\documentclass[twoside,b5paper98pt]{memoir}
	\usepackage{amsmath}
	\usepackage{amsfonts}
	\usepackage{amssymb}
	\usepackage{graphicx}
	\usepackage{color}
	\usepackage{geometry}
        \usepackage{wrapfig}
        \usepackage[utf8]{inputenc}
	\usepackage{booktabs}
\usepackage[colorlinks=true,urlcolor=blue]{hyperref} 
	\newsubfloat{figure}
	\pretitle{\begin{center}\sffamily\huge\MakeUppercase}
	\posttitle{\par\end{center}\vskip 0.5em}
	\maxtocdepth{subsection} 

	\chapterstyle{veelo} 

	\hangsecnum 
	\maxsecnumdepth{subsubsection} 

	\setsecheadstyle{\Large\bfseries\raggedright} 
	\setsubsecheadstyle{\Large\bfseries\raggedright} 
	\setsubsubsecheadstyle{\large\bfseries\raggedright} 

    \setaftersecskip{1.8\baselineskip}
    \setaftersubsecskip{1.5\baselineskip}   
	\newcommand{\degree}{\ensuremath{^\circ}}
\usepackage[font=small,indention=0.2em,labelfont=bf,textfont=it,hang]{caption}

\selectfont
	\title{Memoria M.Sc.}
	\author{Ramiro Checa-Garcia}
\begin{document}

\includegraphics[width=1.00\textwidth,height=1.00\textheight]{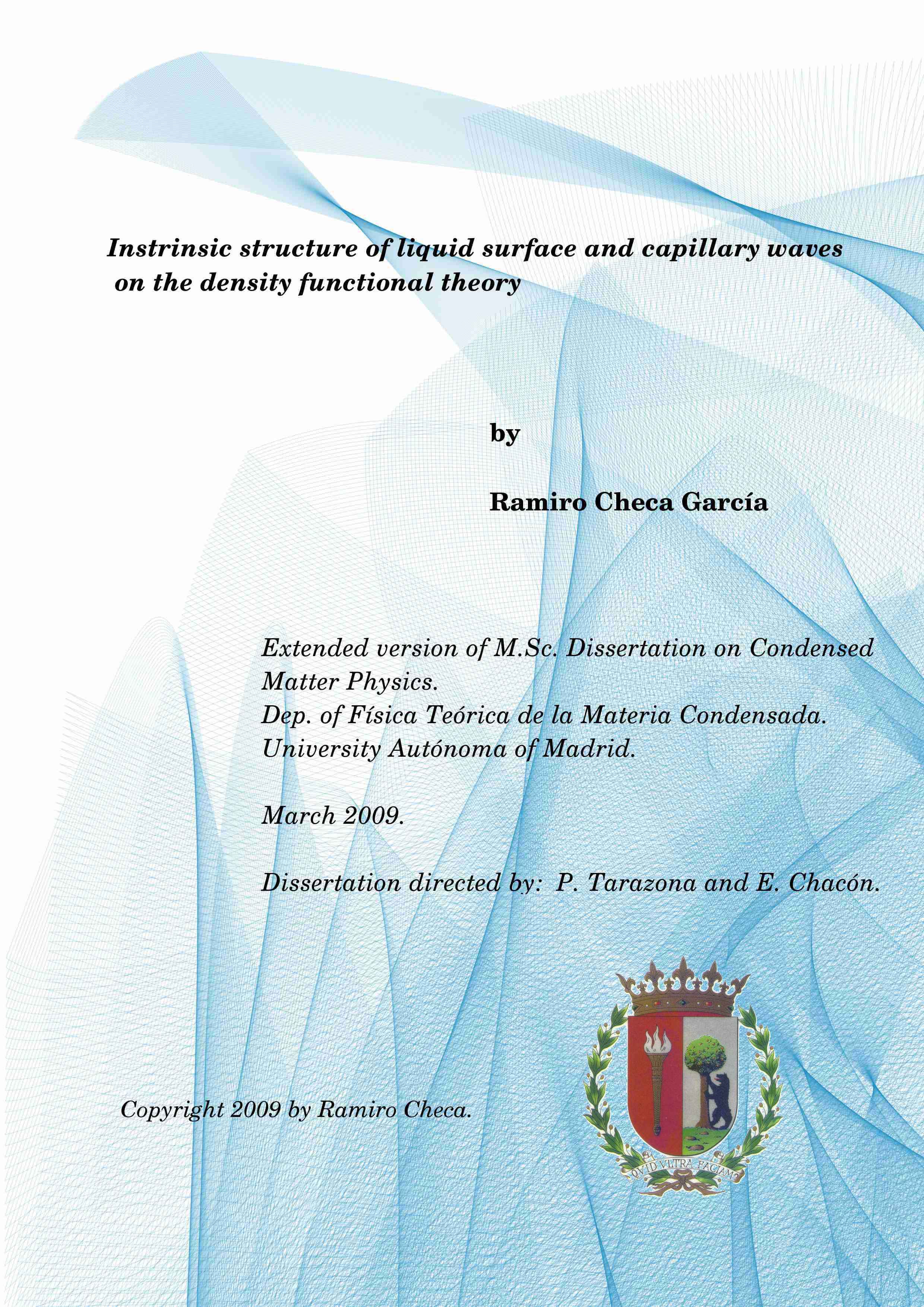}%
\thispagestyle{empty}

	\renewcommand{\listtablename}{Índice de tablas} 
	\renewcommand{\tablename}{Tabla}
	\renewcommand{\bibname}{Bibliografía}
	\renewcommand{\chaptername}{Capítulo}
	\renewcommand{\contentsname}{Índice general}
	\renewcommand{\appendixname}{Apéndices}
	\renewcommand{\partname}{Parte}
	\renewcommand{\figurename}{Figura}
	\renewcommand{\indexname}{Índice de Términos}	
	\renewcommand{\seename}{véase}
	\renewcommand{\preindexhook}{\large\sffamily Índice de definiciones, expresiones y resultados relacionados con la Teoría de Ondas Capilares y la Estructura Intrínseca tratado en el trabajo.\\ \\}	

  \frontmatter

\tableofcontents*

%
\chapter{Agradecimientos}

\vspace{1cm}

\epigraph{And yet, the dying swan's voice is counted the loveliest: he sings without fear.}{Briefe über Gelesenes. \\ \scshape Bertolt Brecht}
\vspace*{0.5cm}

\epigraph{Entendemos que, para el hombre, el mundo es una realidad objetiva, independiente de él, posible de ser conocida. Sin embargo es fundamental partir de la idea de que el hombre es un ser de relaciones y no sólo de contactos, no sólo esta \textit{en} el mundo sino \textit{con} el mundo.}{La sociedad brasileña en transición. \\ \scshape Paulo Freire}
\vspace*{0.5cm}

Existen dos enfoques diferentes cuando analizamos la actividad del ser humano, en uno se reconoce a este como actor, en el otro como autor. La forma del proceder de la actividad científica, las condiciones que el objeto impone al conocimiento adecuado de este, así como, el propio sistema institucional en que esta actividad tiene lugar resaltan la autoría que se constituye, no tanto como un reconocimiento del papel personal en una labor o trabajo concreto, como la mirada con que esta actividad humana es concebida. Empero, en la propia constitución del texto científico, se reserva un espacio donde mínimamente la actividad aparece en su forma histórica y los elementos humanos se descubren como actores en un proceso donde, significativamente, los nombres y los acontecimientos se caracterizan por su singularidad.\\

Como autor se ha de insistir en la imposibilidad de haber realizado este trabajo científico sin la orientación de mis tutores Pedro Tarazona y Enrique Chacón, tanto en la concepción de un tema como problema susceptible de estudio, como en las ideas rectoras que la investigación requiere. Con este fin el autor ha intentado una revisión del programa de investigación para que se aprecien y reconozcan con la nitidez debida estos hechos. Por lo demás, el conjunto de ideas que aparecen en el resto del texto son objetivamente promovidas por la más amplia comprensión que ambos poseen del \textit{corpus} de la mecánica estadística. La ayuda de Enrique Chacón ha sido además importante en la parte final de la escritura de este trabajo.\\

Como actor, el apoyo de mi familia y en especial las experiencias compartidas de mi hermana han resultado cruciales. Agradezco también a David Alba su acogida, a Eva Saldaña su capacidad de transmitir vitalidad y a Inma Alonso su amistad, a\-gra\-de\-ci\-mien\-to extensivo a todos los Ecocampusianos. A Begoña por su paciencia diaria en la vida cotidiana y sus continuas enseñanzas.\\

Como actor, las cuestiones de la vida práctica representan un desafío. Quiero agradecer a César Gonzalez y Yannick Dappe su preocupación y sus consejos aunque no los siguiera y también por lo mismo en esta última etapa a Daniel de la Heras con quien he compartido años de despacho y muchos congresos, ha sido de ayuda continua en las tareas tecnológicas y en los debates sobre el Funcional de la Densidad. También un abrazo para mi otra compañera de despacho, Inés y al resto de miembros del grupo de fluidos. Indudablemente gracias a Laura Cano-Cortés que me ha apoyado de modo continuo en multitud de ocasiones.\\

\chapter{Prólogo}
\vspace{0.5 cm}
\epigraph{El prejuicio consiste en creer que los hechos por si solos, sin libre construcción conceptual, pueden y deben proporcionar conocimiento científico. Semejante ilusión solamente se explica porque no es fácil percatarse de aquellos conceptos que, por estar contrastados y llevar largo tiempo en uso, parecen conectados directamente con el material empírico y en realidad están libremente elegidos.}{\scshape Albert Einstein}

\vspace{2cm}

De las palabras citadas sugiero la importancia de conocer para una teoría física dada no sólo su marco de aplicabilidad y sino además clarificar sus suposiciones básicas, implícitas o explícitas. Teorías con un rango de aplicabilidad mayor (resultados más concretos) e hipótesis menos especificas suelen preferirse. Decidir entre dos teorías conlleva buscar sistemas físicos donde su aplicación determine predicciones contrapuestas y poder decidir así la más adecuada. Por otra parte a la luz de teorías más completas (que engloben a las anteriores como particularizaciones) determinados resultados suelen ser reinterpretados ofreciendo una visión diferente tanto de las suposiciones básicas de la teoría particular como de las consecuencias reales que estas implican pudiendo entonces resultados aparentemente contradictorios verse como diferentes aspectos de una realidad física común. \\

Desde el punto de vista de la filosofía de la ciencia, la afirmación anterior requeriría matizar que se entiende por \textit{visión diferente} y por \textit{reinterpretar}, cuestión que necesitaría un, nada sencillo, posicionamiento epistemológico. Resulta más práctico 
moverse dentro del concepto de \textit{programa de investigación} y desde él analizar que aparece como pro\-ble\-ma, con que metodología hay que afrontarlo, y que formas de respuesta son plausibles desde aquel. En nuestro caso resulta además doblemente cómodo realizar una introducción inicial al programa de investigación sobre interfases fluidas y ondas capilares ya que, además, permite diferenciar los resultados obtenidos en esta memoria del conjunto de resultados previos utilizados dentro del contexto de dicho programa de investigación pero no obtenidos por el autor. De este modo ordenamos el documento en tres partes diferenciadas, una primera constituye una introducción más orgánica y menos formal al problema afrontado en la memoria y fija el contexto del programa de investigación al que pertenece, una segunda parte introduce el formalismo de la teoría de líquidos de un modo general y de un modo particular para el caso de interfases líquidas, la tercera parte aglutina el conjunto de resultados obtenidos por el autor.\\

 \mainmatter
\part{ Contexto: la teoría de ondas capilares como programa de investigación}
\chapter{Introducción}\label{sec:introduccion}


Nuestro objeto de estudio es un sistema de dos fases en equilibrio termodinámico separadas por una interfase, que consideraremos plana, y cuya la posición sabemos no es relevante desde el punto de vista termodinámico mientras se respete la ecuación fundamental $dU=TdS-pdV+\gamma dA+\mu dN$, además, desde el punto de vista mecánico la interfase es consistente con la existencia de una superficie. La imagen puede ser una membrana de anchura y masa cero, donde reside la tensión superficial, mientras que se respete, por ejemplo, la definición mecánica de tensión superficial.\\

Al descender al nivel microscópico la porción de sistema entre las dos fases homogéneas aparece como un sistema físico diferenciado donde, más allá de la definición macroscópica de tensión superficial que ha de respetar, posee una \textit{estructura} susceptible de ser investigada. Bajo esta idea muy general podemos describir el sistema en el caso de una interfase plana mediante dos magnitudes que a priori pueden ser físicamente relevantes la anchura de la interfase y su área $A=L_{x}L_{y}$.\\

En un sistema uniforme cabe la sencilla pregunta de hasta que medida puedo dividirlo en porciones manteniendo estas las propiedades del sistema completo y llamar a esta longitud por $\xi_{B}$. Por analogía, y dadas las simetrías del sistema, podemos definir además otra longitud $\xi_{\parallel}$ que nos indica en que sentido podemos realizar esta misma división en el plano de la interfase\footnote{Por simetría solamente tenemos una longitud característica en las direcciones x e y. En todo caso sí fuera necesario hacemos $L_{x}=L_{y}=L_{\parallel}$.}, mientras que para la anchura de la interfase podemos escribir $\xi_{\perp}$. El origen físico de $\xi_{B}$ en el caso del sistema homogéneo esta relacionado con el alcance de las correlaciones en el volumen y, de modo un poco \textit{naive}, podemos decir que el sistema no siente efectos de tamaño finito hasta que no es reducido a un volumen caracterizado por una longitud del orden de $\xi_{B}$. Podemos intentar dar una definición, en el mismo sentido cualitativa, de la cantidad $\xi_{\parallel}$, determinada por el alcance de las correlaciones en la interfase, es decir en la parte del sistema total donde encontramos una variación de la densidad, o en el mismo razonamiento \textit{naive} preguntarnos cuando el sistema en la interfase siente efectos de tamaño finito al variar A (o $L_{\parallel}$).\\

Basándonos en la posible relevancia de estas tres longitudes de correlación presentamos dos propuestas acerca de la interfase una primera imagen debida a  B. Widom\footnote{La formulación original de Widom perseguía construir una teoría de la interfase líquido-vapor válida en la proximidades del punto crítico, pero sin hacer referencia a algunos aspectos discutidos de la teoría original de van der Waals. Como el controvertido problema de suponer en que la interfase es válida la prolongación analítica de la energía libre a valores de la densidad en los que el sistema homogéneo es metaestable o inestable, o el hecho de encontrar exponentes críticos clásicos. Aquí retomamos la formulación de Widom por ser aparentemente explícita en lo que a las magnitudes definidas anteriormente se refiere, y por hacer hincapié en el problema de las fluctuaciones.} y una segunda que será la teoría de las ondas capilares.\\

\section{Teoría de Widom}
\label{sec:teoriaWidom}
\index{Widom!Teoria}
Dividimos el sistema en elementos de volumen $\xi_{\perp}^{d}$, d es la dimensionalidad del sistema. Ahora en la interfase tienen un volumen $\xi_{\perp}^{d-1}$ microscópico frente a las dos fases en coexistencia pero que en el límite de $T\rightarrow T_{c}$ se hace macroscópica. Widom\cite{widom:3892} consideró que:
\begin{itemize}
\item Las fluctuaciones en la densidad se asientan en estos elementos de volumen. Luego manifiestamente supone que la longitud relevante en la interfase es $\xi_{\perp}$. Y escribe la energía libre de una de estas fluctuaciones como:
\begin{equation}
\phi_{1}\sim\xi_{\perp}^{d}(\Delta n)^{2}/(n^{2}\kappa_{T})
\end{equation}
Es una expresión característica de las fluctuaciones de volumen en el sistema homogéneo.
\item Explícitamente supone que dicha expresión es válida sea cual sea el origen de la fluctuación. En el caso de una fluctuación de volumen las cantidades son las usuales\footnote{Es decir, $\Delta n$ es la fluctuación en la densidad, $n$ la densidad del sistema homogéneo y $\kappa_{T}$ la compresibilidad isoterma.} mientras que en el caso de una fluctuación en la superficie afirma que $\Delta n$ es la diferencia en densidad de las dos fases en coexistencia, mientras que $n^{2}\kappa_{T}$ esta dado por los valores de volumen en cualquiera de las dos fases que serán similares cerca del punto crítico y por tanto nos restringimos a este caso.
\end{itemize} 
Bajo estas dos hipótesis escribir la tensión superficial y su minimización llevaría a $\xi_{\perp}=0$ y tenemos una interfase completamente abrupta, problema que surge de modo análogo en la teoría original de van der Waals. Siguiendo al mismo van der Waals, Widom propuso:
\begin{itemize}
\item Incluir una parte dada por gradientes en la densidad. Así la aproximación a primer orden que respeta la simetría del problema resulta una aproximación de gradiente cuadrado,
\begin{equation}
\phi_{2}\sim\xi_{\perp}^{d}(\Delta n/\xi_{\perp})^{2}
\end{equation}
\end{itemize}
La energía libre total de una inhomogeneidad general contiene las dos contribuciones y la tensión superficial vendrá dada por una suma de ambas.\\

Como consecuencia de las hipótesis anteriores Widom reduce el problema a fluctuaciones que respetan siempre la forma de las fluctuaciones de volumen en cada elemento $\xi_{\perp}$ y por tanto, afirma que esta cantidad es esencialmente $\xi_{B}$. En consecuencia, en esta teoría las tres longitudes antes citadas se reducen a la longitud de correlación del volumen, al menos en las proximidades de la temperatura crítica. Por otra parte, un resultado llamativo es que, además, esta distancia característica $\xi_{\perp}$ esta bien determinada, independientemente de las condiciones gracias a las cuales establezco una interfase con geometría plana o del tamaño superficial determinado macroscopicamente por $L_{\parallel}$.\\

Finalmente las ideas de Widom\footnote{En la formulación original de van der Waals no se sitúa el énfasis en las fluctuaciones en la densidad, como intenta Widom en su propuesta, aunque sus resultados dan lugar al mismo comportamiento crítico de la tensión de la interfase y la anchura de la interfase. Su propuesta\cite{reviewEvans1979,RowlinsonWidom} data de 1894 y es la primera teoría consistente de la descripción microscópica de la interfase líquido-vapor \textemdash Rayleigh llegó a obtener una expresión similar a la de van der Waals para la tensión superficial aunque basándose en argumentos diferentes\textemdash, previa de hecho a la formulación por Gibbs de la mecánica estadística de colectividades. El énfasis en la importancia de las fluctuaciones se debe inicialmente a la formulación Einsteniana (1902-1903) de aquella, y su aplicación al movimiento browniano (1905) en que se mostró que la presencia de las fluctuaciones \textemdash reconocida pero no recalcada por Gibbs\textemdash tenía relevancia en determinados procesos físicos. Einstein utilizó la presencia de fluctuaciones para demostrar la hipótesis atómica y calcular el número de Avogadro y la aplicó con éxito al movimiento browniano. Es más, busco un sistema donde las fluctuaciones de la energía fueran del mismo orden que la energía misma y lo encontró en el caso de la radiación para dimensiones del sistema del orden la longitud de onda de esta.} cristalizan al determinar los exponentes críticos $\mu$ y $\nu$ de $\gamma\sim (\Delta T)^{\mu}$ y $\xi_{\perp}\sim (\Delta T)^{\nu}$, que en su teoría estarán determinados por los exponentes críticos de $\Delta n\sim (\Delta T)^{1/d}$ y $\kappa_{T}\sim (\Delta T)^{-f}$. Esto permite cotejar su teoría con datos experimentales para $\mu$ y $\nu$, y discutir así la viabilidad de las hipótesis anteriores en función de las relaciones de \emph{escalado}\footnote{De \emph{hiper-escalado} si aparece explícitamente la dimensión del sistema.}.\\

\section{Teoría de Ondas Capilares}\label{Sec:Cap1OndasCapilares}
\index{Ondas Capilares!Teoría}
La teoría de ondas capilares\footnote{\textit{Smoluchowsky y Mandelstam} (1913) fueron los primeros en resaltar la importancia de las fluctuaciones térmicas en las superficies líquidas. El primero reformula el problema indicando que el movimiento térmico de las moléculas en una interfase fluido-fluido la debería convertir en \emph{rugosa}, siendo Mandelstam quien aplicando la teoría de las fluctuaciones Einsteiniana describe el problema en términos de \emph{ondas capilares}. Su difusión definitiva no será hasta \textit{F. P. Buff, R. A. Lovett y F. H. Stillinger}\cite{PhysRevLett.15.621} con su conocido artículo, donde rescataron estas ideas para dar forma a la teoría conocida como teoría de ondas capilares (CWT).} aporta una \textit{imagen física} a las fluctuaciones contenidas y características de la interfase, para ello la supone definida como una superficie plana y se cuestiona  el trabajo mínimo, es decir  reversible, que es necesario realizar sobre el sistema para pasar de dicha superficie a una superficie corrugada mediante una función $\xi(\vec{R})$. Fenomenológicamente el cambio en $\xi(\vec{R})$ se hace a expensas de la tensión superficial ($\gamma_{lv}$) que penaliza incrementos de esta. Mientras que, el campo gravitatorio (de intensidad g), necesario para mantener separadas ambas fases y tener una interfase macroscopicamente plana, penaliza también las fluctuaciones sobre una superficie plana\footnote{\label{sec:drumhead}A la expresión que surge se la ha denominado hamiltoniano de \emph{drumhead}, aludiendo la la formulación inicial y su solución realizada por Kac en 1952 y consiste en la ecuación (\ref{eqn:trabajominCWT}) sin la aproximación dada por (\ref{eqn:areaOrden2}).}. En estas condiciones este trabajo mínimo se escribiría como,

\begin{equation}
\mathcal{L}_{cw}[\xi]=\int_{A} d\vec{R}\left[\frac{1}{2}\gamma_{lv}|\nabla\xi(\vec{R})|^{2}+\frac{1}{2}mg\xi(\vec{R})^{2}\right]
\label{eqn:trabajominCWT}
\end{equation}

donde explícitamente suponemos amplitudes $\xi(\vec{R})$ pequeñas para poder representar el cambio de superficie por la expresión indicada. Para ello hemos aplicado que dada una función $\xi(\vec{R})$ sobre un cierto dominio de área A univaluada y diferenciable con continuidad podemos expresar el área $\tilde{A}$ sustentada sobre A por $\xi$ mediante la expresión,
\begin{equation}
\tilde{A}=\int_{A} d\vec{R}\sqrt{1+(\partial_{x}\xi)^{2}+(\partial_{y}\xi)^{2}}
\end{equation}
y para amplitudes $\xi(\vec{R})$ pequeñas podemos aproximar
\begin{equation}
 \Delta A=\tilde{A}-A\sim\int_{A} d\vec{R} \frac{1}{2}|\nabla\xi(\vec{R})|^{2}
\label{eqn:areaOrden2}
\end{equation}

 La aplicación de la teoría de fluctuaciones\cite{Landau_B80} me permite estudiar el efecto de estas sobre la posición de la interfase promediando mediante el factor de Boltzmann la expresión anterior\footnote{Interpretamos como $\mathcal{L}_{cw}$ una energía libre o como una variación de entropía del sistema total cuando dejo libre la ligadura $\xi(\vec{R})=0$.}.\\

La resolución del problema se realiza de modo directo en el espacio de Fourier donde la ecuación (\ref{eqn:trabajominCWT}) queda expresada como una función de distribución gaussiana multidimensional y desacoplamos el problema en modos de oscilación independientes, véase \S\ref{sec:aproxGaussiana}. Una medida de la intensidad local de las fluctuaciones es la fluctuación cuadrática media en la posición de la interfase, que resulta ser,
\index{Ondas Capilares!Fluctuación Cuadrática Media}
\begin{equation}
\Delta^{2}_{cw}=\frac{1}{\beta\gamma_{lv}} ln \left[ \frac{q^{2}_{max}+\xi^{-2}_{cw}}{q^{2}_{min}+\xi^{-2}_{cw}}\right]
\label{eqn:sigmaCWT}
\end{equation}
\index{Ondas Capilares!longitud de capilaridad}
donde $\xi_{cw}$ es la \emph{longitud de capilaridad} definida por
\begin{equation}
\xi_{cw}=\left[\frac{\gamma_{lv}}{mg(\rho_{l}-\rho_{v})}\right]^{1/2}
\end{equation}
con los valores $q_{min}=2\pi/L_{\parallel}$ y $q_{max}=2\pi/l_{min}$. El valor de $q_{max}$ esta relacionado con el nivel máximo de corrugación que creemos describible mediante el funcional $\mathcal{L}_{cw}[\xi]$, sobre el trataremos más adelante. Ahora analizamos las consecuencias de $q_{min}$. Observamos que:
\begin{itemize}
\item Existe una divergencia posible en $\Delta_{cw}^{2}$ provocada por los modos de oscilación con vector de onda q-pequeños que necesitan muy poca energía libre para producirse si $g\rightarrow0$. Luego el campo gravitatorio se convierte en esencial para mantener la interfase definida y acota el valor de la longitud de capilaridad al orden de mm (Argón).
\item La dependencia de $\Delta_{cw}^{2}$ con el campo g y $L_{\parallel}$ es logarítmica así que sus efectos son difícilmente mensurables\footnote{Para incrementar en 10 veces la anchura deberíamos incrementar en $10^{67}$ el valor de la longitud característica $L_{\parallel}$ de la superficie.}. Las cuestiones teóricas que plantea son por contra importantes. 
\end{itemize}

En el contexto antes establecido $\xi_{\parallel}\sim\xi_{cw}$, luego CWT establece un nexo de unión con $\xi_{\perp}$ que podemos identificar como $\Delta^{2}_{cw}$. Si esta identificación es correcta observamos que las correlaciones superficiales se extienden a toda la interfase (para $g\sim0^{+}$) y provocan una divergencia en la anchura de esta. Dividir el sistema en la interfase conlleva obtener dos sistemas que ahora poseen diferentes propiedades a las del sistema total.\\

\section{Separación de escala}\label{sec:propuestaWeeks}

El análisis comparativo de las dos teorías indica que poseen predicciones contrapuestas aunque cabe notar que ambas teorías se aplican a situaciones diferentes dentro del diagrama de fases\footnote{La teoría de Widom la hemos justificado cerca de $T_{c}$, mientras que la hipótesis de superficie plana bien definida es más adecuada cerca del punto triple.} hecho permite lanzar un argumento \textit{heurístico} explicativo\cite{weeks:3106} de sus discrepancias: $\xi_{B}$ es, lejos del punto crítico, una medida microscópica, y de hecho, para encontrar en un sistema uniforme efectos de tamaño finito hemos de aproximarnos significativamente al punto crítico. Por contra $\xi_{\parallel}$ resulta ser casi-macroscópica. La primera separación natural de ambos tipos de fluctuaciones resulta ser de escala\footnote{Idea analizada por Weeks\cite{weeks:3106} y que consiste en dividir el sistema en columnas (más tarde Kayser\cite{PhysRevA.33.1948} sugiere que deberían ser celdas, esquema que \textit{Sengers y van Leeuwen}\cite{PhysRevA.39.6346} aplicaron sobre la base de la teoría fenomenológica de Fisk-Widom) de anchura $l\sim \xi_{B}$ y definir para cada una de ellas una superficie de Gibbs, véase ec. (\ref{eqn:defSupGibbs}). El procedimiento supone un valor similar a la longitud de correlación del sistema uniforme de modo que las fluctuaciones de volumen quedan restringidas a una columna. Así se obtiene, mediante argumentos heurísticos equivalentes a la deducción clásica de la teoría de ondas capilares, la forma en que los grados de libertad restantes, que pueden representarse por los diferentes valores de la superficie de Gibbs local en cada columna, y estimar su contribución a la tensión superficial. Se obtiene para $\mathcal{L}$, esencialmente los mismos resultados que CWT y \emph{explícitamente} se diferencia el régimen donde debería ser válida CWT al menos de modo cualitativo. En cada columna, la descripción realizada por Widom, y en particular las propiedades de la anchura intrínseca que contiene su modelo y el comportamiento de esta anchura y su tensión superficial en las proximidades del punto crítico, es correcta. El efecto de las ondas capilares borrando la definición nítida de la interfase en el límite de campo nulo esta en fluctuaciones que suceden en otra escala de longitud diferente a las del volumen.}.\\

	\begin{figure}[htbp]
	\begin{center}
	\includegraphics[width=5.75in]{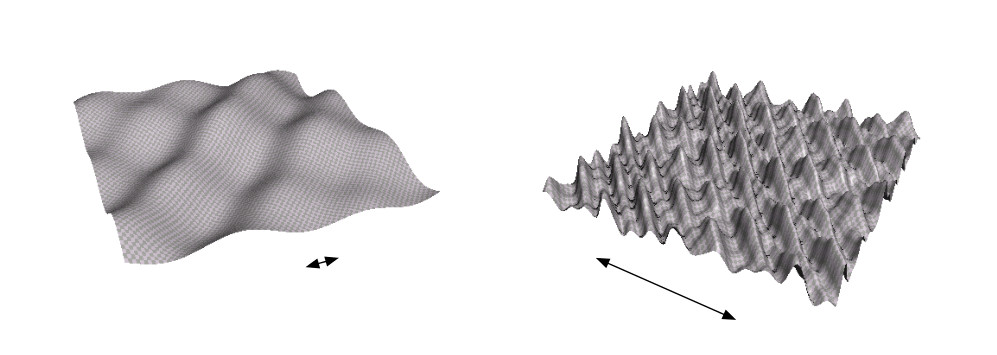}
	\caption{Esquema de dos situaciones opuestas: a la izquierda fluctuaciones superficiales con longitud de escala mayor que $\xi_{B}$, a la derecha fluctuaciones superficiales con longitud de escala menor que  $\xi_{B}$. La longitud de correlación del volumen  $\xi_{B}$ esta representada por una flecha en cada caso. La figura de la derecha esta, por tanto, re-escalada respecto de la izquierda}.
	\label{fig:qminqmax}
	\end{center}
	\end{figure}

Cualitativamente este argumento implica que localmente la anchura de la interfase puede determinarse como $\xi_{B}$, mientras que las ondas capilares deslocalizan la interfase globalmente e introducen la dependencia con $L_{\parallel}$ y $g$.  Llamando $\Delta^{2}_{int}$ a la anchura de la interfase sin la deslocalización que introducen las ondas capilares en el régimen más allá de $\xi_{B}$, la anchura total de la interfase viene determinada por,

\begin{equation}
\Delta^{2}=\Delta^{2}_{int}+\Delta^{2}_{cw}
\label{eqn:WeeksXiPerpen}
\end{equation}

Entonces $\mathcal{L}_{cw}$ se considera una descripción válida de las ondas capilares en el régimen determinado por largas longitudes de onda, donde presumiblemente no interfieren con las propiedades de volumen. Dada una propuesta de estructura interfacial que supongamos ausente de ondas capilares, estructura que llamaremos \emph{intrínseca}, es posible contrastar la ecuación (\ref{eqn:WeeksXiPerpen}). Desde el punto de vista experimental se han realizado medidas de reflectividad óptica\footnote{Seguidamente veremos como la reflectividad de rayos X aporta información relevante.} cerca del punto crítico con este fin. Las medidas que se realizaron en la región próxima a la zona de criticalidad favorecen la hipótesis (\ref{eqn:WeeksXiPerpen}) pero no fueron capaces de diferenciar entre propuestas para la estructura intrínseca diferentes\footnote{Si acaso constataron que $\Delta_{int}^{2}$ podría ser del orden del tamaño molecular lejos del punto crítico.}.

	\begin{figure}[htbp]
	\begin{center}
 	\includegraphics[width=5.5in]{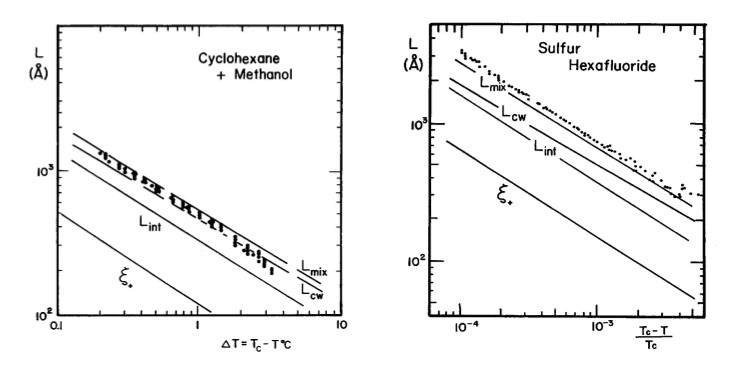}
	\caption{Experimento basado en medidas de reflectividad óptica, cerca del punto crítico de las sustancias indicadas\cite{beysens:3056,beysens:6911}. En la gráfica L representa las anchuras que en el texto denominamos como $\Delta^{2}$.}
	\label{fig:reflectividadTc}
	\end{center}
	\end{figure}

\section{Naturaleza del perfil de densidad}
\index{Estructura Intrínseca}

La teoría de ondas capilares supone que el líquido es de hecho \textit{incompresible}, hipótesis que se manifiesta en que la energía libre $\mathcal{L}_{cw}$ depende únicamente de $\xi(\vec{R})$ y no de la estructura intrínseca sobre la que esta se sustenta. Esto permite que dada una estructura intrínseca descrita por $\tilde{\rho}(z)$ y para una corrugación concreta $\xi$, el perfil de densidad intrínseco sigue rígidamente a la superficie y resulta $\tilde{\rho}(z-\xi(\vec{R}))$ mientras que el promedio sobre las diferentes corrugaciones da lugar al perfil de equilibrio líquido-vapor, y es dependiente de $L_{\parallel}$ y $g$. En consecuencia los perfiles de densidad líquido-vapor no son una propiedad, en este sentido, universal de cada líquido. Tras la teoría de ondas capilares resulta necesario replantearse que significado posee un perfil de densidad con una determinada anchura finita pero que carece de las dependencias funcionales en $L_{\parallel}$ y g expuestas, puede verse \S\ref{sec:funcionesDistribucionAlturas} para los desarrollos analíticos.\\

Los perfiles sin esta dependencia explícita obtenidos mediante diferentes teorías\cite{PhysRevB.30.6666,fisk:3219,Widom1999MolPh1019W} funcionales\footnote{Todas estas propuestas se consideran teorías válidas cualitativamente en las proximidades del punto crítico y dan lugar a perfiles de densidad suaves y de anchura definida que interpolan entre ambas fases, y aunque diferentes propuestas poseen dependencias funcionales diferentes para la estructura intrínseca, desde el punto de vista experimental resulta difícil discriminar entre todas ellas\cite{beysens:3056,PhysRevA.38.2457}, como hemos indicado.} están aparentemente determinados por propiedades \emph{intrínsecas} del fluido y no son dependientes de campos externos o condiciones de frontera particulares, sin embargo, el modo de proceder contenido en ec. (\ref{eqn:WeeksXiPerpen}) no esta exento de debate, algunos autores han utilizado dichos perfiles de densidad como perfil intrínseco sobre el que descongelar las ondas capilares (sugiriendo para ello un $q_{min}$ implícito en dicho perfil) mientras que otros resaltaban que este es ya un perfil de equilibrio aunque bajo una aproximación pudiendo existir un mecanismo de saturación de las ondas capilares dentro de esta teoría haciendo inconsistente este procedimiento\footnote{Al debate sobre la validez de las divergencias de la teoría de ondas capilares se añadió la discusión sobre del papel de siguientes términos del desarrollo, véase ec. (\ref{eqn:areaOrden2}), ordenes $\nabla\xi(\vec{R})^{4}$ cancelando la divergencia presente en CWT.}.\\

Otra vía consiste en obviar inicialmente las presuntas aproximaciones a la estructura intrínseca o de equilibrio líquido-vapor y desarrollar al máximo las implicaciones de la teoría de ondas capilares, volviendo para ello al problema estadístico básico. \textit{Frank H. Stillinger}\cite{StillingerJCPintrinseco} fue el primero en proponer un método diferente para entrever la estructura intrínseca de un fluido en una interfase. Desde las configuraciones moleculares propuso discriminar la función $\xi(\vec{R})$ mediante un criterio percolativo que diferencie las moléculas que pertenecen al líquido de las correspondientes al vapor y a partir de la primera capa de moléculas líquidas determinar el perfil intrínseco, que denominó \textit{inherente}\footnote{Más reciente es la publicación \cite{stillinger:204705}.},  mediante la condición de anular todas las amplitudes $\xi(\vec{q}_{xy})$. El resultado propuesto ha de ser un perfil altamente estructurado lejos de $T_{c}$ con ciertas analogías a la función de distribución de pares g(r). \\
\index{Perfil Inherente}

Si la interpretación de Weeks es correcta y consideramos el perfil propuesto por \textit{Frank H. Stillinger} como el perfil intrínseco entonces la teoría van der Waals y su $\rho_{vdw}$  presentaría una cierto promedio sobre ondas capilares, mediante dos cutoff $q_{min},q_{max}$ desconocidos (que pueden ser o no los sugeridos, $q_{max}\sim2\pi/\sigma$ y $q_{min}\sim2\pi/\xi_{B}$) al igual que la forma de $\mathcal{L}$ en este régimen. Bajo esta imagen perfiles, $\rho(z,q_{min})$, obtenidos por teorías a lo van der Waals pero más elaboradas pueden manifestar, cerca del punto triple donde las ondas capilares sobre una \emph{superficie fría} tienen un papel menor, suaves oscilaciones en la densidad.\\

El hecho de no encontrar dichas oscilaciones en la solución de van der Waals y la larga tradición de perfiles suaves hizo desechar esta hipótesis y también otros resultados que podían indicar una estructura en capas para los perfiles de densidad\cite{0022-3719-4-14-005,0022-3719-4-14-006,0022-3719-4-16-009}. Incluso en las simulaciones sobre típicos sistemas Lennard-Jones se encontraron trazas de esta posible estructuración atribuyéndose a una estadística insuficiente y pasando por alto el papel que el tamaño de las cajas de simulación podría tener a la luz de las dependencias que la teoría de ondas capilares predice.\\

\section{Cuestiones abiertas}

Según la argumentación anterior lo pertinente sería hablar de un perfil de densidad indicando el nivel de corrugación sobre el que ha sido promediado, pero aun queda resolver la cuestión del valor físico de $|q|_{max}$ sobre el cual no tiene sentido la imagen física de ondas capilares, a primera vista podríamos hacerlo del orden del tamaño molecular indicativo de que fluctuaciones con longitud de onda menor no son posibles. En la práctica para de vectores de onda grandes podemos perder la posibilidad de aproximar la superficie por una función suave y en cualquier caso, dada una fluctuación de la densidad en este régimen, no resulta evidente como podemos separar las fluctuaciones de volumen de las de superficie antes diferenciadas por su escala de aplicabilidad en la línea de la argumentación de Weeks\footnote{Si descendemos a niveles menores de $\xi_{B}$ no estaríamos seguros de poder asignar a esta tensión superficial su valor macroscópico y por tanto habríamos de escribir $\gamma(|q|)$  haciéndola depender de la variable que creemos relevante en la descripción de la fluctuación superficial, sin embargo la introducción de esta dependencia manteniendo la forma de $\mathcal{L}$ no es del todo evidente. En las condiciones prácticas en que $2\pi/|q|_{max}>\xi_{B}$ la aproximación  $\gamma(|q|)=\gamma_{lv}$ como la tensión superficial macroscópica es adecuada, y escribir $\mathcal{L}$ una aproximación viable.}.\\

Para definir de modo riguroso la estructura intrínseca y delimitar las condiciones en que la imagen física dada por las ondas capilares posee significado real el objetivo que se nos presenta es separar ambos tipos de fluctuaciones en la escala en que conviven, definir de un modo adecuado un perfil exento de fluctuaciones de superficie y conseguir expresar la energía libre que cuesta descongelar sobre este perfil las ondas capilares. Esta energía libre adquiere la  forma de un funcional\footnote{Necesitamos una definición de $\xi(\vec{R})$ descorrelacionada estadísticamente del perfil intrínseco si esperamos mantener una forma similar pero generalizada de CWT, en el sentido de teoría gaussiana analíticamente resoluble.} de $\xi(\vec{R})$ que debe permitir restituir los grados de libertad colectivos expresados en las ondas capilares y recuperar entonces las propiedades de equilibrio de la interfase líquido-vapor.\\

Concluyendo, si bien hemos determinado el rango de aplicabilidad de CWT no hemos resuelto mucho más. Este era el estado del conocimiento de las fluctuaciones superficiales en los inicios de la pasada década cuando empezaron a realizarse tanto experimentos como simulaciones encaminados a resolver algunas de las disyuntivas que resumimos a continuación:

\begin{itemize}
\item Conocemos el comportamiento de las fluctuaciones de superficie para largas longitudes de onda pero no hemos resuelto como separarlas de modo efectivo de las fluctuaciones de volumen en las escalas en que ambas conviven.
\item No resulta evidente que $\rho_{vdw}(z)$ ha ser el perfil sobre el que descongelar sin más los grados de libertad propuestos por CWT ni, en el caso en que lo fuera, cual sería el valor de $q_{max}$ que incluiría este\footnote{Contrastar con el análisis de \textit{R.Evans}\cite{evansMolecularCWvdw} en la interpretación del análisis de Wertheim aplicado a la teoría de van der Waals.}, el valor de $q_{min}$ viene determinado por el tamaño del sistema.
\item En este punto no tenemos una idea nítida de la forma funcional que tiene el coste en energía libre de las fluctuaciones superficiales en la escala de longitudes de onda pequeñas\footnote{Extensiones fenomenológicas como el \textit{hamiltoniano de Helfrich} que incluye más términos además del incremento de área han de ser interpretados con teorías cuya base sea realmente microscópica\cite{PhysRevE.47.1836}.}. Ni como enlazarlo a la definición rigurosa de un perfil exento de estas.
\end{itemize}

En principio se puede intentar deducir $\mathcal{L}$ desde una teoría más fundamental con base microscópica donde se pueda hacer explícitamente el enlace entre los diferentes niveles de descripción. En el contexto de la teoría del funcional de la densidad ha habido varias aportaciones\cite{PhysRevE.47.1836, PhysRevE.59.6766} intentando evaluar la diferencia de energía libre entre un perfil \textit{interpretado como} un perfil intrínseco plano y un perfil corrugado según una cierta función $\xi(\vec{R})$. La mayoría parten de perfiles intrínsecos que, o bien son funciones paso, o bien son funciones suaves y se realizan tratamientos mediante funcionales locales, limitaciones que impiden indagar en las cuestiones abiertas sobre el perfil intrínseco. Sin embargo en los últimos años las diferentes \textit{imágenes físicas} acerca de la estructura de la interfase sí han podido ser contrastadas mediante simulaciones y experimentos de reflectividad de rayos X, estos últimos notablemente determinantes.

\section{Experimentos de Reflectividad de Rayos X}
\label{sec:experimentosReflectividad}

En la pasada década la aplicación de técnicas experimentales de rayos-X procedentes de fuentes sincrotrón ha tenido un notable éxito en la \emph{caracterización de la estructura superficial de sistemas líquidos}\cite{DaillantRepProgPhys}. Si bien es bastante conocida su aplicación a sustancias sólidas, no lo es tanto a sustancias líquidas en lo que a la influencia de las fluctuaciones superficiales se refiere. Introducimos sus diferencias y las diferentes aproximaciones y métodos utilizados para poder interpretar los resultados experimentales.\\

        \begin{wrapfigure}{r}{7cm}
	\includegraphics[width=2.60in]{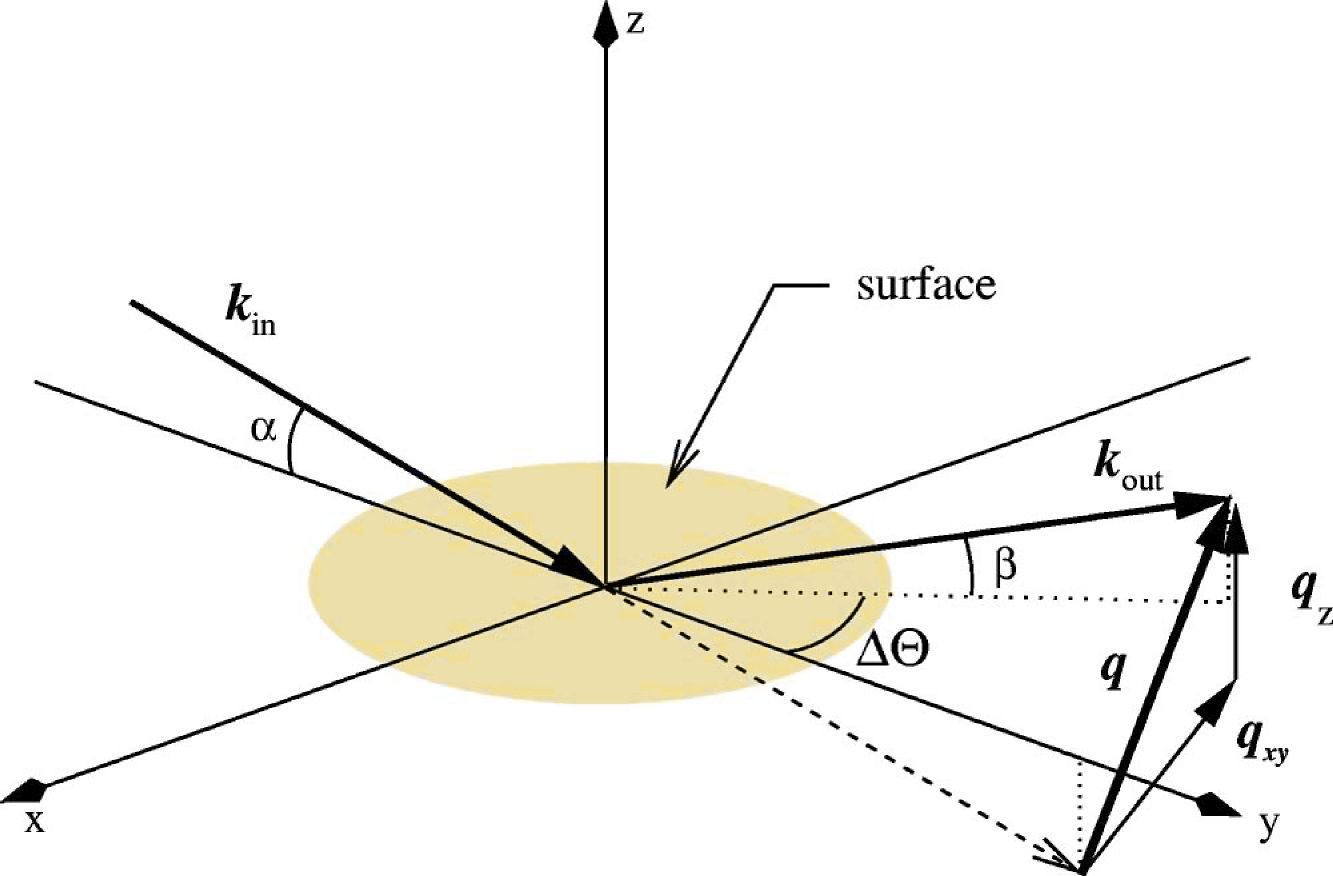}
	\caption{Geometría esencial de los experimentos de reflectividad de rayos X, véase \cite{PhysRevB.67.115405}}
	\label{fig:CinematicaXray}
	\end{wrapfigure}

El índice de refracción de rayos-X en materia es menor de la unidad lo que implica que para ángulos de incidencia pequeños es posible conseguir reflexión total \emph{externa}. La cinemática del proceso de dispersión la podemos ver en la figura anexa (\ref{fig:CinematicaXray}) donde $\alpha=\beta$ e $\Delta\theta=0$ es la \textit{condición especular}, nos interesará esencialmente la dependencia de la reflectividad con $q_{z}$, el momento transferido en la dirección perpendicular a la interfase\footnote{Las técnicas de Rayos X utilizadas por los diferentes grupos experimentales son principalmente tres: reflectividad especular, \emph{scattering difuso} y \emph{Grazing Incidence Diffraction} (GID).}. \\

Por \emph{reflectividad} entendemos la intensidad del haz reflejado respecto del haz incidente sobre una superficie contenida en el plano (x,y).\\

Para una interfase \emph{plana ideal} tanto en el plano de incidencia como en la separación entre dos sistemas de densidad constante en la dirección perpendicular a dicho plano\footnote{En este caso podemos representar la estructura perpendicular mediante una función
\begin{equation}
\rho(z)=\rho_{1}+(\rho(0)-\rho_{1})\theta(z-0)
\end{equation}} el valor de la reflectividad se conoce como \emph{reflectividad de Fresnel}\footnote{Y depende esencialmente del ángulo incidente o para reflexión especular del valor del vector de onda transferido en la dirección perpendicular a la superficie} ($R_{F}$) y puede ser determinado mediante la teoría electromagnética. El resultado es una reflectividad decreciente con el ángulo de incidencia a partir de un ángulo crítico de reflexión total, $\alpha_{c}$, y la absorción jugando solo un papel relevante para ángulos menores que dicho ángulo crítico\footnote{La aproximación más sencilla es $R_{F}(\alpha)\sim (\frac{\alpha_{c}}{2\alpha})^{4}$, donde $\alpha$ es el ángulo de incidencia, para ángulos de incidencia mayores que el crítico esta aproximación es correcta y es el régimen donde encontramos la física que nos interesa.}.\\

Para una interfase \emph{difusa}\footnote{Entenderemos por interfase difusa, un perfil suave, es decir, una interpolación suave hacia la densidad del líquido en la dirección perpendicular al plano de incidencia, esta será plana en el plano de incidencia sino presenta fluctuaciones superficiales o será rugosa si las presenta.} en cambio la reflectividad $R(\alpha)$ presenta un decrecimiento mayor que la reflectividad de Fresnel. La determinación de esta reflectividad es posible realizarla para ángulos suficientemente mayores que un ángulo crítico mediante argumentos análogos a la aproximación de Born, es decir, despreciando dispersión múltiple de rayos-x. La discusión se realiza más adecuadamente en términos de la \emph{sección eficaz diferencial de dispersión} que es proporcional a la intensidad y que se expresa en esta aproximación como\footnote{donde $r_{e}$ es el radio del electrón clásico, para nosotros una constante con unidades de longitud.} ,

\begin{equation}
         \frac{d\sigma}{d\Omega}=r^{2}_{e}\rho^{2}_{bulk}\bigg\vert\frac{1}{\rho_{bulk}}\int e^{-i\vec{q}\vec{r}}\rho(\vec{r})\bigg\vert^{2}
\end{equation}
	
en el caso de una superficie perfectamente plana, no rugosa pero si \textit{difusa}, podemos integrar en el plano (x,y). Si $A_{xy}$ es el área iluminada tenemos,

	\begin{equation}
	\frac{d\sigma}{d\Omega}=r^{2}_{e}\rho^{2}_{bulk}A_{xy}4\pi^{2}\delta^{(2)}(\vec{q}_{xy})\bigg\vert\frac{1}{\rho_{bulk}}\int dz e^{-iq_{z}z}\rho(z)\bigg\vert^{2}
	\end{equation}
	
en el caso de un perfil abrupto podemos integrar también en la coordenada z resultando $\frac{1}{q_{z}^{2}}$, por tanto resulta conveniente expresar lo anterior mediante,

\begin{equation}
	\frac{d\sigma}{d\Omega}=r^{2}_{e}\rho^{2}_{bulk}A_{xy}4\pi^{2}\delta^{(2)}(\vec{q}_{xy})\frac{1}{q_{z}^{2}}\bigg\vert\Phi(q_{z})\bigg\vert^{2}
\end{equation}

donde hemos definido $|\Phi(q_{z})|^{2}$ que se denomina \emph{factor de estructura superficial} y es donde interviene la estructura en la coordenada z de nuestro perfil de densidad de equilibrio. Las medidas se suelen expresar mediante la relación entre la reflectividad medida y la reflectividad de Fresnel toda vez que el resto de contribuciones únicamente me indican que he de medir en las condiciones de reflexión especular y son comunes para una interfase plana y una difusa,
	
	\begin{equation}
	\frac{R(q_{z})}{R_{F}(q_{z})}=|\Phi(q_{z})|^{2}=\bigg\vert\frac{1}{\rho_{bulk}}\int dz \frac{d<\rho(z)>}{dz}e^{-iq_{z}z}\bigg\vert^{2}
	\end{equation}

El valor de $\rho_{bulk}$ es la densidad media electrónica de la muestra lejos de la superficie (valores de z ya en el sistema homogéneo).

	\begin{figure}[htbp]
	\begin{center}
	\includegraphics[width=5.5in]{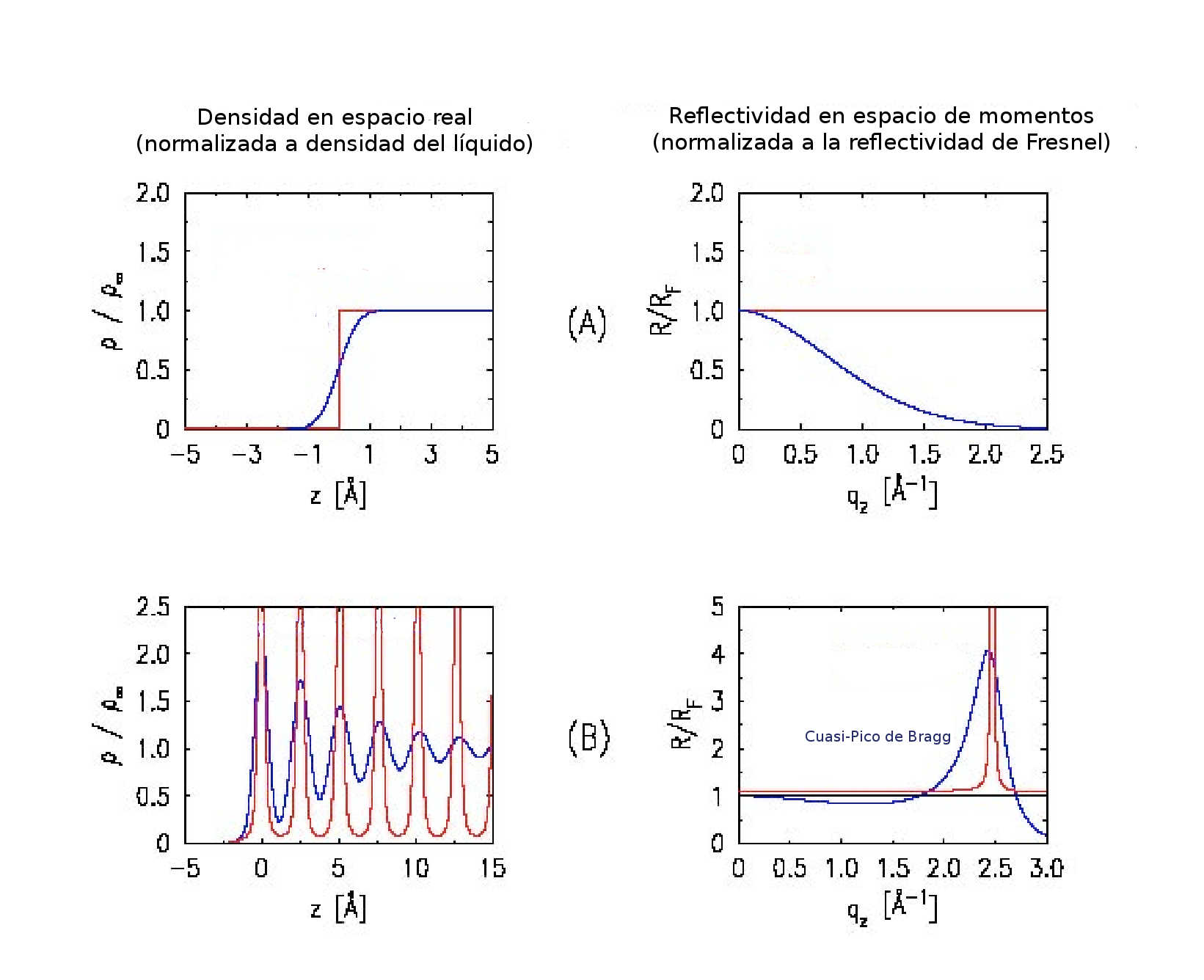}
	\caption{Relación entre los perfiles $\rho(z)$ y el factor de estructura superficial $\Phi(q_{z})$ para varias propuestas. El perfil intrínseco estructurado puede dar lugar a un cuasi-pico de Bragg. Veasé referencia\cite{ShpyrkoPHD}}
	\label{fig:Refletividad1}
	\end{center}
	\end{figure}
	
Nos interesa ver para diferentes propuestas funcionales de $\rho(z)$ que repercusión tienen en $|\Phi(q_{z})|^{2}$, para ello véase la figura (\ref{fig:Refletividad1}). En ella apreciamos como la diferencia en el  factor de estructura superficial de un perfil suave y un perfil oscilante esta en la presencia de un \emph{cuasi-pico de Bragg} en este último. En la discusión que nos ocupa sobre el perfil intrínseco el hecho clave es que ha sido visualizado para metales líquidos gracias a la incorporación de fuentes sincrotrón de rayos-X, con todo aun establecido el carácter estructurado en capas del perfil de densidad de \textit{estos sistemas} han surgido dos cuestiones:

	\begin{itemize}
	\item ?`Es este comportamiento general de todos los sistemas líquidos?
	\item ?`Qué papel juega la teoría de ondas capilares en ellos? ?`Es este papel mensurable?
	\end{itemize}

A la primera cuestión ha recibido diferentes propuestas, inicialmente se ha considerado, en efecto, asociada a la naturaleza metálica de las sustancias utilizadas en los primeros experimentos de reflectividad que induce a pensar en una propiedad exclusiva de los sistemas metálicos, de hecho parte de los experimentos fueron motivados por la predicción de este comportamiento\cite{PhysRevE.56.7033}, la propuesta considera que la diferencia en la densidad de electrones entre las fases vapor y líquido tiene una implicación en la estructura de la interfase debido a que la caída abrupta de la densidad electrónica actúa como una barrera para los iones que responden ordenándose en capas\cite{d'evelyn:5081,d'evelyn:5225} entorno a la superficie. Sin embargo a la vista de los experimentos surge otra propuesta que parte de la idea de determinar que característica de estos sistemas es relevante desde el punto de vista de la segunda cuestión planteada. Vamos a intentar analizar someramente el papel de la teoría de ondas de capilaridad en las medidas de reflectividad para encontrar esta propiedad relevante.\\

	Si deseamos incorporar hipótesis relativas a una superficie no necesariamente plana, además de un perfil estructurado, debemos incorporar una función $\xi(x,y)$ que de cuenta de las diferencias locales respecto de una superficie plana. El resultado es que perdemos la integración en el plano (x,y) y la contribución directa de una función $\delta^{(2)}(x,y)$ quedando (englobamos en C las constantes que no son necesarias en la discusión),
	
\begin{equation}
	\frac{d\sigma}{d\Omega}=C\int_{z_{1}\geqslant0} dz_{1}dz_{2}d^{2}\vec{r}_{xy}<\rho(\vec{r}_{xy},z_{1})\rho(0,z_{2})>e^{-iq_{xy}\vec{r}_{xy}}e^{iq_{z}(z_{1}-z_{2})}
 \end{equation}
	El promedio indica funciones de correlación de la densidad. Resolver como expresar esta integral de forma que explícitamente veamos las contribuciones superficiales conlleva poder separar las fluctuaciones superficiales de las de volumen en todas las escalas. El modo usual de proceder es realizar hipótesis iniciales similares al análisis de Weeks, véase nota en \S\ref{sec:propuestaWeeks}, e imaginar que de algún modo hemos promediado en la estructura más local de la superficie al menos hasta longitudes del orden de las correlaciones de volumen. Bajo esta suposición separamos explícitamente las contribuciones en dos escalas,
	
	\begin{align}
	\frac{d\sigma}{d\Omega}&=C\int_{z_{1}\geqslant0} dz_{1}dz_{2}d^{2}\vec{r}_{xy}e^{-i\vec{q}_{xy}\vec{r}_{xy}}e^{iq_{z}(z_{1}-z_{2})}\\
	&\left[\rho(z_{1}-\xi(\vec{r}_{xy}))\rho(z_{2}-\xi(0))+<\delta\rho(\vec{r}_{xy},z_{1})\delta\rho(0,z_{2})>\right] \notag
 \end{align}
la parte con valores de $\vec{r}_{xy}>\xi_{B}$ lleva incorporada una aproximación presente en la CWT, sobre ella se hará hincapié más adelante. El segundo término da cuenta de la escala en longitudes de onda menores que $\xi_{B}$ y aglutina por tanto dos aspectos diferentes la dispersión difusa procedente de las fluctuaciones de volumen y las fluctuaciones de superficie a la misma escala. La expresión dada por el primer integrando será la aportación relevante en la \textit{condición especular} y puede ser escrita como,

\begin{equation}
   \frac{d\sigma}{d\Omega}=\tilde{C}\frac{|\Phi(q_{z})|^{2}}{q_{z}^{2}}\int_{\vec{r}_{xy}>\xi_{B}} d^{2}\vec{r}_{xy}<e^{-iq_{z}[\xi(\vec{r}_{xy})-\xi(0)]}>e^{i\vec{q}_{xy}\vec{r}_{xy}}
\end{equation}

ahora la definición del factor de estructura superficial queda como,

\begin{equation}
	\Phi(q_{z})=\int dz\frac{\partial\rho(z-\xi(\vec{r}_{xy}))}{\partial z}e^{iq_{z}(z-\xi(\vec{r}_{xy}))}
\end{equation}

Hasta ahora como comentábamos sólo hemos tratado el problema de como la estructura perpendicular a la superficie del sistema influye en la medida de la reflectividad, obviamente tanto en sistemas sólidos como en sistemas líquidos tenemos una estructura superficial que condicionará nuestras medidas de reflectividad. Si podemos caracterizar esta estructura superficial mediante $\xi(\vec{R})$, su relevancia en un experimento conlleva determinar las correlaciones de alturas sobre la superficie $<[\xi(\vec{R}_{1})-\xi(\vec{R}_{2})]^{2}>$.\\

 La principal diferencia entre un sólido y un líquido es que estas correlaciones suelen ser de corto alcance en el primero\footnote{Rigurosamente hablando no es cierto y la temperatura condiciona que podamos hablar o no de correlaciones de corto alcance en un sólido. Los trabajos experimentales de reflectividad de rayos X se realizan en el rango de temperaturas en que esta afirmación es cierta.} y como hemos comentado (y veremos en detalle más adelante) son de largo alcance en el segundo. Esto plantea dos situaciones contrapuestas en ambos sistemas cuando intentamos determinar la sección eficaz diferencial de dispersión en ambos casos, emergen entonces dos situaciones diferentes:
	
\begin{itemize}
	\item Si las correlaciones superficiales son de corto alcance es posible diferenciar las dos contribuciones a la integral en el plano (x,y) una correspondiente a la reflexión especular que estará caracterizada por una expresión similar al caso plano con un factor gaussiano procedente de las correlaciones superficiales pero que esta acotado por su alcance finito, explícitamente para distancias grandes $<[\xi(\vec{R})-\xi(0)]^{2}>\sim 2<\xi(0)^{2}>$ . Y otro término que se corresponde con una dispersión difusa procedente de la rugosidad superficial. La primera condición establece reflexión especular por tanto mediante dos medidas (con y sin $\Delta\theta=0$) es posible determinar la contribución $R(q_{z})/R_{F}(q_{z})\sim e^{\sigma^{2}q_{z}^{2}}|\Phi(q_{z})|^{2}$.\\
	
Para el sólido tendríamos,
	\begin{equation}
	\frac{d\sigma}{d\Omega}=C|\Phi(q_{z})|^{2}q_{z}^{-2}e^{-q_{z}^{2}<\xi(0)^{2}>}\delta^{2}(q_{xy})
	\end{equation}
	
	\item La situación es algo más complicada si las correlaciones superficiales son de largo alcance ya que la diferenciación entre reflexión especular y difusa para ángulos cercanos a la condición especular es difícil y la reflectividad pasa a depender de la \emph{resolución} del espectrómetro.
	
Para un líquido tendríamos,

	\begin{equation}
	\frac{d\sigma}{d\Omega}=\tilde{C}|\Phi(q_{z})|^{2}\frac{4}{\beta\gamma}q_{xy}^{-2}\left[\frac{q_{xy}}{q_{max}}\right]^{\eta}
	\end{equation}
	
donde se ha utilizado para la función de correlación de alturas $<[\xi(\vec{R}_{1})-\xi(\vec{R}_{2})]^{2}>$ la expresión dada por la teoría de ondas capilares\footnote{Veáse \S\ref{sec:funcionesDistribucionAlturas} para su desarrollo analítico.}. Su validez en función de lo comentado anteriormente esta entre $\xi_{B}<r<\xi_{cw}$ donde $\xi_{B}$ como antes determina $q_{max}$. 
\end{itemize}
	
	La medidas de reflectividad implican la integración en el ángulo sólido del detector y por tanto involucran una dependencia explícita con la resolución de este y su geometría, para un detector circular podemos determinarla y tenemos la expresión,
        \begin{equation}
	\frac{R(q_{z})}{R_{F}(q_{z})}=|\Phi(q_{z})|^{2}\left(\frac{q_{res}}{q_{max}}\right)^{\eta}
	\end{equation}
	
Si el factor de estructura superficial determina la presencia de un pico aun tendremos que identificarlo a pesar del factor que en la sección eficaz lo acompaña y separarlo de la difusión que las fluctuaciones de volumen introduzcan. Las condiciones basadas en la expresión anterior en que el cuasi-pico de Bragg es visualizadle se dan solo para $\eta<2$ ya que solo entonces puedo diferenciar ambos tipos de fluctuaciones. En este sentido el parámetro $\eta$ me determina el rango de fluctuaciones que estoy observando y para $\eta>2$ ambas están en la misma escala (en las condiciones de CWT). En consecuencia el parámetro $\eta$ plantea cuestiones importantes acerca de los sistemas en los debo realizar experimentos así como el rango en que es posible visualizar resultados. Este parámetro viene dado por,

	\begin{equation}
	\eta=\frac{q_{z}^{2}}{2\pi\gamma\beta}
	\label{eqn:etapexper}
	\end{equation}

Observamos que la diferenciación de las contribuciones estructurales será más nítida para líquidos con valores altos de la tensión superficial lo que explica que ha sido observada esencialmente en metales líquidos estables a bajas temperaturas. Podemos comparar diferentes líquidos mediante los datos de la tabla  (\ref{tabla:DatosMetalesLiquidos}).

\vspace{0.75cm}
\begin{table}[htdp]
\begin{center}
\begin{tabular}{c c c c c c c} \toprule
\textbf{Elemento} &  $\mathbf{\gamma_{lv}}$ \textbf{(mN/m)}& $\mathbf{T_{p}}$ \textbf{(Kelvin)} &$\mathbf{T_{p}\gamma_{lv}^{-1}}$ &  $\mathbf{q_{pico}\quad\AA^{-1}}$ &$\mathbf{\eta_{pico}}$ \\
\midrule
\textit{Hg}   &   498  & 234.28    & 0.47 & 2.3 & 0.53\\
\textit{Ga}   &   718 & 302.93 & 0.42 & 2.6 & 0.61\\
\textit{In}    &   556 & 429.32   &  0.77& 2.2 & 0.80\\
\textit{K}    &   110 & 336.6    &  3.06 & 1.5 & 1.59\\
\textit{Na}  &   191& 371.0   &  1.94 & 1.9 & 1.54\\
\textit{Ar}   &   13  & 83.80 & 6.44 & 4.1 & 3.83\\
\textit{Au}  &   1169 & 1337.3  & 1.14& 2.5& 1.34\\
\bottomrule
\end{tabular}
\end{center}
\caption{Datos para metales líquidos cerca de punto triple. Recogidos de la referencia \cite{Singh:1997:0034-4885:57}.}
\label{tabla:DatosMetalesLiquidos}
\end{table}
\vspace{0.75cm}

Los valores clave para poder visualizar una estructura superficial son los que indica $\eta_{pico}$, es decir, el valor del exponente $\eta$ en el $q_{z}$ que esperamos característico del cuasi-pico de Bragg, valores de $\eta$ mayores de 2 para este $q_{z,pico}$ imposibilitan detectar la estructura del perfil de densidad contenida en $\Phi(q_{z})$. Los candidatos ideales en la tabla son Hg, Ga, In\cite{PhysRevB.59.783}, Au. Mientras que el Ar resulta inaccesible a las medidas experimentales y los modelos como K, Na se sitúan en el límite accesible ya que limitaciones experimentales\footnote{Además de las condiciones impuestas por el $\eta$ límite teórico  hay otros factores que limitan esta resolución máxima accesible, como el \textit{scattering} difuso por el vapor, la pureza de la muestra en la superficie o la reactividad del líquido en contacto con el aire, algunas de estas limitaciones pueden ser solventadas mediante medidas en UHV(ultra alto vacío). Las características de la ventana del detector son el otro factor limitante.}  hacen difícil llegar de modo fiable a un valor mayor de 1.5. Observar el cuasi-pico de Bragg esta condicionado a tener líquidos con $\frac{T_{p}}{\gamma_{lv}}$ pequeño y un ordenamiento estructural intrínseco tal que $q_{z,pico}$ no quede fuera del alcance experimental. \\

        \begin{figure}
	\begin{center}
	\includegraphics[width=4.2in]{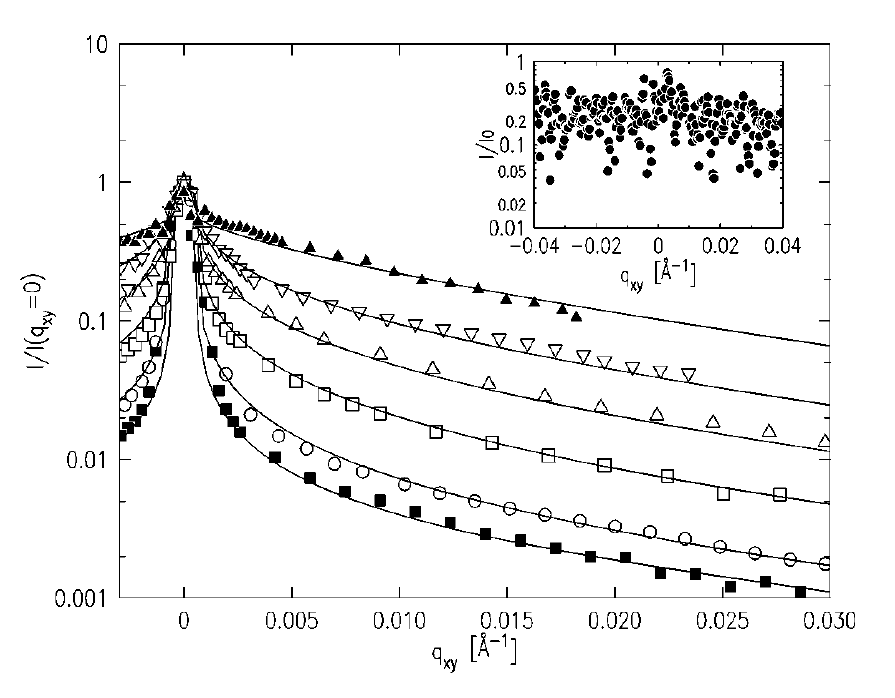}
	\caption{\textbf{Valores de reflectividad para K}, normalizados a 1. Las diferentes gráficas muestran valores diferentes de $q_{z}$, valores mayores implican un mayor valor de $\eta$ y se hace más difícil diferenciar el cuasi-pico de Bragg. En el inset se visualiza un valor de $\eta$ previo al $q_{z,pico}$ esperado lo que imposibilita la resolución completa del factor de estructura. Son medidas de scattering difuso. La anchura del pico entorno a $q_{xy}=0$ esta condicionada por la resolución del detector. Véase \cite{PhysRevB.67.115405}}
	\label{fig:PotasioXrayPapeleta}
	\end{center}
	\end{figure}

        \begin{figure}
	\begin{center}
	\includegraphics[width=0.75\textwidth]{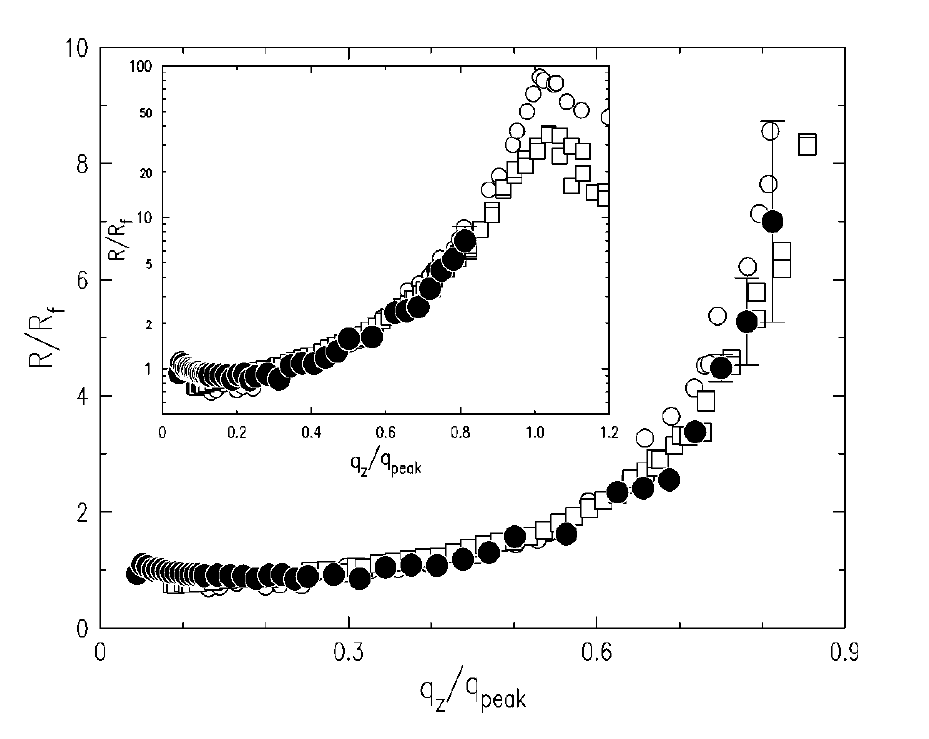}
	\caption{\textbf{Comparación del K con Hg y Ga}. En ellos es posible resolver la estructura a valores de $\eta$ pequeños y por tanto contrastar posibles perfiles de densidad. La gráfica muestra que el K puede seguir siendo un candidato a perfil intrínseco estructurado aunque sea difícil resolver completamente $\Phi(q_{z})$. Véase \cite{PhysRevB.67.115405}}
	\label{fig:ReflectividadPotasium}
	\end{center}
	\end{figure}

\begin{figure}[htbp]
	\begin{center}
	\includegraphics[width=1.10\textwidth]{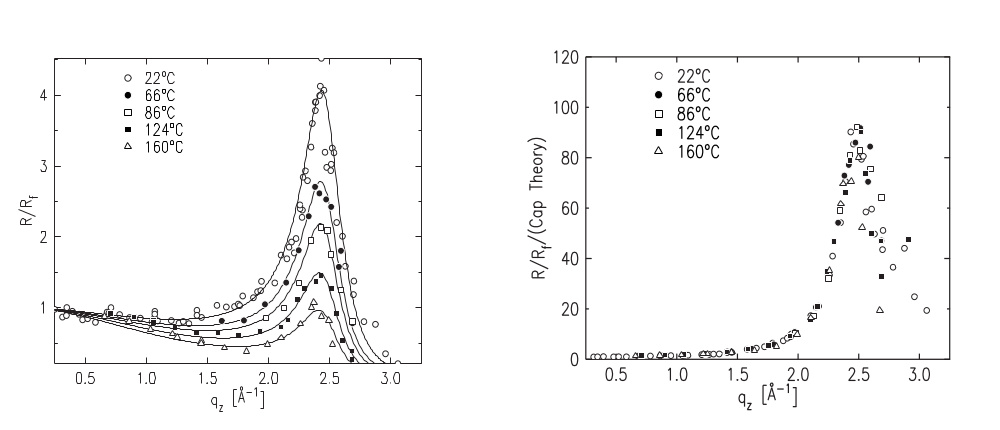}
	\caption{\textbf{Izquierda}: $R/R_{F}$ para \emph{Galium} líquido en función de la temperatura. El mayor pico corresponde a $22\degree C$ y el ultimo a $160\degree C$. \textbf{Derecha}: Substracción de la contribución de las fluctuaciones superficiales predichas por CWT. Véase \cite{PhysRevB.54.9730}}
	\label{fig:Refletividad2}
	\end{center}
	\end{figure}

Esto se puede apreciar de modo nítido en las figuras (\ref{fig:PotasioXrayPapeleta}) y (\ref{fig:ReflectividadPotasium}), la primera de ellas muestra el papel de $\eta$ posibilitando o no la diferenciación de la estructura, mientras que la comparación con dos modelos donde esto es posible se hace en la segunda de ellas. Un sistema particularmente adecuado es el \emph{Galium}, vamos a analizar datos experimentales\cite{PhysRevB.47.6847,PhysRevLett.75.2498,PhysRevB.58.R13419} referentes a él y condensados en la figura (\ref{fig:Refletividad2}). Podemos observar la presencia del cuasi-pico de Bragg y como, mientras su localización en $q_{z}$ permanece constante, su intensidad disminuye al aumentar la temperatura induciendo una correspondencia con perfiles con una estructura progresivamente más amortiguada. Resulta conveniente escribir,

	\begin{equation}
	\frac{R(q_{z},T)}{R_{F}(q_{z})}=|\Phi(q_{z,T})|^{2}=\bigg\vert\frac{1}{\rho_{0}}\int dz \frac{d<\rho(z)>_{T}}{z}e^{iq_{z}z}\bigg\vert^{2}
	\end{equation}

donde explícitamente indicamos la dependencia con la temperatura. Para poder contrastar CWT con diferentes perfiles intrínsecos hemos de medir reflectividades que resultan de la integración de la sección eficaz diferencial, sobre el ángulo sólido sustendido por el detector y la señal medida (por tanto incluso en el caso especular depende de la resolución del espectrómetro.). De la figura (\ref{fig:Refletividad2}) surge la hipótesis de la existencia de un perfil sobre el que las fluctuaciones superficiales son introducidas dando lugar a esta dependencia. La aplicación de la teoría de capilaridad tal y como ha sido introducida previamente lleva a los experimentales a escribir:

	\begin{equation}
	\frac{R(q_{z},T)}{R_{F}(q_{z})}=|\Phi(q_{z},T)|^{2}e^{-\Delta^{2}_{cw}q_{z}^{2}}
	\end{equation}
	
Y la contribución de las ondas capilares sobre un perfil que nominalmente no las tiene viene determinado por el factor indicado donde la dependencia con la resolución del detector participa de $\Delta^{2}_{cw}$ de modo análogo a la dependencia en $L_{\parallel}$ que teníamos en la expresión de CWT . Su aplicación a los resultados experimentales es notable permitiendo obtener los resultados de la derecha en la figura (\ref{fig:Refletividad2}).\\

Al respecto de los modelos para perfiles intrínsecos utilizados para el \textit{Galium} bá\-si\-ca\-men\-te son dos,
	\begin{eqnarray}
	\rho(z)&=&\rho_{bulk}\sum_{i=0}^{\infty}\frac{d}{2\pi\sigma_{i}}e^{\frac{-(z-id-z')^{2}}{2\sigma_{i}^{2}}}\\
	\rho(z)&=&\rho_{bulk}\left[ 1+Ae^{-\frac{z-z'}{\lambda}}sen\left( \frac{2\pi(z-z_{0}-z')}{d}\right) erf\left( \frac{z}{\sigma_{0}}\right) \right] 	
\label{eqn:perfilesIntrinsecosExp}
	\end{eqnarray}
	usualmente el primero es preferido pues tiene menos parámetros libres ($\sigma_{i}^{2}=i\bar{\sigma}^{2}+\sigma_{0}^{2}$).  En ambos casos el procedimiento es realizar la transformada Fourier para obtener $\Phi(q_{z})$ y ajustar los parámetros del modelo a partir de las medidas. De aquí se cotejan los posibles modelos de perfil intrínseco. En el caso del Hg\cite{PhysRevLett.74.4444,PhysRevB.58.R13419} se utiliza el primero de los modelos pero la primera capa es sustituida por una contribución $n_{0}\delta(z)$.\\
	
	\begin{figure}[htbp]
\vspace{0.5cm}
	\begin{center}
	\includegraphics[width=4.2in]{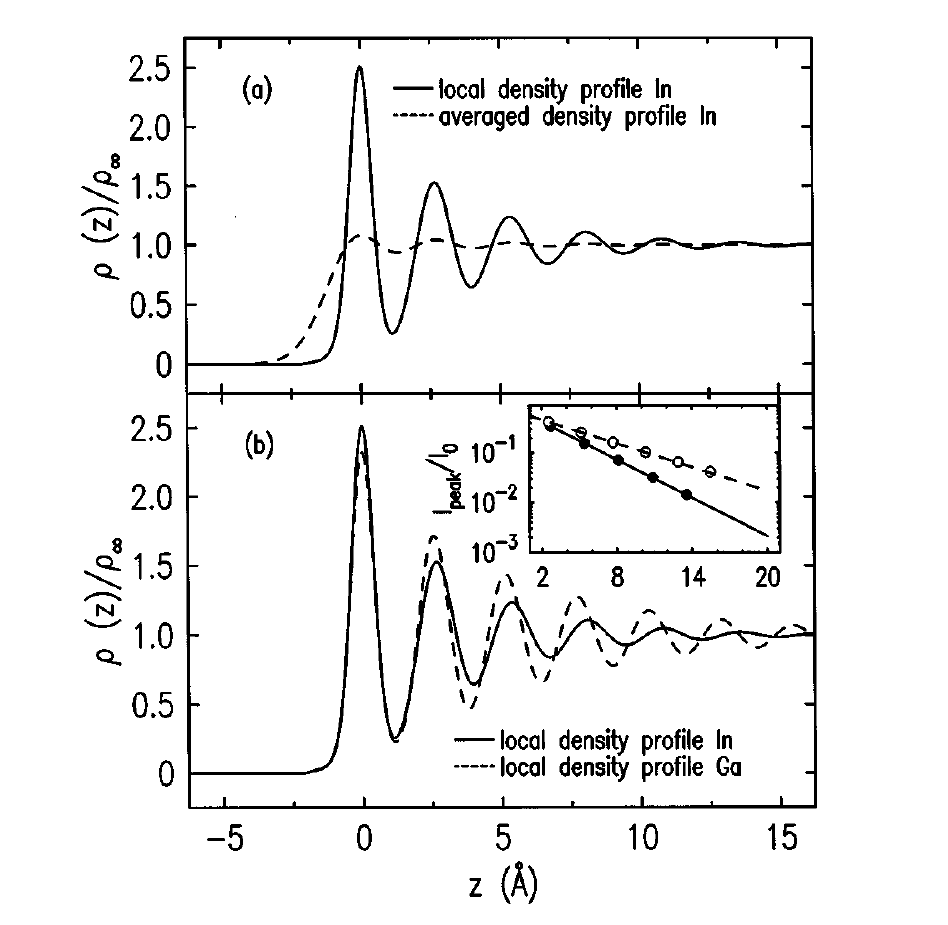}
	\caption{Perfiles intrínsecos. \textbf{(a)} Se compara el perfil intrínseco con el perfil medio que incluye parte del espectro de ondas capilares. \textbf{(b)} Se visualiza las propuestas para perfil intrínseco en los elementos \textit{Galium} e \textit{Indium}.}
	\label{fig:Refletividad3}
	\end{center}
\vspace{0.6cm}

	\end{figure}
	
\begin{figure}[htbp]
\vspace{0.5cm}
	\begin{center}
	\includegraphics[width=4.0in]{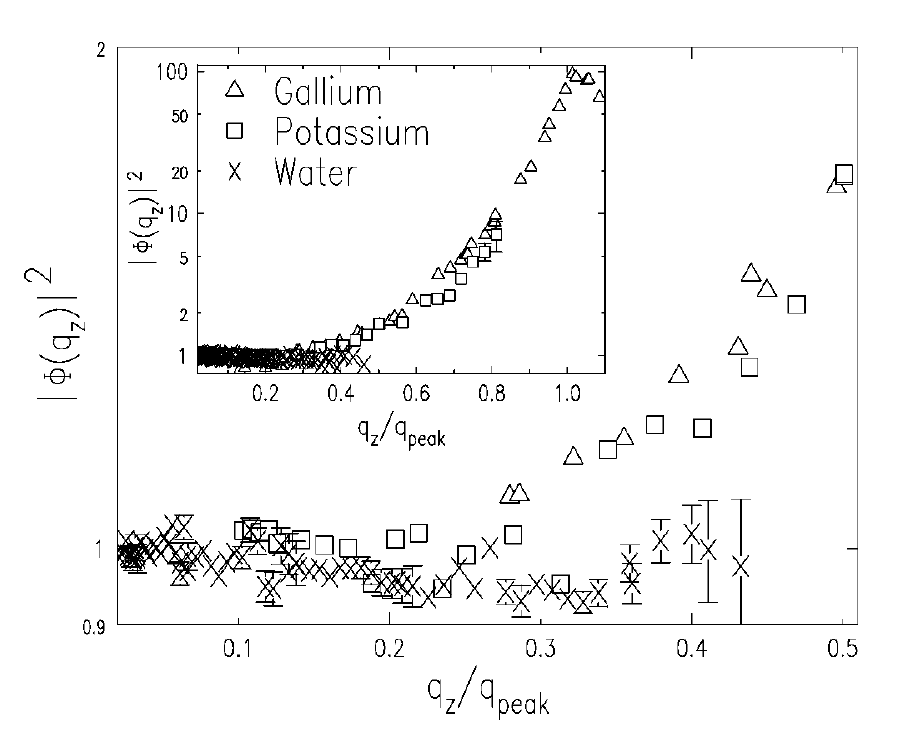}
	\caption{Mediciones para el $H_{2}0$ comparadas con Hg, Ga, K. El inset es más clarificador en lo que respecta a la posibilidad de emitir conclusiones acerca del cuasi-pico de Bragg en este experimento. Las medidas de reflectividad correspondientes a estos tres sistemas muestran que entorno a $q_{z}\simeq 0.8$ mientras que para el \textit{Galium} estan entorno a $10^{-6}$ para el \textit{Potassium} esta entorno a $10^{-10}$ y aun menos para el \textit{Agua} en el límite de capacidad experimental. Véase \cite{shpyrko:245423}}
	\label{fig:RefletividadWater}
	\end{center}
\vspace{0.7cm}

	\end{figure}

La situación respecto de las cuestiones planteadas es, primero la observación en determinados líquidos de un perfil intrínseco estructurado,  y segundo la necesidad para su medición de valores de valores altos de tensión superficial y bajos para el punto triple de modo que podamos enfriar lo suficiente la superficie y en consecuencia hagamos visible la estructura detrás de las fluctuaciones de superficie. Queda el problema de determinar si esta propiedad estructural es universal más allá de su posible medición o es una propiedad concreta de metales líquidos como \textit{Galium} o el \textit{Mercurio}. En el primero casos hemos de encontrar el fenómeno físico subyacente a la estructuración, en el segundo caso la base física de la estructuración podría ser la propuesta por \textit{Rice et al} pero indicaría que otros sistemas no metálicos, como por ejemplo el agua no poseen esta propiedad estructural. En esta línea se han realizado experimentos de reflectividad\cite{PhysRevB.67.115405,PhysRevLett.54.114} para el agua y se han comparado con el K (\textit{Potasio}) cuya tensión superficial es similar podemos observarlo en la figura (\ref{fig:RefletividadWater}). En ella los resultados \textit{no son concluyentes} ya que para $q_{z}/q_{pico}\sim0.4$ deja de haber medidas. Si bien para el caso del K el factor de estructura comienza a desarrollar el cuasi-pico de Bragg en estos valores no es evidente que, a pesar de tener la misma tensión superficial, deba visualizarse en este mismo punto trazas de estructuración. Los experimentos no permiten cerrar la cuestión a no ser que midamos para $q_{z}\sim q_{pico}$ o exista un argumento sólido para esperar que en $q_{z}/q_{pico}\sim0.4$ ya debemos encontrar estructuración.\\

Los dos problemas que hemos dibujado mediante los experimentos son finalmente los que indicábamos como cuestiones abiertas por CWT, a saber, la naturaleza de los perfiles sin fluctuaciones superficiales y la forma de estas en escalas de longitud pequeñas. Para la primera de las cuestiones las simulaciones para modelos dieléctricos parecen indicar la presencia también de las mismas propiedades estructurales que en metales líquidos con, si cabe, mayor riqueza fenomenológica al presentar además de una estructuración en capas en el perfil intrínseco una estructuración orientacional de las moléculas en la superficie fría (agua). La segunda de las cuestiones posee un intenso debate y será tratada a lo largo de la disertación desde el punto de vista de la teoría del funcional de la densidad.

\section{Modelos de interacción}

Hemos visto a partir de los datos extraídos, ver tabla (\ref{tabla:DatosMetalesLiquidos}), de los diferentes elementos junto con los resultados procedentes de los experimentos que una de las propiedades clave en la visualización de estructura en los perfiles de equilibrio líquido-vapor es la posibilidad de tener tensiones superficiales suficientemente altas, propiedad que aparece condensada, véase ec. (\ref{eqn:etapexper}), en el parámetro $\eta$.\\

 	\begin{wrapfigure}{r}{7.4cm}
 	\vspace{-0.8cm}
	\includegraphics[width=2.7in]{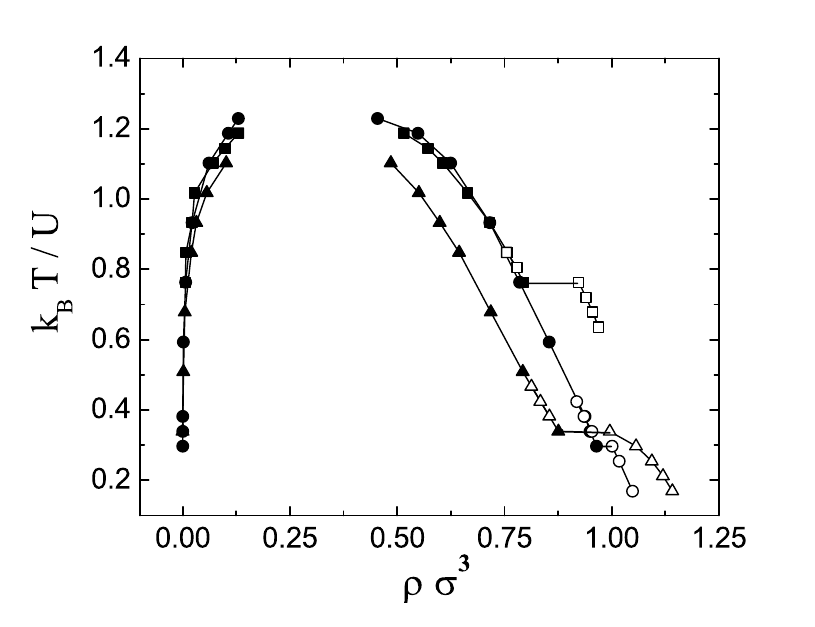}
 	\caption{\small Resultados para los diagramas de fases mediante MC tomados de \cite{PhysRevLett.87.166101}. Círculos Na, Triángulos Hg, Cuadrados Lennard-Jones. Símbolos en negro NPT, en blanco NVT. Simulaciones fueron realizadas en geometría slab.}
 	\label{fig:Montecarlo}
 	\end{wrapfigure}

Tomando como objetivo esta propiedad \textit{Velasco et al} \cite{velasco:10777} determinaron un conjunto de  potenciales a pares que permiten reproducirla y analizaron la posible relevancia de características no superficiales como causantes de la forma estructural de los perfiles líquido-vapor, o desde otro punto de vista si la presencia o no de oscilaciones en $\rho_{lv}(z)$ esta condicionada por una propiedad del sistema uniforme. Esta cuestión surge a raíz del trabajo realizado por \textit{R. Evans}\cite{EvansMolecularPhyscFWline} al señalar la posible relevancia de las propiedades estructurales de la función de correlación total en la estructura de las interfases fluidas. En su planteamiento retoman un problema que surge hace tiempo a partir del estudio realizado para sistemas unidimensionales por \textit{Fisher y Widom}\cite{FisherWidomlinea1969}, y que, motivado por el trabajo de Evans, ha sido abordado posteriormente desde varios enfoques. Mediante simulación Montecarlo se ha analizado extensamente el potencial Lennard-Jones\cite{dijkstra:1449}, mientras que desde de aproximaciones funcionales, esencialmente WDA, se ha analizado el pozo cuadrado\cite{EvansMolecularPhyscFWline}. También se han realizado estudios similares aunque en sistemas diferentes, como mezclas coloide-polímero\cite{PhysRevE.64.041501}, o sistemas no esféricos\cite{savenko:021202} y finalmente con este mismo objetivo pueden abordarse planteamientos basados en la teoría de las ecuaciones integrales que pueden dar una información estructural próxima a las simulaciones\cite{PhysRevE.51.3146}, aunque dependiendo de las aproximaciones utilizadas puede no ser posible resolverlas en el diagrama de fases completo\cite{ferreira:594}.\\

Todos los trabajos indicados al basarse en el uso de potenciales a pares pueden enfocarse como el estudio simplificado de sustancias que podemos encontrar en el laboratorio aunque las propiedades del sistema real pue\-dan ser notablemente más complejas que las representadas por una interacción a pares. Una sustancia sencilla como el Argón puede ser modelizada i\-ni\-ci\-al\-men\-te por un Lennard-Jones pero la interacción a tres cuerpos, que puede ser abordada por un potencial Axilrod-Teller, posee una contribución de hecho finita\cite{PhysRevA.45.3659,PhysRevE.55.2707}. De igual modo las sustancias metálicas como el Mercurio, el Galio o el Sodio poseen interacciones cuyo tratamiento detallado requeriría tener en cuen\-ta el problema estadístico cuántico completo. Los métodos más usuales implican un proceso de \textit{coarse-graining} de modo que un conjunto de grados de libertad puede ser integrado dando lugar a una interacción efectiva, vía que introduce dependencias extra en el potencial de interacción y en general las interacciones \textit{efectivas} resultaran dependientes del estado.\\

Otro método posible de determinar las interacciones es a partir de medidas experimentales del factor de estructura $S(q;\rho)$, el resultado será igualmente un potencial dependiente, en este caso, de la densidad global, $\rho$, del estado. \\

En general una vez determinado un potencial $\phi(\vec{r}_{12};\rho)$, aparecen dos inconvenientes que es preciso notar: un problema de \textit{transferencia} \cite{merabia:054903} debido a que la dependencia funcional en un estado no es necesariamente la misma que en otro estado diferente, y un problema de \textit{representabilidad} \cite{Louis:2002:0953-8984:9187} debido a la posibilidad de que para un estado no exista un único $\phi(\vec{r}_{12};\rho)$. Esta última cuestión es importante ya que puede derivar en la presencia de inconsistencias en la determinación de propiedades termodinámicas y estructurales.\\

Resulta pues preferible retomar los potenciales a pares contenidos en \textit{Velasco et al} \cite{velasco:10777} sin el objetivo de modelizar el diagrama de fases completo\footnote{Que intentaríamos mediante un $\phi(\vec{r}_{12},\rho)$ y que habría de enfrentarse con los problemas comentados y que además nos obligaría en todo momento interpretarlo aludiendo al proceso de integración del conjunto de interacciones \textit{many-body} que intentaría capturar.}, sino de conseguir representar una propiedad concreta, a saber el parámetro $\eta$ adecuado, relacionada con las sustancias de interés.\\

\subsection{Potenciales \textit{tipo Sodio} y \textit{tipo Mercurio}}
\label{sec:potencialesSodioMercurio}
Los dos modelos de interacción a pares que denominamos \textit{Sodio} (Na) y \textit{Mercurio} (Hg) reproducen, por tanto, \textit{algunas} propiedades experimentales de estos metales líquidos \cite{velasco:10777,PhysRevLett.87.166101}. A pesar de no tener en cuenta propiedades complejas del enlace metálico reproducen valores de la relación entre la temperatura triple y la crítica ($T_{t}/T_{c}$) de un modo cualitativo adecuado a nuestros propósitos. En la figura (\ref{fig:Montecarlo}) extraída de \cite{PhysRevLett.87.166101} podemos observar que para el \textit{Mercurio} $T_{t}/T_{c}=0.27$ y para el \textit{Sodio} $T_{t}/T_{c}=0.22$ mientras que para el caso de un Lennard-Jones es apreciablemente mayor, $T_{t}/T_{c}=0.56$.\\

\begin{figure}[htbp] 
   \centering
\vspace{0.01cm}
   \includegraphics[width=0.93\textwidth]{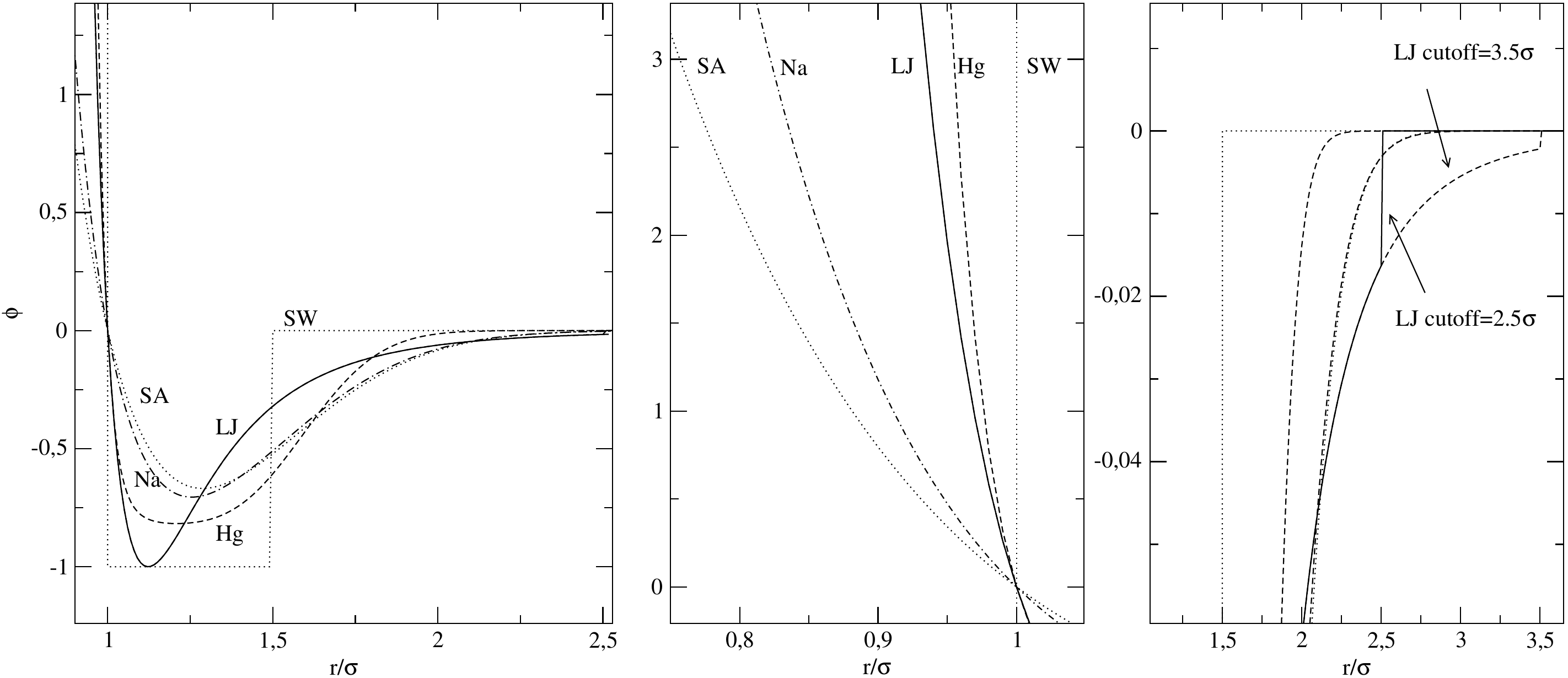} 
   \caption{Potenciales de interacción a pares utilizados a lo largo de la memoria. En el \textbf{centro} observamos el comportamiento de la parte repulsiva (desde la izq son SA,NA,LJ,HG,SW). A la \textit{derecha} podemos ver el comportamiento de la cola atractiva (desde iqz son SW,Hg, SA-Na solapados, Lennard-Jones cutoff en $2.5\sigma$ y Lennard-Jones cutoff en $3.5\sigma$).}
   \label{fig:Potenciales}
\vspace{0.01cm}
\end{figure}

Ambos son construidos mediante una exponencial que describe el término repulsivo a cortas distancia aunque los pa\-rá\-me\-tros que lo constituyen determinan comportamientos cualitativamente diferentes en ambos casos, en el modelo \textit{Mercurio} es muy abrupto, más similar a un modelo de esferas duras o un pozo cuadrado, mientras que el modelo \textit{Sodio} presenta una parte repulsiva más suave. Si comparamos ambos con un modelo Lennard-Jones este presenta un comportamiento intermedio entre ambos. La parte atractiva se construye con un pozo atractivo gaussiano en el caso del \textit{Sodio}, que configura un potencial de corto alcance:

\begin{equation}
\phi_{Na}(r)=A_{0}e^{-\lambda_{0}r}-A_{1}e^{-\lambda_{1}(r-R_{1})^{2}}
\end{equation}
mientras en en el \textit{Mercurio} la parte atractiva es una suma de dos contribuciones gaussianas:
\begin{equation}
\phi_{Hg}(r)=A_{0}e^{-\lambda_{0}r}-A_{1}[e^{-\lambda_{1}(r-R_{1})^{2}}+e^{-\lambda_{1}(r-R_{2})^{2}}]
\end{equation}
\\
ambos son más \textit{planos} en su región atractiva que un Lennard-Jones siendo más acentuado en el modelo \textit{Mercurio}. Los parámetros del \textit{Sodio} permiten enlazar datos de la energía de la fase sólida\cite{PhysRevB.57.15519}. Ajustar las energías de los cristales de Sodio para redes fcc y bcc a un modelo simple de potencial a pares no es posible para un potencial polinómico de la forma $r^{-n}$, incluido un Lennard-Jones, sin embargo para potenciales como el indicado arriba si es factible, los datos son los indicados en la tabla(\ref{tabla:DefinicionPotenciales}). Mientras tanto los valores del Mercurio reproducen datos del factor de estructura del metal en un estado de T=1273 \degree K y densidad $\rho=10.98 gr\cdot cm^{-3}$.\\

También se ha construido otro potencial que denominamos \textit{Soft Alcaline} (SA) que comparte las propiedades de \emph{soft-core} del Sodio pero más acentuadas, para ello se reduce el parámetro $\lambda_{0}$ pero manteniendo las unidades reducidas, que se definen más adelante, básicamente idénticas para lo cual se redefinen las amplitudes $A_{0}$ y $A_{1}$. Las simulaciones Montecarlo muestran que $T_{t}/T_{c}=0.12$. \\

De modo general apreciamos que todos poseen un decaimiento gaussiano a grandes distancias no siendo necesario definir un \emph{cutoff} como suele ser necesario en modelos polinómicos como el Lennard-Jones. Aparecen resumidos los valores que definen los tres potenciales en la tabla (\ref{tabla:DefinicionPotenciales}).

\begin{table}[htdp]
\vspace{0.01cm}
\begin{center}
\begin{tabular}{l l l l}
\toprule
 &  \textbf{Sodio} & \textbf{Soft-Alcaline}  & \textbf{Mercurio} \\
\midrule
$A_{0}$ &  $437.960 \quad eV$ & $37.5500\quad eV$  & $2.06162\cdot 10^{14}\quad eV$ \\
$A_{1}$ &   $ 0.18282\quad eV$ & $0.21054\quad eV$  & $0.075370 \qquad\quad eV$ \\
$\lambda_{0}$ &  $2.3222\quad\AA^{-1}$ & $1.4881\quad\AA^{-1}$  & $12.0237\quad\quad\quad\AA^{-1}$ \\
$\lambda_{1}$ & $0.2140\quad\AA^{-2}$ & $0.2140\quad\AA^{-2}$  & $ 1.22650\quad\qquad\AA^{-2}$ \\
$R_{1}$ &  $3.5344\quad\AA$ & $3.5344\quad\AA$  & $2.95270\quad\qquad\AA$ \\
$R_{2}$ &   &   & $4.12330\quad\qquad\AA$\\
\bottomrule
\end{tabular}
\end{center}
\caption{Datos de los potenciales Sodio, \textit{Soft-Alcaline} y Mercurio.}
\label{tabla:DefinicionPotenciales}
\vspace{0.01cm}

\end{table}%

\begin{table}[htdp]
\vspace{0.01cm}
\begin{center}
\begin{tabular}{c c c c c}
\toprule
- &  \textbf{Na} & \textbf{SA}  & \textbf{Hg} & \textbf{Ar} \\
\midrule
$\sigma $ &  $3.48377 \quad \AA$ & $3.48345\quad \AA$  & $2.94243\quad \AA$ & $3.405\quad \AA$\\
$U$ &   $ 0.188476\quad eV$ & $0.188699\quad eV$  & $0.1211075\quad eV$&$111.9 \quad \degree K$ \\
\bottomrule
\end{tabular}
\end{center}
\caption{Datos unidades reducidas de los potenciales.}
\label{tabla:DatosUnidadesReducidas}
\vspace{0.01cm}
\end{table}%

Como indicábamos ninguno de los modelos anteriores se han construido para reproducir los aspectos de las interacciones que se producen en un metal líquido complejos de describir a un nivel microscópico, tampoco para reproducir fielmente en un rango amplio de densidades y temperaturas propiedades termodinámicas. Pero si que permiten explorar adecuadamente las consecuencias de valores de $T_{t}/T_{c}$ bajos así como sus posibles causas en una representación como potencial a pares\cite{velasco:10777}. Para la comparación entre todos ellos con un potencial de referencia utilizamos unas unidades reducidas tanto en la energía como en la longitud dadas por las relaciones,

\begin{equation}
U=-\frac{9}{8\sigma^{3}}\int dr r^{2}\phi(r)
\label{eqn:UlennardJ}
\end{equation}
\begin{equation}
\phi(\sigma)=0
\end{equation}

que nos permite comparar todos los modelos respecto de un Lennard-Jones con parámetro
$\epsilon=U$ dado por la ecuación,

\begin{equation}
\phi_{LJ}(r)=4\epsilon\left[\left( \frac{\sigma}{{r}}\right)^{6}-\left( \frac{\sigma}{{r}}\right)^{12}\right]
\end{equation}

los valores para $\sigma$ y $U$ los podemos ver en la tabla (\ref{tabla:DatosUnidadesReducidas}) donde se han incluido los datos, como referencia comparativa, típicos de un Lennard-Jones que permiten modelizar el caso del Argón líquido.\\

En la figura (\ref{fig:Potenciales}) encontramos una representación de estos potenciales junto con un Lennard-Jones\footnote{En la literatura el modelo de Lennard-Jones se trata de diferentes maneras, suele ser común realizar un cutoff en $2.5\sigma$ ignorando el resto de la cola ya que a efectos de cálculo se traduce en menos costo computacional. A menudo se complementa con un \textit{shift} del potencial para tener un potencial continuo. En este capítulo representamos el Lennard-Jones sin cutoff tanto en la figura (\ref{fig:Potenciales}) como en las curvas de coexistencia de fases, sin embargo en el cálculo funcional realizado más adelante realizamos un cutoff y consideramos también una $\epsilon$ que hace que comparta unidades reducidas con el resto de potenciales, esto determina diferentes Lennard-Jones con diferentes cutoff que tienen leves diferencias. En lo que respecta a simulación se han utilizado diferentes modelos de Lennard-Jones variando tanto el cutoff como el shift. La comparación del Lennard-Jones con un cutoff en 2.5 respecto de este mismo con un shift muestra que el primero parece preferible al segundo a la hora de realizar simulaciones\cite{trokhymchuk:8510} y es el que usaremos en los experimentos numéricos mediante dinámica molecular.}. A efectos de comparar con la literatura se ha incluido un Potencial Cuadrado de alcance $1.5\sigma$ ya que ha sido analizado extensivamente.

\begin{figure}[htbp] 
   \centering
\vspace{0.01cm}

   \includegraphics[width=0.95\textwidth]{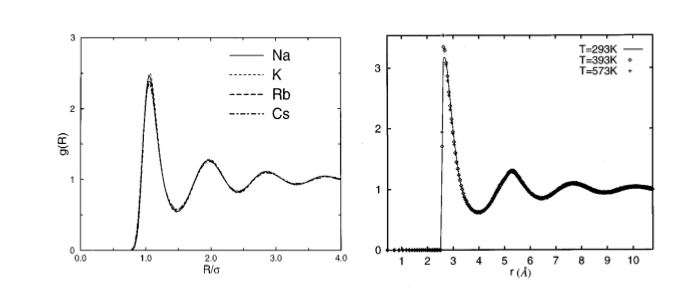} 
   \caption{Funciones de distribución radial obtenidas por \textit{Rice et al} \cite{zhao:1959}, basándose en una representación mediante pseudopotenciales para las interacciones ion-ion e ion-electrón y simulaciones Montecarlo. \textbf{Izquierda}: metales alcalinos, \textbf{Derecha}: Galio. Los potenciales de interacción indicados dan un comportamiento cualitativamente similar en las formas de estructura líquida que se han llamado \textit{Sodio} y \textit{Mercurio}.}
   \label{fig:GdeRmetalesRice}
\vspace{0.01cm}
\end{figure}

\cleartorecto

\section{Simulaciones de Montecarlo y de Dinámica Molecular}

\epigraph{Marco Polo describe un puente, piedra por piedra. -¿Pero cuál es la piedra que sostiene el puente? –pregunta Kublai Kan. –El puente no está sostenido por esta piedra o por aquélla –responde Marco-, sino por la línea del arco que ellas forman. Kublai permanece silencioso, reflexionando. Después añade: -¿Por qué me hablas de piedras? Lo único que me importa es el arco. Polo responde –Sin piedras no hay arco.}{\textit{Las ciudades invisibles}\\ \scshape Italo Calvino}

\vspace{0.5 cm}

Las simulaciones permiten técnicamente, siguiendo indicaciones si\-mi\-la\-res en su fi\-lo\-so\-fía a las propuestas por \textit{Frank H. Stillinger}, determinar el perfil intrínseco y la superficie intrínseca a partir de las configuraciones del sistema. El procedimiento es determinar para cada configuración, de un modo consistente en todas las escalas, $\xi(\vec{R})$ y siguiendo la teoría de ondas capilares tendremos un perfil nominalmente exento de fluctuaciones superficiales mediante $\tilde{\rho}(z)=<\sum_{i}^{N}\delta(z-z_{i}-\xi(\vec{R}))>$. Esta metodología presenta dos sutilezas subyacentes, la primera el nivel de corrugación permitido para $\xi(\vec{R})$ introduce una dependencia implícita en el perfil intrínseco, la seguunda es la posibilidad de establecer conjuntos disjuntos para una definición dada de $\xi(\vec{R})$ en el espacio de configuraciones o dicho de otro modo la independencia estadística de  $\xi(\vec{R})$ y $\tilde{\rho}(z)$.

\subsection{Superficie intrínseca y Perfil intrínseco}
\label{sec:intruSuperIntrinseca}

Como vemos en el caso de las simulaciones la determinación del perfil de densidad intrínseco requiere una definición previa de superficie intrínseca desde las configuraciones moleculares de un sistema, con este objetivo han surgido diferentes metodologías que agrupo genéricamente dos familias de aproximaciones fundamentales.\\

La definición más sencilla corresponde a una \textit{superficie local de Gibbs} y ha sido la propuesta tradicional sobre la que se han realizado la mayoría de las argumentaciones sobre la estructura de la interfase líquido-vapor desde evaluaciones teóricas\cite{weeks:3106} hasta simulaciones recientes de mezclas de polímeros\cite{VinkHorbachBinderJCP} así como la interpretación de diferentes experimentos\cite{DirkAarts05072004}.\\

Esencialmente consiste en dividir el sistema en subsistemas en columnas de un tamaño transversal dado. Sobre cada uno de estos subsistemas se determina la tasa de ocupación de partículas y con ella una superficie intrínseca local. El problema esencial de esta aproximación es que al descender a secciones del subsistema menores que la longitud de correlación de volumen, las fluctuaciones propias del sistema homogéneo tendrán un efecto nítido en los valores de la superficie local de Gibbs y en consecuencia la definición de la superficie intrínseca\cite{stecki:2860} no esta ligada únicamente a propiedades locales de la superficie ya que las tasas de ocupación engloban también propiedades de volumen.\\

\begin{figure}[htbp]
\begin{center}
\vspace{0.5cm}
\includegraphics[width=5.1in]{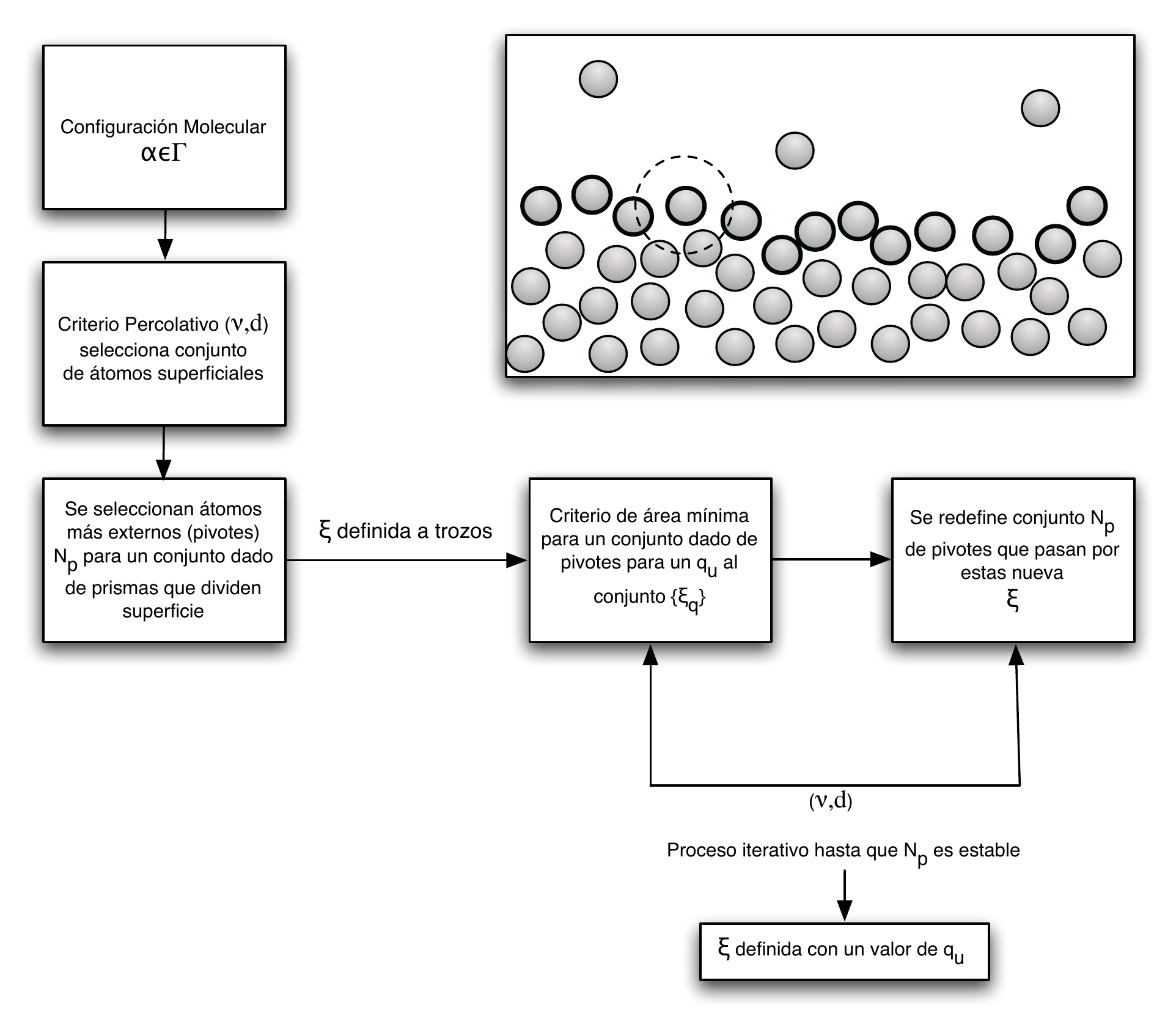}
\caption{Descripción del esquema operacional-percolativo de Tarazona-Chacón\cite{PhysRevB.70.235407,2005JPCM17S3493C}. Los dos parámetros clave son $\nu$, que es el número de vecinos próximos que han de existir en un radio \textit{d} para pertenecer al líquido. Son adecuados $(\nu,d)=(3,1.5\sigma)$.}
\label{fig:PercolativoTCh1}
\end{center}
\vspace{0.9cm}
\end{figure}

Esto motiva la segunda construcción de una definición que en la propuesta inicial de \textit{Tarazona y Chacón} puede verse en la figura (\ref{fig:PercolativoTCh1}). Es un método operacional de carácter percolativo, donde aunque el sistema es igualmente dividido en subsistemas la evaluación de la superficie intrínseca es realizada mediante un \textit{argumento que solamente tiene en cuenta propiedades de distribución molecular en la superficie}\footnote{Su descripción implica a funciones de distribución de partículas de varios ordenes, véase ec. (\ref{eqn:familiaDistribDensidadMacro}), cuando descendemos a escala molecular debido a las correlaciones que involucra un proceso percolativo con optimización. Esta característica no presente en una receta basada en la superficie de Gibbs que sólo involucra a la función de distribución de partículas de orden uno.}. Formalmente, dada una superficie intrínseca,\\
\index{Superficia Intrinseca}
\begin{equation}
\xi(\vec{R})=\hat{\xi}_{0}+\sum_{|\vec{q}|=q_{l}>0}^{q_{u}}\hat{\xi}_{q}e^{-i\vec{q}\vec{R}}
\end{equation}
podemos determinar el perfil intrínseco\footnote{Suponemos que nuestro sistema esta inmerso en una caja de volumen $L_{z}A_{0}$.} a un nivel dado de corrugación de la superficie $\xi(\vec{R})$ especificado por $q_{u}$,
\begin{equation}
\tilde{\rho}(z,q_{u})A_{0}\equiv\left\langle \sum_{i=1}^{N}\delta(z-z_{i}+\xi(\vec{R}_{i}))\right\rangle 
\end{equation}
\index{Perfil Intrínseco}
En el caso de la superficie de Gibbs local cuando el nivel de corrugación, el valor de $q_{u}$, llega a la escala molecular la presencia de fluctuaciones de volumen en la definición de $\xi(\vec{R})$ tiende a eliminar la estructuración en capas mostrando un perfil intrínseco suave en lugar de altamente estructurado. El segundo procedimiento, en cambio, lleva a perfiles intrínsecos progresivamente más estructurados revelando predicciones más acordes con las propuestas experimentales. De modo que ambos métodos dan resultados análogos en el límite mesoscópico $q_{u}\sim 0$ y en ambos en este límite se recupera el perfil de densidad $\rho(z,L_{x}=2\pi/q_{l})$ pero muestran perfiles intrínsecos diferentes a niveles de corrugación mayores incluso aun nítidamente alejados de $q_{max}=2\pi/\sigma$. \\

Los perfiles de equilibrio son determinados mediante,
\begin{equation}
\rho(z,L_{x})A_{0}=\rho(z,q_{l})A_{0}=\left\langle \sum_{i=1}^{N}\delta(z-z_{i}) \right\rangle 
\end{equation}

En cada uno de los esquemas para $\xi$ usados en simulación podemos observar las propiedades estadísticas asociadas a $\xi_{q}$ y los correspondientes perfiles intrínsecos. El método más cómodo de comparación es definir una $\gamma(q)$ y suponer inicialmente que aproximadamente las amplitudes $\xi_{q}$ siguen, tanto en el método percolativo como en la superficie de Gibbs, casos una distribución gaussiana donde los diferentes modos aparecen descorrelacionados y por tanto,
\begin{equation}
\gamma(q)=\frac{1}{\beta A_{0}q^{2}<|\xi_{q}|^{2}>}
\end{equation}
La predicción de la definición de $\xi(\vec{R})$ como una superficie local de Gibbs es una $\gamma(q)$ \textit{decreciente}\footnote{De hecho se obtiene una forma funcional que se puede relacionar con el factor de estructura $S(q)$ del sistema líquido homogéneo, revelando la presencia de fluctuaciones de volumen en el presunto espectro de ondas de capilaridad que de hecho serían predominantes en el comportamiento global de la amplitudes $\xi_{q}$.} mientras que $\gamma(q)$ en Tarazona-Chacón es \textit{creciente}, evitando la necesidad de  introducir un \textit{cutoff} externo $q_{max}$ en el espectro de ondas capilares, lo que es desde el punto de vista físico notablemente sugerente.\\

\begin{figure}[htbp] 
   \centering
   \includegraphics[width=1.00\textwidth]{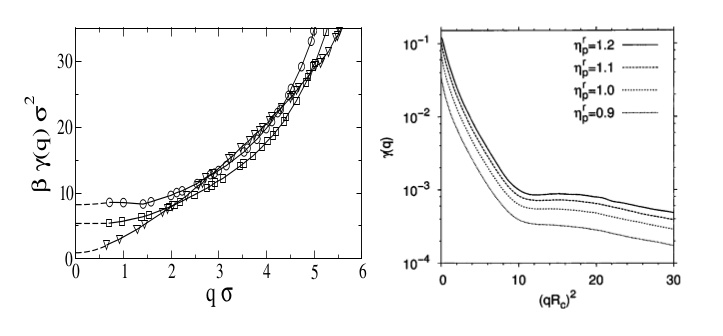} 
   \caption{\textbf{Izquierda}: $\gamma(q)$ obtenida por Tarazona y Chacón\cite{PhysRevLett.91.166103} mediante el método descrito, para diferentes modelos de interacción a pares. \textbf{Derecha}: $\gamma(q)$ obtenida por \textit{R.L.Vink et al mediante un método basado en la superficie de Gibbs para la coexistencia coloide-polímero.}\cite{VinkHorbachBinderJCP}}
   \label{fig:GammaQsimulaciones}
\end{figure}

Los diferentes parámetros que controlan la $\xi(\vec{R})$ tienen una influencia en el perfil intrínseco aunque no esencial, es en $\gamma(q)$ donde si se aprecian con más nitidez las consecuencias de la definición elegida para la superficie intrínseca. Esto nos llevara a que a la hora de buscar un esquema conceptual acerca de la estructura intrínseca sea más conveniente definirla de modo que se obtenga el perfil intrínseco adecuado, mientras que  $\gamma(q)$ respete un conjunto de propiedades razonables.

\section{Otros sistemas: Mezclas coloide-polímero}

Consideremos ahora un sistema constituido por una mezcla binaria de co\-loi\-des y po\-lí\-me\-ros, en él se observa en determinadas condiciones una separación entre una fase con alta concentración de coloides y otra con una baja concentración de estos. Mirando únicamente los coloides no existe una atracción directa entre ellos que permita producir una separación de fases en la imagen tradicional de un líquido de van der Waals, pero existe un mecanismo de origen entrópico que permite establecer una interacción efectiva atractiva entre los coloides (atracción de deplección) y reduce a un modelo sencillo el mecanismo de separación de fases. Desde este esquema es posible tratar este sistema complejo como un sistema simple análogo a los analizados a lo largo de la memoria y estudiar mediante las metodologías utilizadas algunas de las propiedades superficiales de la interfase fluido-fluido\footnote{Siendo posible estudiarlo directamente como mezcla binaria con una extensión directa de los mismos métodos.}.\\

\begin{figure}[htbp] 
   \centering
   \includegraphics[width=0.78\textwidth]{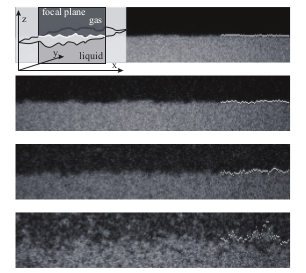} 
   \caption{Imágenes obtenidas mediante LSCM para una mezcla coloide-polímero, tomado de referencia \cite{paperCONmecke}. Conforme se desciende en la figura vemos casos con mayor temperatura.}
   \label{fig:LSCM}
\end{figure}

Por otra parte el rango de tamaños de las partículas coloidales oscila entre los 6nm y 500nm lo que lo sitúa en el rango en que es posible realizar experimentos en el visible.  La sustancia más utilizada es conocida como PMMA y se modeliza adecuadamente por esferas con un radio de unos 70nm y obviando los problemas de polidispersión\footnote{Producidos por variaciones en el radio de los coloides al ser una sustancia creada artificialmente que marginalmente hace que su radio no sea uniforme\cite{PhysRevE.71.066126}.} es un sistema clásico de esferas duras. La introducción de un polímero como el polystyreno, cuyo radio de \textit{gyration} es de unos 40nm, generará la atracción efectiva de deplección haciendo posible ver este sistema mediante la imagen tradicional de van der Waals. Ha sido posible estudiar mediante métodos experimentales\cite{DirkAarts05072004} como LSCM (Laser Scanning Confocal Microscopy) la interfase fluido-fluido como se pueden ver en la figura (\ref{fig:LSCM}).\\

La visualización de la interfase no constituye la resolución del problema ya que en simulaciones también podemos \textit{ver} la interfase y los problemas para la determinación \textit{correcta} de la superficie intrínseca siguen estando presentes. Los autores a partir de las figuras mostradas \cite{paperCONmecke} estiman los valores de $\gamma$ y de $L_{\parallel}$ en función de la concentración de coloides. Es ilustrativo estudiar su metodología ya que esencialmente se enmarca en la aplicación tradicional de la teoría de ondas capilares. La magnitud experimental accesible es la distribución de intensidad de luz en el sistema en el instante de la medición\footnote{Los tiempos de relajación de este sistema respecto del aparato lo permiten.} que denotamos por $I(x,z,t)$. A los coloides creados se les ha dotado de propiedades fluorescentes por tanto la imagen brillante representa la fase rica en coloides y las imágenes mostradas asemejan una interfase líquido-vapor.\\

Su análisis se basa en el concepto de superficie intrínseca que determinan mediante la expresión:
	\begin{equation}
	\int_{0}^{L_{z}} dz I(x,z,t)dz=I_{liq}(x)\xi_{b}(x,t)+I_{gas}(x)(L_{z}-\xi_{b}(x,t))
	\end{equation}
las funciones $I_{liq}(x)$ y $I_{gas}(x)$ son promedios en el volumen de las intensidades y por tanto juegan el papel equivalente a las densidades $\rho_{l}$ y $\rho_{v}$ en una coexistencia líquido-vapor. El criterio es una aplicación local de hipótesis equivalentes a la superficie de Gibbs y por lo mismo en el procesamiento de datos no hay una separación explícita de las fluctuaciones superficiales y las fluctuaciones de volumen. La definición final que encontramos es $\xi(x,t)=\xi_{b}(x,t)-\bar{\xi}_{b}(t)$. Mediante esta se pueden realizar un procedimiento estadístico para calcular la función de distribución de alturas y ajustarla a un modelo gaussiano. Mientras que la función de correlación de alturas esta definida mediante:
	\begin{equation}
g_{h}(x)=<\xi(x',t')\xi(x'+x,t')>
	\end{equation}
De este modo comparan esta función obtenida experimentalmente con un modelo de ondas capilares donde $L_{\parallel}$ que aparece como implícita\footnote{La función de distribución de alturas de la teoría de ondas capilares esta incluida en el apéndice \S\ref{apn:teoriascampoefectivo}, veáse la ecuación (\ref{eqn:defdeSdeR}), que es resoluble en analíticamente mediante funciones de Bessel.}. Los resultados experimentales muestran que la función de distribución de alturas es gaussiana en un rango amplio de temperaturas, razonablemente compatible con la teoría de ondas capilares.\\

Este sistema al contrario que los metales líquidos la interfase es notablemente \textit{blanda} ya que posee unos valores de la tensión superficial muy bajos: $\gamma\sim 1\mu N/m$, mucho menores que los visualizados en la tabla (\ref{tabla:DatosMetalesLiquidos}) y por tanto es un candidato ideal para estudiar otros aspectos de las interfases, posee altos valores de la viscosidad que permiten que el tiempo de relajación de las fluctuaciones superficiales sea del orden de segundos y su dinámica mensurable.\\

Resaltar que en su estudio no intentan investigar en detalle el problema de la separación de las fluctuaciones superficiales de las de volumen ni tampoco la determinación de un perfil intrínseco para este sistema, en cualquier caso este análisis \textit{requeriría métodos análogos a los realizados en simulación}. De hecho se han realizado simulaciones Montecarlo para este sistema presentando los mismos problemas que encontramos en simulaciones de líquidos simples basados en el criterio de Gibbs\cite{VinkHorbachBinderJCP}.\\

\section{Teorías de líquidos acerca de la estructura intrínseca de la interfase}

La teoría de líquidos ha aportado cierta luz en lo referente a la generalidad de un comportamiento oscilante en la densidad ya que los perfiles obtenidos en teorías funcionales también resultan estructurados de modo similar si el funcional usado incorpora una descripción de las interacciones repulsivas (efectos de volumen excluido) adecuados, y toda vez que las teorías de perturbaciones en que se basa el tratamiento de sistemas simples ha tenido notable éxito en la explicación de gran variedad de fenómenos no parece ser un defecto de las aproximaciones funcionales utilizadas. Con todo, como veremos, el tener perfiles estructurados no soluciona la disyuntiva de qué es un perfil de densidad obtenido minimizando un funcional de energía libre pero si abre una vía para lanzar hipótesis falsables acerca de su naturaleza.\\

\section{Sumario}
\begin{itemize}
\item La teoría de ondas capilares forma parte de la física presente en una interfase líquida\cite{PhysRevLett.54.444,PhysRevE.60.6708}. Posee un papel en su termodinámica (tensión superficial) y en su estructura (perfiles de densidad), e introduce una dependencia en las propiedades estructurales relacionada con las condiciones que hacen posible la presencia de la interfase: campo externo y tamaño restringido de la superficie líquida\cite{PhysRevLett.15.621}.
\item Esta teoría se articula sobre dos conceptos que considera relacionados pero independientes desde el punto de vista estadístico. La \textit{superficie intrínseca} que separa las dos fases y el \textit{perfil intrínseco} que se supone no incluye la presencia de ondas capilares (fluctuaciones superficiales) y por tanto carece de las dependencias que el perfil de densidad medio si posee (aunque puede a priori poseer otras)\cite{PhysRevLett.91.166103}.
\item Los resultados experimentales permiten establecer, bajo una teoría de ondas capilares clásica, la presencia en metales líquidos de un perfil intrínseco altamente estructurado. Aunque resultados análogos para otros líquidos como el agua esta actualmente fuera del alcance experimental\cite{PhysRevLett.54.114,chacon014709}.
\item Las simulaciones muestran que esta propiedad del perfil intrínseco es de hecho general y la característica que permite su observación experimental en metales líquidos es ser estables frente al sólido hasta bajas temperaturas.
\item Los resultados anteriores abren cuestiones referentes a posibles extensiones a la teoría de ondas capilares clásica en un régimen donde coexiste con fluctuaciones de volumen. Cuestiones que deben ser tenidas en cuenta por planteamientos teóricos de sistemas no homogéneos\cite{velasco:10777,EvansMolecularPhyscFWline}.
\end{itemize} 

\part{ Introducción a la teoría de líquidos}
\chapter{Formalismo}	
   
Las cuestiones planteadas por la teoría de ondas capilares nos han llevado a recoger interacciones en\-tre partículas que realzan ciertas propiedades fenomenológicas claves que aparecen ligadas a un comportamiento esencialmente colectivo que precisa introducir el formalismo de la mecánica estadística\footnote{En la presente memoria se introduce el formalismo de la mec\'{a}nica estadística en su versi\'on cl\'asica, aunque cabe notar que no es la \'unica posiblidad. En particular, por su aparente simplicidad conceptual, ha tomado relevancia la presentaci\'on del formalismo de colectivades desde la teor\'{i}a de la informaci\'on, m\'as concretamente en la forma propuesta por Jaynes \cite{jaynes1982rationale, PhysRev.106.620}. Desde el punto de vista adoptado en ella se incide en las relaciones que se pueden establecer entre el formalismo de la mecánica estadística y el campo de la inferencia estad\'{i}stica. As\'{i}, del mismo modo que la teor\'{i}a de los fen\'omenos cr\'{i}ticos se ha mostrado fructifera en campos diferentes de su concepci\'on original, el enfoque aportado por el formalismo de m\'axima entrop\'{i}a junto con su interpretaci\'on Bayesiana han mostrado aplicaciones en campos sumamente diversos \cite{PhysRevE.56.4217, checa_phdthesis_2012, phillips2006maximum, PhysRevE.76.031102, 0305-4470-36-3-303} tanto desarrollos te\'oricos como ciencias aplicadas. Finalmente, indicar que incluso dentro de un enfoque cl\'asico es posible incidir en diferentes aspectos del formalismo\cite{AttardBook}.}. Como trataremos con sistemas en equilibrio termodinámico nos centraremos en la teoría de colectividades y la teoría de líquidos basada en ella.\\

Conviene realizar algunas matizaciones para desenvolvernos con soltura en un conjunto amplio de resultados previos. El esquema formal de la física estadística esta cons\-ti\-tui\-do por un conjunto de postulados iniciales considerablemente estables, mientras que por el hecho de tratar con propiedades colectivas, de sistemas a menudo complejos, necesita construirse sobre \emph{modelos} y \emph{aproximaciones} debido a la dificultad matemática de conseguir soluciones analíticas. El material empírico se inyecta principalmente en el sistema formal mediante la contrastación con aproximaciones, no con sus postulados básicos\footnote{Un conjunto amplio de descripciones fenomenológicas relevante puede dar lugar a una teoría completa, como puede haber sucedido con los fenómenos de criticalidad que han rebasado de hecho el marco teórico de la mecánica estadística clásica.}, y mediante las diferentes aproximaciones y modelos se construyen \emph{imágenes} de los sistemas físicos bajo estudio. Los contextos de descubrimiento suelen sustentarse en estas \emph{imágenes} sobre las que se plantean determinadas hipótesis. La elección apropiada de las aproximaciones y modelos no se basa exclusivamente en la completitud entendida como exactitud, sino en una combinación adecuada que sopesa también la sencillez de una aproximación suficiente para responder a la cuestión planteada sin oscurecer con una complejidad innecesaria el razonamiento. Así pues una parte relevante del proceso de descubrimiento es la de clarificación del significado real y posible de dichas imágenes físicas así como el papel que ciertas hipótesis juegan realmente dentro del sistema formal.\\

Seguidamente introducimos tanto los aspectos de base formal como un conjunto suficiente de modelos y aproximaciones usadas en la física de líquidos. El lector que los conozca puede voluntariamente pasar a la parte que describe los resultados, el índice le dará una idea de los contenidos tratados.

\section{Introducción a la teoría de colectividades}

La física estadística se caracteriza por explicar leyes y obtener propiedades a una cierta escala como emergentes de leyes y relaciones presentes en escalas menores. Así en la mecánica estadística aparecen inicialmente dos niveles de descripción diferentes, uno macroscópico que enlaza con la termodinámica y otro microscópico que enlaza con la mecánica
  y la necesidad de explicar la física a estas escalas origina la idiosincrasia particular de esta disciplina. El método clásico de construir este enlace es mediante la mecánica estadística de colectividades que se articula desde concepto de estado en ambos niveles: microestado y macroestado\footnote{En el caso microscópico se denomina \textit{microestado} y en el caso macroscópico \textit{macroestado}.}. Su formalismo requiere, dado un macroestado, identificar cuales serán los \emph{microestados} compatibles con él, y sobre ellos aplicar métodos estadísticos. Para esto habremos de asignar \emph{pesos estadísticos} a los microestados, lo que usualmente se realiza a partir de propiedades dinámicas de transición en\-tre ellos.\\

\paragraph*{Escala macroscópica}

Sus magnitudes físicas suelen ser observables, aunque algunas tienen que ver con propiedades del conjunto de estados microscópicos accesibles al sistema. Se clasifican, incidiendo en que son propiedades de estado y no de proceso, en variables de estado y funciones de estado según la forma funcional con que estemos tratando. Como determinadas relaciones funcionales tienen la deseable propiedad de poder proporcionar toda la información termodinámica a partir de ellas se denominan ecuaciones fundamentales\footnote{De esta manera la entropía en función sus \textit{variables naturales}, $S=S(E,V,N)$ siendo energía interna, E, volumen V, y número de partículas N contiene toda la termodinámica del sistema y análogamente sucede, por ejemplo, con la energía libre de Helmholtz $F=F(T,V,N)$  en las variables T,V y N.} e involucran funciones llamadas \textit{potenciales termodinámicos}. Normalmente la relación funcional entre tres variables termodinámicas observables se denomina \textit{ecuación de estado}, como por ejemplo $P=P(T,V)$, y las derivadas segundas establecen las funciones respuesta como la compresibilidad isoterma o el calor específico. El propio formalismo termodinámico permite deducir algunas relaciones formales entre funciones respuesta de un modo totalmente general. Los postulados de la termodinámica afirman pues las condiciones de la existencia de los potenciales termodinámicos, estableciendo además propiedades matemáticas y su interpretación física, por ejemplo la contenida en las condiciones de estabilidad. Un concepto clave, como vemos, es el de estado de equilibrio, mientras que los conceptos de fase, sistema homogéneo o mezcla, pueden ser definidos convenientemente a este nivel\cite{CallenBook}. Las propiedades que dan lugar a la coexistencia de fases también son establecidas a este nivel macroscópico, pero la forma funcional concreta de las ecuaciones de estado o los potenciales termodinámicos necesita información experimental o microscópica.\\

Un paso intermedio puede ser establecido mediante una teoría termodinámica que permita fluctuaciones en sus variables de estado en la que el concepto de estado es algo más amplio y sus postulados algo más generales\cite{PhysRev.83.1231,kleinTisza,Landau_B80,BOOK-TheorySimpleLiquids}, y en la que por, y solo por, sus características formales se puede incluir la teoría de ondas capilares obviamente sin una interpretación microscópica.

\paragraph*{Escala microscópica}

En nuestro caso nos restringiremos en todo momento a la física estadística basada en la mecánica clásica\footnote{Las simplificaciones involucradas en el tratamiento que aquí se hace son descritas con detalle por \textit{J.A. Barker y D. Henderson}\cite{RevModPhys.48.587}, donde además se comentan diferentes aspectos de las interacciones y las implicaciones que tales simplificaciones suponen.} de \emph{sistemas simples}\footnote{En todo momento supondremos partículas idénticas y esféricas sin grados internos de libertad, aunque muchas de las afirmaciones y metodologías utilizadas poseen un carácter más general, además se suele considerar que en sistema simple esta adecuadamente descrito por interacciones que pueden ser escritas en términos de interacciones entre dos partículas\cite{BOOK-TheorySimpleLiquids}.} donde para un sistema aislado de N partículas el microestado del sistema esta completamente determinando especificando el conjunto de posiciones $\vec{r}^{N}$ y momentos $\vec{p}^{N}$ de todas las partículas, por tanto nuestros microestados pertenecen al espacio de las fases $\Gamma$ dado por $\Gamma=p^{N}q^{N}$ mientras que la dinámica de transiciones entre microestados viene determinada por el hamiltoniano $\mathcal{H}_{N}$ del sistema que, cumpliendo las ecuaciones de Hamilton, podemos escribir en la forma,
\begin{equation}
\mathcal{H}_{N}(\vec{r}^{N},\vec{p}^{N})=\mathcal{K}_{N}(\vec{p}^{N})+\Phi_{N}(\vec{r}^{N})+\mathcal{V}_{N}(\vec{r}^{N})=\mathcal{K}_{N}(\vec{p}^{N})+\mathcal{U}_{N}(\vec{r}^{N})
\label{eqn:Hamiltoniano}
\end{equation}
donde la parte cinética es $\mathcal{K}_{N}$, la interacción entre partículas $\Phi_{N}(\vec{r}^{N})$ y un potencial externo que viene dado por $\mathcal{V}_{N}(\vec{r}^{N})$. La parte no cinética la denominamos con $\mathcal{U}_{N}(\vec{r}^{N})$\footnote{Hemos restringido las posibles dependencias funcionales de los potenciales de interacción del microestado únicamente a las configuraciones.}. En el caso de que podamos describir la interacción entre partículas como interacciones a pares tendremos,
\begin{equation}
\Phi_{N}(\vec{r}^{N})=\frac{1}{2}\sum_{i<j}^{N}\phi(\vec{r}_{i},\vec{r}_{j})
\label{eqn:InteraccionApares}
\end{equation}

\subsection{Colectividad Microcanónica}
El hecho fijar ciertos observables macroscópicos en el sistema se revela a un nivel microscópico mediante una restricción de la región del espacio de las fases accesible. En el caso de $\mathcal{H}_{N}$ que cumple las ecuaciones de Hamilton para un sistema aislado este posee un valor constante que representa su energía, lo que restringe los posibles microestados acordes con la limitación macroscópica a una región concreta del espacio de la fases determinada por $\left\lbrace \alpha\in\Gamma / \mathcal{H}_{N}(\alpha)=E\right\rbrace $\footnote{Otras ligaduras restringirían aun más el espacio de las fases accesible al sistema, cada una delimita más una \textit{hipersuperficie} en el estado de las fases donde se desarrolla la transición entre microestados.}. Esto nos permite definir una \emph{densidad de probabilidad} en el espacio de las fases para un sistema aislado\footnote{En el caso de los sistemas que vamos a estudiar posee tres parámetros extensivos E,V,N que determinan un \emph{macroestado} concreto que queda definido entonces como el conjunto de microestados compatibles con estos parámetros macroscópicos fijados. Dos macroestados son por tanto conjuntos disjuntos de microestados.} junto con una \emph{función de partición} obtenida de normalizar la densidad de probabilidad y a partir de la cual se puede extraer toda la información termodinámica del sistema. La importancia de la densidad de probabilidad radica en que con ella pueden obtenerse los valores promedios correspondientes a la escala macroscópica de funciones dinámicas definidas en la escala microscópica. La densidad de probabilidad, la función de partición y su relación con la termodinámica vienen entonces determinadas respectivamente por,\index{Colectivo!Microcanónico}
\begin{equation}
\mathcal{P}[\Gamma|E,V,N]=\frac{\delta(\mathcal{H}_{N}(\Gamma)-E)}{N!h^{3N}Z(E,V,N)}
\end{equation}
\begin{equation}
Z(E,V,N)=\frac{1}{N!h^{3N}}\int d\Gamma \delta(\mathcal{H}_{N}(\Gamma)-E)
\end{equation}
\begin{equation}
S(E,V,N)=k_{b}lnZ(E,V,N)
\end{equation}
quedando definida la colectividad \emph{microcanónica} que como vemos depende de tres parámetros que definen el macroestado del sistema\footnote{Son tres parámetros extensivos, los parámetros intensivos conjugados son obtenidos por diferenciación de la entropía respecto de E, V y N, obteniendo T, P y $\mu$.}: E, V y N. La entropía se determina por la última de las ecuaciones conocida como \textit{ecuación de Boltzmann}. \\

Nuestro objetivo es introducir colectividades diferentes de la microcanónica que permitan situaciones más generales que la de un sistema aislado y que serán representativas de estados de equilibrio de un sistema con otro sistema mayor que se suele denominar baño, \textit{reservoir} o fuente. El \textit{reservoir} se caracteriza porque sus propiedades se mantienen constantes y determinan de hecho las \emph{condiciones de equilibrio}, esto implica que los macroestados de nuestro sistema, ahora no aislado, vienen representados mediante otros parámetros de estado relacionados con las formas de contacto del sistema y el \textit{reservoir}. Nos interesa construir los colectivos canónico y macrocanónico y especialmente este último por ser particularmente adecuado para obtener resultados teóricos y sobre él introduciremos la teoría del funcional de la densidad. Veamos inicialmente la forma general del método para construir estas colectividades que particularizaremos a continuación.
\subsection{Sistemas en contacto con un \textit{reservoir}}

Dado un sistema aislado dividido en dos subsistemas caracterizados por los subíndices 1 y 2, tendremos que $N_{1}+N_{2}=N$ y $V_{1}+V_{2}=V$ y podemos escribir,
\begin{equation}
\mathcal{H}_{N}(\vec{r}^{N},\vec{p}^{N})=\mathcal{H}_{N_{1}}(\vec{r}^{N_{1}},\vec{p}^{N_{1}})+\mathcal{H}_{N_{2}}(\vec{r}^{N_{2}},\vec{p}^{N_{2}})+\mathcal{H}_{N_{1}N_{2}}(\vec{r}^{N},\vec{p}^{N})
\label{eqn:HamiltonianoDossubsistemas}
\end{equation}

Para sistemas macroscópicos podemos considerar la energía de interacción $\mathcal{H}_{N_{1}N_{2}}$ en\-tre ambos subsistemas despreciable frente a las energías de cada uno lo que permite escribir $\mathcal{H}_{N}=\mathcal{H}_{N_{1}}+\mathcal{H}_{N_{2}}$ y $E_{1}+E_{2}=E$. El reparto de energía, volumen o número de partículas dependerá del tipo de contacto que permitamos entre los dos subsistemas pero la separación de grados de libertad la suponemos posible y en consecuencia es posible restringir la región del espacio de las fases que define el sistema total para valores fijos de uno o varios de los parámetros del macroestado que definen al subsistema 1. Tenemos por tanto funciones de partición y densidades de probabilidad del sistema total aislado para distribuciones concretas de los parámetros extensivos sujetos a intercambio entre ambos subsistemas.\\

El siguiente paso conceptual es suponer que, por ser el subsistema 2 un \textit{reservoir}, para todas estas diferentes distribuciones de valores no cambian los parámetros intensivos conjugados del reservoir y es posible definir tanto una función de distribución como una densidad de probabilidad del subsistema 1 en contacto con un reservoir del que solo nos interesan las propiedades intensivas correspondientes sin importar nada más específico de su naturaleza. Detallamos esto de modo general.\\

Partimos de un sistema aislado compuesto de dos subsistemas en contacto. El macroestado del sistema total queda descrito por los parámetros,
\begin{equation}
\left\lbrace A^{t}_{1}=A^{r}_{1}+A^{s}_{1},...,A^{t}_{l}=A^{r}_{l}+A^{s}_{l}\right\rbrace \end{equation}

Consideramos uno de los subsistemas, que llamo \textit{reservoir}, lo suficientemente grande para que podamos afirmar que sus propiedades intensivas $\left\lbrace a^{r}_{1},...,a^{r}_{l}\right\rbrace $ permanecen constantes\footnote{Suponemos también que la región de contacto es lo suficientemente pequeña para que no juegue un papel en nuestro análisis.}, y el otro será el subsistema cuyas propiedades nos interesa conocer y parte de sus variables de estado (macroestado) pueden variar de acuerdo con el tipo de contacto establecido entre ambos sistemas. Sean este subconjunto el dado por $A^{s}_{1},...,A^{s}_{k}$. De modo general tendremos\footnote{No hay posibilidad de confusión con los superíndices por tanto los suprimimos.}
una densidad de probabilidad
\begin{equation}
\mathcal{P}[A_{1},...,A_{k}|a_{1},..,a_{k},A_{k+1},..,A_{l},T]
\end{equation}
para los parámetros $A_{1},...,A_{k}$ y su normalización determina la función de partición,
\begin{equation}
Z(a_{1},..,a_{k},A_{k+1},..,A_{l},T)
\end{equation}
La propia densidad de probabilidad $\mathcal{P}[A_{1},...,A_{k}|a_{1},..,a_{k},A_{k+1},..,A_{l},T]$ permite obtener unos valores medios $\langle A_{i}\rangle $ así como los valores de las fluctuaciones\footnote{O de las correlaciones.} $\Delta A_{i}=\langle A_{i}^{2}\rangle -\langle A_{i}\rangle ^2$ para los diferentes $A_{1},...,A_{k}$, que como veremos son relacionables con magnitudes con significado físico relevante, tanto a un nivel termodinámico (magnitudes de equilibrio) como un nivel estructural (funciones de distribución). El enlace a nivel estructural se realiza desde la mecánica relacionando la densidad de probabilidad anterior con una densidad de probabilidad en el espacio de las fases $\mathcal{P}[\Gamma|a_{1},..,a_{k},A_{k+1},..,A_{l},T]$ que tomamos como definición para el enlace entre la dinámica y las propiedades de equilibrio. De este modo las propiedades medias\footnote{Y las propiedades medidas vía hipótesis ergódica.} de una propiedad dinámica dada por $f(\Gamma)$ quedan expresadas como,
\begin{equation}
<f>=\int d\Gamma f(\Gamma)\mathcal{P}(\Gamma|a_{1},..,a_{k},A_{k+1},..,A_{l},T)
\end{equation}
Una vez introducido el procedimiento general es aplicable directamente a los colectivos canónico y macrocanónico, lo que permite que únicamente recuperemos los resultados sin demostración en las siguientes secciones\footnote{El esquema descrito arriba no es completamente general pueden existir colectivos que requieran más cuidado, como por ejemplo, el isentálpico, pero las ideas generales son compartidas.}.

\subsubsection{Colectividad Canónica}
\label{sec:canonico}
\index{Colectivo!Canónico}
Representa un sistema en contacto con un baño térmico, de que modo que puede intercambiar energía E con este manteniendo V y N constantes. El baño queda definido por la variable intensiva correspondiente, su temperatura T, que determina las condiciones de equilibrio\footnote{Podemos relacionar tanto la función de distribución como la función de partición canónicas con las análogas de un sistema aislado caracterizado por los parámetros extensión E,V,N.  Mediante las relaciones, 
\begin{equation}
\mathcal{P}[E|T,V,N]=\frac{e^{S(E,V,N)}e^{-\beta E}}{Z(T,V,N)}
\end{equation}
\begin{equation}
Z(E,V,N)=\int dE e^{-\beta E}\int d\Gamma \delta(\mathcal{H}_{N}(\Gamma)-E)
\end{equation}
}. Este sistema posee una función de partición\footnote{La introducción del factor $N!$ en el denominador proviene de la indistinguibilidad de partículas mientras que el factor $h^{3N}$ donde h es la constante de Planck nace de argumentos que permiten relacionar el número de estados calculado en física cuántica con el límite clásico. Su presencia en las expresiones puede predecirse mediante argumentos dimensionales pero su naturaleza y determinación requiere del enlace con la mecánica cuántica, desde el punto de vista de la mecánica clásica podemos prescindir de la determinación de dicha constante mediante un reescalado del cero de potenciales químicos.},
\begin{equation}
Z(N,V,T)=\frac{1}{N!h^{3N}}\int d\vec{p}^{N}\int d\vec{r}^{N}e^{-\beta\mathcal{H}_{N}(\vec{r}^{N},\vec{p}^{N})}
\end{equation}
La distribución de probabilidad en el espacio de las fases para este colectivo se denomina \emph{distribución canónica} y queda expresada de la siguiente manera,
\begin{equation}
\mathcal{P}(\vec{r}^{N},\vec{p}^{N}|N,V,T)=\frac{1}{N!h^{3N}Z(N,V,T)}e^{-\beta\mathcal{H}_{N}(\vec{r}^{N},\vec{p}^{N})}
\label{eqn:DensidadProbabilidadCanonica}
\end{equation}
y por tanto describe la probabilidad de tener N partículas indistinguibles en posiciones $\vec{r}_{1},...,\vec{r}_{N}$ con momentos $\vec{p}_{1},...,\vec{p}_{N}$.\\

La ecuación que relaciona en este caso los dos niveles de descripción es,
\begin{equation}
lnZ(N,V,T)=\beta F(N,V,T)
\end{equation}
donde F(N,V,T) es la energía libre de Helmholtz en sus variables naturales con lo que toda la termodinámica se obtiene de la función de partición. La energía libre introducida verifica la \textit{ecuación fundamental},
\begin{equation}
dF(N,V,T)=-S(N,V,T)dT+P(N,V,T)dV-\mu(N,V,T)dN
\end{equation}
donde las magnitudes entropía S,  presión P y potencial químico $\mu$ son determinadas mediante diferenciación. En particular podemos obtener la \textit{ecuación de estado} $P(T,\rho)$ que relaciona tres parámetros del macroestado mensurables en el laboratorio. Las derivadas segundas determinan la convexidad de la energía libre y la estabilidad del macroestado frente a fluctuaciones, en este colectivo, en la energía.\\

Tratamos con sistemas clásicos y simples luego la forma de $\mathcal{H}_{N}$ permite factorizar la parte cinética tanto en la función de partición como en la densidad de probabilidad.
\begin{equation}
Z(N,V,T)=\frac{1}{N!\Lambda^{3N}}\int d\vec{r}^{N}e^{-\beta\mathcal{U}_{N}(\vec{r}^{N})}=\frac{Z^{id}}{V^{N}}\int d\vec{r}^{N}e^{-\beta\mathcal{U}_{N}(\vec{r}^{N})}
\end{equation}
esta última integral se denomina \emph{integral de configuración}. Esto separa formalmente la función de partición en el producto de una parte ideal (cinética) y una parte de \emph{exceso} sobre la parte ideal que da cuenta de las interacciones y es donde se introducen las diferentes aproximaciones. La separación entre parte ideal y de exceso anterior queda expresada entonces en la energía libre como una suma,
\begin{equation}
F(N,V,T)=F_{id}(N,V,T)+F_{ex}(N,V,T)
\end{equation}

Uno puede definir, de modo análogo a la función de partición, una \emph{distribución de probabilidad configuracional} integrando en momentos de modo que,
\begin{equation}
\mathcal{P}(\vec{r}^{N}|N,V,T)=\frac{1}{Q(N,V,T)}e^{-\beta\mathcal{U}_{N}(\vec{r}^{N})}
\end{equation}

Tanto esta ecuación como la anterior (\ref{eqn:DensidadProbabilidadCanonica}) contienen información completa de la distribución de todas las partículas en el sistema, a pesar de ello resulta preferible trabajar con otras funciones de distribución que aunque contienen una información parcial de la configuración del sistema  permiten una caracterización adecuada de ciertas propiedades del sistema tanto a nivel microscópico como su enlace a un nivel macroscópico. La ventaja de dichas funciones es que pueden ser determinadas directa o indirectamente en el laboratorio lo que permite comparar diferentes aproximaciones mientras que su relación con magnitudes de equilibrio termodinámicas permite a su vez evaluar la consistencia de dichas aproximaciones. Finalmente en ellas se reflejan propiedades estructurales del sistema y por tanto nos serán de gran utilidad para la comprensión y caracterización de las diferentes fases tanto fluidas como sólidas.\\

La \emph{distribución densidad de n-partículas},
\begin{equation}
\rho_{N}^{(n)}(\vec{r}^{n})=\frac{N!}{(N-n)!}\frac{1}{Q(N,V,T)}\int d\vec{r}^{n+1}...d\vec{r}^{N}e^{-\beta\mathcal{U}_{N}(\vec{r}^{N})}
\end{equation}
cuya normalización viene dada por,
\begin{equation}
\int d\vec{r}^{n}\rho_{N}^{(n)}(\vec{r}^{n})=\frac{N!}{(N-n)!}
\end{equation}
representa la probabilidad de encontrar simultáneamente $n$ partículas arbitrarias en las posiciones caracterizadas por $\vec{r}_{1},...,\vec{r}_{n}=\vec{r}^{n}$ independientemente la posición del resto de las $N-n$ partículas.\\

Conviene diferenciar esta familia de funciones de las \emph{funciones de distribución de n-partículas}, estas últimas están normalizadas a la unidad. De este modo escribiríamos,
\begin{equation}
\wp^{(1)}(\vec{u}|N,V,T)=\left<\delta(\vec{u}-\vec{r}_{1})\right>=\int d\Gamma\mathcal{P}(\Gamma;N,V,T)\delta(\vec{u}-\vec{r}_{1})
\end{equation}
mientras que la distribución densidad de una partícula, escrita a partir de una función dinámica se expresará como\footnote{Observar que el sistema refleja en la densidad las simetrías en del hamiltoniano, de modo que si la energía potencial es invariante frente a translaciones lleva a un fluido uniforme. Cuando un campo externo a un cuerpo rompe esa simetría del hamiltoniano, la densidad correspondiente debería mostrar esta ruptura dando lugar aun fluido no homogéneo.},
\begin{equation}
\rho_{N}^{(1)}(\vec{u})=\left<\hat{\rho}_{N}^{(1)}(\vec{u})\right>=\left<\sum_{i=1}^{N}\delta(\vec{u}-\vec{r}_{i})\right>
\label{eqn:operadordensidad}
\end{equation}
y en el caso de la distribución densidad de dos partículas,
\begin{equation}
\rho_{N}^{(2)}(\vec{u},\vec{v})=\left<\hat{\rho}_{N}^{(2)}(\vec{u},\vec{v})\right>=\left<\sum^{N}_{i=0}\sum^{N}_{j\neq i}\delta(\vec{s}-\vec{u}_{i})\delta(\vec{v}-\vec{r}_{j})\right>
\end{equation}

\subsubsection{Colectividad Macrocanónica}
\label{sec:macrocanonico}
\index{Colectivo!Macrocanónico}
\index{Colectivo!Gran-Canónico|see{Colectivo!Macrocanónico}}
Representa un sistema en contacto con un baño de partículas de modo que puede variar su energía E y partículas N, manteniendo V constante. El baño queda definido por su temperatura y potencial químico que determinan las condiciones de equilibrio mutuo\footnote{Al igual que hicimos con la colectividad canónica y la microcanónica se pueden relacionar las funciones de partición canónica $Z(T,V,N)$ y $Z(T,V,\mu)$. Esta relación se extendería en su aplicación al resto de definiciones y por tanto podemos relacionar también las funciones de distribución. Es interesante notar que si tuviéramos un sistema multicomponente habríamos de tener para cada componente i un $\mu_{i}$ que sin embargo no sería una propiedad intensiva solo de i sino que depende de todos los parámetros que determinan el sistema completo.}.\\

La función de partición en este colectivo se define como,
\begin{align}
Z(\mu,V,T) &=\sum_{N=0}^{\infty}e^{\beta\mu N}Z(N,V,T)=\sum_{N=0}^{\infty}\frac{e^{\beta\mu N}}{\Lambda^{3N}N!}Q(N,V,T)\\
&=\sum_{N=0}^{\infty}\frac{e^{\beta\mu N}}{h^{3N}}\int d\vec{q}^{N}d\vec{r}^{N}\frac{1}{N!}e^{-\beta\mathcal{H}_{N}(\vec{r}^{N},\vec{p}^{N})} \notag
\end{align}
que de modo análogo cumple que,
\begin{equation}
lnZ(\mu,V,T)=\beta \Omega(\mu,V,T)
\end{equation}
y
\begin{equation}
d\Omega(\mu,V,T)=-S(\mu,V,T)dT+P(\mu,V,T)dV-N(\mu,V,T)d\mu(N,V,T)
\end{equation}
donde $\Omega(\mu,V,T)$ es la energía libre macrocanónica también en sus variables naturales.\\

Mientras que la densidad de probabilidad se escribiría como,
\begin{equation}
\mathcal{P}(\vec{r}^{N},\vec{p}^{N},N|\mu,V,T)=\frac{e^{-\beta\mu N}e^{-\beta\mathcal{H}_{N}(\vec{r}^{N},\vec{p}^{N})}}{N!h^{3N}Z(\mu,V,T)}
\end{equation}

Las funciones \textit{distribución de densidad de n-partículas} quedan definidas, en completa analogía con las del colectivo canónico, mediante,
\begin{equation}
\rho_{\mu}^{(n)}(\vec{s}^{n})=\frac{1}{Z(\mu,V,T)}\sum_{N=n}^{\infty}\frac{N!}{(N-n)!}e^{\beta\mu N}\int d\vec{r}^{N}d\vec{p}^{N}e^{-\beta\mathcal{H}_{N}(\vec{r}^{N},\vec{p}^{N})}\vec{\delta}^{(n)}(\vec{r}^{n}-\vec{s}^{n})
\label{eqn:familiaDistribDensidadMacro}
\end{equation}
esta función posee dos límites asintóticos bien definidos, a grandes distancias las partículas dejan de estar correlacionadas y la función de distribución de densidad de n-partículas factoriza en el producto de funciones de distribución de orden menor y de modo natural surge otra jerarquía que también se suelen denominar \textit{funciones de distribución de n-partículas} como,
\begin{equation}
g^{(n)}(\vec{r}^{n})=\frac{\rho^{(n)}(\vec{r}^{n})}{\rho^{(1)}(\vec{r}_{1})...\rho^{(1)}(\vec{r}_{n})}
\end{equation}
y que es aplicable también al colectivo canónico. Esperamos además que a separaciones del orden del tamaño de las partículas o menores las funciones obtenidas cumplan ser nulas.\\

Como en el caso de la colectividad canónica escribimos la normalización de las dos primeras funciones de la familia $\rho_{\mu}^{(n)}$,
\begin{equation}
\int d\vec{r}\rho_{\mu}^{(1)}(\vec{r})=<N>_{\mu}
\end{equation}
\begin{equation}
\int d\vec{r}d\vec{s}\rho_{\mu}^{(2)}(\vec{r},\vec{s})=<N^{2}>_{\mu}-<N>_{\mu}
\end{equation}

esta última expresión induce otra definición que surgirá más adelante de modo natural en el contexto de la teoría del funcional de la densidad. 
\begin{equation}
\underbrace{G^{(2)}(\vec{r},\vec{s})}_{Corr.\,densidad-densidad}=\overbrace{\underbrace{\rho_{\mu}^{(2)}(\vec{r},\vec{s})}_{Distr.\,orden\, 2}-\rho_{\mu}^{(1)}(\vec{r})\rho_{\mu}^{(1)}(\vec{s})}^{F.\,corr.\,total\equiv h(\vec{r}_{1},\vec{r}_{2})\rho(\vec{r}_{1})\rho(\vec{r}_{2})}+\underbrace{\rho_{\mu}^{(1)}(\vec{r})\delta(\vec{r}-\vec{s})}_{autocorrelaci\acute{o}n}
\label{eqn:laFuncionG2}
\end{equation}
donde usando la normalización de $\rho_{\mu}^{(1)}$ vemos que su integración da lugar a las fluctuaciones en el colectivo macrocanónico en el número de partículas,
\begin{equation}
\int d\vec{r}d\vec{s}G^{(2)}(\vec{r},\vec{s})=<N^{2}>_{\mu}-<N>^{2}_{\mu}\equiv<(\Delta N)^{2}>
\end{equation}

Esto nos permite establecer una conexión entre una propiedad estructural $G^{(2)}(\vec{r}_{1},\vec{r}_{2})$ y una propiedad termodinámica, la \textit{compresibilidad isoterma}, $\kappa_{T}=\rho^{-1}\frac{\partial \rho}{\partial P}|_{T}$ ya que,

\begin{equation}
<N^{2}>_{\mu}-<N>^{2}_{\mu}=<N>\rho\beta^{-1}\kappa_{T}
\label{eqn:compresibilidad1}
\end{equation}

El hecho de que las funciones de distribución en el caso ideal factorizan en el producto de funciones de distribución de una partícula debido a la ausencia de interacciones motiva definir una \emph{función de correlación total} mediante,
\begin{equation}
h^{(2)}(\vec{r},\vec{s})\rho_{\mu}^{(1)}(\vec{r})\rho_{\mu}^{(1)}(\vec{s})=\rho_{\mu}^{(2)}(\vec{r},\vec{s})-\rho_{\mu}^{(1)}(\vec{r})\rho_{\mu}^{(1)}(\vec{s})
\end{equation}
a la función de distribución densidad de dos partículas le hemos substraído la parte ideal, de modo que en un sistema no-homogéneo pero ideal aunque $\rho^{(2)}(\vec{r},\vec{s})$ pueda presentar estructura la función $h^{(2)}(\vec{r},\vec{s})$ no lo hará.\\

Resulta conveniente para desarrollos posteriores escribir los valores medios en este colectivo del potencial externo y del potencial intermolecular en el caso de potenciales a pares. A saber,
\begin{equation}
\left < V_{ext}(\vec{r}^{N})\right>=\left < \sum_{i=1}^{N}v_{ext}(\vec{r}_{i})\right>=\int d\vec{r}\rho^{(1)}(\vec{r})v_{ext}(\vec{r})
\end{equation}
\begin{equation}
\left < \Phi_{N}(\vec{r}^{N})\right>=\left < \frac{1}{2}\sum_{i<j}\phi(\vec{r}_{i},\vec{r}_{j})\right>=\frac{1}{2}\int d\vec{r}_{1}d\vec{r}_{2}\rho^{(2)}(\vec{r}_{1},\vec{r}_{2})\phi(\vec{r}_{1},\vec{r}_{2})
\label{eqn:promedioPHIN}
\end{equation}

En el caso de potenciales a pares podemos establecer una relación integrodiferencial entre los miembros de la familia $\rho^{(n)}$  de orden n y los de orden n+1, llamada jerarquía\textit{ Yvon-Born-Green}, que para n=1 se expresa como,
\begin{equation}
\nabla_{1}\rho(\vec{r}_{1})=\rho(\vec{r}_{1})\nabla_{1}v(\vec{r}_{1})+\int \rho^{(2)}(\vec{r}_{1},\vec{r}_{2})\nabla_{1}\phi(r_{12})d\vec{r}_{2}
\label{eqn:YBGorden1}
\end{equation}

\subsection{Relaciones entre termodinámica y estructura} 
\label{sec:rutasTermoEstructura}
Para incluir información experimental relacionada con las funciones de distribución, hacemos uso de una definición más, comenzamos con la transformada Fourier de la densidad,
\begin{equation}
\rho^{(1)}_{\vec{k}}=\int d\vec{r}e^{i\vec{k}\vec{r}}\rho(\vec{r}) 
\end{equation}
de modo que su correlación define el \textit{factor de estructura} $S(\vec{k})$
\begin{equation}
S(\vec{k})=\frac{1}{N}\left\langle \rho^{(1)}_{\vec{k}}\rho^{(1)}_{\vec{-k}} \right\rangle=1+\frac{1}{N}\int d\vec{r}\int d\vec{s} e^{i\vec{k}(\vec{r}-\vec{s})}\rho^{(2)}_{N}(\vec{r},\vec{s})
\label{eqn:factorestructura}
\end{equation}
la segunda igualdad de ec.(\ref{eqn:factorestructura}) establece la relación entre el factor de estructura y la función de distribución $g^{(2)}$. El factor de estructura puede determinarse mediante experimentos\footnote{Véase \S\ref{sec:experimentosReflectividad} para el caso de una interfase.} de difracción de rayos X y contiene información de las fluctuaciones contenidas en el sistema, por ejemplo $S(0)$ se relaciona vía ec. (\ref{eqn:compresibilidad1}) con la compresibilidad isoterma. En en el caso homogéneo se simplifican a $S(|\vec{k}|)$ y $g(|\vec{r}_{12}|)$ que se denomina \textit{función de distribución radial}\footnote{De modo general haciendo uso de las definiciones introducidas $S(\vec{k})=1+\rho\delta(\vec{k})+\rho\hat{h}(\vec{k})$, luego aparece relacionado directamente con la transformada de Fourier de $G^{(2)}$.} y por tanto tenemos acceso experimental a esta última.\\

La importancia de la función $\rho^{(2)}(\vec{r}_{1},\vec{r}_{2})$ se debe además a otro hecho, en un sistema donde las interacciones son únicamente entre pares de partículas, véase ec. (\ref{eqn:InteraccionApares}), las propiedades termodinámicas pueden ser expresadas mediante integrales sobre la función de distribución de dos partículas, y cada relación constituye una \textit{ruta} hacia la termodinámica desde la estructura.\\

Las tres ecuaciones más relevantes en el caso de potenciales a pares son la \textit{ecuación de la energía}, la \textit{ecuación del virial} y la \textit{ecuación de la compresibilidad}. En el primer caso la energía del sistema es determinada como el promedio en la colectividad de la función dinámica $\mathcal{H}_{N}$, véase ec.(\ref{eqn:Hamiltoniano}), que nos lleva a
\begin{equation}
E(T,V,N)=E_{id}+\frac{1}{2}\int d\vec{r}_{1}d\vec{r}_{2}\rho^{(2)}(\vec{r}_{1},\vec{r}_{2})\phi(\vec{r}_{1},\vec{r}_{2})+\int d\vec{r}_{1}\rho^{(1)}(\vec{r}_{1})V_{ext}(\vec{r}_{1})
\label{eqn:EcuacionEnergia}
\end{equation}\index{Ecuación!de la Energía}\index{Virial!Ecuación}
para la ecuación del virial,
\begin{equation}
\frac{\beta P(T,V,N)}{\rho}=1-\beta\frac{1}{3N}\int d\vec{r}_{1}d\vec{r}_{2}\rho^{(2)}(\vec{r}_{1},\vec{r}_{2})\vec{r}_{1}\cdot\bigtriangledown_{1}\phi(\vec{r}_{1},\vec{r}_{2})+\int d\vec{r}_{1}\rho^{(1)}(\vec{r}_{1}) \vec{r}_{1}\cdot\bigtriangledown_{1}V_{ext}(\vec{r}_{1})
\label{eqn:EcuacionVirial}
\end{equation}
que determina la presión del sistema esencialmente a partir del promedio de la función dinámica $\beta\sum_{i}\vec{r}_{i}\cdot\bigtriangledown_{i}\mathcal{U}_{N}$. La relación de la compresibilidad ya se cito en ec.(\ref{eqn:compresibilidad1}) y queda expresada en un sistema uniforme como,
\begin{equation}
\kappa_{T}\beta^{-1}=1+\rho\int  d\vec{r}h^{(2)}(\vec{r})
\label{eqn:EcuacionCompresibilidad}
\end{equation}\index{Compresibilidad!Ecuación}

La dificultad práctica de la aplicación de estas relaciones, no solo esta en determinar de un modo más o menos adecuado la estructura del sistema dada por g(r), sino que el modo de obtener esta función permita un cálculo razonablemente consistente por las diferentes \textit{rutas} a la termodinámica, cuestión deseable que una aproximación \textit{a priori} no tiene porque cumplir.\\

Se ha planteado de modo somero la estructura de la mecánica estadística de colectividades. Resolver la función de partición para modelos concretos es una tarea complicada y son contados los casos en que se conoce una solución analítica, por otra parte el objetivo de la teoría de líquidos es comprender porque unas fases son estables frente a otras en determinadas condiciones de presión, temperatura o densidad, analizar la estructura de las fases correspondientes así como propiedades dinámicas de estas, finalmente relacionarlas con las propiedades a un nivel molecular, es decir, como las interacciones y las propiedades geométricas de las moléculas influyen en su comportamiento colectivo. Conviene por tanto, indagar en las condiciones que son necesarias para reproducir fases estables, y si el propio formalismo puede explicar la fenomenología que tiene lugar una transición de fase.

    \vspace{0.2cm}

\section{Transiciones de fase en sistemas simples}

Supongamos resuelto el problema de determinar a partir de los detalles microscópicos de un modelo concreto la función de partición macrocanónica $\Omega(T,V,\mu)$ 
descrita. Aun no hemos respondido a la cuestión de  si los resultados obtenidos son o no consistentes con los postulados de la termodinámica, y si lo con otras colectividades estadísticas. En general la demostración de la consistencia\cite{BalescuBook} es posible sobre sobre hipótesis muy generales\footnote{Se exige al potencial entre partículas $\phi(\vec{r}_{ij})$ ser,
\begin{itemize}
\item \emph{Estable}, $\sum_{i<j}^{N}\phi(\vec{r}_{ij})\geqslant-BN$ con $B\in\mathbb{R}^{+}$
\item \emph{Temperado}, existen $\alpha,\beta>0$ tal que $|\phi(\vec{r}_{ij})|<B|\vec{r}_{ij}|^{-d-\alpha}$ para $|\vec{r}_{ij}|>\beta$
\end{itemize}
Los potenciales de corto alcance y fuertemente repulsivos ejemplarizados por un Lennard-Jonnes cumplen estas propiedades. Con todo hay notables excepciones como el potencial de Coulomb que exige un análisis diferente.} para $\mathcal{H}_{N}$
y se obtiene la deseable propiedad de convexidad de la energía libre resultante de la aplicación del formalismo\cite{BOOK-NigelGoldenfeld}.\\

En la descripción termodinámica de un sistema simple de un componente las propiedades de convexidad tienen su expresión condensada en las condiciones de estabilidad termodinámica, que se concretan en dos expresiones
\begin{subequations}
\begin{equation}
\left( \frac{\partial u}{\partial T}\right)_{V}= c_{v}>0
\end{equation} 
\begin{equation}
-\left( \frac{\partial P}{\partial v}\right)_{T}=\frac{\rho}{\kappa_{T}}>0
\label{eqn:estabilidad2}
\end{equation}
\label{eqn.condestabilidad}
\end{subequations}
y cuyo contenido físico esta expresado en la ley de Le-Chatelier\cite{CallenBook} y si se verifican\footnote{Es una condición necesaria pero no suficiente para tener estados de equilibro.} la descripción es apropiada para un sistema simple monocomponente. Para los valores de los parámetros del macroestado en que no se verifica ec. (\ref{eqn.condestabilidad}) el sistema no puede ser homogéneo\footnote{No puede mantenerse homogéneo frente a perturbaciones externas o fluctuaciones internas.} y se divide en dos o más fases (heterogéneo). Existe además un caso límite en que la relación (\ref{eqn:estabilidad2}) anterior se anula y la compresibilidad isoterma diverge que tiene relación con las fases críticas\footnote{La denominación de funciones respuesta en este contexto viene del \emph{principio de Le Chatelier} que determina como responde el sistema bien a un agente externo (cambio de un parámetro del reservoir por ejemplo) o de modo equivalente como responde el resto del sistema a una fluctuación en una pequeña región del sistema. En general los criterios de estabilidad determinan que el sistema en un estado de equilibrio perturbado responde intentando restablecer el equilibrio, los criterios de estabilidad se expresarían entonces mediante unas funciones respuesta de signos determinados que permiten este restablecimiento.}.\\

En el caso de un fluido simple las transiciones de fase involucran discontinuidades en determinadas funciones e incluso singularidades en las funciones respuesta del sistema y en general comportamientos no analíticos. Antes de pasar a preguntarnos que contenido físico poseen estos cambios de fase nos preguntamos como el formalismo de la mecánica estadística es capaz de reproducirlos, la respuesta a esta cuestión se halla contenida en la siguiente teoría y que forma parte del conjunto de resultados y metodologías que se suelen calificar de rigurosos.\\

\vspace{0.2cm}

\subsection{Teoría de Yang y Lee}

Si analizamos para un sistema de partículas en el colectivo canónico que forma adquiere la energía libre, observamos que para todo valor de N la función de partición posee un valor positivo, real y bien definido para potenciales de interacción razonables (analíticos). Como indicaron \textit{Yang y Lee}\cite{PhysRev.87.404} esto permite dado un sistema de volumen finito escribir la propia función de partición macrocanónica en la forma de un polinomio en $e^{-\beta\mu N}$ y por tanto es expresable como,

\begin{equation}
\Xi(\mu,V,T)= \sum_{N=N_{0}}^{N_{m}}\left( e^{-\beta\mu}\right) ^N\mathcal{Z}_{N}(T,V)=\prod_{i=1}^{N}\left( 1-\frac{e^{-\beta\mu}}{z_{i}} \right)
\end{equation}

donde $z_{i}(V,T)$ son las raíces del polinomio\footnote{Que de hecho no pueden ser reales y positivas.}. La función de partición es una función analítica y lo serán la ecuación de estado $P(T,V,\mu)$ así como la compresibilidad isoterma. Los cambios de fase estan caracterizados por comportamientos no analíticos de algunas de estas magnitudes lo que conduce a la conclusión de que solamente para sistemas en el \textit{límite termodinámico} el formalismo representará, de modo adecuado, los cambios de fase\footnote{El primero en indicar esta importancia fue \textit{Krammer} en 1936, aunque tardo cierto tiempo en conocerse su trabajo y muchos físicos de su época pensaban que la función de partición solo puede describir una fase.}, basándose en el hecho de que el límite de una sucesión de funciones analíticas puede no serlo. Matemáticamente las no analiticidades aparecerán como límites de los puntos de acumulación de ceros de la función de partición en el eje real positivo y por tanto las diferentes fases aparecen como regiones de analiticidad acotadas por dichos puntos. \\

La extensividad de energía libre y de la entropía hace conveniente definir magnitudes molares o magnitudes por unidad de volumen que permanecen bien definidas en el límite termodinámico, su definición aprovecha además la homogeneidad de dichas funciones respecto del número de partículas, $S(U,V,N)=N s(U/N,V/N,1)=Ns(u,v)$ y $F(T,V,N)=Nf_{n}(T,v)=Vf_{v}(T,\rho)$, mientras que directamente $\Omega(T,V,\mu)=N\mu(T,v)$\\

\subsection{Transiciones de fase de primer orden y continuas}
En general las regiones, en el espacio determinado por los parámetros macroscópicos, donde la energía libre sea analítica serán las diferentes \textit{fases} y las regiones que los separen serán las \textit{fronteras de fase}, una representación del diagrama de fases de un sistema simple esta en la figura (\ref{fig:DiagramaPV}b). Nos interesan dos situaciones diferentes compatibles con los teoremas de Yang y Lee que representen situaciones con cambio de fases,
\begin{itemize}
\item Dos regiones con un punto de acumulación de ceros en medio. Tendríamos \textit{transición de fase primer} orden si lleva a discontinuidad en primera derivada.

\item Podemos tener una situación similar pero con continuidad tendríamos una \textit{transición de fase continua}.

\end{itemize}

    \vspace{1.2cm}

\subsection{Interacción entre partículas}

Hemos indicado las propiedades generales de la función de partición y en que condiciones es posible reproducir una transición de fase. Por otra parte observando la mecánica estadística colectividades vemos que si bien su formalismo no depende esencialmente de la forma concreta que poseen las interacciones entre sus constituyentes elementales\footnote{Siempre y cuando conduzcan a funciones de partición que respeten la termodinámica.} la física concreta de un sistema si vendrá determinada por esta. Esto lleva a dos cuestiones\footnote{Entorno a las que giran muchos de los problemas que se tratarán en la memoria.} entrelazadas. Una primera es, que modelo para las interacciones permite reproducir el conjunto de propiedades que tratamos de explicar y la segunda es, dada esta interacción como puedo resolver la función de partición (o funciones de distribución) para ese modelo.\\

	\begin{wrapfigure}{r}{8.4cm}
	\includegraphics[width=2.95in]{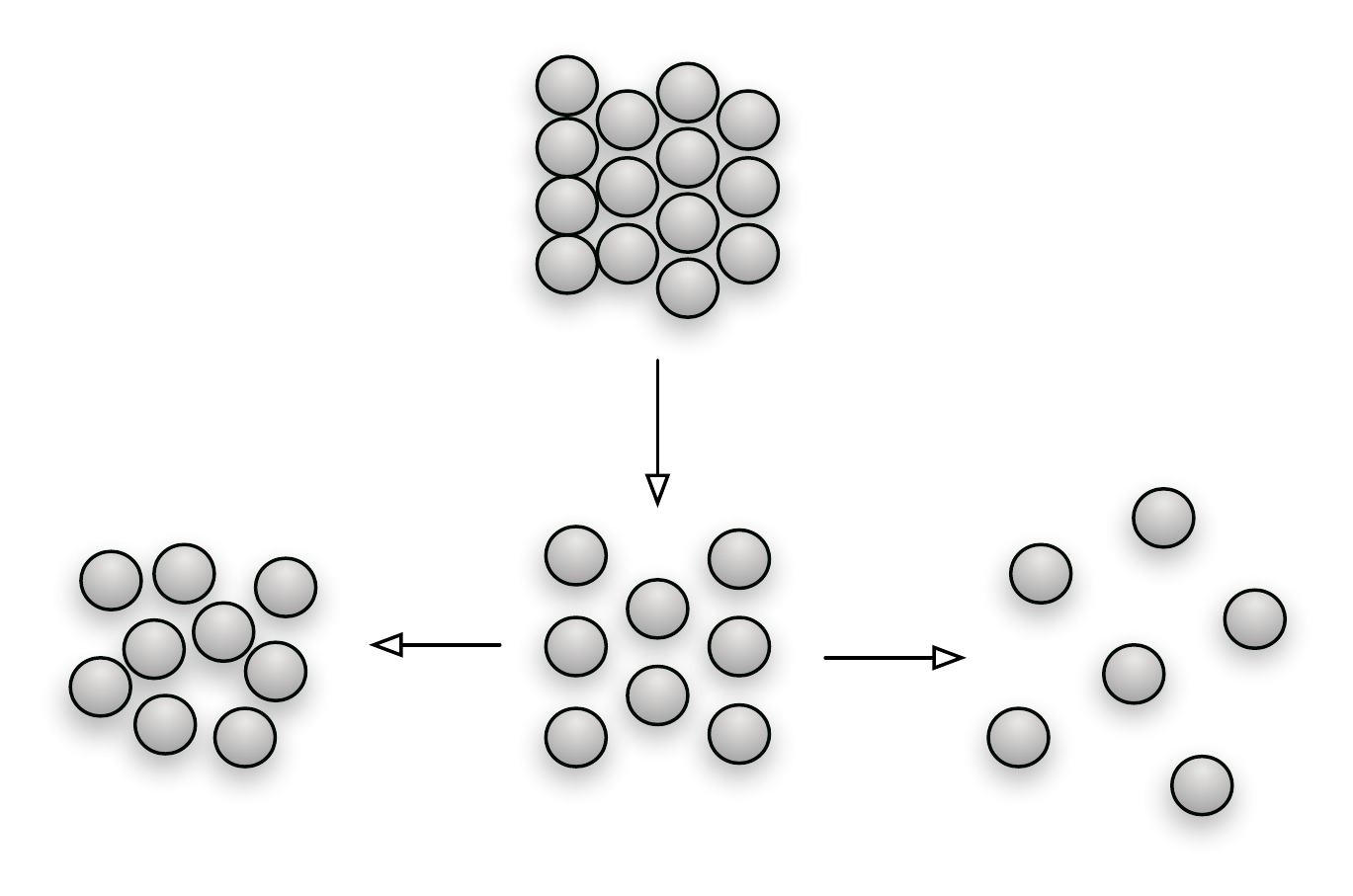}
	\caption{Arriba esquema con empaquetamiento máximo. En el centro expansión en $1/3$ del volumen. Potenciales suficientemente atractivos permiten una configuración como en izq. (líquido) y der. (gas) }
	\label{fig:criterioLindelmann}
	\end{wrapfigure}

Si nuestro objetivo es establecer en que condiciones es posible tener fases sólida y fluida, véase figura (\ref{fig:DiagramaPV}b), así como coexistencia líquido-vapor, la primera cuestión puede ser abordada del siguiente modo. La propia estabilidad de la materia sugiere que a distancias muy cortas la interacción entre las partículas ha de ser repulsiva y hemos de buscar un potencial impenetrable a cortas distancias, de hecho un modelo de esferas duras permite reproducir dos fases estructuralmente diferentes una sólida y otra fluida, y el paso de una a otra viene regido por la densidad del sistema. Si dada una densidad que reproduce la fase sólida ($\eta\sim 0.74$ en sistemas tridimensionales) expandimos el sistema en un tercio de su volumen (criterio de Lindelmann) el sistema pierde las condiciones que permitían su rigidez y pasa a una fase fluida. Imaginemos sin embargo una leve atracción de corto alcance superpuesta la sistema de esfera duras, si esta permite ligar a las esferas, ahora separadas, lo suficiente es posible encontrar una fase líquida estable y en consecuencia es posible encontrar (en función de la agitación térmica) una coexistencia entre líquido y gas (en la medida en que la atracción permita tener a una misma temperatura dos sistemas con diferente densidad estables). \\

Esto determina dos propiedades bá\-si\-cas para las interacciones que reproducen la fenomenología buscada, a\-de\-más los potenciales repulsivos a cortas distancias y atractivos a distancias intermedias son compatibles con la exigencia de estabilidad ter\-mo\-di\-ná\-mi\-ca\footnote{El alcance atractivo del potencial puede ser de hecho crucial para el diagrama de fases\cite{PhysRevLett.73.752}.}. La figura (\ref{fig:Potenciales}) ya mostraba estas pro\-pie\-da\-des. El problema será construir aproximaciones predictivas pa\-ra los modelos de interacción introducidos. En realidad gran parte de la dificultad esta en encontrar métodos de tratamiento adecuados para un potencial de interacción dado. Antes de pasar a describir esta segunda cuestión que abarca gran parte del formalismo podemos analizar la física de coexistencia de fases representada en una ecuación de estado que contiene la imagen física introducida. \\

\subsubsection{Ecuación de van der Waals}
\label{sec:eqvarderWaals}
Convenientemente tratado las propiedades del potencial de interacción se traducen en una ecuación de estado que es la que describirá la física macroscópica de diagrama de fases, como la transición líquido-vapor o la presencia de un punto crítico. La descripción más conocida que contiene en su ecuación de estado las propiedades repulsivas y atractivas comentadas es la \emph{ecuación de estado de van der Waals} que viene dada por la ecuación,
\begin{equation}
P(T,V,N)=\frac{NkT}{V-\beta N}-\alpha\frac{N^{2}}{V^{2}}
\label{eqn:vdwec}
\end{equation}
Parte de la expresión de la ecuación de estado para un gas ideal asumiendo que el principal efecto de la parte repulsiva es reducir el volumen disponible del sistema (covolumen), mientras que la parte atractiva es descrita como un efecto uniforme sobre el que incidiremos más adelante. Los valores concretos de $\alpha$ y $\beta$ son el reflejo de propiedades microscópicas de las interacciones, más allá de esto la ecuación se caracteriza por poseer propiedades universales independientes de estos valores en la escala adecuada.\\

	\begin{figure}
\begin{center}
	\includegraphics[width=5.5in]{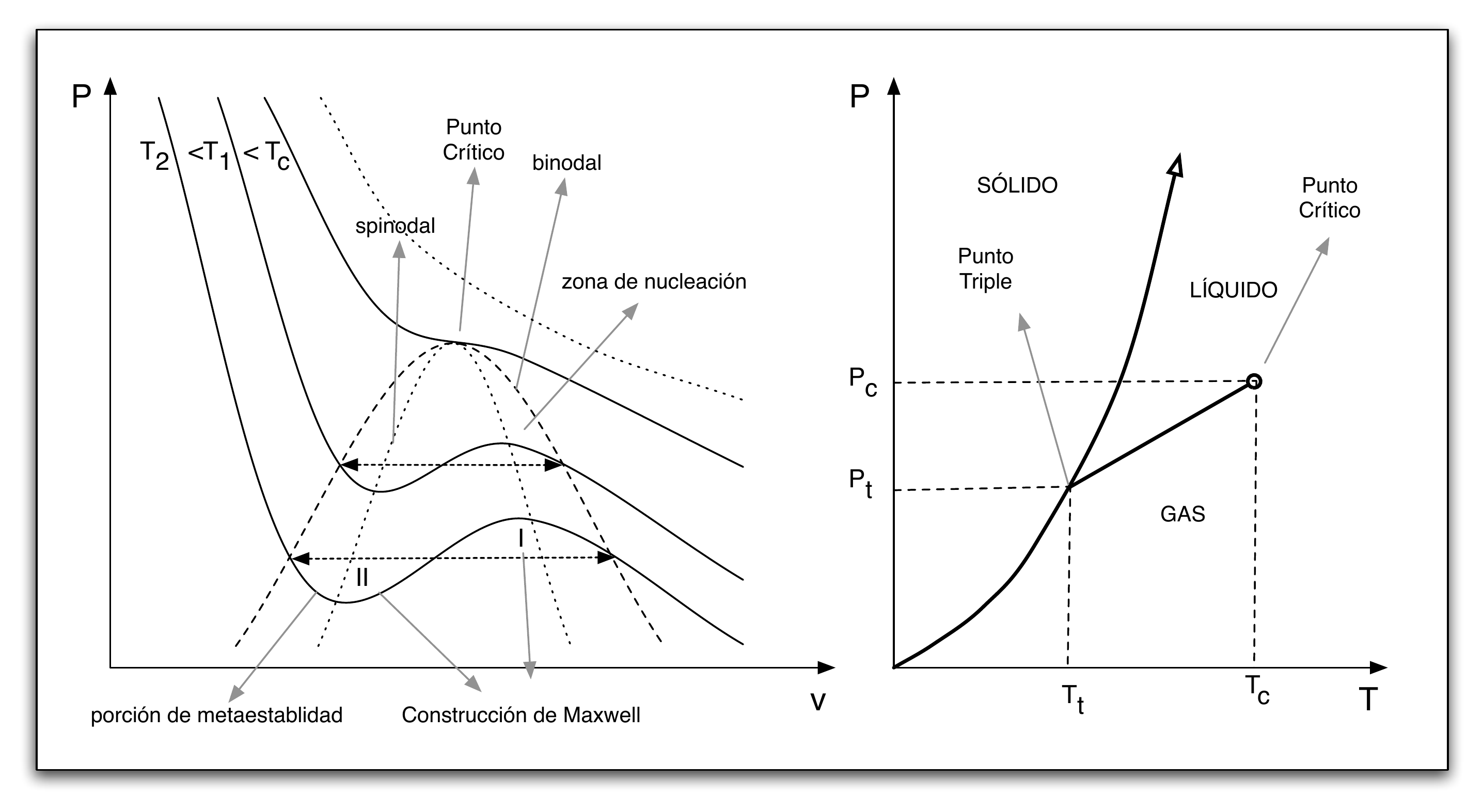}
\end{center}
\caption{\textbf{(a)} Esquema de las isotermas en un diagrama P-V basado en una ecuación de estado tipo van der Waals. Coexistencia de fases o binodal (línea discontinua), dentro esta la curva espinodal (línea de puntos). La región entre ambas es la zona de nucleación y ambas curvas convergen en el punto crítico. \textbf{(b)} Diagrama de fases típico de un sistema monocomponente simple.}
\label{fig:DiagramaPV}
	\end{figure}

Una representación esquemática de $P(v)$ para varios valores de T se halla en la figura (\ref{fig:DiagramaPV}a), como vemos existe una región a $T<T_{c}$ donde posee tres soluciones y es compatible con la separación del sistema en dos fases en equilibrio, cumpliendo por tanto $\mu^{l}(T,P)=\mu^{v}(T,P)$ junto con $T^{l}=T^{v}=T$ y $P^{l}=P^{v}=P$, fruto de que $\mu(P)$ que a $T>T_{c}$ era una curva única ahora posee dos ramas que coinciden en el equilibrio de fases\footnote{El contenido de esta afirmación junto con la integración en la isoterma de la ecuación de Gibbs-Duhem me determina la construcción de Maxwell para la coexistencia líquido-vapor (igualdad de las áreas I y II de la figura (\ref{fig:DiagramaPV})) y fija los valores de las densidades de ambas fases en la coexistencia, es decir los estados de equilibrio correspondientes a los valores dados de temperatura y presión.}.\\

Observando de nuevo la curva $P(v)$, hay valores de los parámetros de estado en que la curva, no correspondiendo a los estados de equilibrio, no violan sin embargo los criterios de estabilidad y representan estados estables solo frente a pequeñas fluctuaciones(metaestables). La zona de la gráfica $P(v)$ de los estados metaestables, es decir, la zona comprendida entonces entre los puntos que verifican la construcción de Maxwell (curva de coexistencia o binodal) y el límite donde dejan de cumplirse los criterios de estabilidad (curva espinodal) se denomina \textit{zona de nucleación} ya que este es el fenómeno físico (la formación de núcleos\footnote{Una estimación de las fluctuaciones necesarias para poder crear estos núcleos, es decir, que sean núcleos estables puede realizarse desde argumentos estadísticos y de hecho desde aplicaciones de la teoría del funcional de la densidad, a este campo se le denomina teoría de la decomposición espinodal.}
de una fase, la más estable, en la otra) que permite al sistema pasar de una fase a otra (saber que existe otra más favorable desde el punto de vista de la energía libre)\footnote{Si volvemos ahora a la construcción de Maxwell y viajamos por la isoterma observamos un cambio brusco en el volumen molar, lo mismo ocurrirá en el caso de la entropía. Esto hace que estas transiciones de fase se denominen discontinuas o de primer orden (en el sentido de \textit{Ehrenfest}). La transición al estado crítico en cambio se denomina continua en la medida en que no aparecen discontinuidades en el volumen molar o la  entropía molar. La señal de la transición de fase se aprecia en los valores del calor especifico y la compresibilidad.}.\\

Como comentábamos el camino desde un modelo de interacción a una ecuación de estado es un proceso analíticamente difícil, hecho que proviene de la dificultad de realizar la integral de configuración. Los sistemas resolubles usualmente son aquellos donde es posible factorizar en el número de partículas dicha integral bien por la ausencia de interacciones, como en un sistema ideal, bien por la posibilidad de diagonalizar apropiadamente el hamiltoniano de interacción como ocurre, por ejemplo, en el oscilador armónico unidimensional.\\

Ahora bien, una primera aproximación del problema puede obtenerse observando que a densidades bajas (sistemas diluidos) las propias características de la interacción dada hacen que dentro del alcance del potencial de interacción encontremos un número limitado de partículas y la integral de configuración pueda ser razonablemente abordable. En la siguiente sección describimos el procedimiento que permite construir un desarrollo en la densidad y lo aplicamos a un sistema de esferas duras, que como hemos sugerido, y veremos en detalle, es parte referencial importante en el estudio de sistemas líquidos. \\
\section{Expansiones en la densidad. Desarrollo del virial}
Resulta conveniente, aunque no estrictamente necesario, restringirnos a interacciones que puedan ser expresadas como una suma de términos de interacción entre dos partículas. Bajo esta suposición definimos dos cantidades,

\begin{equation}
e(r)\equiv e^{-\beta \phi(r)}
\end{equation}
\begin{equation}
f(r)\equiv e(r)-1
\label{eqn:funcionMayer}
\end{equation}
esta última se denomina \emph{función de Mayer} es de corto alcance y permite construir formalmente un desarrollo de la integral de configuración que podemos expresar como,

\begin{equation}
Q(T,V,N)=\int d\vec{r}_{1}...d\vec{r}_{N}\prod_{i<j}^{N}[1+f(r_{ij})]=Q_{0}+Q_{2}+...
\end{equation}

Donde los sucesivos sumandos van involucrando progresivamente más partículas (cero, dos etc). Este desarrollo en el caso de la energía libre queda expresado por,
\begin{equation}
\beta F(T,V,N)=\beta F_{id}+ln[1+Q_{2}/Q_{0}+...]
\end{equation}
diferenciando con respecto al volumen podemos obtener la ecuación de estado con sus correcciones en densidad sobre el caso ideal. En el caso de la función de partición macrocanónica podemos escribir,
\begin{equation}
ln\Xi=V\sum_{i=1}^{\infty}B_{i}\rho^{i}
\end{equation}
los coeficientes $B_{n}$ se denominan \textit{coeficientes del virial}. Y para la ecuación de estado tendremos que,
\begin{equation}
\beta P=\sum_{i=1}^{\infty}B_{i}\rho^{i}
\end{equation}
Dentro de este desarrollo podemos escribir diferentes propiedades, por ejemplo, la compresibilidad isoterma,
\begin{equation}
\chi_{T}^{-1}=-\rho^{-1}\frac{\partial P}{\partial \rho}\bigg\vert_{T}=\beta^{-1}\sum_{k=1}^{\infty}kB_{k}\rho^{k-2}
\end{equation}
El procedimiento formal es general y pueden determinarse para diferentes potenciales los diferentes coeficientes del virial y obtener un desarrollo a bajas densidades. La determinación analítica es progresivamente más complicada y únicamente para ciertos potenciales es posible determinar más allá del tercer coeficiente del virial sin el uso de métodos numéricos. Para un potencial de esferas duras se han determinado analíticamente los cuatro primeros coeficientes del virial y por métodos numéricos se han estimado los siguientes coeficientes, algunos aparecen en la tabla (\ref{Tabla:VirialHS}).

\begin{table}[htdp]
\begin{center}
\begin{tabular}{c c c c c c c}
\toprule
 &  $\mathbf{B_{2}}$ & $\mathbf{B_{3}/B_{2}^{2}}$  & $\mathbf{B_{4}/B_{3}^{3}}$ & $\mathbf{B_{5}/B_{4}^{4}}$ & $\mathbf{B_{6}/B_{5}^{5}}$ & $\mathbf{B_{7}/B_{6}^{6}}$\\
\midrule
\textit{Exacto} &  2.0944 & 0.625  & 0.2869 & 0.1103 & 0.0386 & 0.0138 \\
\textit{SPT}      &  2.0944 & 0.625  & 0.2969 & 0.1211 & 0.0449 & 0.0156 \\
\textit{CS}      &  2.0944 & 0.625  & 0.2813 & 0.1094 & 0.0156 & 0.0132 \\
\midrule
\textit{HNC(C)} &  2.0944 & 0.625  & 0.2029 & 0.0493 & 0.0281 &  \\
\textit{PY(C)}     &  2.0944 & 0.625  & 0.2969 & 0.1211 & 0.0449 & 0.0156 \\
\bottomrule
\end{tabular}
\end{center}
\caption{\textbf{Coeficientes del virial para un sistema de esferas duras}, en unidades del diámetro de esferas duras. Se compara una determinación directa (exacta) con los coeficientes contenidos en diferentes aproximaciones. En las dos últimas se escribe la \textit{vía de la compresibilidad} pues da resultados más ajustados. La ruta del virial en PY se aproxima a los valores exactos por debajo mientras que en la ruta de la compresibilidad se aproxima por encima, en el caso de HNC sucede al contrario.}
\label{Tabla:VirialHS}
\end{table}
\subsection{Ecuación de Carnahan-Starling para esferas duras}
Carnahan y Starling observaron que expresando los coeficientes del virial en un desarrollo en la fracción de empaquetamiento $\eta=\pi\rho d^{3}/6$ los primeros coeficientes eran números enteros o próximo a enteros y que mediante un ajuste de estos coeficientes se podían predecir los siguientes tres coeficientes de modo casi exacto. La suma de la serie resultante dio como resultado lo que se conoce como aproximación de Carnahan-Starling para la ecuación de estado de esferas duras,
\begin{equation}
\beta P=\rho\frac{1+\eta+\eta^{2}-\eta^{3}}{(1-\eta)^{3}}
\label{eqn:CSecuacionestado}
\end{equation}
que se suele tomar como una aproximación semiempírica y de hecho a menudo como un resultado casi exacto\footnote{La forma del denominador que permite estados líquidos a fracciones de empaquetamiento superiores al empaquetamiento máximo responde a la imagen de \textit{líquido ideal} dada por Rosenfeld\cite{PhysRevA.32.1834} en el limite asintótico de altas densidades.}. La aproximación es analítica por tanto es posible integrando y diferenciando la ecuación de estado, conocido que el límite a densidades bajas es el sistema ideal determinar la energía libre de exceso de un sistema de esferas duras,
\begin{equation}
F^{ex}(T,V,N)=N\beta^{-1}\int_{0}^{\rho}d\tilde{\rho}\frac{\beta P^{ex}(\tilde{\rho})}{\tilde{\rho}^{2}}=N\beta^{-1}\frac{\eta(4-3\eta)}{(1-\eta)^{2}}
\end{equation}
\begin{equation}
\beta\mu^{ex}(T,V,N)=\frac{1}{N\beta^{-1}}F^{ex}(T,V,N)+P^{ex}V=\frac{1+5\eta-6\eta^{2}+2\eta^{3}}{(1-\eta)^{3}}
\end{equation}
Existe otra forma complementaria de obtener una ecuación de estado para un sistema de cuerpos duros que resulta ser exacta en el caso unidimensional y que en el caso tridimensional reproduce los primeros coeficientes del virial de modo adecuado y que por otra parte no alude a un desarrollo en densidad.
\section{Teoría de la Partícula Escalada para sistemas duros}
Este método para determinar la ecuación de estado fue desarrollado por \textit{Reiss, Frisch y Lebowitz}\cite{helfand:1379,helfand:1037,reiss:369,1970reiss:4015} basándose en consideraciones diferentes al desarrollo en densidad. Partieron del hecho de que para determinar la ecuación de estado en un sistema de cuerpos duros solo es necesario conocer de manera exacta el valor de la función de distribución radial en el caso de contacto tal y como expresa el teorema del virial, véase ec.(\ref{eqn:EcuacionVirial}),
\begin{equation}
\beta P=\rho+\frac{2}{3}\pi a^{3} \rho^{2}g(a)
\end{equation}
Para determinar $g(a)$, propusieron un método original y diferente para determinar el potencial químico basándose en el trabajo que cuesta crear un hueco de tamaño determinado en un sistema de cuerpos duros de tamaño d, cuando el tamaño del hueco es d, este se comporta exactamente como una esfera más. La cantidad importante es $G(\lambda)$ que tiene relación con la probabilidad media de que un cuerpo duro del sistema este en contacto con el hueco. $G(\lambda)$ puede ser determinada a partir de la probabilidad de que se forme la cavidad y dicha probabilidad es lo que se intenta expresar de un modo sistemático mediante cantidades que tienen que ver con las características geométricas de las partículas. \\

Por este método se puede encontrar la ecuación de estado para sistemas duros de diferente dimesionalidad,\\
\begin{itemize}
\item Segmentos, $\eta=\rho d$
\begin{equation}
\beta P=\rho\frac{1}{1-\eta}
\end{equation}
\item Discos, $\eta=\frac{\pi}{4}\rho d^{2}$
\begin{equation}
\beta P=\rho\frac{1}{(1-\eta)^{2}}
\end{equation}
\item Esferas, $\eta=\frac{\pi}{6}\rho d^{3}$
\begin{equation}
\beta P=\rho\frac{1+\eta+\eta^{2}}{(1-\eta)^{3}}
\label{eqn:EstadoPYescalada}
\end{equation}
\end{itemize}

La comparación con los coeficientes del virial en el caso de esferas duras podemos verlo en la tabla (\ref{Tabla:VirialHS}), el caso unidimensional es exacto y el caso bidimensional es casi exacto. Este desarrollo ha tenido una gran relevancia ya que se puede extender de modo aproximado a cuerpos duros con geometrías convexas diversas. \\

En el caso de \textit{discos duros}, hay correcciones empíricas obtenidas de mejorar la propuesta de SPT, como son la \textit{ecuación de estado de Henderson},
\begin{equation}
\beta P=\frac{\rho}{(1-\eta)^{2}}\left(1+0.128\eta^{2}+\frac{0.043\eta^{4}}{1-\eta} \right)
\label{eqn:Henderson_Hard_Disk}
\end{equation}

 y la de \textit{Baus-Colot}\cite{PhysRevA.36.3912} que fue desarrollada para hiperesferas en dimensión D. Para D=2 se concreta en la ecuación,
\begin{equation}
\beta P=\rho+\rho\left[ \frac{1.001\eta}{(1-\eta)^{2}}+\frac{1.4063\eta}{1-\eta}-\left(0.4073\eta+0.2803\eta^{2}+0.1353\eta^{3}+0.041\eta^{4}\right)\right]
\label{eqn:Baus-Colot_Hard_Disk}
\end{equation}

 Para los propósitos de esta memoria la ecuación de estado dada por SPT será adecuada.

\section{Desarrollo Diagramático}
\label{sec:desarrolloDiagramatico}

El tratamiento sistemático de las expansiones en la densidad se facilita mediante su expresión en forma de diagramas en \textit{clusters} junto con una serie de reglas formales para su manipulación\cite{BOOK-TheorySimpleLiquids}. Estos permiten simplificar simbólicamente el tratamiento de una expansión basada en las funciones de Mayer,
\begin{equation}
\Xi(\mu,V,T)=\sum_{N}\frac{1}{N!}\int d\vec{r}^{N}\prod_{i=1}^{N}z(\vec{r}_{i})\prod_{i<j}^{N}(1+f(\vec{r}_{i},\vec{r}_{j}))
\end{equation}
En nuestro caso nos interesa esencialmente que revela propiedades importantes al nivel de las funciones de distribución de partículas de modo que si escribimos la función de partición en un sistema en un potencial externo $\mathcal{V}_{N}=\sum_{i}v_{ext}(\vec{r}_{i})$ tendríamos,
\begin{equation}
\Xi(\mu,V,T)=\sum_{N}\frac{1}{N!}\int d\vec{r}^{N}\prod_{k=1}^{N}z(\vec{r}_{k})e^{-\beta U_{N}(\vec{r}^{N})}
\end{equation}
donde $z(\vec{r}_{k})=e^{-\beta(\mu-v_{ext}(\vec{r}_{k}))}$. De modo que la \textit{derivada funcional} reiterada de la función de partición sobre esta fugacidad generalizada permite reproducir lo que esencialmente es la $\rho^{(n)}(\vec{s}^{n})$, y en consecuencia definir formalmente la jerarquía de funciones de distribución. Del mismo modo otra jerarquía puede ser definida mediante la diferenciación funcional de la energía libre, es decir, del logaritmo de la función de partición,
\begin{equation}
G^{(n)}(\vec{s}^{n})=\frac{\delta^{n}ln\Xi(\mu,V,T)}{\delta ln z(\vec{s}^{1})...\delta ln z(\vec{s}^{n})}
\end{equation}
Cuyo miembro de orden 2 ya hemos introducido previamente\footnote{También se ha denominado a esta familia funciones de Ursell sobre todo en los años 80 y se solían denotar por $F_{n}$.}. Su inversa funcional permite definir otra jerarquía de funciones $C^{(n)}(\vec{s}^{n})$ y que formalmente pueden ser obtenidas a partir de derivadas funcionales de otro funcional que surgirá de manera natural en el formalismo del funcional de la densidad, y posponemos a entonces su introducción completa.\\

Los desarrollos diagramáticos desde el punto de vista de la teoría de líquidos han contribuido doblemente, por una parte han sido la primera forma en que se ha afrontado la constitución del formalismo del funcional de la densidad y por otra parte a dado lugar a un campo conocido como \textit{Ecuaciones Integrales} para determinar propiedades de los líquidos.  

\section{Ecuaciones Integrales}
\label{sec:ecuacionesIntegrales}

La relación de los segundos elementos de ambas jerarquías tiene una relevancia especial y ha constituido una de las relaciones con más importancia en la teoría de líquidos simples. Fue inicialmente formulada por Ornstein y Zernike y en el caso no-homogéneo queda expresada como,

\begin{equation}
h^{(2)}(\vec{r}_{1},\vec{r}_{2})=c^{(2)}(\vec{r}_{1},\vec{r}_{2})+\int d\vec{r_{3}}h^{(2)}(\vec{r}_{1},\vec{r}_{3})\rho(r_{3})c^{(2)}(\vec{r}_{3},\vec{r}_{2})
\label{eqn:OZ}
\end{equation}
que suele ser tomada como una definición de la \textit{función de correlación directa} $c^{(2)}(\vec{r}_{1},\vec{r}_{2})$. Desde el punto de vista del desarrollo diagramático supone diferenciar un subconjunto de diagramas de $h^{(2)}$ que permite escribir la relación recurrente dada arriba y dicho subconjunto de diagramas es el que define a la función de correlación directa $c^{(2)}$, que posee la relevante propiedad de ser del mismo alcance que el potencial de interacción cosa que no sucede en $h^{(2)}(r)$. La determinación completa de estas dos funciones a partir de la ecuación de Ornstein-Zernique necesita buscar otra relación, que se denomina genéricamente \textit{relación de cierre} y cuya elección dará lugar a diferentes aproximaciones, con el problema a priori de que su carácter aproximado puede hacer dependiente la termodinámica de la ruta elegida para su conexión con la estructura.

\subsection{Teoría de Percus-Yevick}
\label{sec:teoriaPY}
Nos interesa comenzar por sistemas homogéneos donde la posible determinación de las diferentes funciones de distribución y correlación es más sencilla. En este caso la función de distribución radial la expresamos a partir de ec. (\ref{eqn:funcionMayer}) como
\begin{equation}
g(r)=(1+f(r))y(r)
\end{equation}
que define la  llamada \textit{función cavidad} $y(r)$. Esencialmente los primeros términos de desarrollo de $c(r)$ son expresables como $c(r)=f(r)y(r)$ mientras que el resto lo agrupamos en una \textit{función cola} que llamamos $d(r)$.
\begin{equation}
c(r)=f(r)y(r)+d(r)
\label{eqn:cierre1}
\end{equation}
A bajas densidades d(r) sera pequeña y es posible considerar
\begin{equation}
c(r)=f(r)y(r)
\end{equation}
que constituye la \textit{aproximación de Percus-Yevick}.\\

Esta aproximación tiene una relevancia especial en el caso de \textit{esferas duras} puede escribirse como,
\begin{subequations}
\begin{equation}
c(r)=0,\qquad r>d
\end{equation}
\begin{equation}
g(r)=0,\qquad r<d
\end{equation}
\label{eqn:condiccionesHSpercusYevick}
\end{subequations}
que permite obtener una expresión\cite{thiele:474,PhysRevLett.10.321} analítica\footnote{Dadas las condiciones (\ref{eqn:condiccionesHSpercusYevick}) junto con la ecuación de Ornstein-Zernique es posible resolver el problema mediante un método integral, véase \cite{PhysRev.154.170}, alternativamente se puede establecer igualmente desde ec. (\ref{eqn:condiccionesHSpercusYevick}) una ecuación diferencial resoluble para $c_{PY}(r)$, véase \cite{Krienke1998263}.}   para $c_{PY}(r)$ y de hecho obtener la termodinámica del sistema, que en el caso de la ecuación de estado determinada a partir de la ecuación de la compresibilidad permite llegar a la misma ecuación de estado que la teoría de la partícula escalada\footnote{Conviene notar que otras vías diferentes, \S\ref{sec:rutasTermoEstructura},  dan resultados diferentes y no es posible considerar idénticas la teoría de la partícula escalada y la aproximación de Percus-Yevick, aunque como veremos existe un esquema más general que permite desarrollar una imagen más unificada de ambas propuestas, \S\ref{sec:teoriaRosenfeld}.}. Finalmente la ecuación de Carnahan-Starling puede ser obtenida como una combinación de las soluciones de Percus-Yevick obtenidas por medio de la ecuación de la compresibilidad y por medio de la ecuación de la presión.\\

\subsubsection{Extensiones de la aproximación de Percus-Yevick}
Una variante directa de la teoría de Percus-Yevick nace de analizar su significado, $c(r)=f(r)y(r)$ implica que  a grandes distancias el comportamiento predominante es $c(r)=-\beta \phi(r)$.\\

 \textit{Lebowitz y Percus}\cite{PhysRev.144.251} propusieron adoptar esta definición como c(r) en la zona atractiva y el resultado de Percus-Yevick para esferas duras en la zona repulsiva, esta aproximación se conoce como \textit{mean spherical aproximation}.\\

\textit{Andersen y Chandler} propusieron en cambio generalizar la propuesta y sugieren $c(r)=c_{hs}(r)-\beta \phi(r)$ para $r>\sigma$, ideal en el caso de que se tenga una teoría para la función de correlación directa del sistema de esferas duras más precisa que la receta de PY, se conoce como \textit{random phase aproximation}.

\subsection{Aproximación HNC}
\label{sec:teoriaHNC}

Se basa en una expresión alternativa a ec. (\ref{eqn:cierre1}) como,
\begin{equation}
y(r)=e^{h(r)-c(r)}
\label{eqn:cierre2}
\end{equation}
Los resultados mejoran para potenciales atractivos de largo alcance, para los que la aproximación de Percus-Yevick da resultados modestos pero empeoran respecto de esta aproximación para esferas duras sobre todo a altas densidades. Con todo HNC no es una aproximación a densidades bajas ya que en su construcción incluye diagramas a todos los ordenes en la densidad.

    \section{Teoría de perturbaciones}
\label{sec:teoriaPerturbativaIntro}

Hasta el momento el cálculo de las propiedades ha sido planteado como un desarrollo en la densidad pero lógicamente es poco viable a densidades cercanas al punto triple, mientras tanto las funciones de correlación para potenciales genéricos son difíciles de determinar analíticamente. Si hemos visto que la ecuación de estado dada por Carnahan-Starling o la teoría de la partícula escalada, así como la solución analítica de Percus-Yevick para c(r) permiten un buen tratamiento del sistema de esferas duras.\\

Además una extensión\cite{WidomScience} termodinámicamente consistente de la ecuación de van der Waals puede ser escrita como $\beta P=\beta P_{HS}-a\beta\rho^{2}$, que sugiere ver al líquido como estructuralmente similar a un sistema de esferas duras y la parte atractiva jugando esencialmente un papel termodinámico. La \textit{imagen física} asociada que justifica este tratamiento es la de una partícula sometida a fuerzas atractivas de las partículas cercanas que en un sistema homogéneo denso en promedio se cancelan para la mayoría de las configuraciones, en consecuencia, la diferencia estructural con un sistema únicamente de esferas duras es mínima y la parte atractiva puede ser aproximada por un fondo uniforme superpuesto al fluido de esferas duras que no altera ni la estructura ni la dinámica pero aporta una energía de cohesión\footnote{El resultado es que si bien las propiedades de escalado cerca del punto crítico no son las adecuadas y un modelo \textit{lattice-gas} daría exponentes críticos más exactos, las propiedades generales del sistema cerca del punto triple están descritas de este modo correctamente si el sistema de esferas duras es tratado de modo adecuado. La explicación es que cerca del punto crítico las correlaciones que contiene la parte atractiva son más importantes que cerca del punto triple donde la alta compresibilidad impide fluctuaciones de largo alcance que alteran las condiciones en que la imagen de van der Waals resulta efectiva.}.\\

Dos mejoras son posibles en la imagen dada por van der Waals,
\begin{itemize}
\item Determinar la contribución de las fuerzas atractivas teniendo en cuenta las distribuciones de partículas del sistema.
\item Determinar el sistema de esferas duras más adecuado a los pa\-rá\-me\-tros ter\-mo\-di\-ná\-mi\-cos de cada estado.
\end{itemize}

Para ello se han desarrollado una serie de métodos muy útiles al respecto de como representar un líquido simple y que genéricamente se conocen como teorías perturbativas. Esencialmente veremos la expansión $\lambda$ sobre un sistema de referencia genérico, que responde a la primera de las cuestiones y su aplicación a potenciales \textit{blandos} que configura a la segunda de las mejoras.

\subsection{Expansión $\lambda$}
Se considera un sistema de partículas con un potencial de interacción $\phi_{\lambda}$ dependiente de un parámetro $\lambda$, que varía desde un valor inicial $\lambda_{0}$ correspondiente a un potencial $\phi_{\lambda_{0}}$ hasta un valor $\lambda_{1}$ correspondiente a $\phi_{\lambda_{1}}$ y manera que el potencial puede variarse en este parámetro\footnote{Se suele denominar \textit{parámetro de acoplo}.} de modo continuo. La manera usual de expresarlo es mediante,
\begin{equation}
\phi(\lambda,r_{12})=\phi_{0}(r_{12})+\lambda\phi(r_{12})
\end{equation}
el \textit{potencial completo} se recupera en $\lambda_{1}=1$ mientras que en $\lambda_{0}=0$ tenemos únicamente el \textit{potencial de referencia}. La forma de la función de partición permite escribir de modo exacto,
\begin{equation}
\frac{\partial\beta F(\lambda)}{\partial\lambda}=\beta\left\langle \frac{\partial \Phi_{N}(\lambda)}{\partial\lambda}\right\rangle 
\label{eqn:acoploPotencial}
\end{equation}
de esta manera tendremos para la energía libre\footnote{El modo de construir un desarrollo perturbativo es introducir una expansión Taylor dentro de la integral entorno de $\lambda_{0}$, que para la forma de $\phi(\lambda,r_{12})$ se expresa de modo sencillo y se suele denominar expansión en altas temperaturas ya que cada termino aparece multiplicado por potencias crecientes de $\beta^{-1}$, este desarrollo se debe a \textit{Zwanzig}.}.
\begin{equation}
\beta F(\lambda_{1})=\beta F(\lambda_{0})+\int_{0}^{1} d\lambda \beta\left\langle \frac{\partial\Phi_{N}(\lambda)}{\partial\lambda}\right\rangle 
\end{equation}
ya vimos que la cantidad del integrando es posible expresarla mediante la función de distribución $\rho^{(2)}(\vec{r}_{1},\vec{r}_{2};\phi_{\lambda})$ donde indicamos expresamente que esta depende del potencial de interacción del sistema bajo el parámetro de acoplo,
\begin{equation}
\beta F(\lambda_{1})=\beta F(\lambda_{0})+\frac{1}{2}\int_{0}^{1} d\lambda \int d\vec{r}_{1}d\vec{r}_{2} \rho^{(2)}(\vec{r}_{1},\vec{r}_{2};\phi_{\lambda})\phi(r_{12})
\label{eqn:TeoriaPerturbaciones}
\end{equation}
el problema de determinar el valor de la energía libre para un potencial genérico ha sido sustituido por la necesidad de conocer la función de partición en un sistema de referencia definido por el potencial $\phi_{0}$ más el conocimiento de la función distribución de dos cuerpos en toda la familia de potenciales caracterizada por $\lambda\in[0,1]$, que es un problema a priori tan complicado como el inicial. La ventaja de esta expresión es que permite proponer un desarrollo perturbativo en el parámetro de acoplo.\\

El orden en $\lambda$ para la función $\rho^{(2)}(\vec{r}_{1},\vec{r}_{2};\phi_{\lambda})$ equivale al primer término del desarrollo perturbativo en la energía libre y equivale identificar el valor de $\rho^{(2)}(\vec{r}_{1},\vec{r}_{2};\phi_{\lambda})\sim\rho^{(2)}(\vec{r}_{1},\vec{r}_{2};\lambda_{0})$, igual por tanto al sistema de referencia que se supone conocido. Este término puede simplificarse aun más si escribimos $\rho^{(2)}(\vec{r}_{1},\vec{r}_{2};\lambda_{0})\sim\rho^{(1)}(\vec{r}_{1})\rho^{(1)}(\vec{r}_{2})$ que da lugar a una aproximación de campo medio que ignora correlaciones involucradas en el término $\phi(r_{12})$, que para un sistema uniforme recupera la ecuación de estado de van der Waals mejorada con el conocimiento más preciso del sistema de referencia incluido en $\beta F(\lambda_{0})$. \\

\subsection{Modelos de interacción para líquidos simples: Potenciales blandos}
\label{sec:modelosPerturbativos}

Ya hemos comentado la conveniencia de tomar como sistema de referencia un potencial de esferas duras para la descripción de un líquido. La base rigurosa puede verse en los experimentos de \textit{Ashcroft-Lekner}\cite{PhysRev.145.83} donde comparan el factor de estructura de metales líquidos cerca del punto triple con el factor de estructura de un sistema de esferas duras equivalente basándose en la aproximación de Percus-Yevick. Esto fue estudiado por \textit{Verlet}\cite{PhysRev.165.201} para un modelo Lennard-Jones mediante simulaciones buscando el diámetro de esferas duras que permite reproducir los máximos del factor de estructura que encontraba a altas densidades. Los resultados fueron excelentes confirmando la descripción adecuada de un líquido cerca del punto triple como un sistema de esferas duras decorado con un potencial atractivo\footnote{En cambio cerca del punto crítico cuando la compresibilidad diverge el comportamiento a vectores de onda pequeños esta dominado por las fuerzas atractivas y se manifiestan explícitamente los defectos de una aproximación de campo medio.}.\\

Los estudios anteriores implican la elección de un método para determinar adecuadamente el sistema de referencia. La física contenida en el trabajo de Verlet es explicada y reproducida teóricamente por \textit{Weeks, Andersen y Chandler}\cite{Chandler05201983}. Existe además otro tratamiento previo debido a \textit{Barker y Henderson} que permite encontrar adecuadamente el sistema de referencia, no es tan sistemática como la anterior pero destaca por su simplicidad. En ambos casos la separación conlleva métodos diferentes para evaluar perturbativamente la contribución de la cola atractiva, detallamos ambos.

\subsubsection{Propuesta de Barker-Henderson}
\label{ss:BarkerHerderson}
\begin{figure}[htbp]
\begin{center}
\includegraphics[width=4.0in]{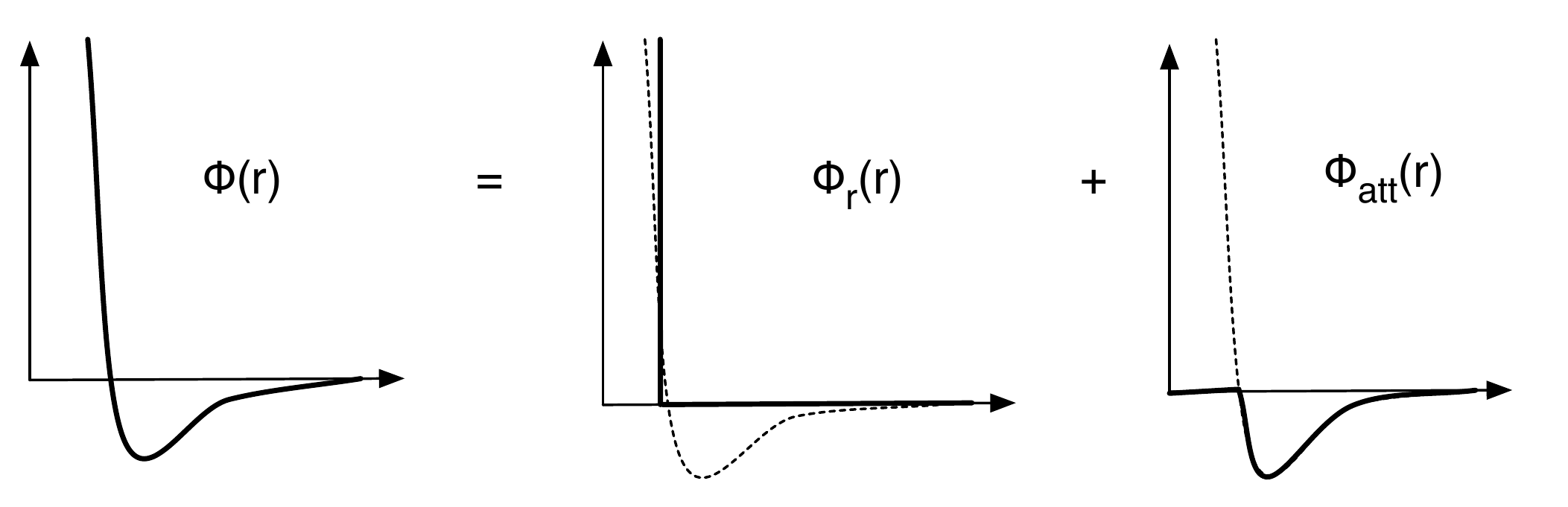}
\caption{División del potencial de interacción en la teoría BH}
\label{fig:TeoriaBH}
\end{center}
\end{figure}

La división del potencial de interacción es la correspondiente a la figura (\ref{fig:TeoriaBH}) donde se separan las contribuciones positivas y negativas de este y se aproxima la primera de ellas. La parte de referencia que se aproxima es,
\begin{eqnarray}
\phi_{R}(r)=\left\{\begin{array}{ll}
\phi(r),&r<\sigma\\0,&r\ge \sigma\end{array}\right.
\label{eqn:BH1}
\end{eqnarray}

La construcción de la teoría perturbativa parte de la ecuación  (\ref{eqn:TeoriaPerturbaciones}) que obtenemos si la particularizamos al caso homogéneo,

\begin{equation}
\frac{\beta F}{N}=\frac{\beta F_{0}}{N}+2\rho\pi\beta\int dr r^{2}g_{0}(\rho,r)\phi(r)+O(\beta^{2})
\end{equation}

Formalmente la expresión arriba indicada, al ser una teoría de perturbaciones, requiere expresar de algún modo los siguientes términos, para evaluar, al menos cualitativamente, que las correcciones a la teoría son pequeñas en el rango de aproximación que nos interesa. Barker y Henderson desarrollaron inicialmente una teoría para el potencial cuadrado\cite{barker:2856} donde expresaban dicha corrección mediante argumentos \textit{mesoscópicos}\footnote{
La expresión en concreto seria,
\begin{equation}
\frac{1}{4}\rho\beta^{2}\int d\vec{r}^{\prime}\left(\phi(\vec{r}^{\prime})\right) ^{2}\beta^{-1}\left.\frac{\partial(\rho g^{(2)})}{\partial P}\right|_{0}
\end{equation}
la expresión involucra la derivada de $g^{(2)}(r)$ que es introducida como una compresibilidad local.}
dando lugar a una descripción adecuada al ser comparada con simulaciones numéricas tanto Montecarlo como de Dinámica Molecular. La desventaja de su argumentación es que su desarrollo sistemático a más ordenes no es del todo evidente y su aplicación menos transparente.\\

En el caso de potenciales como el Lennard-Jones no estrictamente repulsivos, es posible desarrollar el factor de Boltzmann del potencial de interacción en función parámetros uno que calibra la dureza de la parte repulsiva y otro la profundidad de la parte atractiva\cite{barker:4714}. Su desarrollo lleva a una expresión,
\begin{equation}
\beta\frac{F-F_{0}}{N}=-2\pi\rho d^{2}g_{0}(d)\left[d-\int_{0}^{\sigma}1-e^{-\beta\phi(r)}\right] +2\pi\rho\beta\int dr r^{2}g_{0}(r)\phi(r)+O(\beta^{2})
\end{equation}
donde $\sigma$ es el valor en que $\phi(r)=0$. La elección adecuada de $d(T)$ permite anular el primer término del desarrollo y representa el diámetro del sistema de referencia de esferas duras. Así introducen un parámetro $d_{BH}$ como,
\begin{equation}
d_{BH}(T)=\int_{0}^{\sigma}(1-e^{-\beta u(r)})dr
\end{equation}
y en consecuencia $g_{0}(r,\rho)\simeq g_{HS}(r/d_{hs},\eta)$, con $\eta=\rho d_{BH}^{3}\pi/6$. Los resultados son bastante aceptables para un sistema de Lennard-Jones\footnote{Como indiciábamos los siguientes términos involucran funciones de distribución de ordenes mayores y las expresiones y conocimiento necesarios del sistema de referencia se hacen complejos, por ello proponen una corrección por argumentos similares a la propuesta para los modelos duros, que sería en este caso,
\begin{equation}
\frac{1}{4}\rho\beta \kappa^{hs}\int d\vec{r}^{\prime}\left(\phi(\vec{r}^{\prime})\right) ^{2}g^{hs}(d_{BH},\eta)
\end{equation}
Algunos autores\cite{paricaud:154505} han propuesto sustituir $\kappa^{hs}$ por $\kappa^{hs}(1+\chi(\eta))$ con $\chi$ una función ajustable empíricamente basándose en argumentaciones igualmente mesoscópicas.
}. 
\subsubsection{Propuesta de Weeks-Chandler-Andersen}
En el caso de WCA\cite{PhysRevLett.25.149,Chandler05201983} la división del potencial propuesta es diferente y también lo es el método que proponen para estimar el sistema de referencia de esferas duras idóneo para un potencial blando. Su objetivo es que el sistema de referencia sea capaz de reproducir el factor de estructura del sistema completo para vectores de onda grandes\cite{weeks:5237}, partiendo de las observaciones sobre simulaciones de Verlet y los experimentos de Ashcroft y Lekner.\\

En su distinción del papel jugado por las contribuciones atractiva y repulsivas consideran que es la fuerza intermolecular, $-\phi'(r)$, la que juega un papel relevante y proponen una separación que estime la parte de la fuerza entre $\sigma$ y un $r_{0}$ dado por el mínimo del potencial. Esta argumentación lleva a un sistema de referencia repulsivo dado por la expresión,
\begin{figure}[htbp]
\begin{center}
\includegraphics[width=4.0in]{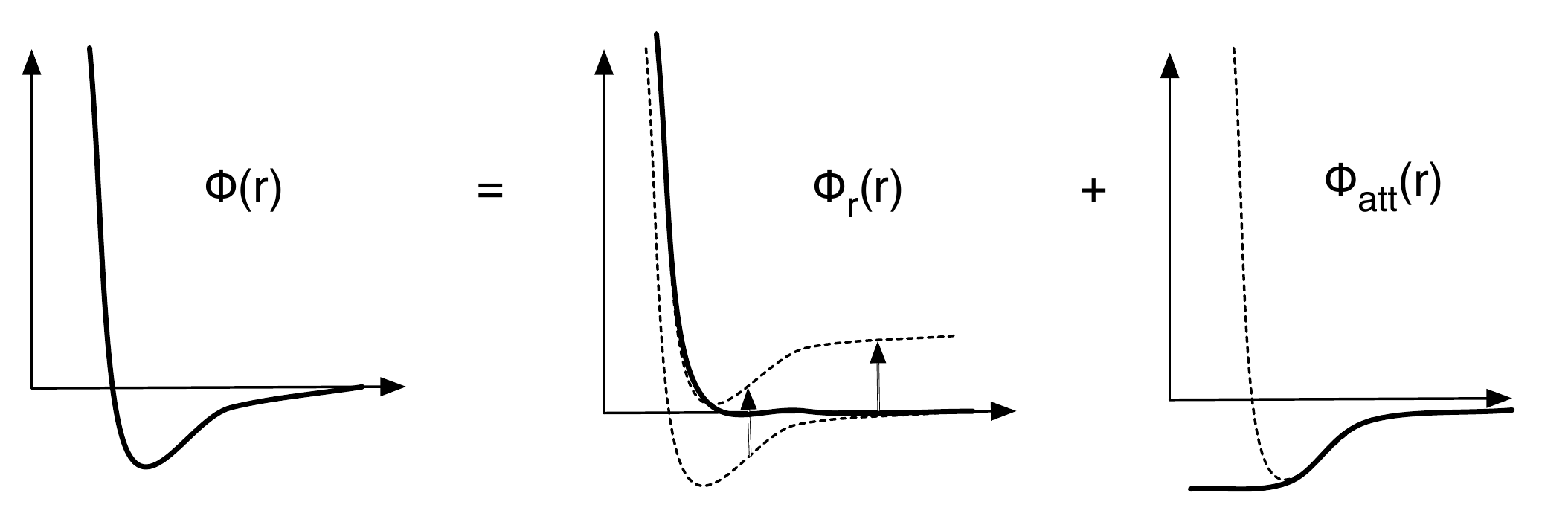}
\caption{División del potencial de interacción en la teoría WCA}
\label{fig:TeoriaWCA}
\end{center}

\end{figure}
\begin{eqnarray}
\phi_{R}(r)=\left\{\begin{array}{ll}
\epsilon+\phi(r),&r<r_{0}\\0,&r\ge r_{0}\end{array}\right.
\label{eqn:WCA1}
\end{eqnarray}
donde $r_{0}$ es el valor que hace mínimo al potencial completo o de modo equivalente que anula la fuerza intermolecular y entonces proponen que la imagen física de la teoría perturbativa se corresponde con $g(r,\phi)\approx g_{0}(r,\phi_{R})$. Igual que en el procedimiento anterior el potencial sobre el que se hace el tratamiento perturbativo es $\phi_{at}(r)=\phi(r)-\phi_{R}(r)$.\\

El modo de determinar el sistema de esferas duras adecuado resulta algo más detallado en esta teoría incorporando un desarrollo del factor de Boltzmann del potencial de referencia dado por ec.(\ref{eqn:WCA1}) entorno del potencial de esferas duras. Esto permite un desarrollo más sistemático que en el caso de Barker-Henderson, aunque al igual que en aquel, se basa en anular la primera corrección con una elección apropiada del diámetro de esferas duras\cite{PhysRevA.4.1597},
\begin{equation}
\beta\frac{F-F_{HS}}{N}=-\frac{T}{2}\rho\int d\vec{r}y_{HS}(r)\bigtriangleup f(r) +2\pi\rho\beta\int dr r^{2}g_{0}(r)\phi(r)+O(\beta^{2})
\end{equation}
donde $\bigtriangleup f(r)=\left[e^{-\beta\phi_{R}}-e^{-\beta\phi_{HS}}\right]$ se denomina \textit{función Blip}. El diámetro de esferas duras $d_{WCA}(T,\rho)$ verifica,
\begin{equation}
\int d\vec{r}y_{HS}(r)\bigtriangleup f(r)=0
\end{equation}
y en este caso depende de T y $\rho$. En la argumentación basada en la similitud necesaria entre la función de distribución radial del sistema completo y la del sistema de referencia, WCA considera por tanto,
\begin{equation}
g(r)\simeq g_{0}(r)=e^{-\beta u_{0}(r)}y_{0}(\vec{r})\simeq e^{-\beta u_{0}(r)}y_{HS}(\vec{r})
\end{equation}

  	\section{Teoría del funcional de la densidad}
\label{sec:DFT}
Hasta el momento la herramienta descrita para un análisis teórico de fluidos es el método de las ecuaciones integrales, centrado en la determinación de la función de correlación total y cuya contrastación experimental esta principalmente en las observaciones del factor de estructura. Además hemos analizado las funciones de distribución de partículas. Observamos dos puntos esenciales:
\begin{itemize}
\item El logaritmo de la función de partición macrocanónica puede ser considerada un funcional de una fugacidad generalizada que \textit{genera} formalmente las diferentes funciones de distribución mediante diferenciación funcional. Pueden existir otros funcionales que sean generadores de otras familias diferentes de funciones de correlación.
\item Se pueden entender las funciones de distribución desde su base mecánica y en consecuencia ser extendidas más allá del equilibrio, de modo que se reduzcan en las condiciones de equilibrio termodinámico a las expresiones conocidas.
\end{itemize}
Estas dos afirmaciones sugieren la posibilidad de construir un formalismo que haga explícitas estas propiedades. Inicialmente el desarrollo de este formalismo se hizo sobre la base de los desarrollos diagramáticos por \textit{Morita-Hiroike}\cite{PTP.23.1003,PTP.24.317,PTP.25.537} y \textit{Frank H. Stillinger-Buff}\cite{StillingerBuffDFT}. Se puede desarrollar además un enfoque replanteándolo como un problema variacional\cite{reviewEvans1979,lebowitz:116}.\\

La teoría del funcional de la densidad se constituye como un formalismo natural para el estudio de sistemas no homogéneos, en ella podemos encontrar cierta analogía con los métodos termodinámicos en el sentido de que las primeras derivadas de un funcional de la energía libre determinarán la situación de equilibrio del sistema mientras que las segundas derivadas se relacionaran con funciones respuesta (y sus criterios de estabilidad). En ella es sencillo obtener relaciones formales y de hecho obtendremos de manera directa la ecuación de Ornsterin-Zernique.

\subsection{Formalismo}
Comenzamos de nuevo con el Hamiltoniano ec. (\ref{eqn:Hamiltoniano}), que a partir de ahora referimos a potenciales externos de la forma,
\begin{equation}
\mathcal{V}_{N}(\vec{r}^{N})=\sum_{i=1}^{N}v_{ext}(\vec{r}_{i})
\end{equation}
que inducen una dependencia en la densidad de probabilidad $\mathcal{P}(\vec{r}^{N},\vec{p}^{N})=\mathcal{P}[v_{ext};\vec{r}^{N},\vec{p}^{N}]$ con el potencial externo, lo que provoca que la función densidad sea, de hecho, un funcional de este $\rho^{(1)}=\rho^{(1)}[v_{ext};\vec{r}]$. Resulta menos trivial\cite{reviewEvans1979} que dada una función densidad del sistema y fijadas las interacciones entre elementos del sistema $\Phi_{N}(\vec{r}^{N})$ en ec. (\ref{eqn:Hamiltoniano}), hay un único potencial externo $\mathcal{V}_{N}(\vec{r}^{N})$ que me la determina, esto permite establecer para una determinada función de densidad $\rho^{(1)}$ una distribución de probabilidad $\mathcal{P}(\vec{r}^{N},\vec{p}^{N})$ y permite también postular la existencia de un funcional de Helmholtz \emph{intrínseco} que escribimos,
\begin{equation}
\mathcal{F}[\mathcal{P}]=\frac{h^{-3N}}{N!}\int d\vec{p}^{N}\int d\vec{r}^{N}\mathcal{P}(\vec{r}^{N},\vec{p}^{N})(K_{N}+\Phi_{N}+\frac{1}{\beta}ln(\mathcal{P}(\vec{r}^{N},\vec{p}^{N}))
\end{equation}
y donde la distribución de densidad $\mathcal{P}(\vec{r}^{N},\vec{p}^{N})$ es la determinada por la función densidad.\\

Podemos definir dos funcionales además del anterior,
\begin{equation}
F[\mathcal{P}]=\mathcal{F}[\mathcal{P}]+\frac{h^{-3N}}{N!}\int d\vec{p}^{N}\int d\vec{r}^{N}\mathcal{P}(\vec{r}^{N},\vec{p}^{N})\mathcal{V}_{N}(\vec{r}^{N})
\end{equation}
que denominaremos funcional de Helmholtz y donde aparece $v_{ext}$ y no es por tanto \emph{intrínseco} al modelo de interacción $\Phi_{N}$ contenido en la dinámica del sistema. La otra definición es el funcional
\begin{align}
\Omega[\mathcal{P}]=F[\mathcal{P}]-\frac{h^{-3N}}{N!}\int d\vec{p}^{N}\int d\vec{r}^{N}\mathcal{P}(\vec{r}^{N},\vec{p}^{N})\mu N
\end{align}
Ahora si utilizamos el hecho de que la densidad de probabilidad esta unívocamente determinada por la función densidad podemos escribir $\mathcal{F}[\rho]$ y por tanto,
\begin{equation}
F[\rho]=\mathcal{F}[\rho]+\int d\vec{r}\rho(\vec{r})v_{ext}(\vec{r})
\end{equation}
así como
\begin{equation}
\Omega[\rho]=\mathcal{F}[\rho]+\int d\vec{r}\rho(\vec{r})v_{ext}(\vec{r})-\mu\int d\vec{r}\rho(\vec{r})
\end{equation}
sobre este funcional se puede demostrar su carácter extremal en la densidad de probabilidad de equilibrio que denotamos por $\mathcal{P}_{eq}$,
\begin{equation}
\Omega[\mathcal{P}]\geqslant\Omega[\mathcal{P}_{eq}]
\label{eqn:VariacionalOmega}
\end{equation}
dándose la igualdad solo en el equilibrio, en tal caso el funcional representa la energía libre en el colectivo macrocanónico. Como hemos señalado el funcional anterior puede ser entendido como un funcional de la densidad $\rho$ y si establecemos que la densidad que determina $\mathcal{P}_{eq}$ es $\rho_{eq}$ podemos escribir,
\begin{equation}
\Omega[\rho]\geqslant\Omega[\rho_{eq}]=\Omega
\end{equation}
esta ecuación indica que evaluando el funcional en la densidad de equilibro obtenemos la energía libre del colectivo macrocanónico. Resulta útil expresar la ecuación anterior que define como determinar la densidad de equilibrio en un problema variacional, mediante una ecuación de Euler-Lagrange, que en vista de las definiciones anteriores queda como,
\begin{equation}
\frac{\delta\mathcal{F}[\rho]}{\delta\rho(\vec{r}_{1})}+v_{ext}(\vec{r}_{1})-\mu=0
\label{eqn:EulerLagrangeMinimizacion}
\end{equation}
en esta ecuación $\mu$ tiene el carácter de un potencial externo constante. Definimos entonces un \emph{potencial generalizado} mediante $u(\vec{r})=\mu-v_{ext}(\vec{r}_{1})$, y definimos un \emph{potencial químico intrínseco} como,
\begin{equation}
\frac{\delta\mathcal{F}[\rho]}{\delta\rho(\vec{r}_{1})}=\mu_{in}[\rho;\vec{r}_{1}]
\end{equation}
Hemos llegado la expresión práctica de la formulación variacional (\ref{eqn:VariacionalOmega}) en el colectivo macrocanónico\footnote{Para definir en el colectivo canónico algo similar  tendríamos que acudir a la definición de problema variacional isoperímetro para llegar a unas ecuaciones en una forma similar a las de Euler-Lagrange de arriba pero con la utilización de multiplicadores de Lagrange. En este caso el multiplicador de Lagrange sería el potencial químico\cite{yang:3732}.}.\\

La demostración rigurosa del problema inverso\footnote{Una vez demostrada la existencia, el problema de la unicidad suele ser sencillo.} de determinar la existencia de un potencial externo que da una densidad dada en el equilibrio puede ser visto en la referencia \cite{Chayes} donde se comenta en que condiciones estrictas puede darse, mientras que un análisis del formalismo desde el punto de vista de la propiedad de convexidad en el funcional esta contenida en \cite{ConvexidadWDAcalliol}, es interesante observar que los funcionales $\Omega[u]$ y $\mathcal{F}[\rho]$ están relacionados por una transformada de Legendre, lo que permite formular como un problema variacional al mismo nivel la interpretación de $\Omega[u]$.  

\subsection{Sistema Ideal y Parte de Exceso}
En el caso de un sistema ideal $\Phi_{N}(\vec{r}^{N})=0$ y  $\mathcal{P}(\vec{r}^{N},\vec{p}^{N})$ podemos factorizarla como producto de funciones que involucran una sola partícula y encontrar una expresión explícita para el funcional de la densidad,
\begin{equation}
\mathcal{F}_{id}[\rho]=\beta^{-1}\int d\vec{r}\rho(\vec{r})[ln(\Lambda^{3}\rho(\vec{r}))-\rho(\vec{r})]
\end{equation}
del que aplicando la condición de equilibro obtenemos la ley barométrica,
\begin{equation}
\rho_{eq}(\vec{r})=e^{-\beta\mu}e^{-\beta v_{ext}(\vec{r})}
\end{equation}
por otra parte la división de la función de partición en parte ideal y de exceso permite escribir nuestro funcional general como,
\begin{equation}
\mathcal{F}[\rho]=\mathcal{F}_{id}[\rho]+\mathcal{F}_{ex}[\rho]
\end{equation}
Cabe hacer dos consideraciones generales:
\begin{itemize}
\item Como es natural es sobre la parte de exceso sobre la que se realizan las diferentes aproximaciones para tratar sistemas con interacciones y conviene definir funciones de correlación basadas en ella.
\item Su construcción no va a necesitar fijar un potencial externo concreto sino que se basa únicamente en las interacciones entre partículas del sistema, y por tanto mediante la relación con el funcional de Helmholtz puede ser aplicado a sistema no-homogéneos de naturaleza muy diferente.
\end{itemize}

\subsection{Jerarquías de funciones de correlación}
\label{ss:jerarquias}
Como comentamos anteriormente, en el caso de la función de partición de un sistema homogéneo de
las primeras derivadas obtenemos propiedades termodinámicas del estado de equilibrio y mediante segundas derivadas
obtenemos criterios de estabilidad y funciones respuesta de mi sistema. En el caso del funcional de la densidad
mediante diferenciación funcional se obtiene una serie de familias o jerarquías de funciones (que ya sugerimos desde un desarrollo diagramático), las dos primeras funciones que se obtienen son claves en la teoría de líquidos, \\

Con la definición anterior de $u(r)$ tenemos que $\Omega[u;V,T]$ es un funcional de potencial externo generalizado, con lo que derivando funcionalmente tenemos,
\begin{equation}
\frac{\delta\Omega}{\delta u(\vec{r})}=-<\hat{\rho}(\vec{r})>=-\rho_{eq}(\vec{r})
\label{eqn:primeraderivadaomegarespdeu}
\end{equation}
\begin{equation}
\frac{\delta^{2}\Omega}{\delta u(\vec{r}_{1})\delta u(\vec{r}_{2})}=G^{(2)}(\vec{r}_{1},\vec{r}_{2})
\end{equation}
Resultado que anunciábamos en \S\ref{sec:desarrolloDiagramatico} y que reproduce de modo sencillo la ecuación (\ref{eqn:laFuncionG2}). Si deseamos interpretar esta función podemos ver que para un sistema en equilibrio en un estado estable un pequeño cambio en el potencial externo vendrá acompañado de un leve cambio en la densidad,
\begin{equation}
\delta\rho(\vec{r})=\int d\vec{s}\frac{\delta \rho(\vec{r})}{\delta u(\vec{s})}\delta u(\vec{s})=\int d\vec{s}\beta^{-1}G^{(2)}(\vec{r},\vec{s})\delta u(\vec{s})=-\int d\vec{s}\chi(\vec{r},\vec{s})\delta u(\vec{s})
\label{eqn:respuestalinealDFT}
\end{equation}
luego la propia definición en este contexto de $G^{(2)}$ permite relacionarla con la \emph{función respuesta lineal estática en la densidad} $\chi$ y por tanto esta última con el factor de estructura del líquido.\\

Igualmente para el funcional $\mathcal{F}[\rho]$ tenemos la siguiente jerarquía de funciones,
\begin{equation}
\frac{\delta F}{\delta \rho(\vec{r})}=\mu_{in}[\rho;r]=C^{(1)}[\rho;\vec{r}]
\end{equation}
\begin{equation}
\frac{\delta^{2}F}{\delta \rho(\vec{r}_{1})\delta \rho(\vec{r}_{2})}=C^{(2)}[\rho;\vec{r}_{1},\vec{r}_{2}]
\end{equation}
A partir de esta jerarquía\footnote{Formalmente podemos definir el primer miembro de la jerarquía como $c^{(0)}=-\beta F$} podemos definir otra de más utilidad directa mediante la separación antes introducida entre la parte ideal y de la parte de exceso,
\begin{equation}
\frac{\delta F_{ex}}{\delta \rho(\vec{r})}=\mu_{in}[\rho;r]=c^{(1)}[\rho;\vec{r}]=C^{(1)}[\rho;\vec{r}]+ln(\lambda^{3}\rho(\vec{r}))
\label{eqn:primeraderivadafexrespderho}
\end{equation}
\begin{equation}
\frac{\delta^{2}F_{ex}}{\delta \rho(\vec{r}_{1})\delta \rho(\vec{r}_{2})}=c^{(2)}[\rho;\vec{r}_{1},\vec{r}_{2}]=
C^{(2)}[\rho;\vec{r}_{1},\vec{r}_{2}]+\frac{\delta(\vec{r}_{1}-\vec{r}_{2})}{\rho(\vec{r}_{2})}
\end{equation}
la segunda de las ecuaciones se denomina \textit{segunda ecuación de Yvon} mientras que la primera de las ecuaciones, haciendo uso de la condición de mínimo del funcional $\Omega$ para la densidad de equilibrio me lleva a expresar la densidad en forma de ecuación auto-consistente:
\begin{equation}
ln\rho_{eq}(\vec{r})=-\beta(\mu+v_{ext}(\vec{r}))+c^{(1)}[\rho_{eq};\vec{r}]
\label{eqn:eqautoconsistenterho}
\end{equation}
mirando las ecuaciones (\ref{eqn:primeraderivadaomegarespdeu})
 y (\ref{eqn:primeraderivadafexrespderho})
 hemos establecido una relación para la densidad de equilibro, entre las primeras derivadas funcionales de dos jerarquías diferentes la segunda de las cuales proviene de separar explícitamente la parte de interacción del potencial de las contribuciones cinéticas (parte ideal). Como consecuencia el termino $c^{(1)}$ puede ser interpretado como un potencial efectivo creado por las interacciones para dar lugar al perfil de densidad en el equilibrio\cite{BOOK-TheorySimpleLiquids}.  Podemos intentar establecer también una relación entre ambas jerarquías al nivel de segundas derivadas funcionales lo que dará lugar a la conocida ecuación de Ornstein-Zernique, que fue introducida antes a partir de un desarrollo diagramático.\\

\subsection{Ecuación Ornstein-Zernique. Relaciones de cierre}
Las dos familias de funciones están relacionadas ya que,
\begin{equation}
\int d\vec{r_{3}}C^{(2)}(\vec{r}_{1},\vec{r}_{3})G^{(2)}(\vec{r}_{3},\vec{r}_{2})=\int d\vec{r}_{3}\frac{\delta \rho(\vec{r}_{3})}{\delta u(\vec{r}_{2})}\frac{\delta u(\vec{r}_{1})}{\delta \rho(\vec{r}_{3})}=\delta(\vec{r}_{1}-\vec{r}_{3})
\end{equation}
la sustitución de las expresiones anteriores nos permite escribir la expresión anterior como la ecuación de Ornstein-Zernique dada anteriormente ec.(\ref{eqn:OZ})
\begin{equation}
h^{(2)}(\vec{r}_{1},\vec{r}_{2})=c^{(2)}(\vec{r}_{1},\vec{r}_{2})+\int d\vec{r_{3}}h^{(2)}(\vec{r}_{1},\vec{r}_{3})\rho(r_{3})c^{(2)}(\vec{r}_{3},\vec{r}_{2})
\end{equation}
aplicada a un fluido no homogéneo. Si expresamos la función de correlación total $h^{(2)}(\vec{r}_{1},\vec{r}_{2})$ como,
\begin{equation}
h^{(2)}(r_{1},r_{2})=-\frac{\delta(\vec{r}_{1}-\vec{r}_{2})}{\rho(\vec{r}_{1})}-\frac{1}{\beta\rho(\vec{r}_{1})\rho(\vec{r}_{2})}\frac{\delta\rho(\vec{r}_{1})}{\delta v(\vec{r}_{2})}
\end{equation}
tenemos la \textit{primera ecuación de Yvon}.\\

En el caso de un sistema homogéneo resultado la ecuación de Ornstein-Zernique usual,
\begin{equation}
h^{(2)}(|\vec{r}_{12}|)=c^{(2)}(|\vec{r}_{12}|)+\int d\vec{r}_{3}h^{(2)}(|\vec{r}_{13}|)\rho(\vec{r}_{3})c^{(2)}(|\vec{r}_{23}|)
\end{equation}
que en el espacio de Fourier permite una interpretación sencilla ya que las convoluciones se transforman en productos quedando,
\begin{equation}
\hat{h}(k)=\hat{c}(k)+\rho \hat{c}(k)\hat{h}(k)=\frac{\hat{c}(k)}{1-\rho \hat{c}(k)}=\hat{c}(k)+\rho \hat{c}(k)^{2}+\rho^{2}\hat{c}(k)^{3}+...
\end{equation}
El desarrollo presentado expresa el hecho de que la correlación total entre dos partículas viene determinada por la correlación directa cuando median sucesivamente más y más partículas.\\

  	\subsection{Aproximación local y de Gradientes Cuadrados}
Hasta el momento no hemos escrito ninguna expresión funcional concreta salvo en el caso de un sistema ideal. La aproximación más sencilla es hacer $\mathcal{F}[\rho]$ igual a la energía libre por unidad de volumen de un sistema homogéneo teóricamente más sencilla de determinar. Tal identificación es factible siempre que el gradiente de la densidad se aproxime a cero en tal caso podemos suponer que se comporta localmente como un sistema homogéneo de densidad $\rho$. Una teoría así construida recuperará por definición el límite homogéneo correctamente, sin embargo tratará igual a regiones con igual densidad aunque en sus proximidades la densidad sea diferente.  Esta aproximación se conoce como \emph{Local Density Aproximation} (LDA) y podemos expresar,

\begin{equation}
\mathcal{F}[\rho]=\int d\vec{r} \psi[\rho(\vec{r})]
\end{equation}

Podemos incluir términos que si den cuenta de las variaciones en la densidad en la proximidades de una partícula si desarrollamos la función densidad de energía libre $\psi$ sobre la función densidad, y aunque no tenemos garantizada, \textit{a priori}, la convergencia de este desarrollo podemos esperar al menos una validez asintótica. Tendremos\cite{reviewEvans1979,RowlinsonWidom},

\begin{equation}
 \psi[\rho(\vec{r})]=\psi^{(0)}(\rho(\vec{r}))+\nabla\psi\nabla\rho+...+\psi^{(2)}(\vec{r})|\vec{\nabla}\rho(\vec{r})|^{2}+O(\vec{\nabla}^4)
\end{equation}

La invariancia rotacional impone una expresión de la forma,

\begin{equation}
\mathcal{F}[\rho]=\int d\vec{r} \left[\psi^{(0)}(\rho(\vec{r}))+\psi^{(2)}(\rho(\vec{r}))|\vec{\nabla}\rho(\vec{r})|^{2}+O(\vec{\nabla}^4)\right]
\end{equation}

Donde las funciones $\psi^{(0)}$ y $\psi^{(2)}$ están sin determinar. Para encontrarlas podemos realizar una expansión en densidad del funcional $\mathcal{F}[\rho]$ general y utilizar las expresiones de las derivadas funcionales anteriores sobre la densidad uniforme $\rho_{u}$, de este modo\cite{RowlinsonWidom} tendríamos expresiones para ambas funciones\footnote{Consistentes con la teoría de la respuesta lineal.},

\begin{equation}
\frac{d\psi^{(0)}(\rho_{u})}{d\rho_{u}}=\mu(\rho_{u})
\end{equation}
\begin{equation}
\psi^{(2)}(\rho_{u})=\frac{1}{12}\beta^{-1}\int dr r^{2}c^{(2)}(\rho_{u};r)
\label{eqn:gradientesCuadradosCoef2}
\end{equation}

Si restringimos aun más el valor de $\psi^{(2)}$ y lo hacemos independiente de la densidad $\rho_{u}$, y aproximamos la función de correlación directa solo por la parte atractiva del potencial obtendríamos, llamado $\tau$ al resultado,

\begin{equation}
\psi^{(2)}=\frac{1}{12}\beta^{-1}\int dr r^{2}\phi(r)\equiv \tau
\label{eqn:definicionTAU}
\end{equation}

Recalcar que hay una suposición implícita en este desarrollo, a saber, que las integrales anteriores que me definen $\psi^{(2)}$ existen lo que es válido para potenciales de corto alcance pero no para cualquier potencial de largo alcance, como los polinómicos de la forma $r^{-n}$.\\

\label{sec:GradCuadrado}
  	\section{Teorías de la densidad promediada}
\label{sec:teoriasDensidadPromediada}
La idea que engloba este conjunto de aproximaciones funcionales es construir el funcional de energía libre a partir de una cierta densidad no local $\bar{\rho}$ definida mediante la convolución de la densidad $\rho$ con una función peso que se suele extender dentro del alcance de las interacciones\footnote{Un ejemplo se ve en la teoría de campo medio donde en este caso la función peso es el propio potencial de interacción
\begin{equation}
\mathcal{F}[\rho]=\frac{1}{2}\int d\vec{r}_{1}d\vec{r}_{2}\rho(\vec{r}_{1})\rho(\vec{r}_{2})\phi(|\vec{r}_{1}-\vec{r}_{2}|)
\end{equation}
}. \\

En este punto emergen una serie de propuestas diferentes conocidas genéricamente como WDA (Weighted Density Aproximation), inicialmente introducidas por \textit{Nordholm}  y desarrolladas por \textit{Tarazona}\cite{PhysRevA.31.2672}, y poco después por \textit{Curtin y Ashcroft}\cite{PhysRevA.32.2909}. Cuyo objetivo es construir un funcional cuya reducción al caso homogéneo obtenga una termodinámica, ecuación de estado y reglas de suma, y una estructura, funciones de correlación, adecuadas al conocimiento existente de sistemas uniformes. \\

Además existen otro conjunto de funcionales de la densidad construidos sin imponer sobre ellos las propiedades del sistema uniforme sino bajo otras premisas y que también reproducen propiedades deseables cuando son evaluados en una densidad uniforme $\rho(\vec{r})=\rho$. A este conjunto de aproximaciones se las suele agrupar con el nombre de \textit{teorías de la medida fundamental} (FMT). Ambos esquemas han tenido un éxito notable describiendo sistemas no-homogéneos.\\

En general las aproximaciones locales y de gradiente cuadrado fallan cuando los perfiles de densidad no son suaves en el rango de las interacciones aunque pueden describir adecuadamente algunos comportamientos donde el perfil de densidad varía poco, como por ejemplo una interfase cerca del punto crítico. En consecuencia las aproximaciones de la densidad promediada (WDA y FMT) intentan resolver las deficiencias que posee cualquier teoría local del funcional de la densidad cuando intentan describir situaciones donde los efectos de exclusión de partículas si son importantes. Desde el punto de vista de las correlaciones representan una mejora sustancial en la descripción de estas en la zona repulsiva del potencial, de hecho parte del éxito de estas aproximaciones es la capacidad para incorporar o reproducir el conocimiento previo, que por medio de la teoría de ecuaciones integrales se tiene, de las correlaciones de un sistema de esferas duras, básicamente la aproximación de Percus-Yevick. Desde un punto de vista más formal en ocasiones representa además una teoría más general y unificada de otras teorías ya existentes y por tanto con un interés teórico intrínseco.

\subsection{WDA}
\label{subsection:WDA}
La idea es partir de la energía libre de exceso escrita de la siguiente manera,
\begin{equation}
\mathcal{F}_{ex}[\rho]=\int d\vec{r}\rho(\vec{r})\Delta\psi(\bar{\rho}(\vec{r}))
\end{equation}
donde hemos definido una densidad promediada usando una función peso, aun por determinar.
\begin{equation}
\bar{\rho}(\vec{r})=\int d\vec{r}'\rho(\vec{r}')\omega(|\vec{r}-\vec{r}'|;\bar{\rho}(\vec{r}))
\label{eqn:pesoWDA-T2}
\end{equation}

La primera recuerda a la aproximación local pero evaluada en la densidad promediada, mientras que la segunda ecuación es una definición de esta ultima por medio de una ecuación integral. De esta manera, mientras que la densidad promediada puede ser lo suficientemente suave como para que la primera ecuación sea una descripción correcta, la segunda permite que la función densidad si que esa altamente no-homogénea en el rango de las interacciones, y podemos seguir utilizando como $\Delta\psi$ la densidad de energía libre de exceso por partícula de un sistema uniforme. Queda por determinar la función peso. Aquí emergen dos aproximaciones diferentes ambas debidas a \textit{Tarazona}. La primera aproximación que denominaremos WDA-T1 consiste en definir una peso independiente de la densidad,

\begin{equation}
\omega(|\vec{r}-\vec{r}'|)=\frac{3}{4\pi\sigma^{3}}\theta(\sigma-|\vec{r}-\vec{r'}|)
\label{eqn:pesoWDA-T1}
\end{equation}

De esta manera ya se puede reproducir de modo cualitativo situaciones como un fluido en contacto con una barrera o un sólido de esferas duras, de hecho es capaz de describir el proceso de \emph{drying} completo de la interfase pared-líquido y observar una estructuración en capas (\emph{layering}) en determinados casos aunque no es capaz de reproducir sus resultados con precisión. A la hora de incorporar la información del sistema homogéneo tenemos dos expresiones,

\begin{equation}
\mathcal{F}[\rho(\vec{r})]\bigg\vert_{( \rho(\vec{r})->\rho)}=F(\rho)
\label{eqn:sistemaHomogeneo}
\end{equation}
\begin{equation}
c^{(2)}(\vec{r},\vec{r'})=-\beta\frac{\delta^{2}\mathcal{F}_{ex}[\rho]}{\delta\rho(\vec{r})\delta\rho(\vec{r}')}
\end{equation}

La primera impone una normalización concreta sobre la función peso $\omega$, y la segunda determina la estructura de correlación incorporada o incorporable al funcional. En el caso de WDA-T1 con la función peso (\ref{eqn:pesoWDA-T1}) y con la necesidad de reproducir para el valor homogéneo $F(\rho)$ la ecuación de estado de Carnahan-Starling ya tenemos determinado el funcional y por tanto las correlaciones que implícitamente hemos incluido en él.\\

Si bien se encuentra un funcional con buenas propiedades realmente tenemos un mayor conocimiento previo de la función de correlación de esferas duras, en particular la aproximación de Percus-Yevick. Por esta razón Tarazona propuso\cite{PhysRevA.31.2672,PhysRevA.32.3148} para (\ref{eqn:pesoWDA-T2}) una definición más general de la función peso. Puesto que la propia función peso depende de la densidad promediada esta ecuación debería resolverse de un modo autoconsistente, sin embargo un camino más sencillo es,

\begin{equation}
\omega(r;\rho)=\omega_{0}(r)+\omega_{1}(r)\rho+\omega_{2}(r)\rho^{2}
\end{equation}

que propone un desarrollo en la densidad para la función peso y donde la ecuación (\ref{eqn:sistemaHomogeneo}) determina la normalización de la función $\omega(r;\rho)$,

\begin{equation}
\int d\vec{r}\omega(\vec{r};\rho)=1
\end{equation}
que dentro del desarrollo propuesto lleva a que,
\begin{equation}
\int d\vec{r}\omega_{0}(\vec{r})=1
\end{equation}
\begin{equation}
\int d\vec{r}\omega_{1,2}(\vec{r})=0
\end{equation}

Si ahora introducimos esta información en $c^{(2)}(\vec{r},\vec{r}')$  y comparamos con el caso homogéneo vemos que tenemos que proponer una ecuación de estado (una $F_{ex}$) y una función de correlación e introducir ambas como un desarrollo del virial. El proceso elegido por \textit{Tarazona} fue introducir Carnahan-Starling como ecuación de estado y Percus-Yevick como función de correlación directa. Como Carnahan-Starling es semi-empírica no se conoce la función de correlación de Carnahan-Starling dentro de un desarrollo analítico, por tanto imponer que se reproduzca Percus-Yevick en el mayor rango posible de densidades es únicamente pedir algo razonable como función de correlación directa para el funcional\footnote{Hay con todo una pequeña inconsistencia en la medida en que fijar un desarrollo del virial para la energía libre me determina un desarrollo dado para la compresibilidad isoterma que no tiene porque coincidir con el determinado a partir de la $c^{(2)}_{PY}$.}.\\

Una propiedad interesante de WDA es que en el límite bidimensional cuando confino el sistema sigue proporcionando una descripción decente del sistema esto es importante y permite una descripción realista de fenómenos de estructuración en una dirección espacial. En general este funcional es capaz de tratar gran variedad sistemas altamente no homogéneos y se ha usado con éxito para determinar desde problemas de solidificación hasta líquidos complejos\cite{1995JPCM7.5753S}.

\subsection{Teorías de la Medida Fundamental}
\label{subsection:FMT}
Como indicábamos esta familia de aproximaciones más que basarse en desarrollos aproximados intentan encontrar expresiones cuyo contenido físico sea sugerente y veremos como el impulso más importante en estas aproximaciones se deberá a las aportaciones de \textit{Rosenfeld} que comenzaron en estudios de sistemas uniformes. El primer funcional de este tipo fue introducido como una generalización directa al caso tridimensional desde la solución exacta al problema unidimensional de segmentos duros por \textit{Percus}\cite{percus:1316}, en el podemos expresar la energía libre de exceso mediante\cite{1981JSPRobledoVarea},
\begin{equation}
F_{ex}[\rho]=\int d\vec{r} \bar{n}_{1}(\vec{r})\phi_{ex}(\bar{n}_{2}(\vec{r}))
\end{equation}
donde $\phi_{ex}$ será la energía libre de exceso del caso uniforme y las densidades $\bar{n}_{1}$ y $\bar{n}_{2}$, son densidades promedio,
\begin{equation}
\bar{n}_{\alpha}(\vec{s})=\int d\vec{r}\sigma_{\alpha}(\vec{r})\rho(\vec{r}+\vec{s})
\label{eqn:densidadespromedio}
\end{equation}
mediante dos pesos dados respectivamente por la expresiones,
\begin{equation}
\sigma_{1}(\vec{r})=\frac{4}{\pi R^{2}}\delta(|\vec{r}|-R)
\end{equation}
\begin{equation}
\sigma_{2}(\vec{r})=\frac{6}{\pi}\frac{8}{R^{3}}\theta(|\vec{r}|-R)
\end{equation}
Sin embargo este funcional da unos resultados bastante modestos\cite{ComprobacionesPerus3D}. Su merito estriba en que abre una vía para introducir funcionales más elaborados. En principio podemos introducir un conjunto de funciones peso mayor pero de manera que el funcional posea la misma estructura,
\begin{equation}
F_{ex}[\rho]=\int d\vec{r} \Phi(\bar{n}_{1},\bar{n}_{2},...)
\label{eqn:estructuraBaseFMT}
\end{equation}
donde tal y como apunta Percus no tenemos porque restringirnos a funciones peso escalares. Una vez propuesto un funcional de esta forma aun tenemos que imponer propiedades para ser un candidato adecuado. Dos propiedades evidentes son, el limite homogéneo que ha de recuperar resultados coherentes y las propiedades de estabilidad adecuadas para tener realmente un funcional convexo, es decir imponemos restricciones sobre la segunda derivada funcional. Esto proporciona criterios razonables para encontrar funciones peso plausibles pero no es una vía que proporcione un funcional único. 

\subsubsection{Teoría original de Rosenfeld}
\label{sec:teoriaRosenfeld}
Hemos descrito tanto la teoría de la partícula escalada para obtener la ecuación de estado en el caso de un sistema de esferas como la función de correlación de Percus-Yevick, ambas resultan ser clave en la descripción y estudio de sistemas uniformes. Del trabajo de Rosenfeld surge una visión unificada de ambas teorías, ya que mediante hipótesis básicas sobre el funcional de energía libre es posible obtener las propiedades del sistema uniforme (y no al contrario), el proceso seguido por \textit{Rosenfeld} se detalla en las referencias\cite{PhysRevE.50.R3318,rosenfeld:4865,rosenfeld:4305,rosenfeld:4272}. Desde el punto de vista de la ecuación básica (\ref{eqn:estructuraBaseFMT}) es necesario determinar el conjunto adecuado de pesos y la forma funcional de $\Phi(\bar{n}_{1},\bar{n}_{2},...)$. Rosenfeld propuso aumentar los pesos del caso unidimensional mediante un peso vectorial,
\begin{equation}
 \vec{\omega}(\vec{r})=\hat{r}\frac{1}{4\pi R^{2}}\delta(R-|\vec{r}|)
\end{equation}
y donde la densidad promedio se construye como en la ec. (\ref{eqn:densidadespromedio}). La construcción de la forma funcional $\Phi(\bar{n}_{\alpha})$ respeta el limite exacto a baja densidad y la función de correlación directa de Percus-Yevick mediante,
\begin{equation}
 \Phi_{Rosenfeld}=\sum_{i=1}^{3}\phi_{i}(\eta(\vec{r}),n(\vec{r}),\vec{v}(\vec{r}))
\end{equation}
donde
\begin{subequations}
 \begin{equation}
 \phi_{1}=-nlog(1-\eta)
 \end{equation}
 \begin{equation}
 \phi_{2}=4\pi R^{3}\frac{n^{2}-\vec{v}\cdot\vec{v}}{1-\eta}
 \end{equation}
 \begin{equation}
 \phi_{3}=8\pi^{2} R^{6}n\frac{n^{2}/3-\vec{v}\cdot\vec{v}}{(1-\eta)^{2}}
 \end{equation}
\end{subequations}

Este funcional contiene ciertas ventajas conceptuales respecto de los funcionales de la densidad promediada previos pero presenta algunos problemas en al descripción del cristal de esferas duras.

\subsubsection{Modelo de Kierlik-Rosinberg}

Conviene preguntarse si es posible determinar unívocamente las funciones peso. \textit{Kierlik y Rosinberg}\cite{PhysRevA.44.5025,PhysRevA.42.3382,PhysRevE.48.618} probaron una definición diferente de este conjunto donde no se usan pesos de carácter vectorial, el precio a pagar es tener derivadas de la función $\delta(|\vec{r}|-R)$ como parte de las definiciones de los pesos\footnote{Este conjunto de funciones peso resulta dar un resultado equivalente a la teoría original de Rosenfeld en el sentido en que $\int d\vec{r}[\Phi_{Rosenfeld}-\Phi_{KR}]=0$.}, esto toda vez que siempre aparecen integradas no supone un formalmente problema. Las cuatro funciones peso diferentes, pero todas ellas escalares, resultan,
\begin{equation}
\omega^{(3)}=\theta(R-r)
\end{equation}
\begin{equation}
\omega^{(2)}=\delta(R-r)
\end{equation}
\begin{equation}
\omega^{(1)}=\frac{1}{8\pi}\delta'(R-r)
\end{equation}
\begin{equation}
\omega^{(0)}=-\frac{1}{8\pi}\delta''(R-r)+\frac{1}{2\pi r}\delta'(R-r)
\end{equation}
La función $\Phi(\bar{n}_{\alpha})$ se escribe como,
\begin{equation}
\Phi_{KR}=-\bar{n}_{0}log(1-\bar{n}_{0})+\frac{\bar{n}_{1}\bar{n}_{2}}{1-\bar{n}_{3}}+\frac{1}{24\pi}\frac{\bar{n}_{2}^{3}}{(1-\bar{n}_{3})^{2}}
\end{equation}
Una consecuencia posible de este funcional es que las hipótesis incluidas a la teoría de Rosenfeld (acerca de la descomposición basándose en la teoría de la partícula escalada) dan lugar a un funcional bien fundamentado ya que diferentes versiones de las funciones peso dan lugar a resultados equivalentes. En general subsiguientes desarrollos prefieren la elección de Rosenfeld, de hecho a partir de ella se han hecho generalizaciones a partículas duras convexas no esféricas.\\

En este funcional de \textit{Kierlik y Rosinberg} es directo ver como se comporta el funcional en el caso de una distribución de densidad $\rho(\vec{r})=\rho^{2d}\delta(z)$ y ver como los resultados son muy similares a los que se obtienen en el caso uniforme desde SPT. El caso de WDA-T2 la coincidencia es razonable aunque en fracciones de empaquetamiento altas sobreestima a la teoría de la partícula escalada. A la cuestión de como mejorar las propiedades del funcional de Rosenfeld para distribuciones confinadas como la anterior responde la siguiente aproximación.

\subsubsection{Interpolación dimensional}
El funcional de Rosenfeld (o equivalentemente KR) resulta predecir adecuadamente muchos fenómenos excepto el fenómeno de solidificación que si resulta bien descrito cualitativamente en WDA. La idea que surge aquí y que explica este hecho es que un funcional que describa bien situaciones donde hemos reducido la dimensión efectiva del sistema debe ser capaz de describir correctamente un sólido de esferas duras\cite{PhysRevLett.84.694}, sin embargo el funcional de Rosenfeld no posee esta capacidad, lo que lleva a intentar definir un funcional que permita reproducir resultados adecuados al reducir la dimensión del sistema\cite{PhysRevE.55.R4873,PhysRevE.55.4245}, en particular reproducir el resultado exacto de segmentos duros.\\

La idea original de FMT es incluir como medida fundamental las pro\-pie\-da\-des geo\-mé\-tri\-cas de las partículas más que reproducir (utilizar como medida fundamental) la ecuación de Mayer. La vía de Rosenfeld fue obtener el funcional a partir de la descomposición similar a la teoría de la partícula escalada, la vía de la \textit{interpolación dimensional} intenta obtener este funcional sometiendo al sistema a confinamientos restringidos, como una cavidad que solo pueda albergar una partícula o una cavidad que reduzca de modo efectivo la dimensionalidad del sistema a uno. La restricción de reproducir los diferentes confinamientos da lugar a un funcional tipo FMT. Si reescribimos la teoría original de Rosenfeld usando una función $\phi_{0}(\eta)$ tendremos tres términos en la energía de exceso de esferas duras a saber,
\begin{equation}
\Phi_{1}=\int d\vec{r}n(\vec{r})\frac{d\phi_{0}(\eta(\vec{r}))}{d\eta}
\end{equation}
\begin{equation}
\Phi_{2}=\pi R^{3}\int d\vec{r}\left[n(\vec{r})^{2}-\vec{v}(\vec{r})\cdot\vec{v}(\vec{r})\right]\frac{d^{2}\phi_{0}(\eta(\vec{r}))}{d\eta^{2}}
\end{equation}
\begin{equation}
\Phi_{3}=8\pi^{2}R^{6}\int d\vec{r}n(\vec{r})\left[\frac{1}{3}n(\vec{r})^{2}-\vec{v}(\vec{r})\cdot\vec{v}(\vec{r})\right]\frac{d^{3}\phi_{0}(\eta(\vec{r}))}{d\eta^{3}}
\label{eqn:FMTrosenfeld3}
\end{equation}
escribimos también las funciones peso explícitamente,
\begin{equation}
\eta(\vec{r})=\int d\vec{s}\rho(\vec{r}+\vec{s})\theta(s-R)
\end{equation}
\begin{equation}
n(\vec{r})=\frac{1}{4\pi R^{2}}\int d\vec{s}\rho(\vec{r}+\vec{s})\delta(s-R)
\end{equation}
\begin{equation}
\vec{v}(\vec{r})=\frac{1}{4\pi R^{2}}\int d\vec{s}\rho(\vec{r}+\vec{s})\delta(s-R)\hat{s}
\end{equation}

Como comentábamos a partir de confinamientos en capas $\rho(\vec{r})=\rho^{2d}(\vec{R})\delta(z)$ o el más exigente  $\rho(\vec{r})=\rho^{1d}(z)\delta(\vec{R})$ se puede analizar esta teoría y ver que mejora a WDA en el primer caso pero presenta una divergencia en el segundo\cite{PhysRevE.55.4245}, pero abre la posibilidad de intentar introducir mejoras sustanciales en el funcional. \textit{Tarazona} propone un nuevo término que sustituye a (\ref{eqn:FMTrosenfeld3}) causante de la divergencia construido mediante
 peso de carácter tensorial que, de un lado generaliza el funcional de Rosenfeld, de otro recupera en el caso uniforme la $c^{(2)}_{PY}(r)$ y de otro soluciona el problema de la divergencia. El término $\Phi_{3}$ junto con la función peso tensorial definen,
\begin{equation}
\Phi_{3}=12\pi^{2}R^{6}\int d\vec{r}f_{3}(\vec{r})\frac{d^{3}\phi_{0}(\eta(\vec{r}))}{d\eta^{3}}
\end{equation}
con
\begin{equation}
f_{3}=\vec{v}\cdot \hat{T} \cdot \vec{v}-n \vec{v}\cdot\vec{v}-Tr[\hat{T}^{3}]+nTr[\hat{T}^{2}]
\end{equation}
donde 
\begin{equation}
\hat{T}_{\alpha,\beta}(\vec{r})=\frac{1}{4\pi R^{2}}\int d\vec{s}\rho(\vec{r}+\vec{s})\delta(s-R)\hat{s}_{\alpha}\hat{s}_{\beta}
\end{equation}

Esto constituye un aproximación que reproduce la ecuación de PY y la función de correlación directa de PY. Podemos reproducir la ecuación de estado de Carnahan-Starling definiendo adecuadamente  $\frac{d^{3}\phi_{0}(\eta(\vec{r})}{d\eta^{3}}$ pero aparece la cuestión de hasta que punto mejorar la descripción del fluido empeora la descripción del sólido. La descripción funcional utilizando este tipo de pesos presenta pues como limitación el uso de una aproximación modesta para la ecuación de estado del líquido y una aproximación buena para la descripción de las propiedades del sólido. En todo caso \textit{Tarazona}\cite{FMTphysicaATaraz} ha introducido una función $\phi^{CS}$ que permite recuperar la ecuación de estado de CS,
\begin{equation}
\phi^{CS}=\frac{2}{3\eta^{2}}\left(\frac{\eta}{(1-\eta^{2})}+log(1-\eta)\right)
\label{eqn:CSparaFMT}
\end{equation}

  	\section{Integración  Termodinámica y Teoría de perturbaciones}
\label{sec:integracionTermodinamica}

Dado un funcional $X[y]$  de una determinada función $y(x)$, podemos construir una jerarquía de funciones que obtenemos mediante diferenciación funcional de X respeto de y, entonces podemos hallar el valor de $X[y_{1}]$ a partir de $X[y_{0}]$ mediante una integración a lo largo de una curva\footnote{Suponemos bien definida X[y] para todas las funciones en las familias de curvas que unen $y_{0}$ e $y_{1}$.} $y(\alpha)$ parametrizada por $\alpha$ de modo que una las funciones $y_{0}=y(\alpha=0)$ e $y_{1}=y(\alpha=1)$, es decir,
\begin{equation}
X[y_{1}]=X[y_{0}]+\int_{0}^{1} d\alpha \int dx \frac{dy}{d\alpha} \frac{\partial X[y]}{\partial y}
\end{equation}

Si observamos la ecuación resultante, conocido el funcional $X[y_{0}]$ podemos determinarlo para $X[y_{1}]$ conocida la derivada funcional $\frac{\partial X[y]}{\partial y}$ en el camino de integración.  En nuestro caso tenemos un funcional de la densidad, o del potencial externo generalizado $u(\vec{r})$ o también podemos verlo como un funcional de la interacción a pares $\phi(\vec{r}_{1},\vec{r}_{2})$. En todos ellos partimos de que el funcional es único para $\rho$, $u(\vec{r})$ o $\phi(\vec{r}_{1},\vec{r}_{2})$ respectivamente luego diferentes caminos deben dar resultados idénticos, elegimos uno particularmente sencillo dado por,
\begin{equation}
y(\alpha)=y_{0}+\alpha(y_{1}-y_{0})
\end{equation}

donde obtenemos que,

\begin{equation}
X[y_{1}]=X[y_{0}]+\int_{0}^{1} d\alpha\int dx (y_{1}-y_{0}) \frac{\partial X[y]}{\partial y}
\end{equation}

Notar que esta integración se puede aplicar reiterativamente de modo que si en lugar de querer incorporar el conocimiento a nivel de primeras derivadas funcionales pretendemos incorporar el conocimiento de la jerarquía de las funciones de correlación al nivel de las segundas derivadas sencillamente aplicamos de nuevo el formalismo anterior donde $X[y]$ es ahora la primera derivada funcional. Esto se puede utilizar, por ejemplo, para construir un funcional a partir del conocimiento de la función de correlación directa.\\

Podemos aplicar estas relaciones para obtener dentro del formalismo del funcional de la densidad una relación explicita entre la función de distribución radial y la función de correlación directa. Comenzamos con la ecuación (\ref{eqn:eqautoconsistenterho}) y la \emph{hipótesis de Percus}. En ella evaluamos la respuesta de un fluido a un campo externo inducido por una partícula fija (en el origen $\vec{r}_{0}=0$) eligiendo dicho campo como el propio potencial de interacción entre las partículas, es decir, $v_{ext}(\vec{r}_{i0})=\phi(\vec{r}_{i0})$. En este caso la densidad del sistema no homogéneo se convierte en,
\begin{equation}
\rho_{eq}(\vec{r})=\frac{\rho^{(2)}(\vec{r})}{\rho_{0}}=\rho_{0}g(\vec{r})
\end{equation}
donde $\rho_{0}$ y $g(\vec{r})$ son la densidad y la función de distribución radial del sistema uniforme. Aplicando los métodos de integración termodinámica que acabamos de ver\footnote{Obviamente podemos obtener también $g(r)$ mediante minimización del funcional de la densidad aplicando la ecuación (\ref{eqn:EulerLagrangeMinimizacion}).} y la hipótesis de Percus la función $g(r)$ se escribe como,

\begin{equation}
ln g(\vec{r}_{1})=-\beta \phi(\vec{r}_{1})+\int_{0}^{1}d\alpha\int d\vec{r}_{2}\rho_{0}(g(\vec{r}_{2})-1)c^{(2)}(\rho_{\alpha};\vec{r}_{1},\vec{r}_{2})
\label{eqn:percuscampoefectivo}
\end{equation}

únicamente expresamos el conocimiento de g(r) mediante la función de correlación directa del sistema no-uniforme.

    \section{Fenómenos críticos en fluidos}
\label{sec:criticalidad}
En la lectura realizada por van der Waals al recibir el novel en 1910, comenta que en su imagen de los estados de agregación líquido y gas implicaba una relación de continuidad entre ambos, hecho que se hace manifiesto en su ecuación de estado y que de hecho se convierte en uno de los fenómenos más relevantes en las transiciones de fase\footnote{En el caso de un sólido y un fluido se argumenta que siempre existe una línea de transición pues hay una diferencia de simetría que no puede ser cambiada continuamente, mientras que esto si es posible entre un líquido y un vapor y en consecuencia su transición puede terminar en un punto crítico.}.La existencia de estados críticos donde las propiedades son notoriamente diferentes es relevante desde el punto de vista de la física estadística y nacen multitud de conceptos que conviene introducir aunque sea muy brevemente. Las diferencias en las propiedades físicas al acercarse al punto crítico dotan a los sistemas de una nueva fenomenología para analizar las diferentes aproximaciones, por otra parte aunque no tratamos con el problema de las fases críticas ciertas propiedades superficiales y los conceptos involucrados se relacionan con conceptos nacidos del estudio de fases críticas, en particular las correlaciones de largo alcance y la relevancia de las fluctuaciones cerca del punto critico tienen cierto parentesco con problemas que aparecen en las interfases fluidas\cite{PhysRevLett.86.2369,PhysRevLett.52.2160}\\

\begin{table}[htdp]
\begin{center}\begin{tabular}{c c c c c}
\toprule
\textbf{Propiedad} & \textbf{Exponente Crítico}   & \textbf{Campo Medio} & \textbf{d=2} & \textbf{d=3} \\
\midrule  $\rho_{l}-\rho_{v}$& $(T_{c}-T)^{\beta}$   &  $\frac{1}{2}$ &$\frac{1}{8}$ & $0.32$ \\
   $\kappa_{T}$& $|T-T_{c}|^{-\gamma}$ &  1 & $\frac{7}{4}$ & $1.24$ \\
   $C_{v}$& $A_{\pm}|T-T_{c})|^{-\alpha}$ &  0 & 0 & 0.11 \\
  $\mu-\mu^{c}$& $(\rho-\rho^{c})|\rho-\rho^{c}|^{\delta-1}$ &  3 & 15 & 4.81 \\
\midrule
  $\xi_{B}$& $|T-T_{c}|^{-\nu}$   &  $\frac{1}{2}$  & 1 & $0.63$ \\
  $h(r)$& $r^{-(d-2+\eta)}$ &  $0$ & $\frac{1}{4}$ & $0.04$ \\

\bottomrule
\end{tabular}
\caption{Definición de los exponentes críticos en un sistema líquido-vapor. Para las funciones de correlación $r/\xi_{B}<<1$.}
\label{tabla:exponCrit1}
\end{center}
\end{table}

La definición de la funciones de correlación total y la ecuación de la compresibilidad llevaba a la relación,
\begin{equation}
1+\rho\int d\vec{r}h(\vec{r})=\beta^{-1}\rho\kappa_{T}
\label{eqn:relacionhderykt}
\end{equation}
y esta definición más la ecuación de Ornstein-Zernique da lugar en el caso homogéneo a
\begin{equation}
1-\rho\int d\vec{r}c^{(2)}(\vec{r})=\frac{\beta}{\rho\kappa_{T}}
\label{eqn:cderTc}
\end{equation}
Ambas ecuaciones relacionan la termodinámica con la estructura. Históricamente la ecuación de Ornstein-Zernique surgió del estudio de fenómenos de scattering cerca del punto crítico, donde las diferentes propiedades de las dos funciones de correlación (la total y la directa) se hacen patentes. La integral de la función de correlación total diverge lo que se interpreta como un aumento progresivo de su alcance mientras que en el caso de la integral de la función de correlación directa no existe tal divergencia e incluso en las proximidades del punto crítico se postula como una función de corto alcance en líquidos simples.\\

En las proximidades del punto crítico la diferencia entre las densidades de ambas fases se anula aproximándose a un valor común $\rho_{c}$ mientras que la compresibilidad isoterma como vimos presenta una divergencia.  Se definen para describir estos comportamientos singulares un conjunto de exponentes críticos\footnote{El exponente $\alpha$ se interpreta como una discontinuidad finita.}, que resumimos en la tabla (\ref{tabla:exponCrit1}). La relación termodinámica-estructura (\ref{eqn:relacionhderykt}), permite relacionar la compresibilidad con la función de correlación total, la propia definición de $h(|\vec{r}|)$ hace que se anule a valores de $|\vec{r}|$ altos y por tanto la contribución a la compresibilidad proviene de un rango que podemos caracterizar por $\xi_{B}$ cuyo significado sería una longitud máxima de coherencia o correlación para las fluctuaciones en la densidad. Cerca del punto critico esta longitud diverge\footnote{La teoría original de Ornstein-Zernique desde el corto alcance de la función de correlación directa obtiene relaciones para el comportamiento de la función h(r) para $r>>\xi_{B}$ y $r<<\xi_{B}$, es última no decae en $d=2$, indicando que para hacer compatibles estas funciones con la existencia real del punto critico en $d=2$ es necesario un parámetro adicional $\eta$ \cite{fisher:944}, los comportamientos que lo incluyen aparecen reflejados en la tabla (\ref{tabla:exponCrit1}). De hecho en $d=2$ resulta $\eta=\frac{1}{4}$ mientras que en $d=3$ la teoría de Ornstein-Zernique original puede ser correcta con $\eta=0$ y de hecho valor es pequeño y se ha estimado como $\eta\sim0.04$.} y la ec. (\ref{eqn:relacionhderykt}) relaciona la divergencia de $\kappa_{T}$ y $\xi_{B}$.\\


En la tabla (\ref{tabla:exponCrit1}) se indican los valores de los exponentes críticos clásicos que se corresponden con una teoría de campo medio, como la ecuación de van der Waals y que es generalizada en el contexto de los fenómenos críticos por la teoría de Landau. Esta teoría se puede desarrollar también en un caso no homogéneo, bien para estudiar la relevancia de un campo externo, bien para analizar formalmente las propiedades de las funciones de correlación. Surge un campo amplio llamado teorías de campo basadas en la construcción de un funcional de un cierto parámetro de orden relevante. Conviene diferenciarlo conceptualmente de la teoría del funcional de la densidad antes introducida, ello se discute en unos de los apéndices .

    \section{Sumario}
Este capítulo ha sido una introducción formal a los métodos usados a lo largo de la memoria: desde teorías perturbativas, hasta los funcionales de la densidad utilizados más adelante. Una introducción a los métodos de teoría basadas en hamiltonianos efectivos se encuentra en el apéndice \S\ref{apn:teoriascampoefectivo}. Nuestro objetivo es el tratamiento de sistemas líquidos y conviene escribir conclusiones importantes al respecto de ellos de utilidad posterior,
\begin{itemize}
\item La función de partición es capaz de dar una descripción conjunta de diferentes fases. El funcional de la densidad además permite tratar las fases sólida, líquida y gaseosa, en situaciones no-homogéneas. Potenciales de interacción fuertemente repulsivos a cortas distancias y levemente atractivos a distancias intermedias permiten describir la coexistencia líquido-vapor.
\item La imagen de van der Waals de un líquido es esencialmente correcta en dos situaciones: potenciales de largo alcance suaves (potenciales de Kac) o cerca del punto triple. De hecho un buen tratamiento de la parte repulsiva permite reproducir bajo una aproximación de campo medio los aspectos estructurales líquidos densos, pero falla al describir el diagrama de fases completo, en particular las propiedades cerca del punto crítico. Las teorías perturbativas permiten una mejora sistemática de esta aproximación pero presentan la necesidad de conocer progresivamente funciones de distribución de partículas de ordenes mayores.
\item Todas las aproximaciones de campo medio presentan los mismos comportamientos clásicos pero no hay una única teoría de campo medio en le sentido de que se puede construir una aproximación de campo medio desde diferentes caminos. En el contexto de la teoría de Landau la bondad como aproximación fenomenológica se suele estimar mediante el criterio de Ginzburg, mientras que su contenido físico indica que es una teoría de modo efectivo adecuada si las correlaciones del sistema que esperamos describir son del alcance de las correlaciones que hemos incluido. Dicho de otro modo, el sistema se siente como si de modo efectivo estuviera limitado en sus correlaciones por un tamaño efectivo del sistema.  
\end{itemize}
\chapter{Aplicación a interfases fluidas}

Se realiza una aplicación del formalismo de la teoría de líquidos a algunos aspectos de las interfases fluidas y se comentan en el proceso procedimientos y resultados que serán analizados y cuestionados en los capítulos posteriores. Se explican diferentes esquemas alrededor de los conceptos de tensión superficial y perfil de densidad líquido-vapor.

\section{Aproximación macroscópica a la tensión superficial}

La termodinámica concibe la \textit{tensión superficial}\footnote{Un análisis exhaustivo de las diferentes definiciones de tensión superficial se puede encontrar en \cite{RowlinsonWidom} y en \cite{NavascuesSurfaceTension}.} únicamente desde la existencia una superficie donde reside dicha tensión de manera que la determinación del trabajo reversible necesario para incrementar el tamaño del sistema en $dV$ cuando incremento el área de dicha superficie en $dA$ queda expresada como $dW=-PdV+\gamma dA$. En consecuencia la descripción correcta requiere como parámetro de estado adicional el área de la interfase y su variable intensiva conjugada será la tensión superficial. La ecuación \textit{ecuación de Gibbs-Duhem}\footnote{Es la ecuación fundamental en forma diferencial en el caso de que todas las variables termodinámicas sean intensivas.} queda,
\begin{equation}
d\gamma=-\frac{S}{A}dT+\frac{V}{A}dP-\frac{N}{A}d\mu
\end{equation}
y $\gamma$ puede ser expresada en los colectivos que hemos definido, \S\ref{sec:canonico} y  \S\ref{sec:macrocanonico},
\begin{equation}
\gamma=\left( \frac{\partial\Omega}{\partial A} \right)_{T,V,\mu}=\left(\frac{\partial F}{\partial A} \right)_{T,V,N}
\label{eqn:tensionsuptermo}
\end{equation}

La posición de dicha superficie en el sistema no aparece como una variable termodinámica\footnote{Nos restringimos al caso de una superficie plana.} mientras que diferenciar la interfase como un sistema independiente no permite encontrar nueva información, aun así la existencia de variaciones en la densidad si puede motivar la definición de una superficie divisoria en la medida estas variaciones pudieran aparecer localizadas. Una vez que elegimos una definición matemática de superficie los volúmenes de ambas fases están fijados y tomando los valores de las fases homogéneas por unidad de volumen de las diferentes magnitudes aparecen de manera natural las contribuciones superficiales como cantidades de exceso sobre las dos fases uniformes. Resulta pues que su valor concreto aparece condicionado por la definición elegida de la superficie divisoria. En el colectivo macrocanónico, donde la ecuación de Euler toma la forma $\Omega=-pV+\gamma A$, la tensión superficial definida como el exceso de $\Omega$ superficial por unidad de área no depende de la elección concreta de la superficie. Se suele considerar como definición de superficie la llamada \emph{superficie equimolar de Gibbs} que equivale, en una interfase líquido-vapor, a anular el valor de la densidad sobre la superficie\footnote{La ecuación de Euler para la energía libre de Helmholtz sería $F=-pV+\gamma A$ solo si hemos determinado una superficie para la interfase de manera que $\mu N^{s}=0$. En el caso del colectivo macrocanónico la relevancia de la definición de superficie se hace más evidente en sistemas multicomponentes donde aparece el concepto de adsorción del componente $i$ como la densidad localizada en la superficie de dicho componente, en cualquier caso es una definición operacional.}.\\

La primera extensión sobre la termodinámica expuesta es la llamada \textit{termodinámica local} que supone que en cada punto del sistema es posible definir valores de las funciones termodinámicas sin ambigüedad, de modo que en una interfase líquido-vapor plana aparecen como funciones de la variable altura. Más allá de la definición posible o su método, en una interfase el carácter de la presión\footnote{Entendida como presión hidrostática.} pasa de ser escalar a ser tensorial. En una interfase plana tendrá dos componentes diferenciadas: tangencial y normal a la superficie, la estabilidad mecánica se suele expresar aludiendo a que la definición del tensor presión debe poseer una divergencia nula o en su caso compensar la presencia de un campo externo. En nuestra geometría planar llamamos $p_{t}=p_{\perp\perp}$ y las componentes normales $p_{n}=p_{\parallel}$. El incremento de energía libre por unidad de área quedaría como,
\begin{equation}
\gamma=\int_{-L}^{L} dz (p_{n}(z)-p_{t}(z))
\end{equation}
donde fuera de la interfase $p_{n}=0$ y $p_{t}=p$ es la presión hidrostática de un fluido en equilibrio. Esta expresión se conoce como \textit{definición mecánica de tensión superficial}.\\

En este contexto es cuando podemos hablar de una interfase $\rho(z)$ y la \emph{superficie equimolar de Gibbs} queda expresada como el valor de $z_{g}$ que verifica la ecuación,
\begin{equation}
\int_{-L}^{z_{g}} dz\left[ \rho^{l}-\rho(z)\right]=\int_{z_{g}}^{L} dz \left[ \rho(z)-\rho^{g}\right] 
\label{eqn:defSupGibbs}
\end{equation}

Los tratamientos anteriores permiten obtener la tensión superficial determinada la energía libre, bien como función de la superficie (ecuación de Gibbs-Duhem a presión y potencial químico constantes) bien como la diferencia en términos de energía libre entre la energía libre total y la contribución de volumen de las dos fases $\gamma=A^{-1}(\Omega-PV)$, esto último se basa en la expresión de Euler para la ecuación diferencial anterior y ambas son la expresión termodinámica de la ecuación fundamental. Es más conveniente realizar un tratamiento para el segundo modo en el colectivo macrocanónico donde este método no depende de la elección concreta de la superficie de Gibbs, en cualquier caso se resalta el hecho de que la tensión superficial aparece como un exceso de presión.\\

\section{Teoría de van der Waals de la interfase líquido-vapor}
\index{van der Waals!Teoría de Interfase}
Hemos visto en el capítulo anterior como en gran medida la imagen actual de un líquido estaba contenida en la ecuación de estado de van der Waals. La extensión de sus ideas al estudio de una interfase líquido-vapor fue realizada por el propio van der Waals, que  puede considerarse como una versión inicial de las teorías del funcional de la densidad locales aplicadas a líquidos y su teoría de la interfase introducida como una aproximación funcional de gradiente cuadrado, aunque originalmente fue formulada desde un punto de vista más fenomenológico\footnote{Presentado ciertas analogías con el hamiltoniano efectivo de una teoría de Ginzburg-Landau, por ejemplo.}.\\

\begin{figure}[htbp]
\begin{center}
\includegraphics[width=3.5in]{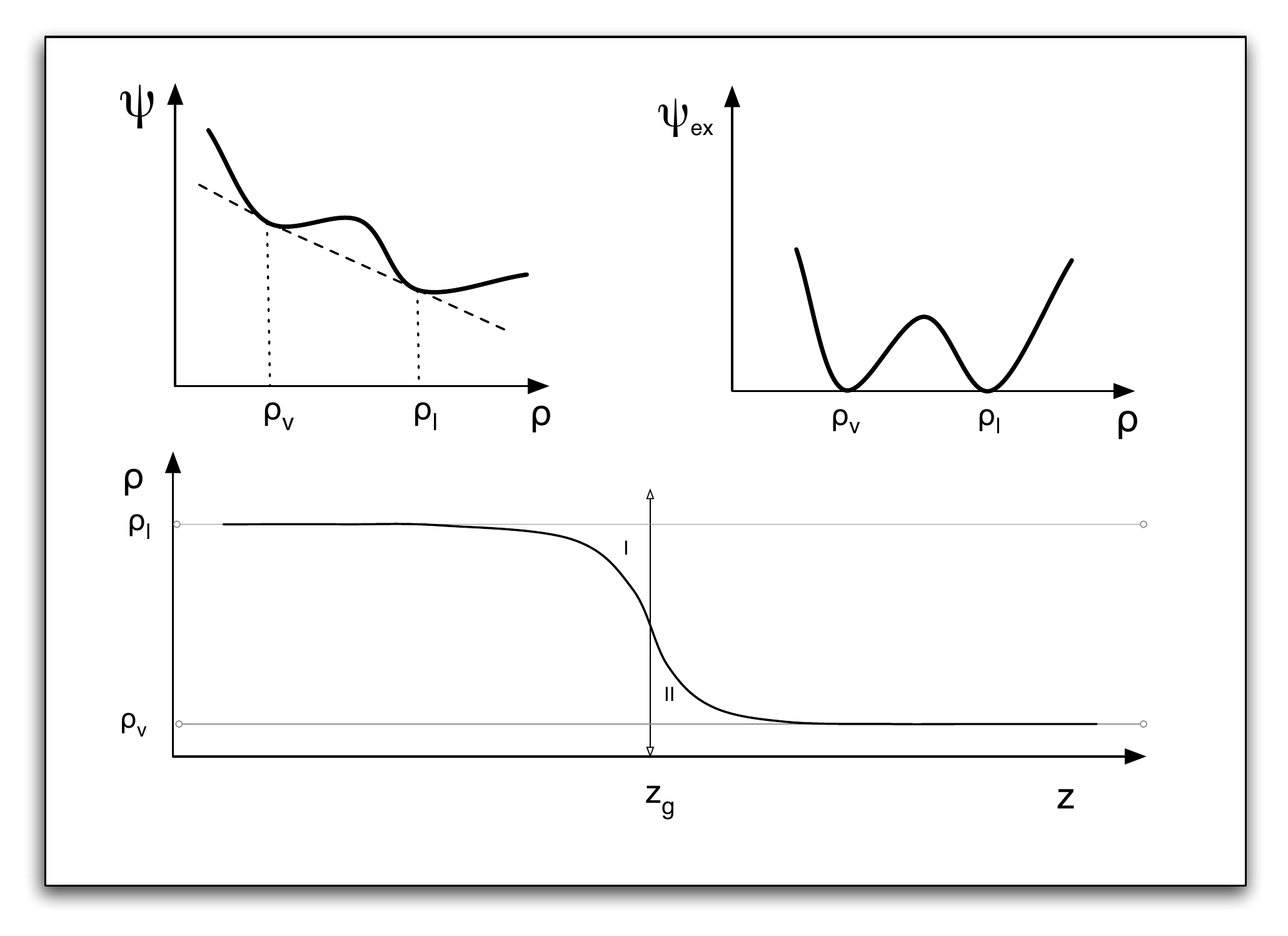}
\caption{\textbf{Teoría de van der Waals}. Densidad de energía libre local del sistema uniforme en el caso de coexistencia de fases. Línea discontinua es la construcción de Maxwell. Línea continua entre las densidades de coexistencia es la continuación analítica de la energía libre. A la derecha una representación de la energía libre de exceso sobre el caso homogéneo. Abajo el perfil de densidad que se obtiene.}
\label{fig:Diagrama-vDW}
\end{center}
\end{figure}

La primera extensión a la densidad de energía libre de un fluido homogéneo a un caso no-homogéneo es considerar una densidad de energía libre local. En su estudio de la interfase líquido-vapor, van der Waals considera relevante, más allá de la definición usada para dicha cantidad o en que ecuación de estado pueda estar basada, el exceso de energía libre sobre el caso homogéneo y para ello considera la diferencia entre una energía libre local y la construcción de tangentes comunes\footnote{Cuyo contenido físico es análogo a la construcción de Maxwell.} y por tanto la energía libre local es \textit{continuada analíticamente} en la región de inestabilidad de la ecuación de estado y la tensión superficial es considerada como un exceso de presión, véase la figura (\ref{fig:Diagrama-vDW}).\\

La integral a lo largo de la interfase de este exceso de energía libre resulta ser independiente de la elección de la superficie de Gibbs pero una aproximación dada únicamente de este modo daría lugar a una interfase completamente abrupta y una tensión superficial nula (como en Widom \S\ref{sec:teoriaWidom}). Esto llevó a van der Waals a completar su energía libre de exceso inicial con un ingrediente que penalizara variaciones rápidas de la densidad del sistema\footnote{Su expresión para $\tau$ esencialmente tiene en cuenta las fuerzas atractivas incluidas en el sistema y de hecho su expresión fue introducida por Rayleigh previamente aunque con una interpretación diferente.},
\begin{equation}
\gamma_{vdw}[\rho]=\int dz \left[ \psi_{ex}(\rho(z))+\frac{\tau}{2}\left( \frac{d\rho_{z}}{dz}\right)^{2}\right] 
\end{equation}

Una vez construida la expresión para la tensión superficial aparece como un funcional $\gamma_{vdw}[\rho,\rho']$ y por tanto es natural incorporar un proceso de minimización variacional con el fin de obtener el perfil de densidad de equilibrio. La forma más inmediata de solución es asumir que el parámetro $\tau$ no depende de la densidad lo que permite una resolución analítica completa del problema.\\

Las posibles soluciones de un funcional de este tipo\footnote{Por otra parte es posible obtener las propiedades esenciales en un campo gravitatorio, aunque en este caso para poder recuperar la degeneración en z debemos incrementar el potencial químico en $mg\delta z$.} fueron analizadas en  detalle por \textit{Yang, P. Fleming y J. H. Gibbs}\cite{yang:3732}, con la salvedad de partir estos de la teoría del funcional de la densidad y por tanto de una ecuación de mínimo análoga a ec. (\ref{eqn:EulerLagrangeMinimizacion}), asi deducen tanto las propiedades de equilibro como la construcción de Maxwell. La forma para los perfiles de densidad puede verse en la figura (\ref{fig:Diagrama-vDW}), son perfiles que interpolan suavemente entre las densidades de las dos fases y poseen una anchura bien definida.\\
 
En este contexto, podemos determinar mediante las dos definiciones iniciales, mecánica y termodinámica, la tensión superficial mostrando que para esta aproximación son consistentes\footnote{Si deseamos ir un poco más allá vemos que en esta aproximación la tensión superficial calculada como diferencia de componentes del tensor presión recupera la forma de \textit{Triezenberg y Zwanzig}\cite{PhysRevE.53.5130}}\\
\begin{equation}
\gamma_{vdW}=2\tau\int dz \rho'(z)^{2}=-2\int dz \psi_{ex}(\rho) 
\end{equation}

La línea de coexistencia líquido-vapor termina en un punto crítico, luego es de esperar que la tensión superficial se anule, con un cierto exponente critico clásico $\mu$, por otra parte la anchura de la interfase también diverge como $\nu_{l}$,
\begin{eqnarray}
\gamma\sim(T_{c}-T)^{\mu}\\
\xi_{\perp}\sim(T_{c}-T)^{-\nu_{l}}
\end{eqnarray}
ambos límites tomados sobre la línea de coexistencia de fases y $T\rightarrow T_{c}$. El exponente critico $\mu$ se espera sea universal y Widom sugiere relacionarlo con el exponente critico de las correlaciones de volumen mediante una relación de hiperescalado. La teoría de van der Waals reproduce los valores de las teorías de campo medio detalladas en el capítulo anterior, dando como resultado valores clásicos. Un intento de extensión a valores no clásicos fue realizado en primera instancia por \textit{Fisk-Widom}\cite{fisk:3219}, que aunque también ha sido criticado\cite{0034-4885-47-9-001ReviewJasnow} preserva una validez fenomenológica.\\

Si analizamos el problema desde la óptica de la teoría de gradientes cuadrados completada con la teoría de la respuesta lineal, véase ec. (\ref{eqn:gradientesCuadradosCoef2}) tenemos que $\tau$ aparece definido como el segundo momento de la función de correlación, esto muestra como la teoría no es capaz de tratar potenciales tipo Lennard-Jones pues se espera que $c(r)\sim\beta\phi(r)$ y su segundo momento puede no existir.\\

Para realizar un análisis más detallado de la física contenida en esta aproximación es conveniente aplicar el formalismo del funcional de la densidad, \S\ref{sec:DFT}, al problema de las interfases fluidas.

\section{Aproximación microscópica a la tensión superficial}\label{sec:LVenDFT}
Gran parte de los conceptos introducidos en el formalismo eran conocidos antes del desarrollo formal de la teoría de colectividades. Los diferentes estados de agregación y la imagen dada por la teoría cinética de los fenómenos térmicos eran herramientas usadas por los físicos ya en la época de Clausius (1857). El concepto de tensión superficial era de uso común en el siglo XIX ya que la fenomenología asociada es fácilmente observable de modo que la representación mecánica como fuerza por unidad de longitud y trabajo por unidad de área eran conocidas en la época de Laplace (1807). Por contra, aunque la idea de fuerzas intermoleculares se afianza en este periodo, una relación cuantitativa con la tensión superficial era desconocida, esencialmente porque una formulación completa de este problema requiere el uso y definición de las funciones de distribución de partículas.\\

En las relaciones macroscópicas previas, ec. (\ref{eqn:tensionsuptermo}), hemos perdido la posibilidad de relacionar la tensión superficial con las propiedades microscópicas, tanto su dependencia en el potencial de interacción intermolecular como la relevancia de las correlaciones del sistema en la energía superficial. Además las expresiones anteriores adolecen del problema de no diferenciar contribuciones de volumen (análogas a las presentes en el sistema homogéneo) y de superficie (características de la interfase) ya que ambas pueden estar presentes en la interfase que realmente que ocupa un cierto volumen.\\
\subsection{Expresión de Kirkwood-Buff}
Desde el punto de vista de la mecánica estadística es posible desde la definición ma\-cros\-có\-pi\-ca, ec. (\ref{eqn:tensionsuptermo}), realizar la diferenciación respecto del área en la expresión de la función de partición, lo cual da paso a una primera definición con información microscópica de la tensión superficial cuya obtención es análoga al cálculo del virial\footnote{En este se expresa $P=-(\frac{\partial F}{\partial V})=-\beta^{-1}(\frac{\partial ln Z_{N}}{\partial V})$ vemos que un cambio en el volumen se traduce en una expresión para la presión que en el caso de un potencial a pares involucra la función de correlación y el potencial de interacción, véase ec. (\ref{eqn:EcuacionVirial}).}. De modo similar manteniendo V fijo pero considerando un cambio de área $\Delta A=\delta x L^{2}$ se puede expresar la derivada de la función de partición respecto al área\cite{reviewEvans1979,RowlinsonWidom},
 \begin{equation}
\gamma=\frac{1}{2}\int dz \int d\vec{r}_{12}\frac{x^{2}_{12}-|\vec{r}^{2}_{12}|}{|\vec{r}_{12}|}\phi^{\prime}(|\vec{r}_{12}|)\rho^{(2)}(\vec{r}_{1},\vec{r}_{2})
\label{eqn:Kirkwood-Buff}
\end{equation}
\index{Tensión superficial!Ecuación de Kirkwood-Buff}
conocida como la ecuación para la \emph{tensión superficial de Kirkwood-Buff}\cite{kirkwood:338}, esta expresión puede ser igualmente obtenida desde el tensor presión. La utilidad práctica depende del contexto donde se utilice esta ecuación, desde el punto de vista del funcional de la densidad su uso presenta el inconveniente de necesitar el conocimiento de $\rho^{(2)}(\vec{r}_{1},\vec{r}_{2})$. Su aplicación directa precisa las aproximaciones usuales: como despreciar correlación en un primer paso e incluir el conocimiento del sistema homogéneo (a través de su función de distribución radial) en un siguiente siguiente paso, véanse los comentarios a la ecuación (\ref{eqn:TeoriaPerturbaciones}). En cambio si que es una expresión útil en el cálculo mediante simulaciones de Dinámica Molecular\cite{0953-8984-13-21-308} y Montecarlo, ya que puede ser escrita como\footnote{En la práctica esta formulación requiere determinar la contribución de la cola del potencial a la tensión superficial que puede ser considerable.},
\begin{equation}
\gamma=\frac{1}{L^{2}}\left\langle \sum_{i<j}^{N}\phi'(r_{ij})\frac{x_{ij}^{2}-z_{ij}^{2}}{r_{ij}} \right\rangle 
\end{equation}

\subsection{Expresión basada en la función de correlación directa}

Desde el punto de vista del funcional de la densidad tiene una utilidad más inmediata una expresión basada en la función de correlación directa sobre la que suelen construirse la mayoría de los funcionales y sobre la que es más viable tener un control de cuan realista es nuestra predicción funcional.\\
\index{Tensión superficial!Ecuación de Triezenberg y Zwanzig}

La expresión que incluimos fue determinada por \textit{Triezenberg y Zwanzig} \cite{PhysRevLett.28.1183}\footnote{E independientemente por Lovett e inicialmente por Yvon.}. Indicamos someramente el proceso que les condujo a esta ya que su metodología contiene información relevante. El procedimiento consta de los siguientes pasos:
\begin{itemize}
\item Expresamos la energía libre que cuesta una desviación $\Delta\rho(\vec{r})$ sobre la \textit{densidad de equilibrio} $\rho_{0}(\vec{r})$.
\item Concretamos esta expresión para una simetría dada por $V_{ext}(z)$.
\item Indagamos la forma que adquiere para desviaciones en la densidad que involucren cambios macroscópicos en el área de la interfase, ya que esto reproduce la definición de tensión superficial (\ref{eqn:tensionsuptermo}) termodinámica.
\end{itemize}

Partimos de un desarrollo de nuestro funcional entorno de la densidad de equilibrio y cuya primera contribución no nula esta dada por el término de segundo orden,
\begin{equation}
\delta \Omega[\rho_{0}+\Delta\rho]=\frac{1}{2!}\int d\vec{r}_{1}\int d\vec{r}_{2}\left.\frac{\delta^{2}\mathcal{F}[\rho]}
{\delta\rho(\vec{r}_{1})\delta\rho(\vec{r}_{2})}\right| _{\rho_{0}}\Delta\rho(\vec{r}_{1})\Delta\rho(\vec{r}_{2})
+O(\delta\rho^{3})
\label{eqn:perturbacionDensidadOrden2}
\end{equation}
el significado de la expresión lo conocemos, ya que dentro de la integral tenemos la función de correlación $C^{(2)}(\vec{r}_{1},\vec{r}_{2};\rho_{0})$, y determina el cambio en energía libre al variar la densidad en dos puntos $\vec{r}_{1}$ y $\vec{r}_{2}$ sobre la densidad de equilibrio.\\

En el caso de la simetría en el plano $xy$ que caracteriza una interfase líquido-vapor plana la ecuación (\ref{eqn:perturbacionDensidadOrden2}) se puede simplificar introduciendo transformadas de Fourier en las coordenadas contenidas en la interfase $\vec{R}$,
\begin{equation}
\delta \Omega[\rho_{0}+\Delta\rho]=\frac{1}{2 A\beta}\int dz_{1}\int dz_{2}\sum_{\vec{q}}C^{(2)}(z_{1},z_{2},\vec{q})\Delta\rho(\vec{q},z_{1})\Delta\rho(-\vec{q},z_{2})
\end{equation}
Podemos proponer una restricción a fluctuaciones que sean expresables como un cambio en el área de la superficie de Gibbs local $\xi_{g}(\vec{R})$ y que se caractericen por ser leves desviaciones de una interfase plana. La forma de la fluctuación se puede escribir como la conocida expresión,\\
\begin{equation}
\Delta\rho(\vec{R}_{1},z_{1})=\rho_{0}(z-\xi_{g}(\vec{R}_{1}))-\rho_{0}(z_{1})=-\xi_{g}(\vec{R})\frac{d\rho_{0}(z_{1})}{dz_{1}}
\label{eqn:perturbacionTipoCWT}
\end{equation}
consideramos que $\xi_{g}(\vec{R})$ es una función univaluada, de manera que el incremento de área sustentado por esta superficie es,
\begin{equation}
\Delta A=\frac{1}{2}\int d\vec{R}|\nabla^{2}\xi_{q}(\vec{R})|^{2}
\label{eqn:cambioArea}
\end{equation}
con este conjunto de suposiciones se pueden escribir los dos primeros ordenes en q \cite{PhysRevLett.28.1183}, 
\begin{equation}
C^{(2)}(z_{1},z_{2},q)=C^{(2)}_{0}(z_{1},z_{2})+q^{2}C^{(2)}_{2}(z_{1},z_{2})+q^{4}C^{(2)}_{4}(z_{1},z_{2})+...
\end{equation}
El término de orden 0 en q es,
\begin{equation}
\int dz_{1}\int dz_{2} \frac{d \rho_{0}(z_{1})}{dz_{1}}\frac{d \rho_{0}(z_{2})}{dz_{2}}C_{0}^{(2)}(z_{1},z_{2})\sum_{q}|\xi_{q}|^{2}
\end{equation}
la proyección de la función $C_{0}^{(2)}(z_{1},z_{2})$ sobre la derivada del perfil veremos que se anula mientras que el término de orden $q^{2}$ viene dado por,
\begin{equation}
\int dz_{1}\int dz_{2} \frac{d \rho_{0}(z_{1})}{dz_{1}}\frac{d \rho_{0}(z_{2})}{dz_{2}}C_{2}^{(2)}(z_{1},z_{2})\sum_{q}q^{2}|\xi_{q}|^{2}
\end{equation}
la suma en q representa el cambio de área introducido en la ec. (\ref{eqn:cambioArea}) expresado en el espacio Fourier y la función $C_{2}^{(2)}(z_{1},z_{2})$ es el segundo momento de $C^{(2)}(z_{1},z_{2},\vec{R}_{12})$ y por tanto la tensión superficial viene dada por,
\begin{equation}
\beta\gamma=\int dz_{1}\int dz_{2} \frac{d \rho_{0}(z_{1})}{dz_{1}}\frac{d \rho_{0}(z_{2})}{dz_{2}}C_{2}^{(2)}(z_{1},z_{2})
\label{eqn:Triezenberg-Zwanzig}
\end{equation}
a esta ecuación la denominamos ecuación de Triezenberg-Zwanzig (TZ).\\

La equivalencia general entre ambas expresiones  (\ref{eqn:Triezenberg-Zwanzig}) y (\ref{eqn:Kirkwood-Buff}) requiere el uso de un formalismo dinámico que no hemos presentado, pero cuyo objetivo es expresar la energía libre superficial en la expresión de Kirkwood-Buff mediante el análogo en la jerarquía de $G^{(n)}$ de la función de correlación directa y utilizar la relación entre estas\cite{RowlinsonWidom}. El paso intermedio es una expresión que ha sido denominada TZW, por ser similar a (\ref{eqn:Triezenberg-Zwanzig}) y basarse en la aportación de Wertheim\cite{wertheim:2377},
\begin{equation}
\delta \Omega[u_{0}+\Delta u]=\frac{1}{2}\int d\vec{r}_{1}\int d\vec{r}_{2}\delta u(\vec{r}_{1})G^{(2)}(\vec{r}_{1},\vec{r}_{2})\delta u(\vec{r}_{2})
\end{equation}
Si suponemos inicialmente un potencial externo $u_{0}(z)$ y una interfase xy que perturbamos mediante una variación del potencial externo con la forma de una onda capilar tendremos $\delta u(\vec{r})=-u'_{0}(z)\xi(\vec{R})$, podemos escribir,
\begin{equation}
\delta \Omega[u_{0}+\Delta u]=\frac{1}{2}\int d\vec{r}_{1}\int d\vec{r}_{2} u'(z_{1})\xi(\vec{R}_{1})G^{(2)}(z_{1},z_{2},\vec{R}_{12})u'(z_{2})\xi(\vec{R}_{2})
\end{equation}
que resulta el análogo directo de la ecuación (\ref{eqn:perturbacionDensidadOrden2}) y del mismo modo el orden 2 en q, permite expresar,
\begin{equation}
\beta\gamma=\int dz_{1}\int dz_{2}u'(z_{1})u'(z_{2})G_{2}^{(2)}(z_{1},z_{2})
\label{eqn:TZWgamma}
\end{equation}
Si aplicamos que,
\begin{equation}
\rho'_{0}(z_{1})=\int d\vec{R}_{12}\int dz_{2} G^{(2)}(z_{1},z_{2},\vec{R}_{12}) u'(z_{2})
\label{eqn:respLinealCampoz}
\end{equation}
y además que $\int dz_{2}\int d\vec{R}_{2}G^{(2)}(z_{1},z_{2},\vec{R}_{12})C^{(2)}(z_{2},z_{3},\vec{R}_{23})=\delta(\vec{R}_{1}-\vec{R}_{3})\delta(z_{1}-z_{3})$, llegamos de hecho a la ecuación ec.(\ref{eqn:Triezenberg-Zwanzig}).\\

Quedaba ver que el orden 1 en el desarrollo ec. (\ref{eqn:perturbacionDensidadOrden2}) bajo ec. (\ref{eqn:perturbacionTipoCWT}), se anulaba pero basta ver que si $u(z)=\lambda z$ la relación equivalente a ec. (\ref{eqn:respLinealCampoz}) me lleva a,
\begin{equation}
\beta\lambda=\int d z_{2}C_{0}^{(2)}(z_{1},z_{2})\rho'_{0}(z_{2})
\end{equation}
que en el límite $\lambda\rightarrow 0^{+}$ debe mantenerse y por tanto como anunciamos previamente se anula.\\

Es interesante analizar en contenido de las relaciones que hemos encontrado, al principio de la memoria indicábamos las dos predicciones opuestas de la teoría de ondas capilares y la teoría de van der Waals. Conviene discernir que significado posee las relaciones exactas dadas en ambos casos lo que requiere determinar, bien las funciones de correlación directa bien las funciones de correlación total. Procedemos de este modo en las siguientes secciones.\\
\section{Análisis de las correlaciones en la interfase}
\label{sec:Wherteim}
¿Cual es la física contenida en la ecuación (\ref{eqn:respLinealCampoz}) aplicada al caso de la interfase líquido-vapor? En el caso de $v_{ext}(z)=mgz$ queda expresada como,
\begin{equation}
\rho'_{0}(z_{1})=-\beta mg\int d\vec{R}_{12}\int dz_{2} G^{(2)}(z_{1},z_{2},|\vec{R}_{12}|)
\label{eqn:Werhteim1}
\end{equation}
que implica que si el perfil de densidad esta bien definido en $g\simeq0^{+}$ necesariamente la integral anterior debe ser divergente como $g^{-1}$, y el sentido físico atribuido a este hecho es la presencia correlaciones de largo alcance en el plano de la interfase. La expresión de esta divergencia fue determinada por \textit{Wertheim} que mostró\cite{wertheim:2377} que,
\begin{equation}
G_{0}^{(2)}(z_{1},z_{2})=\frac{d\rho_{0}(z_{1})}{dz_{1}}\frac{d\rho_{0}(z_{2})}{dz_{2}}\frac{1}{(\rho_{l}-\rho_{v})\beta mg}
\end{equation}
En el caso de un sistema uniforme se da $\rho'_{0}(z_{1})\simeq0$ y esta divergencia no es \textit{requerida}. Del mismo modo que lejos de la interfase la divergencia en las correlaciones transversales desaparece.\\

La función $G_{0}^{(2)}(z_{1},z_{2})$ es el término de orden 0 de la transformada de Fourier de $G^{(2)}(z_{1},z_{2},R_{12})$ para la dependencia en el vector de onda q podemos escribir\cite{wertheim:2377,reviewEvans1979,RowlinsonWidom},
\begin{equation}
G^{(2)}(q,z_{1},z_{2})\simeq\beta^{-1}\frac{d\rho_{0}(z_{1})}{dz_{1}}\frac{d\rho_{0}(z_{2})}{dz_{2}}\frac{1}{(\rho_{l}-\rho_{v})\beta mg+\gamma q^{2}}+O(q^{-4})
\end{equation}
aproximación que es solo válida para el comportamiento a largas longitudes de onda\footnote{Para ello se identifica el término que acompaña a $q^{2}$ como la tensión superficial macroscópica determinada por ec. (\ref{eqn:tensionsuptermo}). Para realizar esta identificación se puede determinar $\Delta\Omega$ para fluctuaciones de la forma dada por la ecuación (\ref{eqn:perturbacionTipoCWT}) y hacer uso de la ecuación (\ref{eqn:Triezenberg-Zwanzig}). Aunque formalmente se puede relacionar con la teoría de ondas de capilaridad que describimos en el capítulo introductorio, y que es válida igualmente en este límite cabe notar que en este análisis de \textit{Wertheim} no se parte de hipótesis que tengan como imagen física las ondas de capilaridad únicamente precisamos una cierta naturaleza para las fluctuaciones en la densidad en la interfase para identificar el factor que acompaña a $q^{2}$ con la tensión superficial macroscópica en el límite $q\rightarrow0$. La única suposición que hemos hecho es que la derivada del perfil de densidad permanece finita en el límite de $g\rightarrow0$. Sobre la posibilidad de análisis un poco más generales vean se los artículos de Weeks\cite{PhysRevLett.52.2160}.} de la función de correlación $G^{(2)}$ pero no permite un análisis a escalas microscópicas.\\

La divergencia de las correlaciones superficiales en la interfase recuerda el problema de la divergencia de las correlaciones de volumen cerca del punto crítico. Podemos analizar mediante argumentos similares a los sugeridos para $h(r)$ en \S\ref{sec:criticalidad} el comportamiento de la función $G^{(2)}(z_{1},z_{2},R_{12})$. Para ello integramos la ecuación (\ref{eqn:Werhteim1}) y obtenemos:
\begin{equation}
\Delta\rho=\rho_{l}-\rho_{v}=\beta mg \int dz_{1}\int dz_{2}\int d^{(d-1)}\vec{R}_{12} G^{(2)}(z_{1},z_{2},|\vec{R}_{12}|)
\label{eqn:WertheimIntegradaWeeks}
\end{equation}
siguiendo a \textit{Weeks}\cite{PhysRevLett.52.2160} podemos proponer, para valores de $r>>\xi_{B}$ donde solo importan los aspectos superficiales y no los de volumen, una forma asintótica para $G^{(2)}$, donde las simetrías imponen dos escalas una dada por $\xi_{cw}$ en la interfase y otra dada por $\xi_{\perp}$ perpendicular a esta,
\begin{equation}
G^{(2)}(z_{1},z_{2},R_{12})\sim R_{12}^{-\theta}G^{(2)}_{S}\left(\frac{z_{1}}{\xi_{\perp}},\frac{z_{2}}{\xi_{\perp}},\frac{|\vec{R}_{12}|}{\xi_{cw}}\right) 
\end{equation}
el exponente $\theta$ deberá darse en el contexto de la aproximación que consideremos para calcular las correlaciones superficiales y no podemos saberlo mediante argumentos de escalado. Si sabemos que su signo ha de ser positivo para que las correlaciones no crezcan con la distancia. Introduciendo esta hipótesis de escala en la ecuación (\ref{eqn:WertheimIntegradaWeeks}) podemos establecer que,
\begin{equation}
\xi_{cw}^{(d-3-\theta)}\xi_{\perp}^{2}\sim C
\label{eqn:scalingWeeks}
\end{equation}
donde $C$ es una constante que no depende de g. El valor de $\theta$ condiciona el caso de dimensión límite en el que aunque $\xi_{cw}$ diverge es posible que la anchura $\xi_{\perp}$ permanezca finita, véase ec.(\ref{eqn:distrualturasSdeR}).

\subsection{Correlaciones en la teoría de van der Waals}
La forma de las correlaciones en una teoría de gradientes cuadrados puede ser determinada mediante diferenciación y, tal y como se vio en el formalismo, obtenemos,
\begin{equation}
C^{(2)}(\vec{r}_{1},\vec{r}_{2})=\beta \left( \frac{d^{2}f_{0}}{d\rho^{2}}-2f_{2}\nabla^{2}\right)\delta(r_{1}-r_{2})
\label{eqn:c2devdw}
\end{equation}
La relación exacta,
\begin{equation}
\int dz_{3}C^{(2)}(q,z_{1},z_{3})G^{(2)}(q,z_{3},z_{2})=\delta(z_{1}-z_{2})
\end{equation}
permite obtener las correlaciones en el modo $q\simeq0^{+}$, \textit{Evans}\cite{evansMolecularCWvdw} obtuvo,
\begin{equation}
G^{(2)}(q\simeq0^{+},z_{1},z_{2})\sim\beta^{-1}\frac{d\rho_{0}(z_{1})}{dz_{1}}\frac{d\rho_{0}(z_{2})}{dz_{2}}\frac{1}{mg(\rho_{l}-\rho_{v})+\gamma_{lv}q^{2}}
\label{eqn:ecMODOq0wdw}
\end{equation}
 que contiene un comportamiento análogo al análisis de Wertheim y a la expresión (\ref{eqn:asintoticoG2cwt}).

\subsection{Correlaciones en la teoría de ondas capilares}
\label{sec:correlacionesCWT}
Las propiedades descritas para hamiltonianos efectivos gaussianos, ec. (\ref{eqn:HeffGaussiano}), nos permiten definir para el hamiltoniano de ondas capilares descrito en la introducción una serie de funciones de correlación y distribución sobre las cuales se pueden analizar las relaciones anteriores y comprobar su significado. Partimos de ec. (\ref{eqn:trabajominCWT}), donde explícitamente escribimos la tensión superficial macroscópica\footnote{Un hamiltoniano de interfase concebido como mejora a CWT, como aparentemente sería \textit{Drumhead}\cite{PhysRevB.32.233} necesitaría una tensión superficial correspondiente a nivel de fluctuaciones microscópicas que pretendiera describir.}. En este contexto la \textit{función de correlación de alturas} (\ref{eqn:FuncionCorrelacionGaussianaOrden2}) aparece definida como:
 \begin{equation}
\langle\xi(\vec{R})\xi(\vec{0})\rangle_{cw}=S(\vec{R})=\frac{1}{\beta\gamma(2\pi)^{d-1}}\int_{|q|<q_{max}} d\vec{q}\frac{e^{i\vec{q}\vec{R}}}{(q^{2}\xi_{cw}^{-2})}
\end{equation}
la medida usual de anchura de la interfase esta dada por $S(0)=\Delta_{cw}^{2}$. Lo más conveniente para caracterizar las propiedades de la interfase es definir \textit{funciones de distribución de alturas},
\begin{equation}
\mathcal{P}^{(1)}(z)=\left\langle\delta(\xi(\vec{R}))-z)\right\rangle_{cw}
\end{equation}
\begin{equation}
\mathcal{P}^{(2)}(z_{1},z_{2},R_{12})=\left\langle \delta(\xi(\vec{R}_{1}))-z_{1})\delta(\xi(\vec{R}_{2}))-z_{2})\right\rangle _{cw}
\end{equation}
Definir fluctuaciones en la densidad y relacionarlas con las distribuciones anteriores requiere dentro del modelo tradicional de ondas capilares definir un perfil de densidad intrínseco cuyo promedio funcional sobre $\xi(\vec{R})$ permite recuperar el perfil de densidad medio $\rho_{0}(z_{1})=\langle \tilde{\rho}_{\xi}(\vec{R}_{1})\rangle_{cw}$,
mientras que la función de correlación de pares aparece como\cite{weeks:6494,bedeaux:972}:
\begin{equation}
G_{cw}^{(2)}(z_{1},z_{2},\vec{R}_{12})=\left\langle \left[\tilde{\rho}_{\xi}(\vec{r}_{1})-\rho_{0}(z_{1})\right] \left[\tilde{\rho}_{\xi}(\vec{r}_{2})-\rho_{0}(z_{2})\right]\right\rangle_{cw}
\end{equation}
a partir de estas expresiones es posible calcular la función de distribución así como la función de correlación y comprobar si la teoría de ondas capilares satisface las relaciones integrodiferenciales comentadas en la teoría del funcional de la densidad, y en consecuencia obtener un forma plausible para $C_{cw}^{(2)}(z_{1},z_{2},\vec{R}_{12})$.\\

Bajo la hipótesis más sencilla que parte de un perfil de densidad intrínseco dado por la función paso se puede expresar a primer orden\cite{bedeaux:972},
 \begin{equation}
G^{(2)}_{cw}(z_{1},z_{2},\vec{R}_{12})\sim S(\vec{R}_{12})\frac{d\rho_{0}(z_{1})}{dz}\frac{d\rho_{0}(z_{2})}{dz}
\label{eqn:asintoticoG2cwt}
\end{equation}
con $\rho_{0}(z)$ dado por ec. (\ref{eqn:perfildensidadmedioCWT}) y una expresión más general que se puede ver en \S\ref{sec:funcionesDistribucionAlturas}. Es una ecuación donde la dimensionalidad del sistema aparece incluida en $S(R_{12})$ siendo más lentamente decreciente en dimensión 2, donde los efectos de las ondas capilares son más nítidos. En este caso \textit{Bedeaux y Weeks}\cite{bedeaux:972} determinaron, a partir de las propiedades de escalado de la función $G^{(2)}_{cw}$, que la correspondiente función $C_{cw}^{(2)}$ cumple la ecuación (\ref{eqn:Triezenberg-Zwanzig}).\\

La física contenida en la ecuación (\ref{eqn:asintoticoG2cwt}) respeta el análisis de Wertheim e incide en que las ondas capilares prevalecen en la superficie en el comportamiento a largas distancias, mientras que derivadas finitas del perfil de densidad medio la confinan en la interfase. Como esta función no contiene información acerca de la compresibilidad del líquido (aproxima el volumen por un líquido incompresible), la función correspondiente $C_{cw}^{(2)}$ no debe mostrar el comportamiento análogo a una función de correlación directa similar a ec. (\ref{eqn:c2devdw}) al alejarnos de la interfase\footnote{La determinación de las funciones de correlación de un modelo de ondas de capilaridad puede parecer directa en principio sin embargo ha generado cierta discusión tanto en si mismo como en el contraste con la teoría de van der Waals como veremos seguidamente. Esencialmente se ha planteado que la teoría de ondas capilares no puede ser una teoría correcta bien por sus predicciones, aparentemente contrarias a van der Waals bien porque la determinación de la función de correlación y su inclusión en TZ parecía dar resultados no convergentes\cite{weeks:6494}.}.\\

Las dos predicciones de CWT, divergencia de la anchura de la interfase y el largo alcance de las correlaciones en la interfase, aparecen relacionadas, véase ec.(\ref{eqn:scalingWeeks}) y nos proporcionan la imagen física de la relación entre ambos comportamientos. En (\ref{eqn:ecMODOq0wdw}) para q=0 encontramos correlaciones superficiales de largo alcance mientras que el perfil $\rho_{vdw}(z)$ permanece con anchura finita, la interpretación correcta de la divergencia contenida en ec.(\ref{eqn:ecMODOq0wdw}) es la presencia de una ruptura de simetría en el colectivo macrocanónico en la dirección z debida a la formación de una interfase $\rho_{vdw}(z)$ y el resultado obtenido refleja la existencia de un \textit{modo de goldstone} originado porque desplazamientos globales de la interfase no cuestan energía libre\footnote{La teoría de van der Waals puede ser considerada desde diferentes puntos de vista, hasta el momento la hemos considerado una aproximación y por tanto esta sujeta a interpretar que física contiene. Una alternativa es encontrar un potencial de interacción para el que sea una teoría exacta, en el caso uniforme es el llamado potencial de Kac. \textit{Weeks, Bedeaux y Zielinska}\cite{weeks:3790} determinaron un modelo asimétrico donde la teoría de la interfase de van der Waals resulta exacta, en este caso solo el modo q=0 resulta ser de largo alcance y aparece la imagen de una interfase con las correlaciones de largo alcance suprimidas pero la fuerte anisotropía del potencial provoca que algunas de la relaciones mostradas pierdan validez ya que por ejemplo la ecuación TZ no obtiene la tensión superficial.}.\\ 

El caso de $q\simeq0$ sugiere no es consistente incorporar sin más la teoría de ondas capilares sobre un perfil obtenido mediante una aproximación funcional, como van der Waals o sus generalizaciones,  sin haber evaluado antes la participación de las correlaciones superficiales contenidas en ella y expresadas en el perfil de densidad, y por otra parte la no divergencia de la interfase en $\rho_{vdw}(z)$ no se debe a un defecto en las predicciones o forma en la que la teoría de ondas capilares es construida a un nivel macroscópico. Esto último se puede ver a partir de resultados exactos obtenidos en modelos de red al compararlas con predicciones de modelos continuos.\\

Onsager en su estudio de un modelo de Ising bidimensional demostró la existencia de una transiciones de fase en el límite termodinámico y también determinó la energía libre interfacial\footnote{El método usado consiste en determinar la energía libre del sistema en tres condiciones de contorno diferentes $(++),(--), y (-+)$ y calculado la energía interfacial como $\mathcal{F}^{+-}-\mathcal{F}^{--}-\mathcal{F}^{++}$, que es similar a la determinación termodinámica comentada al inicio del presente capítulo.}. Los resultados fueron un exponente crítico para $\gamma$ de $\mu=1$. Además el exceso de energía libre por la presencia de la interfase, $\gamma$, es una cantidad bien definida y sin embargo el modelo de Ising bidimensional en ausencia de campo externo no posee una interfase bien definida (localizada). Esta última propiedad es fuertemente dependiente de la dimensionalidad del sistema de modo que en un modelo de Ising tridimensional si que es posible encontrar una interfase bien definida en campo nulo a temperaturas suficientemente bajas, a la temperatura límite se la denomina temperatura de \textit{roughening}.\\

La extensión de esta propiedad a sistemas continuos requiere cautela, ya que si un modelo de red posee una interfase difusa esperamos que el modelo continuo también aunque el recíproco no es cierto y de hecho la temperatura de \textit{roughening} no parece existir en sistema continuos. En dos y tres dimensiones en campo externo nulo no esperamos que exista una interfase localizada.\\

En el caso continuo en ausencia de campo externo podemos suponer que existe una tensión superficial bien definida, como en el resultado de Onsager junto con una interfase localizada, contrariamente a Onsager. La existencia de un perfil de densidad $\rho(z)$ implica\cite{PhysRevLett.54.444} que la función de correlación directa ha de poseer la forma cuando $|\vec{R}|>>1$,
\begin{equation}
c^{(2)}(\vec{R},z_{1},z_{2})\sim \frac{\rho'_{0}(z_{1})\rho'_{0}(z_{2})}{|\vec{R}|^{3+\eta}}
\end{equation}
donde $\eta$ es un exponente positivo menor que la unidad. Ahora basándose en la expresión (\ref{eqn:Triezenberg-Zwanzig}) se llega la conclusión de la divergencia de la tensión superficial, desechando la hipótesis de partida de $\rho'(z)\neq0$. La demostración debida a \textit{M. Robert} se restringe a dimensión $d\leq3$ con lo que estos resultados favorecen las conclusiones de CWT en campo nulo, frente a un perfil de densidad bien definido $\rho_{vdw}(z)$, cuestión además corroborada mediante simulaciones\cite{PhysRevE.60.6708}.\\

La extensión de este argumento a campo externo no nulo y el análisis del comportamiento de $\rho'(z)$ en el límite $g\rightarrow 0^{+}$ no es inmediata. Las ecuaciones expuestas sugieren que para $\xi_{\perp}$ definido como $\Delta^{2}_{cw}$ podemos expresar $G^{(2)}(z_{1},z_{2},\vec{R}\simeq0)\sim \rho'_{0}(z_{1})\rho'_{0}(z_{2})\xi_{\perp}^{2}$. La idea de que el producto $\rho'_{0}(z_{2})\xi_{\perp}$ sea finito para tener un valor finito de la función de correlación densidad-densidad implica una relación entre la derivada del perfil de densidad y la anchura de la interfase, indicando que la divergencia de la segunda de puede compensar si $\rho'_{0}(z_{1})\sim 0$ de modo adecuado\cite{EvansBook}.\\

Desde esta óptica las teorías de campo medio retienen información sobre las ondas capilares en la función de correlación de pares, véase ecuación (\ref{eqn:ecMODOq0wdw}), pero su expresión en el perfil de densidad no es completo, y se hace necesaria otra vía para examinar las fluctuaciones incluidas al menos de modo efectivo en un perfil de densidad líquido-vapor. El enlace concreto entre las propiedades intrínsecas y las fluctuaciones superficiales queda pues abierto.\\

    \section{Modelo de Ondas Capilares Extendido}\label{sec:RRVR}

Se constituye a partir de resultados generales dentro de la teoría del funcional de la densidad concernientes a la expansión funcional entorno de la densidad de equilibrio del sistema. La expresión a segundo orden involucra las funciones de correlación de orden dos accesibles a las teorías funcionales actuales. Sobre esta expresión general se particulariza en las mismas condiciones que TZ y propone un desplazamiento \textit{rígido} del perfil de densidad siguiendo la superficie intrínseca, conceptualizada básicamente como la superficie de Gibbs local,
\begin{equation}
\Delta\Omega=\frac{1}{2\beta}\int d\vec{R}_{1}d\vec{R}_{2}\tilde{C}^{(2)}(|\vec{R}_{1}-\vec{R}_{2}|)\xi(\vec{R}_{1})\xi(\vec{R}_{2})
\end{equation}
donde han considerado con generalidad,
\begin{equation}
\tilde{C}^{(2)}(|\vec{R}_{1}-\vec{R}_{2}|)=\int dz_{1}dz_{2}\frac{d\rho_{0}(z_{1})}{dz_{1}}\frac{d\rho_{0}(z_{2})}{dz_{1}}C^{(2)}(z_{1},z_{2};|\vec{R}_{1}-\vec{R}_{2}|)
\end{equation}
Si identificamos esta función con el término de orden $q^{2}$ de su transformada de Fourier recuperamos la expresión original de TZ que se considera expresión de CWT. Por consiguiente se denomina a este modelo como \textit{modelo de ondas capilares extendido}. Si indagamos en los primeros coeficientes del desarrollo encontramos que,\\
\begin{equation}
\int d\vec{R}\tilde{C}^{(2)}(\vec{R})=\beta\Delta\rho mg=\hat{C}^{(2)}_{0}(0)
\end{equation}
y después usando TZ para la tensión superficial podemos expresar que,
\begin{equation}
\beta\gamma=\hat{C}^{(2)}_{2}(0)
\end{equation}
con lo que obtenemos $\Delta\Omega_{cw}[\xi]$.
El modelo extendido es obtenido incorporando sucesivos momentos de las correlaciones transversales de modo que,
\begin{equation}
\beta\kappa=\hat{C}^{(2)}_{4}(0)
\label{eqn:RRVRkappa}
\end{equation}
se denomina \textit{bending rigidity modulus} y es un coeficiente de rigidez que representa el cambio en energía libre de curvatura elástica de la superficie.\\

De modo general tenemos que,
\begin{equation}
\Delta \Omega=\frac{1}{2\beta}\frac{1}{L^{2}}\sum_{q}\tilde{C}^{(2)}(q)|\xi(q)|^{2}
\end{equation}
y por tanto
\begin{equation}
\langle|\xi(q)|^{2}\rangle=L^{-2}\frac{1}{\tilde{C}^{(2)}(q)}
\end{equation}

Cabe recordar que una vez se han considerado validas las generalizaciones que implica, surge como cuestión natural la relevancia de los siguientes ordenes en las consecuencias que la teoría de ondas capilares clásica da como resultado, en particular las propiedades para la tensión superficial\footnote{Antes y después de incluir fluctuaciones superficiales.} y la fluctuación cuadrática media en la altura, así como las consecuencias que tiene para la propuesta de escalado realizada sobre las correlaciones transversales, ec. (\ref{eqn:scalingWeeks}). Por esta razón \textit{A.Robledo, C. Varea y V.Romero-Rochin} \cite{1992PhyA184367R,1991PhyA177474R} realizan un análisis del posible papel jugado por el coeficiente de rigidez. Abordan también la cuestión, más sutil, de las propiedades que encierra el hamiltoniano efectivo para la interfase que esta teoría implica y los exponentes que podría poseer considerando que respecto del límite $g\rightarrow 0^{+}$ posee un comportamiento crítico. Curiosamente los autores reconocen que su propuesta supone una redefinición de $\gamma$ como $\gamma(q)$ pero desechan esta forma de entender la interfase\cite{PhysRevLett.86.2369}.\\

Siendo una extensión razonable dentro de la teoría del funcional de la densidad surge la cuestión de hasta que punto posee sentido incluir términos progresivamente mayores en un desarrollo en q sobre una hipótesis construida bajo la imagen macroscópica de la interfase, en los últimos capítulos de la memoria haremos incapie en este hecho\footnote{La evaluación mediante simulaciones de dinámica molecular de estas hipótesis ha sido realizada por Stecki\cite{stecki:7574}, definiendo la superficie intrínseca como la superficie local de Gibbs. Stecki encontró una conexión entre dos cantidades aparentemente diferentes $<|\xi_{q}|^{2}>$ y $\chi(q)$, es decir, una propiedad eminentemente superficial con una propiedad definida a partir de fluctuaciones en el volumen del líquido, véase \S\ref{sec:intruSuperIntrinseca}. Esto lleva a considerar que la propuesta de modelo extendido posee fluctuaciones de dos tipos y en consecuencia una verdadera estimación de sus resultados debería ser realizada por un modelo que incluyese fluctuaciones del tipo propuesto arriba junto a fluctuaciones propias del sistema intrínseco.}.\\

Colateralmente uno de los objetivos de \textit{Romero-Rochin, Varea y Robledo} (RRVR) es intentar expresar mediante evaluaciones microscópicas los términos que se encuentran en un hamiltoniano efectivo de Helfrich,\\
\begin{equation}
\Omega_{S}=\int d\vec{S}(\gamma-2\kappa c_{0}J+\kappa J^{2}+\overline{\kappa}K)
\label{eq:Helfrich}
\end{equation}
y por tanto dotar a este en el contexto dado por el funcional de la densidad de una base microscópica\footnote{En este punto solo nos interesa que es una extensión formal de la teoría de ondas capilares clásica donde RRVR dan da la interpretación microscópica de $\kappa$ como ec.(\ref{eqn:RRVRkappa}).}.\\

El modelo de ondas capilares extendido no aclara si la relación presente en CWT entre el perfil intrínseco y la superficie intrínseca puede incluirse en las escalas microscópicas. Con este fin el esquema ofrecido por RRVR puede interpretarse del siguiente modo, el perfil de equilibrio, en el sentido de minimizar la ecuación de Euler-Lagrange (\ref{eqn:EulerLagrangeMinimizacion}), lo llamamos $\rho_{0}(z)$ y verifica que $\rho(z^{x})=\rho^{x}$.  Si extendemos esta definición al plano transversal para todo valor $\vec{R}$ tenemos para cada par $(z^{x},\rho^{x})$ una posible familia de perfiles de densidad y el mínimo de esta familia es nuestro perfil de equilibrio. Si otra vez extendemos esta idea, ahora al caso de una ligadura $(z^{x}=\xi(\vec{R}),\rho^{x})$, de nuevo para una funcion $\xi(\vec{R})$ habremos de realizar un proceso de minimización para encontrar el valor de equilibrio para este sistema \textit{ligado}, la minimización sobre la familia $\xi(\vec{R})$ debe permitir recuperar el perfil de equilibrio líquido-vapor.\\

Para determinar $\Omega_{S}[\xi]$ debemos minimizar el funcional $\Omega[\rho]$ en la familia de perfiles sujetos a la condición $(z^{x}=\xi(\vec{R}),\rho^{x})$ esto nos da una expresión que se puede suponer de la forma dada por ec. (\ref{eq:Helfrich}) e intentar estimar la forma de los coeficientes desde un desarrollo perturbativo en la densidad y que tecnicamente pueden depender de la condicion inicial $(z^{x},\rho^{x})$. En este contexto podemos ampliar aun más la idea del modelo extendido proponiendo una generalización de la ec.(\ref{eqn:perturbacionTipoCWT}) de la forma,
\begin{equation}
\Delta\rho(\vec{R},z)=\int d\vec{R} f(z;|\vec{R}-\vec{R}'|)\xi(\vec{R}')
\end{equation}
que es una convolución con una cierta función peso $f(z,l)$ por determinar. Según esta idea tenemos,
\begin{equation}
\rho(\vec{R},z)=\rho_{0}(z)+\int d\vec{q}e^{i\vec{q}\vec{R}}f(z,|\vec{q}|)\xi(|\vec{q}|)
\end{equation}
Realizando un desarrollo $\hat{f}(z,|\vec{q}|)=\hat{f}_{0}(z)+q^{2}\hat{f}_{2}(z)+...$, comprobamos que si nos restringimos a $\hat{f}_{0}(z)$, debemos recuperar RRVR, y el resto de términos son una extensión formal de RRVR que permiten realizar cierta minimización de la indicada antes como necesaria\footnote{Esta idea parte del \textit{crossing criteria} que se analizará en detalle en \S\ref{sec:capituloHeff}, de ciertas hipótesis implícitas acerca de $\xi(\vec{R})$ y de si realmente es posible expresar $\Omega[\xi]$ mediante este desarrollo en la densidad. Por otra parte una extensión más elaborada de este desarrollo así como de la inclusión formal de las hipótesis sobre la naturaleza del perfil intrínseco en las proximidades del punto triple será indicada en \S\ref{sec:apendiceIntrinseco}.}. La corrección $\hat{f}_{2}(z)$ fue propuesta por \textit{Parry y Boulter}\cite{Parry:1994:7199} y sugieren obtener una expresión aproximada para esta función desde de las correlaciones en una teoría de campo medio\footnote{Véase también \cite{1997JSP89273Robledo}.}, lo usual es tomar un esquema de Landau-Ginzburg-Wilson sencillo sobre el que sea posible encontrar analíticamente la solución a la ecuación de Green para $G_{MF}^{(2)}(z_{1},z_{2},q)$. Como resultado de esta propuesta se obtiene una $\beta\kappa$ acotada superiormente por la obtenida en RRVR.\\

\section{Determinación del perfil de densidad}

Como se indicó en la introducción la forma de los perfiles de densidad de equilibrio ha presentado en la historia de la teoría de líquidos bastante controversia. Una posible vía para resolver el perfil de densidad es aplicar iterativamente las ecuaciones que surgen de (\ref{eqn:eqautoconsistenterho}) y (\ref{eqn:YBGorden1}) de 
tal y como realizaron \textit{Fischer y Methfessel}\cite{PhysRevA.22.2836}, \textit{Toxvaerd}\cite{toxvaerd:2863} y \textit{Croxton}\cite{0022-3719-4-14-006}. Esencialemente se incorpora el conocimiento de la función de correlación, bien directa bien total, del sistema uniforme al caso no-uniforme bajo determinadas hipótesis. La mayoría de los resultados daban perfiles suaves similares a la teoría de van der Waals con diferentes anchuras $\xi_{\perp}$. Se encontró también una solución oscilante que se atribuyo a inestabilidades cerca del punto triple. Croxton fue el primero en sugerir una interpretación diferente, con un modo adecuado de incorporar el sistema uniforme y modo adecuado de realizar el proceso iterativo le fue posible encontrar perfiles oscilantes. Esencialmente la resolución de ec. (\ref{eqn:YBGorden1}) mezcla soluciones iterativamente y en cada paso se redefine la interfase mediante la superficie equimolar de Gibbs, esto según Croxton borra todo rastro de oscilaciones en la solución final. Sus cálculos fueron corroborados por simulaciones sin embargo sus resultados se tomaron como anomalías hasta que dentro del desarrollo del funcional de la densidad aparece la estructuración como una propiedad característica de interfases de diferente naturaleza\footnote{Es resaltable el trabajo realizado por Croxton. Su objetivo fue encontrar expresiones plausibles para las funciones de distribución en la interfase, proponiendo de hecho, una densidad oscilante en la parte densa del perfil de densidad\cite{0022-3719-4-14-005,0022-3719-4-16-009,0022-3719-4-14-006} y constituye por tanto la primera aportación en el contexto de la teoría de líquidos desde el enfoque que al problema de la interfase se da en esta memoria. Croxton observa como bajo la ec. (\ref{eqn:YBGorden1}), tenemos que incluir el conocimiento de la función de distribución a dos cuerpos, difícil de conocer en el caso no-homogéneo y propone incluir un operador que transforme el caso homogéneo en el caso no-homogéneo y resolver así una expresión para la función de distribución a un cuerpo bajo hipótesis plausibles para este operador. En su trabajo de hecho intuye que en los metales líquidos la estructuración debería ser más evidente que en el caso del Argón estudiado por él. Más adelante se han determinado también bajo la ecuación (\ref{eqn:YBGorden1}) estructuras líquidas en contacto con paredes\cite{PhysRevA.22.2836}, sin embargo hasta el desarrollo de funcionales de la densidad que incluían mejores aproximaciones a las correlaciones del sistema de esferas duras no se vislumbra la evidencia de perfiles estructurados y se retoma la imagen dada por Croxton\cite{EvansMolecularPhyscFWline}.}.\\

\begin{figure}[htbp] 
   \centering
   \includegraphics[width=0.70\textwidth]{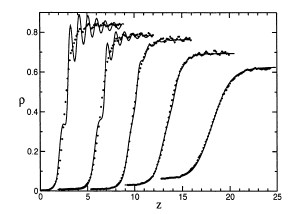} 
   \caption{\textbf{Perfiles de densidad obtenidos por Weeks y Katsov}\cite{KatsovWeeksMWDAliqVap} mediante una propuesta de extensión sobre la teoría de van der Waals. Observar tanto las pronunciadas oscilaciones como la primera pequeña oscilación previa antes de oscilar entorno de $\rho_{l}$. Las curvas de puntos se corresponden con simulaciones de Dinámica Molecular realizadas por Mecke para el modelo Lennard-Jones que también usan Weeks y Katsov.}
\label{fig:WeeksKatsov}
\end{figure}

De modo diferente, desde la imagen de un líquido dada por WCA y la teoría de perturbaciones, \textit{Katsov y Weeks}\cite{KatsovWeeksMWDAliqVap} han encontrado una fuerte estructura en capas para la interfase líquido-vapor y en general de los sistemas no homogéneos. Aunque sugieren la posibilidad de que exista una teoría del funcional de la densidad detrás de esta aproximación no determinan su posible forma, véase figura (\ref{fig:WeeksKatsov}). Obtienen perfiles similares a los perfiles que mostraremos cerca del punto triple aunque en su caso no realizan un análisis que mostraremos necesario para poder comparar perfiles dados por un cálculo teórico con perfiles dados en un esquema de simulación.

    \section{Sumario}
En este capítulo se han analizado la interfase líquido-vapor en el contexto de la teoría de ondas capilares, tanto desde el punto de vista de la tensión superficial como de los perfiles de densidad.
\begin{itemize}
\item La teoría de ondas de capilaridad impone unas predicciones sobre el perfil de densidad y las funciones de correlación que contrastan con los resultados de las aproximaciones funcionales conocidas en la teoría del funcional de la densidad. Con todo dicha teoría de ondas capilares es perfectamente consistente con las reglas de suma del formalismo del funcional de la densidad. Las propiedades del perfil de la densidad y las funciones de correlación están relacionadas por propiedades de escalado.
\item Un análisis de la estabilidad de las aproximaciones funcionales revela propiedades similares en la función de correlación total transversal a las predicciones de la teoría de ondas capilares, sin embargo las funciones de correlación directa incluidas en las aproximaciones funcionales presentan una forma notablemente diferente.
\item Dicho análisis de la estabilidad revela la presencia de ondas capilares en los perfiles de densidad de teorías funcionales pero no la longitud de coherencia que para esta fluctuaciones esta incluida. Diferentes tipos de fluctuaciones en la densidad pueden tener diferentes longitudes de coherencia.
\item La tensión superficial puede determinarse por diferentes caminos, la extensión de los argumentos que llevan a su definición en el funcional de la densidad permite definir teorías que formalmente son extensiones de la teoría de ondas capilares clásica y que pueden enlazarse con teorías de la interfase de carácter también fenomenológico.
\item Los perfiles de densidad pueden ser obtenidos por diferentes vías, un análisis cuidadoso revela el carácter oscilante de estos, y abre la posibilidad de diferentes interpretaciones acerca de la estructura de la interfase.

\end{itemize}

\part{ Planteamiento y Resultados}
\chapter{Diagrama de fases y propiedades estructurales}

En este capítulo se abordan dos objetivos previos al estudio de superficies líquidas mediante métodos funcionales, el primero es un análisis del diagrama de fases para el conjunto de potenciales de interacción descritos en \S\ref{sec:potencialesSodioMercurio} y el segundo es un análisis de las propiedades estructurales del sistema uniforme desde dos cuestiones diferentes: el rango de estabilidad de la fase líquida frente a la fase sólida y las propiedades asintóticas de la estructura de correlación del líquido.\\

Dichos análisis nos van a permitir entender resultados posteriores obtenidos en el marco formal de las \textit{teorías de van der Waals generalizadas} y permiten una primera comparación tanto con los estudios previos desde el punto de vista de las teorías del funcional de la densidad para otros potenciales\cite{EvansMolecularPhyscFWline,dijkstra:1449} como las conclusiones extraídas desde simulaciones Montecarlo\cite{PhysRevLett.87.166101}. Para comprender la física en estas aproximaciones funcionales necesitamos contrastar los resultados de DFT, obtenidos para dos funcionales diferentes basados en FMT y WDA, con otras teorías y diferenciar las propiedades que pertenecen al modelo de interacción y las que pertenecen a las aproximaciones utilizadas, el estudio se amplia por esta razón en dos direcciones, mediante la teoría de las ecuaciones integrales y mediante \textit{experimentos numéricos}\footnote{Los trabajos iniciales con estos potenciales se realizaron mediante simulación Montecarlo, en esta memoria se han realizado simulaciones de dinámica molecular para completar aquellos. La presente memoria amplia los resultados publicados en las referencias \cite{PhysRevE.70.061601,tarazona2007critical}}. Introducimos inicialmente los dos formalismos usados.\\

\section{Teorías de van der Waals generalizadas}
\label{sec:vanDerwaalsGENERAL}
Siguiendo el tratamiento expuesto en \S\ref{sec:integracionTermodinamica} es posible determinar el funcional $F[\rho,\phi]$, entendido como un funcional también del potencial de interacción, siempre que conozcamos la derivada funcional correspondiente, en este caso basándonos en ec. (\ref{eqn:promedioPHIN}) tenemos que $\rho^{(2)}(\vec{r}_{1},\vec{r}_{2})=2\frac{\partial \Omega}{\partial \phi (\vec{r}_{1},\vec{r}_{2})}$ luego conocido el funcional $F[\rho,\phi_{ref}]$ para un potencial de referencia $\phi_{ref}$ y separando el resto del potencial como $\phi_{p}=\phi-\phi_{ref}$, es exacto escribir\cite{Evans92infundamentals,abraham:157,PhysRevA.25.1669},
\begin{equation}
\mathcal{F}[\rho,\phi]=\mathcal{F}_{ref}[\rho,\phi_{ref}]+\frac{1}{2}\int d\alpha\int d\vec{r}_{1}d\vec{r}_{2}\left[\phi(\vec{r}_{1},\vec{r}_{2})-\phi_{ref}(\vec{r}_{1},\vec{r}_{2})\right] \rho^{(2)}(\alpha\phi;\vec{r}_{1},\vec{r}_{2})
\label{eqn:vanDerWaalsGeneralizada}
\end{equation}

donde hemos introducido el potencial de interacción vía ec. (\ref{eqn:acoploPotencial}). La expresión anterior requiere conocer la función de correlación $\rho^{(2)}(\vec{r}_{1},\vec{r}_{2})$ en el rango de potenciales necesario para realizar la integración, información que es obviamente desconocida y que obliga a utilizar diferentes aproximaciones de manera que incluyamos el conocimiento parcial que de esta tengamos. Las aproximaciones básicas son:\\
\begin{itemize}
\item La más sencilla es eliminar por completo las correlaciones entre las partículas y escribir:
\begin{equation}
\rho^{(2)}(\vec{r}_{1},\vec{r}_{2})\approx\rho(\vec{r}_{1})\rho(\vec{r}_{2})
\end{equation}
conocida como \emph{aproximación de campo medio}, de manera que no es necesario conocer la función de distribución en el rango de potenciales\footnote{En la práctica estamos suponiendo que $h^{(2)}(\vec{r}_{1},\vec{r}_{2})=0$ y explícitamente estamos despreciando un término, $\frac{\beta}{2}\int d\alpha\int d\vec{r}_{1}d\vec{r}_{2}\phi_{p}(\vec{r}_{1},\vec{r}_{2})[\rho^{(2)}(\alpha;\vec{r}_{1},\vec{r}_{2})-\rho(\vec{r}_{1})\rho(\vec{r}_{2})]$, aproximación que resulta exacta en el caso del potencial de Kac\cite{PhysRevA.25.1669} y que permite establecer un desarrollo perturbativo sobre el parámetro $\gamma\rightarrow0^{+}$ que define a este potencial $\phi_{p}(\vec{r}_{1},\vec{r}_{2})=\gamma^{3}\omega(\gamma|\vec{r}_{12}|)$.}.
\item La siguiente corrección consiste en incluir ciertas correlaciones partiendo de la definición de las funciones de distribución considerando que dichas correlaciones no varían en el  proceso de integración y por tanto es factible escribir:
\begin{equation}
\rho^{(2)}(\vec{r}_{1},\vec{r}_{2})\cong g_{ref}^{(2)}(\vec{r}_{1},\vec{r}_{2})\rho(\vec{r}_{1})\rho(\vec{r}_{2})
\label{eqn:perturbacionORDEN1}
\end{equation}
\item Mejorar la aproximación anterior complica notablemente el tratamiento y necesita conocer funciones de distribución del sistema de referencia de ordenes superiores\footnote{Esencialmente necesitamos la variación de $\rho^{(2)}$ con el parámetro de acoplo del potencial de la teoría perturbativa evaluada en el sistema de referencia, lo que requiere tanto las funciones $\rho_{ref}^{(3)}$ como $\rho_{ref}^{(4)}$, que se suelen aproximar por su límite de baja densidad donde sabemos factorizan en funciones a dos cuerpos y de hecho el límite asintótico las identifica como las funciones $\rho^{(2)}$.} lo que implica nuevas aproximaciones\cite{BOOK-TheorySimpleLiquids,RevModPhys.48.587}.
\end{itemize}

Para aplicar la ec. (\ref{eqn:perturbacionORDEN1}) necesitamos una aproximación más ya que la función $g_{ref}^{(2)}(\vec{r}_{1},\vec{r}_{2})$ vuelve a ser desconocida, aun así la descripción estructural del líquido respecto de la aproximación de campo medio se mejora notablemente mediante la aproximación,
\begin{equation}
g_{ref}^{(2)}(\vec{r}_{1},\vec{r}_{2})\simeq g_{ref}^{(2)}(|\vec{r}_{12}|;\rho_{eff})
\label{eqn:aproximacionRHOeff}
\end{equation}
La elección de la densidad efectiva $\rho_{eff}$ requiere evaluar de algún modo la densidad en $\vec{r}_{1}$ y $\vec{r}_{2}$. Dos prescripciones razonables se basan en promediar la densidad en una esfera de radio $\tilde{\sigma}$, similar al tamaño molecular, globalmente resulta\cite{sokolowski:5441,wadewitz:2447},
\begin{align}
\rho^{(2)}(\vec{r}_{1},\vec{r}_{2})&\simeq\rho(\vec{r}_{1})\rho(\vec{r}_{2})g_{ref}\left(r_{12}; \bar{\rho}\left(\frac{\vec{r}_{1}+\vec{r}_{2}}{2}\right)\right) \\
&\simeq\rho(\vec{r}_{1})\rho(\vec{r}_{2})g_{ref}\left(r_{12};\frac{\bar{\rho}(\vec{r}_{1})+\bar{\rho}(\vec{r}_{2})}{2}\right)  \notag
\end{align}
La función $g_{ref}(r)$ condensa pues el conocimiento estructural que poseemos del sistema de referencia, mientras que la posible falta de isotropía entre los puntos $\vec{r}_{1}$ y $\vec{r}_{2}$ aparece contenida en el producto de densidades a un cuerpo, véase ec. (\ref{eqn:perturbacionORDEN1}). En el caso de potenciales Lennard-Jones\footnote{Existen alternativas basadas en el sistema de esferas duras como referencia pero en una función $g_{ref}^{(2)}(|\vec{r}_{12}|;\rho_{eff})$, obtenida a partir de un esquema perturbativo dentro de OZ, véase \textit{Tang y Wu}\cite{tang:7388}.} se ha elegido\cite{toxvaerd:3116} el sistema de referencia de esferas duras\cite{TarazonaMarconiEvans} donde $g_{HS}(r)$ que posee descripciones analíticas\cite{wertheim:643} y semiempíricas notables\cite{PhysRevA.5.939} y la determinación del sistema de referencia óptimo se realiza mediante las teorías perturbativas usuales \S\ref{sec:modelosPerturbativos}, mientras que la limitación descriptiva que contiene el caso uniforme hace esperar que las teorías de van der Waals generalizadas así construidas respondan preferentemente a la imagen de un líquido cerca del punto triple.\\

\section{Estudio mediante la teoría de las ecuaciones integrales}

En el estudio de potenciales a pares la termodinámica del sistema se halla contenida en la función de distribución radial y será consistente en la medida en que las diferentes rutas que determinan la relación termodinámica-estructura \S\ref{sec:rutasTermoEstructura}
 den los mismos resultados. El marco ofrecido por la teoría de las ecuaciones integrales, véase \S\ref{sec:ecuacionesIntegrales}, es idóneo para la obtención de propiedades estructurales del sistema uniforme y una elección adecuada de la aproximación en este marco encuentra resultados similares a los ofrecidos por simulación.\\

La teoría de las ecuaciones integrales basada en OZ se completa con diferentes relaciones de cierre, donde las dos consideradas básicas son la de Percus-Yevick \S\ref{sec:teoriaPY}, que funciona adecuadamente en el régimen de cortas distancias, y la de HNC \S\ref{sec:teoriaHNC}, que lo hace en el caso de largas distancias. Como ambas poseen el inconveniente de ser termodinámicamente inconsistentes una primera idea semi-empírica es combinarlas y mediante los parámetros libres de dicha combinación buscar una consistencia más alta entre las relaciones termodinámica-estructura, aprovechando las distancias en que ambas presentan sus ventajas\footnote{La mejora sobre las relaciones de cierre básicas no es necesariamente una combinación en busca de consistencia, basándose en otros criterios \textit{Verlet}\cite{VerletClosure} y \textit{Martynov-Sarkisov}\cite{MartynovSarkisov} han propuesto otras relaciones de cierre con notable mejora en sus resultados y de hecho mayor consistencia, al incorporar propiedades que la función puente se posee, función que seguidamente introduciremos.}. En esta línea ha habido considerables aportaciones como la relación de cierre de Rogers-Young\cite{PhysRevA.30.999}, que interpola entre HNC y PY o HMSA\cite{zerah:2336} que interpola entre HNC y SMSA que a su vez interpola entre PY y MSA.\\

\begin{figure}[htbp] 
   \centering
   \includegraphics[width=1.0\textwidth]{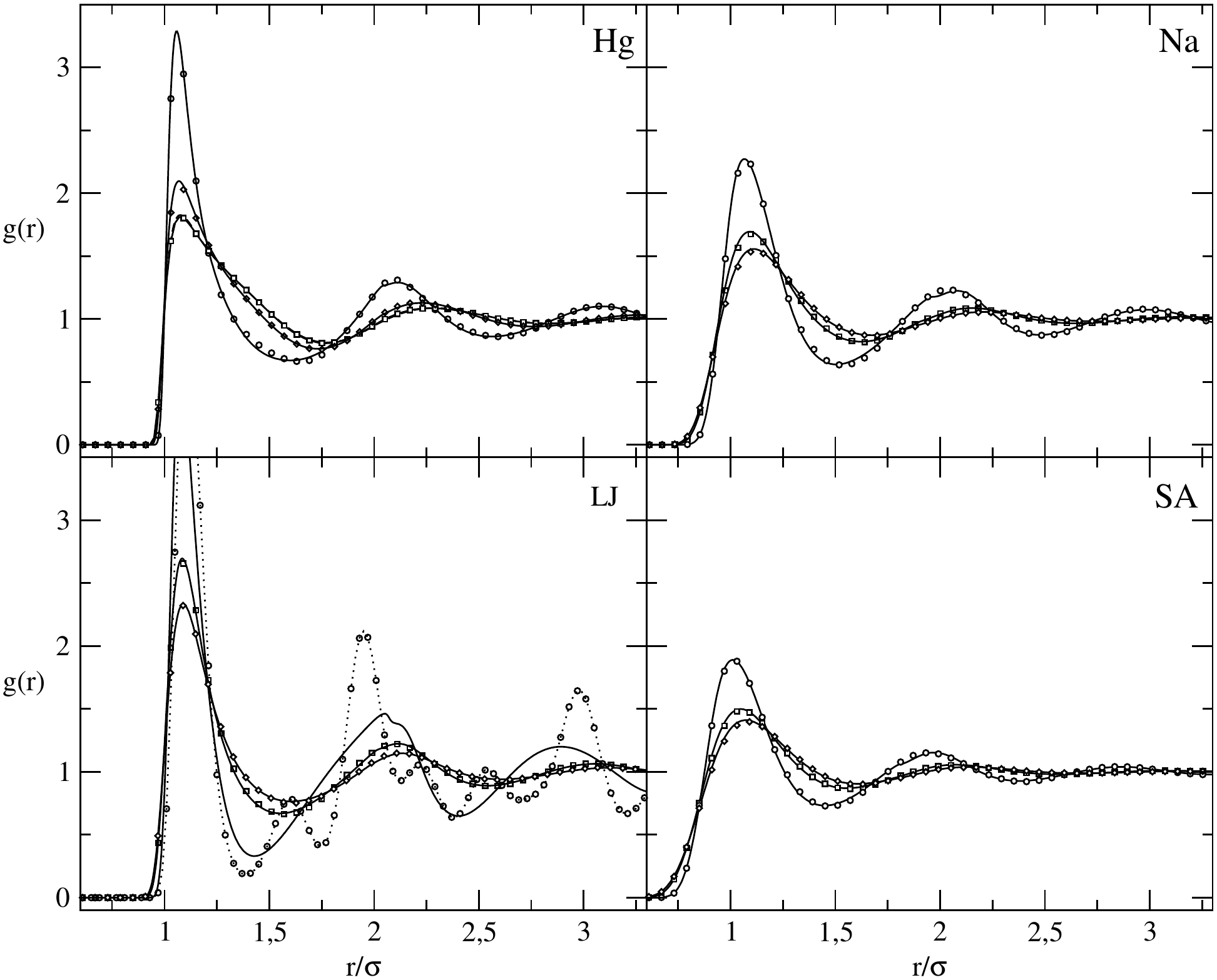} 
   \caption{\textbf{Funciones de distribución radial} para los tres potenciales introducidos y el potencial de Lennard-Jones. Se comparan RHNC-Lado (líneas continuas) con Simulación de dinámica molecular (círculos, cuadrados, rombos) a T/U=0.40, T/U=0.85 y T/U=1.05 sobre la línea de coexistencia de fases en la rama densa de RHNC-Lado. En el caso LJ, a T/U=0.40 la simulación muestra una $g(r)$ característica de una red fcc, mientras RHNC-Lado continua dando una posible solución líquida. Las simulaciones de dinámica molecular han sido realizadas con 4000 partículas y un termostato de Nosé-Hoveer para fijar el colectivo canónico. El lector puede comparar estas figuras con la figura (\ref{fig:GdeRmetalesRice}).}
   \label{fig:FuncionesDistribucionRadial}
\end{figure}

Es posible además introducir otras aproximaciones, y de hecho reescribir varias de las anteriores, desde otra relación exacta que complemente la ecuación de Ornstein-Zernique (\ref{eqn:OZ}), introduciendo la llamada \textit{función puente} b(r) mediante
\begin{equation}
c(r)=h(r)-ln\left[(h(r)+1)e^{\beta\phi(r)}\right]+b(r)
\label{eqn:definicionPuente}
\end{equation}
La relación de $b(r)$ como funcional de $h(r)$ es notablemente compleja\footnote{Involucra diagramas altamente conectados y solamente es posible calcular numéricamente de modo abordable algunos de ellos\cite{AttardBook}, además errores numéricos en g(r) son amplificados en c(r) y más aun en b(r), esto hace que la determinación sistemática sea más complicada para esta función. Aun así para esferas duras el comportamiento a baja densidad es conocido ya que involucra únicamente los primeros diagramas de su desarrollo en densidad. Una vía diferente propuesta por \textit{Labik y Malievsky}\cite{Malijevsky:0026-8976:663} fue parametrizar la función $b_{HS}(r)$ y determinar estos parámetros de modo adecuado basándose en la ecuación de estado de estado de Erpenbeck-Wood y relaciones de suma, es decir, relaciones que integran propiedades estructurales y obtienen propiedades termodinámicas.}. La cuestión se puede resolver razonablemente gracias a la presencia, indicada por \textit{Y. Rosenfeld y N. W. Ashcroft}\cite{PhysRevA.20.1208}, de una cierta universalidad en la función $b(r)$ manifestada independientemente del potencial de interacción $\phi(r)$. Esto abre la posibilidad de analizar el problema incluyendo una función $b_{0}(r)$ basada en un potencial $\phi_{0}(r)$ que sea conocida y que expresada como función ajustable permita, desde un punto de vista variacional, minimizar la energía libre del sistema e introducir requisitos de consistencia en la aproximación\cite{PhysRevA.28.2374}. Específicamente este proceso establece que,
\begin{equation}
4\pi\rho\int dr r^{2}\left[g(r)-g_{0}(r)\right]\frac{\partial b_{0}(r,\sigma)}{\partial\sigma}=0
\label{eqn:criterioLADOfuncionpuente}
\end{equation}
que hace consistente la ruta de la energía \S\ref{eqn:EcuacionEnergia}
 y la ruta del virial \S\ref{eqn:EcuacionVirial}
. El procedimiento necesita de la función puente del sistema de referencia que en lo que sigue tomamos como un sistema de esferas duras a partir de dos funciones aproximadas pero consideradas cuasi-exactas como son la función de distribución radial de \textit{Verlet-Weis}\cite{PhysRevA.5.939} y la función cavidad de \textit{Henderson-Grundke}\cite{henderson:601} en la zona $r<\sigma$. De este modo,
\begin{equation}
b_{0}(r)=ln y_{0}(r)+\gamma_{0}(r)
\end{equation}
la función $\gamma(r)$ se denomina \textit{función serie o nodal} y dada una prescripción para h(r) puede ser obtenida su transformada de Fourier mediante,
\begin{equation}
\hat{\gamma}_{0}(k)=\frac{\rho\hat{h}_{0}^{2}(k)}{1+\rho\hat{h}_{0}(k)}
\end{equation}
ambas funciones, $y_{0}(r)$ y $\gamma_{0}(r)$, aparecen parametrizadas en función de $\sigma$ y es posible la diferenciación de $b_{0}(r;\sigma)$ y en consecuencia ec. (\ref{eqn:criterioLADOfuncionpuente}) es resoluble. Un método de solucionar del sistema de ecuaciones acoplado (\ref{eqn:definicionPuente} y \ref{eqn:OZ})
eficaz numéricamente ha sido descrito con detalle por \textit{Lomba}\cite{1989MolPh..68...87L} y por \textit{Labik, Malijevsky y Vonka}\cite{1985MolPh..56..709L}, mientras que otras propuestas se basan en transformaciones de la secuencia de la serie de soluciones que surge del proceso iterativo y que optimizan sus propiedades de convergencia\cite{Homeier:0010-4655:188}.\\

Una vez resuelta la estructura del líquido uniforme podemos determinar las condiciones de coexistencia de fases líquida y vapor aprovechando que el criterio de Lado permite calcular de modo consistente la presión por la ruta del virial. En lo que sigue denominamos a la aproximación usada RHNC-Lado.\\

Los resultados de resolver el sistema de ecuaciones (\ref{eqn:OZ} y \ref{eqn:definicionPuente}) para la función de distribución radial g(r) del Mercurio\footnote{El modelo Mercurio a T/U=0.40 equivaldría aproximadamente a 570 Kelvin.} comparan \textit{cualitativamente} bien con los resultados \textit{ab initio} de \textit{Zhao et al} para el Galium\cite{PhysRevE.56.7033}. Mientras que para metales alcalinos\cite{chekmarev:768} podemos comparar como en el potencial Sodio la función de distribución radial comparte la forma del primer y segundo pico (más suavizados que un potencial polinómico más repulsivo como el Lennard-Jones) y constituyen una forma muy sencilla de tratar metales alcalinos (Na,K,Rb,Cs),  comparense las figuras (\ref{fig:GdeRmetalesRice}) y (\ref{fig:FuncionesDistribucionRadial}).\\

Podemos entender algo más de estos potenciales a partir de las funciones de distribución radial del Sodio y un Lennard-Jones observando como este último, en el caso de simulación, y partiendo de una configuración líquida se desestabiliza a una estructura fcc, mientras que el modelo de metal alcalino permanece en estado líquido a una densidad y temperatura equivalente\footnote{La curvas de coexistencia de fases en unidades reducidas muestran unos valores de la densidad bastante similares entre ambos potenciales.} (en que veremos que además es estable). La aproximación RHNC-Lado continua siendo una descripción adecuada de la estructura de líquidos \textit{blandos} en el caso uniforme como comprobamos en el \textit{Soft-Alcaline} indicando que las suposiciones de universalidad de b(r) y la determinación realizada del diámetro de esferas duras son adecuadas para un amplio rango de potenciales, lo que es un resultado notable comparando con teorías de carácter perturbativo\footnote{Para verificar nuestro cálculo RHNC-Lado hemos realizado simulaciones de dinámica molecular para los mismo valores de $(T,\rho)$, véase la figura (\ref{fig:FuncionesDistribucionRadial}).}.

\section{Diagrama de fases}
Una vez obtenida la energía libre $f_{V}(T,\rho)$ podemos determinar las condiciones de coexistencia de fases compatibles con la construcción de Maxwell en que se produce la física que nos interesará, mediante el sistema de ecuaciones,
\begin{eqnarray}
\mu(T,\rho_{v})&=&\mu(T,\rho_{l})\\
P(T,\rho_{v})&=& P(T,\rho_{l})
\label{eqn:coexfases}
\end{eqnarray}
que indicábamos en \S\ref{sec:eqvarderWaals} y determina la llamada \textit{línea binodal}, mientras que la \textit{línea espinodal}, que delimita la región inestable de la metaestable, queda expresada en la ecuación,
\begin{equation}
\frac{\partial f_{V}(T,v)}{\partial v}=0
\end{equation}

Diferentes aproximaciones a la energía libre de Helmholtz, darán diferentes resultados para las curvas de coexistencia de fases, pero además será conveniente comparar el contenido estructural y no sólo el termodinámico.\\

En nuestro caso dada la ec.(\ref{eqn:vanDerWaalsGeneralizada}) introducimos aproximaciones funcionales para la coexistencia líquido-vapor construidas sobre un sistemas de referencia de esferas duras y la parte atractiva se incorpora inicialmente mediante una aproximación de campo medio. La elección del funcional de esferas duras se realiza en dos esquemas diferentes introducidos -véase \S\ref{subsection:WDA} y \S\ref{subsection:FMT}- que satisfacen respectivamente las ecuaciones de estado de Carnahan-Starling  ec.(\ref{eqn:CSecuacionestado}) y Percus-Yevick  ec.(\ref{eqn:EstadoPYescalada}). En el caso de la teoría de la medida fundamental \S\ref{subsection:FMT} se puede construir satisfaciendo también la ecuación de Carnahan-Starling, ec. (\ref{eqn:CSparaFMT}), y se ha utilizado más adelante a efectos comparativos. En el sistema de esferas duras introducimos un esquema perturbativo basado en la descomposición del potencial mediante el método de Barker y Henderson \S\ref{ss:BarkerHerderson} y utilizamos su definición de diámetro de esferas duras, $d_{hs}(T)$, que evalúa el sistema de referencia óptimo sin complicar excesivamente la aproximación.\\

Expresamos la energía libre por unidad de volumen para el sistema uniforme contenido en nuestras aproximaciones funcionales de \textit{campo medio},
\begin{equation}
\beta f_{V}(T,\rho)=f_{id}(\rho)+f_{ex-hs}(\rho)-\beta\frac{8}{9}2\pi\sigma^{3}U\rho^{2}=f_{hs}(\rho)-\beta\frac{8}{9}2\pi\sigma^{3}U\rho^{2}
\label{eqn:vanderWaalsgeneralizada}
\end{equation}
donde como vemos el $d_{BH}(T)$ introduce la dependencia que causa las diferencias en el diagrama de fases\footnote{Por tanto esta ecuación de van der Waals generalizada no sigue una ley de estados correspondientes para los diferentes modelos de interacción debido a la presencia de $d_{hs}(T)$.} de los diferentes potenciales que hemos determinado en las unidades reducidas T/U. El análisis no incluye la determinación de la fase sólida en el diagrama de fases por tanto las curvas presentadas corresponden a la coexistencia líquido-vapor solapadas con la continuación de esta en la rama metaestable del líquido frente al sólido en el caso en que sea matemáticamente estable, lo mismo sucede para el caso de RHNC-Lado, que se corresponde casi exactamente con los resultados de simulación Montecarlo\cite{PhysRevLett.87.166101} y además para todos los modelos de interacción.\\

\begin{figure}[htbp] 
   \centering
   \includegraphics[width=1.0\textwidth]{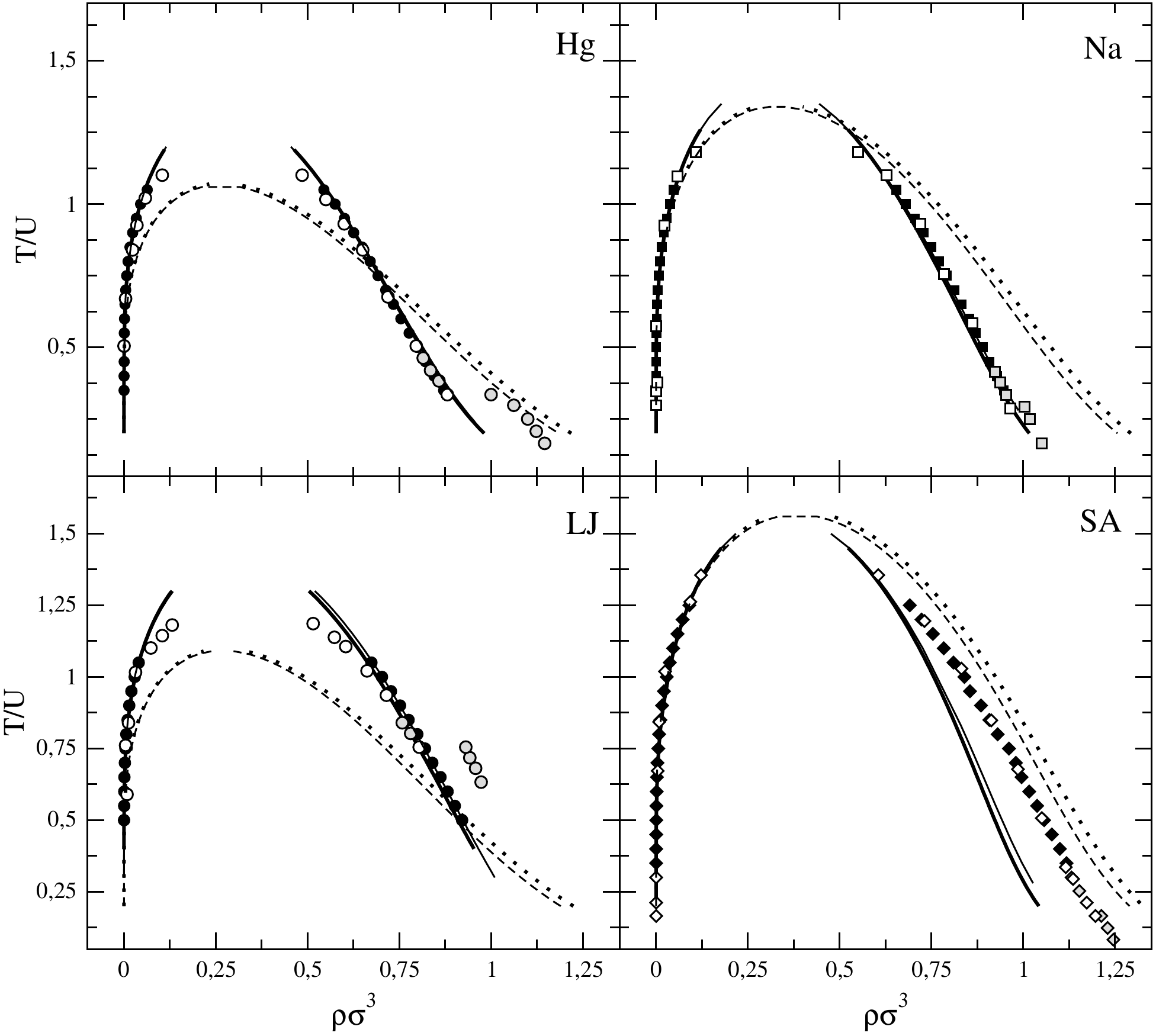} 
   \caption{\textbf{Coexistencia de fases} para los potenciales \textit{Mercurio}, \textit{Sodio},\textit{Soft-Alcaline} y \textit{Lennard-Jones}, para diferentes aproximaciones: $PY+MFA$ (linea discontinua), $CS+MFA$ (linea puntos), $PY+g_{PY}(r)$ (linea continua gruesa), $CS+g_{vw}(r)$ (linea continua fina), RHNC-Lado (figuras en negro), Simulación NVT (figuras blancas), Simulación NPT (figuras grises) estas dos últimas tomadas de \cite{velasco:10777}.}
   \label{fig:DiagramaFASEStodos}
\end{figure}

Los resultados se condensan en la figura (\ref{fig:DiagramaFASEStodos}). En el caso de las temperaturas críticas en campo medio los resultados indican una sobreestimación de $T_{c}$ para modelos \textit{blandos}, mientras que para modelos más próximos al sistema de referencia de esferas duras los valores de $T_{c}$ presentan una estimación menor que los correspondientes a simulaciones Montecarlo. En el caso de RHNC-Lado, existe una región sin solución que abarca la zona de criticalidad y parte de la zona inestable de coexistencia de fases lo que impide obtener soluciones cerca del punto crítico\cite{PhysRevE.51.3146}, comparamos pues con los resultados obtenidos mediante MC en la tabla (\ref{tabla:Tcriticas}).\\

\begin{table}[htdp]
\begin{center}
\vspace{0.25cm}
\begin{tabular}{c c c c c c c c}
\toprule
& & \textbf{SA} & \textbf{Na} & \textbf{Hg} & \textbf{LJ} & \textbf{SW15} \\
 \midrule
& \textit{WDA}  & 1.577 & 1.349   & 1.074 & 1.098 & 1.006 \\

$T_{c}/U$& \textit{FMT-PY}&  1.565 & 1.340   & 1.067 & 1.092 & 1.001 \\

& \textit{MC} &  1.43 & 1.25   & 1.17 & 1.21 & - \\
\midrule
 
& \textit{WDA}  & 0.988 & 0.978   & 0.9345 & 0.92 & 0.94 \\

$T_{FW}/U$& \textit{FMT-PY} &  0.899 & 0.939   & 0.9305 & 0.919 & 0.939 \\

&\textit{RHNC-Lado} &  0.47 & 0.83   & 1.10 & - & - \\
\bottomrule
%
%
%
%
\end{tabular}
\end{center}
\caption{Valores de $T_{c}/U$ y $T_{FW}/U$ para los diferentes modelos y para la aproximación de campo medio, WDA satisface CS+MFA y para FMT se muestra PY+MFA. Comparación con valores procedentes de Montecarlo para $T_{c}/U$ y RHNC-Lado para $T_{FW}/U$. }
\label{tabla:Tcriticas}
\vspace{0.25cm}

\end{table}%

La primera corrección a una teoría de campo medio en el esquema perturbativo que estamos utilizando, necesita la función de distribución radial del sistema de esferas duras que igualmente calculamos con un diámetro $d_{BH}(T)$. La expresión perturbativa a primer orden, que fue ya tratada en ec. (\ref{eqn:TeoriaPerturbaciones}), resulta ser,

\begin{equation}
\beta F=\beta F_{0}+2\pi\rho^{2}\int dr_{12} r^{2}_{12}\phi_{AT}(r_{12})g_{0}(r_{12};\rho_{0})
\label{eqn:EnergiaLibreDesarrollocongder}
\end{equation}

Para la función $g_{HS}(r)$ pueden incluirse diferentes aproximaciones, la más directa es la que surge a partir de la teoría de Percus-Yevick ($g_{PY}(r)$) y que fue expresada\footnote{Su metodología de cálculo se basa en la expresión de su transformada de Laplace, algunas extensiones analíticas se han realizado tanto para sistemas de esferas duras como para otros potenciales\cite{tangLu3,trokhymchuk:024501}.} por Wertheim\cite{wertheim:643,PhysRevLett.10.321}, mientras que la parametrización semi-empírica de Verlet-Weis\cite{PhysRevA.5.939} ($g_{vw}(r)$) permite una descripción más adecuada del valor de contacto $g(d_{HS})$, que en Percus-Yevick es ligeramente menor que en simulación. En la figura (\ref{fig:DiagramaFASEStodos}) se representa un sistema de referencia basado en la ecuación de estado de PY y $g_{PY}(r)$, y en la ecuación de estado de CS con $g_{vw}(r)$.  El resultado es muy similar, mientras que la curva en campo medio con ambas ecuaciones de estado puede verse, más clara, en la figura (\ref{fig:DiagramadeFasesConEstructuras}).\\

Se aprecia la notable mejora para el potencial Mercurio y Sodio de la expresión (\ref{eqn:EnergiaLibreDesarrollocongder}) sobre la aproximación de campo medio lejos del punto crítico. Pero los potenciales considerablemente más blandos que el Lennard-Jones como el \textit{Soft-Alcaline} resultan mejor descritos, desde el punto de la coexistencia de fases, mediante campo medio\footnote{En rigor no siempre es inadecuada una aproximación de campo medio, y de hecho para el caso del potencial gaussiano\cite{stillinger:3968}, en contraste con un potencial repulsivo de esferas duras, resulta notablemente exitosa tanto en propiedades del sistema simple sea o no uniforme\cite{PhysRevE.62.7961} como en mezclas\cite{PhysRevE.64.041501}.}. Los resultados de las densidades de coexistencia líquidas para potenciales blandos en forma de ley de potencias se conoce\cite{paricaud:154505} son estimados por debajo por teorías perturbativas tipo Barker-Henderson y dentro de dicho esquema la siguiente corrección, como comentamos \S\ref{ss:BarkerHerderson},
es de carácter mesoscópico y no soluciona el problema, el primer hecho es corroborado para el \textit{Soft-Alcaline} y es razonable que el segundo también lo sea.\\

Desarrollar una teoría perturbativa adecuada para potenciales repulsivos blandos no es sencillo, se podría argumentar que una mejor elección es utilizar la aproximación de Weeks-Chandler-Andersen en lugar de Barker-Henderson, que se supone afina mejor el sistema de referencia y es más consistente en el desarrollo término a término, sin embargo a densidades altas, en el régimen que más nos interesa, WCA puede fallar en la búsqueda de un diámetro de esferas duras físicamente razonable. También bajo la aproximación dada por ec. (\ref{eqn:EnergiaLibreDesarrollocongder}), se ha intentado estimar variacionalmente\cite{mansoori:4958} $d_{HS}$ basándose en la desigualdad de Gibbs-Bogoliubov\cite{BOOK-TheorySimpleLiquids} que mejora para potenciales blandos\cite{ben-amotz:4844}. En nuestro caso basándonos en los cálculos realizados mediante RHNC-Lado, y el hecho de que incluso para el modelo \textit{Soft-Alcaline} los resultados sean similares a simulación induce a afirmar que el criterio de Lado y su aplicación a una teoría perturbativa con una división del potencial análoga a WCA puede ser un modo adecuado de tratar este tipo de potenciales blandos.\\

 El origen de las discrepancias para potenciales blandos en las teorías perturbativas sobre esferas duras ha sido estudiado por \textit{K.K.Mon}\cite{mon:9392,mon:3245,mon:4766} mediante simulación Montecarlo, y aparecen debidas a la diferencia en el espacio de configuraciones accesible al sistema para potenciales duros, que llamamos $\Gamma_{hs}$, y blandos, que llamamos $\Gamma$. La diferencia denotada por $\Delta_{hs}=\overline{\Gamma\cap\Gamma_{hs}}$, representa la parte accesible al potencial de interacción blando pero inaccesible al sistema de referencia de esferas duras. En consecuencia existe una diferencia intrínseca entre tratar perturbativamente mediante un sistema de referencia de esferas duras y mediante un sistema repulsivo no singular y esta diferencia intrínseca se puede concretar, para $\Phi_{0}$ representando la energía potencial total repulsiva, como,
\begin{equation}
-\frac{\int_{\Delta_{hs}}e^{-\beta\Phi_{0}}}{\int_{\Gamma}e^{-\beta\Phi_{0}}}
\label{eqn:KKmon}
\end{equation}
Esto permite corregir de modo razonablemente eficiente parte de los problemas que una teoría perturbativa (a todos los ordenes) de potenciales blandos sobre el sistema de esferas duras puede generar. Esta formulación del problema presenta la ventaja, como han indicado \textit{Ben-Amotz y Stell}\cite{ben-amotz:4844}, de que en analogía con WCA es posible estimar la contribución más importante a la ec. (\ref{eqn:KKmon}) para potenciales a pares,
\begin{equation}
-\frac{\int_{\Delta_{hs}}e^{-\beta\Phi_{0}}}{\int_{\Gamma} e^{-\beta\Phi_{0}}}\simeq2\pi\rho\int_{0}^{\sigma_{hs}}g_{0}(r)r^{2}dr
\label{eqn:BenAmotzStell}
\end{equation}
 Que al aparecer como una integral dentro de la parte repulsiva del potencial permite a priori, incorporando una parametrización adecuada, como la de \textit{Henderson-Grunke} por ejemplo, ser aplicada a teorías de van der Waals generalizadas\footnote{Bastaría determinar el valor de $d_{hs}$ mediante una aproximación variacional que tenga en cuenta esta expresión junto con el primer orden perturbativo.}. El resultado de la corrección se expresa,
\begin{equation}
-\frac{\int_{\Delta_{hs}}e^{-\beta\Phi_{0}}}{\int_{\Gamma} e^{-\beta\Phi_{0}}}\simeq 2\pi\rho\int_{0}^{\sigma_{hs}}y_{hs}(r)e^{-\beta u_{0}(r)}r^{2}dr
\label{eqn:BenAmotzStell2}
\end{equation}
Esta metodología abre la posibilidad de definir una teoría razonable de van der Waals generalizada\footnote{Como alternativa a una teoría WCA o una teoría WCA-Lado.} para el caso de potenciales blandos, pero para nuestros propósitos se oscurece la interpretación clara de resultados posteriores, por ejemplo como la bondad del funcional resultante fuera del caso homogéneo respecto de la aproximación perturbativa al orden más bajo en diferentes situaciones, y por otra parte los resultados que buscamos no se conocen de hecho en los ordenes más bajos de las teorías perturbativas.\\

\section{Estabilidad de la fase líquida}

Las mismas condiciones (\ref{eqn:coexfases}) son aplicables al caso de la coexistencia sólido-líquido, en cuyo caso necesitaremos determinar mediante algún modelo las propiedades termodinámicas del sólido. Esto determinaría las regiones en que cada fase es estable termodinámicamente pero existen ramas de metaestabilidad en las que una de ellas puede permanecer estable desde el punto de vista mecánico, y surge la cuestión de cual es la relación entre la desestabilización dentro de la fase fluida y la aparición de una fase sólida más estable. A priori la desestabilización del líquido, para una aproximación dada, debería presentarse como una cota al proceso de solidificación. Por otra parte se han desarrollado criterios desde una sola de las fases\cite{PhysRevB.42.2504} para determinar \textit{aproximadamente} los valores termodinámicos para la transición líquido-sólido, criterios como veremos esencialmente estructurales.\\

Desde el punto de vista de los objetivos posteriores el conocimiento de las propiedades de estabilidad de la fase líquida determinan los posibles intervalos en densidad y temperatura en que es posible obtener propiedades de un sistema líquido en que la densidad varíe espacialmente debido a un potencial externo. Es una cuestión además abordable desde el punto de vista de las diferentes aproximaciones comentadas anteriormente.\\

\subsubsection*{Inestabilidad mecánica y soluciones \textit{tipo} sólido}
\label{sec:soluctiposolido}
El problema de la estabilidad se formula de modo sencillo desde la ec. (\ref{eqn:respuestalinealDFT}) en ella la estabilidad mecánica de un fluido indica que una leve perturbación en el potencial externo induce una leve respuesta en la densidad y la inestabilidad se produce cuando la respuesta lineal estática, $\chi$, presenta singularidades, que es lo mismo que encontrar, los ceros de la ecuación $1-\rho c^{(2)}(|\vec{q}|)=0$.\\

La imagen física se completa indagando si es posible que, en ausencia de campo externo, obtengamos junto a la solución líquida uniforme una solución no-uniforme fuertemente oscilante que recuerde distribuciones en la densidad \textit{tipo} sólido. Para ello comenzamos con relaciones generales en sistemas no uniformes de las funciones de distribución y correlación que fueron introducidas independientemente por \textit{Lovett}\cite{lovett:570} y \textit{Wertheim}\cite{wertheim:2377} y constituyen relaciones exactas para los sistemas en equilibro\footnote{Desde el formalismo del funcional de la densidad podemos obtener ambas ecuaciones\cite{reviewEvans1979}.}, la primera de ellas es,
\begin{equation}
\nabla_{1} ln\rho(\vec{r}_{1})=-\beta\nabla_{1} V_{ext}(\vec{r}_{1})+\beta^{-1}\int d\vec{r}_{2} c^{(2)}[\rho;\vec{r}_{1},\vec{r}_{2}]\nabla_{2} \rho(\vec{r}_{2})
\label{eqn:Lovett1}
\end{equation}
que puede ser deducida inmediatamente desde ec.(\ref{eqn:eqautoconsistenterho}). Mientras que bajo la hipótesis adicional de interacción a pares podemos obtener la primera ecuación de la jerarquía de Yvon-Born-Green\footnote{La segunda ecuación de la jerarquía involucra la función $\rho^{(3)}(\vec{r}_{1},\vec{r}_{2},\vec{r}_{3})$ con lo que su aplicación requiere una vez más la incorporación de hipótesis adicionales\cite{BOOK-TheorySimpleLiquids}.},
\begin{equation}
\nabla_{1}\rho(\vec{r}_{1})+\beta\rho(\vec{r}_{1})\nabla_{1}V_{ext}(\vec{r}_{1})=-\beta\int d\vec{r}_{2} \nabla_{2}\phi(\vec{r}_{1},\vec{r}_{2})\rho^{(2)}(\vec{r}_{1},\vec{r}_{2})
\label{eqn:YBG1}
\end{equation}

De ambas la primera de ellas es particularmente sugerente porque no presenta aproximaciones y porque suele ser más sencillo determinar las funciones de correlación directa que las funciones de correlación totales en el funcional de la densidad. Dado nuestro interés en determinar tanto las ramas de inestabilidad de las aproximaciones como la presencia de soluciones no homogéneas resulta conveniente un análisis en el contexto de la teoría de la bifurcación.\\

Si observamos las dos ecuaciones anteriores pueden ser usadas para determinar la forma de las soluciones $\rho(\vec{r})$ para una estructura determinada (incluida en $\rho^{(2)}$ o incluida en $c^{(2)}$). En el caso de ausencia de potencial externo ambas ecuaciones tienen como solución la densidad uniforme, $\rho_{0}$, y además por su no-linealidad es posible que existan otras soluciones cuyo valor medio sigue siendo $\rho_{0}$ pero pueden ser \textit{altamente no homogéneas}\footnote{En el caso de tener la presencia de un potencial externo la bifurcación de las soluciones se produciría desde el perfil de densidad de equilibrio único a densidades bajas pero no necesariamente único (en el sentido anterior) a altas densidades.}. Encontrar estas soluciones y determinar las condiciones en que se produce corresponde a la teoría de la bifurcación\footnote{En el caso de la interfase líquido-vapor es posible utilizar esta misma idea para intentar encontrar restricciones a la forma funcional del perfil de densidad cerca del punto crítico, véase M.Robert\cite{PhysRevB.30.6666} donde se parte de la relación (\ref{eqn:Lovett1}).} y en el caso de líquidos simples, la cuestión inicialmente planteada será ahora si el punto de bifurcación corresponde de manera aproximada o no a una transición entre un líquido y un sólido, es decir, si la desestabilización estructural de las soluciones de la fase líquida se corresponden aproximadamente con una transición termodinámica que formalmente precisa la presencia de la otra fase para la resolución equivalente al sistema de ecuaciones (\ref{eqn:coexfases}).\\

Fijados los parámetros termodinámicos la posible multiplicidad de soluciones emerge en el \textit{límite termodinámico} del hecho de tener diferentes ramas de la energía libre a determinados valores de estos parámetros. Si solamente fijamos dos de estos parámetros podemos observar bifurcación en las soluciones correspondientes a una sola de las ramas de la energía libre, al variar sobre esta el parámetro que queda libre. Si lo denominamos $\lambda$ y las relaciones funcionales (\ref{eqn:Lovett1} y \ref{eqn:YBG1}) las expresamos entonces por $\Upsilon[\lambda,\rho]=0$, en el punto de bifurcación $\lambda_{0}$ se verifica: $\Upsilon[\lambda_{0},\rho]=0$ y $\delta_{\lambda}\Upsilon[\lambda,\rho]|_{\lambda_{0}}=0$. Si dos soluciones bifurcan deben verificar ambas esta última relación funcional y es posible establecer una relación integral\cite{lovett:2425} que para la ec. (\ref{eqn:Lovett1}) resulta\footnote{Como suponemos ausencia de campo externo es posible escribir $c^{(2)}(\vec{r}_{1},\vec{r}_{2})=c^{(2)}(\vec{r}_{12})$.},
\begin{equation}
\nabla_{1}\varphi(\vec{r}_{1})-\rho_{l}\int d\vec{r}_{3}c^{(2)}(\vec{r}_{13})\nabla_{3}  \varphi(\vec{r}_{3})=0
\end{equation}
donde $\varphi$ representa la variación en $\lambda$ de la diferencia entre la posible solución y la homogénea\footnote{Nótese que en el caso de un campo externo la solución no-homogénea respeta la simetría del campo externo responsable de la inhomogeneidad y bajo el cual el sistema se convierte en inestable y surge la bifurcación.}. La solución más sencilla es en el espacio de Fourier donde podemos ver que existen soluciones para un vector de onda $\vec{q}$ si se verifica,
\begin{equation}
1-\rho_{l}\hat{c}^{(2)}(\vec{q})=0
\label{eqn:ecuacionMODOSlovett}
\end{equation}
Es importante recalcar que las soluciones de esta ecuación determinan cuando el sistema se desestabiliza respecto de la solución estable a temperaturas más altas y aparecen soluciones no homogéneas altamente oscilantes pero no necesariamente son indicativo de una transición líquido-sólido\cite{lovett:7353}. Dada una teoría funcional es posible a partir de la función de correlación directa incluida en ella determinar las soluciones de esta ecuación\footnote{Cuestión idénticamente abordable desde la función c(r) obtenida del proceso de convergencia en una aproximación mediante ecuaciones integrales.}.\\

\begin{figure}[htbp] 
   \centering
   \includegraphics[width=1.0\textwidth]{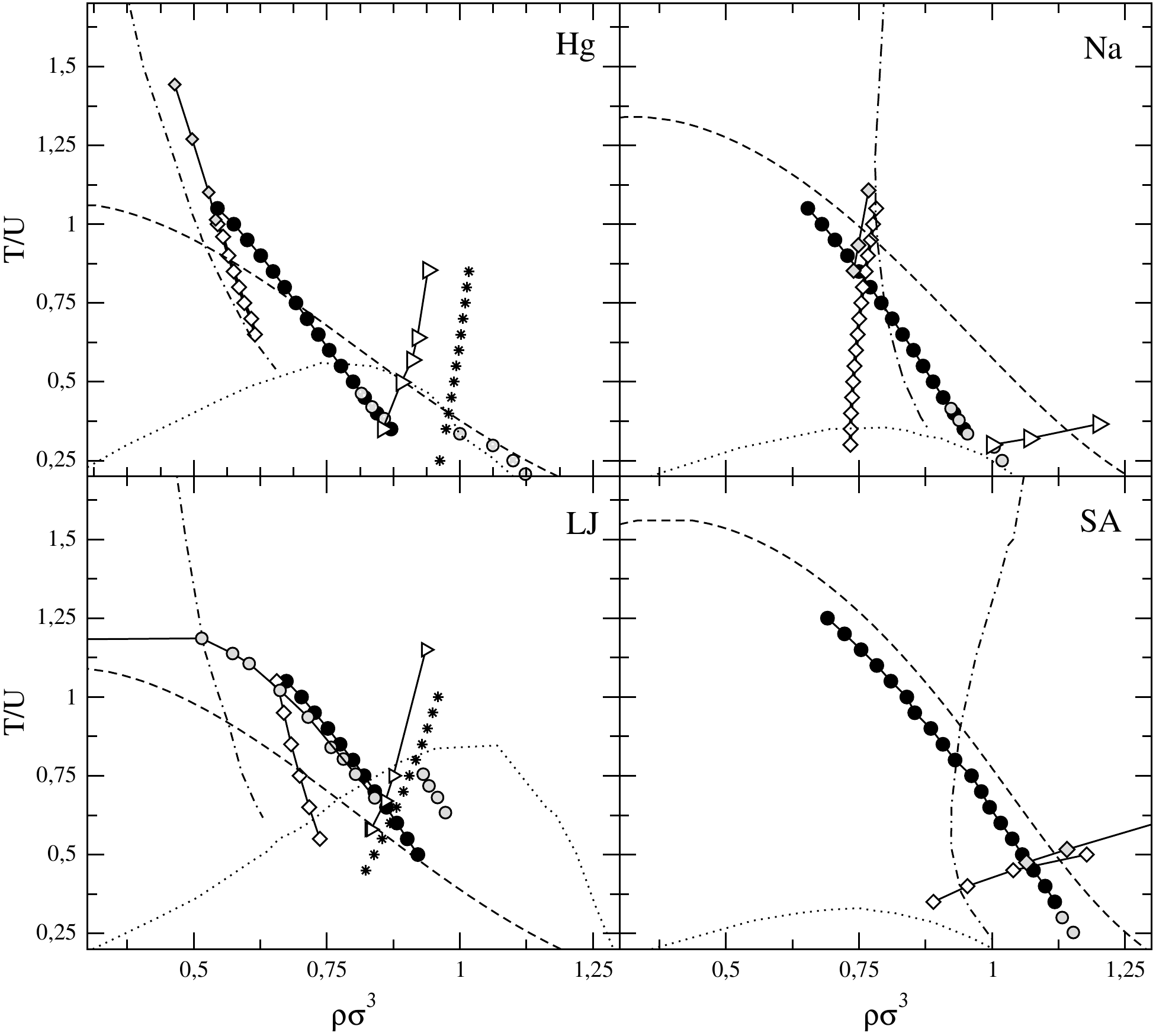} 
   \caption{Comparación entre las Lineas de Fisher-Widom mediante DFT para PY-MFA (líneas discontinuas con puntos) y RHNC-Lado (rombos blancos). Comparativa entre inestabilidades de la solución líquida de DFT (línea de puntos) respecto de punto triple simulación Montecarlo\cite{velasco:10777} y criterios desde la fase líquida para transiciones líquido-sólido: $\Delta s=0$ (asteriscos), Hansen-Verlet con valor $S(q_{0})$, corregido según\cite{velasco:10777} (triángulos blancos). Rama densa de la coexistencia de fases se representa como línea discontinua para PY-MFA y línea continua gruesa para RHNC-Lado. Las lineas con símbolos en gris corresponden a la información procedente de simulaciones Montecarlo\cite{velasco:10777}}
   \label{fig:DiagramaFASESconFW}
\end{figure}

\subsubsection*{Transición líquido-sólido desde aproximaciones al fluido}

Conviene reparar en la diferente descripción que la ecuación de Ornstein-Zernique hace de la fase líquida respecto de la imagen de la fase sólida dentro del funcional de la densidad. En este último la imagen es la de un líquido altamente no homogéneo y por tanto susceptible de ser descrito no por las ecuaciones (\ref{eqn:definicionPuente} y \ref{eqn:OZ}) sino por sus equivalentes no homogéneas\footnote{Sin embargo sobre estas últimas es difícil saber como solucionarlas sin introducir previamente una estructura cristalina, o de otro modo, saber como el sistema de ecuaciones puede diferenciar estructuras con idéntico empaquetamiento y similar energía libre estructural en regiones no macroscópicas del sistema. La teoría del funcional de la medida fundamental construido mediante interpolación dimensional ha sido en este sentido una de las teorías recientes de más éxito en el tratamiento del sólido de esferas duras\cite{PhysRevLett.84.694}.}. Por lo tanto la ecuación de OZ homogénea junto con una relación de cierre puede describir únicamente un líquido metaestable en la región cristalina, véase el caso del Lennard-Jones en la figura (\ref{fig:FuncionesDistribucionRadial}).\\

Aun así, en la búsqueda de una diferenciación de la zona metaestable del líquido y la estable se ha propuesto la existencia de relaciones entre la desestabilización estructural del líquido y la presencia de la transición al sólido también en el contexto de la teoría de las ecuaciones integrales. \textit{Rosenfeld}\cite{PhysRevA.46.4922} mostró que la ausencia de solución a determinadas densidades de la aproximación HNC reproduce de manera fiable las densidades de solidificación para una familia amplia de potenciales. También se ha sugerido un método viable basado en la diferencia\footnote{Donde $s^{ex}$ contiene todos a los ordenes, pero su determinación consistente desde la energía de exceso y la energía libre de exceso puede introducir inconsistencias respecto de la $s_{2}$ obtenida desde g(r), esto hace que las diferentes relaciones de cierre puedan dar diferentes resultados con este criterio, o una inconsistencia entre este criterio y otros similares dentro de una relación de cierre dada.}
\begin{equation}
\Delta s=s^{ex}-s_{2}=\beta(u^{ex}-a^{ex})-\frac{1}{2}\rho\left(\int d\vec{r}h(\vec{r})-\int g(\vec{r})lng(\vec{r})d\vec{r} \right) 
\end{equation}
esencialmente los cambios estructurales de la transición de solidificación se relacionan con su cambio de signo\cite{PhysRevA.45.R6966}. Este método ha sido comprobado para diferentes potenciales mostrando que para ciertas relaciones de cierre permite establecer una correlación aproximada entre $\Delta s=0$ y la cristalización del líquido\cite{PhysRevE.49.5164}. El estudio bajo RHNC-Lado de potenciales no estrictamente duros pero nítidamente repulsivos muestra esta relación conserva al menos su validez cualitativa, más exactamente para el potencial Lennard-Jones \textit{sobreestima} la densidad de solidificación respecto de los experimentos numéricos, y solo de un modo aproximado reproduce resultados de otros métodos. Para ello hemos comparado las predicciones con el criterio de Hansen-Verlet\cite{PhysRev.184.151}, desde la estructura procedente de dinámica molecular\footnote{Dicho criterio se construye sobre la correlación encontrada para el Lennard-Jones entre el valor de la densidad a la que $S(q)=1+\rho h(q)$ alcanza un valor de 2.85 y el estado a esta temperatura en que se produce la solidificación, pero el valor de S(q) puede variar por la naturaleza del potencial de interacción, cuestión tenida en cuenta para el caso del Sodio.}.\\

En el caso de los metales líquidos propuestos,
\begin{itemize}
 \item Para el \textit{Mercurio} ha sido viable determinar los valores de $(T,\rho)$ que verifican que $\Delta s=0$. En la figura (\ref{fig:DiagramaFASESconFW}) vemos que su localización también sobrestima las densidades del punto triple, aun así permite cualitativamente observar la localización de la cristalización para este modelo respecto de Hansen-Verlet\footnote{Si suponemos un comportamiento similar de la sobreestimación al del Lennard-Jones, tendríamos una curva de solidificación que enlaza con el punto triple obtenido mediante simulación Montecarlo.}.
\item Para el \textit{Sodio} solo hemos encontrado una región restringida para valores altos de la densidad y bajos de temperatura ($T/U\lesssim 0.25$) en el limite del diagrama de fases. Si hemos representado, en la figura (\ref{fig:DiagramaFASESconFW}) los resultados para criterio Hansen-Verlet desde la estructura de simulación de Dinámica Molecular.
\item Para el \textit{Soft-Alcaline} no hemos encontrado cambios de signo en $\Delta s=0$, aunque en ese caso no podemos diferenciar si existe una esta sobreestimación que lleva el cambio de signo en $\Delta s$ fuera del rango de densidades reales del fluido.
\end{itemize}
Como consecuencia, el criterio $\Delta s=0$, ha resultado inaplicable a los potenciales más blandos, mientras que para potenciales Lennard-Jones y Mercurio en la aproximación RHNC-Lado hemos encontrado que se aproxima a la línea de \textit{melting}. Globalmente las predicciones son compatibles con los resultados previos acerca de las relaciones $T_{p}/T_{c}$ para estos potenciales\footnote{También hemos comparado nuestros resultados estructurales a la luz de una cuestión que ha generado cierta literatura referente a la posible existencia de \textit{precursores estructurales} del proceso de solidificación. En los modelos de metales utilizados la función de distribución radial encontrada posee un pico residual a altas densidades previo al segundo pico que es asociado a un líquido superenfriado cuyas configuraciones estructurales son similares a las de un sólido amorfo. Este hecho se encuentra en la función de distribución radial de un sistema de esferas duras aunque no es reproducible en la aproximación de Percus-Yevick. Al respecto de como responde estructuralmente hablando una teoría como RHNC-Lado a altas densidades y como la función de distribución g(r) en zonas metaestables puede ser vista como un intento de reproducir algunas propiedades de la función g(r) de una fcc, véase figura    (\ref{fig:FuncionesDistribucionRadial}) del LJ a T/U=0.40.}, véase \S\ref{sec:potencialesSodioMercurio}.\\

\section{Línea de Fisher-Widom}

Cerca del punto crítico la función de distribución de pares g(r) posee un comportamiento asintótico de largo alcance dado por un decaimiento exponencial monótono\cite{RowlinsonWidom} con un parámetro de decaimiento relacionado con la longitud de correlación de volumen del sistema $\xi_{B}$. Cerca del punto triple la parte repulsiva del potencial y en consecuencia los efectos de empaquetamiento molecular son dominantes y esperamos que esto sea reflejado en un comportamiento diferente de g(r). Nuestro objetivo de diferenciación termodinámica y estructural entre propiedades de volumen y de superficie requiere conocer la localización de los diferentes comportamientos estructurales del sistema uniforme en el diagrama de fases, cuestión abordada por \textit{Fisher y Widom}\cite{FisherWidomlinea1969} que evaluaron el comportamiento asintótico de g(r) para sistemas unidimensionales con interacciones de corto alcance\footnote{De hecho si bien su introducción formal es más general el cálculo concreto lo realizaron con potenciales de interacción únicamente hasta primeros vecinos.}.\\

En sistemas unidimensionales el análisis de la transformada de Laplace de la función de partición permite expresar las propiedades termodinámicas mediante polos simples de esta y de modo análogo el comportamiento asintótico de las función de distribución radial a partir de los polos de su transformada de \emph{Laplace} $\tilde{g}(q)$. De este modo es expresable g(r) como una suma de las contribuciones de los polos de la función $\tilde{g}(q)$, es decir, $g(r)=\sum \gamma_{n}e^{-q_{n}r}$ con $q_{n}$ los polos de $\tilde{g}(q)$ y $\gamma_{n}$ sus residuos asociados\footnote{Las expresiones formales son sencillas ya que la única función clave a determinar es la transformada de Laplace del factor de Boltzmann del potencial de interacción, es decir, esencialmente la transformada Laplace de la función de Mayer.}. \\

Así la presencia de los dos modos de comportamiento en g(r) asociados inicialmente y a \textit{grosso modo} a la predominancia del carácter repulsivo o atractivo del potencial\footnote{\textit{Kirkwood et al}\cite{kirkwood:1040} obtuvieron una solución asintótica oscilatoria para el fluido de esferas duras tridimensional, mientras que \textit{Martynov} fue quien sugirió que una presencia atractiva predominante transforma dicho comportamiento en monótono. } es formalizada por \textit{Fisher y Widom} y asociada \emph{explícitamente} a dos tipos diferentes de polos en $\tilde{g}(q)$, uno real, que da lugar aun comportamiento monótono y otro complejo, que da lugar a un comportamiento oscilatorio. En el caso general propusieron una superposición de ambos comportamientos resultado del carácter dual de la interacción. Los valores de $q_{n}$ dependen del estado termodinámico luego en diferentes estados es posible que el término predominante sea bien el monótono si el $q_{n}$ real asociado tiene mayor valor (menor valor negativo) o bien el oscilatorio si $q_{n}$ es complejo y es su parte real la de mayor valor. Surgen entonces dos cuestiones genéricas de interés:\\
\begin{itemize}
\item Donde están localizados ambos comportamientos en un diagrama de fases. Y por tanto donde esta localizada en el diagrama de fases la transición entre ambos comportamientos que  denominaremos \emph{línea de Fisher-Widom}.
\item La relevancia que tiene sobre toda la forma funcional de g(r) la \emph{línea de Fisher-Widom} y relación que existe entre ambas en las diferentes aproximaciones de la teoría de líquidos y los diferentes potenciales de interacción.
\end{itemize}

\begin{figure}[htbp] 
   \centering
   
   \includegraphics[width=1.00\textwidth]{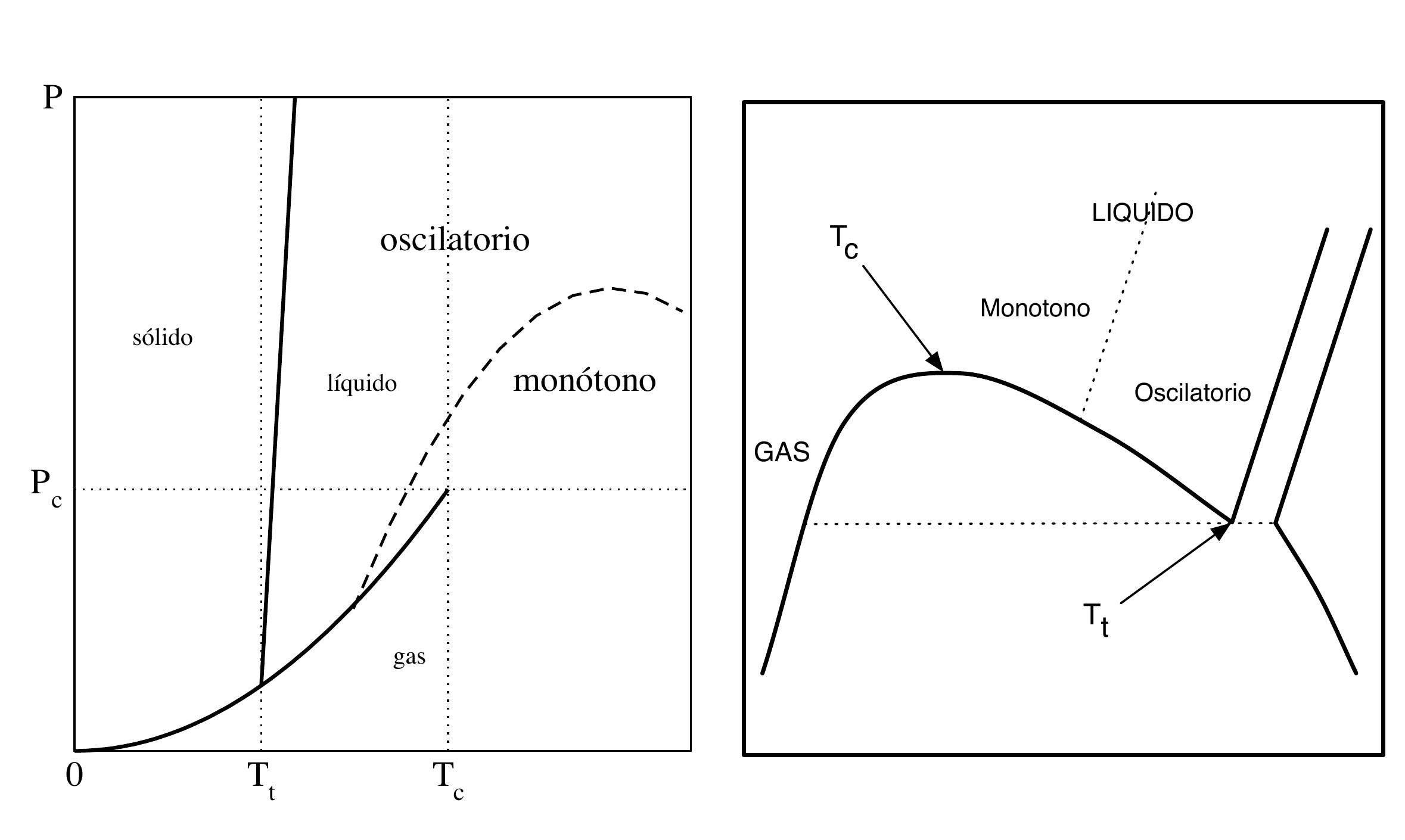} 
   \caption{\textbf{Conjetura de Fisher y Widom} para los diagramas de fase (P,T) y (T,$\rho$) dividiendo las zonas según el comportamiento asintótico de g(r). Línea discontinua representaría la línea de Fisher-Widom. Una propiedad que consideran relevante es que la curva en el diagrama (T,P) siempre esta acotada, de manera que dada una T siempre existe una presión tal que el sistema posee un comportamiento oscilatorio en g(r).}
   \label{fig:ConjeturaFW}
\end{figure}

Los sistemas unidimensionales estudiados por Fisher y Widom carecen de transiciones de fase y lógicamente la interpretación de estos resultados en un diagrama de fases no es inmediata\footnote{El salto de propiedades del sistemas unidimensionales a sistemas de mayor dimensionalidad es un proceso que requiere cautela, recuerde la afirmación de Ising y Lenz acerca del modelo de Ising en dimensiones mayores de uno a partir de sus cálculos en una dimensión.}. El interés en situar de modo cualitativo la línea de Fisher-Widom en dicho diagrama de fases y poder evaluar si es factible, y como realizar, la búsqueda de la transición entre ambos modos en un sistema real lleva a comparar sus resultados con una ecuación de estado de van der Waals generalizada\cite{WidomScience}, véase ec.     (\ref{eqn:vanderWaalsgeneralizada}) o \S\ref{sec:teoriaPerturbativaIntro}. Intentaron situar en sus resultados para este cambio de comportamiento en los diagramas (P,T) y $(\rho,T)$ argumentando donde quedaría el punto crítico y el punto triple respecto de su línea de transición de forma cualitativa.  Concluyeron que la mayoría de la zona de coexistencia de fases debería de ser monótona habiendo un comportamiento oscilatorio solo cerca de la transición líquido-sólido, véase figura (\ref{fig:ConjeturaFW}). La línea de Fisher-Widom en la coexistencia de fases la situaron entorno de $0.7\,T_{c}$ \textit{independientemente del modelo}.\\

En el caso de sistemas de mayor dimensionalidad el análisis realizado por Fisher-Widom basado en la transformada de Laplace no es tan conveniente, pero la propiedad que indica que los polos de una transformación integral de la función de distribución radial puede determinar el comportamiento asintótico de esta si puede ser generalizado. Para ello hacemos uso de la ecuación de Ornstein-Zernique, ec.(\ref{eqn:OZ}). En el caso de la teoría del funcional de la densidad tenemos acceso directo a la función de correlación directa c(r) del sistema uniforme y resultará un modo adecuado para obtener el comportamiento asintótico de la función de correlación total h(r), mientras que en la teoría de las ecuaciones integrales accedemos a ambas y mediante simulación podemos determinar h(r).

\section{Comportamiento asintótico de la función de correlación total}
Partiendo de la relación entre $h(\vec{r})$, y la función de correlación directa $c(\vec{r})$, que viene determinada por la ecuación de Ornstein-Zernique, tenemos,
\begin{equation}
h(\vec{r})=c(\vec{r})+\rho\int d\vec{r}^{\prime} h(\vec{r}^{\prime})c(|\vec{r}-\vec{r}^{\prime}|)
\end{equation} 
la convolución de dos funciones queda expresada como un producto de sus transformadas de Fourier por tanto se expresa de manera más sencilla como,
\begin{equation}
\hat{h}(\vec{q})=\frac{\hat{c}(\vec{q})}{1-\rho \hat{c}(\vec{q})}
\end{equation}
haciendo uso de que en el caso de un fluido uniforme la dependencia es únicamente con el modulo de r, podemos expresar utilizando la función de Bessel,  $J_{0}(qr)$,
\begin{equation}
rh(r)=\int_{0}^{\infty} dq\;q\;J_{0}(qr)\hat{h}(q)=\int_{0}^{\infty} dq\;q\;J_{0}(qr)\frac{\hat{c}(q)}{1-\rho \hat{c}(q)}
\end{equation}
la resolución de la ecuación $1-\rho \hat{c}(q)=0$ determina los polos de la función $\hat{h}(q)$ y por tanto el comportamiento asintótico de la función $rh(r)$. Esta ecuación ya fue estudiada en el contexto de la teoría de la bifurcación, asociada a la inestabilidad del líquido frente al sólido\cite{kirkwood:514,lovett:2425} y relacionada con la divergencia del factor de estructura en el estudio de la estabilidad. Aquí se generaliza el problema \cite{henderson:6750,evans:591}, recordando el trabajo de Fisher-Widom, afirmando que podemos analizar el comportamiento de sistemas estables si generalizamos q a valores no únicamente reales. Resolvemos por tanto para $q=\alpha_{0}+i\alpha_{1}$ complejo la ecuación anterior utilizando que,
\begin{equation}
\hat{c}(q)=4\pi\int_{0}^{\infty}dr r^{2}c(r)\frac{sen(qr)}{qr}
\end{equation}
por tanto
\begin{equation}
1-4\pi\rho\int dr r^{2}c(r)\frac{sen(qr)}{qr}=0
\end{equation}
que equivale a resolver el par de ecuaciones,
\begin{eqnarray}
\alpha_{1}=4\pi\rho\int dr r^{2}c(r)\frac{senh(\alpha_{1} r)}{r}cos(\alpha_{0} r)\\
\alpha_{0}=4\pi\rho\int dr r^{2}c(r)\frac{cosh(\alpha_{1} r)}{r}sen(\alpha_{0} r)
\label{eqn:IntegralesModosDecaimiento}
\end{eqnarray}
Si escribimos haciendo uso del teorema de los residuos la función h(r) en función de las soluciones anteriores por medio de una integral de contorno en el plano complejo\footnote{Formalmente escribiríamos,
\begin{equation*}
rh(r)=\frac{1}{2\pi^{2}}\int_{0}^{\infty}dq\,J_{0}(qr)\hat{h}(q)=\frac{1}{4\pi^{2}i}\int_{-\infty}^{\infty}e^{iqr}\hat{h}(q)
\end{equation*}
lo que permite expresar  (\ref{eqn:desarrolloPOLOShr}) a partir del teorema de los residuos. Necesitamos construir un contorno en el plano complejo lo cual es obviamente posible para un número finito de ellos, el caso de un número infinito como esperamos en un fluido con una parte de esferas duras \cite{perram1980rps} se puede demostrar que están aislados y el contorno de integración puede ser construido.},

\begin{equation}
rh(r)=\sum_{n}e^{iq_{n}}\frac{R_{n}}{2\pi}
\label{eqn:desarrolloPOLOShr}
\end{equation}
vemos que el comportamiento asintótico vendrá regido por los valores de $\alpha_{1}$ mayores y de la solución anterior tendremos tres posibilidades para el primer término.
\begin{enumerate}
\item \textit{Solución imaginaria pura}: $\alpha_{0}=0$ que dará una exponencial monótona decreciente y esperamos $rh(r)\sim e^{-\alpha_{1} r}$. En este caso $\alpha_{1}^{-1}$ esta relacionado con la longitud de correlación del volumen que presenta una divergencia cerca de $T_{c}$.
\item \textit{Solución real pura}: $\alpha_{1}=0$ que dará una función oscilante y podemos relacionarlo con el análisis de la estabilidad y de la teoría de la bifurcación\cite{henderson:6750} como indicábamos antes.
\item \textit{Dos soluciones} dadas por $q_{1}=\alpha_{0}+i\alpha_{1}$ y $q_{2}=-\alpha_{0}+i\alpha_{1}$ lo que da una función oscilante amortiguada por una función exponencial decreciente\cite{evans:591} y por tanto esperamos que $rh(r)\sim e^{-\alpha_{1} r}cos(\alpha_{0}r-\theta)$.
\end{enumerate}

Si nos movemos en un amplio rango de valores de la temperatura podemos esperar una superposición de ambos comportamientos, y por tanto,

\begin{equation}
h(r,T)\sim A_{h}(T,\rho)\frac{e^{-\alpha_{1} r}}{r}cos(\alpha_{0}r-\theta)+B_{h}(T,\rho)\frac{e^{-\tilde{\alpha}_{1} r}}{r}
\label{eqn:asintoticogder}
\end{equation}

La generalidad de la expresión anterior será puesta de manifiesto en el siguiente capítulo\footnote{Hay extensiones de los argumentos anteriores para fluidos moleculares ver por ejemplo la referencia\cite{savenko:021202} para su aplicación a esferocilindros.}. Las amplitudes que acompañan a la solución pueden ser determinadas por medio de teorema de los residuos para $rh(r)$ del mismo modo que las $\gamma_{n}$ en el esquema introducido por Fisher-Widom.\\

Un análisis completo ha sido realizado por \textit{Martynov y Sarkisov}\cite{martynov:3445} directamente en el espacio real para un sistema de esferas duras, desde la ecuación de Ornstein-Zernique en el contexto de las ecuaciones integrales y para diferentes relaciones de clausura. Obtiene las mismas expresiones para el cálculo de $\alpha_{1}$ y $\alpha_{0}$, este método no involucra transformadas de Fourier para las amplitudes, sin embargo algunas de sus expresiones aproximadas para estas no son completamente adecuadas\cite{evans:591} en su aplicación al caso general.\\

La línea de Fisher-Widom es la condición  $\alpha_{1}=\tilde{\alpha}_{1}$, en ec. (\ref{eqn:asintoticogder}). En lo que resta denominaremos a $\tilde{\alpha}_{1}$, $\alpha_{1}$ y $\alpha_{0}$  \textit{modos de decaimiento asintótico} y conviene determinarlos en detalle para los diferentes potenciales de interacción introducidos y notar que dependen del estado termodinámico $(T,\rho)$.

\begin{figure}[htbp] 
   \centering
   \vspace{0.45cm}
   \includegraphics[width=1.0\textwidth]{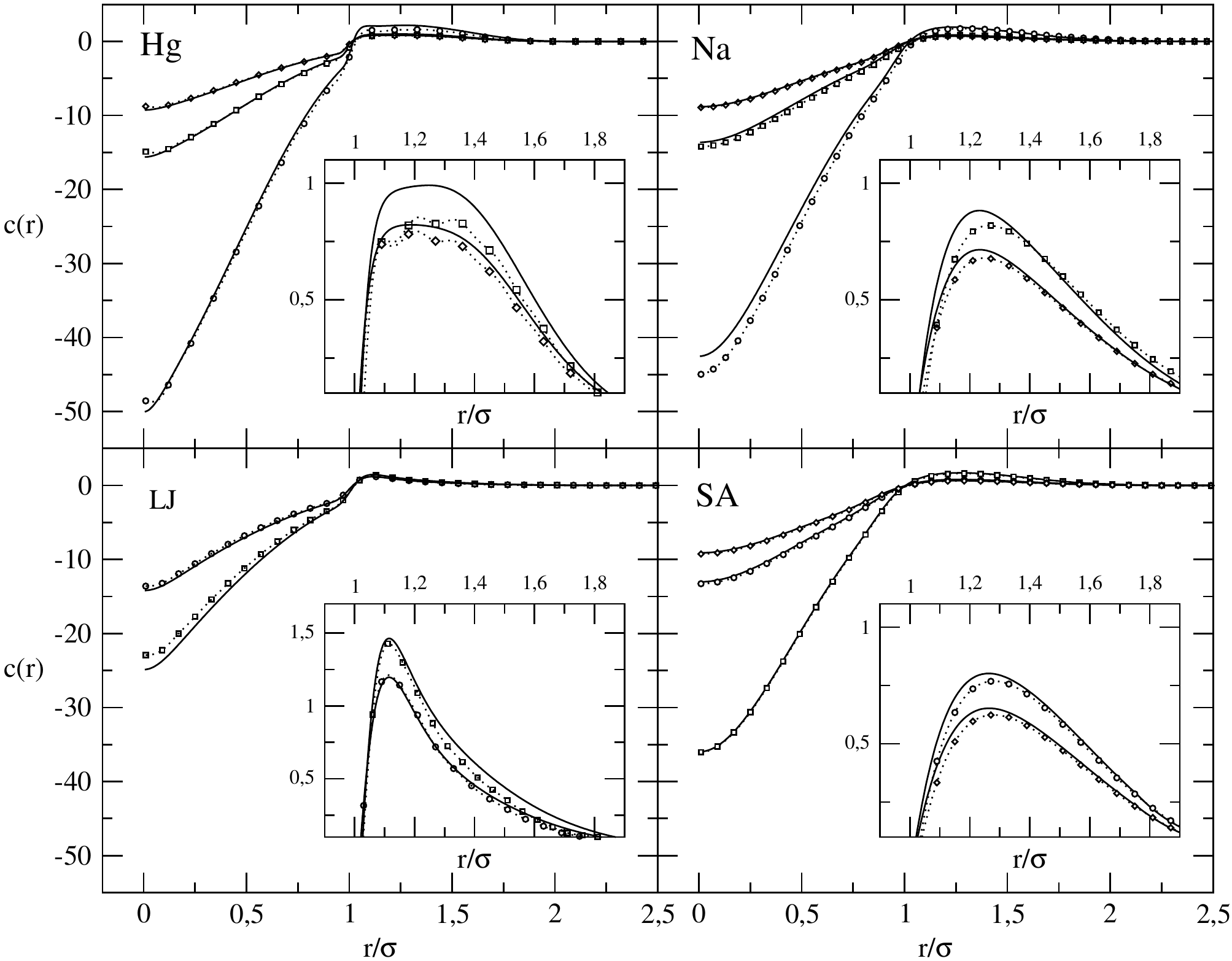} 
   \vspace{0.25cm}

   \caption{\textbf{Funciones de correlación directa} para los tres modelos analizados y su comparación con el Lennard-Jones, en la aproximación RHNC-Lado (líneas continuas) y simulación dinámica molecular (línea de puntos con círculos/cuadrados/rombos). Valores de T/U=0.40, T/U=0.85 y T/U=1.05 para los valores de densidad de la rama densa de la coexistencia de fases de RHNC-Lado. En el caso del LJ T/U=0.40 no se muestra en consonancia con el comportamiento de g(r) de simulación. Los inset corresponden a los valores T/U=0.85 y T/U=1.05.}
   \label{fig:FuncionesCdeR}
   \vspace{0.45cm}

\end{figure}

\subsection{Función de correlación directa}

Conviene puntualizar que para la determinación de los polos de $\hat{h}(q)$ la presencia de partes positivas y partes negativas en la función c(r) es clave y la competición entre ambas partes de la función de correlación directa expresa el contenido físico del comportamiento asintótico, ya que el sistema (\ref{eqn:IntegralesModosDecaimiento}) que pretendemos resolver involucra un cálculo integral que puede amplificar las diferencias leves de la función $c(r)$, recordemos de corto alcance, en la función h(r). Resolvemos el problema estructural para las dos recetas funcionales del sistema de referencia y analizamos su relevancia en los parámetros de decaimiento\footnote{La importancia de la cola de la función de correlación directa en la determinación de propiedades generales de la función de correlación total fue ya indicada por Stell, ver por ejemplo \cite{1999cond.mat.12033K}.}. Analizamos también la función de correlación directa c(r) de RHNC-Lado y la comparamos con los resultados de las teorías funcionales.\\

\subsubsection*{Función $c(r)$ desde RHNC-Lado}
\label{sec:cderRHNC-Lado}
En la figura (\ref{fig:FuncionesCdeR}) representamos la función de c(r) de nuestros modelos,  obtenida desde simulación de dinámica molecular a partir de g(r) y mediante RHNC-Lado. La coincidencia no es tan buena como en g(r) pero esperable por las limitaciones de tamaño del sistema en ambos métodos. La semejanza es notable sobre todo en el caso del \textit{Soft-Alcaline}, donde la limitación en tamaño se deja sentir menos, mientras que en el Mercurio las diferencias aparecen esencialmente en la parte positiva de c(r) más sensible a limitaciones de cálculo. Observamos que para los tres modelos los valores de c(0) para temperaturas altas son muy similares (aunque algo diferentes del Lennard-Jones \textit{sin cutoff}, la simulación si posee un cutoff en $2.5\sigma$) y solamente se aprecian diferencias al bajar la temperatura. La diferencia en dureza de los potenciales se refleja en c(r) en lo abrupto del cambio de signo entorno a $c(\sigma)$, y vemos como la aproximación RHNC-Lado representa decentemente las diferencias de los potenciales en este aspecto, exagerándolo en el caso del Mercurio y que como veremos se reflejará en la línea de Fisher-Widom.\\

 La forma de c(r) respeta razonablemente el comportamiento asintótico esperado $c(r)\simeq-\beta\phi(r)$ en ambos métodos, aunque en el caso del Mercurio las diferencias en la parte positiva entorno a $r\gtrsim\sigma$ se compensan con una rama oscilante a largas distancias, que no se aprecia en la figura, este hecho esta presente aunque más atenuado en el Sodio y prácticamente ausente en el \textit{Soft-Alcaline}. Este hecho es además responsable parcial de las diferencias en los modos de decaimiento para RHNC-Lado a altas densidades y posiblemente de las leves diferencias en la predicción de Fisher-Widom para el Mercurio respecto de los datos de simulación. Aun así teniendo en cuenta las notables diferencias estructurales de los potenciales tratados demostradas mediante simulación resalta nuevamente la predicción del tratamiento mediante ecuaciones integrales que hemos hecho. Además las intuiciones acerca de la relevancia de las interacciones en la estructura asintótica es entendida a partir de la función de correlación directa, cuestión que analizamos comparando también con las teorías de van der Waals generalizadas.\\

\subsubsection*{Función $c(r)$ desde Teorías de van der Waals generalizadas}
\label{sec:VderWallsCdeR}

Las expresiones funcionales que hemos utilizado pueden ser condensadas en la expresión,
\begin{equation}
\mathcal{F}[\rho]=\mathcal{F}_{hs}[\rho]+\frac{1}{2}\int d\vec{r}_{1}\vec{r}_{2} \rho(\vec{r}_{1})\rho(\vec{r}_{2})\phi_{AT}(\vec{r}_{1},\vec{r}_{2})
\end{equation}
donde $\phi_{AT}(\vec{r}_{12})$ es determinado como en \S\ref{ss:BarkerHerderson}. Luego la función de correlación directa incluida en los dos modelos funcionales usados: WDA-CS y FMT-PY puede ser determinada mediante la segunda derivada funcional de la parte de exceso de nuestros funcionales, tal y como se vio en \S\ref{ss:jerarquias}. Los cálculos aquí mostrados son en aproximación de campo medio con lo que la expresión para la función de correlación directa queda del siguiente modo,

\begin{equation}
c(|\vec{r}_{12}|)\equiv\beta\frac{\delta^{2}F_{ex}[\rho]}{\delta\rho(\vec{r}_{1})\delta\rho(\vec{r}_{2})}\bigg\vert_{\rho_{B}}=c_{HS}(|\vec{r}_{12}|)-\beta\phi(|\vec{r}_{12}|)
\end{equation}

que se corresponde con la llamada \emph{Random Phase Aproximation}(RPA), y refleja el hecho antes comentado de no tener en cuenta correlaciones en el término funcional que contiene la cola atractiva. Es una aproximación sencilla que supone que la contribución de la parte atractiva equivale al comportamiento asintótico de la función de correlación directa, ya que mediante desarrollos diagramáticos se puede identificar como $\beta\phi(r)$. A pesar de su simplicidad, dentro del marco del desarrollo asintótico algunos de sus resultados puede ser más generales siempre que las correlaciones de la parte atractiva no incluidas no tengan un papel notable en el régimen de estudio de la función h(r).

\begin{center}
\begin{figure}[htbp] 
\vspace{0.45cm}
\centering
   \includegraphics[width=1.00\textwidth]{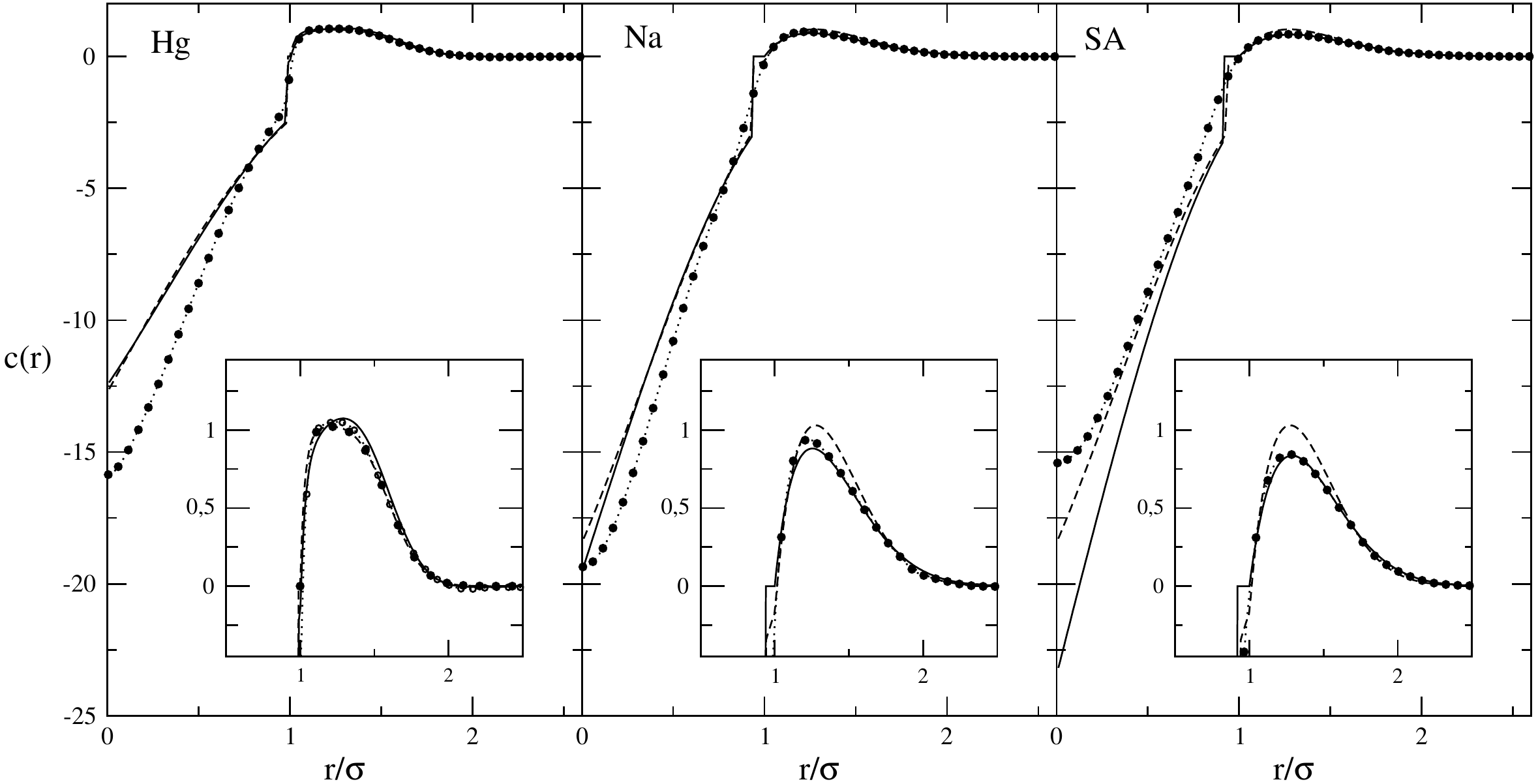} 
   \vspace{0.45cm}
   \caption{\textbf{Funciones de correlación directa}. De izquierda a derecha tenemos Mercurio, Sodio y Soft-Alcaline. Líneas continuas representan FMT-PY-MFA a $T/U=0.80$ y líneas discontinuas representan WDA-CS-MFA para la misma densidad $\rho_{liquid}$, siendo $Hg=0.82$, $Na=0.84$, $SA=0.87$. Estos valores de $(\rho,T)$ se pueden ver representados en el diagrama de fases (\ref{fig:DiagramadeFasesConEstructuras}). Las líneas con círculos negros representan la c(r) obtenida para los mismo valores de $(T,\rho)$ de FMT-PY-MFA pero bajo RHNC-Lado.}
   \label{fig:EsquemaCdeR}
   \vspace{0.45cm}

\end{figure}
\end{center}

Ambos funcionales tratan de reproducir la función de correlación de Percus-Yevick, en el caso de FMT-PY es exacta, en el caso de WDA-CS no se anula para valores mayores de $\sigma$ sino que presenta una pequeña cola oscilante, véase apéndice \S\ref{sec:apendiceCwda}.\\

\begin{figure}[htbp] 
   \centering
   \includegraphics[width=1.25\textwidth,angle=270]{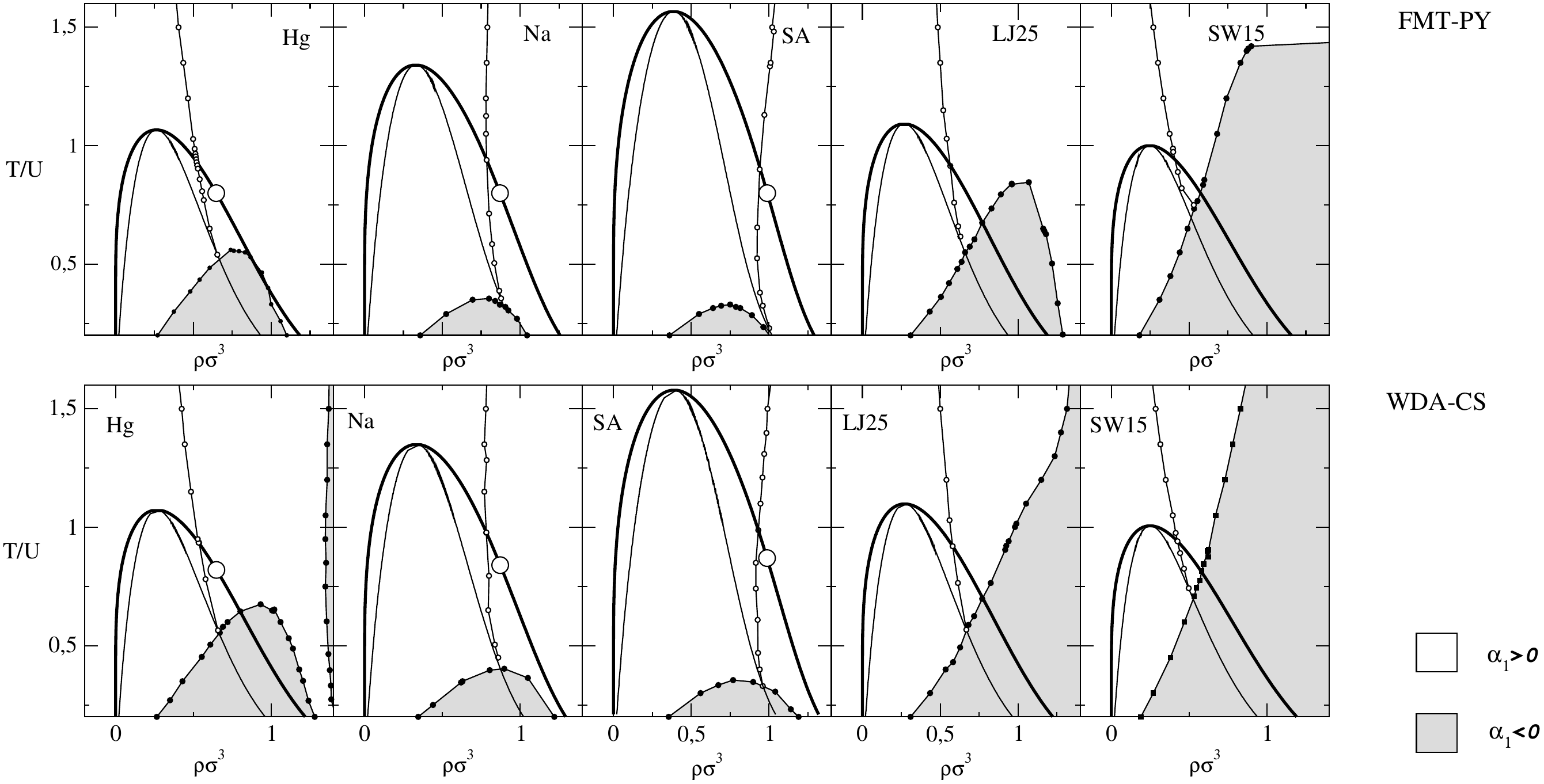} 
   \caption{Se muestra la línea de coexistencia de fases (binodal) así como la línea espinodal, para los modelos estudiados, y para ambas recetas funcionales FMT y WDA. En el diagrama se incluye las líneas $\alpha_{1}=\tilde{\alpha}_{1}$ junto con la línea $\alpha_{1}=0$. Los círculos blancos grandes indican los valores para los que se ha representado c(r).}
   \label{fig:DiagramadeFasesConEstructuras}
\end{figure}

La función de correlación de Percus-Yevick es solo negativa y por tanto no esperamos que posea soluciones reales, una primera consecuencia es que las soluciones serán siempre típicas de un líquido para la función h(r) independientemente de la densidad siempre que la ecuación integral este bien definida. Por otra parte a densidades muy altas la aproximación de $c_{WDA}(r)$ para la función de correlación de Percus-Yevick al presentar valores positivos puede técnicamente presentar una solución tipo sólido\footnote{Se puede comparar la mayor inestabilidad de WDA respecto de FMT viendo la figura (\ref{fig:DiagramadeFasesConEstructuras}) en los modelos Lennard-Jones con un cutoff en $2.5\sigma$ y el pozo cuadrado en $1.5\sigma$.} en un sistema de esferas duras\cite{henderson:6750}.\\

En la siguiente figura (\ref{fig:EsquemaCdeR}) se muestran estas funciones c(r) en las dos aproximaciones para dos densidades del líquido y para diferentes modelos\footnote{Que obviamente posee consistencia con un cálculo numérico directo dentro de nuestro funcional.} comparadas con RHNC-Lado. Vemos como el tratamiento mediante el sistema de esferas duras presenta casos opuestos en el caso del Mercurio y el Soft-Alcaline, la parte abrupta del potencial Mercurio es representada correctamente por el salto que presenta las teorías basadas en la ecuación de Percus-Yevick, pero falla en lo que a una descripción de potenciales blandos se refiere. El potencial Sodio presenta un comportamiento estructural bien descrito por RPA, esencialmente debido a que compensa las deficiencias de la descripción simple dada por Barker-Henderson y la también simple, $c(r)=-\beta\phi(r)$ para $r>\sigma$.\\

\section{Modos de Decaimiento Asintótico}

\begin{figure}[htbp] 
  \centering
  \vspace{0.85cm}

   \includegraphics[width=1.09\textwidth]{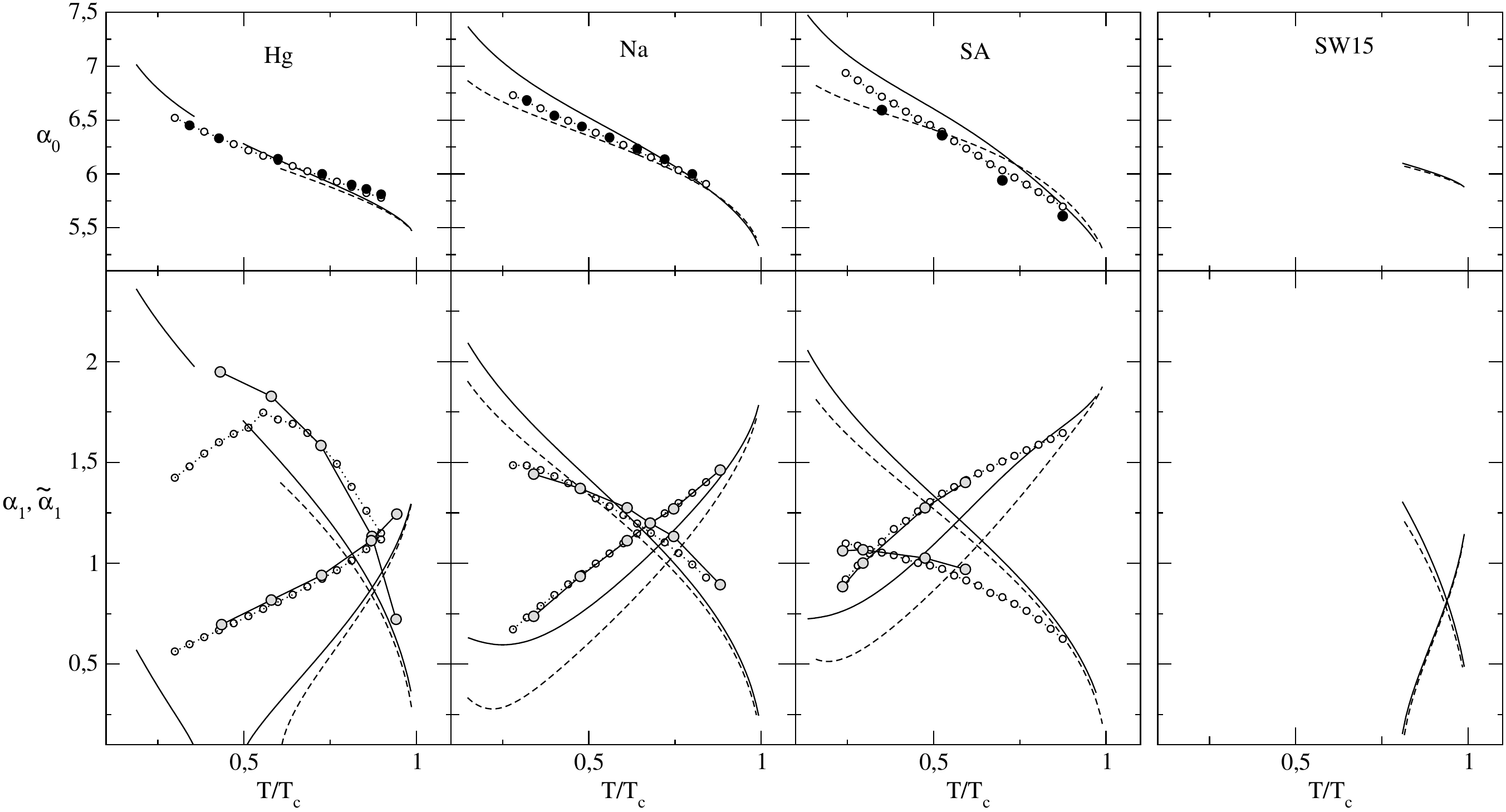} 
   \vspace{0.65cm}
   \caption{\textbf{Modos de decaimiento a lo largo de la curva de Coexistencia}. De izquierda a derecha tenemos Hg, Na, SA y SW15. líneas continuas representan FMT-PY-MFA y líneas discontinuas representan WDA-CS-MFA. Arriba valores de $\alpha_{0}$. Abajo las líneas decrecientes representan el parámetro $\tilde{\alpha}_{1}$ y las líneas crecientes el parámetro $\alpha_{1}$. Los círculos blancos pequeños representan los resultados RHNC-Lado escalados con $T_{c}$ obtenida de simulación Montecarlo. Los círculos negros representan los valores que para esos mismos valores de $(T,\rho)$ se obtienen desde simulación de Dinámica Molecular. Los círculos grises representan los resultados de simulación Montecarlo obtenidos en\cite{PhysRevLett.87.166101}.} 
   \label{fig:DecayModesHgNaSASW}
   \vspace{0.85cm}

	\end{figure}

\begin{wrapfigure}{r}{8.9cm}
	\includegraphics[width=3.0in]{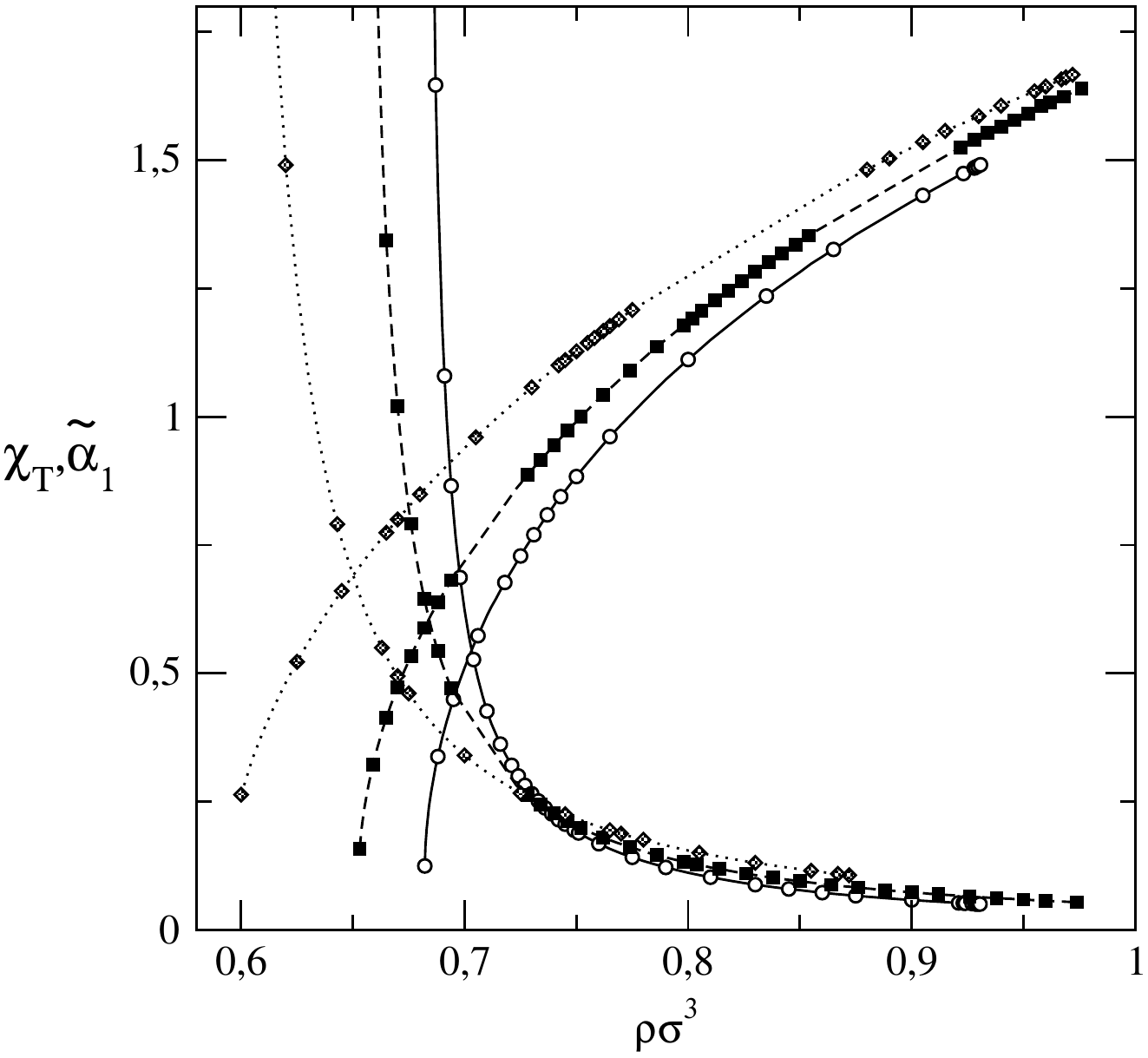}
	\caption{Correlación entre límite $\tilde{\alpha}\rightarrow0^{+}$ y la divergencia de $\chi_{T}$ hacia la línea espinodal desde la fase líquida. Aproximación RHNC-Lado. Modelo Na. líneas correspondientes a tres temperaturas: Continua T/U=0.40, Discontinua T/U=0.55, Puntos T/U=0.80.}
	\label{fig:CorrelacionCompresibilidadBeta}
	\end{wrapfigure}

Los resultados a lo largo de la curva de coexistencia en la rama líquida para $\alpha_{1}$,$\tilde{\alpha}_{1}$ y $\alpha_{0}$ a partir de las funciones c(r) aparece mostrado en la figura (\ref{fig:DecayModesHgNaSASW}), para los modelos \textit{Mercurio}, \textit{Sodio}, \textit{Soft-Alcaline} así como un Pozo Cuadrado de alcance $1.5\sigma$, usado por otros autores\cite{EvansMolecularPhyscFWline} y que incluimos a efectos comparativos.
\begin{itemize}
\item Término oscilatorio ($\alpha_{0}$ y $\alpha_{1}$)
\begin{itemize}
\item \textit{Estructuración intrínseca del líquido} ($\alpha_{0}$), determina la anchura de las diferentes capas de líquido entorno de una partícula fija y los diferentes modelos poseen valores entorno a $\alpha_{0}\sim 2\pi/\sigma$. En el modelo del \textit{Mercurio} la dependencia con T es menor, presentando un limite determinado por una anchura de $\sigma=1.005$ por debajo de la cual el líquido se hace inestable (léase que el otro modo $\alpha_{1}$ se anula), aunque en el caso de FMT-PY aparece una nueva rama a densidades altas. Esto no ocurre en los modelos del \textit{Sodio} y \textit{Soft-Alcaline} donde podemos obtener resultados a temperaturas más bajas y observar anchuras con valores del orden de 0.83 ya en la zona metaestable. Las diferencias entre las aproximaciones WDA y FMT son leves en el caso de potenciales fuertemente repulsivos pero más relevantes en el caso de potenciales blandos donde las densidades de coexistencia a T/U baja son mayores y las diferencias entre $c_{WDA}(r)$ y $c_{PY}(r)$ se acentúan.
\item \textit{Longitud de correlación modo oscilatorio} ($\alpha_{1}^{-1}$): Determina la longitud efectiva del término oscilatorio. El caso del Mercurio en la aproximación de Percus-Yevick posee la propiedad de recuperar a densidades suficientemente altas una rama estable (metaestable), en el sentido de valores de $\alpha_{1}>0$, cosa que no se produce de hecho en el caso de la $c_{WDA}(r)$, realzando una vez más el hecho de que las diferencias entre ambas recetas funcionales se relevan en la estructura a densidades altas, del mismo modo, las curvas de los modelos \textit{blandos} presentan diferencias mayores a valores de T bajos (densidades mayores en la curva de coexistencia) llegando a ser notables, pero compartiendo la propiedad de metaestabilidad.
\end{itemize}
\item Término monótono ($\tilde{\alpha}_{1}$)
\begin{itemize}
\item \textit{Longitud de correlación del término monótono} ($\tilde{\alpha}_{1}^{-1}$): Determina el alcance efectivo del término monótono. Las diferencias entre los diferentes modelos son sistemáticas pero menos acentuadas, en todos ellos las propiedades de $\tilde{\alpha}^{-1}_{1}$  cerca del punto critico revelan los exponentes de una teoría de campo medio, $\xi_{B}\simeq\tilde{\alpha}^{-1}_{1}$. Por otra parte la línea $\tilde{\alpha}_{1}\simeq0$ se correlaciona con la línea espinodal, tanto para el líquido como para el vapor, en la figura 	(\ref{fig:CorrelacionCompresibilidadBeta}) se muestra para el caso de RHNC-Lado.
\end{itemize}
\end{itemize}

\newpage

Al respecto de la línea de Fisher-Widom, los valores $\alpha_{FW}=\alpha_{1}=\tilde{\alpha}_{1}$ se ven condicionados por la dureza del potencial, que para el Pozo Cuadrado presenta el valor menor, mientras que para el Soft-Alcaline presenta el  mayores valores de $\alpha_{FW}$, esto sucede en ambas recetas funcionales pero no es una propiedad extensible a las simulaciones Montecarlo.\\

Cualitativamente RPA presenta similitudes con los resultados de Montecarlo especialmente para el potencial Sodio. En el caso de \textit{Mercurio} la pre\-di\-cci\-ón para \textit{Fisher-Wi\-dom} es similar pero las curvas de $\tilde{\alpha}_{1}$ presentan valores sis\-te\-má\-ti\-ca\-men\-te me\-no\-res que los obtenidos por simulación. En el caso de \textit{Soft-Alcaline} la situación es e\-xac\-ta\-men\-te la contraria y las aproximaciones funcionales utilizadas sobrestiman $\tilde{\alpha}_{1}$, que proviene de los diferentes defectos que en cada modelo se presentan en c(r). En el caso del pa\-rá\-me\-tro $\alpha_{1}$  los resultados fun\-cio\-na\-les para todos los modelos dan valores me\-no\-res que los obtenidos mediante si\-mu\-la\-ci\-ón\footnote{En simulación la magnitud que se obtiene es g(r) o equivalentemente h(r), se pueden analizar en ella los valores de los modos de decaimiento mediante ajustes a su forma funcional o utilizar Ornstein-Zernique para recuperar la función de correlación directa.}.\\

Los potenciales blandos po\-seen diferencias en $\alpha_{1}$ debido a que la incorporación de los efectos de repulsión es reducida a un $d_{hs}$ que en el caso del \textit{Sodio} parece ser eficiente mientras que en el caso del \textit{Soft-Alcaline}, más blando, la aproximación deja ser ser eficiente mostrando más diferencias con simulación y entre las dos recetas funcionales. El \textit{Mercurio} al ser fuertemente repulsivo puede estar modelizado adecuadamente por $d_{hs}$, sin embargo es un pozo notablemente más ancho y plano que un Lennard-Jones como se aprecia en la figura (\ref{fig:Potenciales}) esto permite bajar la relación $T_{p}/T_{c}$ pero necesita un modo eficiente de incluir las correlaciones de la parte atractiva para reproducir adecuadamente la estructuración real creada por el potencial, cuestión no captada adecuadamente por una aproximación de campo medio pero si por RHNC-Lado.\\

De la comparación entre las dos aproximaciones funcionales WDA y FMT, esta última reproduce mejor los resultados de simulación siendo bastante próximos en el Sodio especialmente en la localización de cruze entre la línea de Fisher-Widom y la curva binodal densa.\\

\subsubsection*{Resultados fuera de la línea binodal}
Las diferencias entre los diferentes funcionales sugieren la determinación fuera de la coexistencia tanto de la línea de Fisher-Widom como de la línea de inestabilidad del líquido. En la figura (\ref{fig:DiagramadeFasesConEstructuras}) podemos ver dichas propiedades junto con las curvas binodal y espinodal, mientras que en la figura (\ref{fig:DiagramaFASESconFW}), tenemos el caso RHNC-Lado.\\

Como observamos los modelos Mercurio, Lennard-Jones y Potencial Cuadrado presentan todos una pendiente negativa para la línea de Fisher-Widom, mientras que el potencial Soft-Alcaline presenta una pendiente positiva. Calentar el líquido a temperatura constante permite en el primer caso pasar de un sistema monótono a uno oscilante mientras que en el segundo caso permite pasar de un comportamiento oscilante a uno monótono. Esto es un reflejo nuevamente de las propiedades repulsivas de los diferentes modelos, la forma de las correlaciones implica que un aumento de la temperatura disminuye la contribución atractiva y por tanto favorece efectos de empaquetamiento propios del comportamiento oscilatorio pero esto solo es posible si este efecto de disminución de la relevancia atractiva no se ve compensado por una disminución del diámetro de esferas duras que hace menos relevante la parte oscilante y favorece un comportamiento monótono. Este es el fenómeno que ocurre precisamente en el Soft-Alcaline.  El modelo Sodio presenta un comportamiento intermedio entre ambos. Esto matiza otras de las posibles propiedades generales de la línea de Fisher-Widom. La misma situación se presenta en simulación Montecarlo\cite{FWlineMolPhysNaHgSA} y RHNC-Lado, aunque en estos casos las diferencias están más acentuadas entre los diferentes modelos. \\

En la línea espinodal ha de verificarse trivialmente que la solución $1-\rho \hat{c}(q)=0$ es q=0 y por tanto en ella se cruzan la línea $\alpha_{1}(T,\rho)=0$ y la línea de Fisher-Widom. El lugar geométrico de puntos en $(T,\rho)$ que verifica $\alpha_{1}(T,\rho)=0$ nos permite comparar la línea de inestabilidad frente al sólido de los diferentes modelos. En el caso de los modelos Lennard-Jones y Potencial Cuadrado vemos que atraviesa la línea de coexistencia de fases más cerca del punto critico respecto de las mismas curvas en los otros modelos que significativamente presentan un máximo y decrecen progresivamente a altas densidades. En el caso del Mercurio cruza la línea de coexistencia presentando también una inestabilidad frente al sólido pero a una temperatura significativamente menor pudiendo por tanto observarse más nítidamente en él, respecto del Lennard-Jones o el Potencial Cuadrado, los efectos de empaquetamiento molecular en la función de distribución radial. En el caso de los potenciales Sodio y Soft-Alcaline el máximo se encuentra fuera de la metaestabilidad y permite mantener el sistema líquido estable a temperaturas muy bajas, en ningún caso en el rango de temperaturas analizado hemos podido observar una desestabilización de estos modelos aunque en el Sodio podría ser posible a temperaturas muy bajas, fuera del rango donde podemos realizar una interpretación con sentido físico.\\

\subsubsection*{Excursus: Relevancia del alcance del potencial}
Los potenciales tratados hasta el momento han sido de corto alcance\footnote{Los resultados de simulación corresponden a un Lennard-Jones con \textit{cutoff} de $2.5\sigma$, mientras que hasta el momento los resultados en RPA para este potencial se corresponden con este mismo cutoff. En el caso de RHNC-Lado hemos tratado un potencial Lennard-Jones con un alcance de $20\sigma$.}, sin embargo existen sistemas reales que poseen interacciones de largo alcance, sean interacciones electroestáticas, dipolares o fuerzas de London (o fuerzas de dispersión).\\

Fuera de la región crítica, el estudio del desarrollo diagramático establece\footnote{Información que también esta contenida en la sencilla RPA.} que $c(r)\sim -\beta\phi(r)$ luego para $\phi(r)\sim r^{-6}$ los momentos de orden 2 o mayores de $c(r)$ dejan de ser finitos, como consecuencia las ecuaciones (\ref{eqn:IntegralesModosDecaimiento}) suponen una convergencia que no se da en modelos polinómicos como el Lennard-Jones y para la determinación y relevancia de los polos las expresiones utilizadas no son adecuadas\footnote{Básicamente bajo corto alcance existe un desarrollo Taylor de $\hat{c}(q)$ entorno a $q=0$, pero potenciales de largo alcance rompen esta propiedad.} y en consecuencia el análisis llevado acabo anteriormente para el comportamiento asintótico de la función de correlación deja de ser válido.\\

\begin{figure}[htbp] 
   \centering
   \includegraphics[width=1.00\textwidth]{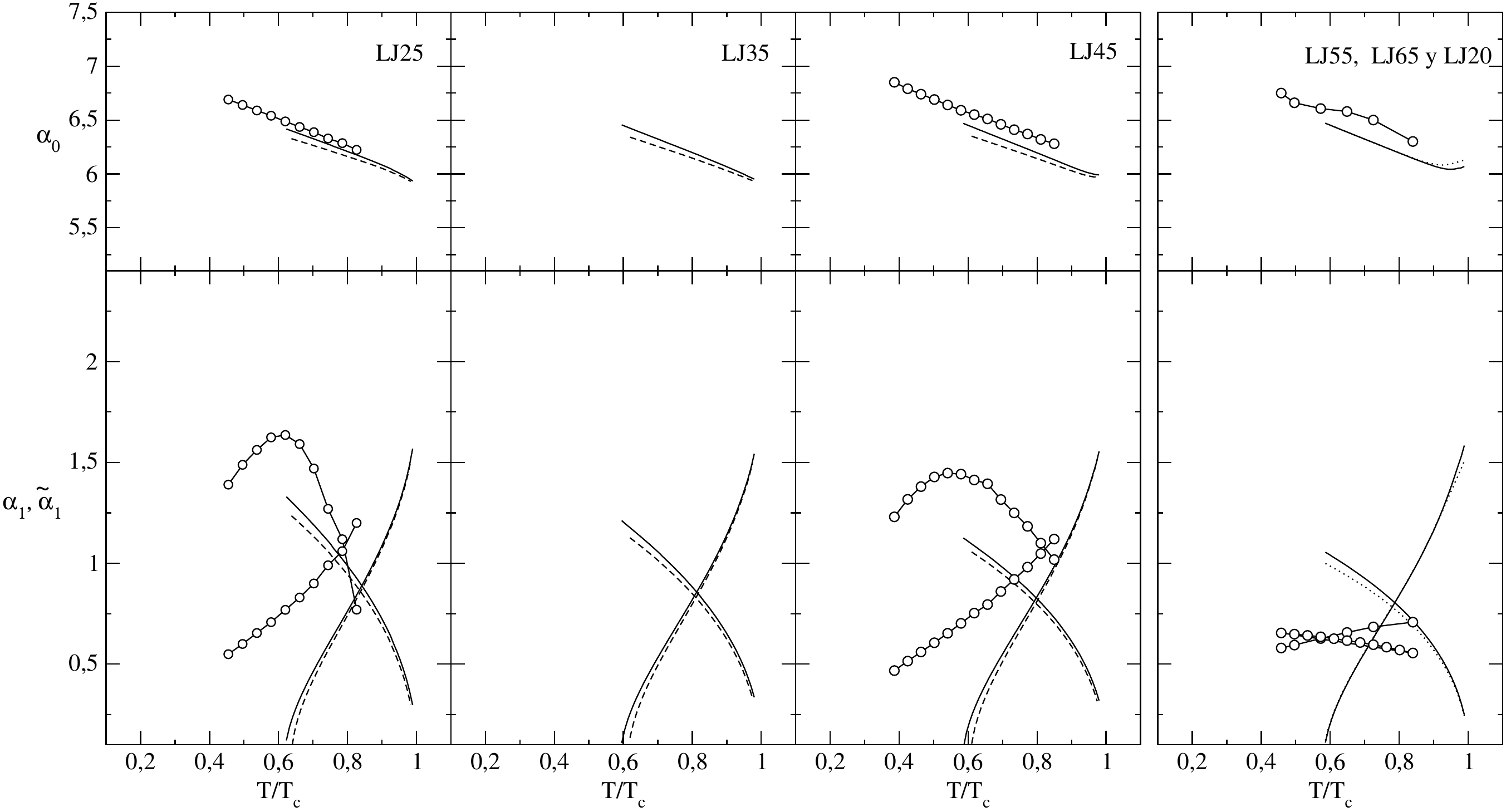} 
   \caption{\textbf{Comparación de los modos de decaimiento del Lennard-Jones para diferentes cutoff}. Las tres primeras cajas corresponden a $2.5\sigma$,$3.5\sigma$ $4.5\sigma$ en las aproximaciones FMT (línea continua),   WDA (línea discontinua) y RHNC-Lado (círculos). A la derecha se incluyen también para FMT los alcances $5.5\sigma$ (línea continua) y $6.5\sigma$ (línea de puntos) para FMT, además de $20\sigma$ en RHNC-Lado (círculos). La escala en ordenadas y abscisas coincide con figura (\ref{fig:DecayModesHgNaSASW}).}
   \label{fig:DecayModesLJ}
\end{figure}

Por otra parte para potenciales polinómicos con un cierto \textit{cutoff} si es valida y los estudios realizados mediante ecuaciones integrales para un modelo Lennard-Jones completo revelan una estructura oscilante análoga a distancias intermedias, esto sugiere la posibilidad de combinar un análisis asintótico exponencial a rangos intermedios con uno polinómico que domina a largas distancias, esta argumentación permite separar la parte repulsiva de los efectos polinómicos de la cola atractiva, método más obvio en RPA pero extensible a otras relaciones de cierre de la ecuación de Ornstein-Zernique, y realizar una extensión que se sintetiza en\cite{Sarkisov:2002},
\begin{equation}
rh(r)\sim A_{h}e^{-\alpha_{1}r}cos(\alpha_{0}r-\theta)+B_{h}\left(\rho\chi_{T}\right) ^{2}\beta^{-1}\phi(r)
\end{equation}
donde nos hemos restringido a un primer término oscilante y una corrección de carácter polinómico. Como primera consecuencia no nos es posible determinar la línea de Fisher-Widom.\\

El camino analítico que da como resultado la expresión anterior fue realizado por \textit{R.J.F. Leote de Carvalho, R. Evans, D.C. Hoyle y J.R. Henderson}\cite{decarvalho1994dpc} añadiendo, al desarrollo mediante el teorema de los residuos, la expansión en serie que surge de la cola polinómica de un modelo Lennard-Jones y analizaron su propuesta en el contexto de RPA. Dentro del concepto de potenciales de corto alcance también ha sido analizado el comportamiento de la línea de Fisher-Widom al incrementar el valor del \textit{cutoff} del potencial Lennard-Jones\cite{PhysRevE.51.3146}. Nosotros hemos comparado la variación de este alcance en RPA respecto de la variación del alcance en RHNC-Lado y su papel en los diferentes modos de decaimiento. Los resultados aparecen en la figura (\ref{fig:DecayModesLJ}), en esta figura hemos preservado las escalas de la figura (\ref{fig:DecayModesHgNaSASW}) para que el lector pueda comparar las curvas de todos los modelos tratados.\\

Tanto en el caso de RPA como en RHNC-Lado el punto de Fisher-Widom se verifica a valores de $T_{FW}$ progresivamente menores al incrementar el alcance del potencial, igualmente sucede con la inestabilidad $\alpha_{1}=0$ para RPA\footnote{Aunque en estos dos casos pueden ser reducidas las diferencias si escalamos no usando $\epsilon_{\infty}$ sino el $\epsilon$ obtenido del propio potencial, véase ec.(\ref{eqn:UlennardJ}), no así en la curva completa $\alpha_{1}$ que no colapsa en una única curva.}. En esta última aproximación además vemos que las discrepancias al variar el alcance son mínimas y se dejan sentir más en $\tilde{\alpha}_{1}$ que disminuyen al incrementar el alcance presentando un comportamiento progresivamente más suave\footnote{Esto se verá más nítido en los perfiles de densidad, y es una traza de la progresiva desaparición del término exponencial monótono en favor del término polinómico en el la inclusión completa de la cola atractiva.}. En el caso de RHNC-Lado las diferencias son más claras, el desplazamiento del punto de FW a temperaturas más bajas al incrementar notablemente el cutoff y las diferencias en $\tilde{\alpha}_{1}$ que muestran que en efecto la presencia del término monótono cambia notablemente. Los valores de $\alpha_{0}$ se mantienen similares aunque para los valores más altos del cutoff encontramos una anomalía en RPA cerca de $T_{c}$\footnote{Posiblemente una traza de lo inadecuado del procedimiento de resolución del sistema de ecuaciones que hemos de encontrar en ausencia de cutoff.}.\\

Terminamos indicando la correlación entre en $\alpha_{1}\simeq0$ de RPA para todos los modelos en que se anula y el comportamiento anómalo en $\tilde{\alpha}_{1}$ en RHNC-Lado que presenta un máximo justo en la presencia de dicha inestabilidad indicando posiblemente la presencia de un zona de metaestabilidad hasta el valor $\tilde{\alpha}_{1}\simeq0$ en que aparece la línea espinodal.\\

\section{Conclusiones}
Resumimos el conjunto de resultados obtenidos:
\begin{itemize}
\item Mediante la aproximación de ecuaciones integrales basada en RHNC, junto con un criterio para determinar adecuadamente el sistema de referencia incluido en $b_{hs}(r)$, hemos descrito en un solo esquema los tres potenciales de interacción para metales líquidos con resultados cuantitativamente aceptables para la descripción de coexistencia de fases líquido-vapor. La descripción mediante una teoría perturbativa, y su extensión a sistemas no homogéneos dada por las teorías de van der Waals generalizadas más usuales, de los potenciales introducidos para metales líquidos aunque presenta limitaciones en su descripción de sistemas blandos capta cualitativamente las diferencias estructurales que nos interesa analizar.
\item La línea de Fisher-Widom no posee un carácter universal ya que su localización y propiedades en el diagrama de fases depende del potencial de interacción. No presenta relevancia en cuanto a la termodinámica pero pudiera presentar una relevancia estructural diversa en el caso general de líquidos no homogéneos.
\item Reproducir la descripción estructural contenida en los modos de decaimiento obtenidos mediante simulación requiere una teoría que describa correctamente la función de correlación directa. Del modelo Soft-Alcaline vemos que una descripción similar a simulación de las propiedades termodinámicas de coexistencia no indica una descripción igualmente precisa de propiedades estructurales.
\item La inestabilidad del líquido frente a perturbaciones \textit{tipo} sólido depende tanto del potencial de interacción como de la receta funcional utilizada, información que se encuentra codificada en la función de correlación directa, siendo un balance entre propiedades atractivas y repulsivas. En el caso de los modelos duros como el Mercurio o el Lennard-Jones la inestabilidad se aproxima a la zona de solidificación mostrando ser una cota aproximada. Otros criterios estructurales poseen una validez semi-cuantitativa.
\item En RPA, los potenciales Sodio y Soft-Alcaline permiten extender el rango de estabilidad del líquido frente a dichas perturbaciones, esto se consigue debido a que la línea binodal que representa el líquido no cruza la rama de inestabilidad aunque esta existe igualmente. Dos procesos involucran este fenómeno, las mayores densidades de la fase líquida en modelos blandos y la menor relevancia de la zona inestable en sistemas progresivamente menos duros. En el caso del potencial Mercurio el cruce se da aunque existen para la función de correlación de Percus-Yevick valores estables a temperaturas muy bajas y donde una teoría de campo medio y la simulación Montecarlo coinciden. Una teoría más completa como RHNC-Lado muestra la presencia de una inestabilidad en un decrecimiento del modo de decaimiento $\tilde{\alpha}_{1}$.

\end{itemize}

\chapter{Relevancia de Fisher-Widom y las Ondas Capilares}
\label{sec:capituloFWyCWT}

Las propiedades estructurales del sistema uniforme tienen una generalidad mayor de la introducida en el capítulo anterior esto nos permite primero, establecer la existencia de oscilaciones en los perfiles de densidad líquido-vapor resultantes de teorías de van der Waals generalizadas, segundo relacionar propiedades de estas con la presencia de ondas capilares y tercero determinar la manera \textit{efectiva} en que están incluidas estas últimas en los perfiles de densidad líquido-vapor.\\

\section{Universalidad del comportamiento asintótico}

En el capítulo anterior hemos estudiado las propiedades asintóticas de la función de correlación total desde la ecuación de Ornstein-Zernique mediante un procedimiento que arranca desde el trabajo de Fisher-Widom y que de modo general consiste en determinar la forma asintótica de una distribución que describe la estructura del sistema bajo un cierto potencial externo gracias a una transformación integral que permite expresar esta distribución en función de los polos de su transformada\footnote{En este sentido la ec. (\ref{eqn:Lovett1}) puede ser un punto de partida idóneo para estudiar líquidos bajo diferentes potenciales externos que impliquen diferentes inhomogeneidades y observar sus \textit{a priori} diferentes comportamientos asintóticos.}.\\

Este método puede ser incluso más general y aplicable a, por ejemplo,  mezclas binarias de moléculas basándose también en la expresión de la ecuación de Ornstein-Zernique para este sistema\cite{evans:591,Salmon0953-8984:11443,PhysRevA.45.7621}. Las relaciones en este caso en el espacio de Fourier son más complicadas ya que para una disolución de n componentes, que podamos representar como fluidos simples, tenemos (n+1) ecuaciones,
\begin{equation}
h_{ij}(r)=c_{ij}(r)+\sum_{k}\rho_{k}\int d\vec{s} h_{ik}(s)c_{kj}(|\vec{r}-\vec{s}|)
\end{equation}
En el caso de una mezcla binaria donde denotamos \textit{s} para el soluto y \textit{d} para el disolvente tendremos $h_{sd}(r)$,$h_{dd}(r)$ y $h_{ss}(r)$ y sus transformadas de Fourier son\footnote{La expresión $\hat{h}_{dd}(k)$ se obtiene desde $\hat{h}_{ss}(k)$ sustituyendo el soluto por el disolvente.},
\begin{align}
\hat{h}_{ss}(k)= &\frac{\hat{c}_{ss}(k)+\rho_{d}\left(\hat{c}_{ds}(k)^{2}-\hat{c}_{aa}(k)\hat{c}_{bb}(k)\right)}{D(k)}\\
\hat{h}_{sd}(k)= &\frac{\hat{c}_{sd}(k)}{D(k)}
\end{align}
donde $D(k)=\left[1-\rho_{s}\hat{c}_{ss}(k)\right]\left[1-\rho_{d}\hat{c}_{dd}(k)\right]-\rho_{s}\rho_{d}\hat{c}_{sd}(k)^{2}$ es común a las tres funciones y por tanto comparten sus polos y en consecuencia sus modos de decaimiento\footnote{Así el parámetro $\alpha_{0}\sim 2\pi/\sigma$ que determina el ordenamiento intrínseco del fluido y podríamos pensar determinado en cada $\hat{h}_{ij}$ por el valor de $\sigma$ de cada uno de los componentes de la mezcla, esta determinado por el mayor diámetro de los componentes.} a\-sin\-tó\-ti\-cos\cite{evans:591}, pero esta propiedad no es compartida por las amplitudes $A_{ss}$,$A_{dd}$ y$A_{sd}$, determinadas de modo análogo al caso monocomponente por los residuos de los polos, que resultan en general diferentes\footnote{Aunque es posible establecer una relación funcional entre ellas tanto para polos complejos como para polos reales.}. De modo que en este sistema la universalidad de los modos de decaimiento contrasta con la particularidad de las amplitudes con las que dichos modos participan en la estructura.\\

\begin{figure}[htbp] 
   \centering
   \includegraphics[width=1.05\textwidth]{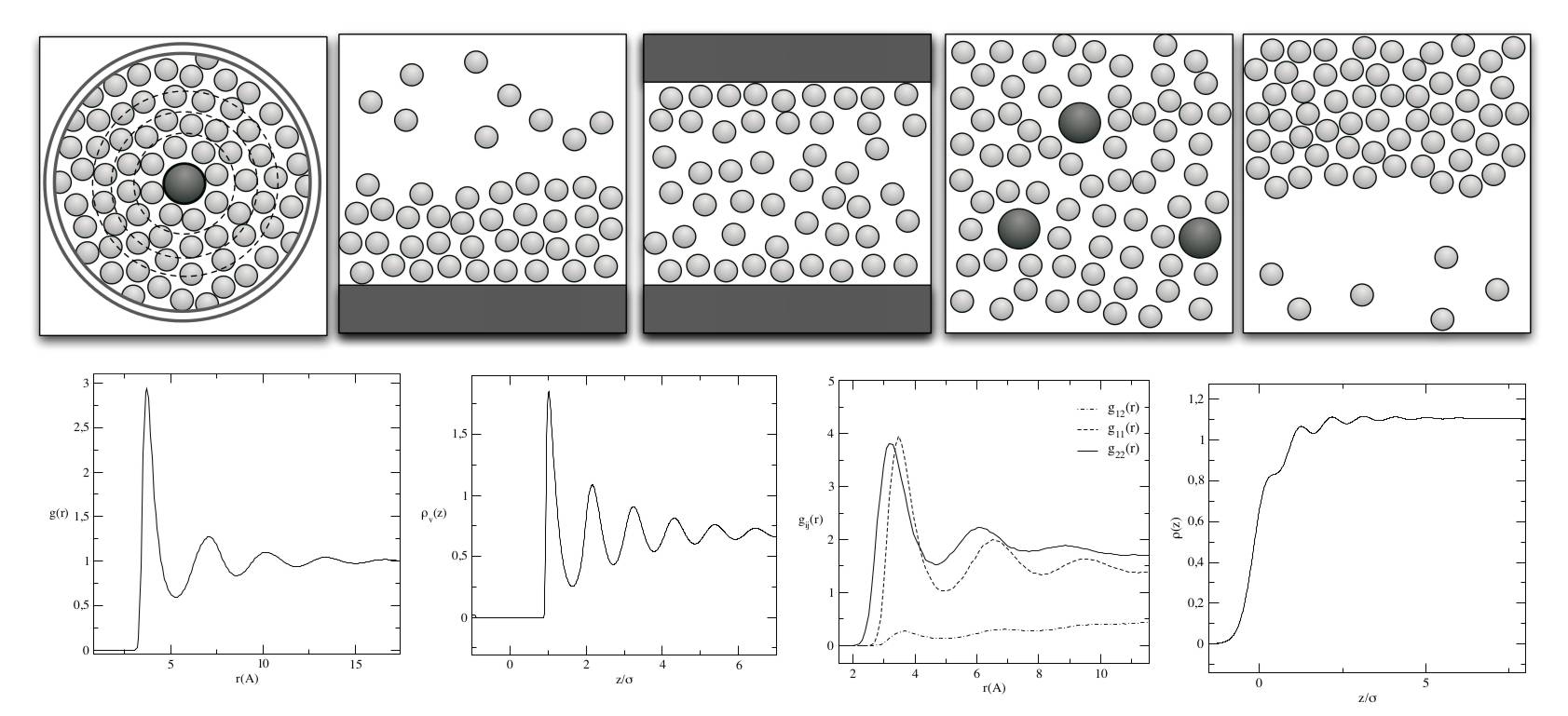} 
   \caption{Comparación del mismo fluido en contacto con una partícula, una pared, confinado, una mezcla y una interfase líquido-vapor.\textbf{ Figuras de arriba} representan una configuración concreta en cada caso. Las \textbf{figuras de abajo} representan la forma de las subsiguientes funciones que surgen de los promedios estructurales sobre configuraciones, el primer caso un perfil radial creado por una partícula mediante simulación, después un potencial externo local similar al segundo caso de arriba, una mezcla basada en los potenciales \textit{Soft-Alcaline} y \textit{Sodio} realizada mediante dinámica molecular (representamos $g_{ss}$, $g_{dd}$ y $g_{ds}$) y un perfil líquido-vapor para el potencial \textit{Sodio} a bajas temperaturas en DFT.}
   \label{fig:EjemplosESTRUCTURA}
\end{figure}

Esto mismo puede ser extendido al problema de un fluido en contacto con una pared ya que se corresponde con el una mezcla binaria donde se introduce una partícula de un componente arbitrariamente grande, y permite establecer una ecuación de OZ que queda expresada en el espacio de Fourier mediante, 
$\hat{h}_{wf}(q)=\hat{c}_{wf}(q)+\rho\hat{h}_{wf}(q)\hat{c}(q)$, con lo que
\begin{equation}
\hat{h}_{wf}(q)=\frac{\hat{c}_{wf}(q)}{1-\rho\hat{c}(q)}
\end{equation}
donde hay que resaltar que la función de correlación directa $wf$ (pared-fluido) no es idéntica a $\hat{c}(q)$ del fluido uniforme.\\
 
En el capítulo anterior hemos indicado que las amplitudes del desarrollo (\ref{eqn:desarrolloPOLOShr}) son determinadas mediante los residuos de cada uno de los polos. Si consideramos estos como \textit{polos simples} tenemos para las amplitudes en el estudio de un fluido uniforme\footnote{Como propiedad general de la aproximación de polo simple podemos decir que, en el caso de g(r), obtiene valores en orden de magnitud adecuados para los polos que representan el comportamiento asintótico comparados con los valores reales obtenidos del comportamiento de h(r) mediante ajustes que reproduzcan \textit{exactamente} los mismos valores de los modos que ec.(\ref{eqn:IntegralesModosDecaimiento}). Así $A_{g}(T)$ reproduce valores del orden $\rho_{l}$ y $B_{g}(T)$ dos ordenes de magnitud menor. La dependencia en la temperatura también es reproducida correctamente, mientras que en ambos casos son cotas interiores de los valores reales con un error relativo aproximado de 0.15. Una aplicación del cálculo analítico de estas amplitudes se puede ver en la figura (\ref{fig:detallesgder}).}
\begin{equation}
A_{n}=-q_{n}\frac{\hat{c}(q_{n})}{\rho\hat{c}^{\prime}(q_{n})}
\end{equation}
mientras que en el numerador aparece $\hat{c}_{wf}(q_{n})$ para un fluido en contacto con una pared. En consecuencia si bien los polos, siguiendo el análisis del capítulo anterior, son idénticos en ambos casos no lo serán las amplitudes ya que en general $\hat{c}_{wf}(q)\neq\hat{c}(q)$.\\

Si comparamos este problema de un fluido en contacto con una pared con la función de distribución radial por la hipótesis de la partícula test de Percus, véase (\ref{eqn:percuscampoefectivo}), apreciamos que la diferencia es la presencia de potenciales externos diferentes que crean la inhomogeneidad del fluido, mientras las propiedades de ordenación intrínseca lejos de la perturbación son las mismas\footnote{Por otra parte esta propiedad de ordenación intrínseca del fluido también permite explicar otras propiedades como las oscilaciones presentes en la fuerza de solvatación $f(L)$ en un fluido confinado entre dos paredes separadas una distancia L, donde la periodicidad de esta función esta determinada por el valor del modo de decaimiento $\alpha_{0}$.}. Estas ideas se representan en la figura (\ref{fig:EjemplosESTRUCTURA}).\\

Desde DFT este hecho es directo, sea $v_{ext}(x)$ un potencial local (bien sea una pared, bien sea una fase fluida de diferente densidad, bien otro potencial externo) que determina un perfil de densidad $\rho(x)$ tal que si $x>>1$ se de $\rho(x)\rightarrow\rho_{0}$, conforme nos acercamos al sistema uniforme y nos alejamos de $v_{ext}(x)$  debe poseer una propiedades determinadas por la función de correlación directa del fluido uniforme y sucede igualmente para la función de distribución radial conforme nos alejamos de la partícula que consideramos fija en el sistema\footnote{Este último problema resulta claramente análogo al anterior desde el punto de vista de la hipótesis de la partícula test de Percus y fue como tal fue tratado dentro de una aproximación del funcional de la densidad\cite{lebowitz:248} en sus albores. Como indicamos en el capítulo anterior podemos realizar también un análisis dentro de las aproximaciones determinadas mediante las ecuaciones integrales, o como detalla Evans\cite{reviewEvans1979} en una ecuación de \textit{Yvon-Green-Born}, véase (\ref{eqn:YBGorden1}) obtener un decaimiento exponencial asintótico hacia la fase uniforme desde la interfase líquido-vapor universal en su forma aunque determinado por un longitud de decaimiento que depende de las aproximaciones utilizadas en la aplicación de cada una de estas ecuaciones. La analogía formal determinaría que este problema ha de tener un solución similar para la función $g(r)$, y de hecho, una vez más, de otros problemas similares como el perfil de densidad de una pared en contacto con un líquido o el modelo de perfil intrínseco que será mostrado más adelante en esta memoria.}.\\

Igualmente dado un fluido uniforme que perturbamos levemente por un potencial externo las características del fluido determinan su respuesta expresada en los modos de decaimiento y las características de la perturbación condicionan únicamente las amplitudes de cada modo intrínseco de respuesta, es decir, estamos en las condiciones de una teoría de la respuesta lineal.
\subsection{Teoría de la Respuesta Lineal}
Empezamos entonces con un sistema sujeto a un cierto potencial externo $v(\vec{r})$. Bajo la teoría del funcional de la densidad existe un perfil de equilibrio $\rho(\vec{r})$ que de hecho minimiza el funcional $\Omega[\rho]$, un perfil que puede ser determinado la ecuación integral para $\mu-v(\vec{r})$, véase ec.(\ref{eqn:EulerLagrangeMinimizacion}). Imaginemos una perturbación sobre el perfil de equilibrio $\rho(\vec{r})$ de manera que el perfil de densidad venga determinado por $\tilde{\rho}(\vec{r})$, podemos evaluar la diferencia en términos de energía libre de esta nueva solución. Escribimos un desarrollo entorno de la densidad de equilibrio, que por satisfacer la ecuación integral de Euler-Lagrange mencionada queda expresado como,
\begin{equation}
\Delta\Omega=\Omega[\tilde{\rho}(\vec{r})]-\Omega[\rho(\vec{r})]=
\int d\vec{r}_{1}d\vec{r}_{2}\left.\frac{\delta \Omega}{\delta \rho(\vec{r}_{1})\delta \rho(\vec{r}_{2})}\right|_{\rho_{0}}\Delta\rho(\vec{r}_{1})\Delta\rho(\vec{r}_{2})
\label{eqn:RespuestaLinealOmega}
\end{equation} 
donde el término de orden 1 del desarrollo no aparece por estar evaluado en la densidad de equilibrio, véase ec. (\ref{eqn:EulerLagrangeMinimizacion}), mientras que $\Delta\rho(\vec{r})=\tilde{\rho}(\vec{r})-\rho_{0}(\vec{r})$.\\

Siguiendo \S\ref{ss:jerarquias}, la segunda derivada funcional aparece relacionada con las correlaciones incluidas en el \textit{sistema uniforme} por tanto la expresión en el espacio de Fourier permite escribir la derivada funcional como $1-\rho_{0}\hat{c}^{(2)}(q)$. Luego si partimos de la solución homogénea, que siempre existe, tenemos más soluciones que verifican la ecuación de Euler-Lagrange si hacemos  $1-\rho_{0}\hat{c}^{(2)}(q)=0$, análisis equivalente al indicado en ec. (\ref{eqn:ecuacionMODOSlovett}).\\

En el caso de tener un potencial $V_{ext}(z)$ o unas condiciones de contorno equivalentes, los perfiles de densidad cerca de la fase homogénea tendrán un comportamiento exponencial determinado por la solución de la ecuación anterior también para posibles valores complejos de q.

\begin{equation}
\rho[z,T;V_{ext}(z)]\sim A_{v}(T)e^{-\alpha_{1} z}cos(\alpha_{0}z-\theta)+B_{v}(T)e^{-\tilde{\alpha}_{1} z}
\label{eqn:ajusteperfildensidad}
\end{equation}

Los valores de las amplitudes $A_{v}$ y $B_{v}$ dependen explícitamente de $V_{ext}(z)$, por tanto encontrar las amplitudes del comportamiento asintótico requiere analizar las soluciones concretas obtenidas mediante la ecuación de Euler-Lagrange de nuestro problema funcional. A esta tarea se dedica el resto del capítulo y extraemos conclusiones generales acerca de la relevancia de las amplitudes en el perfil de densidad así como de la línea de Fisher-Widom en estas\footnote{En el comienzo de la teoría de líquidos, antes de hecho de la formulación de la mecánica estadística, el cociente entre el momento de orden cero y el momento de orden dos de la fuerza atractiva era tomado como una posible medida del alcance la interacción entre las moléculas, o al menos en que medida un sistema podía ser tomado como uniforme, en el contexto de la teoría de las funciones de distribución aparecen las magnitudes $\alpha_{0}$ y $\alpha_{1}$ del mismo modo como medidas del alcance, en un cierto sentido, de las correlaciones en los sistemas líquidos. De este modo la teoría de la respuesta lineal aplicada al funcional de la densidad, tal y como es indicado por \cite{1985MolPh.56.557T}, revela precisamente la interpretación de $\alpha_{i}^{-2}$ como el cociente mencionado y avala la notable intuición de algunos físicos a finales del siglo XIX.}.

\begin{figure}[htbp] 
   \centering
   \includegraphics[width=1.00\textwidth]{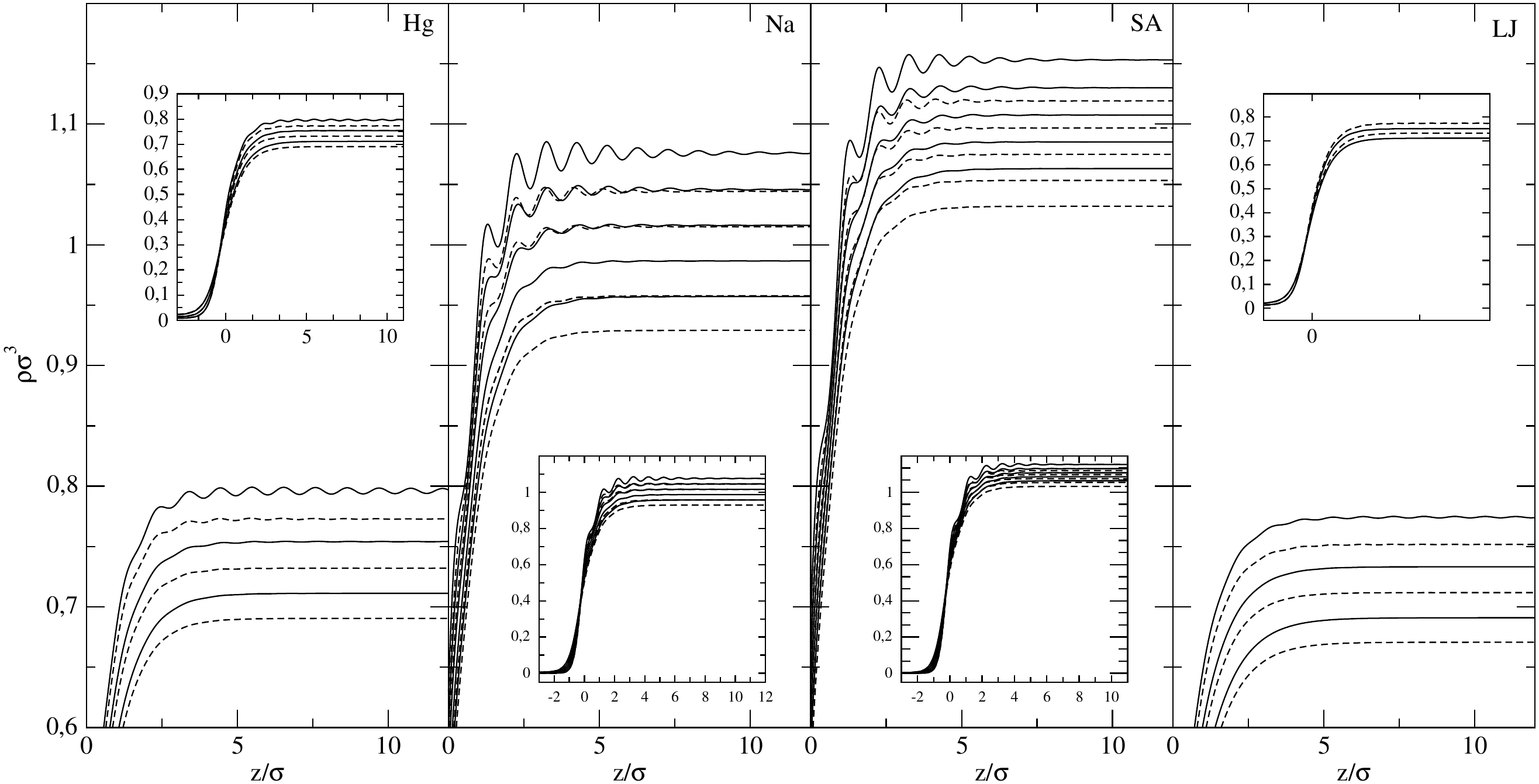} 
   \caption{\textbf{Perfiles de densidad líquido-vapor} $\rho(z/\sigma,T/U)\sigma^{3}$. Líneas continuas WDA-CS, líneas discontinuas FMT-PY. Los modelos de izq. a der. son \textit{Mercurio}(Hg), \textit{Sodio}(Na),  \textit{Soft-Alcaline}(SA) y un \textit{Lennard-Jones}(LJ) con cutoff en $2.5\sigma$. Para el Mercurio(Hg) los valores de T/U=0.65,0.70,0.75, para los modelos \textit{Sodio} y \textit{Soft-Alcaline} tenemos T/U= 0.50,0.55,0.60,0.65,0.70, valores más bajos de temperatura corresponde a valores mayores en $\rho_{liquid}$. Para el Lennard-Jones representamos T/U=0.70, 0.75 y 0.80. En el inset las dos primeras.}
   \label{fig:perfilesLV}
\end{figure}
\section{Perfiles de densidad de equilibrio líquido-vapor}
\label{sec:perfilesLV}
Nos basamos en las teorías de van der Waals generalizadas dadas por ec. (\ref{eqn:vanDerWaalsGeneralizada}), para obtener perfiles de densidad líquido-vapor realizando una minimización de la energía libre macrocanónica $\Omega[\rho]$ respecto de perfiles con simetría plana mediante la \textit{condición de frontera} dada por las densidades $\rho_{l}$ y $\rho_{v}$. El uso del colectivo macrocanónico implica que el perfil de densidad presenta una indeterminación en la posición de la interfase\cite{weeks:3106}, sin embargo fijada como condición inicial para la densidad la dada por la solución de van der Waals no homogénea\footnote{Explícitamente $\rho_{vdw}(z)=c_{1}+c_{2}tanh(\lambda z)$}, la convergencia a la solución es directa sin degeneración y además la superficie de Gibbs resultante es compatible con $z_{g}\sim 0$ en la resolución espacial utilizada\footnote{Por tanto en la presentación de los perfiles de densidad representamos directamente los resultados obtenidos como $\rho(z)$ aunque por lo dicho pueden de ser vistos como $\rho(z-z_{g})$, ambas representaciones serian equivalentes.}. De esta manera el funcional expresado a la presión y potencial químico de coexistencia permite encontrar tanto el perfil de densidad $\rho(z,T)$ como la tensión superficial $\gamma=(\Omega[\rho]-\Omega_{uniforme})/A$ en la condición de mínimo.\\

Hemos obtenido perfiles de densidad líquido-vapor para todos los modelos de interacción y aproximaciones funcionales analizadas en el capítulo anterior, podemos ver una sinopsis de algunos\footnote{El caso del Pozo Cuadrado nuestros son similares a \cite{EvansMolecularPhyscFWline} aunque los valores de los modos de decaimiento difieren levemente seguramente por las diferencias que un alcance diferente en el peso $\omega_{1}$ de WDA-CS puede inducir en la c(r).} para \textit{Mercurio}, \textit{Sodio}, \textit{Soft-Alcaline} y Lennard-Jones de $2.5\sigma$ en la figura (\ref{fig:perfilesLV}).\\

\subsubsection*{Propiedades generales}

Las primeras observaciones de estos perfiles a lo largo de todo el diagrama de fases indican que:

\begin{itemize}
\item Es posible obtener perfiles líquido-vapor para modelos blandos, \textit{Sodio} y \textit{Soft-Alcaline}, a temperaturas bajas y por consiguiente densidades más altas\footnote{Hemos podido constatar que para los potenciales \textit{Soft-Alcaline} y \textit{Sodio} podemos obtener perfiles de coexistencia a temperaturas arbitrariamente bajas aunque el proceso de minimización se dificulta notablemente para densidades del vapor próximas a cero.} que para el resto de potenciales.
\item Para los modelos Mercurio, Lennard-Jones y Pozo Cuadrado aparece una solución \textit{tipo sólido} a T/U por debajo de la línea de inestabilidad $\alpha(T,\rho_{l})=0$ tal y como se indicaba en \S\ref{sec:soluctiposolido}.
\item Los perfiles $\rho_{lv}(z)$ del \textit{Sodio} y el \textit{Soft-Alcaline} son similares, lo mismo ocurre entre el \textit{Mercurio} y Lennard-Jones.
\end{itemize}

Las propiedades del diagrama de fases obtenido por \textit{E. Chacón,  Reinaldo-Falagán, E. Velasco y P. Tarazona}\cite{PhysRevLett.87.166101} explican parte de las propiedades de los perfiles de coexistencia, las curvas de coexistencia del \textit{Soft-Alcaline} y del \textit{Sodio} son similares pero la relación $T_{m}/T_{c}$ es significativamente diferente y ello se traduce en un aspecto notablemente diferente a valores similares de T/U, mientras que las semejanzas se encuentran de hecho entre los potenciales \textit{Sodio} y \textit{Mercurio} cuya relación $T_{m}/T_{c}$ es notablemente similar a pesar de las diferencias en la curva de coexistencia de fases.\\

Esta comparación es más difícil de realizar en nuestros cálculos funcionales ya que carecemos de los valores de $T_{m}$. Los resultados del capítulo anterior indicaban una mayor semejanza estructural entre el \textit{Sodio} y el \textit{Soft-Alcaline}, semejanza que también observamos en los perfiles de densidad (a pesar de que sus curvas de coexistencia son diferentes), mientras que los perfiles del Mercurio y del Lennard-Jones aparecen similares, debido tanto a la similitud del diagrama de fases como al posible comportamiento frente a la fase sólida sugerido por la línea de inestabilidad. Su semejanza, por contra no se debe a los valores similares de $T_{FW}/T_{c}$ en la medida en que para el \textit{Sodio} y el \textit{Soft-Alcaline} son notablemente diferentes. Los resultados parecen pues indicar una relevancia notable de las propiedades de potencial duro frente a potencial blando y su capacidad de representación mediante un $d_{hs}(T)$, mientras que las similitudes en simulación entre el \textit{Sodio} y el \textit{Mercurio} se pierden debido a la aproximación de campo medio que empeora las predicciones de este último, cuestiones ya observadas en el estudio estructural previo.\\

Es llamativo que todos los perfiles obtenidos poseen un primer pico que representa una oscilación sobre una densidad menor que los siguientes (a algunas temperaturas extensible al segundo pico), hecho compatible con un perfil intrínseco fuertemente estructurado suavizado por ondas capilares que dejan un pico residual que es más evidente al bajar la temperatura. En el caso del \textit{Mercurio} y el Lennard-Jones, véase figura (\ref{fig:perfilesLV}) aparece suavizado en la escala utilizada cosa que no sucede en el \textit{Sodio} y el \textit{Soft-Alcaline} a bajas temperaturas, por tanto en el funcional FMT-PY hemos realizado comprobaciones a T/U=0.35 donde vuelve a ser estable la fase líquida del \textit{Mercurio} y observamos de nuevo un comportamiento parecido al \textit{Soft-Alcaline} en el primer pico aunque con amplitudes mayores. La diferencia de la tensión superficial notablemente mayor a la misma T/U en el \textit{Mercurio} explicaría la mayor amplitud de las oscilaciones precisamente por las hipótesis manejadas en el capítulo introductorio \S\ref{sec:introduccion}. Como primera conclusión un elemento crucial en la interpretación estructural de los perfiles de densidad es la tensión superficial y su influencia en las amplitudes de oscilación. Intentamos establecer una serie de relaciones cuantitativas.

\subsubsection*{Validez de la teoría de la respuesta lineal}
\label{sec:validezTRLenrhoz}

Análisis previos\cite{EvansMolecularPhyscFWline} realizados sobre $\rho_{lv}(z)$ para teorías de van der Waals generalizadas suelen partir de una teoría de campo medio para la parte atractiva y considerar por esta razón que no hay fluctuaciones superficiales que invaliden el análisis lineal incluso en campos nulos\footnote{El proceso seguido para incorporar las condiciones de frontera y el método de minimización seguido es diferente, las condiciones de contorno utilizan \textit{drying} completo del líquido en contacto con una pared mientras que el proceso de minimización utiliza interacción de Picard y el método de elementos finitos de Galerkin\cite{henderson:6750}, en nuestro caso hemos usado el método de gradientes conjugados no lineales (GCNL)\cite{R.Fletcher02011964}.}. Interpretan en consecuencia este perfil como \textit{intrínseco} y la dependencia de las amplitudes\footnote{Las amplitudes se notaran por $A_{lv},B_{lv}$, para los perfiles líquido-vapor o simplemente por A y B sino hay lugar a confusión.} $A_{lv}$ y $B_{lv}$  en el diagrama de fases debe ser la misma que en potenciales externos donde explícitamente no existan ondas capilares, es decir, la dependencia funcional la entienden ligada a las propiedades del sistema uniforme\footnote{En el caso de la función g(r) los análisis realizados mediante simulaciones\cite{velasco:10777,FWlineMolPhysNaHgSA} indican que cruzar la línea de Fisher-Widom no anula el término oscilatorio de la función de correlación de pares estando presente a ambos lados de la línea de Fisher-Widom, la generalidad del problema sugiere que es posible que también suceda, que no suceda no liga necesariamente la hipótesis de perfil intrínseco al perfil líquido-vapor, en $\rho_{lv}(z)$ y es por tanto una cuestión abierta la posible correlación entre $A_{lv}(T)$ y presencia de la línea de Fisher-Widom.}.\\

Más allá de la interpretación del perfil $\rho_{lv}(z)$ la posible presencia de ondas capilares\cite{EvansMolecularPhyscFWline}, si bien es importante en las amplitudes $A_{lv}(T)$, no altera la estructuración en capas de los perfiles de densidad (sus modos de decaimiento) y no esperamos que la presencia \textit{parcial} de ondas capilares anule la respuesta lineal sino que explique aspectos clave de la variación de los perfiles de densidad con la temperatura. La expresión (\ref{eqn:ajusteperfildensidad}) permite evaluar de modo concreto la validez de  (\ref{eqn:RespuestaLinealOmega}) mediante el ajuste de nuestros perfiles de equilibrio en el máximo rango de temperaturas posibles, la coincidencia de los valores de los modos de decaimiento es condición necesaria para que se verifique la respuesta lineal, y las amplitudes $A_{lv}$ y $B_{lv}$, que son las que determinan más significativamente el aspecto de este, también son determinadas en el proceso.\\

\begin{figure}[htbp] 
   \centering
\includegraphics[angle=-90,width=0.60\textwidth]{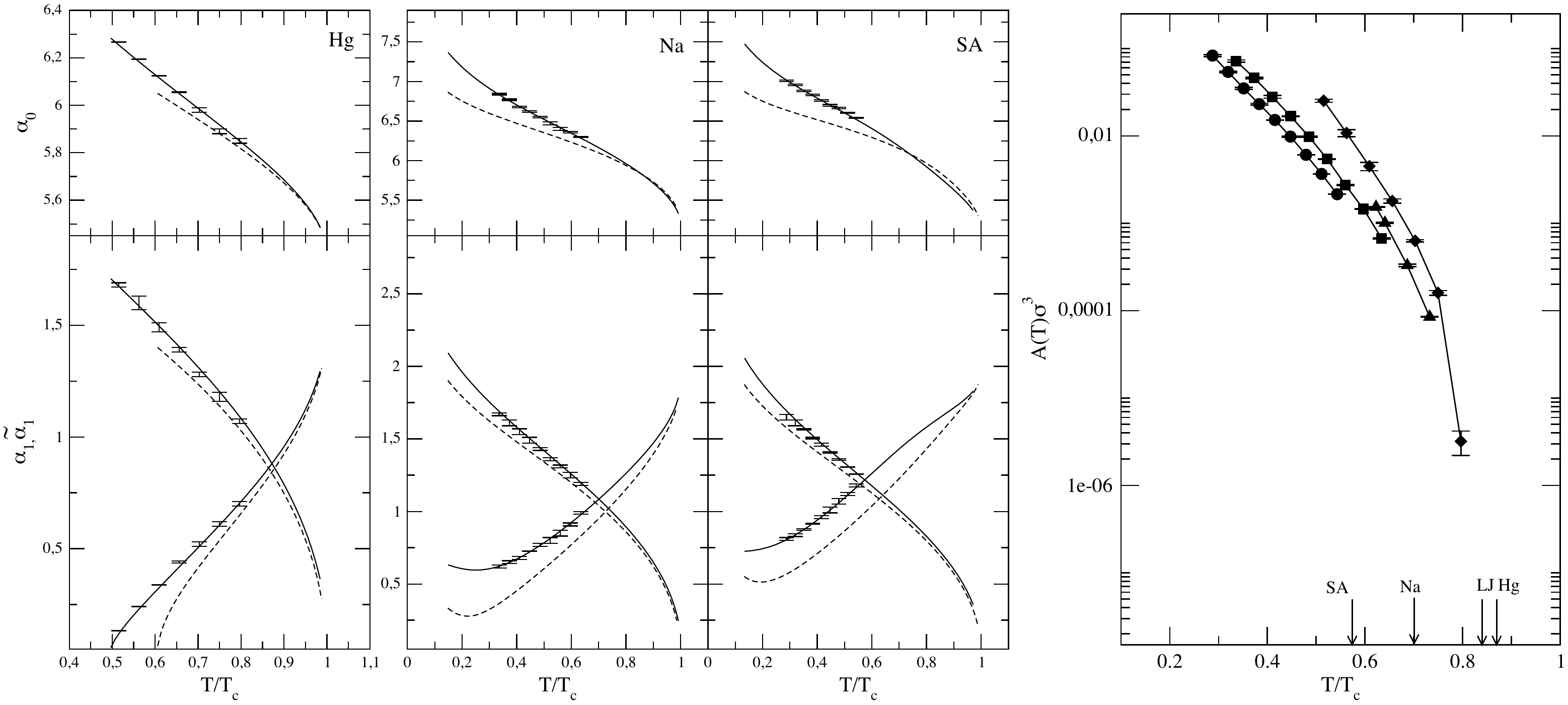} 
   \caption{\textbf{Arriba} Consistencia de la teoría de la respuesta lineal. Presentamos resultado ajustes para modos de decaimiento, se compara con la figura (\ref{fig:DecayModesHgNaSASW}) para FMT-PY. Caso WDA-CS es análogo no se presenta por claridad. Se muestra un punto en Mercurio a $T/T_{c}=0.75$ para apreciar diferencia entre doble precisión y cuádruple precisión. \textbf{Abajo}. A(T). Círculos \textit{Soft-Alcaline}(SA), Cuadrados \textit{Sodio}(Na), Rombos \textit{Mercurio}(Hg) y Triángulos \textit{Lennard-Jones}. Con flechas se representan los valores de $T_{FW}/T_{c}$ para FMT-PY.}
   \label{fig:decaysAmplitudesFMT}
\end{figure}

\begin{figure}[htbp] 
   \centering
   \includegraphics[angle=-90,width=0.60\textwidth]{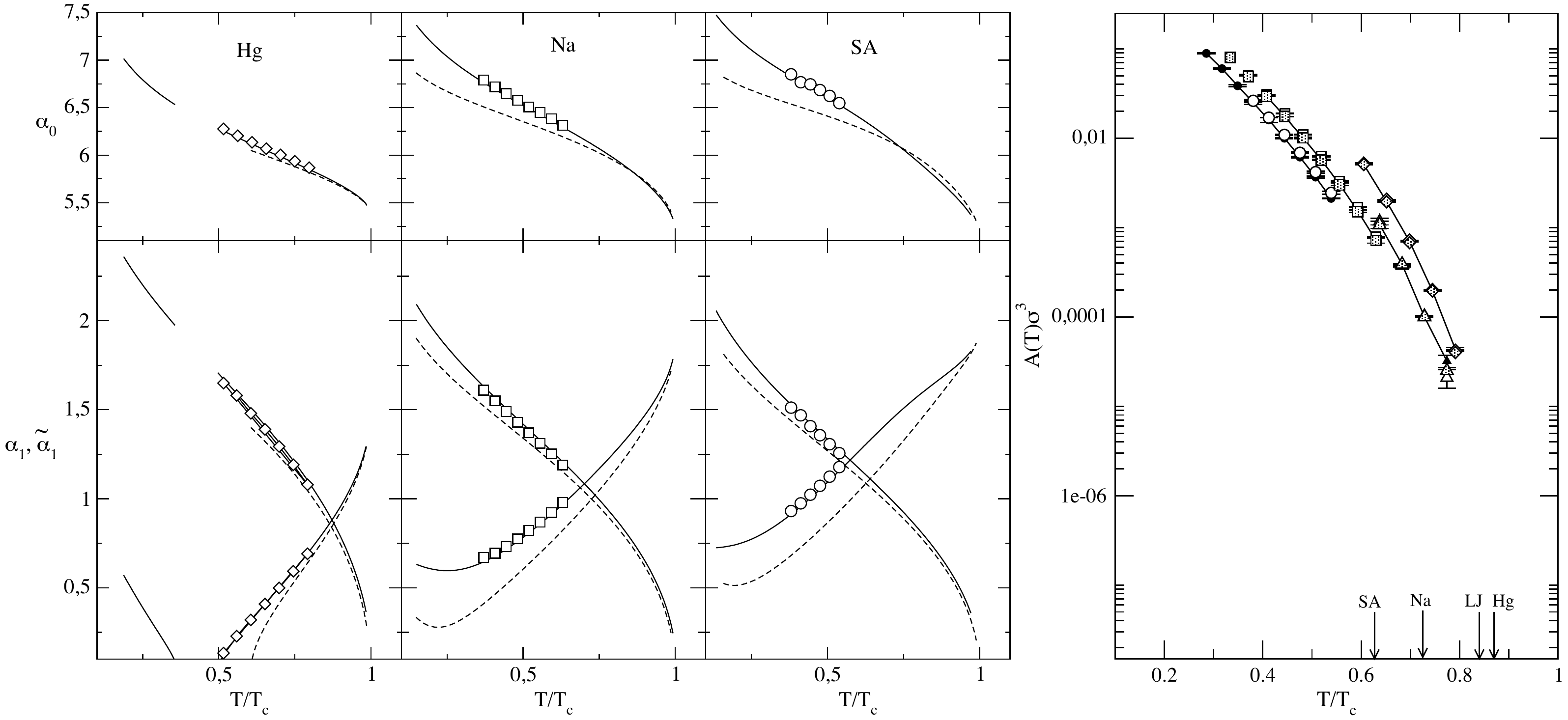} 
   \caption{\textbf{Arriba}. Consistencia de la teoría de la respuesta lineal. Presentamos los valores de $\alpha_{1}$,$\tilde{\alpha}_{1}$ y $\alpha_{0}$ para FMT-CS. \textbf{Abajo}. A(T) para todas las recetas funcionales. Negro: FMT-PY, Blanco.FMT-CS, relleno puntos WDA-CS. Algunas T/U no se representan para mostrar superposición curvas.  Círculos \textit{Soft-Alcaline}(SA), Cuadrados \textit{Sodio}(Na), Rombos \textit{Mercurio}(Hg) y Triángulos \textit{Lennard-Jones}. Con flechas se representan los valores de $T_{FW}/T_{c}$ para WDA-CS.}
   \label{fig:decaysAmplitudesWDA}
\end{figure}

Como muestran las figuras (\ref{fig:decaysAmplitudesFMT}) y (\ref{fig:decaysAmplitudesWDA}), la teoría de la respuesta lineal aplicada a nuestros perfiles de densidad es capaz de reproducir toda la fenomenología asociada a la función de correlación directa. En el \textit{Mercurio} observamos como $\alpha_{1}\simeq0$ a T/U baja reflejando la existencia de la inestabilidad frente al sólido, esto se refleja en una oscilaciones muy pronunciadas en el perfil de densidad que se extienden hasta el volumen. En el \textit{Sodio} vemos nítidamente que el perfil de densidad reproduce las propiedades de existencia de un mínimo en $\alpha_{1}$ y encontramos de hecho dos perfiles con idéntico valor de $\alpha_{1}$ pero diferente distancia entre capas dada por $\alpha_{0}$. Los resultados más significativos son para las $A_{lv}(T)$ que decrecen exponencialmente con T/U aunque muestra aparentemente que podría ser cruzada la línea de Fisher-Widom en el caso de \textit{Soft-Alcaline} y determinar más allá de esta línea valores de $A_{lv}$ finitos, pareciendo este decrecimiento de $A_{lv}(T)$ descorrelacionado\footnote{Representar $A(T)$ frente a $(T-T_{fw})/T_{fw}$ no encuentra un comportamiento en las amplitudes unificado para los diferentes potenciales y por tanto $T_{fw}$ no es la única propiedad física que condiciona las amplitudes.}  de $T_{FW}/U$.\\

Un análisis cuidadoso 
 requiere determinar la dependencia en z (desde la interfase) de los términos que aparecen en la expresión (\ref{eqn:ajusteperfildensidad}) indicándonos el rango de aplicación del análisis lineal así como la consistencia entre los resultados de minimización funcional y de las propiedades de la función de correlación directa. En consecuencia realizamos un ajuste general para todas T para ambos modos y comprobamos su dependencia en z. De modo general los valores de $\tilde{\alpha}_{1}$, $\alpha_{1}$ y $\alpha_{0}$ obtenidos del ajuste coinciden con un error relativo pequeño con los valores obtenidos de la función de correlación. Nuestros valores presentan una indeterminación en z con algunas propiedades sistemáticas:
\begin{itemize}
\item Para valores de z cercanos a la interfase el ajuste monótono es adecuado (reproduce valores razonables de $\tilde{\alpha}_{1}$) pero el oscilatorio no. Esta región cae fuera del análisis asintótico.
\item Para valores muy altos de z es posible realizar solo un ajuste a la parte oscilante reproduciendo adecuadamente la cola del perfil de densidad pero suelen mejorar levemente si incluimos el término monótono. Región idónea para el análisis asintótico.
\item El valor de z a partir del cual los ajustes se corresponden mejor con los esperados es \emph{creciente con la T}. Localizado cerca de la primera oscilación.
\end{itemize}
Los dos últimos resultados al caer en el régimen asintótico son los más significativos, uno nos indica la mayor relevancia del término oscilante\footnote{Cuestión que esperábamos, marginalmente siempre están presentes ambos términos por otra parte que la presencia del término monótono mejore levemente los ajustes respecto del término oscilante no es extraño, añadir nuevos términos al ajuste permite mejorar la aproximación al problema. Por otra parte diferencias para z grandes son esperables debido a las dificultades de minimizar una estructura adecuadamente a valores muy grandes de z. Con este fin hemos realizado dos minimizaciones una en doble precisión y otra en cuádruple precisión, permitiendo incrementar así el rango de temperaturas en que nuestro análisis resultaba válido.}, el otro la zona donde es posible encontrar esta relevancia mayor.\\

Respecto de la línea de Fisher-Widom que a temperaturas crecientes el papel mo\-nó\-to\-no sea progresivamente más relevante y que no hemos encontrado perfiles que consideremos de equilibrio con oscilaciones más allá de ella parece indicar una correlación entre Fisher-Widom y $A_{lv}(T)$, sin embargo veremos que entender la última de las propiedades indicadas explica porque no hemos encontrado término oscilante más allá de Fisher-Widom, es decir $A_{lv}(T)>0$ para $T>T_{fw}$. 

\section{Relevancia de Fisher-Widom}

Una explicación del aspecto diferente entre un perfil de densidad y la función de correlación total puede encontrarse en las propiedades comentadas de nuestros ajustes y el hecho de que estos se desplazan a z crecientes con la temperatura cosa que no sucede en el caso de la función de distribución de pares o del perfil de un líquido en contacto con una pared que presentan oscilaciones desde la primera capa. Este hecho puede ser explicado por la diferencia importante que introducen las amplitudes en uno y otro caso.\\

En el caso de la función g(r) tenemos\footnote{Esta propiedad la hemos verificado para la aproximación RHNC-Lado, véase la figura (\ref{fig:FuncionesDistribucionRadial}) tanto mediante ajustes directos a la cola de la función h(r) como mediante la aproximación de polos simples para los residuos de los términos relevantes del comportamiento asintótico, véase referencia\cite{evans:591}, y la consiguiente determinación analítica. En ambos casos los ordenes de magnitud se mantienen en $A_{g}\approx 1.5$, algo mayor en los ajustes directos, mientras que $B_{g}$ varía de $\approx0.001$ a $\approx0.01$ , la dependencia con la temperatura es mayor en $B_{g}(T)$ que presenta un comportamiento significativamente creciente.} que $A_{g}>>B_{g}$ mientras que en el caso de $\rho_{lv}(z)$ obtenemos $A_{lv}<<|B_{lv}|$. La aplicación de ambos casos a las ecuaciones de ajuste y el signo relativo de $\tilde{\alpha}_{1}(T)$ y $\alpha_{1}(T)$ antes y después de Fisher-Widom indican que la predominancia del término monótono en el caso de g(r) es solo para temperaturas mayores de Fisher-Widom y valores de z  (o r) altos, mientras que en $\rho_{lv}(z)$ para $T<T_{FW}$ siempre esta presente el término oscilatorio pero a valores crecientes de z presentando una divergencia en z al alcanzar $T_{FW}$, esto explica porque encontrábamos la posición en que se verificaba el comportamiento asintótico en valores de z progresivamente más alejados de la interfase.\\ 

El aspecto cualitativo del perfil de densidad en diferentes situaciones viene mediado por los valores de las amplitudes de oscilación, y por tanto por la forma de la perturbación, más que por los valores de sus modos. Las dos situaciones antes expuestas pueden ser caracterizadas mediante expresiones obtenidas de la ecuación de ajuste. Denotamos $\alpha_{c}=\alpha_{0}+i\alpha_{1}$\\
\begin{itemize}
\item Si $A<<B$ entonces para $T<T_{FW}$ se tiene que $\alpha_{1}(T)<\tilde{\alpha}_{1}(T)$ y
\begin{equation}
z_{min}^{osc}(T)=\frac{1}{\tilde{\alpha}_{1} -\alpha_{1}}\ln \left[\frac{\tilde{\alpha}_{1} B}{\vert\alpha_{c}\vert A} \right] 
\label{eqn:maxdensidadlv}
\end{equation}
que indica la localización de la presencia de la primera oscilación en $\rho_{lv}$

\item Si $A>B$ entonces para $T>T_{FW}$ se tiene que $\alpha_{1}(T)>\tilde{\alpha}_{1}(T)$ y
\begin{equation}
z_{max}^{osc}(T)=\frac{1}{\alpha_{1}-\tilde{\alpha}_{1}}\ln \left[\frac{\vert\alpha_{c}\vert A}{\tilde{\alpha}_{1} B} \right] 
\label{eqn:maxdensidadgder}
\end{equation}
Que localiza la presencia de la última oscilación visible para g(r).
\end{itemize}

\begin{figure}[htp]
  \begin{center}
    \subtop[Figura extraída de la referencia \cite{dijkstra:1449} ]{\label{fig:dijkstra}\includegraphics[scale=0.70]{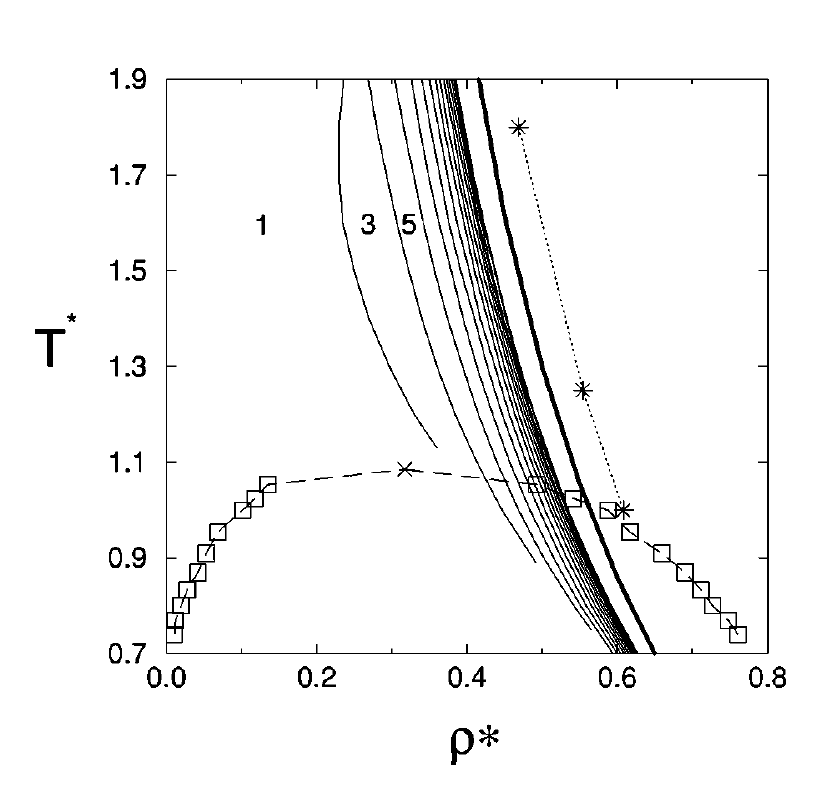}}
    \subtop[Resultados ec.(\ref{eqn:maxdensidadgder}) para $h(r)$ desde la aproximación RHNC-Lado. Modelos SA y Na.]{\label{fig:maximosGder}\includegraphics[scale=0.45]{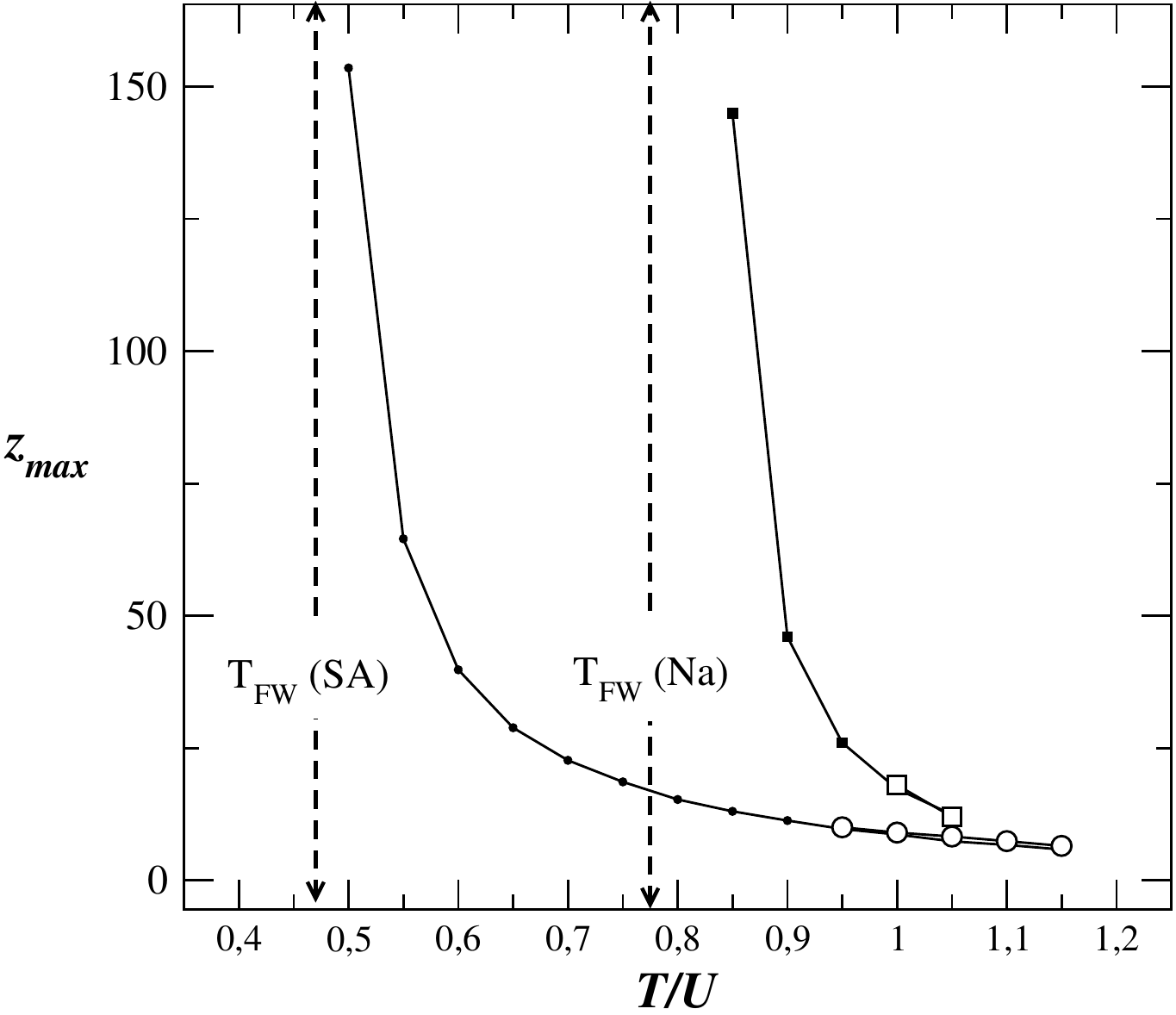}} 
  \end{center}
  \caption{\textbf{(a)}. Muestra las transiciones entre funciones h(r) con n ceros y n+2 ceros, para un Lennard-Jones $2.5\sigma$ y RPA. Como se puede apreciar $n\rightarrow\infty$ al aproximarse a la línea de Fisher-Widom.
  \textbf{(b)}. Cálculo de la última posición de la oscilación de h(r) para los Modelos \textit{Soft-Alcaline} (representado por círculos) y \textit{Sodio} (representado por cuadrados). Línea con figuras negras resultados mediante (\ref{eqn:maxdensidadgder}) donde introducimos $A_{g}(T)$ y $B_{g}(T)$ calculadas analíticamente como los residuos de los primeros polos bajo la suposición de polos simples. Figuras blancas representan posición real de la ultima oscilación en h(r).}
  \label{fig:detallesgder}
\end{figure}
	
La localización de un número de oscilaciones finito para $g(r)$ después de la línea de Fisher-Widom es compatible con los resultados de simulación Montecarlo de \textit{Marjolein Dijkstra} y \textit{R.Evans}\cite{dijkstra:1449} que determinan el número de ceros después de FW\footnote{El motivo de este cálculo proviene, de hecho, de una sugerencia del propio B.Widom.}. Esto se aprecia en la figura (\ref{fig:dijkstra}) que comparamos para dos potenciales en RHNC-Lada con la relación propuesta (\ref{eqn:maxdensidadgder}) que mostramos en la figura (\ref{fig:maximosGder}).\\

\begin{figure}[htbp] 
   \centering
   \includegraphics[width=1.00\textwidth]{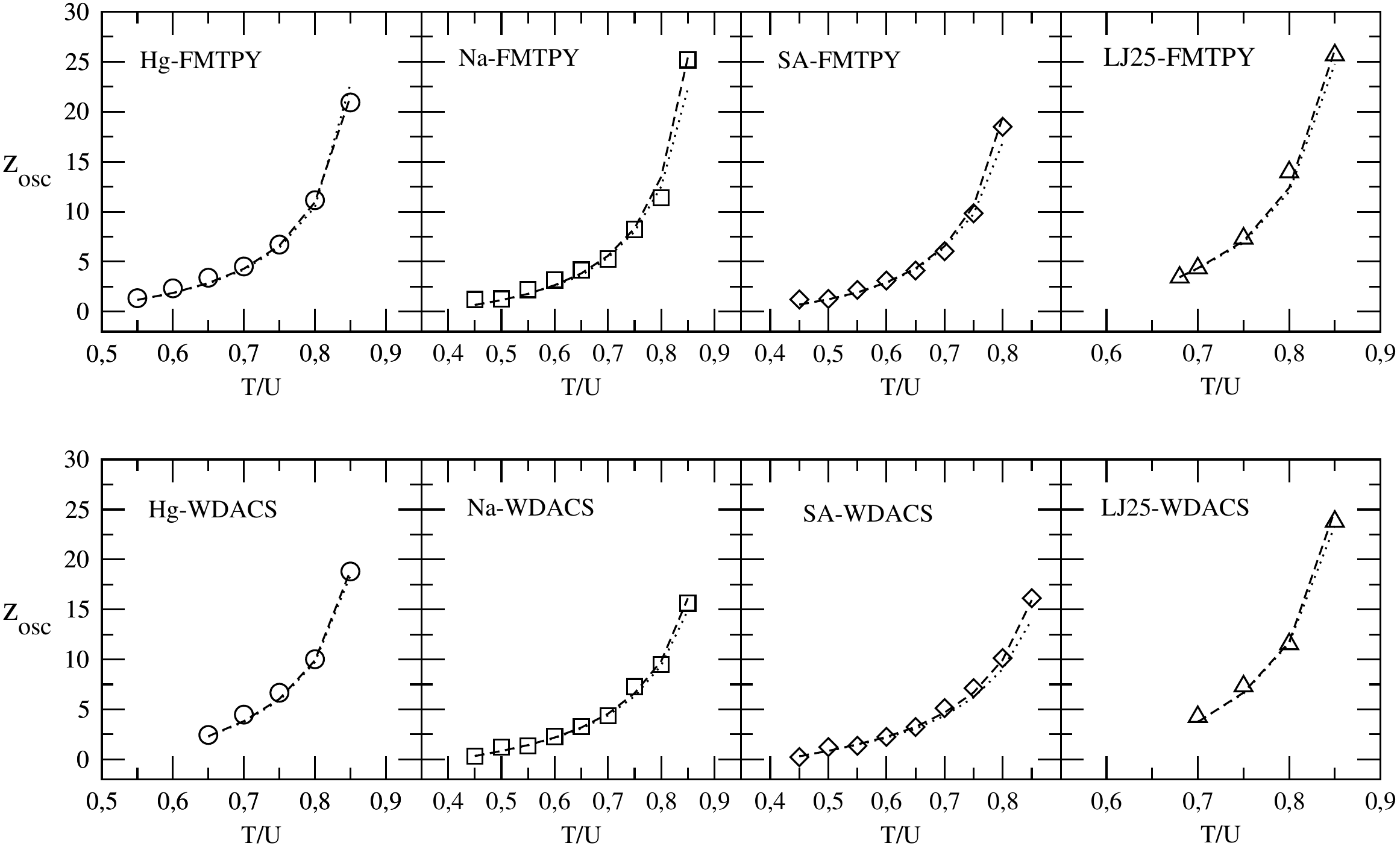} 
   \caption{Estimación de la posición de los máximos mediante la ec. (\ref{eqn:maxdensidadlv}). Figuras son valores de $z_{min}^{osc}(T)$ tomados directamente del perfil de densidad. Se comparan con el cálculo usando primero unicamente los parámetros del ajuste (línea discontinua a trazos) y segundo usando las amplitudes del ajuste con los modos procedentes c(r) (línea de puntos). \textbf{Arriba} funcional FMT-PY. \textbf{Abajo} funcional WDA-CS.}
   \label{fig:DistribucionMax}
\end{figure}

Para el caso de $\rho(z)$ hemos comprobado que nuestro resultado (\ref{eqn:maxdensidadlv}) y la posición real de la primera oscilación coinciden para todas las temperaturas, véase la figura (\ref{fig:DistribucionMax}). El caso del perfil de densidad en contacto con una pared posee un análisis análogo a g(r), de modo que variando las propiedades del potencial de la pared se pueden variar las amplitudes $A_{v}$ y $B_{v}$, a efectos de completitud será analizado más adelante \S\ref{sec:excursusFWintrinseco}, donde veremos que de hecho es posible anular el término monótono $B_{v}\simeq 0$ para determinados $V_{ext}(z)$ y en consecuencia la posición de la última oscilación se adentra completamente en el volumen después de la línea de Fisher-Widom.\\

\section{Relación entre ondas capilares y perfiles de densidad}
\label{sec:relacionCWyDEN}

Comenzamos analizando la posible relación existente entre las ondas capilares y los perfiles obtenidos mediante teorías del funcional de la densidad  evaluando la influencia que la presencia de las ondas capilares implica en la dependencia $A_{lv}(T)$. Podemos determinar, de manera \textit{naive}, dicha relevancia basándonos directamente en las siguientes hipótesis inspiradas en los capítulos precedentes\footnote{Y consistentes con la hipótesis hidrostática para los resultados de las siguientes secciones.},
\begin{enumerate}[\hspace{0.5cm}\bfseries H 1. ]
\item Consideramos válidas las hipótesis de la teoría de ondas capilares que permiten obtener un perfil $\rho(z,L_{1})$ a partir de un perfil $\rho(z,L_{0})$ donde $L_{1}>L_{0}$ mediante la convolución gaussiana con una $\Delta^{2}_{cw}$ obtenida mediante el hamiltoniano efectivo $\mathcal{H}_{CWT}$. En particular lo consideramos válido para $L_{0}$ en el rango del diámetro molecular.\label{H1}
\item Existencia de un perfil intrínseco fuertemente estructurado que puede ser asociado a tamaños transversales del orden del diámetro molecular y presentan una amplitud en las oscilaciones del orden de la densidad del líquido.\label{H2}
\end{enumerate}
Bajo la primera de las hipótesis podemos relacionar los perfiles de densidad para dos valores del tamaño transversal caracterizado por L mediante la expresión, véase ec. (\ref{eqn:perfildensidadmedioCWT}),
\begin{equation}
\rho(z,L_{1})=\int ds \mathcal{P}_{cw}(s,L_{1},L_{0})\rho(z-s,L_{0})
\end{equation}
donde la función  $\mathcal{P}_{cw}(s,L_{1},L_{0})$ es una función de distribución gaussiana y   posee dependencia en los dos cutoffs\footnote{Para su descripción detalla ver \S\ref{sec:funcionesDistribucionAlturas} y la referencia\cite{bedeaux:972}.}. De esta manera dado el comportamiento asintótico del perfil de densidad para un tamaño transversal $L_{0}$. Podemos escribir,

\begin{equation}
\rho(z>z_{1osc},L_{0})=\rho_{liq}+B(L_{0})e^{-\tilde{\alpha}_{1}z}+A(L_{0})e^{-\alpha_{1}(z-z_{0})}cos\left(\alpha_{0}(z-z_{0})\right)
 \label{eqn:Asintotico}
\end{equation}
donde análogamente a ec. (\ref{eqn:ajusteperfildensidad}) consideramos la amplitud $A(L_{0})$ que corresponde a la primera capa del perfil de densidad, manteniendo para ello $z_{0}\lesssim2\pi/\alpha_{0}$. Podemos expresar mediante la convolución gaussiana\footnote{Aplicamos que:
\begin{equation}
\int dz e^{\frac{-x^{2}}{2\Delta}}e^{-k x}=\sqrt{\frac{\Delta\Pi}{2}}e^{\frac{ck^{2}}{2}}erf{\frac{ck+x}{\sqrt{2c}}}
\end{equation}
y que
\begin{equation}
\int dz e^{\frac{-x^{2}}{2\Delta}}e^{-i k x}=i\sqrt{\frac{\Delta\Pi}{2}}e^{\frac{ck^{2}}{2}}erfi{\frac{ck-ix}{\sqrt{2c}}}
\end{equation}
y los comportamiento asintóticos $Re[erf(z)]\longrightarrow \pm 1$ cuando $Re[z]\longrightarrow \pm \infty$ con Im(z) acotado y $erf(z)=-ierfi(z)$.} de la expresión anterior obteniendo,
\begin{align}
\rho(z>z_{1osc},L_{1})=&\rho_{liq}+B(L_{1})e^{-\tilde{\alpha}_{1} z}\nonumber\\
+&A(L_{1})e^{\frac{1}{2}\Delta_{cw}^{2}(\alpha_{0}^{2}-\alpha_{1}^{2})}e^{-\alpha_{1}(z-z_{0})}cos\left(\alpha_{0}z-\alpha_{0}(z_{0}-\Delta_{cw}^{2}\alpha_{1})\right) 
\label{eqn:convucionAsintotico}
\end{align}

Las restricciones $z>>\sigma$ pueden ser aproximadas por $z\gtrsim z_{1osc}$ como vimos en \S\ref{sec:validezTRLenrhoz}. Además fijadas las condiciones termodinámicas y los tamaños $L_{1}$ y $L_{0}$ tendremos que $\Delta_{cw}^{2}\alpha_{1}$ esta determinado y por tanto haciendo uso de la expresión dada por la ec.(\ref{eqn:DeltaCW}) y definiendo $\hat{z}_{0}=z_{0}-\Delta_{cw}^{2}\alpha_{1}$ podemos comparando la expresión (\ref{eqn:convucionAsintotico}) con la anterior (\ref{eqn:Asintotico}) evaluada en $L_{1}$ escribir,
\begin{equation}
A(L_{1})=A(L_{0})\left(\frac{L_{0}}{L_{1}}\right)^{\eta(T)}
\label{eqn:AmplitudOsciConTemp}
\end{equation}
donde hemos definido un parámetro $\eta$ por la ecuación,
\begin{equation}
\eta(T)=\frac{\alpha_{0}^{2}+\alpha_{1}^{2}}{4\pi\beta\gamma}
\end{equation}
esta relación es válida en ausencia de gravedad\footnote{En las referencias\cite{evans:591},\cite{PhysRevE.64.041501} aparece un $\alpha_{0}^{2}-\alpha_{1}^{2}$, el modo consistente de definir las amplitudes en nuestro caso implica la redefinición a $\hat{z}_{0}$ e involucra un cambio en el signo de $\alpha_{1}^{2}$.}.\\

Desde la ecuación que determina el comportamiento asintótico del perfil de densidad vemos que la información que nos interesa $\Delta_{cw}^{2}$ esta igualmente contenida como una fase en la función periódica, a partir de las diferencias entre $\hat{z}_{0}=z_{0}-\Delta_{cw}^{2}\alpha_{1}$ es posible determinar teóricamente el valor de $\Delta_{cw}$.
 El problema que presenta esta elección es que aunque pudiéramos determinar de modo suficientemente preciso los valores de $\hat{z}_{0}$ quedaría aun el problema de si dichos valores pueden ser predictivos para $\Delta_{cw}^{2}$, las amplitudes no dependen, a priori, drásticamente de la aproximación de fluido incompresible implícita en CWT\footnote{Como se vio antes pueden ser determinadas de modo consistente desde las primeras capas del fluido.} pero este desplazamiento $z_{0}$ puede tener una dependencia más importante en esta suposición y su no validez condicionaría los resultados de un modo más difícilmente predecible\footnote{De hecho los valores de A(T) y A(T,mg) resultan más estables que los de $z_{0}$ del que extraer $\Delta_{cw}^{2}$ resultaría más complicado.}.\\

La viabilidad de explicar $A(T)$ mediante ondas capilares se reduce a determinar $L_{0}$ y $L_{1}$ que dependerían de nuestro funcional. Además escrito como $\Omega[\rho,\phi]$ resalta su dependencia implícita con el potencial de interacción usado además de la aproximación funcional para $\Omega[\rho]$.\\

Si suponemos que la hipótesis segunda (H\ref{H2}) es compatible con la expresión deducida tendremos que $L_{0}\sim \sigma$ y $A(L_{0}=A_{0})\sim \rho_{l}$. Para que la ecuación anterior sea cierta sin más debería ser extrapolable la función de distribución de alturas, véase (\ref{eqn:defdeSdeR}), a todas las escalas involucradas en el paso de $L_{0}$ a $L_{1}$. En este sentido obtenemos un valor de $L_{1}$ orientativo pero condicionado a los valores reales de $L_{0}$. Y donde técnicamente la elección de $L_{0}$ esta atada a reproducir en esta escala amplitudes dadas por $A_{0}=\rho_{l}$. \\

\begin{figure}[htbp] 
   \centering
   \includegraphics[width=1.00\textwidth]{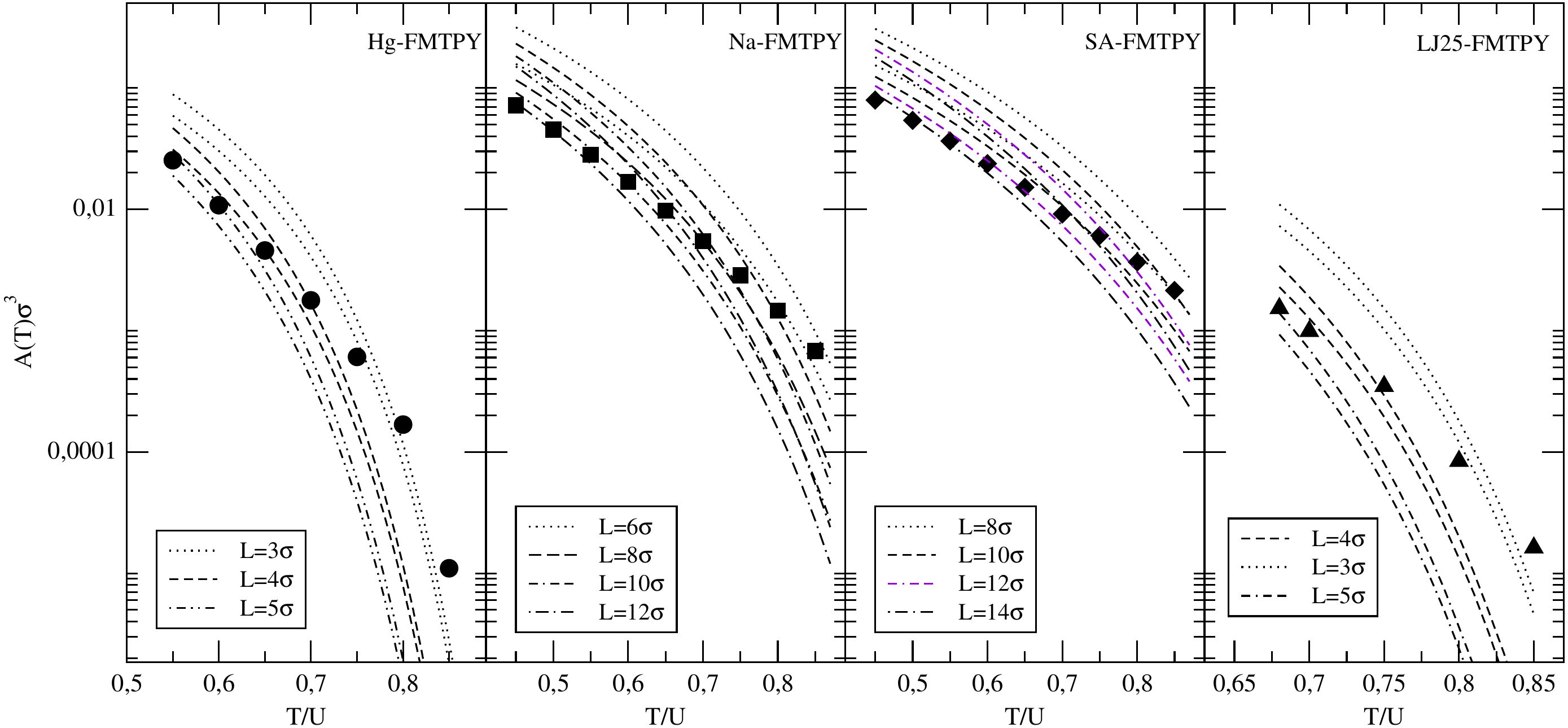} 
   \caption{\textbf{Estimación de la curva A(T)} mediante la expresión (\ref{eqn:AmplitudOsciConTemp}). Se determina para $A(L_{0})=1.2\rho_{l}$ y $A(L_{0})=0.8\rho_{l}$.  La razón $L_{1}/L_{0}$ viene indicada en cada cuadro como L. Para cada valor de $L_{1}/L_{0}$ la línea superior se corresponde con el mayor valor de $A(L_{0})$. \textbf{Funcional FMT-PY}}
   \label{fig:CurvasAdeT-FMT}
\end{figure}

\begin{figure}[htbp] 
   \centering
   \includegraphics[width=1.00\textwidth]{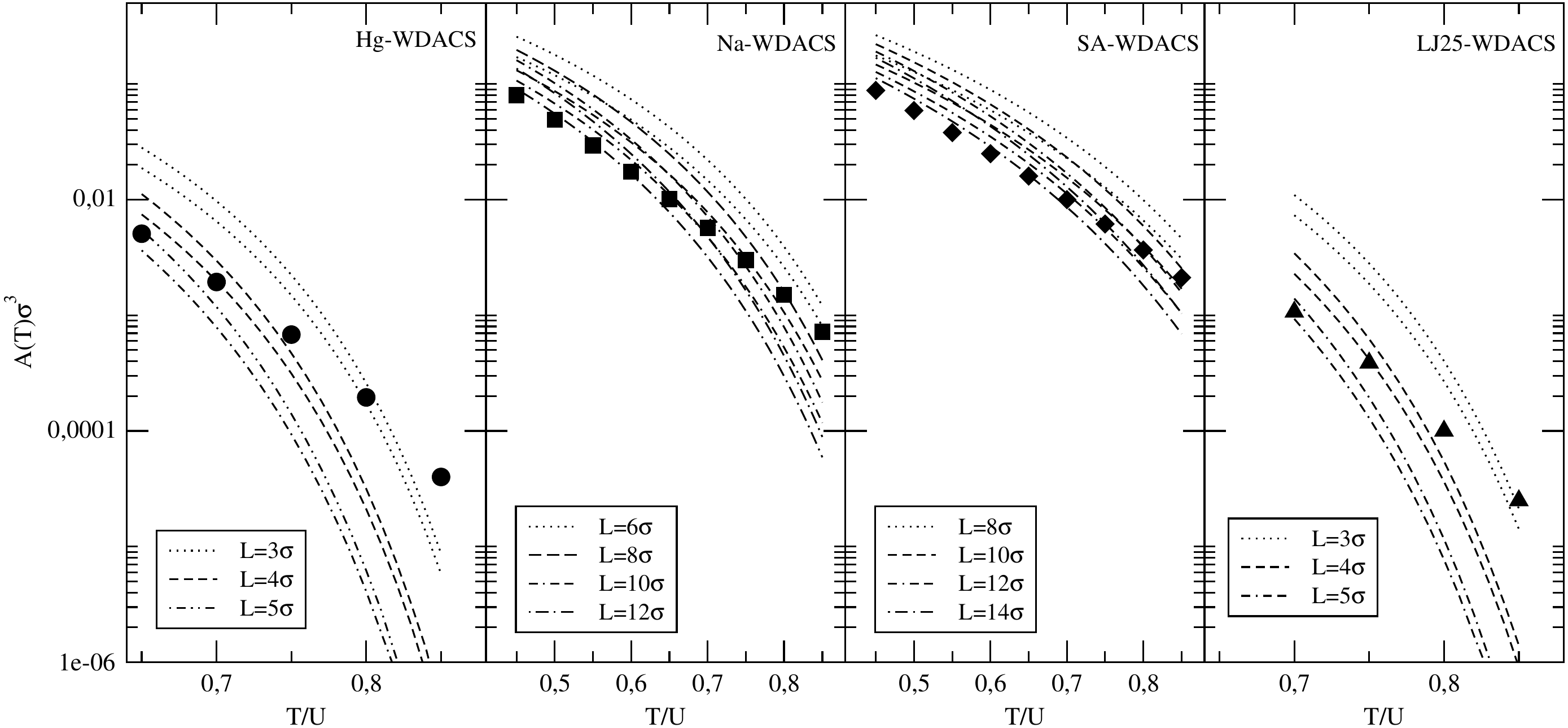} 
   \caption{\textbf{Estimación de la curva A(T)} mediante la expresión (\ref{eqn:AmplitudOsciConTemp}). Se determina para $A(L_{0})=1.2\rho_{l}$ y $A(L_{0})=0.8\rho_{l}$.  La razón $L_{1}/L_{0}$ viene indicada en cada cuadro como L. Para cada valor de $L_{1}/L_{0}$ la línea superior se corresponde con el mayor valor de $A(L_{0})$. \textbf{Funcional WDA-CS}}
   \label{fig:CurvasAdeT-WDA}
\end{figure}
Los resultados de las hipótesis precedentes podemos observarlos en la figuras (\ref{fig:CurvasAdeT-FMT}) y (\ref{fig:CurvasAdeT-WDA}) para los modelos \textit{Soft-Alcaline}, \textit{Sodio}, \textit{Mercurio} y Lennard-Jones. Inicialmente observamos que la determinación de $L_{1}$ independiente del resto de parámetros no es posible de un modo cuantitativo, en especial independiente del valor de $L_{0}$, aunque podemos acotar un posible rango para estos en el rango de temperaturas indicado. Fijar $L_{0}$ independiente del modelo lleva a obtener valores de $L_{1}$ bastante diferentes entre los distintos potenciales. Por otra parte relacionar $L_{1}$ con una longitud característica de las fluctuaciones de volumen como $\alpha_{1}^{-1}$ tampoco es viable.\\

Es interesante la comparación entre las dos recetas funcionales, WDA y FMT-CS, las amplitudes $A(T)$ son similares entre las dos recetas funcionales sin embargo esperamos que la estructuración en WDA sea levemente mayor para $A(L_{0})$, cuestión que observamos se da en $T\simeq0^{+}$ donde $A(T\simeq0)$ es mayor en el caso de WDA ya en la zona de metaestabilidad.

\section{Perfiles de densidad en un campo externo}

Si introducimos el potencial externo $V_{ext}(z)$ en el estudio anterior podemos observar los mismos fenómenos que teníamos en el caso anterior aunque ahora la relación de las propiedades estructurales con los modos de decaimiento es algo más sutil. El análisis tanto de la posible relevancia de Fisher-Widom en las amplitudes como el papel de las ondas capilares en los perfiles de densidad lleva a conclusiones análogas a las anteriores pero las propiedades que obtendremos en campos externos nos permitirán contrastar las hipótesis, introducidas en la sección anterior, tanto cualitativamente como cuantitativamente en la medida en que el parámetro que defina la intensidad del potencial externo permitirá determinar valores de $L_{1}$ independientemente de los valores concretos de la segunda hipótesis que habíamos introducido.\\

Las condiciones de contorno, que definían nuestros perfiles de densidad sin campo externo que permitían identificar los valores del sistema uniforme y definir nítidamente una perturbación de este y aplicar así la teoría de la respuesta lineal, ahora son diferentes. Las condiciones en un campo externo serán en el líquido situado en $z>0$
\begin{equation}
z_{1}>z_{2}>0 \Rightarrow \rho(z_{1})\gtrsim\rho(z_{2})
\end{equation}
y la situación inversa en el vapor. Bajo estas condiciones en campos suficientemente intensos y tamaños del sistema suficientemente grandes se alcanza el empaquetamiento máximo y puede aparecer una fase sólida en la fase líquida en el comportamiento asintótico\footnote{En la práctica es necesario campos mucho más intensos que la gravedad terrestre para observar este fenómeno o bien tamaños de caja muy grandes.}. La forma de introducir por tanto las condiciones de frontera de nuestro cálculo variacional requiere cierto cuidado ya que el requerimiento $\frac{d\rho}{dz}\xrightarrow[z\rightarrow+\infty]\quad 0$ no se cumple para cualquier campo externo. Hemos comprobado diferentes formas de introducir las  condiciones de contorno especificas y obtenemos propiedades de los perfiles de densidad que solo dependen del parámetro usado para definir convenientemente la intensidad del campo externo $V_{ext}(z,mg)$. Se llega al hecho de que el comportamiento asintótico se alcanza a distancias intermedias bien determinadas en función de la temperatura y la intensidad del campo externo.\\

Para el estudio que vamos a realizar hemos introducido $V_{ext}(z)=mgz$ y los perfiles de densidad resultantes podemos verlos  para diferentes modelos en la siguiente figura en función, tanto de la temperatura, como de la intensidad del campo gravitatorio\footnote{En el caso del Lennard-Jones visualizar un perfil de densidad para un rango aceptable de temperaturas en valores de intensidades del campo gravitatorio es difícil por ser el rango de densidades donde el sistema es estable más limitado que en los otros modelos, problema aun más acentuado en un pozo cuadrado, nos restringimos por tanto a los tres potenciales en que el rango de estabilidad frente al sólido es lo suficientemente amplio \textit{en las aproximaciones funcionales que estamos usando}.}.

\begin{figure}[htbp] 
   \centering
   \includegraphics[width=1.0\textwidth]{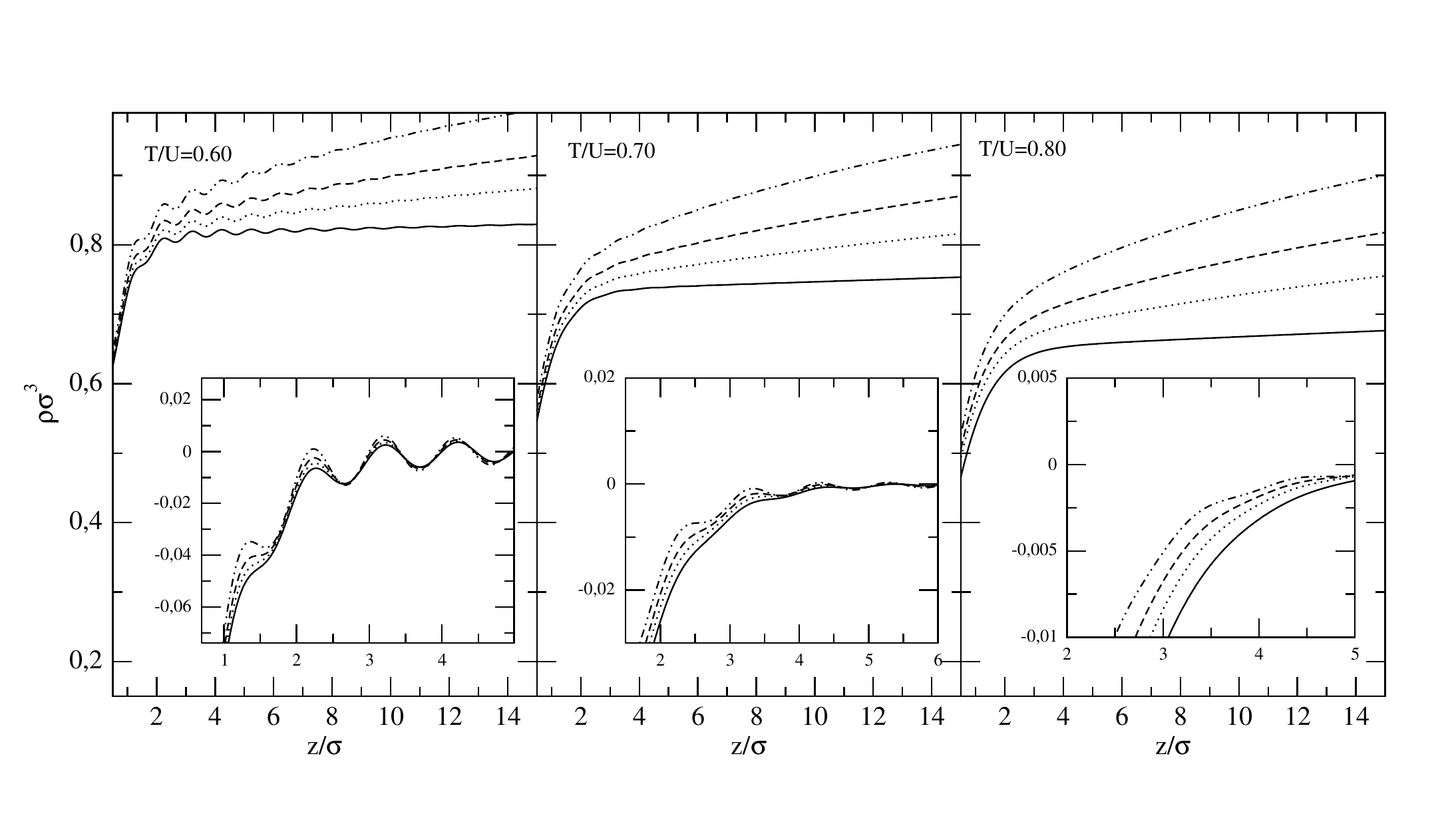}
   \caption{\textbf{Perfiles en un campo externo $V_{ext}(z)=mgz$} con cutoff en $z=60\sigma$.  Aproximación funcional FMT. \textit{Potencial de Interacción Mercurio, \textit{Hg}}. Valores de la intensidad: mg=0.01 (línea continua), mg=0.05 (línea de puntos), mg=0.1 (línea discontinua), mg=0.2 (línea discontinua con puntos).}
   \label{fig:PerfilesGravedadHgPY}
\end{figure}
\begin{figure}[htbp] 
   \centering
   \includegraphics[width=1.0\textwidth]{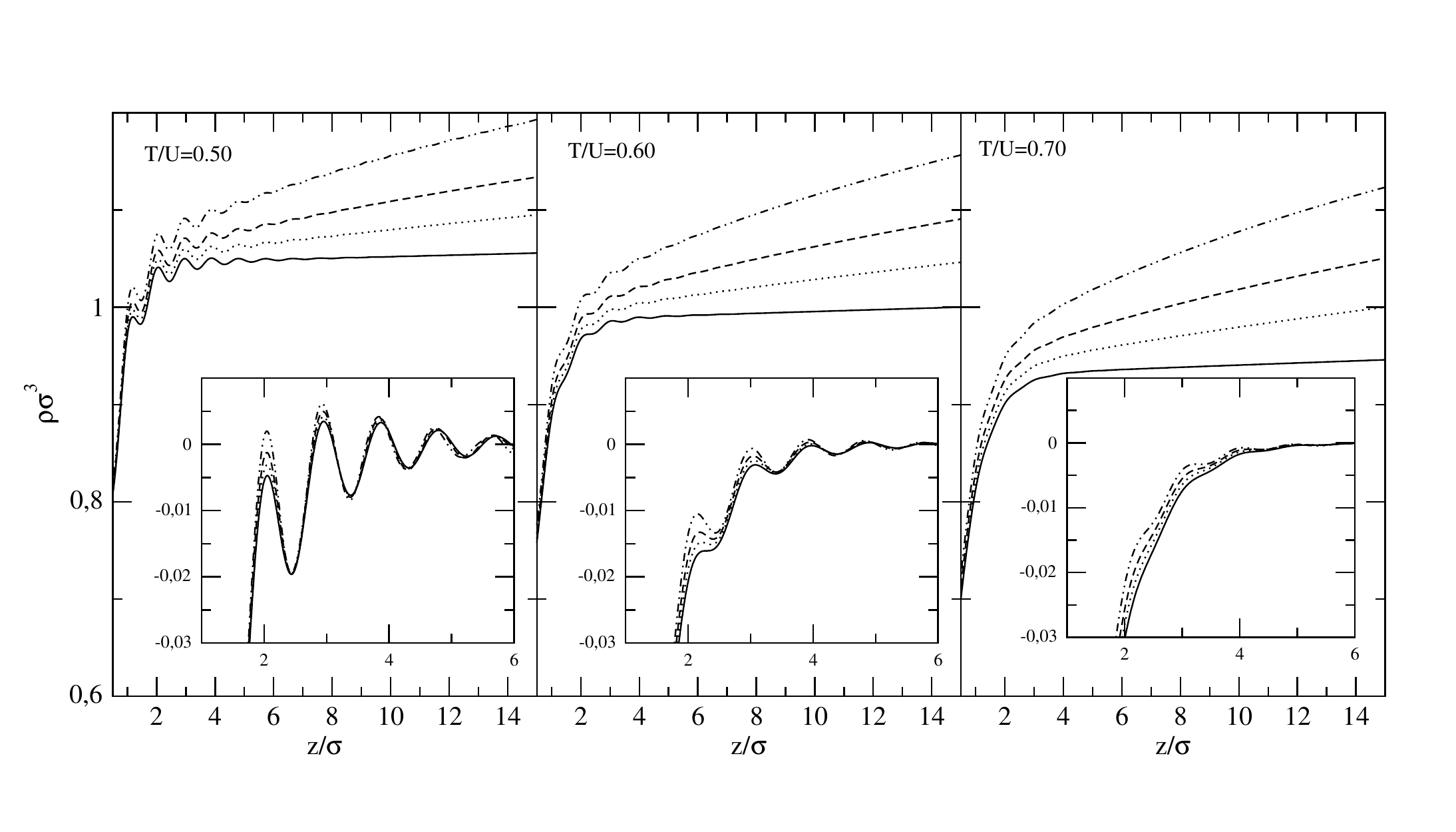}
   \caption{\textbf{Perfiles en un campo externo $V_{ext}(z)=mgz$} con cutoff en $z=60\sigma$. Aproximación funcional FMT. \textit{Potencial de Interacción Sodio, \textit{Na}}. Valores de la intensidad: mg=0.01 (línea continua), mg=0.05 (línea de puntos), mg=0.1 (línea discontinua), mg=0.2 (línea discontinua con puntos).}
   \label{fig:PerfilesGravedadNaPY}
\end{figure}
\begin{figure}[htbp] 
   \centering
   \includegraphics[width=1.05\textwidth]{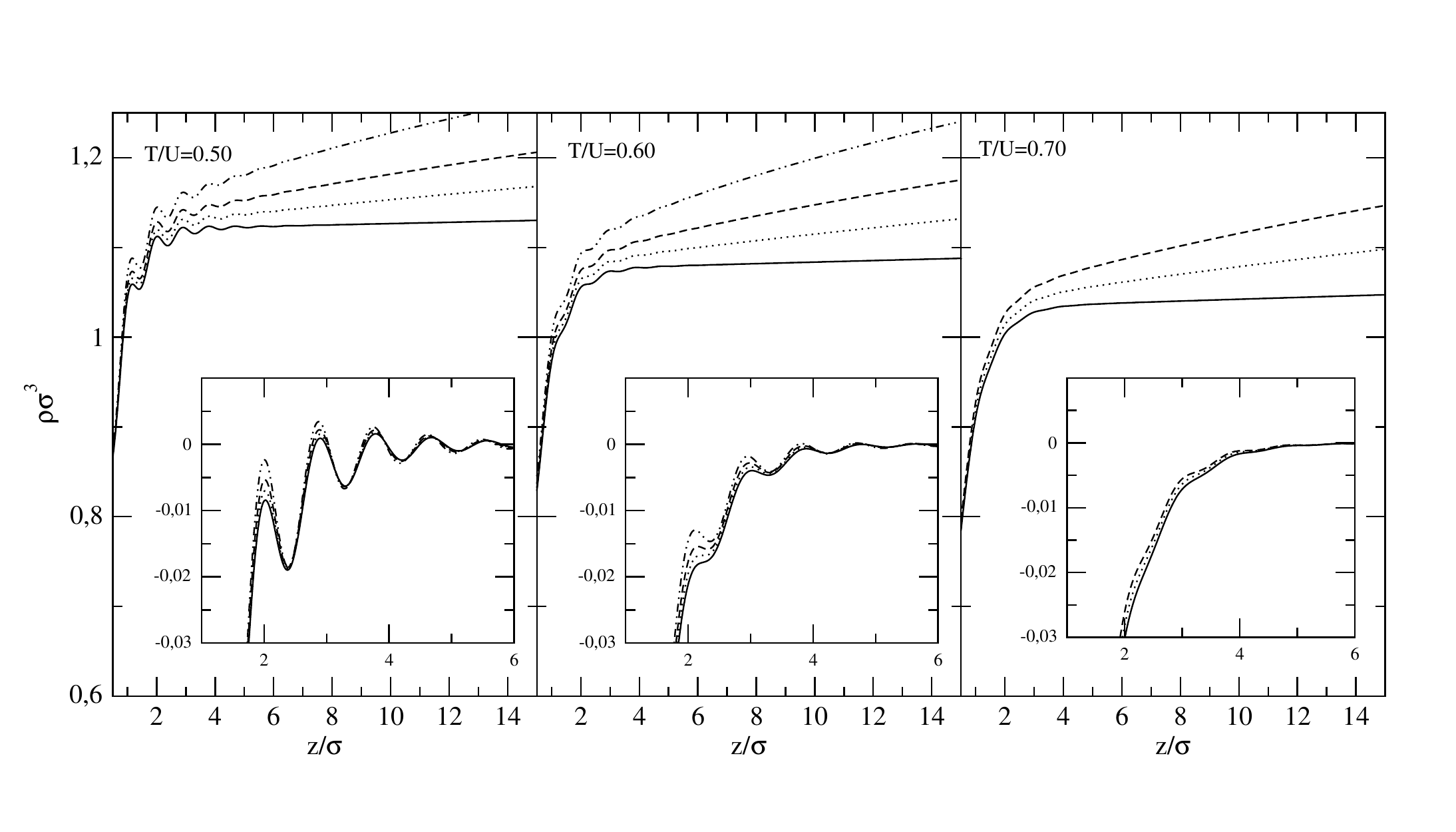}
   \caption{\textbf{Perfiles en un campo externo $V_{ext}(z)=mgz$} con cutoff en $z=60\sigma$. Aproximación funcional FMT. \textit{Potencial de Interacción Soft-Alcaline, \textit{SA}}. Valores de la intensidad: mg=0.01 (línea continua), mg=0.05 (línea de puntos), mg=0.1 (línea discontinua), mg=0.2 (línea discontinua con puntos).}
   \label{fig:PerfilesGravedadSaPY}
\end{figure}

\begin{figure}[htbp] 
  \centering
   \includegraphics[width=1.00\textwidth]{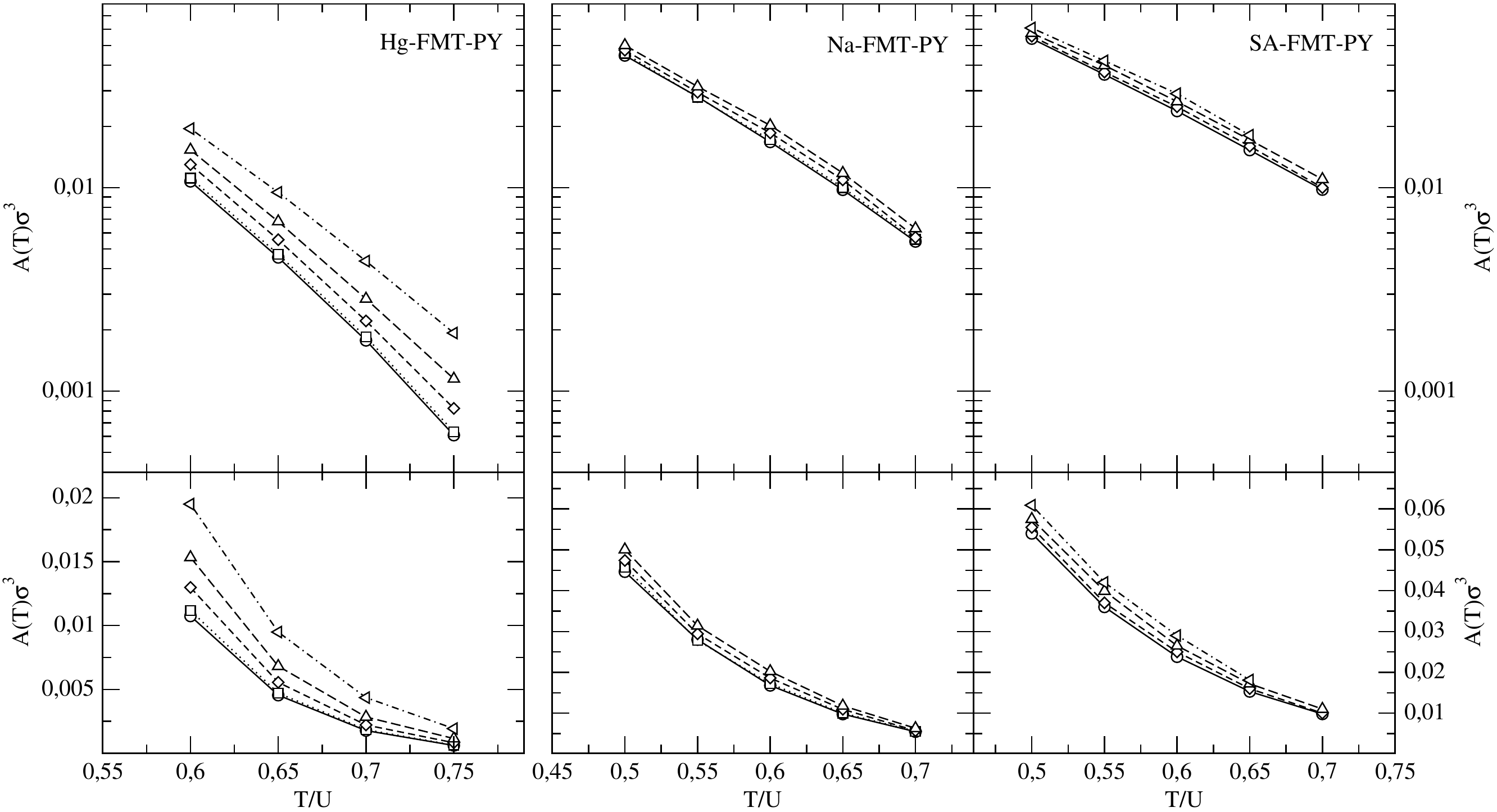}
   \caption{Valores de las Amplitudes A(T,mg) para los perfiles mostrados en el inset figuras          (\ref{fig:PerfilesGravedadHgPY}, \ref{fig:PerfilesGravedadNaPY} y \ref{fig:PerfilesGravedadSaPY}). \textbf{Arriba} escala logarítmica. \textbf{Abajo} escala lineal.}
   \label{fig:AmplitudesFMTPYmg}
\end{figure}

\begin{figure}[htbp] 
   \centering
   \includegraphics[width=1.00\textwidth]{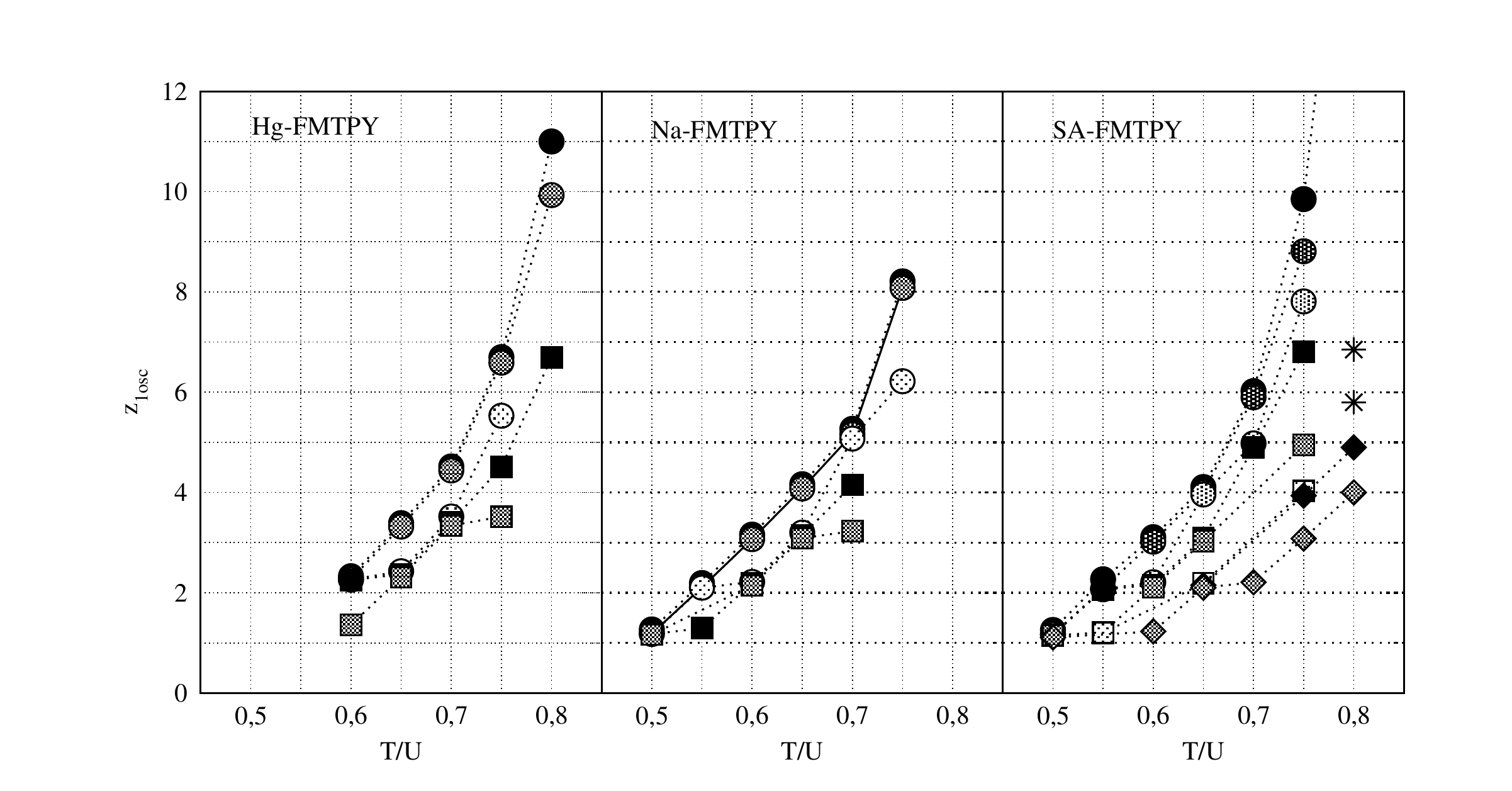}
   \caption{La diferentes curvas representan la distribución de máximos al aumentar la intensidad $mg$. círculos negros $mg=0$, círculos grises $mg=0.01$, círculos blancos $mg=0.05$. Cuadrados negros $mg=0.1$, cuadrados grises $mg=0.2$, cuadrados blancos $mg=0.3$, Diamantes negros $mg=0.4$ y Diamantes grises es $mg=0.5$.
           }
 \label{fig:posicionMaxMG}
\end{figure}

Como indicábamos la teoría de la respuesta lineal desarrollada en los capítulos anteriores y el estudio tanto de propiedades de volumen como de las ondas capilares a partir de esta no es posible para estos perfiles de densidad. Si suponemos válida la \textit{aproximación hidrostática} podemos obtener un perfil de densidad sobre el que aplicar la teoría de la respuesta lineal. Consideramos que el efecto del potencial externo es producir una variación del potencial químico que si suponemos local permite determinar $\rho_{HD}(z)$ mediante la expresión,

\begin{equation}
\mu(\rho_{HD}(z))=\mu_{B}-\phi_{ext}(z)
\label{eqn:aproximacionHD}
\end{equation}

resuelta punto a punto. Los perfiles resultantes pueden verse en los inset de las figuras (\ref{fig:PerfilesGravedadHgPY}, \ref{fig:PerfilesGravedadNaPY} y \ref{fig:PerfilesGravedadSaPY}). Ajustamos $\rho_{HD}(z)$ a la misma forma funcional que $\rho(z)$ en ausencia de campo externo para obtener los valores de las amplitudes y modos de decaimiento\footnote{Para determinar el papel de la aproximacin introducida, lo evaluamos dos tipos diferentes campos externos ambos relacionados con al situación física que tratamos, pero variando algunos parámetros incluidos en ellos.
\begin{itemize}
\item $V_{ext}(z)=-mg\lambda tanh(z/\lambda)$
\item $V_{ext}(z)=-mgz$
\end{itemize}
Para el primero jugamos con el parámetro $\lambda$ y para el segundo diferentes valores del cutoff en el líquido a partir del cual desenganchamos el campo externo. Los resultados son perfiles considerablemente similares cerca de la interfase pero ligeramente diferentes incrementando el valor de z, estas diferencias se matizan con T y mg. Nos interesa esencialmente encontrar que el método es consistente en la determinación de las amplitudes $A_{\rho}(T,mg)$. A este fin evaluamos el siguiente ajuste para el perfil obtenido mediante aprox. LDA-HD, ec.(\ref{eqn:aproximacionHD}),
\begin{equation}
\rho_{HD}(z)=A_{0}+Be^{-\tilde{\alpha}_{1} z}+Ae^{-\alpha_{1}(z-z_{0})}cos(\alpha_{0}(z-z_{0}))+Cz+Dz^2
\label{eqn:ajusteparamg}
\end{equation}
evaluándolo en diferentes ventanas $[z_{1},z_{2}]$ pero donde el intervalo contiene al menos 10 capas. De manera sistemática obtenemos que:
\begin{itemize}
\item Los valores de los parámetros libres $\tilde{\alpha}_{1},\alpha_{1},\alpha_{0},z_{0}$ de ajuste dependen solo de T, mg y z (y solo muy marginalmente de la ventana de ajuste).
\item Los parámetros $A_{0},C,D$ dependen manifiestamente de la forma de campo externo y la ventana de ajuste pero dejan entrever un perfil subyacente a ellos que es común a todos los modos de introducir el campo externo.
\item Para las amplitudes observamos que los valores de A y B dependen esencialmente de T y mg. En el caso  de B como ocurría en $mg=0$ no es estable al variar z para T baja.  La dependencia en z es marginal. La ventana de ajuste presenta relevancia en B debido a la interferencia estadística con C y D, pero no en A salvo a T/U próxima a Fisher-Widom, donde es necesario aproximarse mucho al volumen o introducir campos muy intensos.
\item Los ajustes mediante la ecuación (\ref{eqn:ajusteparamg}) son mas consistentes en la variación en z y permiten ajusta en un rango de T y mg mayor de modo fiable.
\end{itemize}
La introducción de C y D es conveniente para estabilizar los valores de A(T,mg) así como de los modos de decaimiento asintótico, observando la figura (\ref{fig:posicionMaxMG}) permiten además diferenciar los máximos en las posiciones predichas por ec. (\ref{eqn:maxdensidadlv}), véase los puntos marcados con asteriscos que muestran la posición de los máximos tras substraer $Cz-Dz^{2}$.}. Los valores de $A(T,mg)$ pueden verse en la figura (\ref{fig:AmplitudesFMTPYmg}).\\

Al igual que en el capitulo precedente analizamos la posición de los diferentes máximos al variar $(T,mg)$. Que aparece indicado en la figura (\ref{fig:posicionMaxMG}).

\subsection{Papel de las ondas capilares en las amplitudes de las oscilaciones}

Todas las propiedades de estructuración en capas de los perfiles de densidad en un campo externo pueden ser explicadas aludiendo a las dos hipótesis introducidas previamente \S\ref{sec:relacionCWyDEN}, concebidas desde la teoría de ondas capilares clásica. Además la presencia de un campo externo permite la determinación cuantitativa de las magnitudes que permiten interpretar los perfiles de densidad en ausencia de campo externo como perfiles de densidad $\rho(z,L_{eff})$.

\subsubsection{Resultados para los parámetros de la teoría de ondas capilares}

En el caso de introducir un campo externo gravitatorio podemos utilizar el resultado previo del \S\ref{Sec:Cap1OndasCapilares} para $\Delta_{cw}^{2}$ que me implicaría que,
\begin{equation}
A(L_{1},mg)=A(L_{0},mg)\left[\left(\frac{L_{0}}{L_{1}}\right)^2+\frac{(\rho_{l}-\rho_{v})mg}{(2\pi)^{2}\gamma}L_{0}^{2}\right]^{\eta(T)/2}\left[ 1+\frac{L_{0}^{2}}{(2\pi \xi_{cw})^{2}}\right]^{\eta(T)/2}
\label{eqn:basicaLAmplitudesMG}
\end{equation}
En el caso de introducir el sistema en campos intensos el factor mg antes macroscópico se convierte en relevante en la escala esperada para $L_{1}$, pudiendo evaluar los efectos del campo externo en las amplitudes, esto implica llevar la intensidad $mg\sigma$ a valores entorno de $0.1$ que es $10^{13}$ veces mayor que la gravedad terrestre.\\

La ecuación precedente involucra dos cuestiones diferentes, una primera como determinar $A(L_{1},mg)$ a partir de los perfiles de densidad y otra segunda como determinar $L_{1}$ a partir de esta expresión. 
El análisis de la segunda de ellas permite afrontar el objetivo de interpretación del perfil $\rho_{DF}(z)$ planteados en la introducción y se detalla aquí, la primera cuestión involucra la posibilidad de aplicar la teoría de la respuesta lineal a este caso incluso en campos intensos y ha sido corroborada en todo el rango de intensidades utilizadas.\\

Si nos limitamos a valores de $mg\sigma\sim 1$ el último factor de ec. (\ref{eqn:basicaLAmplitudesMG}) puede ser aproximado por 1 para un intervalo aceptable de temperaturas que coincide con el intervalo en que analizamos los resultados\footnote{Nos sera difícil ir más allá de $mg\sim0.5$ con lo que podemos llegar a un valor de $\eta\sim 5$ y la discrepancia es únicamente del orden de milésimas, en cualquier caso tenemos presente que la aproximación acota por encima a las amplitudes tanto más cuando mayor es la temperatura.}. En el caso de mg pequeño 
podemos realizar un desarrollo de Taylor de la expresión incluyendo solamente el primer corchete\footnote{Desarrollar a orden lineal la expresión completa de arriba lleva a realizar la misma aproximación y obtener el mismo comportamiento lineal.} despreciando términos de orden $(mg)^{3}$,
\begin{equation}
\frac{A(L_{1},mg)}{A(L_{0},mg)}=\left(\frac{L_{0}}{L_{1}}\right) ^{\eta}\left[ 1+\frac{(\rho_{l}-\rho_{v})L\eta}{4(2\pi)^{2}\gamma}mg+\frac{1}{8}L_{1}^{4}\left( \frac{\rho_{l}-\rho_{v}}{(2\pi)^{2}\gamma}\right)^{2}[\eta^{2}-2\eta](mg)^{2}\right]
\end{equation}
Conviene observar la aproximación lineal entorno a $\eta=2$ es de hecho cierta a orden $(mg)^{3}$ lo que podemos tener presente al evaluar los resultados. Mientras que a orden lineal\footnote{En la práctica evaluamos para $mg\lesssim0.5$ con lo que la aproximación lineal debe ser correcta la evaluación estricta de consistencia se hará tomando los valores para las magnitudes obtenidas en la aproximación lineal y comprobando el valore del siguiente término para verificar que no es de un orden mayor.} definimos un parámetro $(L\eta)_{eff}$ efectivo responsable de la variación A(T),
\begin{equation}
(L\eta)_{eff}=\left[ \frac{2\pi (\alpha_{0}^{2}+\alpha_{1}^{2})}{\beta(\rho_{l}-\rho_{v})}\frac{A(L_{1},mg)-A(L_{1},0)}{mgA(L_{1},0)}\right]^{1/2}
\label{eqn:ecLetaeff}
\end{equation}
donde por la propia definición de las amplitudes del perfil intrínseco altamente estructurado esperamos que $A(L_{0},mg)\simeq A(L_{0},0)$ en el rango de valores de mg que estamos manejando\footnote{Esta hipótesis por lo demás podría ser comparada tanto con resultados para el modelo Sodio en la zona de metaestabilidad como en el apéndice \S\ref{sec:apendiceIntrinseco} al definir los perfiles intrínsecos altamente estructurados como perfiles de equilibrio bajo un campo externo de intensidad $mg$.}. Mientras que lo relevante de esta expresión es que permite determinar dicho $(L\eta)_{eff}$ a partir únicamente de amplitudes determinadas de los perfiles obtenidos mediante DFT y en particular \textit{no} involucra hipótesis sobre $L_{1}$ salvo que en el rango que usemos consideremos válidas las suposiciones iniciales \S\ref{sec:relacionCWyDEN}.\\

\begin{figure}[htbp] 
   \centering
   \includegraphics[width=1.00\textwidth]{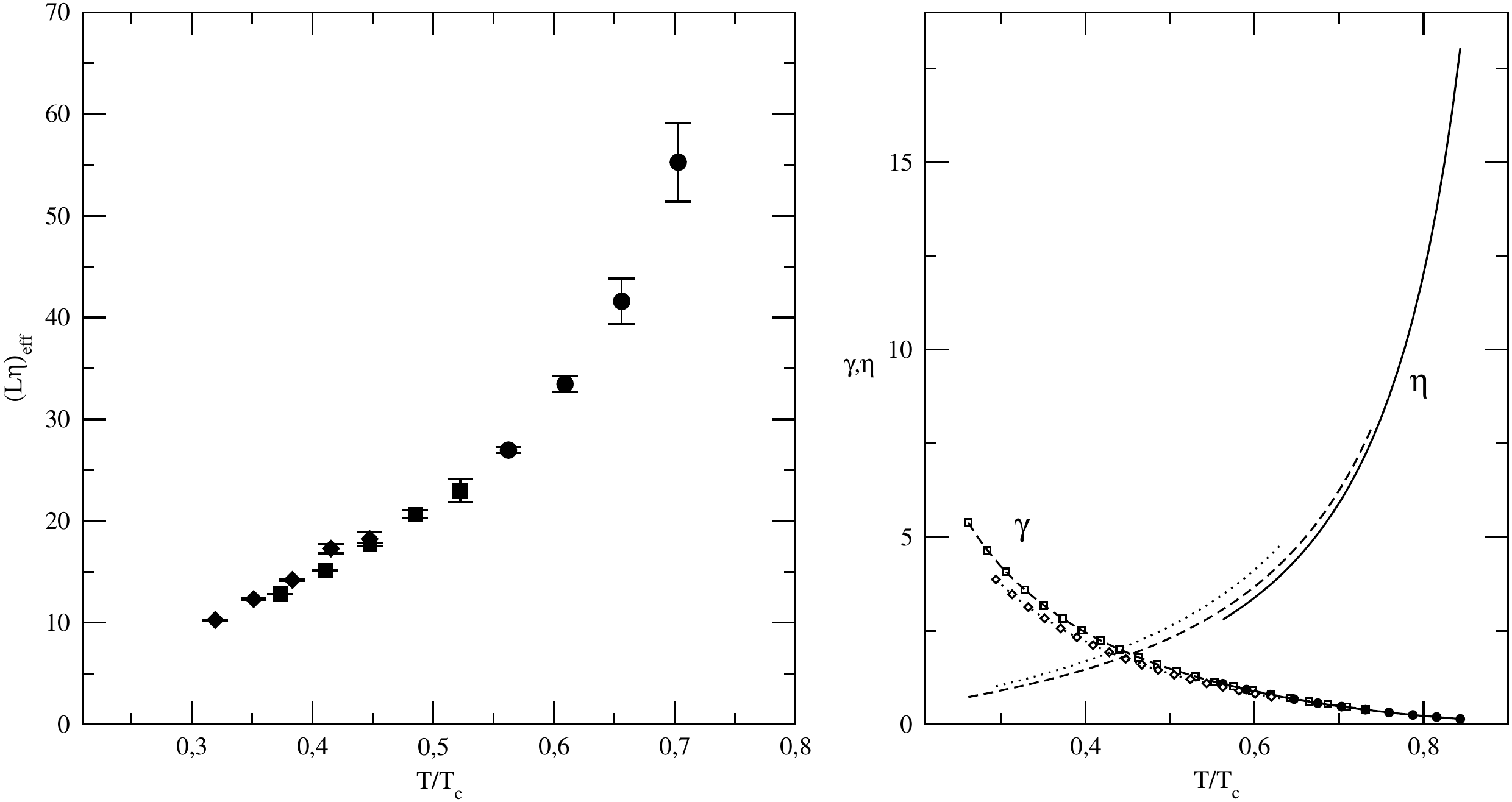}
   \caption{Valores de $(L\eta)_{eff}$ en la escala $T/T_{c}$. A la derecha curvas de $\beta\gamma_{lv}$ y $\eta$.}
   \label{fig:curvaLetaeff}
\end{figure}
\begin{figure}[htbp] 
   \centering
   \includegraphics[width=1.00\textwidth]{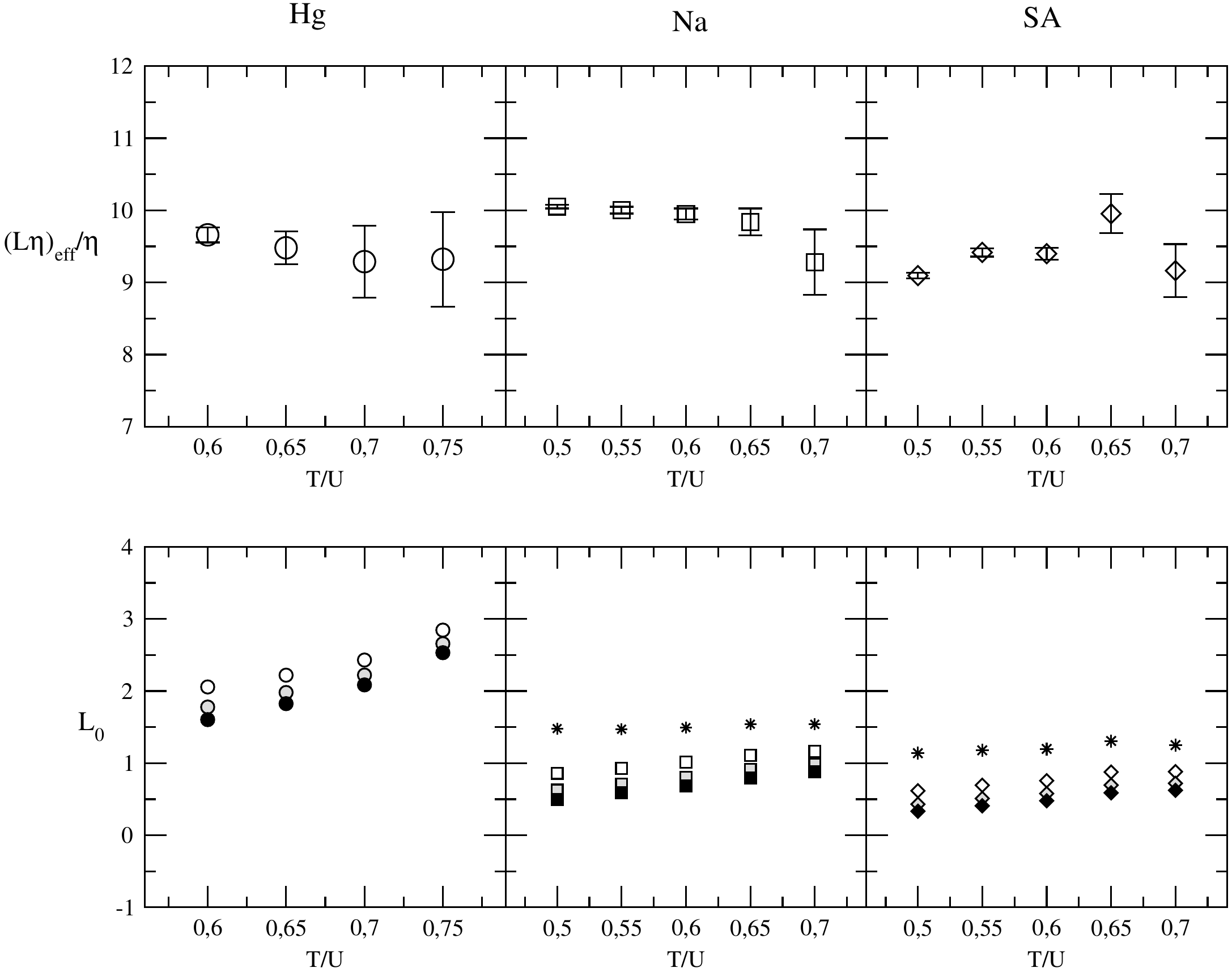}
   \caption{Valores de $(L\eta)_{eff}/\eta$ para los diferentes modelos. Se determinan los errores por la indeterminación del ajuste lineal de las amplitudes, vía ec. (\ref{eqn:ecLetaeff}). La indeterminación en la ventana de ajuste en z no esta incluida y provoca diferentes valores de dicho ajuste ya que diferentes valores de $A(T,mg)$ lleva a diferencias en ec. (\ref{eqn:ecLetaeff}). La que se presenta es una elección de ventana con unos valores del ajuste de regresión óptimos, aunque los resultados no se ven condicionados de modo relevante por esta elección. En la segunda figura representa valores de $L_{0}$ hallados mediante $L_{eff}$ de arriba y diferentes valores de $A_{0}$. \textit{Asteriscos}:  $A_{0}=0.5\rho_{l}$,\textit{Símbolos blancos}: $A_{0}=1.0\rho_{l}$, \textit{Símbolos grises}: $A_{0}=1.5\rho_{l}$, \textit{Símbolos negros}: $A_{0}=2.0\rho_{l}$}
\label{fig:metodo1Leff}
\end{figure}

La curva con los resultados para $(L\eta)_{eff}$ se incluye en la figura (\ref{fig:curvaLetaeff})  para diferentes valores de la temperatura, verificando una \textit{ley de estados correspondientes}, similar a la encontrada para $\eta$, aunque en aquel caso tenemos una desviación respecto de una perfecta curva debida a la diferencia que introducen $d_{hs}$ en la parte repulsiva del potencial, véase la misma figura (\ref{fig:curvaLetaeff}) a la derecha. Nuestro objetivo es determinar que predicciones podemos realizar suponiendo un perfil intrínseco con $L_{0}$ del orden de $\sigma$ y $A_{0}$ del orden de $\rho_{liq}$. Para ello contrastamos valores de ambos parámetros compatibles con la primera hipótesis citada en \S\ref{sec:relacionCWyDEN} mediante la expresión (\ref{eqn:AmplitudOsciConTemp}), donde dos métodos equivalentes son posibles.
\begin{itemize}
\item A partir de los valores de $\eta$ basados en los modos de la respuesta lineal y la tensión superficial sin gravedad definimos $L_{eff}\equiv(L\eta)_{eff}/\eta$. Fijando alternativamente $L_{0}$ o $A_{0}$ obtenemos mediante (\ref{eqn:AmplitudOsciConTemp}) los valores de todos los parámetros, véase figura (\ref{fig:metodo1Leff}).
\item Para cada par ($L_{0},A_{0}$) determinamos un valor de $\eta_{eff}(T)$ mediante la relación (\ref{eqn:AmplitudOsciConTemp}), y el correspondiente $L_{eff}=(L\eta)_{eff}/\eta_{eff}$.
\end{itemize}

\begin{figure}[htbp] 
   \centering
   \includegraphics[width=1.00\textwidth]{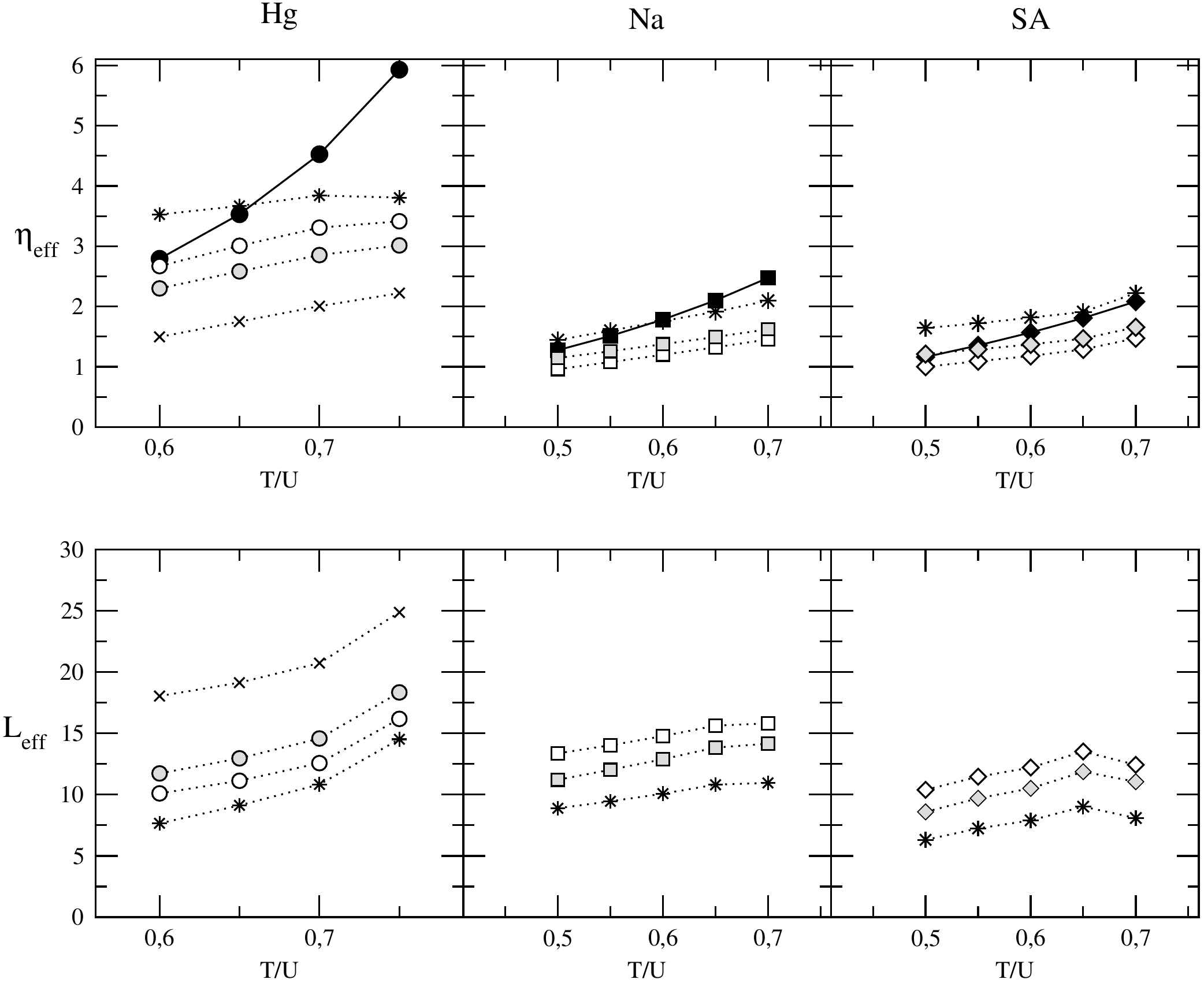}
   \caption{Valores de $\eta_{eff}$ para los diferentes modelos. Figuras negras son los valores de $\eta$. Las otras figuras representan pares $(L_{0},A_{0})$ diferentes. Para el \textit{Mercurio}: asteriscos  $(2\sigma,1.5\rho_{l})$, círculos blancos $(2\sigma,\rho_{l})$, círculos grises $(1.5\sigma,1.5\rho_{l})$, y el aspa $(1\sigma,\rho_{l})$. Para el \textit{Sodio}: asteriscos $(1\sigma,\rho_{l})$, cuadrado blanco $(0.5\sigma,\rho_{l})$, cuadrado gris $(0.5\sigma,1.5\rho_{l})$. En el \textit{Soft-Alcaline}: asteriscos $(1\sigma,\rho_{l})$, rombo blanco $(0.5\sigma,\rho_{l})$, rombo gris $(0.5\sigma,1.5\rho_{l})$  En la parte inferior estarían los valores de $L_{eff}$ correspondientes a $(L\eta)_{eff}/\eta_{eff}$.}
\label{fig:metodo2etaeff}
\end{figure}

Los resultados de ambos métodos deben ser consistentes ya que ambos se basan en ec. (\ref{eqn:AmplitudOsciConTemp}), de modo que si imponemos los valores de $L_{0}(T)$ y el $A_{0}(T)$ resultado del primer método en el segundo volvemos a recobrar los valores de $L_{eff}$ y $\eta$. Conviene pues introducir valores constantes $(L_{0},A_{0})$, como vemos en la figura (\ref{fig:metodo2etaeff}), de modo que la dependencia en T este en el par $(L_{eff},\eta_{eff})$. Los valores de $\eta_{eff}$ resultan aproximadamente constantes con un leve incremento con la temperatura, mientras que la comparación de los valores razonablemente similares entre ambos métodos sugieren valores que verifican la relación  $\eta_{eff}\lesssim\eta$ para bajas temperaturas con diferencias mayores al aumentar la temperatura\footnote{Los resultados se pueden ver en la figura (\ref{fig:metodo2etaeff}) y muestran los valores de $\eta_{eff}$ para diferentes valores de $(L_{0},A_{0})$ junto con valores de $\eta$. Vemos que los valores para los que $\eta_{eff}>\eta$ llevan a valores de $L_{eff}$ menores que los del método anterior. Valores de $\eta_{eff}<<\eta$ llevan por contra a valores de $L_{eff}$ significativamente mayores. El valor óptimo de $\eta_{eff}$ es aquel que aproxima a $\eta$ por debajo para valores de T/U pequeños.}. La definición de $\eta$ sugiere la presencia de $\gamma_{eff}(T)\gtrsim\gamma_{lv}(T)$. Esta situación se corresponde además con parámetros $L_{0}$ y $A_{0}$ en concordancia con el perfil de densidad intrínseco propuesto por los experimentales, y reflejado en la figura (\ref{fig:Refletividad3}), ya que ellos también se basan en la validez de la teoría de ondas capilares clásica. Sin embargo desde el punto de vista del perfil intrínseco real poseen un carácter a priori semi-cuantitativo por lo naive de la segunda hipótesis en \S\ref{sec:relacionCWyDEN}.\\


Los valores correspondientes al primero de los métodos aparecen reflejados en la figura (\ref{fig:metodo1Leff}) que conviene interpretar. Los resultados para los diferentes modelos parecen compatibles de modo global con un valor de $L_{1}\sim 9.5\sigma$ con dos fuentes de error posibles, una primera nacida de la propia indeterminación en el conocimiento del valor de z a partir del cual el comportamiento asintótico esta presente en un campo externo intenso, mientras que fijado un valor de este z aun los valores de las amplitudes están sujetos a cierta indeterminación a la hora de determinar un comportamiento lineal. En la parte de arriba de la figura (\ref{fig:metodo1Leff}) se muestra una curva para un z determinado y la segunda de las fuentes de error, que como vemos solo es relevante a altas temperaturas. La primera fuente de error puede interpretarse 
como la causa de la dispersión de los puntos que observamos en, por ejemplo, el \textit{Soft-Alcaline} donde la indeterminación en el momento en que el comportamiento asintótico se produce es mayor. De modo general establecemos\footnote{De modo que un grado conveniente de consistencia funcional nos ha permitido estimar un rango para $A_{0},L_{0},L_{eff}$, mientras que las limitaciones predictivas del método se presentan para $\eta>5$ para el cual el término de orden 2 en mg puede superar al término lineal incluso en campos no muy intensos.} que 
para $L_{1}\sim (9.5\pm1.5)\sigma$.\\
\begin{figure}[htp]
  \begin{center}
    \subtop[Figura extraída de la referencia \cite{PhysRevB.70.235407} ]{\label{fig:qeffPRB}\includegraphics[scale=0.80]{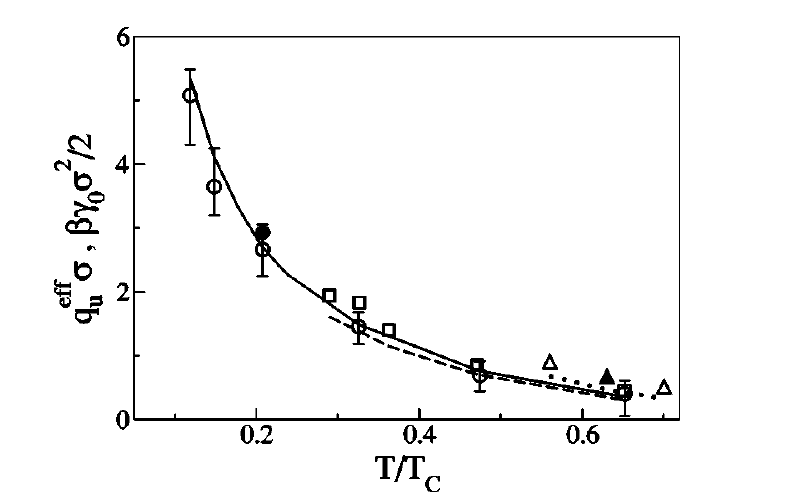}}
    \subtop[Resultados para $L_{0}$ y $1/L_{0}$]{\label{fig:leyEstadosL0}\includegraphics[scale=0.35]{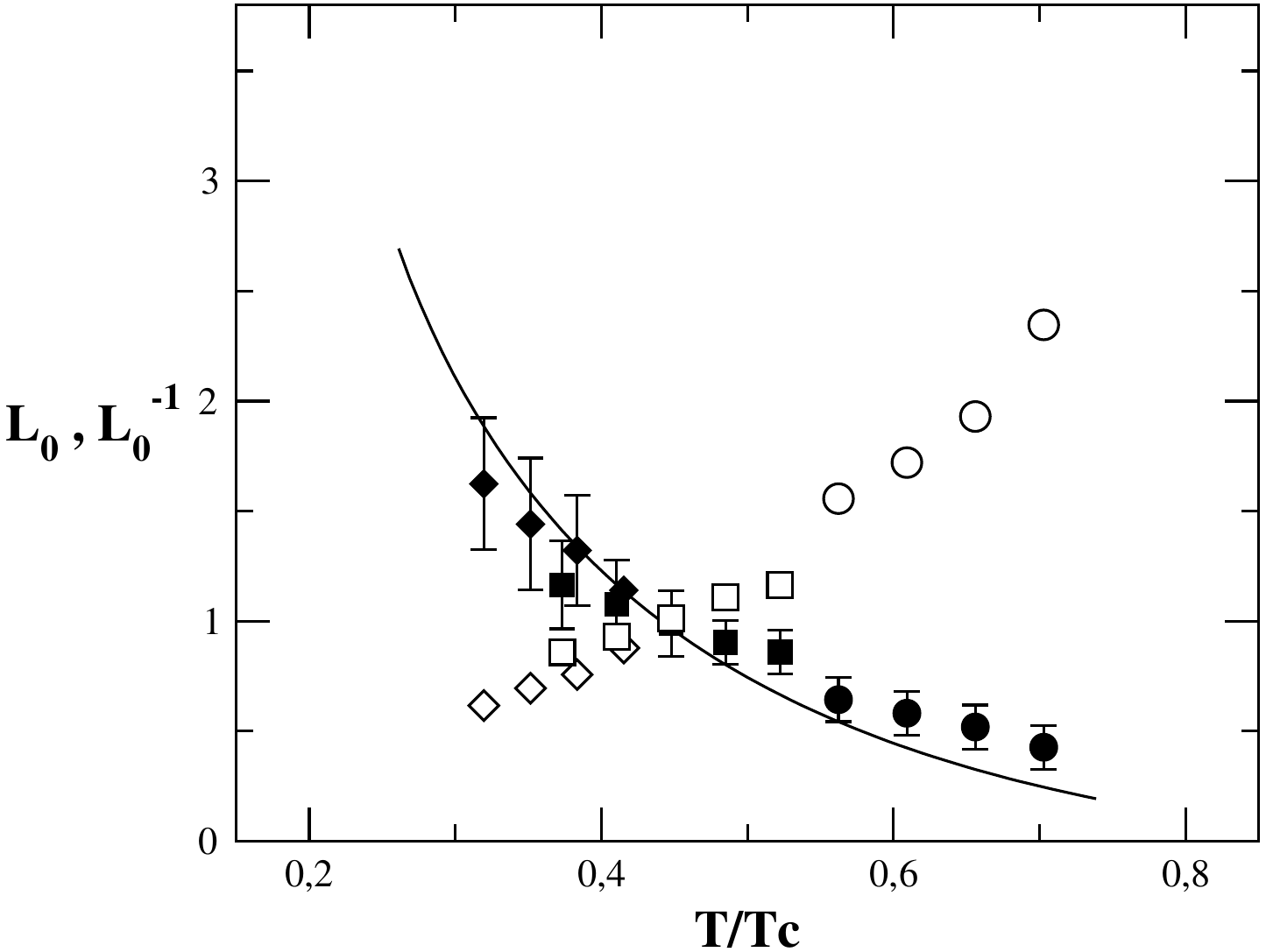}} 
  \end{center}
  \caption{\textbf{(a)}. Muestra el valor de $q_{eff}$ para el que imponiendo $\mathcal{H}_{CWT}$ se obtiene como resultado $\Delta_{CW}$ obtenido del método descrito en figura (\ref{fig:PercolativoTCh1}) desde las configuraciones moleculares. línea continua $\beta\gamma/2$ para \textit{Soft-Alcaline} y línea discontinua para Sodio.
  \textbf{(b)}. Bajo la hipótesis de $A_{0}\simeq 1.25\rho_{l}$ y $L_{eff}=(L\eta)_{eff}/\eta$ descrito en figura (\ref{fig:metodo1Leff}), se representa el resultado para $L_{0}$ (figuras blancas) y $1/L_{0}$ (figuras negras) con una indeterminación debida a $\Delta A_{0}=0.25\rho_{l}$. La línea continua es $\beta\gamma/2$ para \textit{Sodio}.}
  \label{fig:leyestadosCorrespondientesQeff}
\end{figure}

Por último podemos discutir el primer método en que determinamos $L_{0}(T)$ que permite una comparación cualitativa con simulación, ya que el parámetro $(L\eta)_{eff}/\eta\simeq L_{eff}$ es aproximadamente constante para todos los modelos y fijando un $A_{0}\simeq (1.25\pm0.25)\rho_{l}$ podemos determinar el valor de $L_{0}(T)$ a partir de la relación de amplitudes (\ref{eqn:AmplitudOsciConTemp}). Esto permite obtener  para $L_{0}^{-1}$ un comportamiento correlacionado con $\beta\gamma/2$, véase la figura (\ref{fig:leyestadosCorrespondientesQeff}).


\section{Conclusiones}

\begin{itemize}
\item Los perfiles de densidad líquido-vapor obtenidos mediante teorías de van der Waals generalizadas responden al comportamiento asintótico dado por las dos primeras soluciones de la teoría de la respuesta lineal, una oscilatoria y otra monótona. Además este comportamiento es adecuado, también cuantitativamente, para representar los perfiles de densidad desde la primera oscilación completa.

\item La imposibilidad de determinar amplitudes más all de Fisher-Widom se debe que la localización de la primera oscilación sigue la ley (\ref{eqn:maxdensidadlv}) en contraposición a (\ref{eqn:maxdensidadgder}) que permite encontrar oscilaciones más allá de la línea de Fisher-Widom para g(r). El papel de la línea de Fisher-Widom no esta en condicionar las amplitudes del comportamiento asintótico, sino que son las amplitudes de este comportamiento las que dan un papel u otro a la línea de Fisher-Widom.
\item La variación $A_{\rho}(T)$ no se presenta correlacionada con la presencia de Fisher-Widom sino debida a la presencia parcial o efectiva de ondas capilares.
\item Introduciendo campos moderadamente intensos, en el sentido en que la aproximación hidrostática pueda ser valida en un rango que incluya el comportamiento asintótico, es posible determinar de modo efectivo como están incluidas las ondas capilares bajo las hipótesis de CWT y de este modo se postula un perfil intrínseco que no incluye de ondas capilares y resulta ser fuertemente estructurado. Los perfiles de densidad líquido-vapor se corresponden entonces con perfiles caracterizados por una longitud efectiva transversal del orden de $10\sigma$ nítidamente alejada de $\xi_{B}$ en el rango de temperaturas explorado.
\item Hemos diferenciado pues dos conceptos: un perfil intrínseco fuertemente oscilante sin presencia de ondas capilares aun por determinar, y un perfil líquido-vapor, obtenido como perfil de equilibro desde teorías de van der Waals generalizadas, que incluye parcialmente el espectro de ondas capilares.
\end{itemize}

\chapter{Hamiltonianos de interfase en la teoría del funcional de la densidad}

\label{sec:capituloHeff}

\epigraph{\textit{Er hat den archimedischen Punkt gefunden, hat ihn aber gegen sich ausgenutzt, offenbat hat er ihn nur unter dieser Bedingung finden dürfen.} \\ \vspace{0.3 cm}
Encontró el punto de Arquímedes, pero lo usó contra sí mismo; parece que sólo se le permitió encontrarlo con esta condición.
}{\scshape Frank Kafka}
\vspace{0.3 cm}

Los resultados anteriores sugieren la existencia de un hamiltoniano de interfase, del que hemos extraído la información correspondiente a un $\gamma_{eff}>\gamma_{lv}$, pero no hemos determinado. Procedemos a realizar un análisis crítico de las propuestas existentes para este posible hamiltoniano de interfase y extendemos algunos resultados existentes a teorías funcionales no locales como las contenidas en las teorías de van der Waals generalizadas utilizadas hasta el momento.\\

\section{Introducción}

La teoría de ondas capilares inicialmente fundamenta su metodología (como vimos en \S\ref{Sec:Cap1OndasCapilares}) en una propuesta de carácter fenomenológico de la que obtiene un hamiltoniano descriptivo de propiedades de la interfase, $\mathcal{H}_{I}$, y realiza el tratamiento estadístico basándose en la teoría de fluctuaciones termodinámica \cite{CallenBook} a partir del factor de Boltzmann de este. Una ampliación dentro de la misma filosofía fenomenológica puede verse en el \textit{Hamiltoniano de Helfrich} y que se introducirá brevemente en el primer apartado.\\

	\begin{wrapfigure}{r}{7.1cm}
	\includegraphics[width=2.45in]{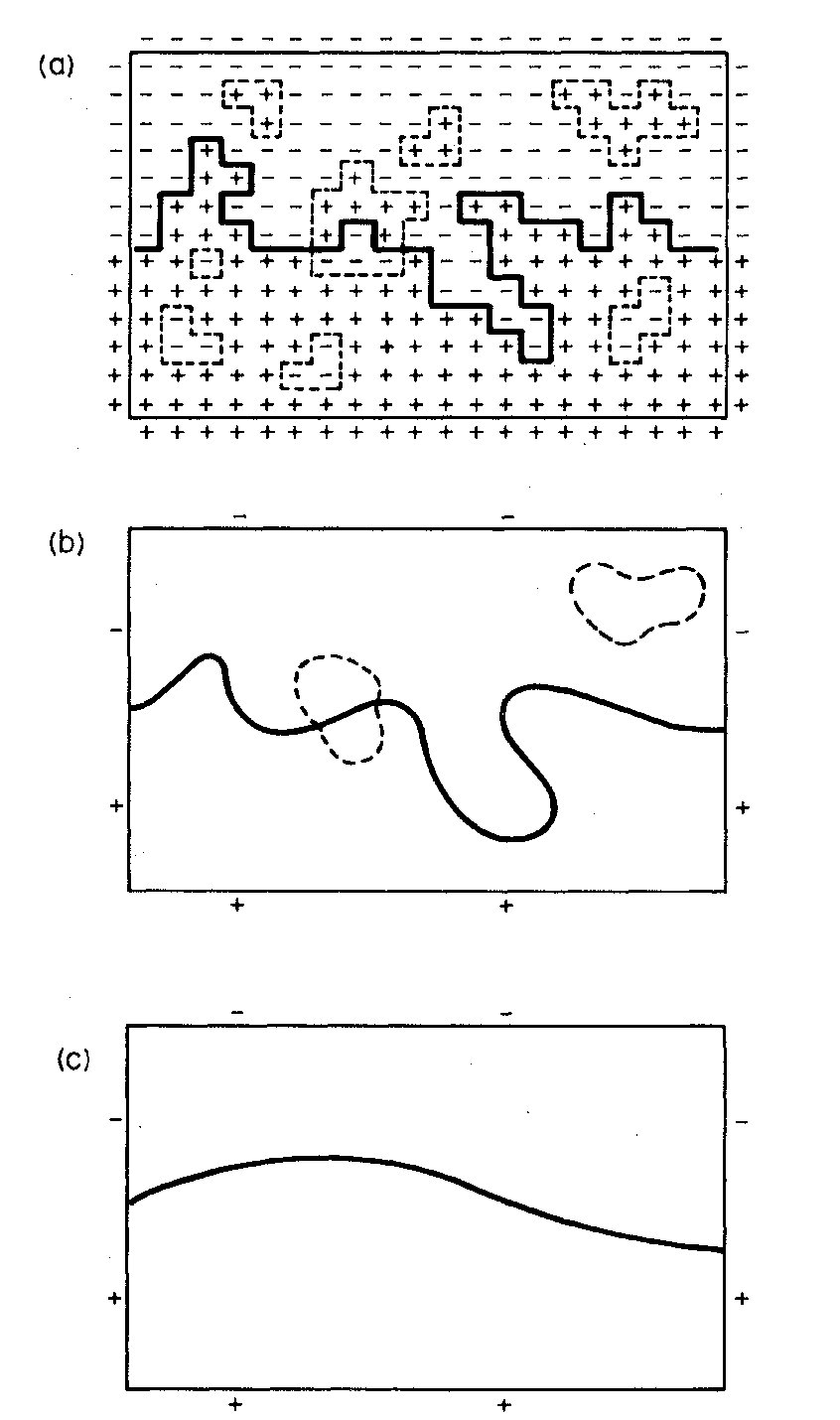}
	\caption{Esquema detalle microscópico, detalles $\mathcal{H}_{LGW}$, detalles $\mathcal{H}_{I}$}
	\label{fig:integracionenHlgwPaperWeeks}
	\end{wrapfigure}

La \textit{imagen física} compartida por ambas propuestas es una mem\-bra\-na elástica y la descripción de los grados de libertad que contiene, aun así conviene insistir que esta no es una interfase entre dos fluidos (o dos fases homogéneas) y aunque algunas propiedades sean análogas no constituye una forma adecuada de construir un modelo \textit{completo}. Permite, no obstante, discutir pro\-pie\-da\-des emergentes o\-ri\-gi\-na\-das en los grados de libertad propios de la interfase a un nivel macroscópico pero han de sugerirse propuestas que busquen en una fun\-da\-men\-ta\-ción estadística más firme, esencialmente recuperaremos dos de ellas.\\

La primera determina un $\mathcal{H}_{I}$ desde Hamiltonianos de Landau-Ginzburg-Wilson donde \textit{a priori} integrando todos los grados de libertad excepto los correspondientes a la interfase podemos obtener un posible hamiltoniano efectivo para esta. El resultado suele ser un ha\-mil\-to\-nia\-no donde aparecen diferentes invariantes geométricos relacionados con las diferentes curvaturas de la superficie y que finalmente se intenta encajar en una expresión análoga al hamiltoniano de Helfrich para reinterpretar los coeficientes fenomenológicos a partir de propiedades con base microscópica. Este camino se ha explorado ampliamente en el contexto del problema del \textit{wetting} donde la superficie es considerada como \textit{sujeta} a un sustrato en lugar del problema que estudiamos de interfases \textit{libres} aun así determinados aspectos del formalismo obtenido en un contexto es trasladable al otro.\\

Finalmente otro camino parte de teorías propias de sistemas no homogéneos pero con una base directamente microscópica como la teoría del funcional de la densidad que presenta la ventaja ya de partida de no restringirse necesariamente a una expresión en gradientes de lo que en el caso anterior de $\mathcal{H}_{LGW}$ sería el parámetro de orden. En este contexto ya se han comentado algunos resultados rigurosos \S\ref{sec:Wherteim} y de hecho extensiones al modelo de ondas capilares desde DFT permiten también relacionar los parámetros fenomenológicos con propiedades microscópicas contenidas en la función de correlación del sistema no homogéneo \S\ref{sec:RRVR}.\\

Mayor alcance poseen aquellas propuestas que intentan encontrar un $\mathcal{H}_{I}$ aplicable a todas las escalas de longitud ya que permitiría responder a las preguntas planteadas en la introducción \S\ref{sec:introduccion} de un modo riguroso dentro de DFT. Un imagen visual se ha intentado en la figura anexa, en ella se parte (a) de una configuración a nivel microscópico,  después (b) vemos las teorías basadas en $\mathcal{H}_{LGW}$ parten de un nivel de escala determinado por $\xi_{B}$ y el promedio correspondiente sobre longitudes menores puede perder características microscópicas de la interfase, la última escala (c) es la de acceso fenomenológico donde la integración progresiva de grados de libertad deja solo aspectos de la interfase relevantes macroscópicamente. En este capítulo se realiza un resumen de las diferentes aportaciones a la construcción de $\mathcal{H}_{I}$ en particular se analizarán críticamente los resultados más extendidos acerca de la determinación de Hamiltonianos de interfase efectivos dentro de la teoría del funcional de la densidad.\\
\section{Modelos fenomenológicos}
Partimos de una superficie definida por $\xi(\vec{R})$, para fijar ideas podemos imaginar una membrana elástica (sus componentes microscópicos en este nivel nos resultan indiferentes), y nos preguntamos esencialmente por la energía superficial construida como un funcional de $\xi$. En ausencia de un campo externo esperamos que:
\begin{equation}
\mathcal{H}[\xi(\vec{r}+\vec{a})]=\mathcal{H}[\xi(\vec{r})]=\mathcal{H}[\xi(\mathcal{R}\vec{r})]
\end{equation}

el vector $\vec{a}$ define una translación y la matriz $\mathcal{R}$ una rotación, expresando su invarianza en el conjunto de transformaciones euclídeas y donde $\vec{r}$ es un vector d-dimensional que define en el espacio la superficie $\xi$ (d-1) dimensional\footnote{Y puede ser determinado paramétricamente mediante d-1 componentes superficiales.}. En cambio transformaciones que deformen la superficie si son relevantes y su costo en energía libre puede ser establecido a partir de propiedades locales de la superficie que pueden ser caracterizadas mediante la \textit{curvatura media} y la \textit{curvatura gaussiana}\footnote{Por introducir intuitivamente estas cantidades, dada una superficie $\xi(\vec{R})$, para cada punto en ella puedo trazar diferentes curvas que pasan por él y están contenidas en $\xi$. Cada una de estas ellas posee una curvatura, que responde al concepto intuitivo de curva más o menos pronunciada, las curvaturas que poseen valores maximal y minimal de este conjunto son las \textit{curvaturas principales}. Con ellas podemos definir tanto la curvatura gaussiana que resulta ser el producto de ambas como la curvatura media que es la media aritmética. La curvatura gaussiana se puede definir solo a partir de distancias medidas en la superficie y por tanto es intrínseca a ella, no así la curvatura media. En cuanto al significado de la primera, su signo positivo indica una superficie localmente convexa y su signo negativo indica una superficie tipo punto de silla. Si dos superficies poseen idéntica curvatura gaussiana podemos decir que localmente podemos transformar una en la otra únicamente \textit{curvando} lo suficiente una de ellas pero sin estirarla, el costo de energía sería entonces cero excepto por la presencia de la curvatura media que si podría ser diferente de cero. Por esta razón se suele llamar al coeficiente que acompaña a K y H, coeficientes de rigidez ya que determina la resistencia de la superficie a ser curvada sin ser estirada.}, respectivamente denotadas por H y K, y que obedecen a las expresiones\footnote{Denotamos $\xi_{x}=\partial_{x}\xi(\vec{R})$ y $\xi_{xy}=\partial_{x}\partial_{y}\xi(\vec{R})$.}:
\begin{equation}
H(\vec{R})=\frac{1}{2g^{2/3}}\left( \xi_{xx}(1+\xi_{y}^{2})+\xi_{yy}(1+\xi_{x}^{2})-2\xi_{xy}\xi_{x}\xi_{y} \right) 
\end{equation}
\begin{equation}
K(\vec{R})=\frac{1}{2g^{2}}\left( \xi_{xx}\xi_{yy}-\xi_{xy}\xi_{yx} \right) 
\end{equation}
Donde g es una métrica definida por $g=(1+\xi_{x}^{2}+\xi_{y}^{2})$. Imponiendo una \textit{forma cuadrática} en las curvaturas se suele expresar:
\begin{equation}
\mathcal{H}[\xi(\vec{R})]=\int d^{2}\vec{R}\left[ 2\kappa(H-c_{0})^{2}+\bar{\kappa}K+\gamma \right] 
\end{equation}
donde $c_{0}$ se introduce para definir una posible \textit{curvatura media espontánea}. Se ha incluido la tensión superficial que penaliza incrementos globales del área superficial y se denominan rigidez  a $\kappa$ y rigidez gaussiana a $\bar{\kappa}$. Para determinar la superficie de equilibrio se realiza mediante el planteamiento usual como problema variacional de modo que la configuración considerada de equilibrio minimiza el funcional definido anteriormente.\\
Sobre esta configuración extremal aparecen fluctuaciones cuyo estudio es relevante para multitud de fenómenos físicos. El estudio de las fluctuaciones de la superficie y las propiedades que esto conlleva suele plantearse a partir de un desarrollo en gradientes de la superficie respecto de la que hemos determinado como la configuración de equilibrio anteriormente.\\

Si la configuración extremal es el caso plano el planteamiento anterior lleva a una expresión extendida respecto del caso de la teoría de ondas capilares, ec.(\ref{eqn:trabajominCWT}). Aparece un término relacionado con la curvatura media dependiente de $(\nabla^{2}\xi)^{2}$ que implica en el desarrollo Fourier un término $q^{4}$ que adquiere la forma:
 \begin{equation}
F[\lbrace \xi_{q}\rbrace ]=\frac{1}{2L^{2}}\sum_{q}\left[\gamma q^{2}+\kappa q^{4} \right]  |\xi_{q}|^{2}=\frac{1}{2L^{2}}\sum_{q}\gamma q^{2}\left[1+\frac{\kappa}{\gamma} q^{2} \right]  |\xi_{q}|^{2}
\label{eqn:extensionFenomeCWT}
\end{equation}
Este modelo ha sido aplicado con éxito al estudio de membranas biológicas\cite{PhysRevLett.59.2486,PhysRevA.39.5280,Deuling1976861}.\\

Para las superficies líquidas, la expresión (\ref{eqn:extensionFenomeCWT}) es útil al igual que en la teoría de ondas capilares clásica ya que permite la determinación de propiedades estadísticas como la anchura cuadrática media\footnote{En este caso las conclusiones son matizadas por el papel jugado por $\kappa$.}. Por otra parte definiendo $\gamma(q)/\gamma\equiv(1+\kappa q^{2})$ tenemos una tensión superficial dependiente de q, que en el caso en que $\kappa$ sea positiva es una función monótonamente creciente.

\section{Hamiltoniano efectivo de interfase basado en $\mathcal{H}_{LGW}$}
\label{sec:HIsobreLGW}

Esencialmente parten de un nivel de descripción basado en un parámetro de orden que podemos denominar $m(\vec{r})$, de esta manera la descripción del sistema aparece regida por un hamiltoniano $\mathcal{H}_{LGW}$ 
y por tanto desde el inicio partimos de una descripción \textit{mesoscópica} que podemos condensar en la expresión\footnote{En el caso del problema del \textit{wetting} se complementa con un término extra que da cuenta de la interacción entre el sustrato y la fase de \textit{wetting}.},
 \begin{equation}
\mathcal{H}_{LGW}=\int dzd\vec{R}\left[ \frac{1}{2}(\nabla m(z,\vec{R}))^{2}+\Phi(m)\right] 
\label{eqn:LGWcap6}
\end{equation}
donde $\Phi(m)$ describe la densidad de energía libre de volumen fuera de la región crítica.\\

La construcción desde $\mathcal{H}_{LGW}$ de $\mathcal{H}_{I}$ se basa en la introducción de una \textit{variable colectiva} $\xi(\vec{R})$ bajo la cual restringimos los estados accesibles al sistema completo, descritos por $m(\vec{r})$, al subconjunto de estos que reproduce $\xi(\vec{R})$ bajo algún criterio. La definición formal queda como,
\begin{equation}
e^{-\beta\mathcal{H}_{I}[\xi(\vec{R})]}=\underset{\{m(\vec{r})\}}{Tr}^{\xi(\vec{R})}[e^{-\beta\mathcal{H}_{LGW}[m(\vec{r})]}]
\label{eqn:defHeff}
\end{equation}
donde hemos querido significar la traza definida es sobre $m(\vec{r})$ atada a los parámetros de orden que reproduzcan $\xi(\vec{R})$. La traza completa sobre $m(\vec{r})$ puede expresarse mediante,
\begin{equation}
\underset{\{m\}}{Tr}=\underset{\{\xi\}}{Tr}\underset{\{m\}}{Tr}^{\xi}
\end{equation}

El problema nace de que resolver la expresión (\ref{eqn:defHeff})
 es tan difícil como resolver el problema general y por tanto su dificultad analítica implica la necesidad de realizar algún tipo de aproximación. En este punto \textit{Fischer-Jin}\cite{PhysRevB.47.7365,PhysRevLett.69.792,PhysRevB.44.1430} supusieron que fijada $\xi(\vec{R})$ el resto de fluctuaciones son irrelevantes al situarse alejados del punto crítico de la transición líquido-vapor y consideraron válido,
\begin{eqnarray}
\mathcal{H}_{I}[\xi(\vec{R})]\simeq \mathcal{H}_{LGW}[m_{\Xi}(\vec{r};\xi(\vec{R}))]=\nonumber\\ =min_{m(\vec{R})}^{\xi(\vec{R})}{\mathcal{H}_{LGW}[m(\vec{R})]}
\label{eqn:defHIdesdeLGW}
\end{eqnarray}
$m_{\Xi}(\vec{r};\xi(\vec{R}))$ es el parámetro de orden que hace mínimo el hamiltoniano dado por ec. (\ref{eqn:LGWcap6}) bajo la ligadura de reproducir $\xi(\vec{R})$.

Además de la validez de esta aproximación resta encontrar un modo efectivo de incluir la restricción de la traza. El criterio más utilizado se denomina \textit{crossing criterion}\footnote{No es el único utilizado y ha habido en el contexto de \textit{wetting} otras propuestas diferentes esencialmente \textit{integral constrains} pero estas se fundamentan igualmente en una $\xi(\vec{R})$ basada en el \textit{crossing criterion} mientras que en el esquema de obtención de $\mathcal{H}_{I}$ se utiliza una versión integral que suaviza los perfiles resultantes y condiciona no tanto la forma funcional como los parámetros incluidos en la definición de $\mathcal{H}_{I}$}, y consiste en  proponer,
\begin{equation}
m[z=\xi(\vec{R}),\vec{R}]=m_{0}\quad\forall\vec{R}
\label{eqn:crossingCriteria}
\end{equation}
donde $m_{0}$ es un valor de referencia que esta comprendido entre los valores de volumen de las dos fases\footnote{En modelos de Ising suele tomarse $m_{0}=0$ en sistemas líquidos $\rho_{0}=\frac{\rho_{l}+\rho_{v}}{2}$ pero no tiene porque restringirse a estos valores concretos, de hecho muchos de los resultados no dependen explícitamente de esta elección y su arbitrariedad es manifiesta al nivel del hamiltoniano efectivo aunque requiere matizaciones al nivel de la funciones de correlación del parámetro de orden $m(\vec{r})$.}. Bajo estas premisas se suelen determinar la forma concreta de los hamiltonianos $\mathcal{H}_{I}[\xi(\vec{R})]$. Y si realizamos una minimización de (\ref{eqn:defHIdesdeLGW})
en ${\xi(\vec{R})}$ permite encontrar el perfil óptimo\footnote{Una primera evaluación del hamiltoniano efectivo de interfase puede ser entonces determinada a partir de un método perturbativo que desde $\xi(\vec{R})\sim\xi_{\pi}+\delta\xi(\vec{R})$ lo relacione con variaciones del parámetro de orden.} que poseerá una interfase plana $\xi_{\pi}$.\\

Es importante notar que el método se basa en que congelada la configuración de la interfase las fluctuaciones restantes son únicamente de volumen correspondientes a las dos fases en cuestión, por tanto el método es consistente únicamente si la condición por la cual restringimos la traza permite realmente determinar unívocamente que estados corresponden a que configuración de la interfase.\\

Siguiendo a \textit{Rejmer-Napiórkowski}\cite{PhysRevE.53.881} este procedimiento aplicado a una interfase libre permite reescribir de nuevo el Hamiltoniano de Helfrich que expresan como:
\begin{equation}
\mathcal{H}_{I}[\xi]=\int d^{2}\vec{R} h(\vec{R},[\xi])
\end{equation}

donde suponemos reescrito $h(\vec{R},[\xi])$ en función de una serie de parámetros que equivalen a los parámetros fenomenológicos del modelo anterior, es decir, nacen de identificar las curvaturas gaussiana y media a partir de la expresión funcional de $\mathcal{H}_{I}$. El resultado final queda reescrito como,
\begin{equation}
h(\vec{R},[\xi])=\Delta A[\xi]\gamma+A[\xi](c_{H}H+c_{K}K+c_{H^{2}}H^{2}+...)
\label{eqn:Helfrich}
\end{equation}
obviamente hemos partido inicialmente de una descripción no completa del sistema dada por $\mathcal{H}_{LGW}$ y una aproximación fijada para $\Phi(m)$, que usualmente consiste en una doble parábola. Las expresiones para $\gamma,c_{H},c_{H^{2}}$ estarán relacionadas con las propiedades de este $\Phi(m)$ de partida\footnote{$c_{K}$ permanece arbitrario.} que recordemos es de carácter mesoscópico. Como indicábamos al principio sería deseable poder extender este tipo de procedimientos a teorías con una base directamente microscópica como pueda ser la teoría del funcional de la densidad. Aportaciones en esta línea son descritas en la siguientes secciones y como veremos las ideas contenidas en el análisis de \textit{Rejmer-Napiórkowski} son las que motivan la aportación de \textit{Napiórkowski-Dietrich} (NaD).

\section{Hamiltonianos efectivos de interfase dentro del funcional de la densidad}

Anteriormente ha sido comentada la aplicación de la teoría del funcional de la densidad, a la interfase fluidas encontrando algunas relaciones y propiedades, véase \S\ref{sec:LVenDFT} y \S\ref{sec:RRVR}. Un análisis más detallado ha sido realizado por \textit{Napiórkowski-Dietrich}\cite{PhysRevE.47.1836} y \textit{Mecke-Dietrich}\cite{PhysRevE.59.6766,hiester:184701,mecke:204723} en la determinación desde la teoría del funcional de la densidad de una expresión para $\mathcal{H}_{I}$ basado en una definición semejante a la utilizada en la teoría que surge desde hamiltonianos de \textit{Landau-Ginzburg-Wilson}.\\

\subsection{Propuesta de Napiórkowski-Dietrich}
Parten de una definición de hamiltoniano efectivo como en la sección anterior que señalan no restringido al incrementos en el área sino extendido a la energía libre en llevar una interfase plana a una interfase rugosa dada por una función $\xi(\vec{R})$. Esencialmente definen,
\begin{equation}
\mathcal{H}_{NaD}[\xi]\equiv\Omega[\rho_{int}(z-\xi(\vec{R}))]-\Omega[\rho_{int}(z)]
\label{eqn:HeffdeNapiorkowski-Dietrich}
\end{equation}
mientras que su propuesta para $\rho_{int}$ viene determinada por:
\begin{equation}
\rho_{int}(z)=\rho_{g}\Theta(z)+\rho_{l}\Theta(-z)
\label{eqn:sharp-kinkmodel}
\end{equation}
la primera ecuación parte de la hipótesis de la teoría de ondas capilares clásica\footnote{La primera aportación de \textit{Napiórkowski-Dietrich} es previa a los estudios de \textit{Fischer-Jin} en que desarrollan extensamente el \textit{crossing criterion}. La hipótesis $\rho(z-\xi(\vec{R}))$ fue analizada en \S\ref{sec:RRVR}. Esencialmente suponemos que la deformación de $\xi(\vec{R})$ no condiciona a $\rho(z)$, es decir, repetimos la hipótesis clásica de CWT.} y supone un desplazamiento rígido del perfil intrínseco siguiendo la superficie intrínseca. Para el perfil de densidad toman la versión más sencilla posible. La identificación correspondiente a $\mathcal{H}_{I}[\xi]$ esta dada por ($\vec{r}=(z,\vec{R})$, $\vec{s}=(z',\vec{S})$, y $\Delta\rho=\rho_{l}-\rho_{v}$),
\begin{equation}
h_{NaD}(\vec{S},[\xi])=-\frac{1}{2}\Delta\rho\int d^{2}\vec{R}\int dz\int_{0}^{\xi(\vec{R})-\xi(\vec{S})} dz' c^{(2)}(|\vec{r}-\vec{s}|)
\label{eqn:Napiorkowski-Dietrich}
\end{equation}
aunque en la referencia original parten de un funcional en aproximación local para esferas duras y campo medio para la cola atractiva con lo que aparecía la función $\omega_{at}$ en lugar de la función de correlación directa\cite{1996JPCMDietrich} $c^{(2)}$. Si el sistema esta inmerso en un campo externo gravitatorio aparecerá un término extra dado por $[\xi(\vec{R})]^{2}$. Conviene resumir las propiedades que este hamiltoniano efectivo posee:
\begin{itemize}
\item Las necesarias $\mathcal{H}_{I}[\xi(\vec{R})+C]=\mathcal{H}_{I}[\xi(\vec{R})]\quad;\quad\mathcal{H}_{I}[C]=0$ que expresan que no cuesta energía libre mover la interfase \textit{globalmente} y que en el caso plano esta energía es cero.
\item Si $\xi(\vec{R})$ es una función suave es posible desarrollar la diferencia $\xi(\vec{R})-\xi(\vec{R'})$ en serie Taylor. A orden lineal se recupera la forma del hamiltoniano \textit{drumhead}, véase la nota (\ref{sec:drumhead}) en \S\ref{Sec:Cap1OndasCapilares} y la referencia \cite{PhysRevB.32.233}, mientras que a orden cuadrático permite reescribir el hamiltoniano como la extensión fenomenológica de Helfrich, donde los coeficientes $\kappa$ y $\gamma$ aparecen como momentos de $c^{(2)}$.
\item El desarrollo anterior no es analítico para potenciales diferentes de los de corto alcance. En particular fuerzas de dispersión llevan a $\kappa$ divergente mientras que efectos de retardo pueden hacerlo finito pero fuertemente dependiente de momento en que la interacción $r^{-7}$ aparezca en sustitución de $r^{-6}$. Este tipo de divergencia la encontrábamos también en otros desarrollos en gradientes del funcional de la densidad \S\ref{sec:GradCuadrado} y por esta razón no lo consideran un defecto de su propuesta.
\item Es no local, y no gaussiano aunque puede ser aproximado de modo que encaje en ec. (\ref{eqn:hamiltonianoGaussiano}).

\end{itemize}

Más allá de las predicciones de este $\mathcal{H}_{I}$ que serán analizadas más adelante hace uso de dos aproximaciones que, \textit{desde su interpretación}, sería deseable eliminar,
\begin{itemize}
\item La primera es partir de un perfil de densidad intrínseco completamente abrupto en lugar de un perfil de densidad que interpole\footnote{Mejorar esta aproximación en este sentido se basa en la interpretación del perfil de densidad en las teorías funcionales como un perfil intrínseco, en la imagen física del perfil intrínseco fuertemente estructurado sugerida en \S\ref{sec:relacionCWyDEN} la aproximación del perfil intrínseco como una función paso es igualmente equivocada pero su mejora habría de realizarse en otra dirección diferente.} de algún modo entre $\rho_{l}$ y $\rho_{v}$.
\item La segunda supone que las propias características de $\xi(\vec{R})$ no condicionan la forma de $\rho_{int}$ y supone (como en la teoría de ondas capilares clásica \S\ref{Sec:Cap1OndasCapilares} en que se cumple ec. (\ref{eqn:perturbacionTipoCWT}) en todas las escalas) que ambos están descorrelacionados. 
\end{itemize}

Estas dos limitaciones intentan ser mejoradas en la propuesta de \textit{Mecke-Dietrich}(MeD) que desarrollan un modo más elaborado de construir un $\mathcal{H}_{I}(\vec{R})$, que denominamos $\mathcal{H}_{MeD}(\vec{R})$.\\

\subsection{Propuesta de Mecke-Dietrich}

Esencialmente incorporan una condición similar a la que encontrábamos en el análisis de $\mathcal{H}_{LGW}$ para restringir a microestados compatibles con una configuración de la interfase dada bajo una aproximación funcional análoga a la usada por \textit{Napiórkowski-Dietrich}. La analogía se establece por un lado entre el parámetro de orden $m(\vec{r})$ que veíamos en \S\ref{sec:HIsobreLGW} y el perfil de densidad $\rho(\vec{r})$ considerado como intrínseco y por otro entre el $\mathcal{H}_{LGW}[m(\vec{r})]$ y la energía libre $\Omega[\rho(\vec{r})]$. Vamos a analizarlo en detalle.\\

El perfil de densidad intrínseco dado para una configuración $\xi(\vec{R})$ se considera determinado por el perfil de densidad que \textit{minimiza} $\Omega[\rho(\vec{r})]$ bajo la ligadura 
$\rho(\vec{r}=(z,\vec{R}))=\rho(z=\xi(\vec{R}),\vec{R})\equiv\bar{\rho}$, por tanto escribimos dicho mínimo por, véase ec. (\ref{eqn:crossingCriteria}),
\begin{equation}
\rho_{\xi}(\vec{r})\equiv\rho_{int}(\vec{r};{\xi(\vec{R})},\bar{\rho})
\end{equation}
mientras que en el caso de una superficie plana la dependencia es solamente en la coordenada z y escribimos,
\begin{equation}
\rho_{0}(z)\equiv\rho_{int}(\vec{r};{\xi(\vec{R})=0},\bar{\rho})
\end{equation}
que es determinado mediante Euler-Lagrange con la condición $\rho_{0}(0)=\bar{\rho}$. Esto supone ya de partida que el perfil de densidad usual obtenido en las diferentes aproximaciones funcionales $\rho_{DF}(z)$ se corresponde con el \textit{perfil intrínseco} asociado a una interfase plana. En estas condiciones definimos el hamiltoniano efectivo de la interfase como,
\begin{equation}
\mathcal{H}_{MeD}[\xi(\vec{R})]=\Omega[\rho_{\xi}(\vec{r})]-\Omega[\rho_{0}(z)]
\label{eqn:HeffdeMecke-Dietrich}
\end{equation}
Conviene resaltar que el perfil de densidad $\rho_{\xi}(\vec{r})$ continua siendo desconocido, en particular su dependencia explícita en $\xi(\vec{R})$. Suponen que el proceso de minimización restringida  da como resultado un perfil que localmente se similar al perfil intrínseco para una configuración superficial plana siempre y cuando sea medido en la dirección normal a la variedad $\xi_{R}$.  Esta hipótesis permite aproximar el perfil desconocido $\rho_{\xi}(\vec{r})\simeq\rho_{\xi}(u)$, donde la variable $u$ aparece definida como la distancia mínima a la interfase. El segundo paso consiste en proponer un desarrollo del perfil de densidad $\rho_{\xi}(u)$ en una serie basada en las curvaturas de la interfase\footnote{Asume que la densidad en la proximidad de la interfase depende de $u$ así como de la geometría de la interfase y esta se supone descrita por las dos curvaturas ya introducidas, la curvatura media de la interfase $H(\vec{r})$ y la curvatura gaussiana $K(\vec{r})$. Ambas contribuciones se separan en un desarrollo en potencias de las curvaturas que por orden serán $\nu=H,K,H^{2},H^{3},HK...$ y la dependencia en $u$ aparece únicamente en los coeficientes $\rho_{\nu}(u)$ que tienen que ser determinados. Conviene señalar que en rigor $u$ depende de la propia superficie $\xi(\vec{R})$.}:
\begin{equation} \rho_{\xi}(\vec{r})\simeq\rho_{\xi}(u)=\rho_{0}(u)+2H\rho_{H}(u)+K\rho_{K}(u)+(2H)^{2}\rho_{H^{2}}(u)+...
\end{equation}
 de este modo proponen incluir las propiedades de $\xi(\vec{R})$ explícitamente en $\mathcal{H}_{I}$.\\

La imagen conceptual de un perfil de densidad intrínseco que no es exclusivamente una translación rígida siguiendo $\xi(\vec{R})$ supone una mejora \textit{necesaria} sobre la aproximación anterior sin embargo conviene notar que el procedimiento seguido después no realiza una minimización explícita\footnote{Volvemos a insistir al lector, en un sistema macroscópico donde la longitud de correlación sea pequeña es posible escribir $\Omega[s]$, donde $s$ describe la geométrica del contorno del sistema, como una suma de propiedades geométricas del recipiente, donde los coeficientes solo no dependen de este,  sin embargo a longitudes de escala del orden de la longitud de correlación del sistema perdemos esta identificación\cite{PhysRevLett.93.160601}.}.\\

Tras un notable esfuerzo analítico consiguen una expresión explícita del hamiltoniano que contiene contribuciones locales y no locales (en ellas aparecen convoluciones de la función peso atractiva que da un carácter no local al funcional $\mathcal{H}_{MeD}[\xi]$) que particularizado a un perfil intrínseco en coordenadas normales dado por una función paso permite reducir sus expresiones a la forma dada por \textit{Napiórkowski-Dietrich}. Conviene notar que hay dos cuestiones esenciales en este procedimiento. Una primera que el \textit{crossing criterium } permita diferenciar adecuadamente configuraciones microscópicas y una segunda que el procedimiento de minimización que permite hallar $\mathcal{H}_{MeD}[\xi(\vec{R})]$ sea adecuado, es decir, que el paso de la definición implícita a la explicita realmente permita minimizar, bajo la ligadura, el funcional.\\
\subsection{Tensión superficial en la aproximación Gaussiana}
Los resultados esenciales se producen dentro de una aproximación gaussiana donde podemos estudiar esta propuesta y la anterior. En ella como es usual se desacoplan los diferentes modos correspondientes a la interfase $\xi(\vec{R})$.  La aproximación gaussiana viene determinada por $\mathcal{H}_{I}[\xi]=\mathcal{H}^{(g)}[\xi]+O^{3}[\xi]$. Una transformación Fourier permite expresarlo de modo análogo a ec.(\ref{eqn:hamiltonianoGaussiano}) donde $h(q)=q^{2}\gamma(q)$,
\begin{equation}
\mathcal{H}_{I}[\xi_{q}]=\frac{1}{2}\int d^{2}q\frac{1}{(2\pi)^{d-1}}q^{2}\gamma(q)|\hat{\xi}_{q}|^{2}+O(\xi^{3})
\label{eqn:GaussianoCONgammaq}
\end{equation}
y $\gamma(q)$ permite analizar las implicaciones de un hamiltoniano no-local (particularizamos en todo momento a campo externo cero) comparando los resultados para, por ejemplo $\mathcal{H}_{NaD}$ en una aproximación gaussiana, con una una aproximación estrictamente bilineal basada en el desarrollo directo en $\mathcal{H}_{NaD}$ de la diferencia $\xi(\vec{R})-\xi(\vec{S})$ previo a la obtención del Hamiltoniano en el espacio de Fourier, véase \cite{PhysRevE.47.1836}.\\

\subsubsection{Aproximación Gaussiana en NaD}

Sus conclusiones son básicamente un comportamiento diferente entre un hamiltoniano efectivo de interfase mediante su procedimiento y un hamiltoniano fenomenológico, en el primero $\gamma(q)<\gamma(q=0)$ mientras que en el segundo se espera que  $\gamma(q)>\gamma(q=0)$.\\

 Esta propiedad de la aproximación fenomenológica es deseable ya que permite eliminar la necesidad de un \textit{cutoff} en una teoría de la forma dad por la ec. (\ref{eqn:GaussianoCONgammaq}), en efecto, si vemos la ecuación (\ref{eqn:valorEsperadoXIq2}) que representa el $<|\hat{\xi}_{q}|^{2}>$ vemos que una predicción para $\gamma(q)$ suficientemente creciente da lugar, para los modos con q alto, a amplitudes cuadráticas medias cercanas a cero, en particular elimina la necesidad de un cutoff $q_{max}$ en la determinación de $\Delta_{CW}^{2}$, véase (\ref{eqn:DeltaCW}). De modo equivalente observando (\ref{eqn:promedioestadiscoGaussiano}) una función  $\gamma(q)$ creciente permite que el propio peso estadístico $e^{-\mathcal{H}_{I}[\xi]}$ muestre para q alto una probabilidad casi nula y la irrelevancia de q fuera de un rango físico razonable. En consecuencia el resultado más sugerente resulta contrario a su propuesta desde el funcional de la densidad, lógicamente este resultado además de la mejora en el procedimiento motivaron el estudio de MeD.
 
\begin{figure}[htp]
  \begin{center}
    \subtop[Diferentes teorías para $\gamma(q)$ ]{\label{fig:Mecke-Dietrich-vs-otras}\includegraphics[scale=0.40]{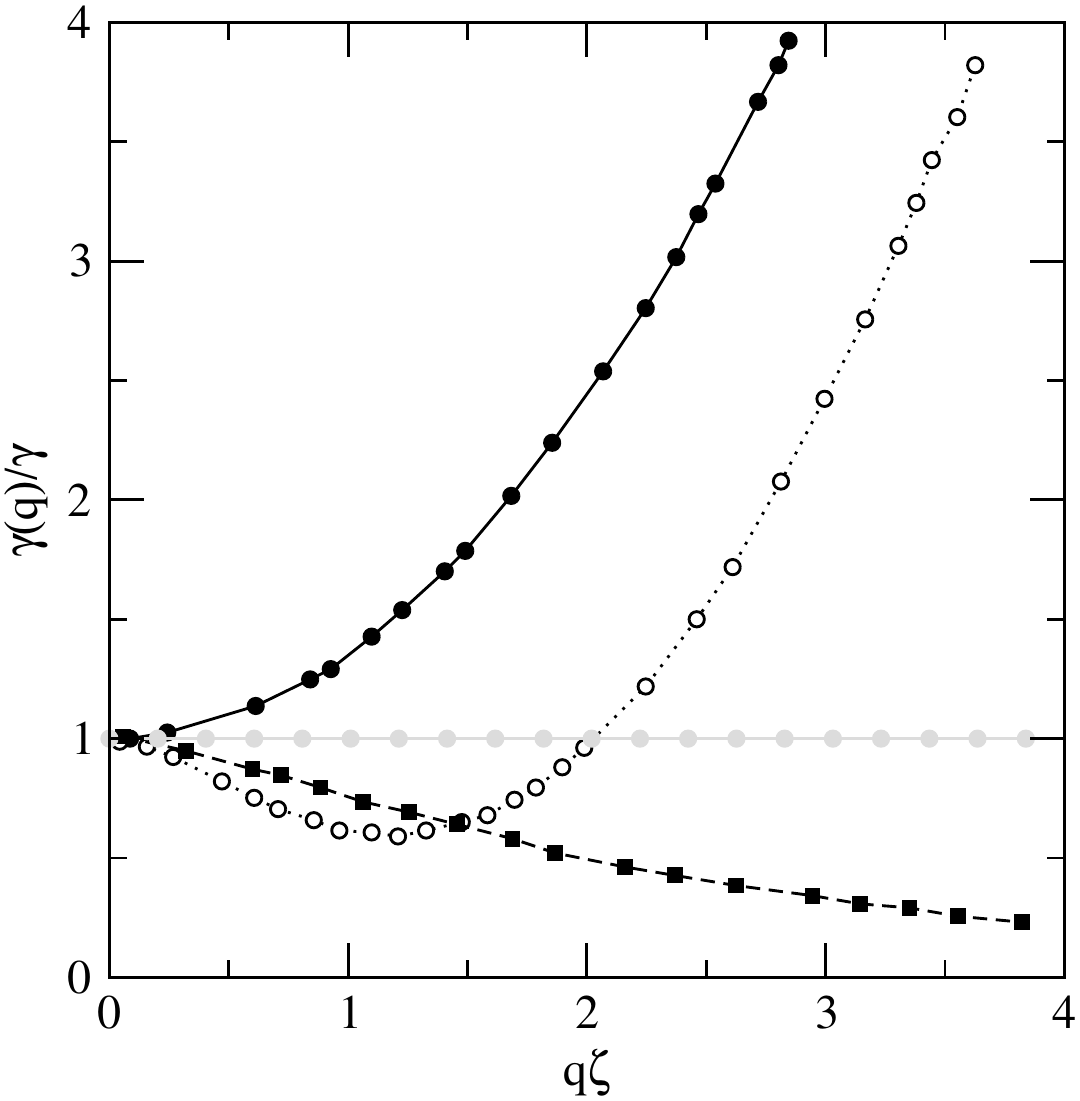}}
    \subtop[Diferentes regímenes en la teoría de Mecke-Dietrich]{\label{fig:RegimenesMeD}
\includegraphics[scale=0.35]{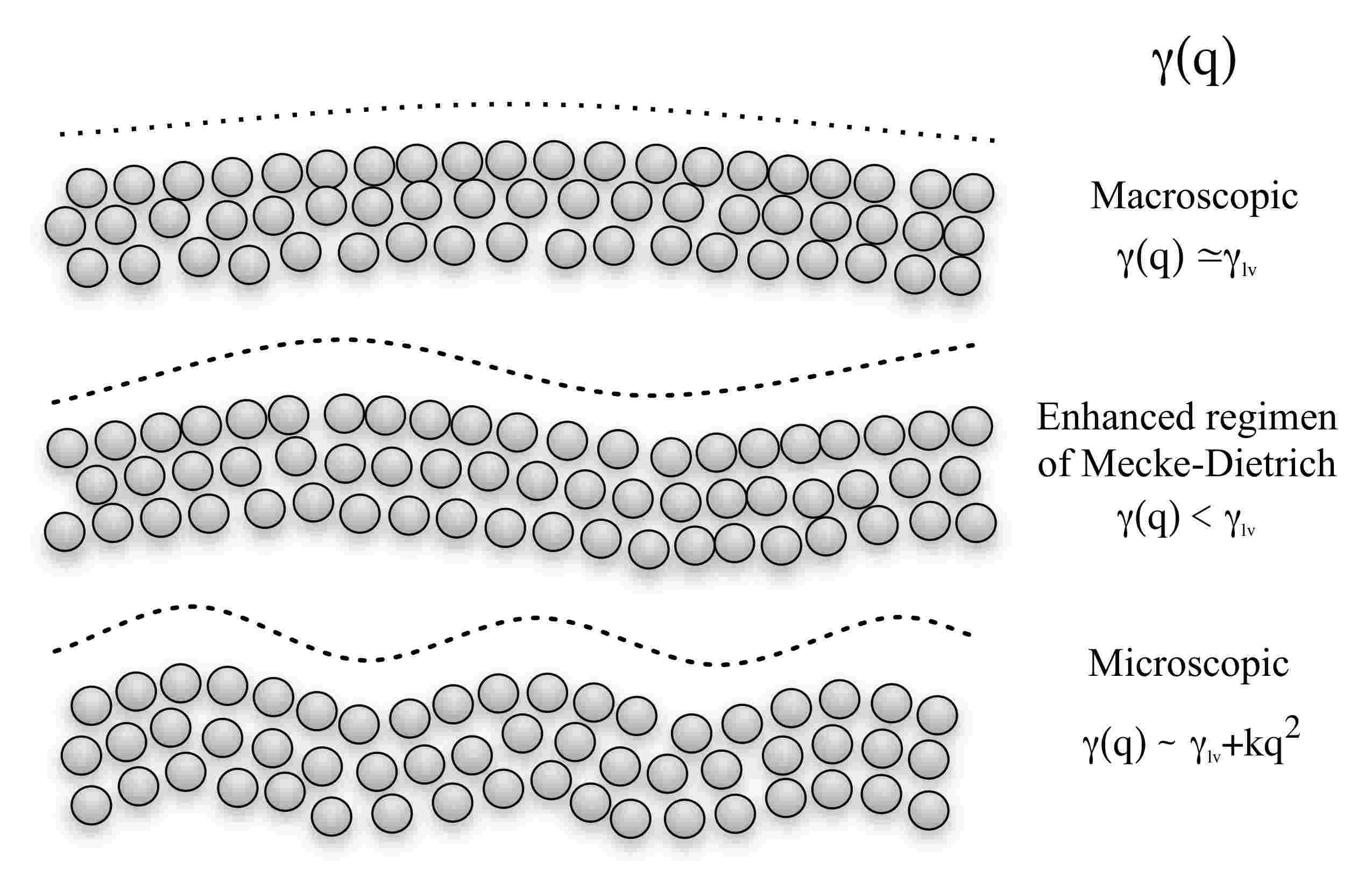}} 
  \end{center}
  \caption{\textbf{(a)}. Comparación diferentes teorías para $\gamma(q)$. CWT corresponde una linea tal que $\gamma(q)/\gamma=1$. Cuadrados negros corresponde a Napiórkowski-Dietrich, Círculos blancos a la propuesta de Mecke-Dietrich y Círculos negros a la aproximación fenomenológica. En el eje de abscisas se ha incluido un valor de $q\zeta$, $\zeta$ aquí únicamente representa una escala considerada adecuada para representar el tamaño asignado por las interacciones a las partículas.
  \textbf{(b)} Predicción de la propuesta de Mecke-Dietrich, existe un régimen de vectores de onda en que la tensión superficial es menor que la macroscópica, y a vectores de onda intenta recuperar las predicciones de una teoría fenomenológica.}
  \label{fig:TeoriaMeDyOTRAS}
\end{figure}

\subsubsection{Aproximación Gaussiana en MeD}
La definición (\ref{eqn:GaussianoCONgammaq})
 puede ser trasladada aquí igualmente para definir $\gamma^{(g)}_{MeD}(q)$, y analizar algunas de sus propiedades. 
En esta aproximación gaussiana encuentran en ausencia de campo externo,
\begin{equation}
\gamma_{MeD}^{(g)}(q)=\frac{1}{q^{2}}\left[ h^{g}_{00}(q)-h^{g}_{00}(0)\right] +2\left[ \kappa^{g}_{H0}(q)-\kappa^{g}_{H0}(0)\right] +q^{2}\left[ \kappa^{g}_{hs}(0)+\kappa^{g}_{HH}(q)\right] 
\label{eqn:gammaqMeD}
\end{equation}

Las constantes provienen de las convoluciones del peso atractivo con el perfil de densidad $\rho_{H}(u)$ o  $\partial\rho_{0}(u)/\partial u$, esto se indica con los subíndices\footnote{Por tanto en su teoría los parámetros pueden ser dependientes de las propiedades de la parte atractiva en su comportamiento asintótico, es decir, difieren en una interacción tipo van der Waals, un potencial que decaiga exponencialmente con la distancia o en potenciales de soporte finito como el Pozo Cuadrado.}. El caso de $\kappa^{g}_{hs}(0)$ proviene de la convolución de $\rho_{H}(u)^{2}$ con $\partial\mu_{hs}/\partial\rho$. Vemos previamente las propiedades generales:

\begin{itemize}
\item $\gamma_{MeD}^{(g)}(q\rightarrow\infty)=\kappa q^{2}+O(q^{4})$ donde de la expresión de $\kappa$ la consideran positiva e independiente de q. La definición de $\kappa$ hace que esta propiedad este determinada por la contribución de esferas duras\footnote{En el cálculo de MeD un funcional local.}.
\item  $\gamma_{MeD}^{(g)}(q \rightarrow 0)=\gamma_{lv}+\kappa_{0} q^{2}+O(q^{4})$ donde de la expresión de $\kappa_{0}$ contiene otra serie de contribuciones extra a $\kappa$ dependientes de la forma de las interacciones atractivas de hecho con contribuciones anómalas para fuerzas de dispersión como ocurre en NaD. Resaltan Mecke y Dietrich que $\kappa_{0}$ puede ser \textit{negativa} y de hecho indican que así es.
\end{itemize}

\vspace*{0.3cm}

\begin{figure}[htbp]
\begin{center}
\includegraphics[width=4.0in]{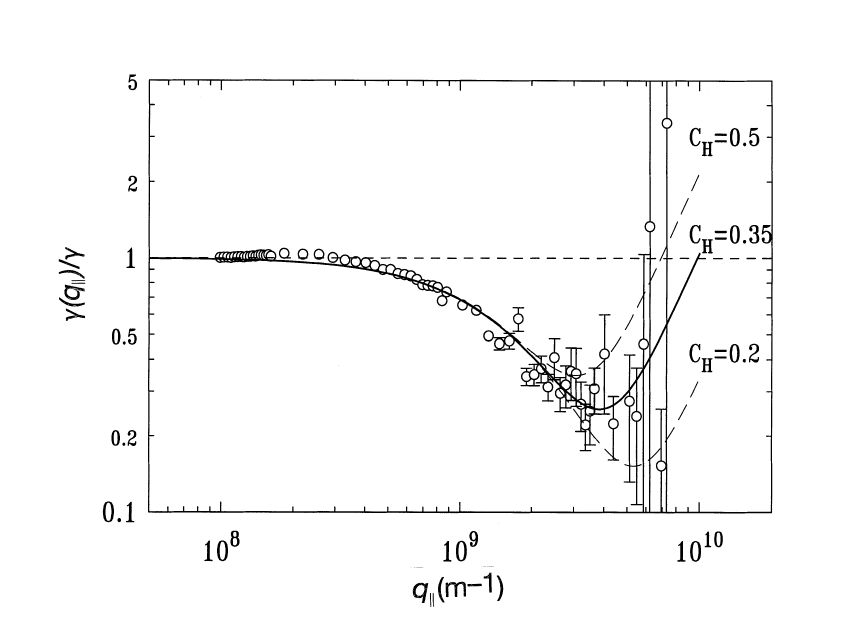}
\caption{Resultados para $\gamma(q)$ obtenidos por el grupo experimental de \textit{Fradin-Daillant}\cite{2000Natur.403..871F}. En la figura observamos que sus resultados son también compatibles con $c_{H}$ dependiente de q progresivamente menor, a q grande la incertidumbre en las medidas es demasiado alta para ser concluyente y el propio método que intenta separar en la sección eficaz diferencial diferentes contribuciones cuestionable\cite{Pershan:0927-7757:149}, mientras que la posibilidad de que en líquidos como el agua usados en el experimento\cite{2000Natur.403..871F} pueden ver disminuida su tensión superficial en determinadas escalas debido a fuentes intensas de Rayos-X ha sido sugerida también recientemente\cite{weon:217403}.}
\label{fig:ExperimentoNature}
\end{center}
\end{figure}

Ambos comportamientos implican la presencia de un \textit{mínimo} en $\gamma(q)$ y en consecuencia una región de modos que penaliza menos la pre\-sen\-cia de ondas capilares que el propio régimen macroscópico. Esta pre\-di\-cci\-ón acerca de la existencia de unos valores mesoscópicos en que $\gamma(q)<\gamma_{lv}$ ha tenido implicaciones en plan\-tea\-mi\-en\-tos de ca\-rác\-ter ex\-pe\-ri\-men\-tal que han intentado una confirmación, véase figura (\ref{fig:ExperimentoNature}).\\

	\begin{wrapfigure}{r}{6.92cm}
	\includegraphics[width=2.25in]{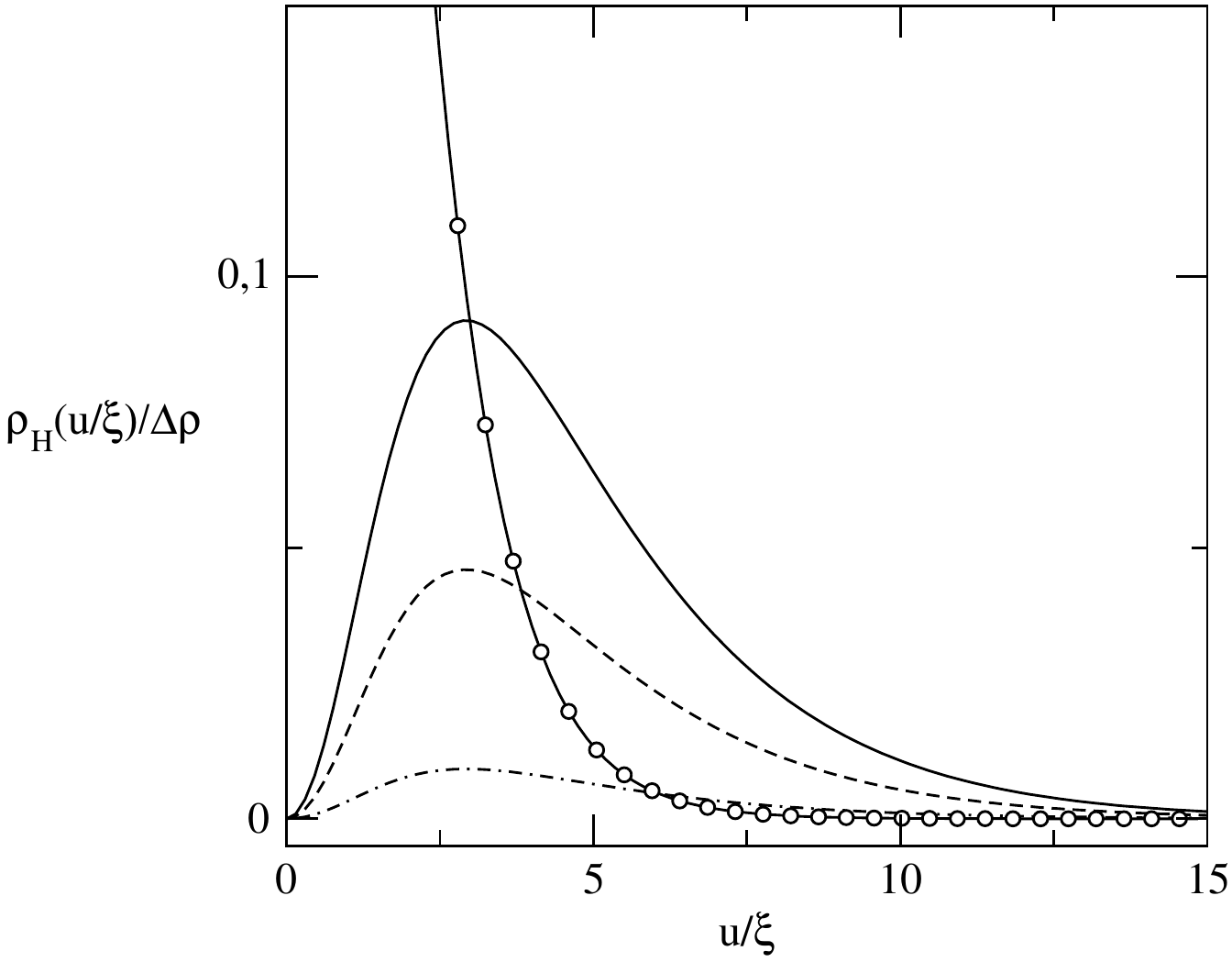}
	\caption{$\rho_{H}(u/\xi)$ variando $C_{H}$, $C_{H}=1$ línea continua, $C_{H}=0.5$ línea discontinua y $C_{H}=0.1$ línea discontinua y de puntos. La línea con círculos representa $\rho_{0}'(u/\xi))$ y es la función con la que convolucionamos $\rho_{H}(u/\xi)$.}
	\label{fig:rhoH}
	\end{wrapfigure}

La evaluación completa de las expresiones del modelo gaussiano requieren utilizar un método nu\-mé\-ri\-co. A\-de\-más los coe\-fi\-ci\-en\-tes $\rho_{\lambda}$ per\-te\-ne\-cien\-tes al desarrollo en curvaturas de la densidad son desconocidos\footnote{En el caso de una pared corrugada mesoscópicamente han sido hallados algunos de estos coeficientes pa\-ra una forma cilíndrica y esférica de la pared\cite{CurvatureProfiles}.}. Pa\-ra indagar en la física subyacente a su solución proponen una forma funcional basándose en una hipótesis de escalado tanto pa\-ra $\rho_{0}(u)$ como pa\-ra $\rho_{H}(u)$, que pa\-ra temperaturas altas debe ser no solo cualitativamente sino también cuantitativamente correcta y su com\-pa\-ra\-ción con los resultados nu\-mé\-ri\-cos de su teoría así parece atestiguarlo.\\

\begin{figure}[htbp]
\begin{center}
\includegraphics[width=4.0in]{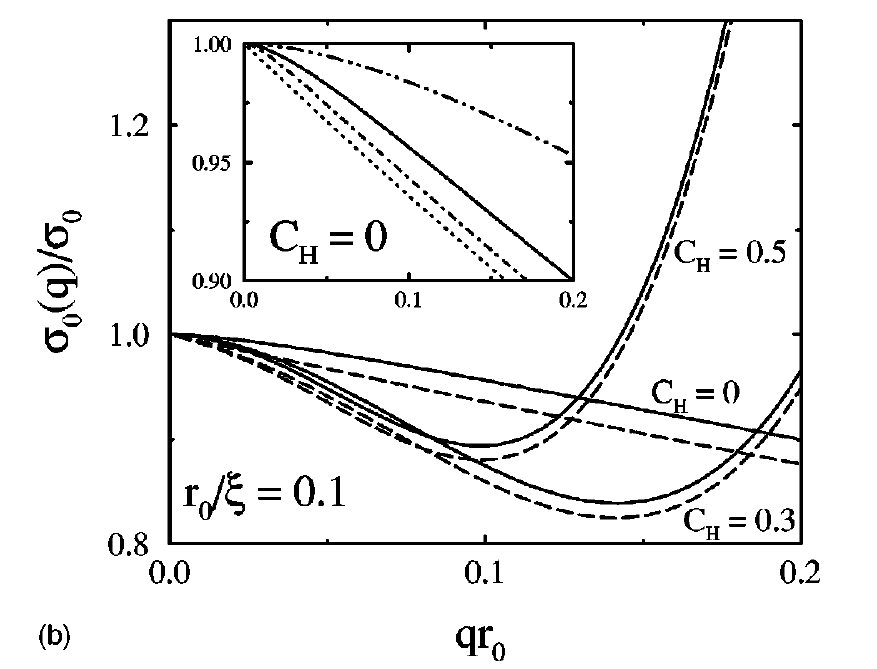}
\caption{Relevancia de $c_{H}$ en la forma de $\gamma(q)$, se puede observar como una dependencia en q de $c_{H}$ puede permitir recuperar una forma similar a NaD, mientras que la posición del mínimo depende nítidamente del valor asignado a $C_{H}$. Figura tomada de referencia\cite{PhysRevE.59.6766}, conservamos en la figura su notación para $\gamma$ como $\sigma$ y $r_{0}$ es el $\zeta$ que hemos usado en figura (\ref{fig:TeoriaMeDyOTRAS}).}
\label{fig:TeoriaMeDdiferentesCH}
\end{center}
\end{figure}

Básicamente parten de propiedades de escalado para los dos perfiles de densidad relevantes en la aproximación gaussiana, y suponen que el perfil intrínseco corrugado depende entonces de dos propiedades la \textit{longitud de correlación} en el volumen dado por $\xi$ y un \textit{parámetro de rigidez} dado por $C_{H}$ que describe la capacidad de distorsión del perfil intrínseco debido a deformaciones de la interfase, es decir,
\begin{equation}
\rho_{0}(u)-\bar{\rho}=\frac{1}{2}\Delta{\rho}\bar{\rho}_{0}(\frac{u}{\xi})
\end{equation}
\begin{equation}
\rho_{H}(u)=C_{H}\Delta{\rho}\xi\bar{\rho}_{H}(\frac{u}{\xi})
\end{equation}

Una de las claves es asumir que esta constante $C_{H}$ es independiente\footnote{Y de la temperatura, por otra parte se espera que $0<C_{H}<1$. Una vez introducidas estas dos expresiones para $\rho_{0}$ y $\rho_{H}$ los valores de las constantes de rigidez presentes en ec. (\ref{eqn:gammaqMeD}) pasan a depender de $C_{H}$ y con ellos la tensión $\gamma(q)_{MeD}^{(g)}$.} de q y obtienen diferentes $\gamma(q)$ en función de $C_{H}$. Concluyendo estas suposiciones les permiten encontrar el comportamiento a q grandes creciente, basándose en el comportamiento para q pequeño. Tomando $C_{H}=0$ ten\-drí\-a\-mos una predicción sin la presencia de un mínimo, análoga a la de \textit{Napiórkowski-Dietrich} donde el \textit{feedback} entre la superficie intrínseca y el perfil intrínseco era ignorado. La forma funcional de $\gamma(q)$, en particular los coeficientes de rigidez que rigen los comportamientos en q  dependen de $C_{H}^{2}$. Los resultados de esta teoría aparecen condensados en la figura (\ref{fig:TeoriaMeDdiferentesCH}).\\

\section{Análisis de los resultados}
\label{sec:analisis}

Resulta conveniente analizar las recetas de NaD y MeD globalmente ya que ambas definen un hamiltoniano efectivo como el coste de energía libre entre una configuración plana y una configuración corrugada según $\xi(\vec{R})$. NaD sencillamente desplaza el perfil localmente según esta función $\xi$ pero no se plantea obtener un perfil que realmente minimice el funcional $\Omega$ para esta superficie intrínseca. MeD introducen un desarrollo que permite de una parte introducir perfiles suaves y de otra permite una cierta optimización del perfil de densidad en su desarrollo en curvaturas de la función $\xi$ pero no realizan una minimización \textit{completa}.  Desde este punto de vista resulta que $C_{H}$ determinada en el régimen $q\sim0$ es usada también en $q\rightarrow\infty$ sin observar que el proceso de optimización de la única amplitud relevante que ha quedado en este nivel de aproximación pueda ser dependiente\footnote{De hecho físicamente parece más sugerente que $\rho_{H}$ pueda depender del grado de corrugación de la interfase a través de la amplitud $C_{H}$, en la figura (\ref{fig:rhoH}) se han representado las funciones $\rho_{H}$ utilizadas por Mecke-Dietrich para diferentes valores de $C_{H}$.} del grado de corrugación de la superficie $\xi$. Si tal dependencia existe no es de esperar que el comportamiento a q grandes este correctamente representado por su $\gamma^{(g)}_{MeD}(q)$. De hecho si comparamos el exceso de energía libre de NaD y MeD este último es menor para q pequeño y sin embargo mayor para valores de q grandes lo que puede estar asociado a una inconsistencia al incorporar información de una zona del espectro de ondas capilares a otra.\\

 Esto puede ser comprobado incluyendo un criterio variacional para la amplitud relevante $C_{H}(q)$ que permita una optimización dentro de propio esquema de la teoría de Mecke-Dietrich\footnote{Si el conjunto de aproximaciones realizadas es el adecuado las predicciones no deberían peligrar en un proceso de optimización y no altera el procedimiento de desarrollo en curvaturas de la interfase para el perfil de densidad, toda vez que el valor de $\rho_{H}$ no ha sido determinado dentro de la teoría a este orden la teoría solo puede técnicamente mejorar.}.\\

\begin{figure}[htbp]
\begin{center}
\includegraphics[width=4.0in]{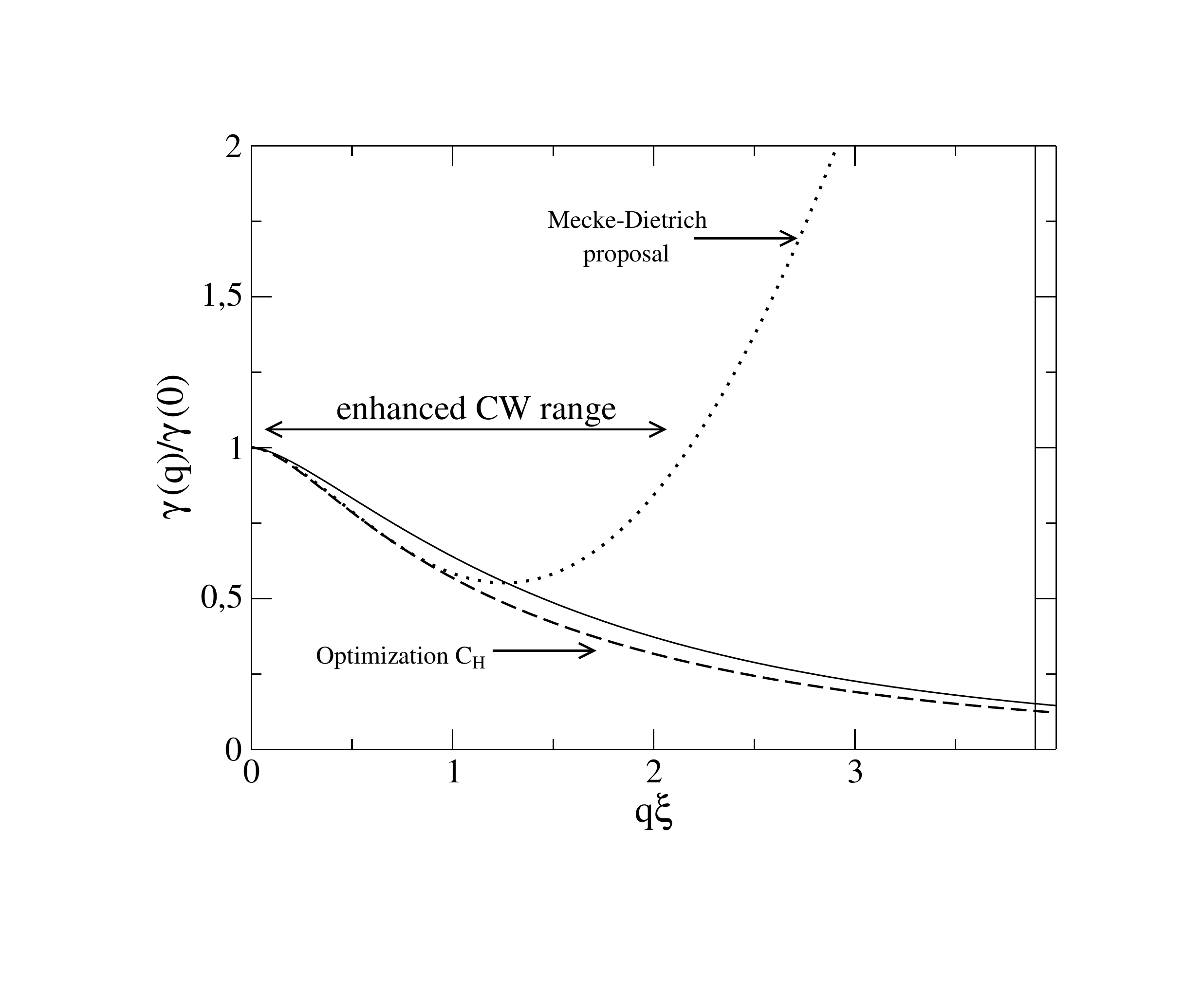}
\caption{Comparación de $\gamma(q)$ entre el resultado de Mecke-Dietrich\cite{PhysRevE.59.6766} y el proceso de optimización basado en $C_{H}(q)$. Línea de puntos, Mecke-Dietrich. Línea discontinua, optimización mediante $C_{H}(q)$, línea continua resultados de Napiórkowski-Dietrich\cite{PhysRevE.47.1836}.}
\label{fig:OptimizacionCH}
\end{center}
\end{figure}

El resultado puede observarse en la figura (\ref{fig:OptimizacionCH}) e indica una variación de $C_{H}\sim q^{-2}$ lo que elimina el crecimiento espúreo de $\gamma(q)$ dentro de la aproximación de MeD revelándose similar a NaD y obteniendo un comportamiento cierto de $\gamma^{(g)}(q\rightarrow\infty)=\kappa q^{2}+O(q^{4})$ pero con $\kappa \sim q^{-4}$.\\

Como consecuencia el desacuerdo entre NaD y las aportaciones fenomenológicas persiste más acentuado en MeD. De este modo ni la incorporación de perfiles suaves ni subsiguiente desarrollo en curvaturas locales permiten obtener un resultado para $\gamma^{(g)}(q)$ acorde con lo que fenomenológicamente es esperable.\\

Podemos completar el análisis observando que un desplazamiento rígido de la interfase, como en la predicción básica de la teoría de ondas capilares, solo es sentido en $\mathcal{H}_{I}$ por la parte contenida en $\mathcal{F}_{ex}[\rho_{\xi}]$, la energía libre de exceso del funcional donde están contenidas las interacciones y que esta construida a su vez mediante medidas no-locales que poseen un alcance dado por $\zeta$, por lo tanto si solo corrugamos en los modos con  $q\gg \zeta^{-1}$ el perfil correspondiente a minimizar el funcional debe ser similar a promediar en la coordenada transversal y por tanto en realidad para valores de q altos equivale de hecho a la ec. (\ref{eqn:perfildensidadmedioCWT}), siempre que $\Delta^{2}_{CW}$ en ec. (\ref{eqn:DeltaCW}) incluya solo estos modos, si realizamos un desarrollo de Taylor de la convolución dada en ec. (\ref{eqn:perfildensidadmedioCWT}). Si aplicamos que $<\xi(\vec{R})>=0$ podemos estimar una expresión para $\mathcal{F}_{ex}[\rho_{\xi}]$ si $q\gg \zeta_{0}^{-1}$, es decir, se cumple,
\begin{equation}
\mathcal{F}_{ex}[\rho_{\xi}(\vec{r})]\simeq\mathcal{F}_{ex}[\langle\rho_{\xi}\rangle_{\vec{R}}]
\end{equation}
y este mismo perfil estaría dado por,
\begin{equation}
\langle\rho_{\xi}\rangle_{\vec{R}}=\left\langle \rho_{DF}(z-\xi(\vec{R}))\right\rangle _{\xi}
\end{equation}
y
\begin{equation}
\langle\rho_{\xi}\rangle_{\vec{R}}=\rho_{DF}(z)+\frac{1}{2}\frac{d^{2}\rho(z)}{dz^{2}}\Delta_{G}^{2}
\end{equation}
donde $\Delta^{2}_{G}=\sum_{q}\xi_{q}\xi^{*}_{q}$ la dispersión cuadrática media en el rango $q>>\zeta^{-1}$.\\

Podemos expandir el funcional de exceso responsable de las contribuciones esenciales (como se manifiesta explícitamente en la aproximación NaD)  y usando la expansión termodinámica de la parte de exceso escribir,
\begin{equation}
\mathcal{F}_{ex}[\langle\rho_{\xi}\rangle_{\vec{R}}]\simeq \mathcal{F}_{ex}[\rho_{DF}]+\int d\vec{r}\frac{\delta^{2}\mathcal{F}_{ex}[\rho]}{\delta\rho(\vec{R})^{2}}\frac{1}{2}\frac{d^{2}\rho(z)}{dz^{2}}\Delta_{CW}^{2}
\end{equation}
donde las propiedades de $\rho_{DF}(z)$ en los limites del volumen permite escribir:
\begin{equation}
\mathcal{H}_{DF}[\xi(\vec{R})]\simeq \frac{A}{2}\int dz \frac{1}{\beta\rho_{DF}(z)}[\rho'_{DF}(z)]\sum_{q}\xi_{q}\xi^{*}_{q}
\end{equation}
gracias a que al escribir la función de correlación de primer orden (primera derivada de la energía libre) hemos hecho uso implícitamente la ecuación de Euler-Lagrange que representa la condición de equilibrio para la densidad. Obviamente en el limite $q\gg \zeta^{-1}$ la parte de exceso relacionada con la cola atractiva no es relevante\footnote{De hecho $\hat{w}(q\rightarrow\infty,z)\simeq0$.} y podemos en un funcional local aproximar $c^{(1)}(\rho)\simeq f'_{HS}(\rho)$ y de hecho expresar por tanto:
\begin{equation}
\mathcal{H}_{DF}[\xi]\sim \frac{A}{2}\int dz f''_{hs}(\rho)\frac{d^{2}\rho_{DF}}{dz^{2}}\sum_{q}|\xi_{q}|^{2}
\end{equation}
En ambas situaciones la dependencia en el \textit{área de la interfase intrínseca} se ha perdido quedando únicamente una dependencia en el \textit{área macroscópica} de la interfase lo que explica el comportamiento a q grande como:
\begin{equation}
\gamma^{(g)}(q)\sim\gamma_{DF}q^{-2}
\end{equation}
ya que el resto de factores incluidos no poseen dependencia en q. 

%
%
\section{Inclusión de una parte no-local en el sistema de referencia de esferas duras}
\label{sec:Cap6parteNOLOCAL}

Hasta el momento todas las determinaciones de $\mathcal{H}_{eff}$ han surgido dentro del contexto de funcionales locales en la bibliografía existente\footnote{El crossing criterium es solo efectivo en funcionales locales, esta restricción sobre un funcional no-local que se defina mediante promedios sobre el perfil de densidad usando convoluciones como el que se introdujo en \S\ref{sec:teoriasDensidadPromediada} carece de viabilidad ya que no tiene efecto sobre las funciones densidad promediadas.}, ¿qué papel juega restringirnos a funcionales locales? Hemos sugerido que los problemas no están relacionados con la aproximación concreta usado sino con la hipótesis de partida que identifica $\mathcal{H}_{I}$ con diferencias entre la energía libre plana y corrugada. El objetivo ahora es obtener bajo las hipótesis de la propuestas de \textit{Napiorkowski-Dietrich} o \textit{Mecke-Dietrich} resultados en el caso de un funcional no-local e insistir por tanto en que el problema de estas propuestas no esta relacionado con la localidad del funcional utilizado, determinamos por tanto la función $\gamma(q)$ dentro de un esquema funcional \textit{no-local} que en consonancia con los funcionales utilizados en los apartados anteriores será un funcional FMT-MFA, entendido como una teoría de van der Waals generalizada. \\

\begin{figure}[htbp]
\begin{center}
\includegraphics[width=1.05\textwidth]{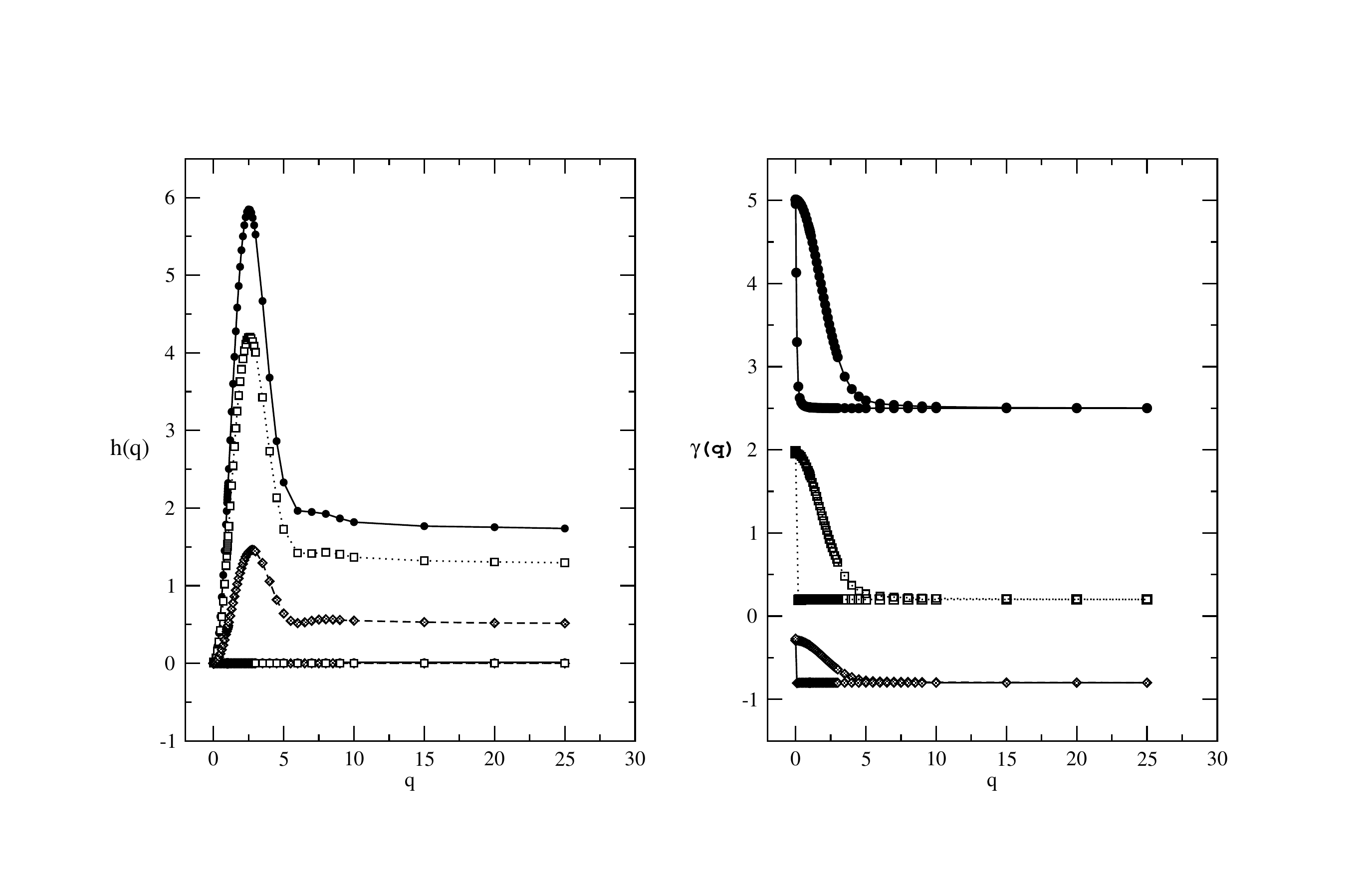}
\caption{\textbf{Solución para $h(q)$ en diferentes modelos}. De arriba a abajo. SA-PY T/U=0.60 Círculos. Na-PY T/U=0.637 Cuadrados Hg-PY T/U=0.75 Diamantes. A la izquierda la curvas visibles representan el costo en energía libre de un desplazamiento rígido de perfil mientras que las curvas compatibles con $h(q)\sim0$ son las anteriores incluyendo la relajación en el sistema. A la derecha podemos ver estos mismos resultados con la hipotética $\gamma(q)=h(q)/q^{2}$, análogamente vemos que incluir la relajación es compatible con $\gamma(q)\sim0$ excepto en $q\sim0$ en que se recupera la $\gamma_{lv}$, tanto en la parte sin relajación como en la parte relajativa de un modo adecuado. En la parte de la derecha se han desplazado verticalmente las figuras para su correcta visualización.}
\label{fig:n0iguala0}
\end{center}
\end{figure}

Comenzamos desde ec. (\ref{eqn:HeffdeNapiorkowski-Dietrich}) aplicando las aproximaciones funcionales introducidas en \S\ref{sec:vanDerwaalsGENERAL}, además al igual que Mecke-Dietrich planteamos $\rho_{DF}(z)$ como perfil intrínseco\footnote{Esta hipótesis es de partida incompatible con los resultados vistos en \S\ref{sec:relacionCWyDEN} que consideraban que el perfil $\rho_{DF}(z)$ incluía parte del espectro de ondas capilares, sencillamente intentamos ver a donde nos conduce.}, por lo visto anteriormente los resultados que obtengamos serán directamente comparables con los obtenidos optimizando $C_{H}$ para funcionales locales. Por tanto realizamos una expansión del funcional de la densidad como para determinar la diferencia,

\begin{equation}
\mathcal{F}_{ex}[\rho(z-\xi(\vec{R}))]-\mathcal{F}_{ex}[\rho(z)]
\label{eqn:Heffnolocal}
\end{equation}
los detalles del cálculo son dados en \S\ref{sec:desarrolloParateExceso} donde el lector puede consultar las expresiones recordando que $\rho^{in}$ aquí utilizado es el perfil líquido-vapor presentado en \S\ref{sec:perfilesLV} y términos de la forma $\rho_{q}(z)$ no son tenidos en cuenta\footnote{También puede consultar las expresiones en \S\ref{sec:determinacionEspectroOndasCapilares} si allí sustituye $\rho_{v,\delta}$ por $\rho_{DF}$, en cualquier caso es posible que el lector volver a este punto tras leer el apéndice \S\ref{sec:apendiceIntrinseco}, ya que operacionalmente hablando este es un caso particular de aquel.}. Los resultados se pueden ver en la figura (\ref{fig:n0iguala0}) que confirma una curva, para la función $\gamma(q)$ igualmente decreciente a la mostrada en (\ref{fig:OptimizacionCH}) tanto cualitativa como cuantitativamente.\\

Como hemos comentado en el caso de MeD bajo sus hipótesis  y en otros previos, véase (\ref{sec:RRVR}) un cálculo propiamente realizado requiere tener en cuenta la posible \textit{relajación} del sistema, es decir el perfil intrínseco plano en presencia de una corrugación $\xi(\vec{R})$ se ve alterado por la reacomododación de las partículas a la nueva imposición de una superficie $\xi$, y por tanto es conveniente extender el argumento dado anteriormente a un caso más general. Extendemos la forma usual de la teoría de ondas capilares que supone que el desplazamiento rígido anula la dependencia en la coordenada transversal y expresamos correcciones debido a la posibilidad de tener un líquido compresible por la expresión:
\begin{equation}
\rho_{\xi}(z,\vec{R})=\rho(z-\xi(\vec{R}),\vec{R})=\sum_{q}\rho_{q}(z-\xi(\vec{R}))e^{-i\vec{q}\vec{R}}
\end{equation}

Si el lector vuelve a los detalles dados en \S\ref{sec:desarrolloParateExceso} ahora incluimos los términos $\rho_{q}(z)$ que determinamos pidiendo que hagan extremal a la expresión indicada (\ref{eqn:Heffnolocal}), que recuerde no es sino ec. (\ref{eqn:HeffdeMecke-Dietrich}), para cada q. Esto permite formalmente determinar las funciones $\rho_{q}$ así como el costo en energía libre de corrugar el perfil de densidad obtenido como perfil medio en los capítulos anteriores en ausencia de potencial externo.\\

En cuanto a la relación con resultados previos, notar que eliminando la relajación ($\tilde{\rho}_{q\neq0}=0$) reproducimos el modelo de ondas capilares extendido \S\ref{sec:RRVR} mientras que incluyendo únicamente el término $\sum_{q}e^{i\vec{q}\vec{R}}\xi_{q}\tilde{\rho}_{q}(z)$ tenemos la pro\-pu\-es\-ta de \textit{Parry y Boulter}\cite{Parry:1994:7199}.\\ 

\begin{figure}[htbp]
\begin{center}
\includegraphics[width=1.05\textwidth]{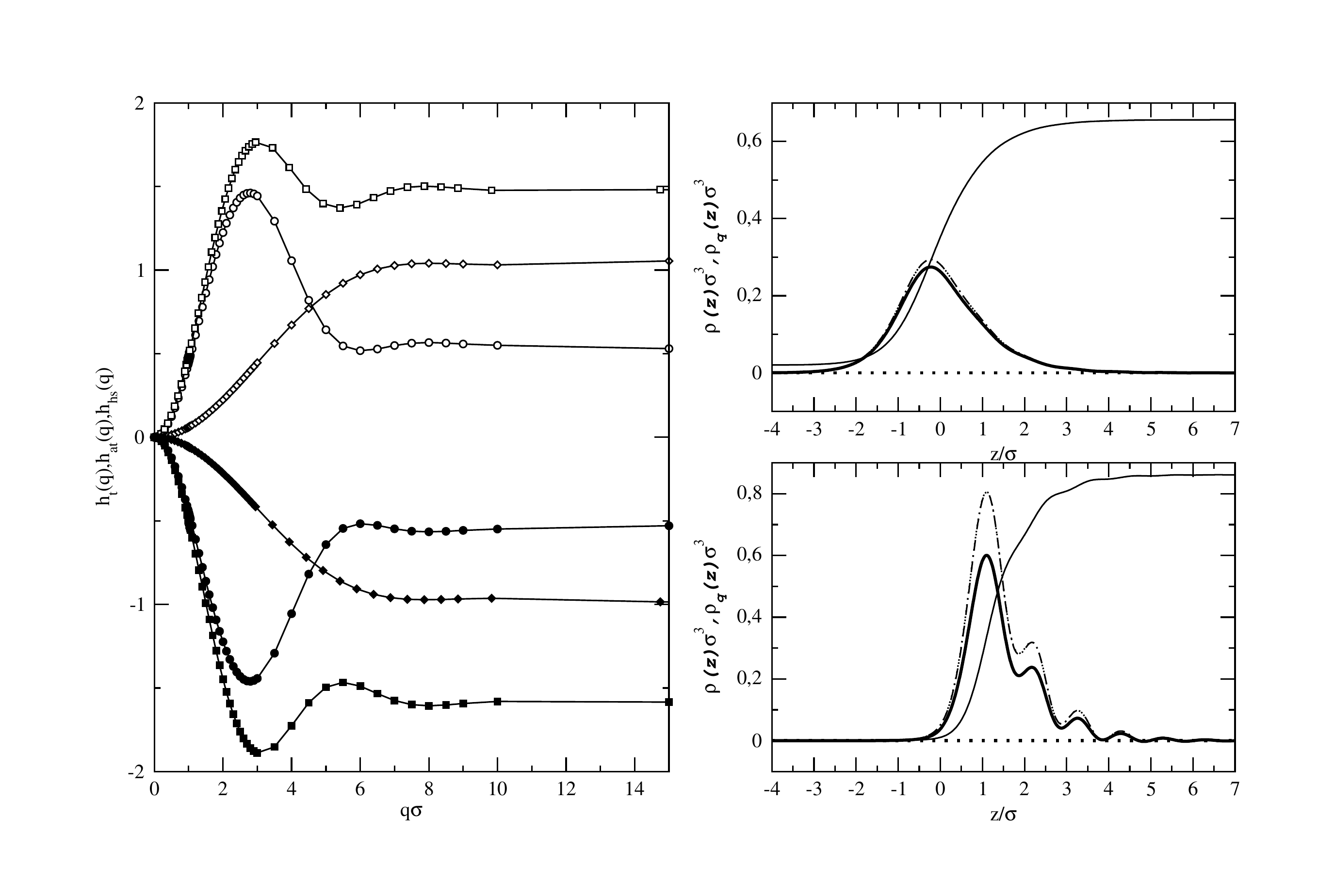}
\caption{\textbf{Izquierda}. Hg-FMT-PY T/U=0.75. Cuadrados HS+AT, círculos HS, rombos AT. Símbolos blancos representa contribuciones a $\Delta_{q}$ y símbolos negros los correspondientes a $R_{q}$, calculados como se indica en el texto, vemos las dos escalar diferentes en el comportamiento en q de HS y AT. A la \textbf{derecha}, línea gruesa representa $\rho'(z)$ mientras que en línea punteada es $\rho_{q\sim0}(z)$ que es proporcional también a $\rho'(z)$ pero con un factor q hace que solo sea relevante la parte de desplazamiento en $q\sim0$. Las líneas discontinua y discontinua-punteada aparecen superpuestas y se corresponden con las funciones $\rho_{q}(z)$ para $q=1$ y $q=5$, arriba un caso suave dado por Hg-FMT-PY T/U=0.75, abajo un caso con estructura dado por SA-FMT-PY T/U=0.60}
\label{fig:RelajacionSINno}
\end{center}
\end{figure}

En la parte derecha de la figura (\ref{fig:RelajacionSINno}) podemos ver la forma de las funciones $\rho_{q}(z)$, que se complementa con la parte izquierda en que vemos las diferentes contribuciones a $h(q)=R_{q}+\Delta_{q}$, como vemos en ella los valores de la relajación son tales que los términos relajativos, cuya suma llamamos $R_{q}$, y los términos de desplazamiento, cuya suma llamamos $\Delta_{q}$, son idénticos para cada parte del funcional, esto lo podemos ver del siguiente modo, dado el valor de $\rho_{q}(z)$ obtenido de nuestra optimización la contribución relajativa es:

\begin{equation}
R_{q}=-\int dz \mathcal{B}(z)\rho_{q}(z)=-\int dz \left[ \mathcal{B}_{HS}(z)+\mathcal{B}_{AT}(z)\right] \rho_{q}(z)
\end{equation}

a valores de q altos tenemos $h(q)\sim0$, y si de hecho sustituimos $\rho_{q}(z)$ por $\rho'_{0}(z)$ vemos que las expresiones son iguales término a término formalmente. Las diferencias a q alto entre $\rho_{q}(z)$ y $\rho'_{0}(z)$ se deben a las inconsistencias numéricas propias de un cálculo de la parte repulsiva donde se incluyen integrales dobles sobre dos kernels con variación en q implícita en funciones de Bessel sobre una amplia región de integración\footnote{Por esta razón hemos indicado $h(q)\sim 0$ ya que en rigor es diferente pero muy próximo a 0, podemos estimar a grosso modo nuestro error del siguiente modo, donde tengamos integrales de orden 1, $\partial_{\nu}\phi$ lo expresamos mediante $\int dz\sum_{\mu}\partial_{\nu\mu}\phi\rho'_{0}(z)$ ambas deberían ser idénticas en el caso exacto, esto permite entender que la regla de suma $h(q\sim0)\simeq\gamma_{lv}q^{2}$ se de solo en sentido aproximado: una aproximación la acotará por arriba y la otra por abajo, con todo damos preferencia a los cálculos con $\partial_{\nu}\phi$ ya que leves discrepancias en la normalización de $\omega_{q}^{\nu}$ tendrán menos relevancia en el cálculo a valores de q grandes por este método. En cierto sentido la regla de suma que impone TZ ha de verificarla todo cálculo consistente mientras todo cálculo con una parte repulsiva no-local permite dotar a este límite de un significado físico más allá de la simple contribución de la parte atractiva que realmente ya estaba incluida en van der Waals el problema numérico en $q\simeq0$ es menos relevante ya que la verdadera dificultad esta en idear una teoría con sentido en q alto. En nuestro cálculo el comportamiento es $\gamma q^{2}$ y aunque la parte repulsiva posee un error del treinta al cuarenta por ciento la parte atractiva sigue siendo la de mayor peso en este régimen.}.\\

El resultado es compatible con el hecho de que formas más generales de corrugación que suponer un líquido incompresible ya están incluidas en un perfil que realmente esta promediado sobre valores de q altos. 

Una vez obtenemos el resultado que hace mínima la forma cuadrática descrita en el apéndice tenemos la expresión para la relajación como:

\begin{equation}
R_{q}=-\int dz_{1}dz_{2}\rho'_{0}(z_{1})c^{(2)}(z_{1},z_{2},q)\rho^{0}_{q}(z_{2})
\end{equation}
el hecho de que sea igual a la expresión de $\Delta_{q}$ se debe a que $\rho^{0}_{q}(z_{2})=\rho'_{0}(z_{1})$ de modo que $\Delta_{q}+R_{q}=h(q)\sim0$.\\

La expresión de $c^{(2)}(z_{1},z_{2},q)$ no incluye la parte ideal aunque para la determinación de las funciones $\rho'_{0}(z_{1})$ y $\rho^{0}_{q}(z_{2})$ ha sido necesaria. Conviene un estudio completo de $C^{(2)}(z_{1},z_{2},\vec{R}_{12})$.

\subsection{Excursus: Espectro de autovalores de $C^{(2)}(z_{1},z_{2},q)$}

La física contenida en el cálculo\cite{WidomMolecularPhysics} puede verse a partir del espectro de autovalores de la función $C^{(2)}(z_{1},z_{2},q)$ que si incluye la parte ideal.

\begin{figure}[htp]
  \begin{center}
    \subtop[Autovectores de $C^{(2)}(q)$ similares a $\rho'(z)$ desde $q\simeq 0$. Modelo Na. Aprox. FMT-PY. T/U=1.15]{\label{fig:C2qNaautovectores}\includegraphics[scale=0.45]{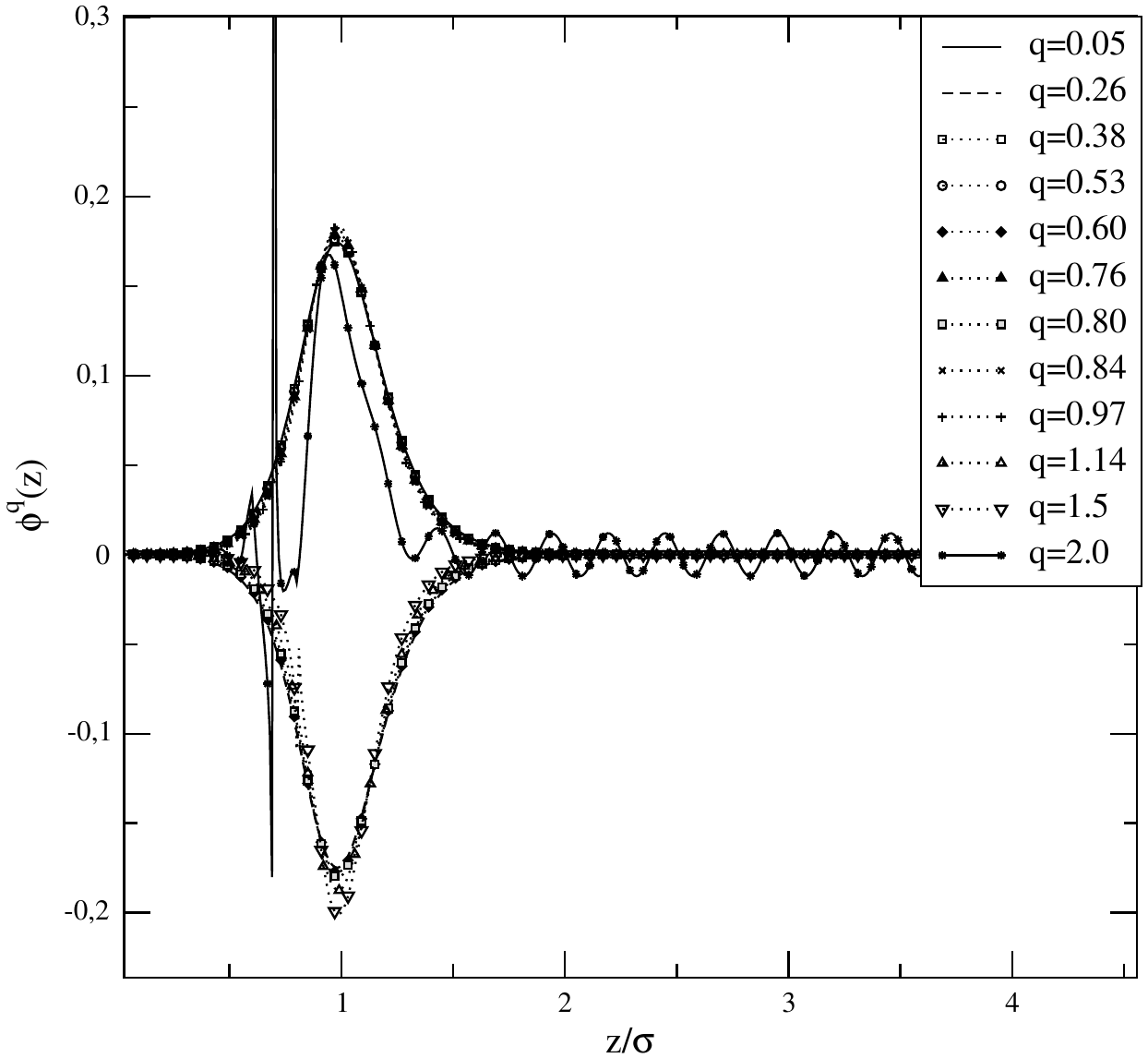}}
    \subtop[Autovectores de $C^{(2)}(q)$ similares a $\rho'(z)$ desde $q\simeq 0$. Modelo LJ$2.5\sigma$. Aprox. FMT-PY. T/U=0.68]{\label{fig:C2qLJautovectores}
\includegraphics[scale=0.45]{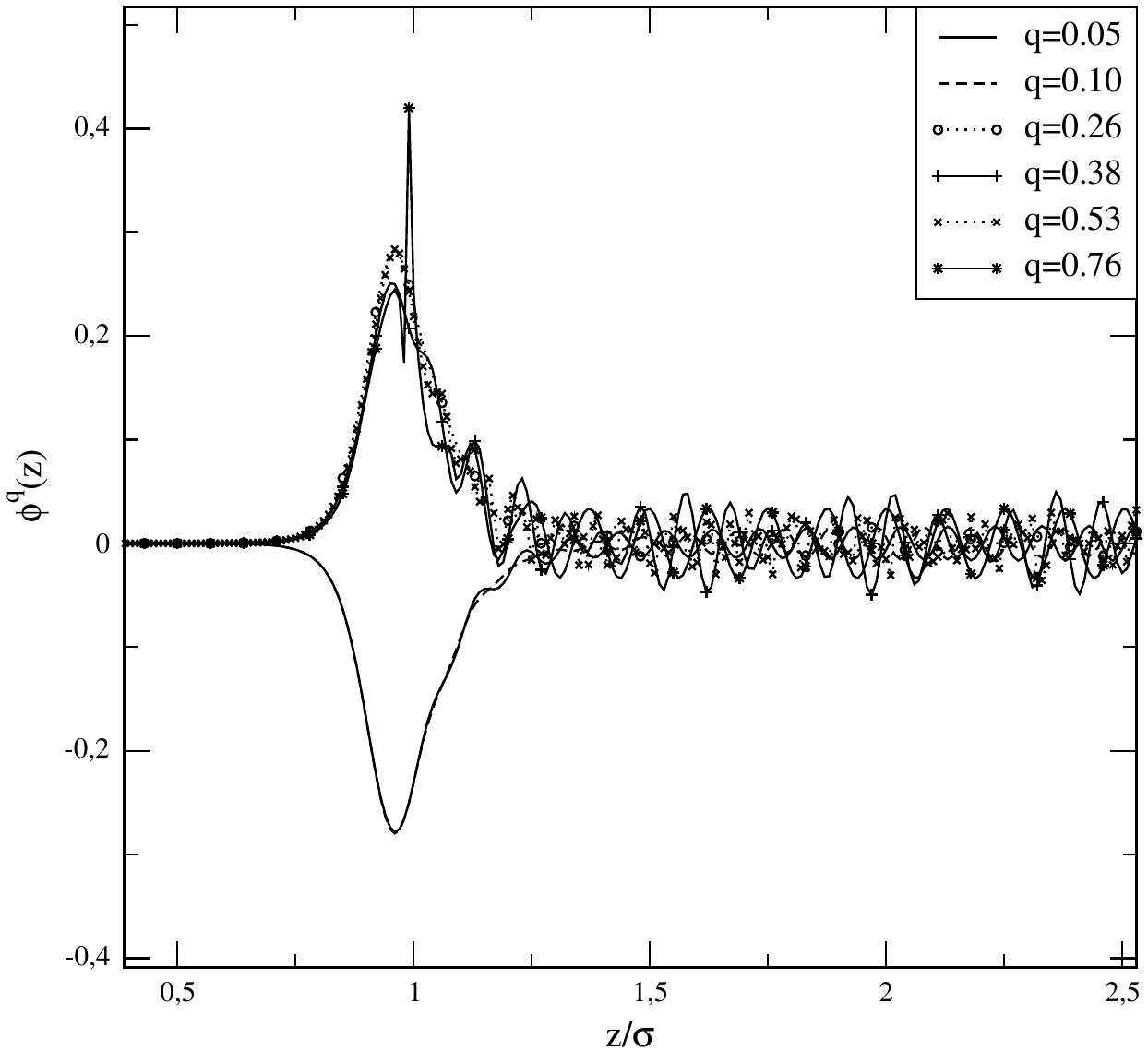}} 
  \end{center}
    \begin{center}
    \subtop[Proyección autovectores de $C^{(2)}(q)$ similares a $\rho'(z)$ desde $q\simeq 0$. Modelo Na. Aprox. FMT-PY. Varias temperaturas.]{\label{fig:C2qNaautovectores}\includegraphics[scale=0.45]{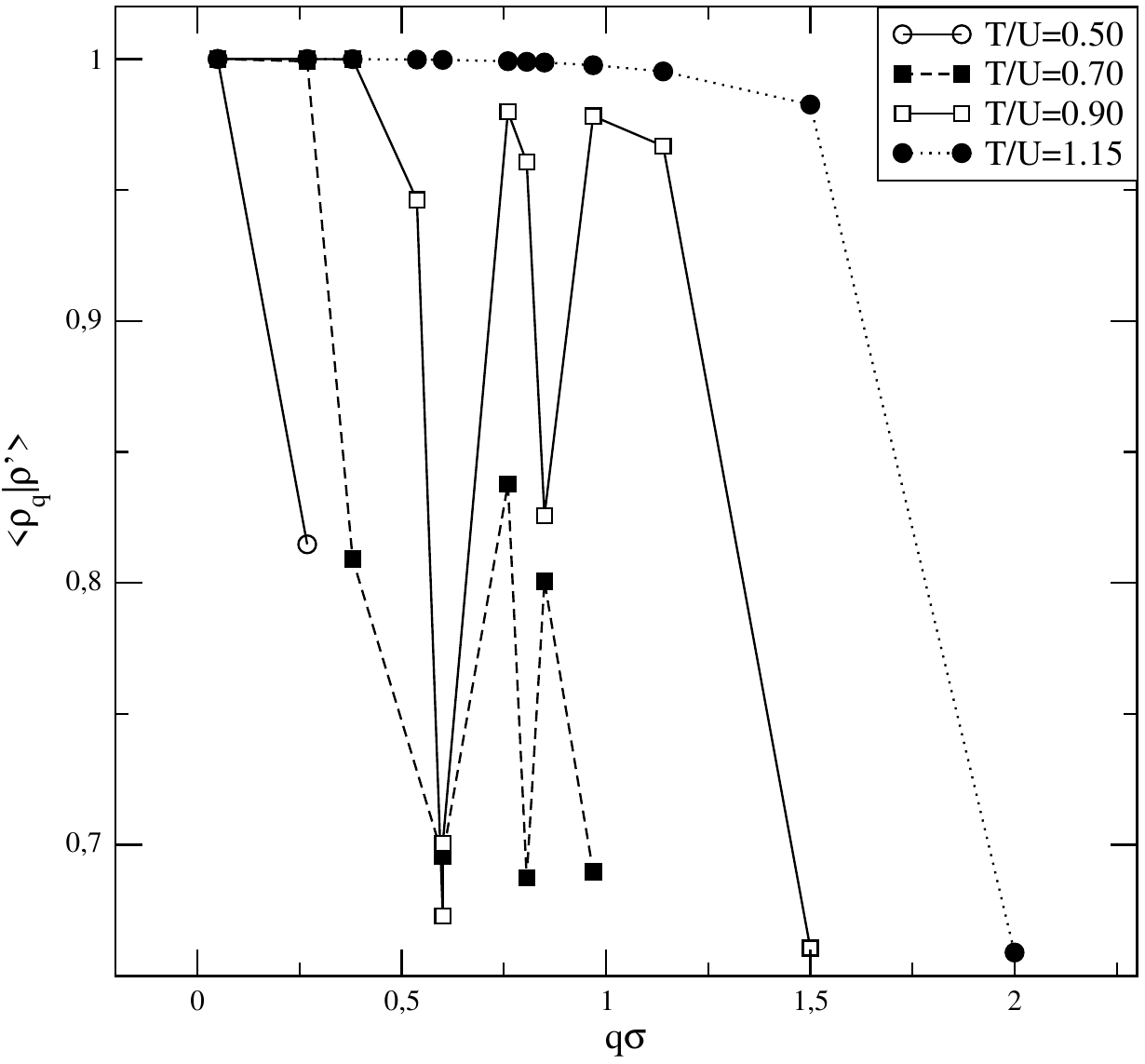}}
    \subtop[Proyección autovectores de $C^{(2)}(q)$ similares a $\rho'(z)$ desde $q\simeq 0$. Modelo LJ$2.5\sigma$. Aprox. FMT-PY. Varias temperaturas.]{\label{fig:C2qLJautovectores}
\includegraphics[scale=0.45]{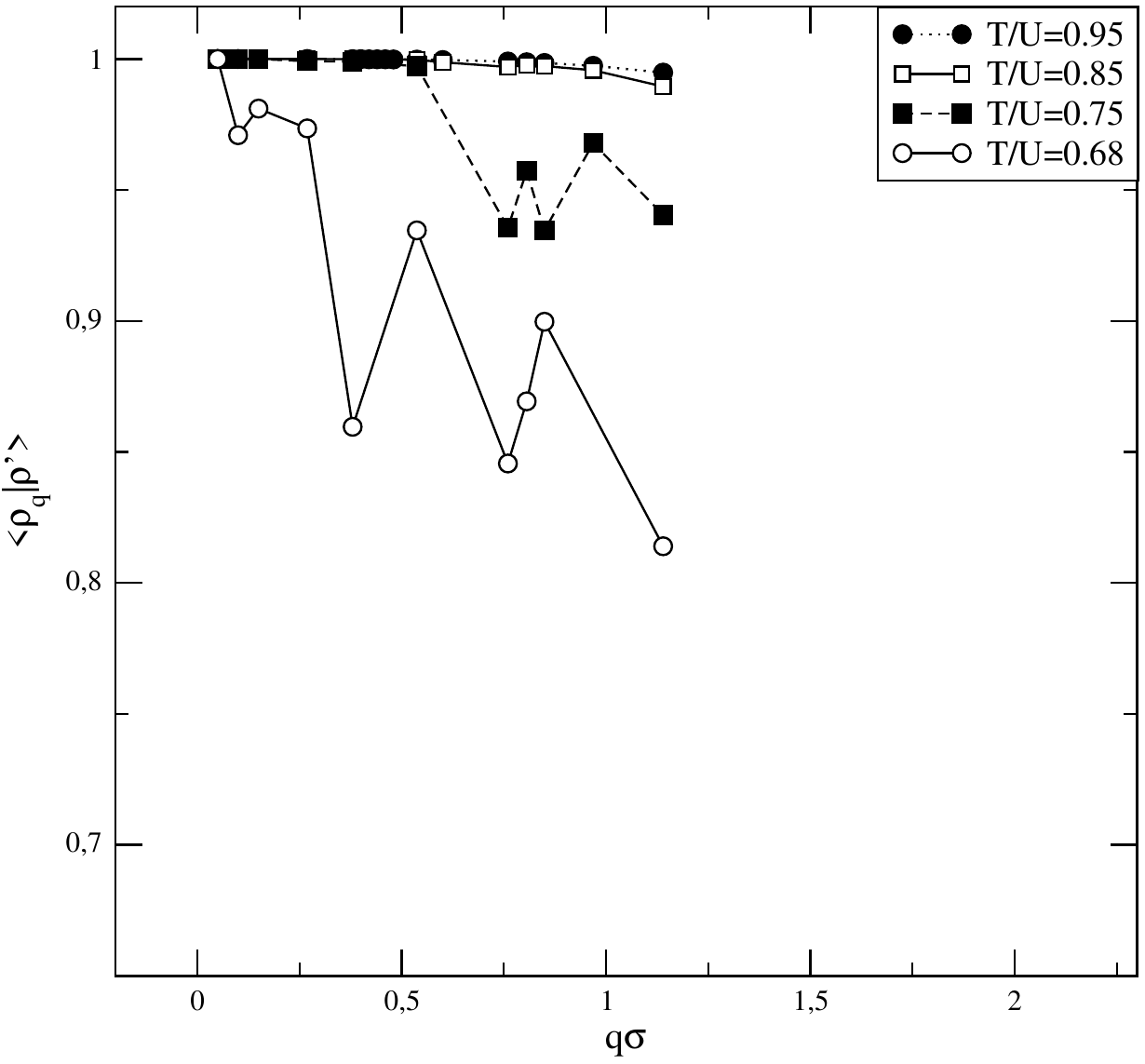}} 
  \end{center}
  \caption{Autovectores de $C^{(2)}(q)$ similares a la derivada del perfil de densidad $\rho_{DF}(z)$ determinado en \S\ref{sec:perfilesLV}, para valores de q mayores de los presentados las diferencias de la proyección de estas funciones con el caso q=0.05 presenta discrepancias mayores. Se muestra el valor de la proyección al variar q sobre el caso q=0.05}
  \label{fig:AutovectorePhiq}
\end{figure}

\begin{figure}[htp]
  \begin{center}
    \subtop[Análisis autovalores $E(\lambda_{cw})$ del autovector similar a $\rho'(z)$ desde $q\simeq 0$ hasta $q\simeq 1$, en escala $q^{2}\sigma^{2}$. Modelo LJ$2.5\sigma$. Aprox. FMT-PY. T/U=0.68, 0.75, 0.85, 0.95]{\label{fig:C2qNaautovectores}\includegraphics[scale=0.45]{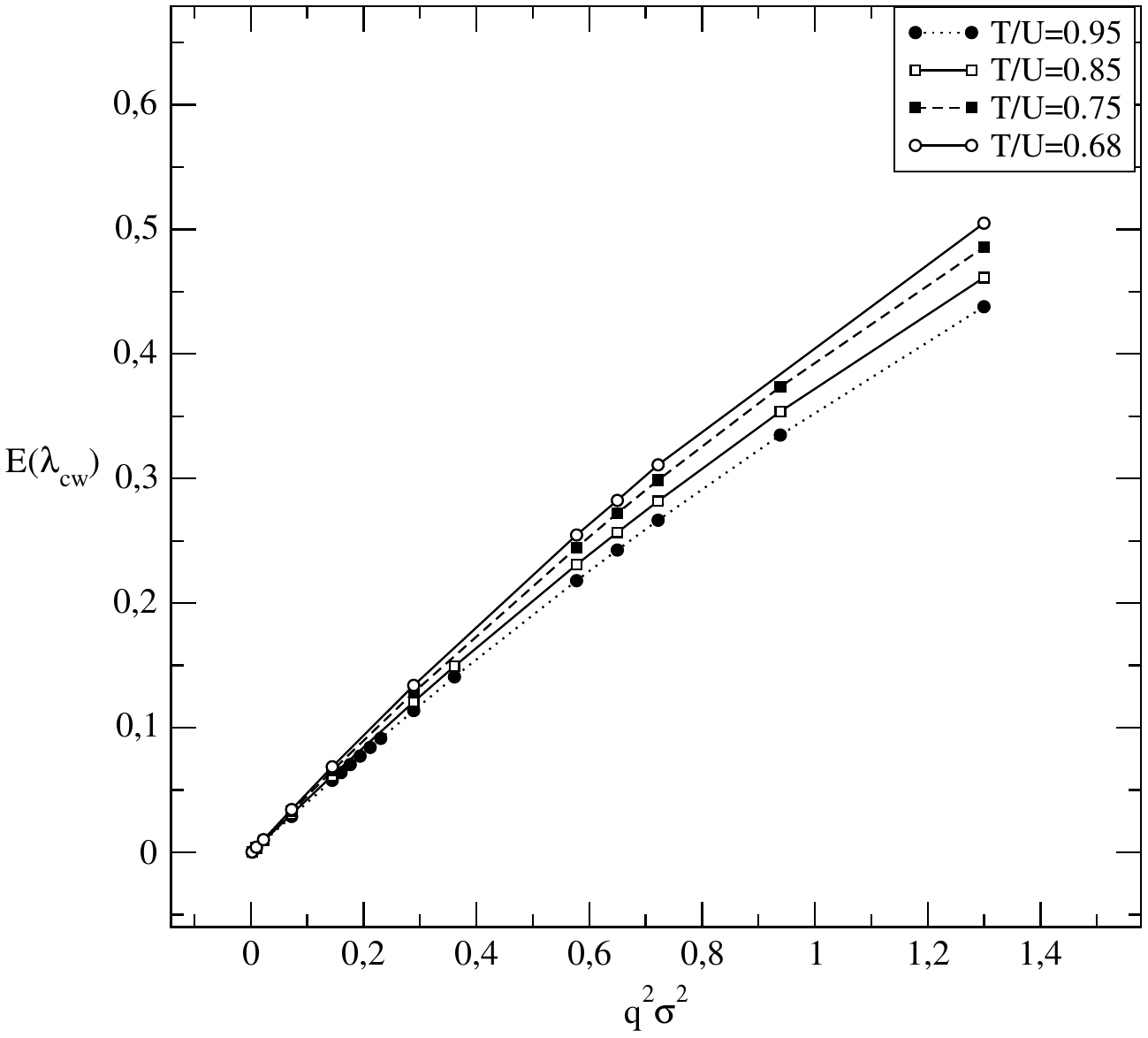}}
    \subtop[Posic. autovectores de $C^{(2)}(q)$ similares a $\rho'(z)$ en el espectro al variar q desde $q\simeq 0$. Modelo LJ$2.5\sigma$. Aprox. FMT-PY. T/U=0.68]{\label{fig:C2qLJautovectores}
\includegraphics[scale=0.45]{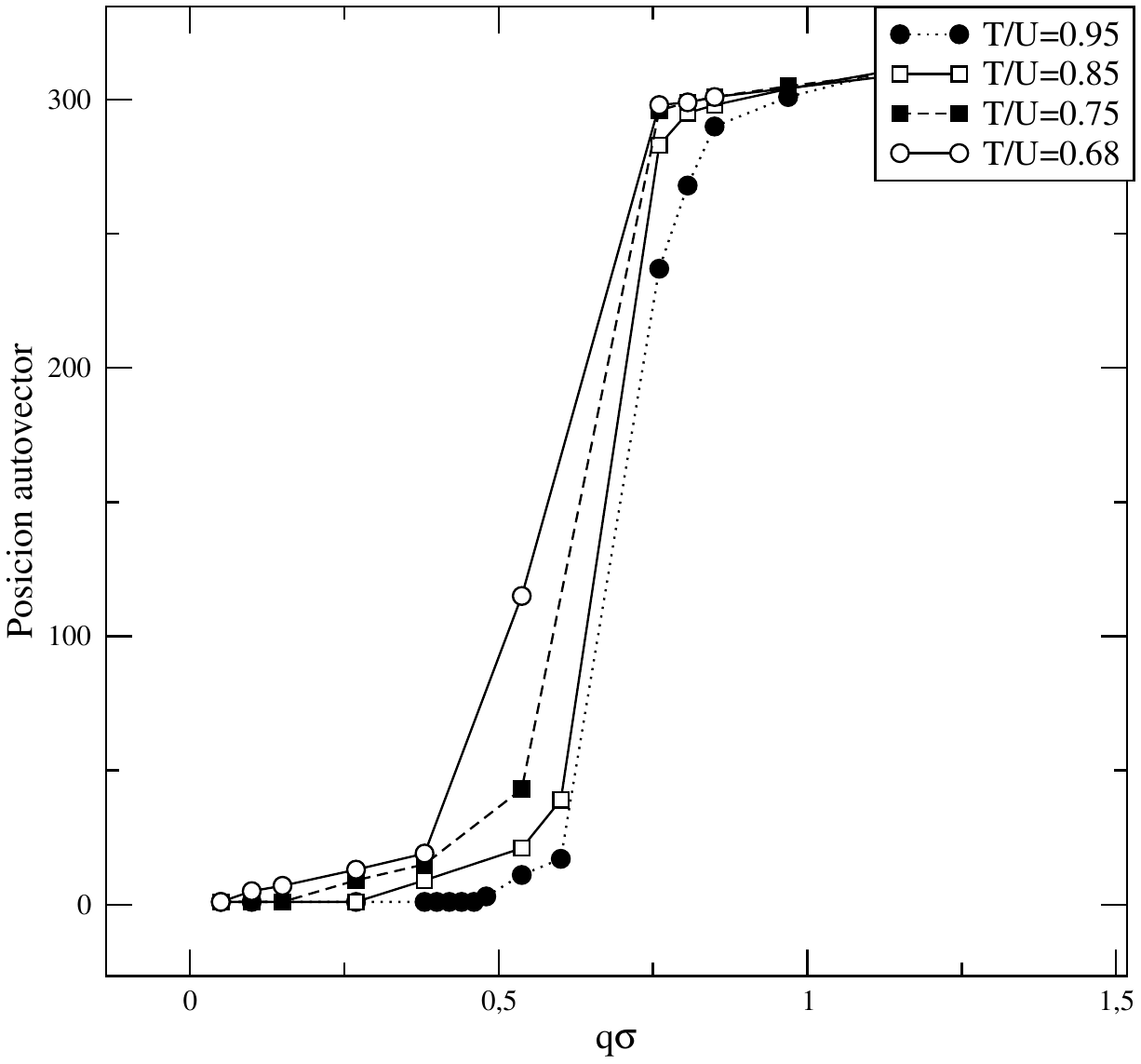}} 
  \end{center}
\caption{\textbf{(a)} Autovalores del autovector representativo de $\rho'(z)$, para cuatro valores de T/U.\textbf{(b)} Posición en el espectro de autovalores completo de autovalor $E(\lambda_{cw})$ para la cuatro temperaturas en (a) representando su variación en q. Se comprueban las dos regiones comentadas en el texto donde se pueden encontrar determinados autovectores.}
  \label{fig:AutovaloreElambda}
\end{figure}

Para analizar las correlaciones incluidas en en los funcionales no-locales hemos determinado el espectro de la función $C^{(2)}(z_{1},z_{2},q)$ diagonalizando esta matriz para cada q, la descomposición espectral subsiguiente para cada modo queda como,
\begin{equation}
\hat{C}(z_{1},z_{2},q)=\sum_{\lambda}E(\lambda,q)\phi_{\lambda}^{q}(z_{1})\phi_{\lambda}^{q}(z_{2})
\end{equation}
La forma concreta de los autovectores permite indagar la forma en que están presentes las ondas capilares en la función de correlación. Para ello retomamos las expresiones que daban lugar a la ecuación TZ,
\begin{equation}
\delta \Omega[\rho_{0}+\Delta\rho]=\frac{1}{2 A\beta}\int dz_{1}\int dz_{2}\sum_{\vec{q}}C^{(2)}(z_{1},z_{2},\vec{q})\Delta\rho(\vec{q},z_{1})\Delta\rho(-\vec{q},z_{2})
\end{equation}
introducimos la descomposición espectral, reagrupamos términos e imponemos fluctuaciones de la forma de ondas capilares, recuérdese la expresión (\ref{eqn:perturbacionTipoCWT}), 
\begin{equation}
\delta \Omega[\rho_{0}+\Delta\rho]=\frac{1}{2 A\beta}\sum_{q}\sum_{\lambda}E(\lambda,q)|\xi_{q}|^{2}\int dz_{1}\phi_{\lambda}^{q}(z_{1})\rho_{0}'(z_{1})\int dz_{2}\phi_{\lambda}^{q}(z_{2})\rho_{0}'(z_{2})
\end{equation}
el conjunto de funciones $\phi_{\lambda}^{q}$ para cada $\lambda$ es ortonormal luego si $\rho_{0}'(z_{2})\in \{ \phi_{\lambda}^{q}\} $ esperamos que para cada q solo exista una posible función cuya proyección con $\rho_{0}'(z)$ sea próxima a la unidad. En la figura (\ref{fig:AutovectorePhiq}) podemos ver para diferentes valores de q el autovector que cumple $\int dz\phi_{\lambda}^{q}(z)\rho_{0}'(z)\simeq 1$ donde hemos permitido variaciones de este valor como se aprecia en la misma figura. Para valores de q alta el autovector característico desarrolla oscilaciones propias de las soluciones de volumen más o menos acopladas a modos particulares según el modelo de interacción, observamos una fuerte dependencia con la temperatura, a temperaturas altas donde la interfase es más blanda y más ancha el modo similar a $\rho_{0}'(z)$ persiste en el espectro sin variar significativamente al aumentar q. Cerca del punto triple, decrece aunque no de modo monótono. Si llamamos al autovalor correspondiente $E(\lambda_{CW})$ y al autovector $\phi_{\lambda_{cw}}^{q}(z)$ podemos expresar que en primera aproximación,
\begin{equation}
\delta \Omega[\rho_{0}+\Delta\rho]=\frac{1}{2 A\beta}\sum_{q}E(\lambda_{CW},q)|\xi_{q}|^{2}\int dz_{1}\phi_{\lambda_{CW}}^{q}(z_{1})\rho_{0}'(z_{1})\int dz_{2}\phi_{\lambda_{CW}}^{q}(z_{2})\rho_{0}'(z_{2})
\end{equation}
los valores de $E(\lambda_{CW},q)$ se pueden ver en figura (\ref{fig:AutovaloreElambda}), donde se representa frente a $q^{2}$ y para $q<<1$ presenta un comportamiento lineal que se pierde para $q\gtrsim 1$ como es de esperar\footnote{Donde además la hipótesis $\int dz\phi_{\lambda}^{q}(z)\rho_{0}'(z)\lesssim1$ se da en un sentido solo aproximado al incrementar q.} por lo visto en el análisis \S\ref{sec:analisis}.\\

Respecto de la posición en el espectro completo de autovalores $E(\lambda_{CW})$ si ordenamos de modo creciente el conjunto $E(\lambda)$ como aparece reflejado en la figura (\ref{fig:AutovaloreElambda}), vemos que hay dos regiones estabilizadas en las que el autovector $\phi_{\lambda_{cw}}^{q}(z)$ se posiciona estando la transición localizada en $q\simeq 0.7$ que equivale a $L\simeq 10\sigma$. Estas dos regiones se corresponden con la zona del espectro donde se localizan los modos correspondientes a fluctuaciones superficiales, analizamos esta cuestión comparando con resultados de simulación.

\subsubsection{Comparación con resultados de simulación}

La función $G^{(2)}$ es accesible mediante experimentos numéricos lo que permite comparar con los autovalores encontrados desde propuestas del funcional de la densidad.\\

\begin{figure}[htp]
  \begin{center}
    \subtop[Espectro de $G^{(2)}$ para LJ de \textit{Steki}]{\label{fig:C2qNaautovectores}\includegraphics[scale=0.70]{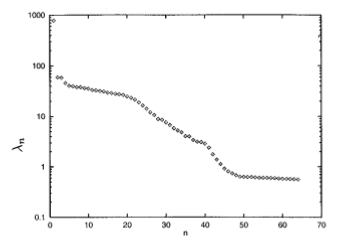}}
    \subtop[Espectro de $G^{(2}$ desde FMT-PY]{\label{fig:C2qLJautovectores}
\includegraphics[scale=0.40]{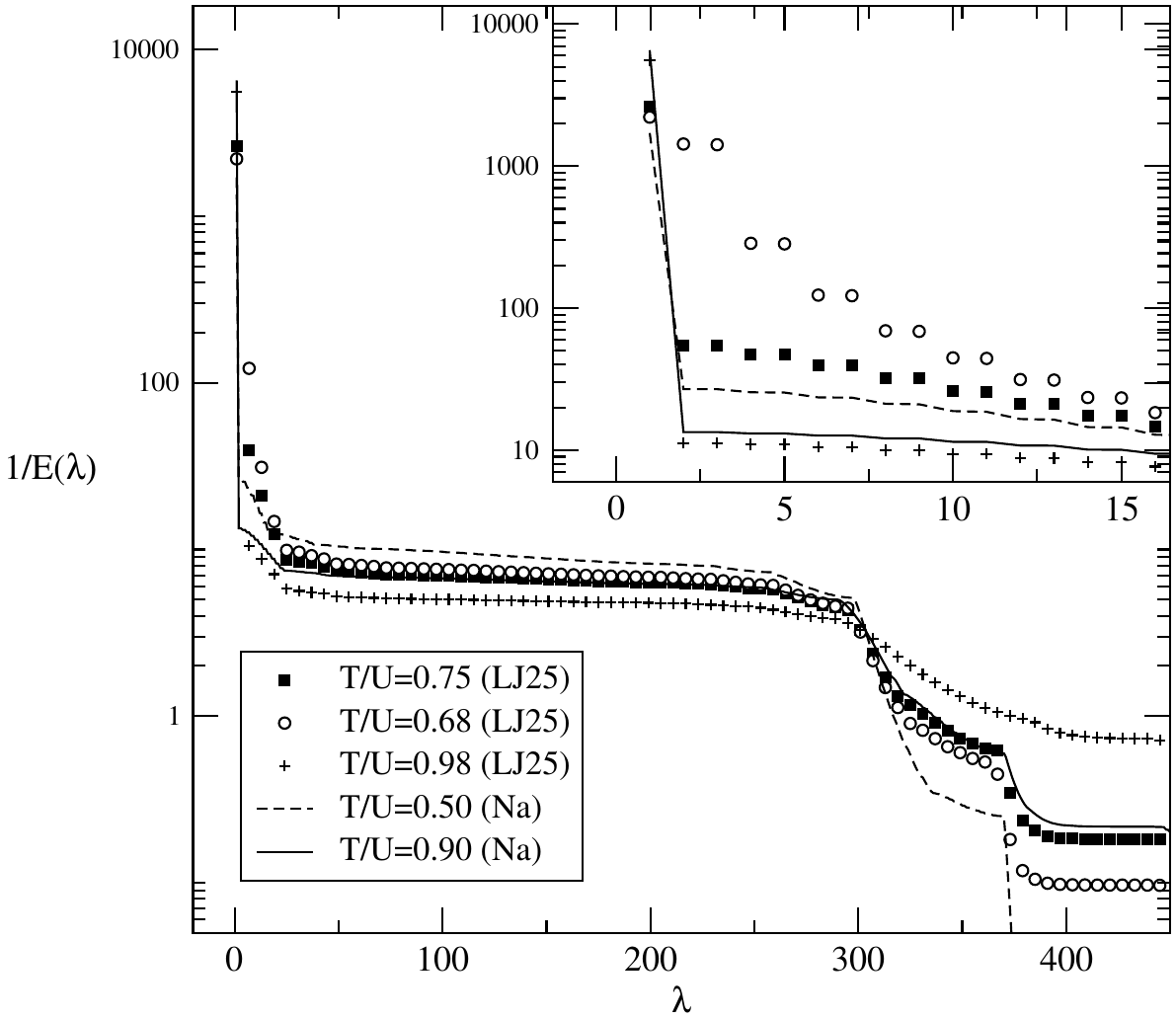}} 
  \end{center}
  \caption{\textbf{(a)} Espectro de $G^{(2)}$ obtenido por \textit{Steki}\cite{steckiJCP7967} para LJ $2.5\sigma$ a T/U=0.75. \textbf{(b)} Espectros de $G^{(2)}$ obtenidos para FMT-PY desde espectro $C^{(2)}$ para un modelo LJ $2.5\sigma$ a T/U=0.68 , 0.75, 0.95 y Sodio a T/U=0.50 y T/U=0.90}
  \label{fig:EspectroCONsimulacion}
\end{figure}

\begin{figure}[htp]
  \begin{center}
    \subtop[Autovectores de $C^{(2)}(q)$ similares a $\rho''(z)$ desde $q\simeq 0$. Modelo LJ$2.5\sigma$. Aprox. FMT-PY. T/U=0.95]{\label{fig:C2qNaautovectores}\includegraphics[scale=0.75]{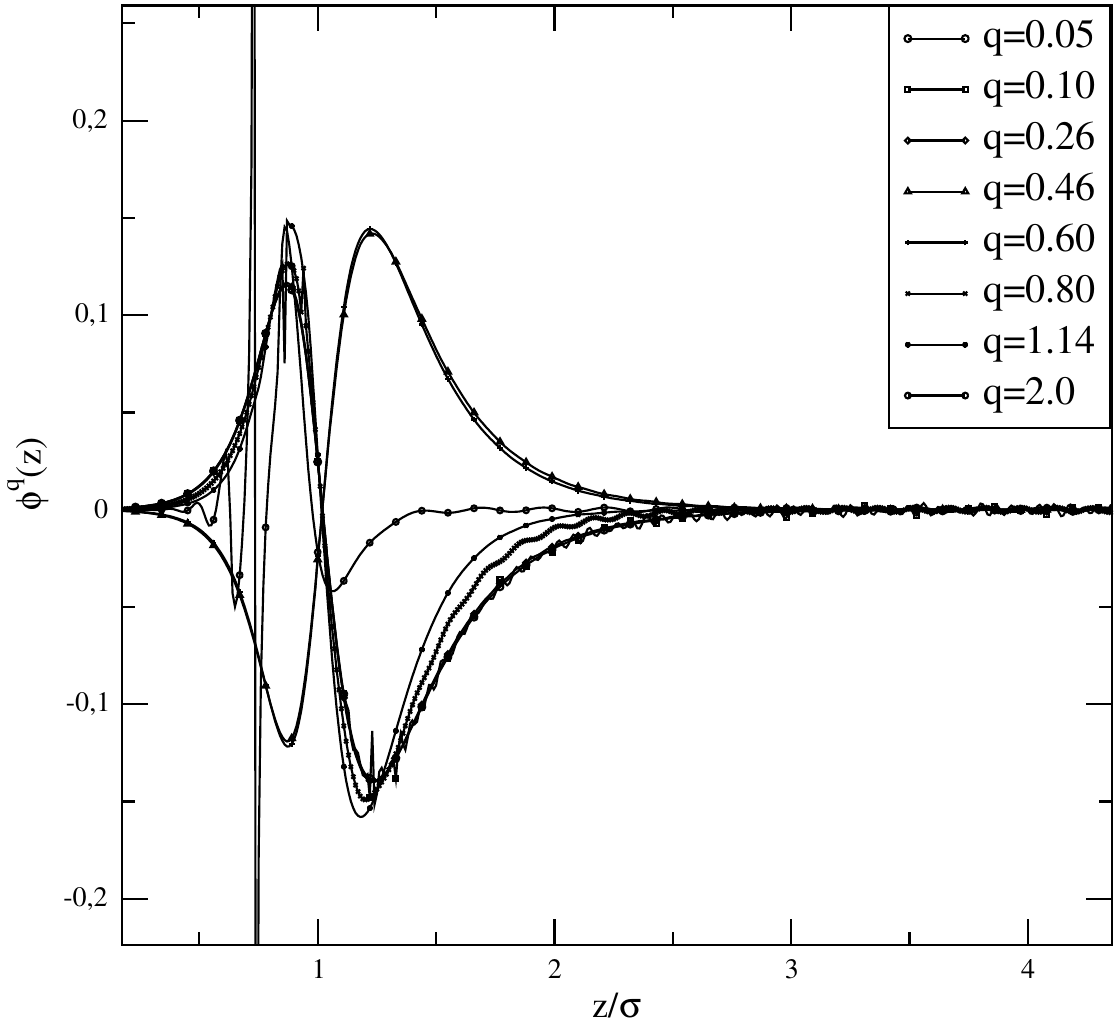}}
    \subtop[Autovectores de $C^{(2)}(q)$ similares a $\rho''(z)$ desde $q\simeq 0$. Modelo LJ$2.5\sigma$. Aprox. FMT-PY. T/U=0.75]{\label{fig:C2qLJautovectores}
\includegraphics[scale=0.75]{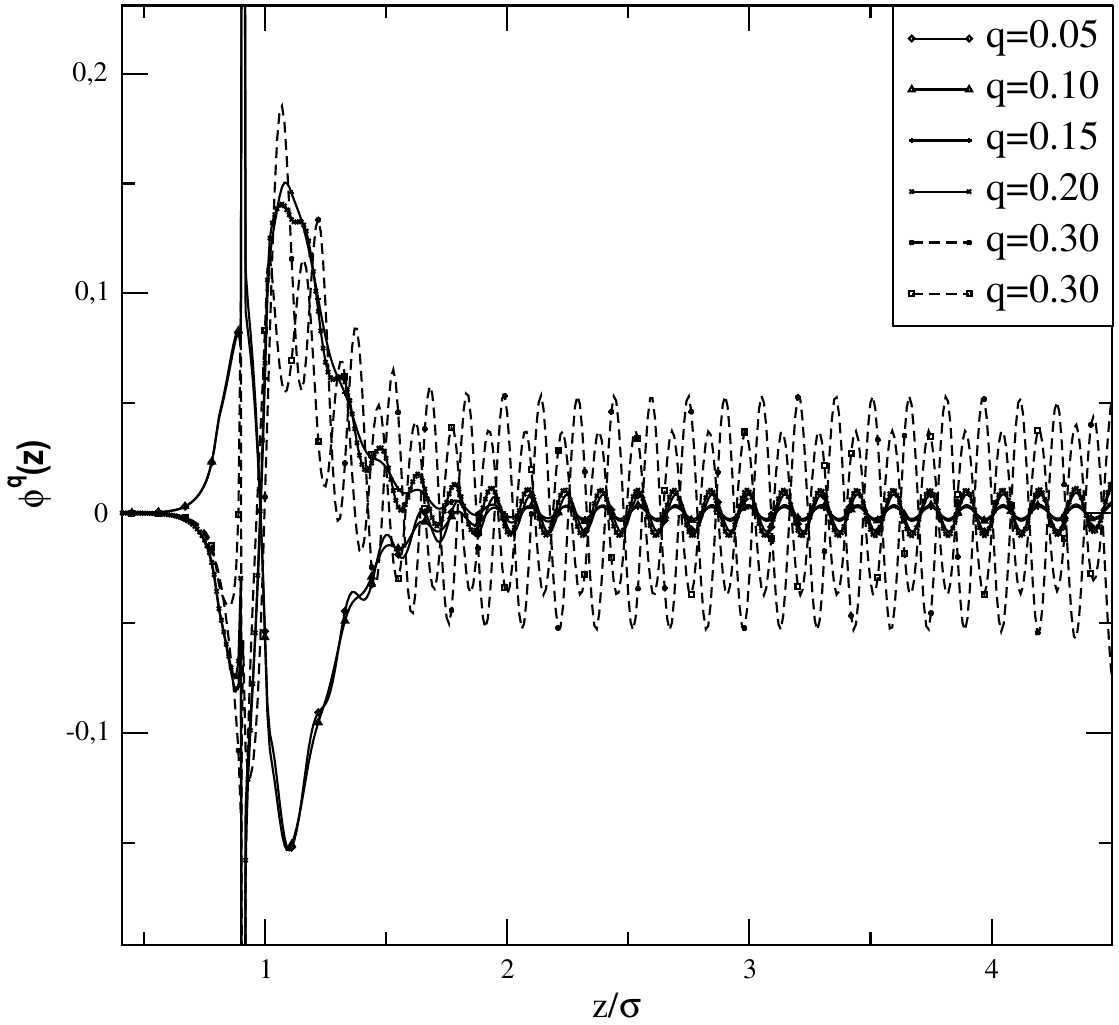}} 
  \end{center}
  \caption{\textbf{(a)} Autovectores de $C^{(2)}(q)$ similares a $\rho''(z)$ desde $q\simeq 0$. Modelo LJ$2.5\sigma$. Aprox. FMT-PY. T/U=0.95 \textbf{(b)} Autovectores de $C^{(2)}(q)$ similares a $\rho''(z)$ desde $q\simeq 0$. Modelo LJ$2.5\sigma$. Aprox. FMT-PY. T/U=0.75, podemos apreciar la relevancia de la temperatura. En esta T/U vemos dos valores a $q=0.30$ cuyas proyecciones son similares y que presentan una zona notablemente oscilante cerca del volumen.}
  \label{fig:AutovectoresEigen2}
\end{figure}

En la figura (\ref{fig:EspectroCONsimulacion}) comparamos los resultados para el espectro en $q\simeq 0$ determinados mediante simulación de Dinámica Molecular para un modelo LJ25 y $T/U=0.75$ obtenido por \textit{J.Steki}\cite{steckiJCP7967} con nuestros resultados en $q=0.05$ para el modelo Lennard-Jones y en el modelo Sodio donde hemos variado la temperatura. Como en el resultado de Steki aparece separado del espectro un autovalor correspondiente a la teoría de ondas capilares clásica y su divergencia en $q\rightarrow0^{+}$, aunque la separación entre este valor y el resto del espectro depende notablemente de la temperatura, para T alta la separación es de hasta tres ordenes de magnitud, para T cercana al punto triple o la inestabilidad frente al sólido la separación es nítida (escala logarítmica) pero mucho menor. El espectro se divide en \textit{plateau} y zonas de decrecimiento, es en estas zonas donde es localiza el modo $\phi_{\lambda_{cw}}^{q}(z)$ y son las zonas donde se presenta mayor discrepancia en $E(\lambda,q)$ al variar la temperatura. Uno de los objetivos de \textit{Stecki} fue localizar los modos con mayor dependencia en q, encontrando la presencia de otro modo, que identificó como característicos de la interfase, equivalente a $\rho''(z)$, en el caso del funcional FMT-PY tenemos los resultados que se visualizan en la figura (\ref{fig:AutovectoresEigen2}), donde buscamos autovectores similares a $\frac{d\phi_{\lambda_{cw}}^{0}(z)}{dz}$.\\

En dicha figura observamos como se comporta este modo en su variación con q, el lector puede ahora comparar con la figura (\ref{fig:rhoH}), y apreciar como los dos perfiles que la teoría de MeD convoluciona\footnote{En el caso de un funcional local bajo aproximación de gradientes cuadrados se puede resolver analíticamente el problema de autovalores para un $f(\rho)=\tau(\rho(z)-\rho_{l})^{2}(\rho(z)-\rho_{v})^{2}$ y resultan ser polinomios de Legendre asociados $P_{2}^{m}(x)$ cuya variable es $x=tanh(z/\xi_{B})$, de modo que cualitativamente los dos primeros autovectores se corresponden con las dos funciones que hemos encontrado aunque en nuestro caso no son completamente simétricas respecto de la interfase como si lo son en este caso analítico. Para otras funciones $f(\rho)$ no es general la obtención de dos modos aislados de un espectro continuo con esta forma funcional.}  son aproximadamente $\phi_{\lambda_{cw}}^{0}(z)$ que se corresponde con $\frac{\rho_{0}(z)}{dz}$ y $\frac{d\phi_{\lambda_{cw}}^{0}(z)}{dz}$ que se corresponde con $\rho_{H}(z)$, el primero persiste en un análisis de los autovectores de $C^{(2)}$ mientras que el segundo no persiste salvo en q bajos y para temperaturas altas. En cuanto a los valores de las proyecciones sabemos que $|<\phi_{\lambda_{cw}}^{0}(z)|\phi_{\lambda_{cw}}^{q}(z)>|\simeq 1$ y $|<\phi_{\lambda_{cw}}^{0}(z)|\phi_{\lambda_{cw}}^{q}(z)>|\simeq 5$ con una variación mayor y sin ser el único autovector con proyección significativamente diferente de cero.\\

\begin{figure}[htp]
  \begin{center}
    \subtop[Espectro de autovalores de $C^{(2)}(q)$, relativo a q=0.05, Modelo LJ$2.5\sigma$. Aprox. FMT-PY. T/U=0.68]{\label{fig:C2qNaautovectores}\includegraphics[scale=0.65]{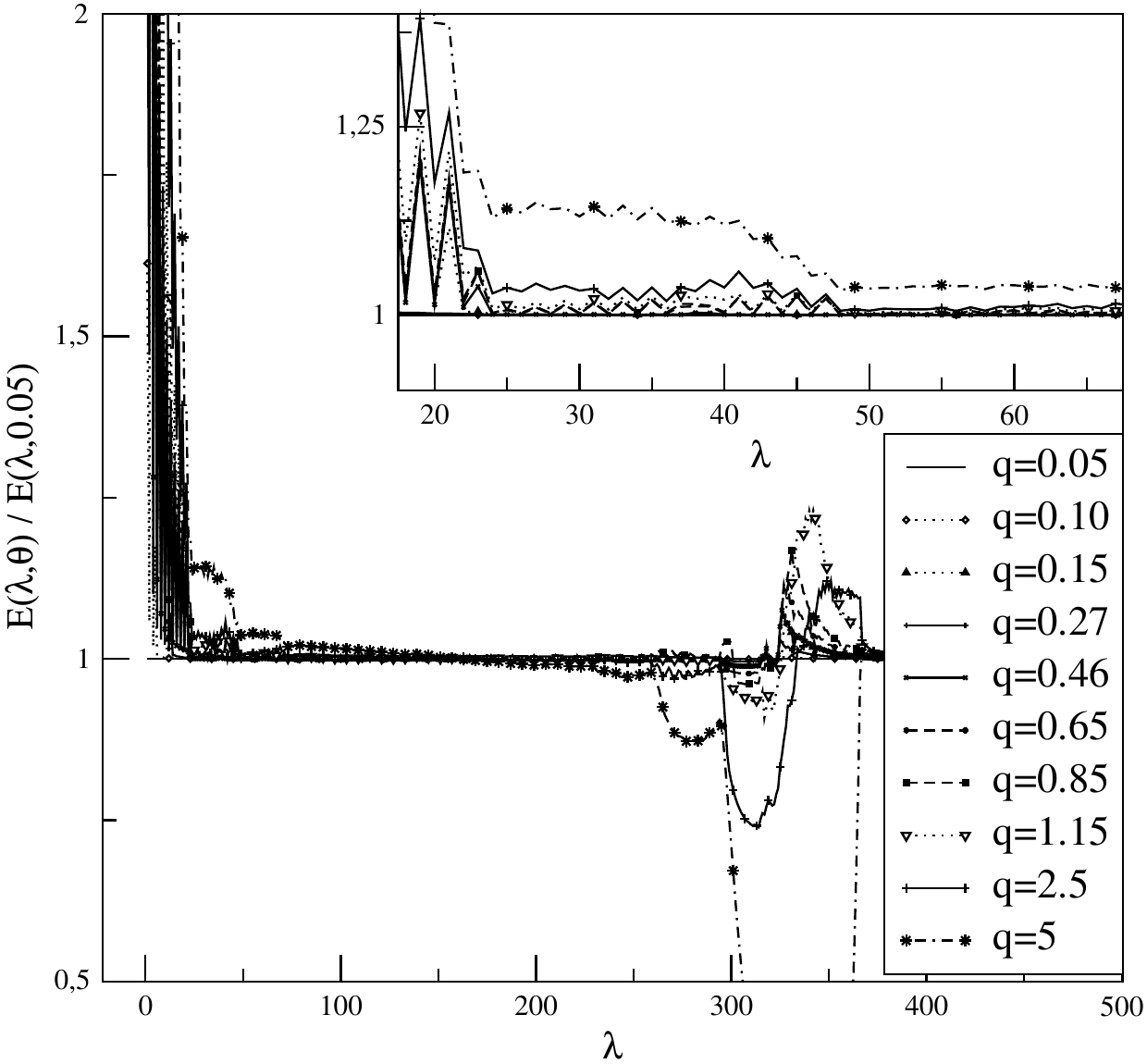}}
    \subtop[Espectro de autovalores de $C^{(2)}(q)$, relativo a q=0.05, Modelo LJ$2.5\sigma$. Aprox. FMT-PY. T/U=0.95]{\label{fig:C2qLJautovectores}
\includegraphics[scale=0.65]{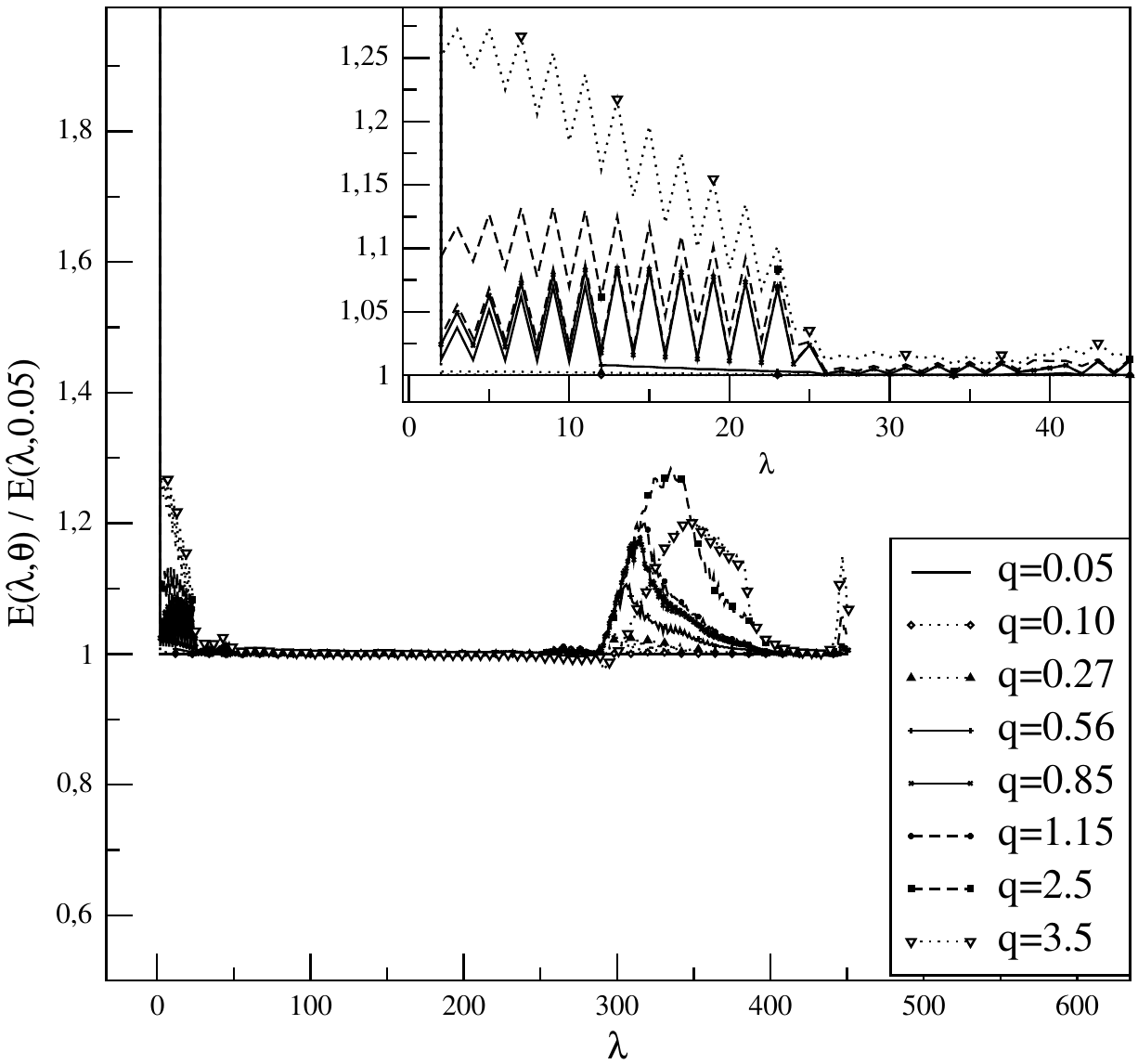}} 
  \end{center}
  \caption{Variación del espectro de autovalores al variar el valor de q. Se determina comparando con el valor en $q=0.05$. Se muestra para dos T/U diferentes y apreciar la relevancia de T/U.}
  \label{fig:EspectroAutovaloresRelativo}
\end{figure}

El análisis completo del espectro de autovalores de $C^{(2)}(q)$ puede verse en la figura 
  (\ref{fig:EspectroAutovaloresRelativo}) que como vemos muestra las zonas donde esta localizada la variación en q y donde deben estar localizados los modos de la interfase, y comparándola con la figura (\ref{fig:EspectroCONsimulacion}) ver que en efecto los valores más estables del espectro correspondientes a autovectores propios de las fluctuaciones de volumen se hallan contenidas en los \textit{plateau} y las mayores variaciones corresponden a fluctuaciones en la superficie.\\
  
En cuanto a un modo de la forma $\rho'''(z)$ estabilizado en un cierto rango de q no nos ha sido posible encontrarlo sugiriendo que dentro de las teorías generalizadas de van der Waals un análisis que incluya las dos primeras derivadas como contribuciones superficiales puede ser suficiente, la proyección $\rho'''(z)$ con funciones $\phi_{\lambda}(z)$ aparece repartida entre multitud de autovectores y aunque ninguno preserva la forma funcional de $\rho'''(z)$ si aparecen siempre localizados en la zona del espectro correspondiente a modos que varían con q.\\

Los resultados de \textit{Stecki} hacen sugerir que para conocer las contribuciones superficiales a la función $G^{(2)}$ es preciso incluir más términos que el usual dado por ec. (\ref{eqn:asintoticoG2cwt}), y propone incluir términos de la forma $<\xi(\vec{R}_{1})\xi(\vec{R}_{2})>\rho^{n}(z_{1})\rho^{m}(z_{1})$ sin embargo no evalúa la presencia o no de estos términos con la temperatura cuestión que a la luz de los resultados mostrados para teorías de van der Waals generalizadas parece necesario.\\ 

Los resultados que obtuvo Steki también son relevantes al respecto de la discusión acerca de $\gamma(q)$, en su estudio usando una definición de superficie intrínseca como una superficie local de Gibbs encontró $<|\xi_{q}|^{2}>\sim \chi(q)$, expresando el hecho de que esta definición de $\xi(\vec{R})$ no esta correlacionada con la superficie intrínseca a nivel microscópico lo que se refleja en fluctuaciones de volumen presentes en $<|\xi_{q}|^{2}>$. Si comparamos el comportamiento de $\gamma(q)$ con el método percolativo-interpolativo, véase la figura (\ref{fig:GammaQsimulaciones}), en el primer caso $<|\xi_{q}|^{2}>\sim q$ para q grande mientras que nuevo criterio muestra un decrecimiento mayor de la predicción de CWT. En consecuencia este último reproduce el limite macroscópico dado por CWT y propone una $\gamma(q)$ monótonamente creciente. En simulaciones numéricas el origen de la dificultad para definir un $\mathcal{H}_{I}$ esta en la necesidad de especificar la superficie intrínseca $\xi(\vec{R})$ para cada configuración del sistema  \S\ref{sec:intruSuperIntrinseca}.

\section{Conclusiones}
\begin{itemize}
\item En el trabajo de \textit{Mecke y Dietrich} se insiste en la separación a todas las escalas de ambos tipos de fluctuaciones para lograr una evaluación correcta del la energía contenida en las ondas capilares, sin embargo, una determinación más completa dentro de su propia teoría lleva a un decrecimiento espúreo en la energía asociada a las ondas capilares a escalas microscópicas indicando el aparentemente contradictorio hecho de que a escalas menores la superficie es progresivamente más rugosa y no posee un \textit{cutoff} propio que impida las divergencias que afectan a CWT.
\item El problema nace de la definición de hamiltoniano de interfase como $\mathcal{H}_{I}[\xi(\vec{R})]=\Omega[\rho_{\xi}(\vec{r})]-\Omega[\rho_{0}(z)]$ ya que la densidad $\rho_{\xi}(\vec{r})$ no puede ser utilizada de modo univoco en todas las escalas para determinar la superficie intrínseca, y el \textit{crossing criterium} se convierte en insuficiente\footnote{Esto se ha visto en una teoría local para la parte de esferas duras en un funcional no-local las posibilidades de definir de este modo $\xi(\vec{R})$ son aun más difíciles.}. Y consecuentemente la separación de las fluctuaciones de superficie y volumen no se realiza correctamente.
\item Resulta por tanto necesario construir una teoría que parta de un perfil de densidad donde han sido primero diferenciadas adecuadamente todas las fluctuaciones superficiales presentes para en un segundo paso construir un esquema que diferencie correctamente fluctuaciones de volumen y superficie y así introducir solo las fluctuaciones superficiales no presentes. Un avance a las propiedades de una teoría con estos objetivos se desarrolla en el primer apéndice.
\item La diagonalización de la función de correlación directa muestra unos resultados para el espectro de autovalores similar al obtenido mediante simulaciones numéricas por Steki. El autovector $\rho'(z)$ esta presente en todo el espectro a altas temperaturas cuando la interfase más difusa y las tensiones superficiales menores, a temperaturas bajas persiste hasta q altos mezcladose levemente con autovectores más propios del volumen. Este fenómeno de mezcla se produce más claramente en $\rho''(z)$ que a valores altos de q ha desaparecido en los autovectores usuales de volumen. 
\end{itemize}

\appendix
\chapter{Perfil intrínseco en la teoría del funcional de la densidad}

\label{sec:apendiceIntrinseco}

Al comenzar, véase \S\ref{sec:experimentosReflectividad}, hemos visto como los resultados experimentales para metales líquidos son compatibles con un esquema interpretativo donde la estructuración en capas del perfil de densidad nace de un perfil intrínseco altamente estructurado sobre el que la teoría de ondas capilares clásica da lugar al perfil observado. Los resultados en \S\ref{sec:relacionCWyDEN} indican que este mismo esquema es adecuado para los perfiles que surgen en la teoría del funcional de la densidad para aproximaciones funcionales que reproduzcan adecuadamente al menos parte de las correlaciones incluidas en el sistema.\\

Dentro de las simulaciones, véase \S\ref{sec:intruSuperIntrinseca}, es posible determinar tanto el perfil de densidad medio como un perfil intrínseco, este último necesita un procedimiento para determinar la superficie intrínseca a partir de las configuraciones moleculares y acabará presentando una dependencia en el nivel de corrugación dado de la superficie intrínseca. Si el procedimiento que determina la superficie intrínseca es adecuado permite obtener resultados otra vez compatibles tanto con los resultados experimentales como con la reinterpretación realizada en \S\ref{sec:relacionCWyDEN} de $\rho_{DF}(\vec{r})$.\\

En este apéndice los objetivos son por tanto, recuperar las principales conclusiones obtenidas dentro de los esquemas de simulación y constituirlas en conceptos sobre los que se articula una aproximación que permita el análisis de la estructura intrínseca dentro la teoría del funcional de la densidad. Comenzamos con la definición de Hamiltoniano efectivo de superficie que es equivalente a la presentada en ec. (\ref{eqn:defHeff}) y en \S\ref{sec:Heff} pero que aquí utilizamos para realizar dos definiciones más que serán de utilidad, un \textit{operador proyección} y una \textit{función de distribución}.

\section{Hamiltoniano efectivo de superficie}
\label{sec:definicionHeffyProyector}
Partimos del espacio de configuraciones del sistema que denominamos $\Gamma$, suponemos que cada configuración permite determinar \textit{una} superficie intrínseca $\xi(\vec{R})$, de modo que tenemos una partición del espacio $\Gamma$ en subespacios $\Gamma_{\xi}$, y tal que que $\cup\Gamma_\xi=\Gamma$. Esto permite formalmente restringir la traza usual del colectivo macrocanónico, que llamaremos traza clásica, y extraer las propiedades de volumen para definir un Hamiltoniano efectivo de superficie $\mathcal{H}[\xi]$,
\begin{eqnarray}
\Xi[T,V,\mu] &=&Tr_{cl}\left[ e^{-\beta(\mathcal{H}_{N}-\mu N)}\right] =\sum_{N=N_{0}}^{\infty}\frac{e^{\beta\mu N}}{\Lambda^{3N}N!}\sum_{\vec{r}^{N}}e^{-\beta U_{N}}\nonumber \\
&= &\underset{\{\xi\}}{Tr}\left[ \sum_{(N,\vec{r}^{N})\in \Gamma_{\xi}}\frac{e^{-\beta(U_{N}-\mu N)}}{\Lambda^{3N}N!}\right] =e^{\beta pV}\underset{\{\xi\}}{Tr}\left[ e^{-\beta\mathcal{H}[\xi]}\right] 
\end{eqnarray}
ya que sabemos que\footnote{Consideramos por tanto despreciables términos superficiales del sistema en contacto con posibles paredes.} $log \Xi=-\beta pV-\beta\gamma_{lv} A_{0}$. Hemos escrito la presión de coexistencia y la tensión superficial macroscópica líquido-vapor así como el área también macroscópica de la interfase. La traza clásica en el subespacio $\Gamma_{\xi}$ la notaremos por $Tr_{cl}^{(\xi)}$ y se respeta la propiedad $Tr_{cl}=Tr_{(\xi)}Tr_{cl}^{(\xi)}$. De modo que hemos definido,
\begin{equation}
e^{-\beta\mathcal{H}[\xi]}\equiv \sum_{\Gamma_{\xi}}\frac{e^{-\beta(U_{N}-\mu N+pV)}}{\Lambda^{3N}N!}=Tr_{cl}^{(\xi)}\left[ e^{-\beta(\mathcal{H}_{N}-\mu N)}\right] 
\end{equation}

Introducimos un \textit{operador proyección} del espacio $\Gamma$ al subespacio $\Gamma_{\xi}$ que permita simplificar y clarificar tanto la notación como la definición de $\mathcal{H}[\xi]$. Dado dicho operador $\Lambda(\xi,\vec{r}^{N})$ el hamiltoniano efectivo queda definido por la relación\footnote{Notar la analogía con los métodos del grupo de renormalización\cite{RevModPhys.55.583,BOOK-NigelGoldenfeld}.},
\begin{equation}
e^{-\beta\mathcal{H}[\xi]}\equiv Tr_{cl}\left[ \Lambda(\xi,\vec{r}^{N})e^{-\beta (\mathcal{H}_{N}-\mu N)} \right]\equiv\mathcal{Z}[\xi]
\end{equation}
Como consecuencia este operador ha de tener necesariamente las siguientes dos pro\-pie\-da\-des,
\begin{itemize}
\item $\Lambda(\xi,\vec{r}^{N})\geq0$ de esta manera $e^{-\beta\mathcal{H}[\xi]}$ será positivo.
\item $\underset{\{\xi\}}{\sum}\Lambda(\xi,\vec{r}^{N})=1$, esto nos garantiza que $\underset{\{\xi\}}{Tr}\left[e^{-\beta\mathcal{H}[\xi]}\right]=Tr_{cl}\left[e^{-\beta (\mathcal{H}_{N}-\mu N)}\right]$
\end{itemize}
\section{Función de distribución densidad en $\Gamma_{\xi}$}
Definimos en este espacio restringido $\Gamma_{\xi}$ una función de distribución dada por,
\begin{equation}
\rho_{\xi}(\vec{r})=\frac{1}{\mathcal{Z}[\xi]}Tr_{cl}^{(\xi)}[\hat{\rho}_{op}e^{-\beta\mathcal{H}_{N}(\vec{r}^{N})}]\equiv \frac{1}{\mathcal{Z}[\xi]}\sum_{(N,\vec{r})\in\Gamma_{\xi}}e^{\beta U_{N}-\mu N+pV}\hat{\rho}_{op}(\vec{r})
\end{equation}
donde $\hat{\rho}_{op}(\vec{r})$ es el usual,
\begin{equation}
\hat{\rho}_{op}(\vec{r})=\sum_{i=1}^{N}\delta^{(3)}(\vec{r}-\vec{r_{i}})=\sum_{i=1}^{N}\delta^{(1)}(z-z_{i})\delta^{(2)}(\vec{R}-\vec{R_{i}})
\end{equation}
resulta ahora cómodo expresar que:
\begin{equation}
\rho_{\xi}(\vec{r})=\frac{1}{\mathcal{Z}[\xi]}Tr_{cl}[\Lambda(\xi,\vec{r}^{N})\hat{\rho}_{op}e^{-\beta\mathcal{H}_{N}(\vec{r}^{N})}]
\end{equation}

\section{Estudio bajo el Funcional de la Densidad}

En el funcional de la densidad el tratamiento que hasta el momento se utilizado consiste en escribir,

\begin{equation}
 e^{-\beta(\Omega[\rho_{\xi}]-\Omega[\rho])}\equiv e^{-\beta\mathcal{H}_{I}[\xi]}\equiv Tr_{cl}\left[ \Lambda(\xi,\vec{r}^{N})e^{-\beta(\mathcal{H}_{N}-\mu N)}\right] 
\end{equation}
lo que teniendo en cuenta que,
\begin{equation}
e^{-\beta\Omega[\rho_{\xi}]}=Tr_{cl}\left[ \left( e^{-\beta V_{ext}(\vec{r}^{N};\xi)}\right) e^{-\beta(\mathcal{U}_{N}+\mathcal{K}_{N}-\mu N)}\right] 
\label{eqn:hipotesisDFTdesdeDEN1}
\end{equation}

equivale a afirmar que el proyector que esperamos diferencie adecuadamente los subespacios $\Gamma_{\xi}$ es un potencial externo a un cuerpo, que se define vía superficie de Gibbs local o vía crossing criterium para funcionales locales. Vamos a analizar con detenimiento sus implicaciones, para ello utilizamos una notación compacta para $\Xi[T,V,\mu;u]$ donde $u(\vec{r})=\beta\mu-\beta v_{ext}(\vec{r})$, que explícitamente diferencie los términos expresables como un potencial externo\footnote{Es la notación usada en\cite{Chayes} y \cite{ConvexidadWDAcalliol}.}.\\

Escribimos  $\mathcal{H}[\vec{r}^{N}]=U_{N}[\vec{r}^{N}]$ y la parte cinética se integra para dar la longitud de onda térmica de \textit{de Broglie} $\Lambda$, definimos una medida en el espacio de la fases $\Gamma$, la medida macrocanónica, mediante,

\begin{equation}
d\mu(\omega)=e^{-\beta U_{N}[\omega]}d\omega
\end{equation}
\begin{equation}
\int d\omega=\sum_{N}\frac{1}{\Lambda^{dN}N!}\int d\vec{x}_{1}... d\vec{x}_{N}
\end{equation}
de modo que si expresamos el término correspondiente al potencial externo a un cuerpo, es decir $\sum_{i}v_{ext}(\vec{r}_{i})$ como $<\hat{\rho}^{(1)}(\omega)|u)>=\int_{V}d\vec{r}^{N}\hat{\rho}^{(1)}(\vec{r};\omega)u(\vec{r})$, recordando ec. (\ref{eqn:operadordensidad}), tenemos de modo compacto,
\begin{equation}
\Xi[u]=\int d\mu(\omega) e^{-<\hat{\rho}|u>}
\end{equation}
Diferenciamos en el formalismo las partículas $\vec{x}_{0},...,\vec{x}_{k}$, es decir, $\omega=(\tilde{\omega},\vec{x}_{0},...,\vec{x}_{k})$ el resultado explicito es,
\begin{eqnarray}
\Xi[u] &=&\int d\vec{x}_{0}...d\vec{x}_{k}\int d\mu(\tilde{\omega}) e^{-<\hat{\rho}(\omega)|u>}= \nonumber\\
         &=&\int d\vec{x}_{0}...d\vec{x}_{k}e^{-<\hat{\rho}(\vec{x}_{0}...\vec{x}_{k})|u>}\int d\mu(\tilde{\omega}) e^{-<\rho(\tilde{\omega})|u>}\nonumber \\
         &=&\int d\vec{x}_{0}...d\vec{x}_{k}e^{-<\hat{\rho}(\vec{x}_{0}...\vec{x}_{k})|u>}e^{-\beta\tilde{U}(\vec{x}_{0}...\vec{x}_{k})}\cdot \nonumber\\
&&\cdot\int d\mu(\tilde{\omega}) e^{-<\hat{\rho}(\tilde{\omega})|u>}e^{-\beta U[\vec{x}_{0}...\vec{x}_{k};\tilde{\omega}]} \\
\end{eqnarray}
hemos definido $\tilde{U}(\vec{x}_{0},...,\vec{x}_{k})=\frac{1}{2}\sum_{i,j}\phi(x_{i},x_{j})$ para potenciales a pares, el índice i recorre $\omega$ y el índice j se restringe a $(\vec{x}_{0},...,\vec{x}_{k})$. Y por tanto podemos definir una función de partición $\Xi^{*}[\tilde{u}|\vec{x}_{0}...\vec{x}_{k}]$ que surge de un potencial externo definido como $\tilde{u}(\tilde{\omega})=u(\tilde{\omega})+\phi(\vec{x}_{1},\tilde{\omega})+...+\phi(\vec{x}_{k},\tilde{\omega})$
La última igualdad permite escribir,
\begin{equation}
\Xi[u]=\int d\vec{x}_{0}...d\vec{x}_{k}e^{-<\rho(\vec{x}_{0}...\vec{x}_{k})|u>}e^{-\beta\tilde{U}(\vec{x}_{0}...\vec{x}_{k})}\Xi^{*}[\tilde{u}|\vec{x}_{0}...\vec{x}_{k}]
\end{equation}
la función $\Xi^{*}[\tilde{u}|\vec{x}_{0}...\vec{x}_{k}]$ esta bien definida bajo el potencial externo que incluye el original $u(\omega)$ junto con el creado por el conjunto de k partículas fijas.  El formalismo del funcional de la densidad permite obtener $\Xi^{*}[\tilde{u}|\vec{x}_{0}...\vec{x}_{k}]$ así como $\rho(\vec{x}|\vec{x}_{0}...\vec{x}_{k})$ y los proyectores que restringen el espacio $\Gamma$ son funciones $\delta$, funciones que son compatibles con las propiedades dadas en \S\ref{sec:definicionHeffyProyector}.\\
\begin{figure}[htbp]
\begin{center}
\includegraphics[width=0.95\textwidth]{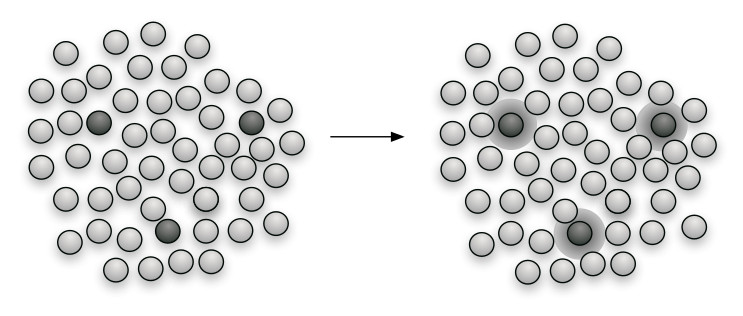}
\caption{Se ilustra la diferencia indicada en el texto entre mirar un conjunto de partículas y permitir un proceso de \textit{coarse-graining} en sus posiciones.}
\label{fig:FigureVext1}
\end{center}
\end{figure}

En la figura (\ref{fig:FigureVext1}) contiene junto al caso bien descrito, como acabamos de ver, por el funcional de la densidad (izquierda) un caso diferente (derecha) en el que dada una configuración $\tilde{\omega}$ se permite al conjunto $(\vec{x}_{0}...\vec{x}_{k})$ una cierta relajación en sus posiciones, un ejemplo puede ser un sistema homogéneo denso con impurezas que permitimos difundir dentro del resto de partículas, esto introduce inmediatamente correlaciones entre el conjunto de grados de libertad $\tilde{\omega}$ y el conjunto $(\vec{x}_{0}...\vec{x}_{k})$. En el caso de que nos interesen las funciones de distribución en $\tilde{\omega}$ permitiendo este proceso de relajación, el problema ya no es representable mediante un potencial externo $\tilde{u}$ causado por k-partículas y tenemos que describir el sistema mediante un $e^{-\beta \tilde{W}}$ que no es un potencial a un cuerpo respecto de las variables $\tilde{\omega}$ y obliga a tener en cuenta correlaciones inducidas entre las propias $\tilde{\omega}$.\\

El ejemplo más clásico consiste en diferenciar una partícula, la función de partición macrocanónica \textit{restringida} a tener una partícula en $\vec{x}_{0}$ permite obtener, mediante el formalismo del funcional de la densidad, la función de distribución condicional $\rho(\vec{x}|\vec{x}_{0})$ de modo que si suponemos el sistema total en ausencia de campo externo $\rho(\vec{x}|\vec{x}_{0})=\rho_{u}g(|\vec{x}-\vec{x}_{0}|)$ y en consecuencia recuperamos estructuralmente la función de distribución radial. Si relajamos levemente el sistema en la coordenada $x_{0}$, resta integrar una \textit{variable colectiva} $\lambda_{0}$ en su lugar,
 
\begin{equation}
\Xi[u]=\sum_{\lambda_{0}}\int d\tilde{\omega}e^{-W(\tilde{\omega})}e^{-<\hat{\rho}(\tilde{\omega})|u>}\int_{R_{0}}e^{-<\hat{\rho}(x_{0})|u>}e^{-W[x_{0},\tilde{\omega}]}
\end{equation}
la última integral representa un promedio, dada la configuración $\tilde{\omega}$ de la variable $\vec{x}_{0}$ sobre una cierta región $R_{0}$, e integrando sobre la variable colectiva $\lambda_{0}$ recuperamos la función de partición y con ella la termodinámica del sistema.\\

En un líquido denso la función $g(|\vec{x}-\vec{x}_{0}|)$ puede ser una buena descripción del objeto que realmente aparece en el formalismo $\rho(\vec{x}|\lambda_{0})$ y también aun en el límite macroscópico la descripción del funcional de la densidad puede ser adecuada para la regla de suma que me permite obtener $pV$, pero en escalas en que la diferencia entre el tratamiento de $\vec{x}_{0}$ y $\lambda_{0}$ se hace relevante la aproximación falla al no diferenciar adecuadamente configuraciones que deben ser diferenciadas por un proyector  $\Lambda(\lambda_{0}|\vec{x}^{N})$ y no un proyector $\Lambda(\vec{x}_{0}|\vec{x}^{N})$ lo que implica por tanto que su energía libre efectiva $\mathcal{H}[\lambda_{0}]$ aproximada por el hamiltoniano efectivo $\mathcal{H}[\vec{x}_{0}]$ es inadecuada al describir el sistema en \textit{estas escalas}.\\

\begin{figure}[htbp]
\begin{center}
\includegraphics[width=0.95\textwidth]{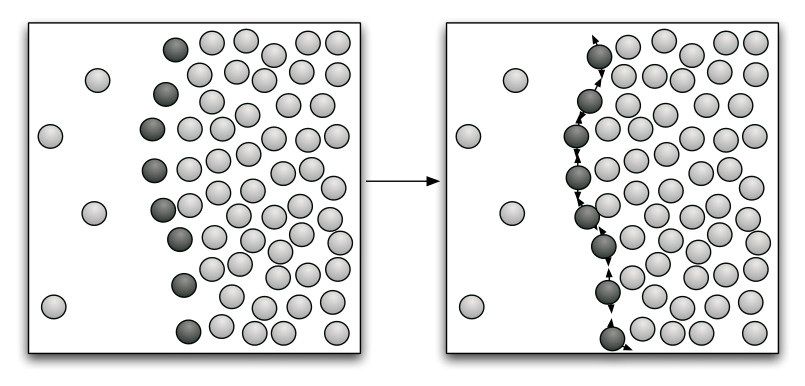}
\caption{Interfase conceptualizada por una capa de partículas oscuras, a la izquierda sus posiciones están fijas, a la derecha representamos la existencia cierta relajación. El problema dado a la izquierda es representable por un potencial externo creado por las partículas oscuras, en el problema de la derecha la existencia de correlaciones efectivas impide dicha representabilidad.}
\label{fig:FigureVext2}
\end{center}
\end{figure}

Esta idea es aplicable al caso de proyectores que intenten separar el espacio de las fases en regiones $\Gamma_{\xi}$,  véase la figura (\ref{fig:FigureVext2}). Si definimos dicho proyector únicamente aludiendo a un potencial externo sobre el resto de las partículas tenemos una descripción conceptualmente diferente de la de un proyector construido de modo que es posible relajar la posición de las partículas de dicha superficie intrínseca, este último caso se relaciona con el criterio operacional-percolativo donde la optimización en la elección de la superficie intrínseca no puede ser descrita con el proyector implícito en el formalismo del funcional de la densidad.\\

Concretando más, estamos diciendo que cada definición de $\mathcal{H}_{I}[\xi]$ posee una definición implícita o explicita de $\Lambda(\xi,\omega)$,
\begin{eqnarray}
\Xi[u] &=& \int d\mu(\omega) e^{-<\hat{\rho}|u>}=\nonumber \\
         &=& \int d\mu(\omega)\left\lbrace \int D\xi \Lambda(\xi,\omega) \right\rbrace  e^{-<\hat{\rho}|u>}=\nonumber\\
         &=& \int D\xi e^{-\beta\mathcal{H}[\xi]}
\end{eqnarray}

Para hallar el proyector correspondiente a la superficie de Gibbs, procedemos a dividir el sistema en $\mathit{l}$ columnas, $C_{0},...C_{l}$ y definir $\xi_{G}=(\xi_{0},...,\xi_{l})$ una configuración $\omega$ es compatible con $\xi_{G}$ si y solo si el numero de partículas contenidas en cada una de las columnas viene dado por $\tilde{N}_{i}=\xi_{i}\Delta\rho A_{0}$ en consecuencia el operador proyección se expresa como:
\begin{eqnarray}
\Lambda(\xi_{g},\omega)&=&\delta\left(\tilde{N}_{0}-\int dz\int_{C_{0}} d\vec{s} \int d\mu(\omega) \hat{\rho}(z,\vec{s};\omega)\right)\cdots \nonumber \\
&\cdots&\delta\left(\tilde{N}_{l}-\int dz\int_{C_{l}} d\vec{s} \int d\mu(\omega) \hat{\rho}(z,\vec{s};\omega)\right)
\end{eqnarray} 
 el hecho de que se exprese como producto de funciones $\delta$  que incluyen solo densidades $\rho^{(1)}$ señala que es representable mediante un potencial externo a un cuerpo, del mismo modo que diferenciar la posición de k-partículas\footnote{Si lo expresamos de modo continuo puede ser expresado como $\Lambda(\xi_{q}-\int dz \rho_{q}(z))$. Si además \textit{imponemos} que el hamiltoniano efectivo resultante sea gaussiano podemos escribir:
\begin{equation}
e^{-\beta h(q)|\xi_{q}|^{2}}=\int d\mu(\omega)\delta(\xi_{q}-\int dz \rho_{q}(z))
\end{equation}
el problema es aplicar esta relación en el caso de $q \simeq \pi/\sigma$ donde la descripción mediante $\rho_{q}(z)$ resulta insuficiente.}.
\\

 En el caso del \textit{crossing criterium} definimos el perfil intrínseco mediante el mínimo del funcional $\Omega[\rho(\vec{r})]$ tal que $\rho(z,\xi(\vec{R}))=\rho^{*}$, y por tanto suponemos la existencia de un potencial externo único $u_{ext}(\vec{r},\xi(\vec{R}),\rho^{*})$ tal que:
\begin{equation}
\rho_{int}(\vec{r};\xi(\vec{R}),\rho^{*})=\left. \frac{\delta ln\Xi[u|\xi(\vec{R}),\rho^{*}]}{\delta u}\right|_{eq}
\end{equation}
en el equilibrio, que como hemos discutido no es cierta para una imagen a escala molecular de la superficie intrínseca\footnote{Más allá de que como indicamos en el capítulo anterior el crossing criterium solo es efectivo en funcionales locales y su desarrollo en gradientes pero para no funcionales más elaborados basados en convoluciones.}.\\

Si aun deseamos seguir entendiendo el problema como representable implícitamente por dicho potencial externo, este dependerá realmente de un conjunto amplio de funciones $\xi(\vec{R})$ ya que no es capaz de diferenciarlas adecuadamente y representar únicamente una de ellas, sin embargo no todas las propiedades reflejaran dependencia del mismo modo. Aquellas donde realmente la mezcla de funciones $\xi(\vec{R})$ sea completa dicha teoría será incapaz de dar un resultado coherente reflejado el \textit{problema de representabilidad} en dicha propiedad. Así la determinación de perfil intrínseco para una $\xi(\vec{R})$ plana, por ejemplo, no debería tener un problema de representabilidad determinante, la energía que cuesta llevar dicho perfil a un perfil corrugado por una función $\xi(\vec{R})$ con relevancia microscópica será, por contra, imposible\footnote{Un caso que nos recuerda este se puede encontrar en la determinación del potencial de interacción en una sustancia mediante el factor de estructura de un material, el resultado será realmente un potencial de interacción dependiente del estado del sistema, propiedades que no dependan de modo importante en esta dependencia pueden ser representadas de modo adecuado por un $\phi_{eff}(z,\rho_{0})$ construido a partir de un estado concreto, pero otras propiedades que impliquen sumar a conjunto completo de estados pueden no ser representables adecuadamente por $\phi_{eff}(z,\rho_{0})$.}.\\

En conclusión vamos a definir de modo aproximado dentro del funcional de la densidad una función de distribución densidad intrínseca que estará cualitativamente bien descrita del mismo modo que, como comentábamos, una aproximación adecuada al problema de la distribución de densidad de una partícula fija en el sistema puede ser una buena aproximación para la función de distribución real en que su posición puede relajarse levemente. Sin embargo una vez definida esta aproximación para $\rho_{\xi}(\vec{r})$ la determinación de $\mathcal{H}_{I}(\xi)$ puede mostrar un grado altamente variable de representabilidad como veremos.\\

\section{Distribución de densidad intrínseca}

Vamos a denominar \textit{distribución de densidad intrínseca} a la función de distribución definida mediante,
\begin{equation}
\tilde{\rho}_{\xi}(z,\vec{R})\equiv\rho_{\xi}(z+\xi(\vec{R}),\vec{R})
\end{equation}
el subíndice $\xi$ indica que la definición es en el espacio restringido de la fases pero puede verse como una dependencia funcional implícita en la función de distribución de la densidad. Esta definición se inspira en la definición de $\tilde{\rho}(z,q_{u})$ que vimos en \S\ref{sec:intruSuperIntrinseca}.\\

La teoría de ondas capilares clásica supone que esta definición descorrelaciona completamente la distribución de densidad intrínseca con la superficie intrínseca y por tanto es posible expresar\footnote{Dentro de superficies intrínsecas donde el cutoff es $q_{u}$, es decir, con una dependencia en el grado de corrugación implícito. Por otra parte la ausencia de la dependencia en $\xi(\vec{R})$ dentro de CWT implica que el perfil intrínseco en dicha teoría podría ser obtenido promediando en el espacio $\Gamma$ completo.} 
\begin{equation}
\tilde{\rho}_{\xi}(z,\vec{R})\overset{CWT}{=}\tilde{\rho}(z)
\end{equation}

Sin embargo los resultados de simulación resaltan la existencia de una correlación entre la superficie intrínseca y esta función de distribución intrínseca. Es conveniente partir de una expresión que permite tratar con perfiles intrínsecos donde exista un cierta dependencia tanto en el plano de la interfase como en la superficie intrínseca a modo de extensión sobre la teoría de ondas capilares clásica. Suponemos por tanto la existencia de una dependencia en las coordenadas transversales que podemos escribir mediante un desarrollo de Fourier en el plano xy.  A primer orden dicha dependencia la suponemos lineal respecto de la amplitudes $\hat{\xi}_{q}$ que caracterizan a $\xi(\vec{R})$ definida implícitamente. De este modo podemos escribir,
\begin{equation}
\rho_{\xi}(z+\xi(\vec{R}),\vec{R})=\tilde{\rho}_{\xi}(z,\vec{R})\simeq\tilde{\rho}(z)+\sum_{q\neq0}\hat{\xi_{q}}e^{i\vec{q}\vec{R}}\tilde{\rho}_{q}(z)
\label{eqn:intrinsecodesarrollo}
\end{equation}

Las conclusiones centrales de esta extensión de CWT son que el perfil de densidad ahora no es la convolución gaussiana del perfil intrínseco sino que posee correcciones de orden $|\xi_{q}|^{2}$,
\begin{equation}
\rho(z,L_{x})\equiv \frac{1}{\mathcal{Z}_{I}}\underset{\{\xi\}}{Tr}\left[\rho_{\xi}(z,\vec{R})e^{-\beta\mathcal{H}[\xi]}\right] =\frac{1}{\mathcal{Z}_{I}}\underset{\{\xi\}}{Tr}\left[ \tilde{\rho}_{\xi}(z-\xi(\vec{R}),\vec{R})e^{-\beta\mathcal{H}[\xi]}\right]
\end{equation}
incluyendo el desarrollo anterior tenemos que\footnote{Podemos escribir de modo explícito ya que que $Tr_{(\xi)}$ equivale a una integral funcional sobre $\xi$,
\begin{equation}
\rho(z,L_{x})\equiv  \frac{1}{\mathcal{Z}_{I}}\int \mathcal{D}[\xi] e^{-\beta\mathcal{H}[\xi]}\rho_{\xi}(z,\vec{R})=\frac{1}{\mathcal{Z}_{I}}\int \mathcal{D}[\xi]  e^{-\beta\mathcal{H}[\xi]}\tilde{\rho}_{\xi}(z-\xi(\vec{R}),\vec{R})
\end{equation}
\begin{equation}
\rho(z,L_{x})\simeq\frac{1}{\mathcal{Z}_{I}}\int \mathcal{D}[\xi] e^{-\beta\mathcal{H}[\xi]}\left[ \tilde{\rho}(z-\xi(\vec{R}))+\sum_{q}\xi_{q}e^{i\vec{q}\vec{R}}\tilde{\rho}_{q}(z-\xi(\vec{R}))\right]
\end{equation}
Donde con $\mathcal{D}[\xi]$ resaltamos dicha integración funcional, y suponemos una normalización adecuada dada por $\mathcal{Z}_{I}$, véase el apéndice \S\ref{sec:funcionesDistribucionAlturas}.},

\begin{eqnarray}
\rho(z,L_{x}) \simeq &&\frac{1}{\mathcal{Z}_{I}}\underset{\{\xi\}}{Tr}\left[\tilde{\rho}_{\xi}(z-\xi(\vec{R})e^{-\beta\mathcal{H}[\xi]}\right]+\nonumber\\ &+&\sum_{|\vec{q}|>0}\frac{1}{\mathcal{Z}_{I}}\underset{\{\xi\}}{Tr}\left[|\xi_{q}|^{2}e^{-\beta\mathcal{H}[\xi]}\right] \frac{d}{dz}\frac{1}{\mathcal{Z}_{I}}\underset{\{\xi\}}{Tr}[\tilde{\rho}_{q}(z-\xi(\vec{R}))e^{-\beta\mathcal{H}[\xi]}]
\label{eqn:denLxDESDEdenINT}
\end{eqnarray}

Para resolver esta expresión en la práctica procedemos como en la teoría de ondas capilares usuales a partir la función de correlación de alturas, los pasos formales se indican en \S\ref{sec:funcionesDistribucionAlturas}.\\

La definición que realicemos del proyector $\Lambda(\xi,\vec{r}^{N})$ permite enlazar de modo continuo los conceptos de perfil medio $\rho(z,L_{x})$ y perfil intrínseco, comparando el primero con $\tilde{\rho}(z,q_{u})$ para un valor de $q_{u}$ nítidamente alejado del nivel molecular $q_{m}$, véase las referencias \cite{PhysRevB.70.235407,2005JPCM17S3493C} y \S\ref{sec:intruSuperIntrinseca}, si partimos de una definición de $\xi(\vec{R})$ sujeta a un nivel de máxima corrugación dado por $q_{u}$.\\

Esto provoca la necesidad de establecer como punto de partida un perfil intrínseco al máximo nivel de corrugación sobre él que sea posible definir adecuadamente un resultado para $\gamma(q)$ que asegure que el procedimiento de obtención de perfiles $\tilde{\rho}(z,q_{u})$ para diferentes $q_{u}$ no procede inconsistentemente incorporando el espectro de ondas capilares, es decir, que para una parte del espectro de ondas capilares no incluyamos artificialmente determinados modos posiblemente ya incluidos.\\

En la teoría de \textit{Mecke-Dietrich} definían el perfil intrínseco mediante una propiedad de equilibrio obtenida al minimizar una aproximación funcional, que hemos visto ya incluye parte del espectro de ondas capilares por tanto el perfil intrínseco por ellos considerado es un perfil $\tilde{\rho}(z,q_{u})$ que interpretan como $\tilde{\rho}(z,q_{m})$ y les lleva a una recorrugación artificial del primero.\\

El siguiente paso es pues determinar bajo el funcional de la densidad $\tilde{\rho}(z,q_{m})$ aproximado como una propiedad de equilibrio, pero bajo una la imagen estructural completamente diferente de la propuesta de Mecke-Dietrich que si elimine completamente la presencia de ondas capilares.

\section{Perfil intrínseco $\tilde{\rho}(z,q_{m})$ como estructura de equilibrio}
\label{sec:perfilesIntrinsecos}
La definición más abrupta de perfiles intrínsecos\footnote{Como los observados en simulación y postulados para en Mercurio líquido en experimentos, por ejemplo.} definida por los experimentales estaba, recuérdese la ec. (\ref{eqn:perfilesIntrinsecosExp}), constituida por función $n_{0}\delta(z)$ que representa la moléculas de la primera capa del líquido, mediante una densidad bidimensional $n_{0}$. Sobre tales perfiles intrínsecos se incorporaría el espectro de ondas capilares completo mientras que el valor de $q_{m}\simeq 2\pi/\sigma$ podría ser entendido como la resolución molecular pero que en el caso de poseer un método adecuado de la determinación del espectro de ondas capilares estaría implícito y no explicito en la forma funcional de dicho espectro\cite{PhysRevLett.91.166103} como vimos en el capitulo anterior.\\

Desde los resultados de simulación esta definición de \textit{primera capa líquida} aparece como un concepto \textit{amplio} sobre el que varias definiciones concretas son posibles, su significado físico sin embargo esta condicionado a valores restringidos del parámetro $n_{0}$ que además permite reproducir propiedades para $\gamma(q)$ con sentido físico. Con todo la condición $\gamma(q)>>\gamma_{lv}$ deja el comportamiento a modos con vector de onda q grande abierto a cierta ambigüedad.\\

El procedimiento físico por el que definimos la superficie intrínseca y consecuentemente obtenemos el perfil intrínseco corresponde a una representación concreta de la estructura intrínseca de la interfase, en ella, la fuerte estructuración proviene de una distribución de probabilidad condicional en la que promediamos sobre la coordenada transversal cuando sobre una determinada capa de moléculas imponemos una densidad y localización determinadas. Su analogía puede ser la interpretación de la función de distribución de pares desde el método de la partícula test de Percus. En este caso $\rho^{(2)}(\vec{r}_{1},\vec{r}_{2})=\rho(\vec{r}_{1})\rho(\vec{r}_{2}|\vec{r}_{1})$ mientras que para el perfil intrínseco $\tilde{\rho}_{\xi}(z)$ buscamos una analogía con $\rho^{(1)}(z,\vec{R}|N_{0}\in\xi(\vec{R}))$ y esta densidad si que la promediamos en $\Gamma$. Si volvemos al razonamiento indicado en anteriormente, hay un promedio sobre la coordenada transversal y en consecuencia no es posible de modo riguroso representar como un campo externo sobre el resto del sistema a efectos de energía libre, aunque a efectos de estructura es una aproximación viable.\\

En el caso del método de la partícula test de Percus representamos la partícula test por un potencial externo en el origen, idéntico al potencial intermolecular, mientras que en nuestro esquema la superficie intrínseca es aquella que pasa por las posiciones moleculares de la primera capa. Esto implica que ec. (\ref{eqn:intrinsecodesarrollo}) posee componentes de Fourier dadas por,
\begin{equation}
\tilde{\rho}_{q}(z)=n_{q}\delta(z)+\rho^{(in)}_{q}(z)
\end{equation}
\begin{equation}
\tilde{\rho}(z)=n_{0}\delta(z)+\rho^{(in)}_{0}(z)
\end{equation}
Dos esquemas diferentes puede ser ideados para representar esta imagen:
\begin{itemize}
\item  Uno en que la primera capa líquida es considerada como un potencial externo donde las corrugaciones de la superficie intrínseca no condicionan la densidad de esta primera capa, de este modo únicamente tendremos una parte $\rho^{(in)}(z)$ y una densidad para la primera capa de $n_{0}$ y por tanto:
\begin{equation}
V_{0}(z_{1},\vec{R}_{1};n_{0},\xi)=n_{0}\int d^{2}\vec{R}_{2}\phi\left( \sqrt{|\vec{R}_{1}-\vec{R}_{2}|^{2}+(z-\xi(\vec{R}_{2}))^{2}}\right) 
\end{equation}
\item Otro donde la primera capa es incluida directamente en la propia estructura de correlación del funcional y donde si que incluimos los términos $n_{0}\delta(z)$ y $n_{q}\delta(z)$ dentro del funcional y en particular estan correlacionados con el resto de las capas de partículas del sistema. 
\end{itemize}
Ambas representaciones pueden ser desarrolladas en la teoría del funcional de la densidad conviene previamente a su introducción aclarar la posición de esta imagen respecto de los resultados previos de simulación y en consecuencia respecto de las definiciones antes realizadas.\\

Nos interesa contrastar esta definición de perfil intrínseco como estructura de equilibrio con la definición dada por $\tilde{\rho}(z,q_{u})$ usual en simulaciones. El primer caso debería ser un perfil aun más estructurado que la segunda definición en su propagación hacia el volumen, ya que la segunda definición presenta un promedio en las fluctuaciones térmicas dadas por diferentes $\xi(\vec{R})$ que para capas moleculares alejadas de la superficie debe producir un cierto amortiguamiento de la estructuración impuesta por la primera capa. Además si nuestras hipótesis son de hecho correctas la diferencia entre un perfil siguiendo una superficie restringida a un modo q posee correcciones de orden lineal de modo que su promedio sobre la distribución de probabilidad de esta $\xi_{q}$ debe anularse. Esto sin embargo no debe suceder si se incluye el espectro completo de ondas capilares y es posible encontrar un amortiguamiento efectivo en las capas más internas haciendo valida la hipótesis que hemos propuesto y haciendo compatibles los resultados para nuestras representación y la forma de los resultados de $\tilde{\rho}(z,q_{u})$ conocidos previamente de simulación.\\
\subsection{Aproximación a $\tilde{\rho}(z,q_{m})$ en la teoría del funcional de la densidad}
Reproducimos la forma que ambas representaciones toman dentro de una teoría del funcional de la densidad,
\begin{itemize}
\item Dentro de la aproximación la estructura intrínseca al equilibrio determinada por la primera de la imágenes, las capas internas de la distribución de densidad son aquellas que minimizan la funcional macrocanónico para el potencial externo definido anteriormente:
\begin{equation}
\Omega_{v}[\tilde{\rho}^{(in)};n_{0},\xi]=\mathcal{F}[\tilde{\rho}^{(in)}(z,\vec{R};\xi)]+\int d^{2}\vec{R}\int dz \tilde{\rho}^{(in)}(z,\vec{R};\xi)\left[ V_{0}(z,\vec{R};n_{0},\xi)-\mu\right] 
\end{equation}
con las condiciones de frontera establecidas por las fases homogéneas en coexistencia, un área transversal dada por $A_{0}$ y un volumen total $V=A_{0}L$.\\

\item La segunda de las aproximaciones separa la contribución ideal y escribe la parte de exceso incluyendo tanto el perfil $\rho^{(in)}(z,\vec{R};\xi)$ como una primera capa correlacionada por tanto con el resto de la distribución de densidad y condicionada a una densidad superficial dada por $n_{0}$.
\begin{equation}
\Omega_{\delta}[\tilde{\rho}^{(in)};n_{0},\xi]=\mathcal{F}_{id}[\tilde{\rho}^{(in)}]+\mathcal{F}_{ex}[\tilde{\rho}^{(total)}]-\mu\int d^{2}\vec{R}\int dz [\tilde{\rho}^{(total)}] 
\end{equation}
Donde hemos escrito\footnote{Para clarificar más la notación se ha escrito un $\rho^{(total)}$ y el $\rho^{(in)}$, se puede ver de que modo se mantiene en lo que resta}:
\begin{equation}
\tilde{\rho}^{(total)}=\tilde{\rho}^{(in)}+n_{0}\delta(z-\xi(\vec{R}))
\end{equation}
\end{itemize} 
Sobre este esquema es conveniente precisar que no toda aproximación funcional permite a priori obtener unos resultados adecuados. El caso más sencillo, que consistente en aplicar una aproximación local para la parte de esferas duras más una aproximación de campo medio para la parte atractiva, no permite diferenciar de modo adecuado la forma en que se puedan correlacionar las capas internas con la primera capa líquida ya que la parte atractiva de campo medio y un potencial externo efectivo dado por las interacciones entre las partículas encierra la misma física. Es el tratamiento de la parte repulsiva quien condiciona, en una aproximación de campo medio, dos formas diferentes de equilibrar la primera capa líquida con el resto de la función de distribución de densidad . Además necesitamos una aproximación en el funcional de la densidad que permita la incorporación de distribuciones de densidad de diferente dimensionalidad como pueda ser una capa bidimensional corrugada junto con una distribución de densidad tridimensional. Un funcional idóneo sería el funcional de la medida fundamental obtenido mediante reducción dimensional y por tanto los resultados concretos que presentemos a partir de ahora estarán referidos esencialmente a este funcional\footnote{Únicamente se compararan algunos resultados con una teoría de la densidad promediada que también da resultados aceptables para confinaciones bidimensionales del sistema, en cualquier caso como veremos solo presentan diferencias notables en algunos casos.}.\\

La comparación de los valores de la energía libre minimizada en ambos procedimientos debe ser convenientemente comparada lo que requiere sumar al valor de $\Omega_{v}$ para el perfil de equilibrio las energías de interacción dentro de la primera capa no incluidas en este método pero si tenidas en cuenta en $\Omega_{\delta}$, esencialmente las correspondientes a un líquido bidimensional de esfera duras con una cola atractiva tratado en campo medio\footnote{Es importante indicar el hecho de que la inclusión de la primera capa líquida en la parte ideal conlleva una divergencia si es evaluada dentro de un funcional en el que se incorporan distribuciones de volumen, esta divergencia nace de fijar los grados de libertad de las moléculas de la capa intrínseca a un plano dentro de un esquema en que el potencial químico es el de volumen. Estos grados de libertad no los incorporamos en el procedimiento de cálculo del perfil intrínseco y serán incorporados al hacer la traza sobre configuraciones de la superficie intrínseca.}.\\

\subsection{Determinación de $\tilde{\rho}(z-\xi_{0})$ perfil intrínseco plano}

\begin{figure}[htbp]
\begin{center}
\includegraphics[width=0.95\textwidth]{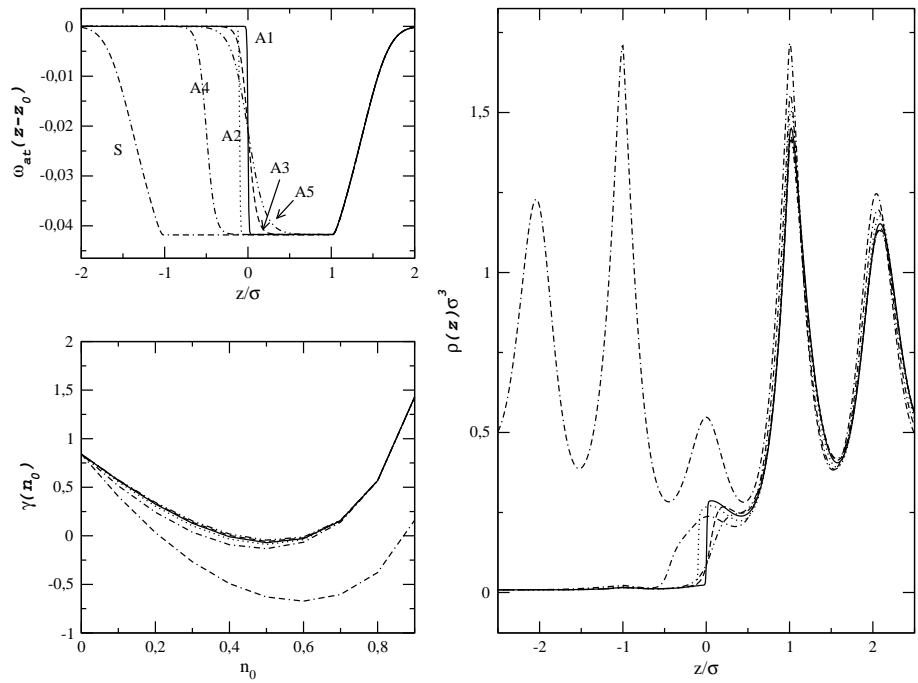}
\caption{Se comparan para un modelo Mercurio T/U=0.65 varios esquemas diferentes de asimetrización. Arriba a la derecha aparece la forma de la interacción entre la capa $n_{0}$ y el resto del perfil $\tilde{\rho}(z)$. Abajo los valores de la energía obtenida al minimizar sin tener en cuenta ni es costo en energía libre de fijar una capa, ni la energía libre contenida en la capa de ocupación $n_{0}$. A la derecha se ven los perfiles para $n_{0}=0.4$, para $n_{0}$ mayores las diferencias son menores.}
\label{fig:ASIMDelta}
\end{center}
\end{figure}

El primer paso en la determinación de los perfiles intrínsecos es su determinación para una configuración plana de la superficie intrínseca, con todo la obtención directa de resultados dentro de un esquema en el colectivo macrocanónico presenta una dificultad para tasas de ocupación altas de la primera capa intrínseca ya que el sistema encuentra favorable el crecimiento de sucesivas capas líquidas en la parte del vapor de la superficie intrínseca. Esto no sucede mediante una definición operacional de $\xi(\vec{R})$ en la que fijamos la capa intrínseca como la primera capa del líquido. Para imponer una condición análoga en nuestro esquema funcional dotamos de una asimetría en la parte atractiva de la primera capa de modo que solo sea atractiva hacia el líquido y sea de modo efectivo la frontera con el vapor. Para introducirla en el primero de los casos, $\Omega_{v}$ , es suficiente con asimetrizar la parte atractiva del potencial externo mientras que en el segundo de los métodos, $\Omega_{\delta}$ , hay que diferenciarla en la expresión de campo medio de la interacción atractiva entre la primera capa, ($n_{0}$ y $n_{q}$), y la parte interna del perfil $\rho^{(in)}(z)$. Lógicamente la parte de interacción dentro de la capa no se ve modificada y tampoco entre las interacciones en $\rho^{(in)}(z)$. A efectos de evaluar su relevancia en el perfil intrínseco se proponen diferentes asimetrizaciones que se puede ver en la figura (\ref{fig:ASIMDelta}).\\

\begin{figure}[htbp]
\begin{center}
\includegraphics[width=1.20\textwidth ,angle=-90]{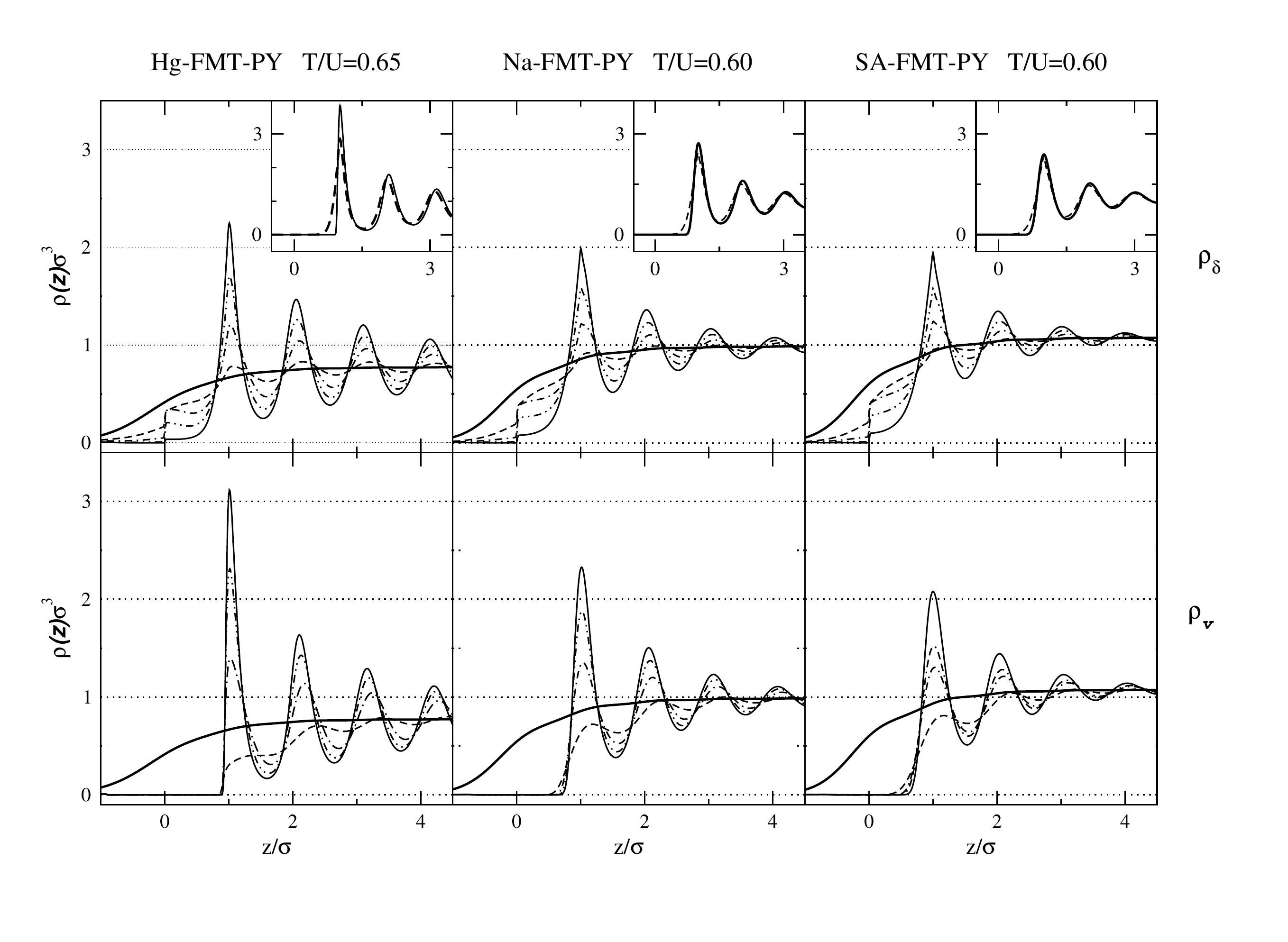}
\caption{Comparación entre perfiles intrínsecos $\rho_{\delta}(z)$ y $\rho_{v}(z)$. Se indica para valores de $n_{0}=0.1$,$n_{0}=0.3$, $n_{0}=0.5$ y $n_{0}=0.7$. La receta asimétrica usada para caso $\rho_{\delta}(z)$ es A1. En los inset se comparan $\rho_{\delta}(z)$ con línea a trazos y $\rho_{v}(z)$ con línea continua, para ocupaciones de $n_{0}=0.9$. Los perfiles son obtenidos para el funcional FMT-PY-MFA.}
\label{fig:DeltavsVextASIM}
\end{center}
\end{figure}

\begin{figure}[htbp]
\begin{center}
\includegraphics[width=1.20\textwidth ,angle=-90]{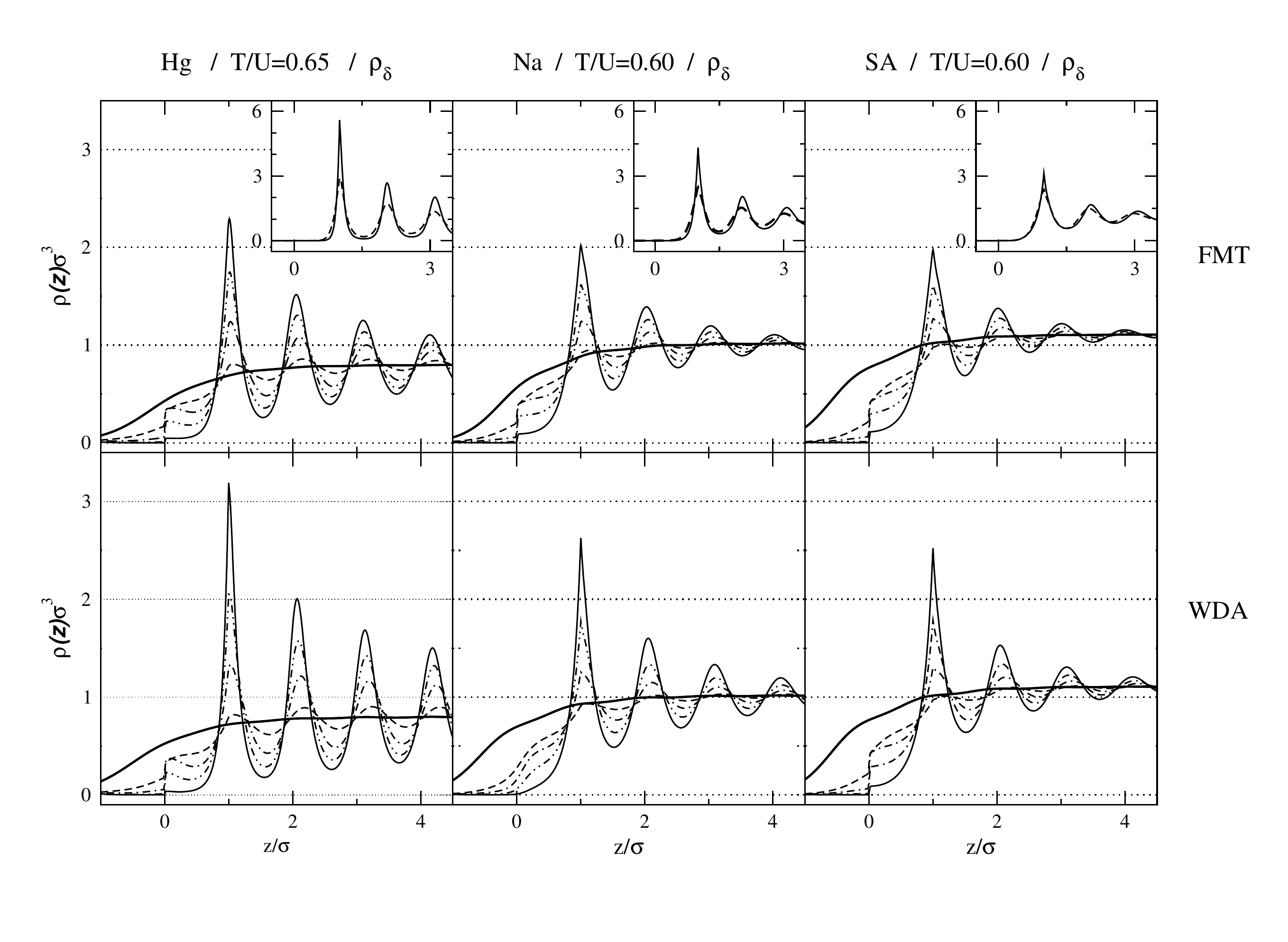}
\caption{Comparación entre perfiles intrínsecos $\rho_{\delta}(z)$ en los funcionales FMT y WDA. Se indica para valores de $n_{0}=0.1$,$n_{0}=0.3$, $n_{0}=0.5$ y $n_{0}=0.7$. La receta asimétrica usada para caso $\rho_{\delta}(z)$ es A1 excepto en WDA para Na que se muestra A5 para comparar suavidad entorno a z=0. En los inset se comparan tasas de  ocupación de $n_{0}=0.9$, WDA línea continua, FMT línea a trazos. La ecuación de estado para ambos funcionales es Carnahan-Starling.}
\label{fig:DeltaASIMwdafmt}
\end{center}
\end{figure}

Evaluamos directamente las expresiones $\Omega_{v}[\rho^{(in)};n_{0},\hat{\xi}_{0}]$ y $\Omega_{\delta}[\rho^{(in)};n_{0},\hat{\xi}_{0}]$  por ambos métodos donde la comparación físicamente relevante se realiza para valores de $n_{0}$ altos.  A efectos comparativos además de FMT, véase la figura (\ref{fig:DeltavsVextASIM}), hemos calculado ambos casos en WDA, véase la figura (\ref{fig:DeltaASIMwdafmt}), en que los perfiles $\rho^{(in)}$ presentan mayor estructuración que FMT para todos los modelos.
 
\subsubsection{Correcciones a la energía libre para $\Omega_{v}$ y $\Omega_{\delta}$.}

La energía libre introducida en el funcional FMT-PY en el caso de una distribución bidimensional corresponde en un modo muy aproximado a la ecuación de estado de discos duros en la teoría de la partícula escalada que podemos escribir como:
\begin{equation} 
\beta f^{2D}_{N}=-\log(1-\eta^{2D})+\frac{\eta^{2D}}{1-\eta^{2D}}
\end{equation}
y obtenemos la energía libre por numero de partículas donde hemos escrito que $\eta^{2D}=\pi R^{2}\rho^{2D}$ con un radio $R=d_{hs}/2$. Por tanto $\beta F_{HS}[\rho^{2D}]=\rho^{2D}\beta f^{2D}_{N}$. Las diferencias entre introducir esta energía libre y la incluida en el propio funcional son mínimas \cite{PhysRevA.42.3382,PhysRevA.44.5025}.\\

\begin{figure}[htbp]
\begin{center}
\includegraphics[width=0.95\textwidth]{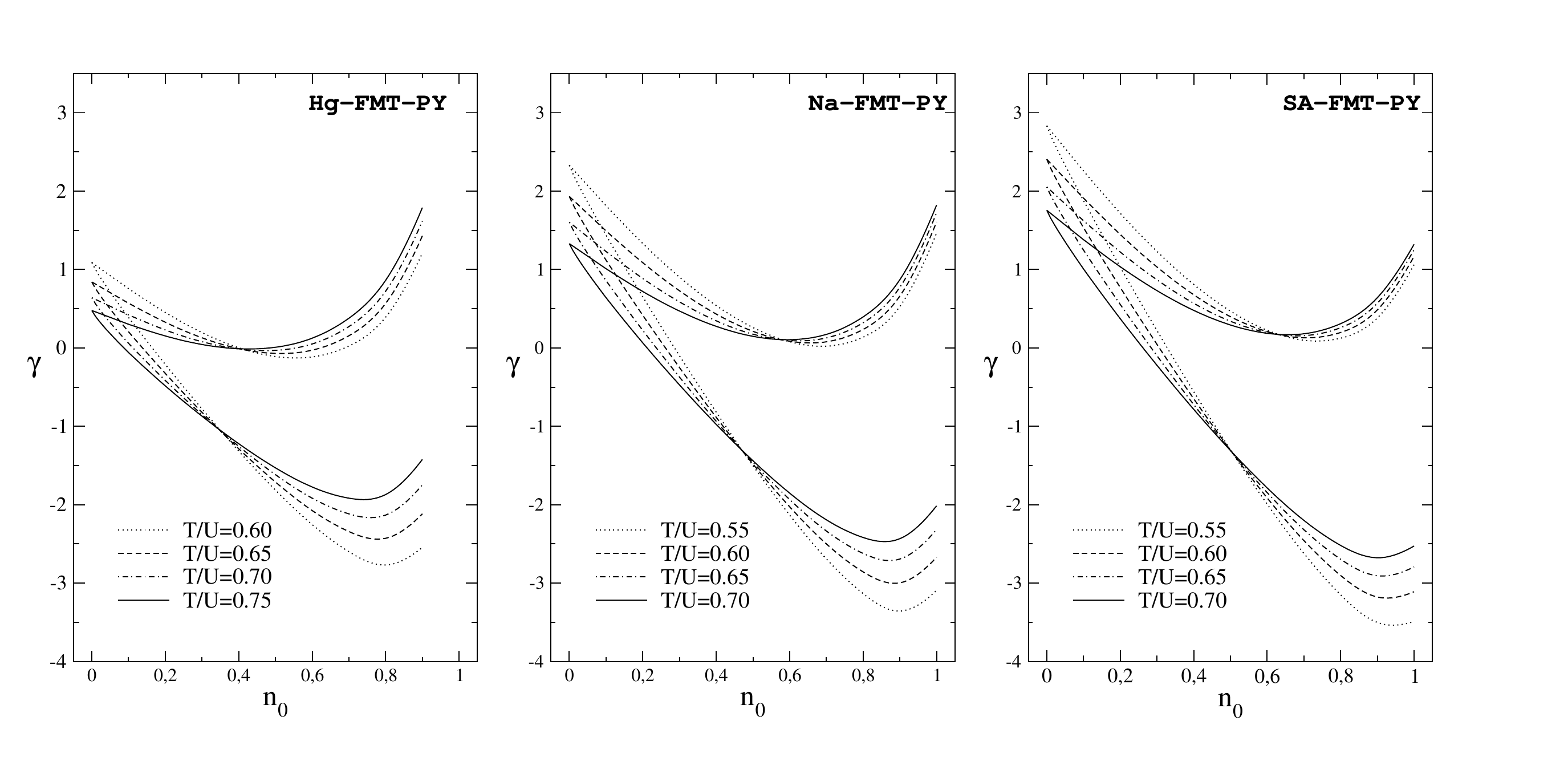}
\caption{Para Mercurio (Hg), Sodio (Na) y \textit{Soft Alcaline} (SA). Funcional FMT-PY. Se indican las tensiones superficiales de minimización del perfil de equilibrio $\rho_{\delta}(z)$ junto con las curvas incluyendo los términos que corresponden a las interacción dentro de la primera capa.}
\label{fig:EnergiasDeltaFMTPY}
\end{center}
\end{figure}

\begin{figure}[htbp]
\begin{center}
\includegraphics[width=0.95\textwidth]{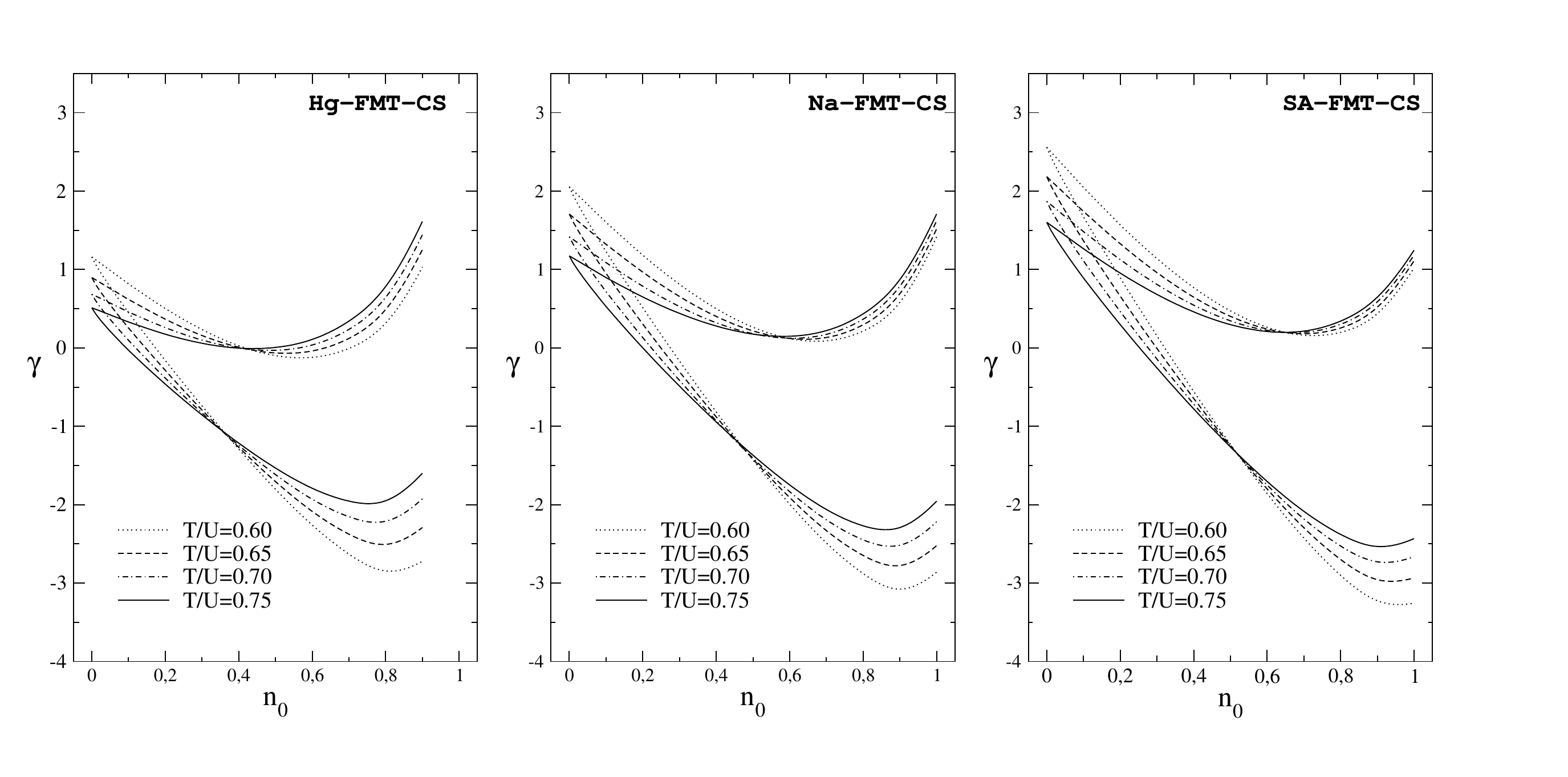}
\caption{Para Mercurio (Hg), Sodio (Na) y \textit{Soft Alcaline} (SA). Funcional FMT-CS. se indican las tensiones superficiales de minimización del perfil de equilibrio $\rho_{\delta}(z)$ junto con las curvas incluyendo los términos que corresponden a las interacción dentro de la primera capa.}
\label{fig:EnergiasDeltaMFTCS}
\end{center}
\end{figure}

\begin{figure}[htbp]
\begin{center}
\includegraphics[width=0.95\textwidth]{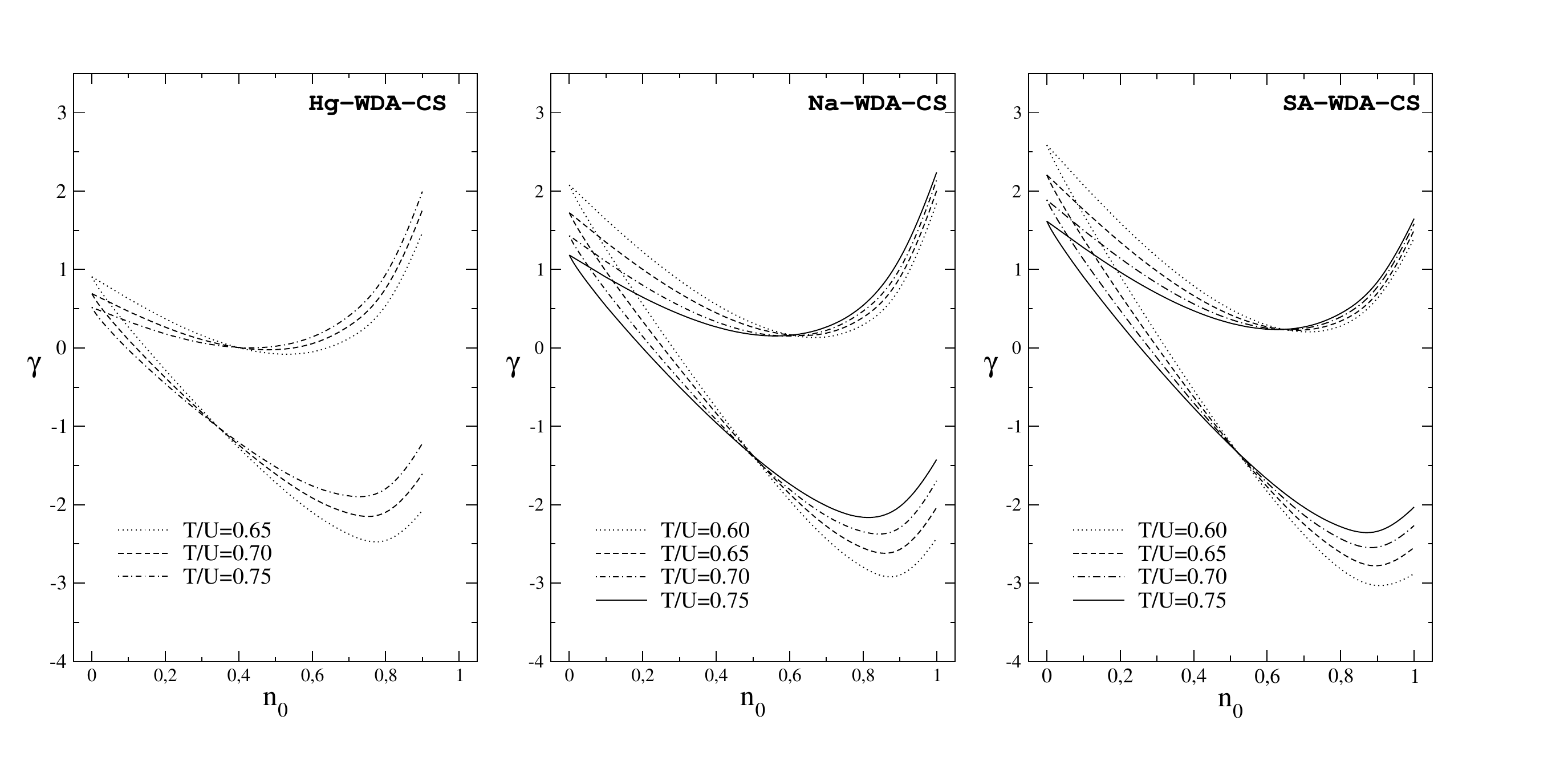}
\caption{Para Mercurio (Hg), Sodio (Na) y \textit{Soft Alcaline} (SA). Funcional WDA-CS. Se indican las tensiones superficiales de minimización del perfil de equilibrio $\rho_{\delta}(z)$ junto con las curvas incluyendo los términos que corresponden a las interacción dentro de la primera capa. En el caso del Hg no se indica T/U=0.60 como en FMT pues aquí se desestabiliza a este valor de la temperatura.}
\label{fig:EnergiasDeltaWDACS}
\end{center}
\end{figure}

\begin{figure}[htbp]
\begin{center}
\includegraphics[width=0.95\textwidth]{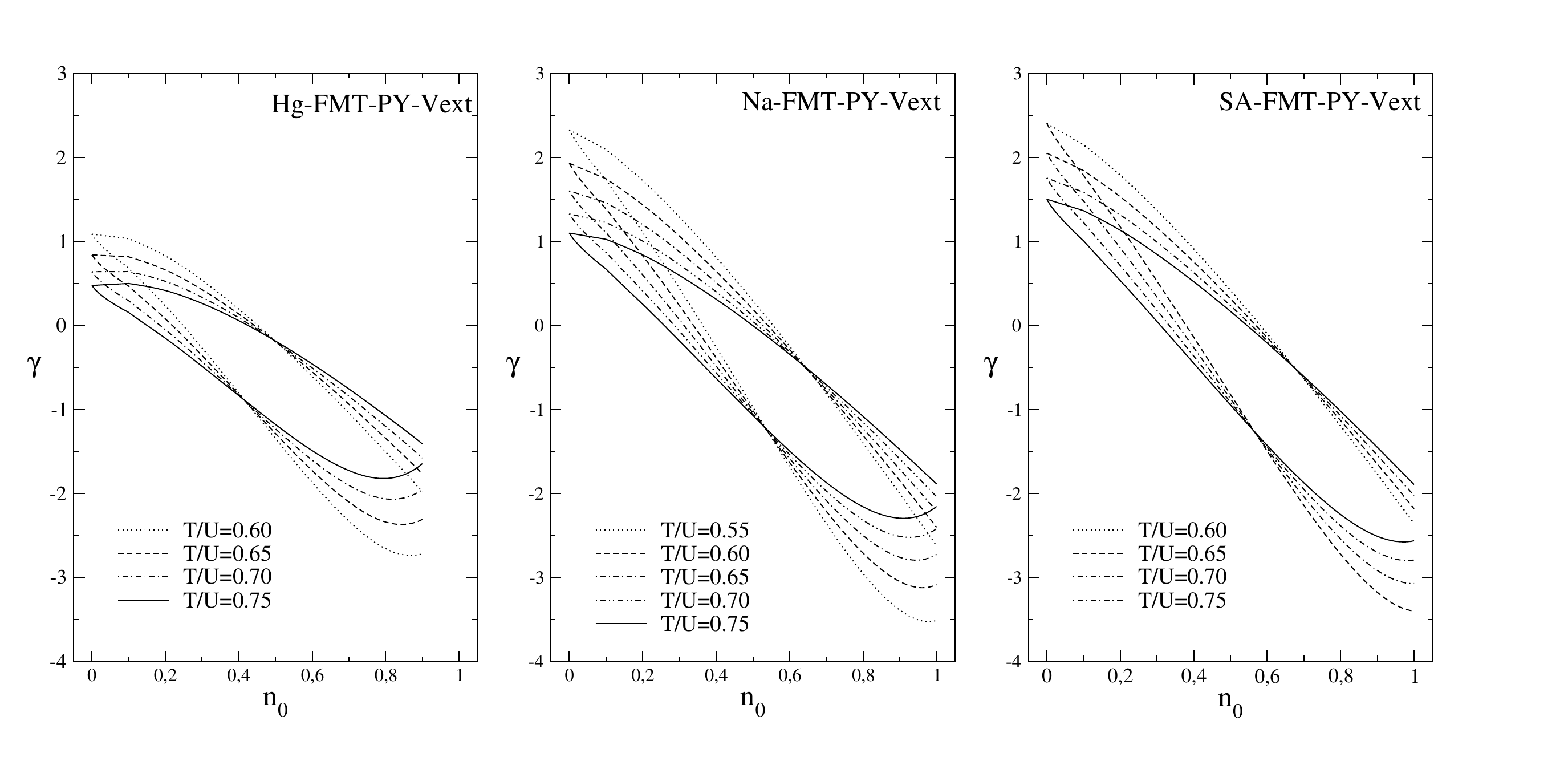}
\caption{Para Mercurio (Hg), Sodio (Na) y \textit{Soft Alcaline} (SA). Funcional FMT-PY. Se indican las tensiones superficiales de minimización del perfil de equilibrio $\rho_{v}(z)$ junto con las curvas incluyendo los términos que corresponden a las interacción dentro de la primera capa.}
\label{fig:EnergiasVextFMT}
\end{center}
\end{figure}

Para la parte atractiva hemos de añadir,
\begin{equation} 
\beta F_{AT}[\rho^{2D}]=\frac{1}{2}\beta\left[\rho^{2D}\right]^{2}\omega_{AT}(\xi_{0})
\end{equation}
mientras que la parte ideal será:
\begin{equation} 
\beta F_{ID}[\rho^{2D}]=\rho^{2D}\left(\log(d_{hs}^{2}\rho^{2D})-1\right) 
\end{equation}
introducimos las energías superficiales de la primera capa líquida para poder comparar las energías por ambos métodos. En las figuras (\ref{fig:EnergiasDeltaFMTPY}),(\ref{fig:EnergiasDeltaMFTCS}),(\ref{fig:EnergiasDeltaWDACS}) y (\ref{fig:EnergiasVextFMT}) se puede apreciar el resultado para los diferentes potenciales de interacción, los diferentes funcionales utilizados y para ambas metodologías $\rho_{\delta,v}$.

\subsubsection{Contraste con los resultados para $L_{eff}$}

Un primer análisis de la viabilidad de las definiciones anteriores como perfiles intrínsecos en un caso plano, puede analizarse a partir del desarrollo asintótico de estos, ya que del mismo modo que en el caso de los perfiles de equilibrio podemos ajustar a la forma funcional caracterizada por los parámetros de decaimiento $\alpha_{1}$, $\tilde{\alpha}_{1}$ y $\alpha_{0}$.\\

\begin{figure}[htbp]
\begin{center}
\includegraphics[width=1.25\textwidth,angle=-90]{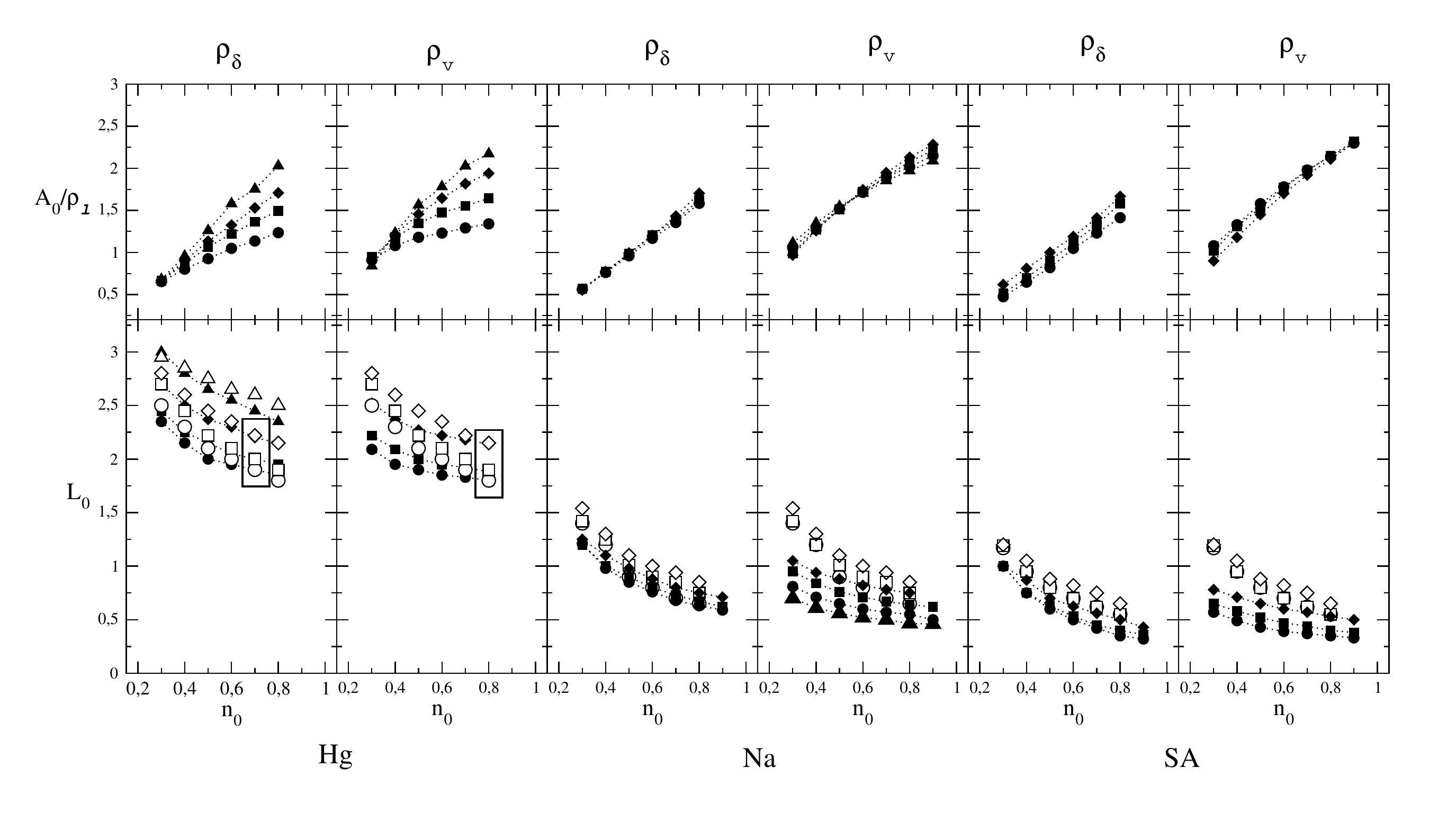}
\caption{\textbf{Arriba}: Amplitudes perfil intrínseco para los tres modelos de interacción y las dos aproximaciones funcionales $\delta$ y $v_{ext}$. Círculos: T/U=0.60, Cuadrados: T/U=0.65, Rombos T/U=0.70. Triángulos representa T/U=0.75 para Hg, T/U=0.55 para Na. \textbf{Abajo} dada $L_{eff}$ de \S\ref{sec:relacionCWyDEN} se representa el valor de $L_{0}$  obtenido del ajuste asintótico óptimo entre $<\rho(z)>_{\xi}$ y el perfil de equilibrio líquido-vapor. Se muestran con figuras blancas los valores de $L_{0}$ correspondientes a las $A_{0}/\rho_{l}$ que surgen del análisis de realizado en el \S\ref{sec:relacionCWyDEN}. Para caso Hg se encuadra caso de mayor consistencia.}
\label{fig:L0asintotico}
\end{center}
\end{figure}

En la figura siguiente (\ref{fig:L0asintotico}) hemos representado los valores de las amplitudes $A_{0}$ para el caso de los perfiles $\rho_{\delta}(z)$ en función de $n_{0}$ mostrando un comportamiento notablemente lineal, los valores de $\alpha_{1}$ y $\alpha_{0}$ son idénticos a los valores de $\rho(z)$ y por tanto similares a los que obteníamos del análisis de la respuesta lineal de modo consistente con el desarrollo mostrado en el capitulo anterior, véase por ejemplo la figura (\ref{fig:decaysAmplitudesFMT}). En el estudio de los perfiles de equilibrio en un campo externo obteníamos dos valores clave\footnote{En lo que sigue denominamos al $L_{0}$ de \S\ref{sec:relacionCWyDEN} como
 $L_{0}^{(mg)}$} en la reinterpretación de los perfiles de densidad $L_{eff}$ y $L_{0}^{(mg)}$, este último era determinado fijando un valor de $A_{0}$. Ahora podemos proceder de modo contrario, y es dado el perfil intrínseco y por tanto $A_{0}$ determinar para $L_{eff}$ el valor de $L_{0}^{(in)}$ que permite reproducir de modo exacto el comportamiento asintótico, salvo el valor de $z_{0}$ que es diferente en $\rho(z)$ y $\tilde{\rho}_{\delta,v}(z)$. En la figura (\ref{fig:L0asintotico}) se puede apreciar que los valores de este $L_{0}^{(in)}$ son similares aunque sistemáticamente menores.\\

En el caso del Mercurio vemos que hay un punto en que coinciden $L_{0}^{(in)}$ y $L_{0}^{(mg)}$ que corresponde a $n_{0}\simeq 0.7$ que podemos considerar el punto de mayor consistencia entre ambas propuestas y que se ha encuadrado en la figura (\ref{fig:L0asintotico}).\\

En el caso de Sodio y \textit{Soft Alcaline} (SA) esta diferencia es más sistemática y no es posible encontrar un punto de mayor consistencia en el rango de $n_{0}$ que hemos indagado, con todo el problema puede estar en el valor elegido para $L_{eff}$, ya que especialmente en el caso del SA hay cierta dispersión como vimos en la figura (\ref{fig:metodo1Leff}), también de las figuras que representan la energía libre vemos que el mínimo de la energía esta en valores de $n_{0}$ más altos, el régimen para el comportamiento asintótico es más incierto y la variación de $A_{0}$ con z mayor, de lo que la determinación univoca de $n_{0}$ varía según en que punto consideremos que el comportamiento es el asintótico.\\

\begin{figure}[htbp]
\begin{center}
\includegraphics[width=0.95\textwidth]{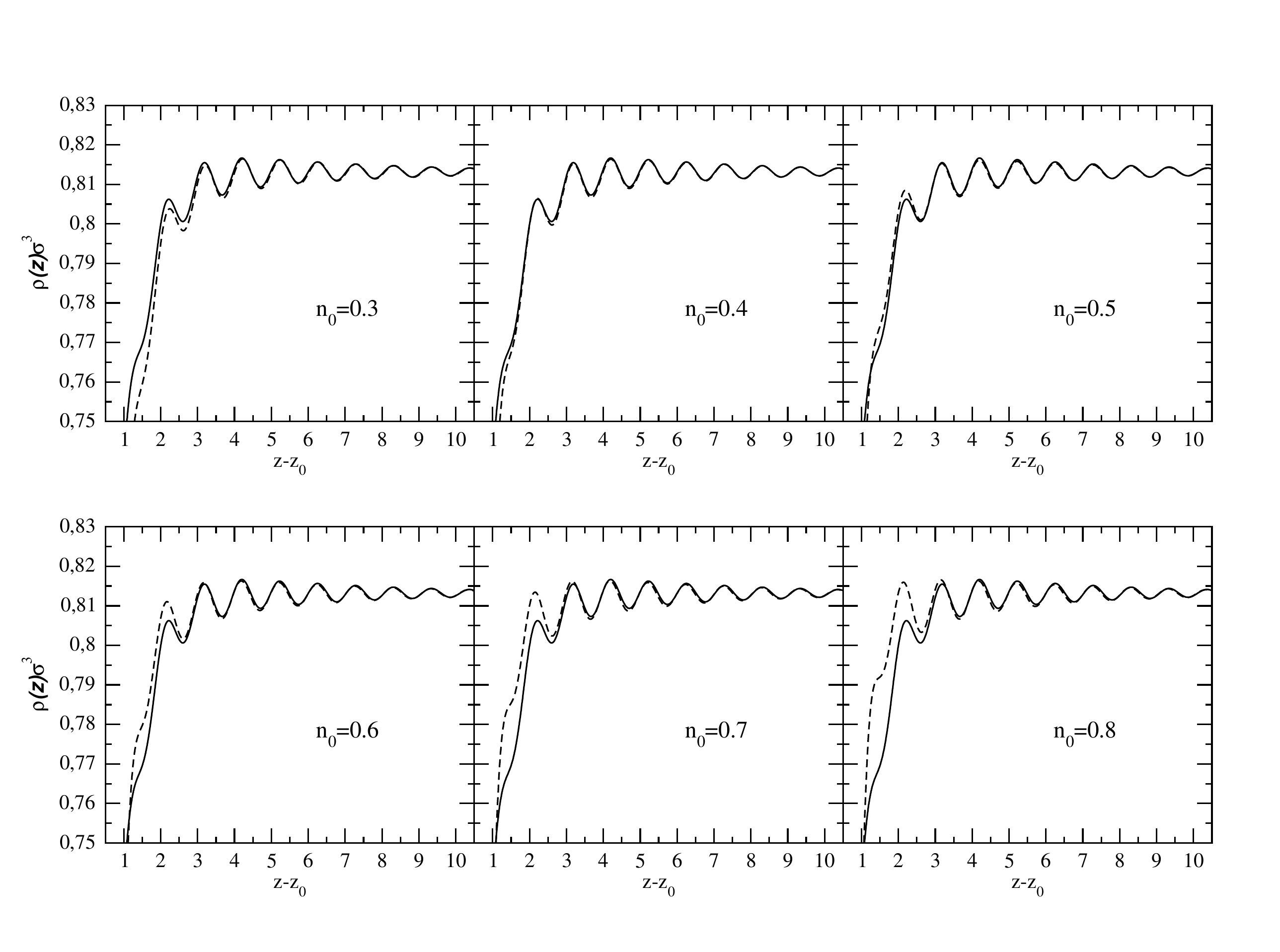}
\caption{Resultado de la convolución del perfil intrínseco $\rho_{\delta}(z)$ con $\mathcal{P}_{\xi}(L_{eff},L_{0})$, con $L_{eff}=9.5\sigma$ y $L_{0}$ como parámetro libre, resultados figura ( \ref{fig:L0asintotico}). T/U=0.60. Hg. FMT-PY}
\label{fig:DeltaHgL0asintoticoT60Hg}
\end{center}
\end{figure}

Respecto a la interpretación de estos resultados cabe indicar que todos los valores de $n_{0}$ son equivalentes desde el punto de vista del comportamiento cerca del volumen, las figuras mostradas más adelante expresan el resultado de la convolución gaussiana de los perfiles intrínsecos para el par $L_{eff}$ junto con $L^{(in)}_{0}$, mostrando que en el caso de $\rho_{\delta}$ la similitud fuera del comportamiento asintótico se extiende a rangos intermedios. Esto indica que desde la simple hipótesis de:
\begin{equation}
\rho(z)=\int dz_{1} \mathcal{P}_{L_{eff},L_{0}}(z_{1})\rho^{(in)}(z-z_{1})
\label{eqn:convolution}
\end{equation}
los $n_{0}$ que en un rango entorno de 0.4 en el caso de $\rho_{\delta}$ y algo mayor en el caso de $\rho_{v}$ reproducen $\rho(z)$ más allá del régimen asintótico, estos valores de $n_{0}$ coinciden casualmente con el mínimo en la energía libre mostrado en la figura (\ref{fig:EnergiasDeltaFMTPY}) sin incluir la energía de la primera capa. Aun así la coincidencia para estos valores de $n_{0}$ no es realmente significativa ya que parámetros de ocupación pequeños acercan el perfil medio al intrínseco sin que sea indicativo de la verdadera ocupación de la primera capa. Si es más sugerente que cuando la incluimos contribuciones superficiales el mínimo se sitúa en parámetros de ocupación entorno a $n_{0}\sim0.7$ que coincide cualitativamente con lo que encontramos en el caso $\rho_{v}$, véase la figura (\ref{fig:EnergiasVextFMT}). La descripción de los perfiles intrínsecos mediante simulación Montecarlo\cite{PhysRevLett.91.166103} coincide en atribuir este rango de densidades a $n_{0}$. Por último la comparación de los de $\rho_{DF}(z,L_{eff})$ y y $\rho(z)$ de la ecuación (\ref{eqn:convolution}) para estos valores de $n_{0}$, véase la figura (\ref{fig:DeltaHgL0asintoticoT60Hg}), puede indicar la existencia de términos extra relevantes más allá de la aproximación de ondas capilares de la convolución anterior que se muestran en las discrepancias de las primeras capas.

\begin{figure}[htbp]
\begin{center}
\includegraphics[width=1.30\textwidth,angle=-90]{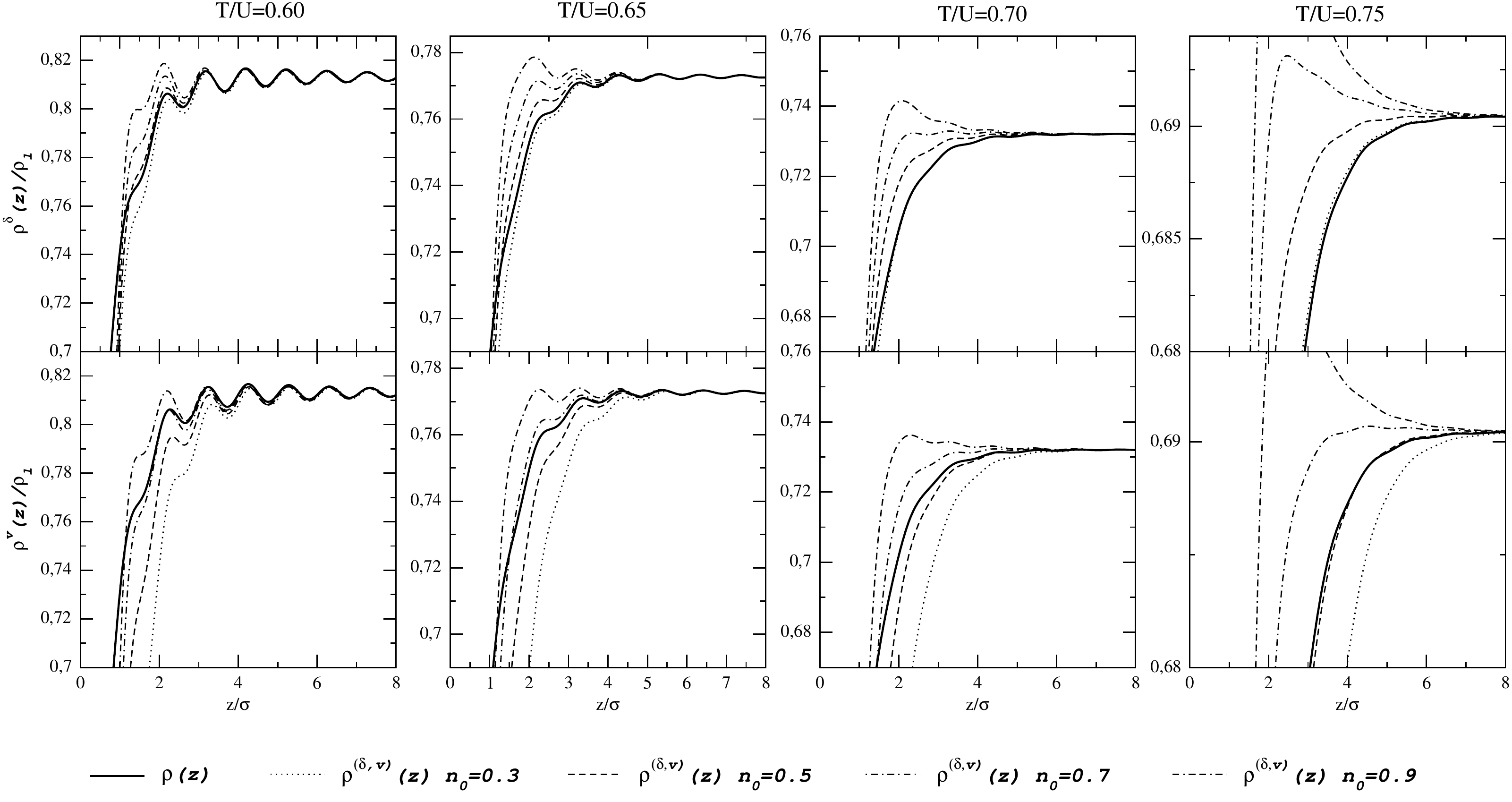}
\caption{Resultado de la convolución del perfil intrínseco $\tilde{\rho}(z)$ con $\mathcal{P}_{\xi}(L_{eff},L_{0})$, con $L_{eff}=9.5\sigma$ y $L_{0}$ como parámetro libre, resultados figura ( \ref{fig:L0asintotico}). Modelo Hg Funcional FMT-PY. Se comparan $\rho_{\delta}(z)$ con $\rho_{vext}(z)$ a la misma T/U. Los valores de $n_{0}$ representados son 0.3, 0.5, 0.7 y 0.9}
\label{fig:DeltaHgL0asintotico}
\end{center}
\end{figure}

\begin{figure}[htbp]
\begin{center}
\includegraphics[width=1.30\textwidth,angle=-90]{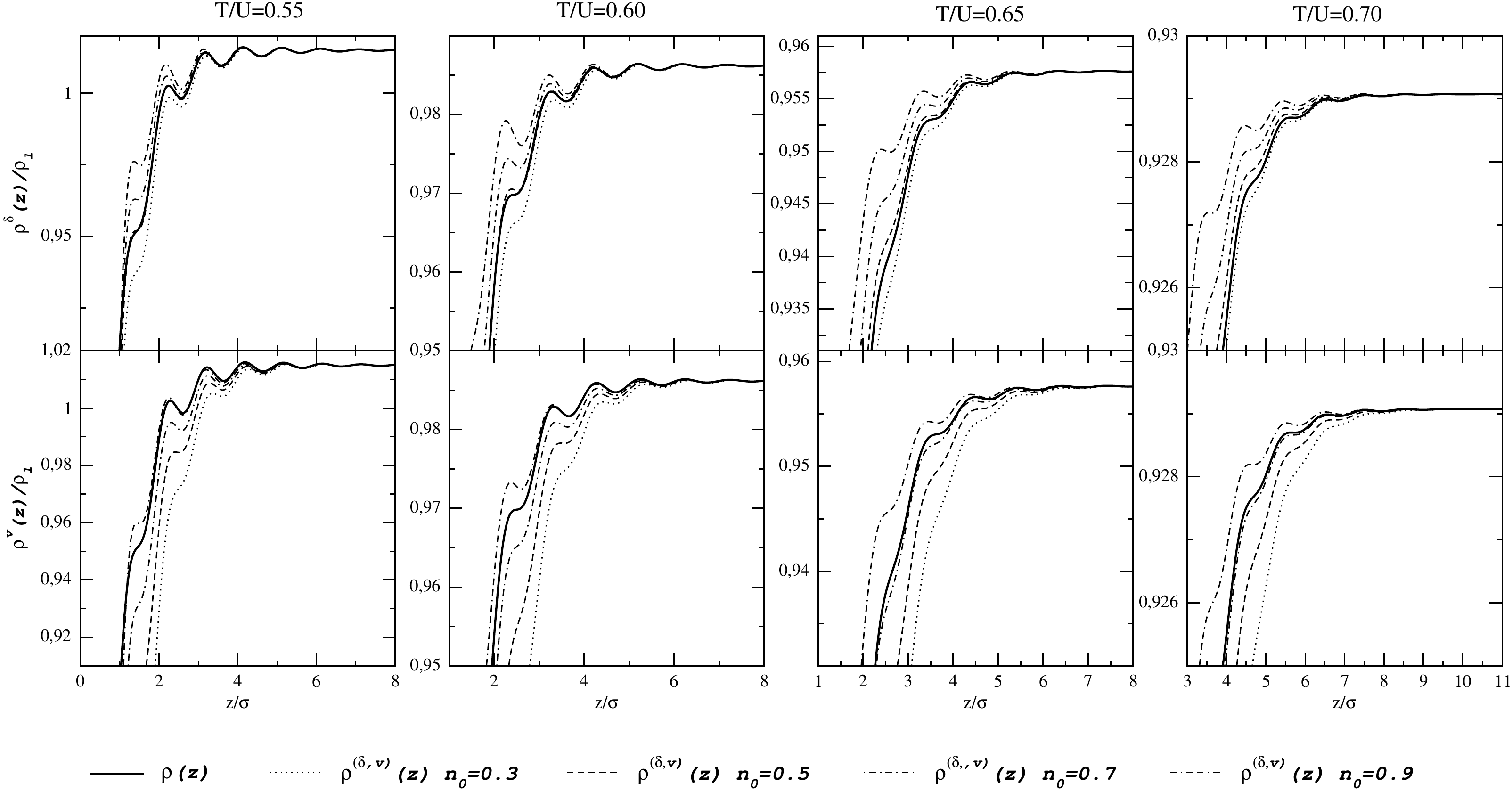}
\caption{Resultado de la convolución del perfil intrínseco $\tilde{\rho}(z)$ con $\mathcal{P}_{\xi}(L_{eff},L_{0})$, con $L_{eff}=9.5\sigma$ y $L_{0}$ como parámetro libre, resultados figura ( \ref{fig:L0asintotico}). Modelo Na. Funcional FMT-PY. Se comparan $\rho_{\delta}(z)$ con $\rho_{vext}(z)$ a la misma T/U.  Los valores de $n_{0}$ representados son 0.3, 0.5, 0.7 y 0.9}
\label{fig:DeltaNaL0asintotico}
\end{center}
\end{figure}

\begin{figure}[htbp]
\begin{center}
\includegraphics[width=1.30\textwidth,angle=-90]{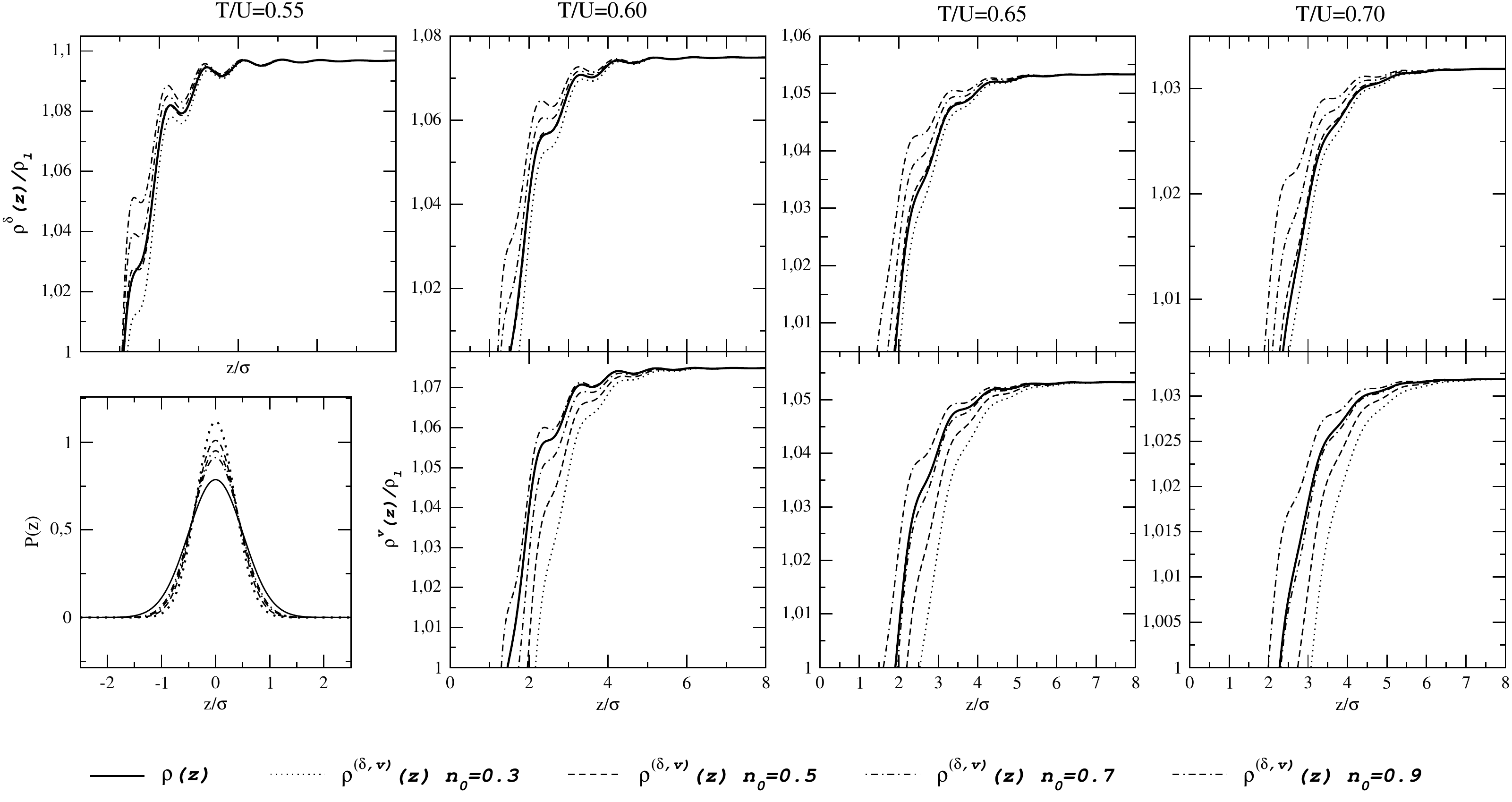}
\caption{Resultado de la convolución del perfil intrínseco $\tilde{\rho}(z)$ con $\mathcal{P}_{\xi}(L_{eff},L_{0})$, con $L_{eff}=9.5\sigma$ y $L_{0}$ como parámetro libre, resultados figura (\ref{fig:L0asintotico}). Modelo Na. Funcional FMT-PY. Se comparan $\rho_{\delta}(z)$ con $\rho_{vext}(z)$ a la misma T/U.  Los valores de $n_{0}$ representados son 0.3, 0.5, 0.7, 0.9. La figura abajo a la izquierda representa la función $\mathcal{P}(z)$, la línea continua es $n_{0}=0.9$ con $T/U=0.70$, el resto son $T/U=0.55$ con $n_{0}=0.3, 0.5, 0.7$ y $0.9$}
\label{fig:DeltaSAL0asintotico}
\end{center}
\end{figure}

\section{Excursus: Línea de Fisher-Widom para $\rho^{(\delta,v)}$}
\label{sec:excursusFWintrinseco}
En el caso del perfil intrínseco los valores de $A_{in}$ y $B_{in}$ están condicionados por $n_{0}$, mientras que los valores de $\alpha_{1}$,$\tilde{\alpha}_{1}$ y $\alpha_{0}$ están determinados por el análisis de la respuesta lineal, la determinación desde el perfil de densidad intrínseco de estos últimos es compatible al igual que lo era desde el  perfil de densidad líquido-vapor.\\

\begin{figure}[htbp]
\begin{center}
\includegraphics[width=0.95\textwidth]{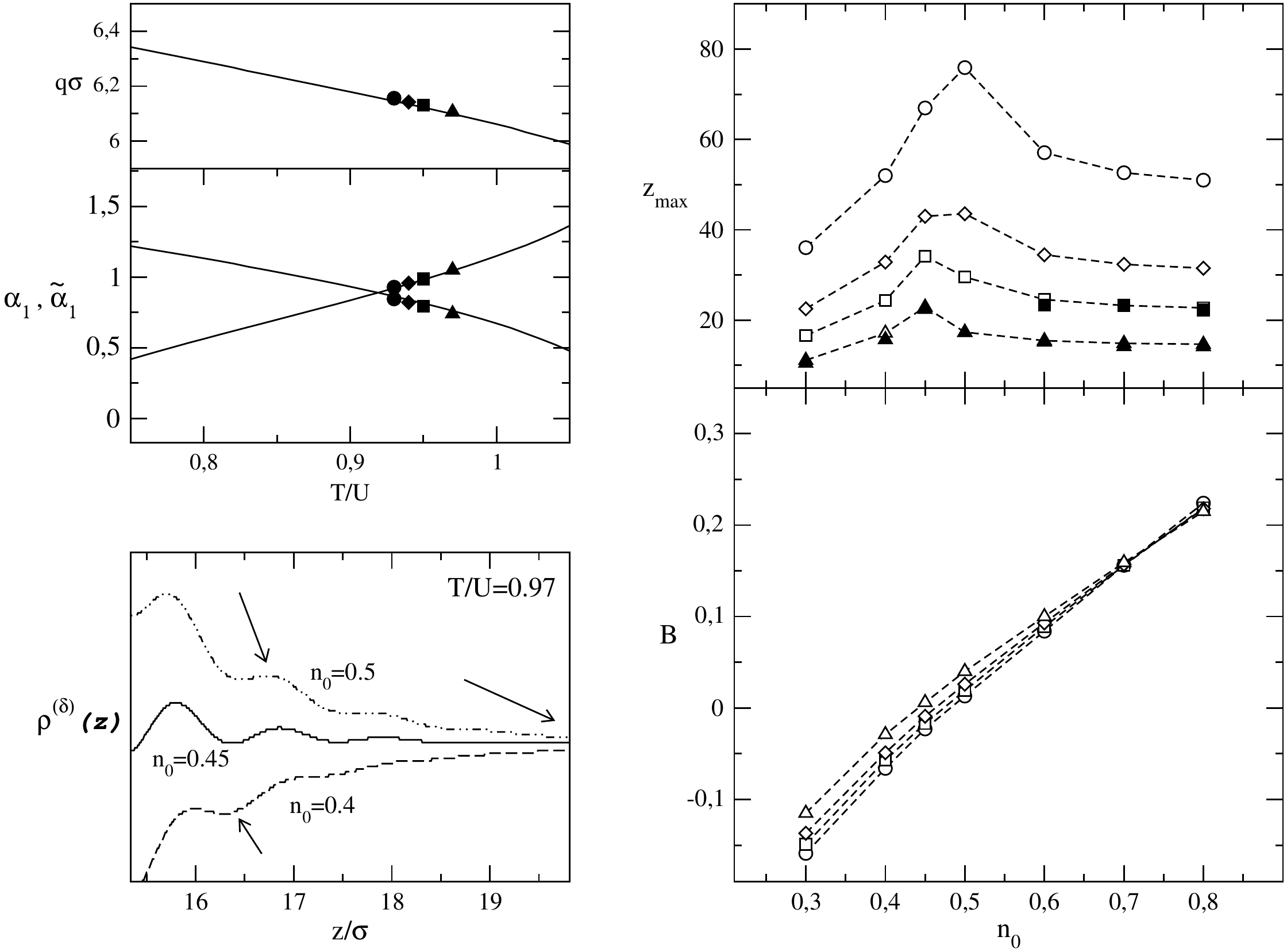}
\caption{Se analizan los modos de decaimiento para temperaturas mayores que Fisher-Widom, para temperaturas menores se han visto antes. Se comprueba que para los valores de $B_{in}$ y $A_{in}$ procedentes del ajuste la forma introducida en ec. (\ref{eqn:ajusteperfildensidad}) es válida. Incluso cuando $B_{in}\simeq 0$ predice correctamente la presencia de oscilaciones propagándose hacia el volumen más allá de la línea de Fisher-Widom. Se muestran tres perfiles donde se ve el fenómeno una flecha indica la posición de la última oscilación completa presente en perfil.}
\label{fig:EstudioLJcasodelta}
\end{center}
\end{figure}

Hemos analizado el comportamiento general con la temperatura y con $n_{0}$ de los perfiles intrínsecos planos. Nos interesa comprobar la relación entre las propiedades estructurales y la línea de Fisher-Widom tal y como sucedía con el perfil de densidad líquido-vapor. Los resultados para el caso Lennard-Jones aparecen condensados en la figura (\ref{fig:EstudioLJcasodelta}), tanto de la consistencia de los perfiles con los valores de la respuesta lineal como los valores de $B_{in}$ variando con $n_{0}$, ya que los valores de $A_{in}$ presentan un comportamiento análogo a los modelos anteriores. \\

Los ajustes de las amplitudes son relacionados con la posición del último máximo como se puede ver en la figura (\ref{fig:EstudioLJcasodelta})
mostrando que la ley que obteníamos si $A_{in}>B_{in}$ para $T>T_{FW}$ cuando se tiene que $\alpha_{1}(T)>\tilde{\alpha}_{1}(T)$ cumple ciertamente que,
\begin{equation}
z_{max}^{osc}(T,n_{0})=\frac{1}{\alpha_{1}(T)-\tilde{\alpha}_{1}(T)}\ln \left[\frac{\vert\alpha_{c}\vert A_{in}(T,n_{0})}{\tilde{\alpha}_{1} B_{in}(T,n_{0})}\right] 
\end{equation}

Como se indica en la figura donde representamos curvas de $z_{max}$ para diferentes valores de T/U 
(cada curva posee diferentes valores de $n_{0}$ y en consecuencia diferentes valores de $A_{in}$ y $B_{in}$, aunque la ley para $z_{max}$ se cumple en todos los casos, con el matiz de que en el caso $B_{in}\simeq 0$ la posición 
$z_{max}^{osc}(T,n_{0}) >> 1$). Los puntos representan los valores de $z_{max}^{osc}(T,n_{0})$
que se encuentran en el perfil. En el caso de $\rho_{v}(z)$ los valores de $n_{0}$ que hacen el valor de $B(n_{0})\simeq 0$ están entorno de 0.6 donde es factible diferenciar oscilaciones en el perfil de densidad m\'as allá de la capa $z/\sigma \simeq 30$ con T/U m\'as allá de Fisher-Widom, indicando que el valor de las amplitudes puede hacer irrelevante a la línea de Fisher-Widom, véase figura (\ref{fig:EstudioLJcasodelta}).
\section{Determinación del espectro de ondas capilares para $\Omega_{v,\delta}$}
\label{sec:determinacionEspectroOndasCapilares}
Los conceptos de perfil intrínseco a un nivel dado de corrugación antes introducido pueden ser evaluados inicialmente mediante un cálculo a segundo orden en las amplitudes de las ondas capilares, $|\xi_{q}|^{2}$, donde se introduzca la hipótesis de la aproximación de los perfiles intrínsecos como perfiles de equilibrio que minimizan un funcional de la densidad. Por lo comentado al inicio del capítulo no esperamos que las características del espectro de ondas capilares que obtengamos sea adecuada ya que estamos diferenciando los subespacios $\Gamma_{\xi}$ mediante la hipótesis dada en ec. (\ref{eqn:hipotesisDFTdesdeDEN1}). El interés esta en que las dos propuestas $\Omega_{v,\delta}$ y sus dos perfiles intrínsecos $\rho^{(v,\delta)}$ incorporan de modo diferente las correlaciones y de la comparación de ambos casos podemos analizar la verosimilitud de los problemas que (\ref{eqn:hipotesisDFTdesdeDEN1}) posee.\\

Para ello hemos determinado un modo de construir $\tilde{\rho}(z-\xi(\vec{R}))$  desde una aproximación como un perfil de equilibrio para una función de distribución condicional. Sin embargo, más que una minimización directa, es conveniente un primer modo de ver la viabilidad dentro del conjunto de aproximaciones realizadas tomando un desarrollo hasta segundo orden en las amplitudes $\xi_{q}$ sobre un caso plano $\tilde{\rho}(z-\hat{\xi}_{0})$. Este último puede ser obtenido mediante minimización directa de los funcionales anteriores donde las simetrías me permiten reducir el problema a una densidad dependiente de una sola coordenada, mientras que la incorporación hasta segundo orden de las amplitudes de las ondas capilares es tratable teóricamente.\\

De este modo si expresamos $\tilde{\rho}^{v,\delta}[z-\xi(\vec{R}),\vec{R};\xi]$ como un desarrollo en $\xi(\vec{R})$ sobre un caso plano y expresamos:
\begin{equation}
\tilde{\rho}^{v,\delta}[z-\xi(\vec{R}),\vec{R};\xi]\simeq\tilde{\rho}^{v,\delta}(z-\xi(\vec{R}))+\sum_{|\vec{q}|>0}\xi_{q}\tilde{\rho}^{v,\delta}_{q}(z-\xi(\vec{R}))e^{-\vec{q}\vec{R}}+O(\xi_{q}^{2})
\end{equation}
como
\begin{equation}
\tilde{\rho}^{v,\delta}(z-\xi(\vec{R}))\simeq\tilde{\rho}^{v,\delta}(z-\hat{\xi}_{0})-\frac{d\tilde{\rho}^{v,\delta}(z-\xi_{0})}{dz}\bigg\vert _{z+\xi_{0}}\xi(\vec{R})+\frac{1}{2}\frac{d^{2}\tilde{\rho}^{v,\delta}(z-\xi_{0})}{d^{2}z}\xi(\vec{R})^{2}+O(\xi_{q}^{3})
\label{eq:DesarrolloIntrinsecosobreplano}
\end{equation}
y
\begin{equation}
\sum_{|\vec{q}|>0}\xi_{q}\tilde{\rho}^{v,\delta}_{q}(z-\xi(\vec{R}))e^{-\vec{q}\vec{R}}=\sum_{|\vec{q}|>0}\xi_{q}\tilde{\rho}^{v,\delta}_{q}(z-\xi_{0})e^{-\vec{q}\vec{R}}-\sum_{|\vec{q}|>0}\frac{d\tilde{\rho}^{v,\delta}_{q}(z-\xi_{0})}{dz}\xi(\vec{R})e^{-\vec{q}\vec{R}}+O(\xi_{q}^{3})
\end{equation}

podemos expresar $\tilde{\rho}^{v,\delta}[z-\xi(\vec{R}),\vec{R};\xi]\simeq\tilde{\rho}^{v,\delta}(z-\hat{\xi}_{0})+\delta\tilde{\rho}^{v,\delta}[z,\vec{R},\xi]$. Esto nos permite incluir un desarrollo de modo consistente mediante un desarrollo en funcional entorno a la densidad $\tilde{\rho}(z-\hat{\xi}_{0})$, de nuestro funcional $\Omega[\tilde{\rho}[z-\xi(\vec{R}),\vec{R};\xi]]$\footnote{En el caso del desarrollo de la primera capa líquida a orden 2 en $\xi_{q}$ debemos considerar las funciones resultantes como distribuciones y aplicar la definición de estas dentro de nuestro funcional que únicamente las define a partir de su aplicación mediante la integración por partes.}.\\

Esto puede ser expresado de modo completo y en ambas imágenes mediante:
\begin{eqnarray}
\Omega_{v,\delta}[\rho+\delta\rho] &=& \Omega_{v,\delta}[\rho] \nonumber\\
&+&\frac{1}{2}\sum_{q}|\xi_{q}|^{2}\left[ \Delta_{v,\delta}(q)+\mathcal{D}_{v,\delta}(\rho_{q})\right] 
\end{eqnarray}
tenemos una forma cuadrática $\mathcal{D}_{v,\delta}(\rho_{q})$ para las funciones $\rho_{q}(z)$ que en el caso de $\Omega_{v}$ serán $\rho^{(in)}(z)$ mientras que en el caso de $\Omega_{\delta}$ incluyen las funciones $n_{q}$. En ambas imágenes tenemos la expresión\footnote{La expresión global es una forma cuadrática pero a efectos de hallar extremales para la energía libre no es relevante la constante $\Delta_{q}$ en la definición de $\mathcal{D}$.}:
\begin{equation}
\mathcal{D}_{v,\delta}(\rho^{v,\delta}_{q})\equiv\frac{1}{2}\int dzdz'\rho^{v,\delta}_{q}(z)\mathcal{A}_{v,\delta}(z,z')\rho^{v,\delta}_{q}(z')-\int dz \mathcal{B}_{v,\delta}(z)\rho^{v,\delta}_{q}(z)
\end{equation}
Las expresiones para teorías de la forma FMT-MFA son dadas en \S\ref{sec:desarrolloAnaliticoDEOmega}, donde se aprecia que la matriz $\mathcal{A}_{v,\delta}(z,z')$ es simétrica y definida positiva, por tanto el mínimo de la forma cuadrática $\mathcal{D}_{v,\delta}$ se encuentra a partir de la solución del sistema lineal
\begin{equation}
\int dz'\mathcal{A}_{v,\delta}(z,z')\rho^{v,\delta}_{q}(z')=\mathcal{B}_{v,\delta}(z)
\end{equation}
esto permite expresar 
\begin{equation}
\Omega_{v,\delta}[\rho+\delta\rho]=\Omega_{v,\delta}[\rho]+\frac{1}{2}\sum_{q}|\xi_{q}|^{2}[\Delta_{v,\delta}(q)+R_{v,\delta}(q)]
\end{equation}
donde aparece una parte relajativa que debe condicionar la forma de $\gamma(q)$ y donde hemos expresado para una solución $\varrho^{v,\delta}_{q}(z)$ dicha parte relajativa por:\\
\begin{equation}
R_{v,\delta}(q)=-\frac{1}{2}\int dz \mathcal{B}_{v,\delta}(z)\varrho^{v,\delta}_{q}(z)
\end{equation}
el significado tanto de $\Delta_{q}$ como $R_{q}$ fue analizado  en \S\ref{sec:Cap6parteNOLOCAL} ya que en el caso de $n_{0}=0$  el perfil intrínseco se haría el perfil líquido-vapor $\rho_{DF}(z)$ usado entonces.\\

Antes de presentar los resultados comparamos con las propuestas ya analizadas, el desarrollo de la parte atractiva no relajativa, $\Delta_{q}^{AT}$ tiene la forma:
\begin{equation}
\Delta_{q}^{AT}=\frac{\beta}{2}\int dz_{1}\rho'(z_{1})\int dz_{2}\rho'(z_{2})\left[ \omega^{at}_{-q}(z_{12})-\omega^{at}_{0}(z_{12})\right] 
\end{equation}
mientras que la predicción de la ecuación TZ será,
\begin{equation}
\beta\gamma=-\int dz_{1}\int dz_{2}\rho'(z_{1})\rho'(z_{2})\frac{c^{(2)}(z_{1},z_{2};\delta q)-c^{(2)}(z_{1},z_{2};0)}{(\delta q)^{2}}
\end{equation}
La propia forma en que las correlaciones estan incluidas en los funcionales de teorías de van der Waals generalizadas hace que la función de correlación directa adquiera la forma dada por \textit{random phase aproximation} y por tanto hace compatible nuestra expresión y TZ donde,
\begin{equation}
c_{2}(z_{1},z_{2})=-\beta\frac{\left[ \omega^{at}_{-q}(z_{12})-\omega^{at}_{0}(z_{12})\right] }{(\delta q)^{2}}
\end{equation}
ya que en el límite $q\rightarrow0$ la parte atractiva es la parte numéricamente relevante. Como consecuencia nuestra aproximación para $n_{0}=0$ en la parte no relajativa resulta equivalente a la expresión que \textit{Mecke-Dietrich} obtienen  como:
\begin{equation}
\beta\gamma=\frac{1}{2}\int dz_{1}\int dz_{2}\rho'(z_{1})\rho'(z_{2})\frac{d^{2}\omega^{at}_{-q}(z_{12})}{dq^{2}}
\end{equation}
obviamente en la siguiente corrección a orden $q^{2}$ también la parte atractiva contribuye a la tensión superficial $\gamma(q)$ pero en este caso los otros términos son relevantes.\\

\begin{figure}[htbp]
\begin{center}
\includegraphics[width=0.95\textwidth]{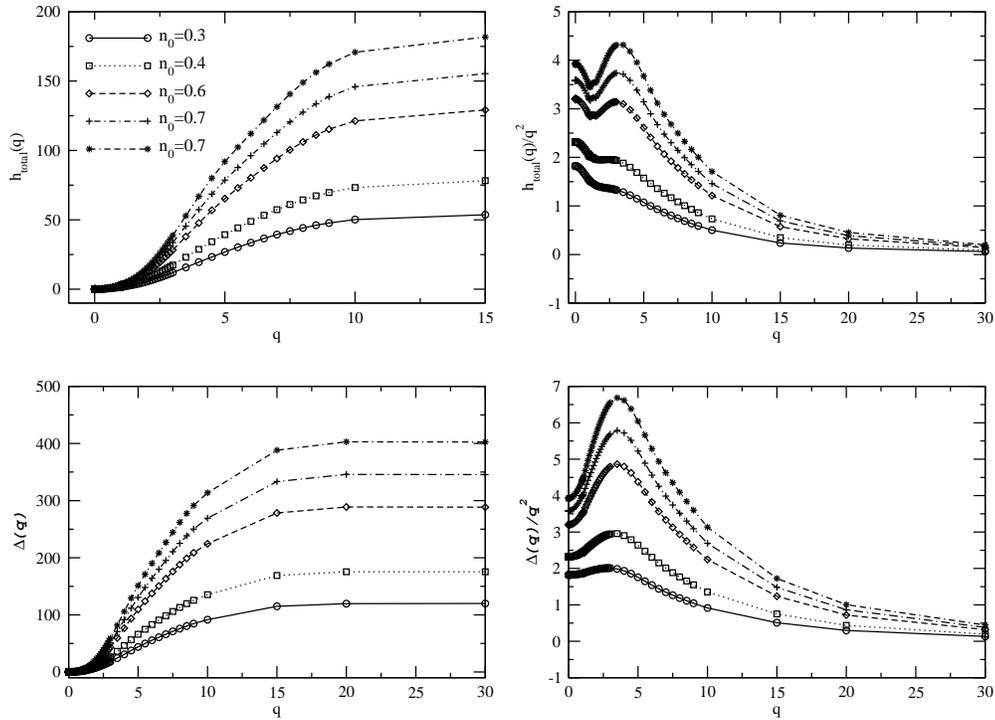}
\caption{Valor de $\delta\Omega^{v}$ en el caso Hg-FMT-PY y T/U=0.65, se muestran los valores para diferentes tasas de ocupación de la capa $n_{0}$. En el caso de $\delta\Omega^{\delta}$ las curvas para $\gamma(q)$ aunque difieren entre sí para las diferentes recetas de asimetrización resultan monótonamente decrecientes y similares a las obtenidas en el capítulo anterior, el valor de $h(q)$ no crece lo suficientemente rápido debido a los problemas comentados en el texto.}
\label{fig:CorrugacionHgPY65Vext}
\end{center}
\end{figure}

Existen diferencias notables entre esta propuesta para $\Omega_{v,\delta}$ y la forma de la teoría de Dietrich y Mecke en la parte relajativa $R_{q}$. 
En la figura (\ref{fig:CorrugacionHgPY65Vext}) mostramos los valores de $\Delta^{v}_{q}$ y de $h^{v}(q)=\Delta^{v}_{q}+R^{v}_{q}$ en el caso de $\delta\Omega_{v}$. También mostramos $h^{v}(q)/q^{2}$ que sería la predicción para $\gamma^{v}(q)$ y el caso sin relajación $\Delta^{v}_{q}/q^{2}$.\\

Podemos ver como en el caso de $\delta\Omega^{v}$, los valores $\Delta^{v}_{q}$ son monótonamente crecientes aunque menores que en los resultados mediante MonteCarlo\cite{PhysRevB.70.235407}, comparese con figura (\ref{fig:GammaQsimulaciones}). La relajación hace disminuir aun más los valores de $\gamma^{v}(q)$ e induce un comportamiento no monótono en la propia forma de $h^{v}(q)/q^{2}$, debido a que cada uno de los términos involucrado posee una escala en q de relajación diferente, la parte de esferas duras relaja monótonamente y más lentamente, la parte atractiva relaja algo más rápido aunque satura entorno a $q\simeq 4$, el acoplo entre las funciones $\rho_{q}$ y el potencial externo relaja muy rápidamente y es el causante del mínimo relativo que observamos en las curvas de $h^{v}(q)/q^{2}$.\\

En el caso de $\delta\Omega^{\delta}$, los valores de $h^{\delta}(q)/q^{2}$ son monótonamente decrecientes debido a que $h^{\delta}(q) << h^{v}(q)$ y los resultados son similares los presentados en el capitulo anterior. Y por tanto observamos como siendo los perfiles intrínsecos $\rho^{in}_{v}$ y $\rho^{in}_{\delta}$ \textit{cualitativamente similares} los correspondientes $h^{v}(q)$ y  $h^{\delta}(q)$ son \textit{significativamente diferentes} lo que se debe a la diferencia en la capacidad de representabilidad que un proyector $\Lambda(\xi,\vec{R})$ construido como un potencial externo posee\footnote{El lector al ver la figura (\ref{fig:CorrugacionHgPY65Vext}) no debe compararla directamente con las figuras (\ref{fig:n0iguala0}) y (\ref{fig:RelajacionSINno}). La estructura sobre la que se asientan es diferente y el caso en que ahora nos ocupa no cumple que $\gamma(q=0)=\gamma_{lv}$ mientras que la presencia del mínimo se debe a que la problemática de DFT comentada en el apéndice se deja sentir más en la parte relajativa que baja $\gamma(q)$ de modo inconsistente en las diferentes partes del espectro.}.

\subsection{Determinación de la Tensión superficial}

Las definiciones usadas permiten escribir que:
\begin{equation}
e^{-\beta\gamma_{lv}A_{0}}=Tr_{(\xi)}e^{-\beta\mathcal{H}[\xi(\vec{R})]}
\end{equation}
por tanto
\begin{equation}
\gamma_{lv}=-\beta^{-1}A_{0}^{-1}\log[Tr_{(\xi)}e^{-\beta\mathcal{H}[\xi(\vec{R})]}]
\end{equation}
Si consideramos la aproximación gaussiana donde las amplitudes $\hat{\xi}_{q}$ poseen probabilidades independientes podemos resolver la traza resultado:
\begin{equation}
\gamma_{lv}=-\beta^{-1}A_{0}^{-1}\log \left[\prod_{|q|=q_{l}}^{q_{u}} \int d\xi_{q} e^{-\beta\mathcal{H}_{q}(\xi_{q})}\right] =\gamma(0)-\beta^{-1}A_{0}^{-1}\sum_{|q|=q_{l}}^{q_{u}}\log \left[ \frac{2\pi}{\beta A_{0}q^{2}\gamma(q)} \right]^{1/2}
\end{equation}
donde $\gamma(q)=\gamma(0)+K_{v,\delta}(n_{0})q^{2}+...$.\\

Como indicábamos y acabamos de ver por las diferencias entre $h^{v}$ y $h^{\delta}$ no será posible hacer consistentes dos propiedades deseables: el rango de valores de q que permite recuperar $\gamma_{lv}$ desde los valores obtenidos en el perfil intrínseco sin ondas capilares, con el rango aceptable que permite hacer compatible la corrugación de este con el comportamiento asintótico de los perfiles líquido-vapor. Mientras que esto último es posible de un modo razonable el recuperar las tensiones superficiales líquido-vapor no.

\section{Excursus: Sobre la definición de perfil intrínseco}
La definición antes utilizada consiste esencialmente en afirmar,
\begin{equation}
\tilde{\rho}(z,\vec{R};\xi)=\tilde{\rho}(z)+\sum_{|\vec{q}|>0}\hat{\xi}_{q}e^{-i\vec{q}\vec{R}}\tilde{\rho}_{q}(z)+O^{2}(\hat{\xi}_{q})
\end{equation}
la definición se complementa con:
\begin{equation}
\tilde{\rho}(z,\vec{R};\xi)\equiv \rho(z-\xi(\vec{R}),\vec{R};\xi)
\end{equation}
donde esta densidad es la distribución densidad condicionada por la ligadura $\xi(\vec{R})$, es decir, calculada en el subespacio restringido a configuraciones que reproducen $\xi(\vec{R})$. En estas condiciones,
\begin{equation}
\tilde{\rho}(z,\vec{R};\xi)=n_{0}\delta(z-\xi(\vec{R}))+\rho_{0}(z-\xi(\vec{R}))+\sum_{|\vec{q}|>0}\hat{\xi}_{q}e^{i\vec{q}\vec{R}}\rho_{q}(z-\xi(\vec{R}))
\end{equation}
con $\rho_{q}(z-\xi(\vec{R}))=\sum_{q}n_{q}\delta(z-\xi(\vec{R}))+\sum_{q}\rho_{q}(z-\xi(\vec{R}))$ y hemos separado explícitamente la primera capa. En esta expresión observamos que todo el perfil sigue a la superficie intrínseca aunque las cantidades $\rho_{q}$ aun están por determinar y por tanto no es únicamente un desplazamiento sin más de un caso plano.\\

Como es necesario incorporar un esquema que determine para las cantidades $\rho_{q}$ otra imagen física es viable, una primera capa en que se asienta la superficie intrínseca y el resto de perfil de densidad únicamente determinado mediante un esquema de optimización. En este caso obtendríamos:

\begin{equation}
\tilde{\rho}(z,\vec{R};\xi)=n_{0}\delta(z-\xi(\vec{R}))+\sum_{|\vec{q}|>0}n_{q}e^{-i\vec{q}\vec{R}}\delta(z-\xi(\vec{R}))+\rho_{0}(z)+\sum_{|\vec{q}|>0}\hat{\xi}_{q}e^{-i\vec{q}\vec{R}}\rho_{q}(z)
\end{equation}
Podemos escribir  ambas propuestas mediante una representación general que incluya ambos como casos particulares. Para ello podemos definir una $C_{q}(k)=\xi_{q}^{k}\equiv c_{q}(k)\xi_{q}$ donde k representa cada capa de partículas. De modo que $c_{q}(k)=1$ y $c_{q}(k)=\delta_{k0}$ representan los dos casos límite expuestos y esperamos que $0\leq c_{q}(k)\leq1$.

En esta forma de escribir las ecuaciones tendríamos una expresión genérica como,
\begin{equation}
\Omega_{v,\delta}[\rho+\delta\rho]=\Omega_{v,\delta}[\rho]+\frac{1}{2}\sum_{q}|\xi_{q}|^{2}\left[ \mathcal{D}^{(1)}_{v,\delta}(\rho_{q})+\mathcal{D}^{(2)}_{v,\delta}(c_{q})+\mathcal{D}^{(3)}_{v,\delta}(\rho_{q},c_{q})\right] 
\end{equation}
Si definimos  $c_{q}(z)=1$ tenemos la expresión usual para la forma cuadrática ya escrita suponiendo que $\rho_{q}(z-\xi(\vec{R}))$. En este caso la suma de $\mathcal{D}^{(i)}$ pueden ser expresado como una sola forma cuadrática, donde $\mathcal{D}^{(3)}_{v,\delta}(\rho_{q},c_{q})=\mathcal{B}_{v,\delta}$ y  $\mathcal{D}^{(2)}_{v,\delta}(c_{q})=\Delta_{v,\delta}$ es la parte no relajativa. Si hacemos $\rho_{q}=0$ anulamos la parte relajativa de nuestro cálculo funcional pero si permitimos que exista $c_{q}(z)$ tenemos una función a determinar que todavía podemos optimizar.
\chapter{Desarrollo analítico de $\Omega_{v,\delta}$}
\label{sec:desarrolloAnaliticoDEOmega}

En este apéndice se detallan las expresiones analíticas pertenecientes al \S\ref{sec:apendiceIntrinseco}. Partimos de un funcional FMT-PY sobre el que se realiza un desarrollo perturbativo en la densidad, incorporando una condición relajativa sobre este. Recordamos las expresiones del desarrollo,

\begin{eqnarray}
\Omega_{v,\delta}[\rho+\delta\rho] &=& \Omega_{v,\delta}[\rho] \nonumber\\
&+&\frac{1}{2}\sum_{q}|\xi_{q}|^{2}\left[ \Delta_{v,\delta}(q)+\mathcal{D}_{v,\delta}(\rho_{q})\right] 
\end{eqnarray}
y
\begin{equation}
\mathcal{D}_{v,\delta}(\rho^{v,\delta}_{q})\equiv\frac{1}{2}\int dzdz'\rho^{v,\delta}_{q}(z)\mathcal{A}_{v,\delta}(z,z')\rho^{v,\delta}_{q}(z')-\int dz \mathcal{B}_{v,\delta}(z)\rho^{v,\delta}_{q}(z)
\end{equation}
Las funciones que aquí vamos a expresar corresponden a $\Delta^{v,\delta}_{q}$, $\mathcal{A}_{v,\delta}(z,z')$ y $\mathcal{B}_{v,\delta}(z)$. En el caso de $\Omega_{v}$, se introduce un potencial externo que participará en $\Delta^{v,\delta}_{q}$ y $\mathcal{B}_{v,\delta}(z)$. En el caso de $\Omega_{\delta}$ los términos correspondientes a la interacción entre la \textit{primera capa} y $\rho^{(in)}$ están contenidos en la misma $\mathcal{F}_{ex}$.

\section{Potencial Externo}
\begin{equation}
\int dzd\vec{R}\rho^{(in)}_{\xi}(z,\vec{R})n_{0}\int d\vec{R}'\phi \left( \sqrt{|\vec{R}-\vec{R}'|^{2}+|z-\xi(\vec{R}') |^{2})}\right)
\end{equation}
Aplicando la definición de $\rho^{(in)}_{\xi}(z,\vec{R})=\tilde{\rho}^{(in)}(z-\xi(\vec{R}),\vec{R}))$ , utilizando un desarrollo Fourier para las funciones peso, e introduciendo el desarrollo (\ref{eq:DesarrolloIntrinsecosobreplano}) y hasta orden 2 en $\xi_{q}$ podemos llegar a,
\begin{equation}
n_{0}\sum_{|q|>0}|\xi_{q}|^{2}\left[ \frac{1}{2}\int dz\rho''^{(in)}_{0}(z)\left(2\Phi_{0}(z)-\Phi_{-q}(z)-\Phi_{q}(z)\right)+\int dz\rho'^{(in)}_{-q}(z)\left(\Phi_{q}(z)-\Phi_{0}(z)\right)\right] 
\end{equation}
o de modo equivalente,
\begin{equation}
n_{0}\sum_{|q|>0}|\xi_{q}|^{2}\left[ \frac{1}{2}\int dz\rho^{(in)}_{0}(z)\left(2\Phi''_{0}(z)-\Phi''_{-q}(z)-\Phi''_{q}(z)\right)+\int dz\rho^{(in)}_{-q}(z)\left(\Phi'_{q}(z)-\Phi'_{0}(z)\right)\right] 
\end{equation}
donde hemos de introducir sobre estas funciones la asimetrización correspondiente. Las expresiones finales son,
\begin{equation}
\Delta_{q}^{v}=n_{0}\int dz\rho^{(in)}_{0}(z)\left(2\Phi''_{0}(z)-\Phi''_{-q}(z)-\Phi''_{q}(z)\right)
\end{equation}
\begin{equation}
\mathcal{B}^{v}_{v}(z)=n_{0}\left(\Phi'_{0}(z)-\Phi'_{q}(z)\right)
\end{equation}
\section{Parte de Exceso}
\label{sec:desarrolloParateExceso}
La expresión formal del desarrollo a orden 2 del funcional de exceso es expresada de modo general mediante,
\begin{equation}
\begin{split}
\mathcal{F}_{ex}[\rho+\delta\rho]&=\mathcal{F}_{ex}[\rho]+\\
&+\sum_{\alpha=0}^{4}\sum_{\nu=0}^{8}\int dz\left( \partial_{\nu}\phi_{\alpha}[\rho]\right) \int d\vec{R}\delta  n^{\nu}(z,\vec{R})+\\
&+\frac{1}{2}\sum_{\alpha=0}^{4}\sum_{\nu,\mu=0}^{8}\int dz \left( \partial_{\nu,\mu}\phi_{\alpha}[\rho]\right) \int d\vec{R}d\vec{R}'\delta n^{\nu}(z,\vec{R})\delta n^{\mu}(z,\vec{R}')
\end{split}
\end{equation}
que esencialmente se basa en que la aproximación funcional utilizada es \textit{local} en el conjunto de funciones $n^{\nu}[\rho]$ donde el índice $\nu$ recorre los índices que definen las funciones peso en FMT, los términos $\phi_{\alpha}$ expresan los tres términos de que consta FMT. La parte atractiva puede ser incluida como un término más $\phi_{4}$ en una aproximación de campo medio.\\

\subsection{Interacción atractiva}
Puede ser obtenida directamente de las expresiones que más adelante introducimos para la parte de esferas duras pero previamente expresamos el resultado explícito para establecer una analogía con el potencial externo,
\begin{equation}
\mathcal{F}_{at}^{v,\delta}=\frac{1}{2}\int dz_{1}d\vec{R}_{1}\rho^{(in)}_{\xi}(z_{1},\vec{R}_{1})\int dz_{2}d\vec{R}_{2}\rho^{(in)}_{\xi}(z_{1}+z_{2},\vec{R}_{1}+\vec{R}_{2})\omega_{AT}(z_{2},\vec{R}_{2})
\end{equation}
que podemos escribir para (\ref{eq:DesarrolloIntrinsecosobreplano})
\begin{subequations}
\begin{align*}
& \frac{1}{2}\int dz_{1}dz_{2}\rho_{0}(z_{1})\rho''_{0}(z_{1}+z_{2})\left(2\omega_{0}(z_{2})-\omega_{-q}(z_{2})-\omega_{q}(z_{2})\right)+\\
&\frac{1}{2}\int dz_{1}dz_{2}\rho_{0}(z_{1})\rho'_{-q}(z_{1}+z_{2})\left(\omega_{-q}(z_{2})-\omega_{0}(z_{2})\right) +\\
&\frac{1}{2}\int dz_{1}dz_{2}\rho_{q}(z_{1})\rho'_{0}(z_{1}+z_{2})\left(\omega_{-q}(z_{2})-\omega_{0}(z_{2})\right) +\\
&\frac{1}{2}\int dz_{1}dz_{2}\rho_{q}(z_{1})\rho_{-q}(z_{1}+z_{2})\omega_{q}(z_{2}) 
\end{align*}
\end{subequations}
Tenemos
\begin{equation}
\Delta_{q}^{at}=\rho_{0}(z_{1})\rho''_{0}(z_{1}+z_{2})\left(2\omega_{0}(z_{2})-\omega_{-q}(z_{2})-\omega_{q}(z_{2})\right)
\end{equation}
\begin{equation}
\mathcal{A}^{at}(z_{1},z_{2})=\omega_{q}(z_{1}-z_{2})
\end{equation}
el resto de los términos se corresponden con la función $\int dz_{1}\rho_{q}(z_{1})\mathcal{B}_{at}(z_{1})$\\

Hemos indicado derivadas de la función $\rho'_{0}$ y $\rho'_{q}$. En el caso de $\Omega_{\delta}$ aparecen funciones $\delta$ luego supone un desarrollo que involucra derivadas en la función distribución $\delta$ pero que siempre aparecen bajo convoluciones. Este mismo desarrollo, aprovechando la presencia de convoluciones, puede expresarse mediante derivadas en las funciones peso que es más natural en un desarrollo de la primera capa como una función $\delta$ aunque ambos casos son  equivalentes, escribimos bajo este último modo el desarrollo de la parte de esferas duras, para ilustrar el procedimiento.\\

\subsection{Exceso de Esferas Duras}
Como indicábamos desarrollamos las funciones $n^{\nu}_{\xi}(z,\vec{R})$. Siguiendo la analogía con el caso atractivo construimos,
\begin{equation}
n_{\xi}^{\nu}(z_{1}+\xi(\vec{R}_{1}),\vec{R}_{1})=\int dz_{2}d\vec{R}_{2}\rho_{\xi}(z_{1}+z_{2}+\xi(\vec{R}_{1}),\vec{R}_{1}+\vec{R}_{2})\omega^{\nu}(z_{2},\vec{R}_{2})
\end{equation}
de esta manera podemos expresar,
\begin{equation}
\begin{split}
\Delta_{v}(q)=&\frac{1}{2}\sum_{\nu}\int dz(\partial_{\nu}\phi)(z)\int dz_{1}\rho_{0}^{(in)}(z+z_{1})\left[ 2\omega''^{\nu}_{0}(z_{1})-\omega''^{\nu}_{-q}(z_{1})-\omega''^{\nu}_{q}(z_{1})\right]+\\
+&\frac{1}{2}\int dz\sum_{\nu,\mu}(\partial_{\nu,\mu}\phi)(z)\left[ \int dz_{1}\rho_{0}^{(in)}(z+z_{1})\left[ \omega'^{\nu}_{-q}(z_{1})-\omega'^{\nu}_{0}(z_{1})\right]\right. \\
&\left. \int dz_{2}\rho_{0}^{(in)}(z+z_{2}) \left[ \omega'^{\mu}_{q}(z_{2})-\omega'^{\mu}_{0}(z_{2})\right]\right]  
\end{split}
\end{equation}
los términos involucrados en la parte relajativa aparecen como,
\begin{equation}
\mathcal{A}_{v}(z_{1},z_{2})=\int dz\sum_{\nu,\mu}(\partial_{\nu,\mu}\phi)(z)\omega^{\nu}_{-q}(z_{1}-z)\omega^{\mu}_{q}(z_{2}-z)
\end{equation}
\begin{equation}
\begin{split}
\mathcal{B}_{v}(z_{1})=-&\sum_{\nu}\int dz \int dz(\partial_{\nu}\phi)(z)\left[\omega''^{\nu}_{q}(z_{1}-z)-\omega''^{\nu}_{0}(z_{1}-z)\right]+\\
+&\int dz\sum_{\nu,\mu}(\partial_{\nu,\mu}\phi)(z)\left[\omega^{\nu}_{q}(z_{1}-z)\right]\int dz_{2}\rho_{0}^{(in)}(z+z_{2})\left[\omega'^{\mu}_{q}(z_{2})-\omega'^{\mu}_{0}(z_{2})\right]
\end{split}
\end{equation}
si sustituimos en las expresiones $\rho^{(in)}$ por $\rho^{(total)}$ podemos reproducir directamente el caso $\Omega_{\delta}$. En ambos casos en el desarrollo es entorno al perfil plano sobre el que se evalúan las funciones $\phi$ que caracterizan al funcional FMT. Las funciones peso $\omega^{\nu}$ están incluidas en el apéndice \S\ref{sec:funcionesPesoq}.

\section{Parte Ideal}

La parte ideal se escribe como,
\begin{eqnarray}
\mathcal{F}_{id}^{v,\delta} &=& \int dzd\vec{R}\rho^{(in)}_{\xi}(z,\vec{R})\left[ ln \left( \rho^{(in)}_{\xi}(z,\vec{R})\right)-1\right] \nonumber \\
&=& \int dzd\vec{R}\rho^{(in)}(z-\xi,\vec{R})\left[ ln \left( \rho^{(in)}_{\xi}(z-\xi,\vec{R})\right)-1\right]
\end{eqnarray}
expresamos $\rho^{(in)}_{\xi}(z-\xi,\vec{R})=\rho_{0}(z)+\delta\rho(z-\xi,\vec{R})$. Esta función $\delta\rho$ posee términos de orden 1 y de orden 2 en $\xi_{q}$, es decir, en las siguientes expresiones llamamos $\delta\rho$ a,
\begin{eqnarray}
\delta\rho &=&\sum_{|q|>0}\xi_{q}(\rho_{q}(z)-\rho'_{0}(z))e^{i\vec{q}\vec{R}} \nonumber \\
&+&\sum_{|q|,|q'|>0}\xi_{q}\xi_{q'}(\frac{1}{2}\rho''_{0}(z)-\rho_{q}(z))e^{i\vec{q+q'}\vec{R}}
\end{eqnarray}
En el funcional tendremos,
\begin{equation}
\mathcal{F}_{id}^{v,\delta} = \int dzd\vec{R}\left(\rho_{0}(z)+\delta\rho\right)\left[ ln\left(\rho_{0}(z)+\delta\rho\right)-1\right]
\end{equation}
Definimos $\Delta\mathcal{F}_{id}^{v,\delta}=\mathcal{F}_{id}^{v,\delta}-\mathcal{F}_{id}^{0}$, donde $\mathcal{F}_{id}^{0}$ es el funcional del sistema ideal para el perfil dado por $\rho_{0}(z)$ entonces tras un desarrollo en el término logarítmico podemos escribir, 
\begin{equation}
\Delta\mathcal{F}_{id}^{v,\delta} =\int dzd\vec{R}(\delta\rho)\ln (\rho_{0}(z))+\int dzd\vec{R}\frac{1}{2}\frac{(\delta\rho)^{2}}{\rho_{0}(z)}
\end{equation}
De todos los términos que incluye la parte ideal solo sobrevive el que incluye $\rho_{q}(z)$ y aparece en la matriz $\mathcal{A}(z_{1},z_{2})$,
\begin{equation}
\mathcal{F}_{id}^{v,\delta} =\frac{1}{2}\int dz\frac{1}{\rho_{0}(z)}\rho_{q}(z)\rho_{-q}(z)
\end{equation}
y por tanto,
\begin{equation}
\mathcal{A}_{id}(z_{1},z_{2})=\frac{1}{\rho_{0}(z_{1})}\delta(z_{1}-z_{2})
\end{equation}
que como es de esperar busca hacer más homogéneo el sistema y en consecuencia solo aparece en la diagonal de la matriz $\mathcal{A}$.
\chapter{Funciones peso para el cálculo de $\Omega_{v,\delta}$}
\label{sec:funcionesPesoq}

Para completar la descripción dada en \S\ref{sec:desarrolloAnaliticoDEOmega} son necesarias las funciones peso del funcional FMT-MFA utilizado, se detallan en este apéndice.

\section{Funciones peso para FMT-interpolación dimensional}
Se escriben las funciones peso en la simetría adecuada para el desarrollo dado en el apéndice \S\ref{sec:apendiceIntrinseco}. Se comienza con dos definiciones para el desarrollo Fourier realizado que escribiremos para fijar notación,
\begin{equation}
\hat{f}(z,\vec{q})=\int d\vec{R} f(z,\vec{R})e^{-i\vec{q}\vec{R}}
\end{equation}
\begin{equation}
f(z,\vec{R})=2\pi \int d\vec{q} \hat{f}(z,\vec{q})e^{i\vec{q}\vec{R}}
\end{equation}
Las expresiones para los pesos son determinadas en un sistema de coordenadas $(\hat{\mu}_{\shortparallel},\hat{\mu}_{\perp},\hat{\mu}_{z})$ donde el desarrollo en el plano xy es realizado mediante un vector unitario paralelo $\hat{\mu}_{\shortparallel}$ y perpendicular $\hat{\mu}_{\perp}$ al vector $\vec{q}$.\\

Haciendo uso de las funciones de Bessel y notando $y=\sqrt{R^{2}-z^{2}}$ con $R=\sigma/2$ el radio molecular, el conjunto de pesos puede escribirse como,
\begin{equation}
\omega^ {0}(z,-q)=\frac{2\pi}{q^{2}}qyJ_{1}(qy)\Theta(R-\vert z \vert)
\end{equation}
\begin{equation}
\omega^ {1}(z,-q)=\frac{1}{2R}J_{0}(qy)\Theta(R-\vert z \vert)
\end{equation}
\begin{equation}
\vec{\omega}(z,-q)=\left[ \frac{1}{2R^{2}}yJ_{1}(qy)\hat{\mu}_{\shortparallel}+zJ_{0}(qy)\hat{\mu}_{z}\right] \Theta(R-\vert z \vert)
\end{equation}
\begin{equation}
\widehat{\omega}(z,-q)=\frac{1}{2R^{3}}
    \left( \begin{array}{lcr}
            y^{2}\frac{J_{0}(qy)-J_{2}(qy)}{2}       & 0                                                       & zy^{2}J_{1}(qy)  \\
            0                                                       & y^{2}\frac{J_{0}(qy)+J_{2}(qy)}{2}      & 0  \\
            zy^{2}J_{1}(qy)                            &  0                                                      &  z^{2}J_{0}(qy)
           \end{array}
    \right )
\Theta(R-\vert z \vert)
\end{equation}

Todos los pesos poseen una misma estructura $\omega^{\nu}(z,-q)=W^{\nu}(z,-q)\Theta(R-\vert z \vert)$ por tanto el cálculo sistemático de las diferenciaciones en la coordenada z de las funciones peso tendrá contribuciones tanto de la derivada de $W$ como de la función paso de Heaviside. Como la última es conocida escribimos, haciendo uso de las propiedades de diferenciación de las funciones de Bessel las derivadas primera y segunda de los factores $W^{\nu}(z,-q)$,
\begin{equation}
\frac{d}{dz}W^{0}(z,-q)=-2\pi J_{0}(qy)z
\end{equation}
\begin{equation}
\frac{d}{dz}W^{1}(z,-q)=\frac{zq^{2}}{2Rqy}J_{1}(qy)
\end{equation}
\begin{equation}
\frac{d}{dz}\vec{W}(z,-q)=\left[ \frac{1}{2R^{2}}yJ_{1}(qy)\hat{\mu}_{\shortparallel}+zJ_{0}(qy)\hat{\mu}_{z}\right]
\end{equation}
\begin{equation}
\frac{d}{dz}\widehat{W}(z,-q)=\frac{1}{2R^{3}}
    \left( \begin{array}{lcr}
            y^{2}\frac{J_{0}(qy)-J_{2}(qy)}{2}       & 0                                                       & zy^{2}J_{1}(qy)  \\
            0                                                       & y^{2}\frac{J_{0}(qy)+J_{2}(qy)}{2}      & 0  \\
            zy^{2}J_{1}(qy)                            &  0                                                      &  z^{2}J_{0}(qy)
           \end{array}
    \right )
\end{equation}
la suma de los elementos de la diagonal vuelve a resultar el peso escalar $\frac{d}{dz}W^{1}(z,-q)$.
Mientras que las derivadas segundas se expresan como,

\begin{equation}
\frac{d^{2}}{dz^{2}}W^{0}(z,-q)=-2\pi \left[ J_{0}(qy)+\frac{z^{2}q^{2}}{qy}J_{1}(qy)\right] 
\end{equation}
\begin{equation}
\frac{d^{2}}{dz^{2}}W^{1}(z,-q)=\frac{1}{2R}\frac{q^{2}}{qy}\left[ J_{1}(qy)+\frac{z^{2}q^{2}}{qy}J_{2}(qy) \right]
\end{equation}
\begin{equation}
\frac{d^{2}}{dz^{2}}\vec{W}(z,-q)= \frac{1}{2R^{2}}\left\lbrace \left[-qJ_{0}(qy)+\frac{z^{2}q^{2}}{qy}\right]\hat{\mu}_{\shortparallel}+\left[\frac{3zq^{2}}{qy}J_{1}(qy)+\frac{z^{3}q^{4}}{(qy)^{2}}\right]\hat{\mu}_{z}\right\rbrace
\end{equation}
indicamos las componentes del tensor,
\begin{equation}
\frac{d^{2}}{dz^{2}}\widehat{W}^{(\shortparallel,\shortparallel)}(z,-q)=
\frac{1}{2R^{3}}\left[(qy)J_{1}(qy)-J_{0}(qy)-\frac{z^{2}q^{2}}{qy}(J_{1}(qy)+(qy)J_{0}(qy) \right] 
\end{equation}
\begin{equation}
\frac{d^{2}}{dz^{2}}\widehat{W}^{(\perp,\perp)}(z,-q)=
\frac{1}{2R^{3}}(-)\left[J_{0}(qy)+\frac{z^{2}q^{2}}{qy}J_{1}(qy) \right] 
\end{equation}
\begin{equation}
\frac{d^{2}}{dz^{2}}\widehat{W}^{(z,z)}(z,-q)=
\frac{1}{2R^{3}}\left[2J_{0}(qy)+\frac{5z^{2}q^{2}}{qy}J_{1}(qy)+\frac{z^{4}q^{4}}{(qy)^{2}}J_{2}(qy)\right] 
\end{equation}
\begin{equation}
\frac{d^{2}}{dz^{2}}\widehat{W}^{(\shortparallel,z)}(z,-q)=
\frac{1}{2R^{3}}\frac{1}{q}\left[ (qy+z^{2}q^{2})J_{1}(qy)+(-2z)q^{2}(qy)J_{0}(qy)\right] 
\end{equation}
la propiedad de la traza del tensor vuelve a mantenerse en virtud de la relación de recurrencia $J_{n+1}(x)=\frac{2n}{x}J_{n}(x)-J_{n-1}(x)$.\\

Como el lector puede comprobar en \S\ref{sec:desarrolloAnaliticoDEOmega} aparecen derivadas segundas de las funciones peso y por consiguiente funciones $\delta'$ como acabamos de ver. El cálculo detallado de los factores que acompañan los reduce a las expresiones de la forma,
\begin{equation}
\sum_{\mu,\nu}(\partial_{\nu,\mu}\phi)W^{\nu}_{0}(\pm R)W^{\mu}_{0}(\pm R)
\end{equation}
y sustituyendo los valores para los pesos obtenidos la suma anterior demuestra ser cero y por tanto las contribuciones a la energía libre están bien definidas.\\

\section{Parte Atractiva en campo medio}
En el caso del peso en la parte atractiva\footnote{No hemos incluido el cutoff en esta expresiones, entre los potenciales utilizados solo puede ser necesario para el Lennard-Jones mientras que en la práctica los cálculos de corrugación los hemos extendido lo suficiente para que no sea relevante.}, 
\begin{equation}
\omega_{at}(z,-q)=2\pi\int_{0}^{\infty}dy y J_{0}(qy)\phi_{at}(\sqrt{z^{2}+y^{2}})
\end{equation}
para la primera derivada,
\begin{equation}
\frac{d}{dz}\omega_{at}(z,-q)=2\pi\int_{0}^{\infty}dy y J_{0}(qy)\phi'_{at}(\sqrt{z^{2}+y^{2}})\frac{z}{\sqrt{z^{2}+y^{2}}}
\end{equation}
que también puede ser expresada utilizando funciones de Bessel de orden 1, desde la expresión $\omega_{at}(z,-q)=2\pi\int_{|z|}^{\infty}dy y J_{0}(q\sqrt{z^{2}+y^{2}})\phi_{at}(|y|)$.
\\
En el caso de la asimetrización del peso atractivo hemos usado,
\begin{equation}
\omega_{at}^{asim}(z,-q)=\omega_{at}^{sim}(z,-q)\frac{1}{2}\left[1+tanh[(z+\alpha)\beta]\right]
\end{equation}
Representa la interacción atractiva de la primera capa con el resto del perfil. Un ejemplo de diferentes parámetros se puede ver en la figura (\ref{fig:EstudioLJcasodelta}).

\chapter{Teorías de campo efectivo}
\label{apn:teoriascampoefectivo}

Las teorías de campo efectivo intentan describir propiedades del sistema en una escala mesoscópica salvando los detalles microscópicos, para lo que se define un parámetro de orden que refleja los cambios fenomenológicos de interés y con el que se construye un funcional, a menudo definido de modo heurístico, que permite obtener la física de interés. Usualmente la conexión microscópica es argumentada más que demostrada y los parámetros que aparecen en este funcional pueden contener dependencias en parámetros termodinámicos cosa que no sucede en un hamiltoniano microscópico. Si este funcional posee un campo externo acoplado al parámetro de orden puede describir situaciones no homogéneas al igual que en las aproximaciones del funcional de la densidad. En consecuencia las teorías de campo y los funcionales de la densidad son dos modos de tratar formalmente sistemas no homogéneos que presentan similitudes pero son aproximaciones de carácter diferente.\\

En la teoría del funcional de la densidad si partimos del funcional exacto, el perfil de densidad es una cantidad de equilibrio (una vez realizada la minimización correspondiente) que ha promediado las fluctuaciones consistentes con el estado de equilibrio estudiado. En la práctica la situación difiere ya que las diferentes aproximaciones a dicho funcional de la densidad exacto permiten obtener perfiles de densidad similares al que suponemos de equilibrio pero no constituyen funcionales exactos y de aquí resulta obvio que los efectos de determinadas fluctuaciones pueden perderse según la naturaleza de la aproximación, en los casos macroscópicos no suele tener gran relevancia pero en sistema pequeños o interfases resulta importante determinar \textit{que} fluctuaciones y \textit{de que modo} están incluidas en el funcional.\\

Las teorías de campo efectivo al partir de un parámetro de orden\footnote{La idea original parte de Landau quien cerca del punto crítico propuso una teoría general a partir de las aproximaciones de campo medio conocidas. Esta teoría partía de la existencia de un parámetro de orden $\eta$ que daba cuenta del comportamiento de sistema cerca del punto critico. Esta teoría se puede generalizar introduciendo el sistema en un campo externo $V_{ext}(\vec{r})$ con lo que habremos de utilizar un parámetro de orden $\eta(\vec{r})$ que representa el promedio de una región de cierto tamaño $\Lambda^{-3}$ la definición de $\eta$ esta por tanto ligada a valor de esta constante, con todo en la teoría de Landau esperamos que al final no tengamos en las predicciones de la teoría una dependencia con el parámetro $\Lambda$.}
 que determina el estado del sistema, expresan la energía libre del sistema cuando lo \emph{obligamos a estar en dicho estado}, mientras que la función de partición que tiene en cuenta todas las posibles configuraciones del sistema es recuperada tras una integración funcional de un hamiltoniano efectivo, $\mathcal{H}_{eff}$ sobre dicho parámetro de orden. 

\section{Propiedades termodinámicas desde $\mathcal{H}_{eff}$}\label{sec:Heff}

Los hamiltonianos efectivos son funcionales del parámetro de orden\footnote{Y que permiten formalmente obtener este para un estado termodinámico mediante su minimización.} pero no representan la energía libre del sistema en equilibro, para determinarla a partir del $\mathcal{H}_{eff}[\eta(\vec{r})]$ procedemos mediante la expresión\cite{BOOK-NigelGoldenfeld},\\
\begin{equation}
Z(T,V,N)=\int D\eta(\vec{r})e^{-\beta H_{eff}[\eta(\vec{r})]}
\label{eqn:ZdesdeHeff}
\end{equation}
y la energía libre termodinámica se calcula mediante\footnote{Usamos el colectivo canónico pero igualmente podemos hacer referencia al macrocanónico.},
\begin{equation}
F(T,V,N)=\beta lnZ(T,V,N)
\end{equation}

Para describir sistemas no-homogéneos hemos introducido un parámetro de orden dependiente de las coordenadas espaciales y un hamiltoniano efectivo que incluye un potencial externo acoplado, $V_{ext}(\vec{r})\eta(\vec{r})$.\\

Desde el punto de vista de los microestados cada \emph{estado}, determinado por un parámetro de orden, corresponderá a un conjunto de microestados relativamente amplio, mientras que la conexión microscópica para $\mathcal{H}_{eff}(\vec{r})$ es,
\begin{equation}
e^{-\beta H_{eff}[\eta(\vec{r})]} =Tr_{\eta}e^{-\beta \mathcal{H}(r_{N})}
\end{equation}
donde $Tr_{\eta}$ es una traza parcial restringida a los microestados compatibles con el parámetro de orden $\eta$ y $\mathcal{H}_{N}$ el hamiltoniano microscópico. Por esto la función de partición, ec. (\ref{eqn:ZdesdeHeff}), se convierte en una integral funcional sobre los posibles parámetros de orden $\eta$ del factor de Boltzmann del $\mathcal{H}_{eff}$ y justifica que a esta última se la denomine energía libre de grano grueso. Los requerimientos termodinámicos están sobre $Z(T,V,N)$ y no sobre $H_{eff}$.\\

El cálculo de la integral funcional se realiza de modo práctico en el espacio de Fourier\footnote{En el es más fácil incorporar la presencia del cutoff $\Lambda$ que realce el carácter mesoscópico de $\eta$.} donde la integral funcional indicada queda,
\begin{equation}
\int D\eta(\vec{r})=\int \prod_{|\vec{k}|<\Lambda}d\eta_{\vec{k}}
\end{equation}
y $\eta_{\vec{k}}$ posee una parte real y una parte compleja. Si $\eta(\vec{r})$ es real tendremos que las partes real y compleja están relacionadas por,
\begin{equation}
Re(\eta_{\vec{k}})=Re(\eta_{\vec{-k}})
\end{equation}
\begin{equation}
Im(\eta_{\vec{k}})=-Im(\eta_{\vec{-k}})
\end{equation}

En el formalismo del funcional de la densidad, las funciones de distribución y correlación obtenidas mediante diferenciación funcional así como sus relaciones han sido la base sobre la que se han construido la mayoría de los resultados. En este contexto también se definen una serie de funciones a partir de derivadas funcionales de la energía libre respecto del potencial externo. Las más relevantes para nosotros serán,\\
\begin{equation}
<\eta(\vec{r})>=\frac{\delta F}{\delta V_{ext}(\vec{r})}
\end{equation}
\begin{equation}
G(\vec{r},\vec{r}')=\beta^{-1}\chi_{T}(\vec{r},\vec{r}')=\beta^{-1}\frac{\delta <\eta(\vec{r})>}{\delta V_{ext}(\vec{r'})}
\label{eqn:G2desdeHeff}
\end{equation}
los promedios son en el espacio de parámetros de orden $\eta(\vec{r})$.

\subsection{Aproximación de campo medio}
Realizar el procedimiento de integración funcional de modo analítico es complicado y la primera aproximación razonable es determinar el parámetro de orden de equilibrio como aquel que minimiza el hamiltoniano efectivo\footnote{Se suele denominar aproximación de punto de silla, y es viable ya que aunque $\mathcal{H}_{eff}$ ha de ser analítico, un proceso de minimización puede dar lugar a comportamientos no-analíticos y puede por tanto describir transiciones de fase y fenómenos críticos que suele ser el objetivo final.} y se determinan en una segunda etapa posibles correcciones debidas a fluctuaciones sobre el parámetro de orden anterior. De este manera la imposición de que $\mathcal{H}_{eff}$ sea \textit{estacionario} lleva a una ecuación de Euler-Lagrange,\\
\begin{equation}
\frac{\delta H_{eff}[\eta]}{\delta\eta}=0
\end{equation}
me permite obtener\footnote{Bajo estas definiciones es posible encontrar haciendo uso de la ecuación (\ref{eqn:G2desdeHeff})
 una relación para la función de correlación de dos puntos que permite establecer su comportamiento asintótico para $r>>\xi_{B}$,\\
\begin{equation}
G(r)\sim\frac{e^{|\vec{r}|/\xi}}{|\vec{r}|^{(d-1)/2}\xi^{(d-3)/2}}
\end{equation}
en el caso de estudiar las proximidades del punto crítico en un sistema uniforme el parámetro de orden puede ser la densidad y la G(r) estaría relacionado con la función de distribución radial.
} un $\eta_{MF}(\vec{r})$.

\subsection{Aproximación Gaussiana}
\label{sec:aproxGaussiana}
\index{Gaussiana!Teoría}
La forma más extendida de hamiltoniano efectivo viene dada por la expresión de \textit{Landau-Ginzburg-Wilson} que condensa propiedades generales de simetría, localidad, homogeneidad y analiticidad que se imponen heurísticamente sobre $\mathcal{H}_{eff}$.\\

Su forma es,
\begin{equation}
\mathcal{H}_{LGW}[\eta(\vec{r})]=\int d^{d}\vec{r}\left[ a\eta(\vec{r})^{2}+b\eta(\vec{r})^{4}+\frac{\gamma}{2}(\nabla\eta(\vec{x}))^{2}-V_{ext}(\vec{r})\eta(\vec{r})\right] 
\end{equation}
a la que hemos acoplado el campo externo $V_{ext}(\vec{r})$ y la exigencia de estabilidad impone restricciones sobre los signos de los parámetros fenomenológicos ($a,b,\gamma$).\\

Podemos ir más allá de la aproximación de campo medio, ya que la integral funcional que define la función de partición si es resoluble analíticamente en el caso de poseer la forma de una gaussiana\footnote{Se puede resolver el problema analíticamente ya que la característica fundamental de esta aproximación es suponer que las distribuciones de probabilidad de las fluctuaciones siguen una distribución normal lo que permite que las diferentes variables aleatorias se conviertan en independientes.}, por tanto se puede estudiar las propiedades de $\mathcal{H}_{LGW}$ si ignoramos los términos no cuadráticos en el parámetro de orden.\\

La expresión gaussiana en ausencia de campo externo es,

\begin{equation}
\mathcal{H}^{(g)}[\eta(\vec{r})]=\int d^{d}\vec{r}\left[\frac{1}{2}\gamma(\nabla\eta(\vec{r}))^{2}+at\eta(\vec{r})^{2}\right]+a_{0}V
\label{eqn:HeffGaussiano}
\end{equation}

Una transformación que lleve $\mathcal{H}^{(g)}$ a una forma diagonal permite realizar las integraciones en $D\eta(\vec{r})$, y la manera más sencilla\footnote{Son posibles transformaciones en el espacio real.} es una transformada Fourier donde la expresión directamente es diagonal\footnote{
Las expresiones que usamos son:
\begin{eqnarray}
\eta(\vec{x})=\frac{1}{L^{d}}\sum_{\vec{k}}\eta_{\vec{k}}e^{-i\vec{k}\vec{x}}\\
\eta_{k}=\int d^{d}\vec{x}\eta(\vec{x})e^{i\vec{k}\vec{x}}
\end{eqnarray}
las relaciones se completan con los siguientes resultados
\begin{eqnarray}
\sum_{k}e^{i\vec{k}(\vec{x}-(\vec{x}')}=L^{d}\delta(\vec{x}-\vec{x'})\\
\int d^{d}\vec{x}e^{i(\vec{k}-\vec{k}')\vec{x}}=L^{d}\delta_{\vec{k},\vec{k}'}
\end{eqnarray}
donde hemos escrito el caso d-dimensional de volumen $L^{d}$.}
 de $\eta(\vec{r})$ tenemos,
\begin{equation}
\mathcal{H}^{(g)}[\{\eta_{k}\}]=\frac{1}{V}\sum_{k}\frac{1}{2}|\eta_{k}|^{2}\left[2at+\gamma k^{2}\right]+a_{0}V
\end{equation}
el cálculo de la función de partición se realiza mediante,
\begin{equation}
e^{-\beta F}=\int D\eta(\vec{r})e^{-\beta \mathcal{H}[\eta(\vec{r})]}=\int \prod_{|\vec{k}|<\Lambda}d\eta_{\vec{k}}e^{-\beta \mathcal{H}}=\int \prod_{|\vec{k}|<\Lambda}d\,Re(\eta_{\vec{k}})d\, Im(\eta_{\vec{k}})e^{-\beta \mathcal{H}}
\end{equation}
donde en $\mathcal{H}^{(g)}$ tenemos $|\eta_{\vec{k}}|^{2}=Re(\eta_{\vec{k}})^{2}+Im(\eta_{\vec{k}})^{2}$.\\

Si $\eta(\vec{r})$ es real su partes real e imaginaria esta relacionadas luego hemos de restringir la suma en vectores $\vec{k}$, e incluimos el factor $\frac{1}{2}$ de modo explicito. El resultado final para la energía libre es,
\begin{equation}
F=a_{0}V-\frac{1}{2}\sum_{|\vec{k}|<\Lambda}ln\frac{2\pi V\beta^{-1}}{2at+\gamma k^{2}}
\label{eqn:elibreGaussiana}
\end{equation}
Para el caso de las función de correlación entre dos puntos podemos podemos hacer uso de que de modo directo,
\begin{equation}
<|\eta_{k}|^{2}>=\frac{V\beta^{-1}}{2at+\gamma k^{2}}=V\hat{G}(\vec{k})
\end{equation}
y 
\begin{equation}
<\eta(\vec{r})\eta(\vec{r}')>=\frac{1}{V}\sum_{\vec{k}<\Lambda}\frac{\beta^{-1}}{2at+\gamma k^{2}}e^{i\vec{k}(\vec{r}-\vec{r}')}
\label{eqn:FuncionCorrelacionGaussianaOrden2}
\end{equation}
que nos servirá en la teoría de ondas de capilares para establecer una medida de la anchura de la interfase.\\
\subsection{Determinación de fluctuaciones sobre campo medio}
El cálculo anterior puede ser utilizado del siguiente modo, partimos de un hamiltoniano efectivo general $L_{eff}$ caracterizado por ser local en el parámetro de orden salvo un término que como en $\mathcal{H}_{LGW}$ depende del cuadrado del gradiente del parámetro de orden. Sea $\phi_{0}(\vec{x})$ el parámetro de orden que hace estacionario a $L_{eff}$ y buscamos la relevancia de las fluctuaciones entorno de dicho parámetro de orden $\eta(\vec{x})=\delta\phi(\vec{x})=\phi(\vec{x})-\phi_{0}(\vec{x})$. Si desarrollamos hasta orden $\eta^{2}$ nuestro $L_{eff}$ y aplicamos que $\phi_{0}(\vec{x})$ hace estacionario al funcional podemos expresar,
\begin{equation}
L_{eff}[\phi]=L_{eff}[\phi_{0}]+L^{(g)}_{eff}[\eta]+O[\eta^{2}]
\end{equation}
donde $L^{(g)}_{eff}[\eta]$ posee una forma gaussiana dada por
\begin{equation}
L^{(g)}_{eff}[\phi]=\int d^{d}\vec{r}\left[\frac{1}{2}\gamma(\nabla\eta)^{2}+\mathcal{L}''_{eff}(\eta) \eta^{2}\right]
\end{equation}
que podemos diagonalizar.  Es interesante notar que este procedimiento es consistente en el proceso de incorporar fluctuaciones (más allá de que solo incluimos términos a ordenes bajos), en el sentido de que $\phi_{mf}=\phi_{0}$ no incorpora previamente fluctuaciones\footnote{Insistimos esto no ocurre así a priori en una aproximación de van der Waals generalizada donde primero debemos determinar que fluctuaciones hemos incluido en  $\phi_{DFT}$.}.\\

Supongamos la siguiente situación, en ausencia de potencial externo $\phi_{mf}$ es uniforme pero sus fluctuaciones si presentan variaciones espaciales. Si existen varios grados de libertad $(\phi_{1},...,\phi_{n})$ tal que uno de ellos representa $\phi_{mf}$ mientras que el resto son nulos, las fluctuaciones sobre estos últimos son posibles sin costo de energía\cite{BOOK-NigelGoldenfeld}. Los modos \textit{colectivos} que se producen son conocidos como \textit{modos de Goldstone} y se corresponden con la ruptura de la simetría del sistema presente en el hamiltoniano. Estos modos, al igual que las ondas capilares, no tienen una fuerza restauradora en ausencia de campos externos y presentan ciertas analogías. El análisis de los modos de Goldstone a ordenes superiores al gaussiano muestra que es posible reexpresar el término gaussiano mediante una constante $\gamma$ redefinida, este hecho hizo sugerir que la interacción de las ondas capilares a distintas escalas hace subir la tensión superficial y sobre argumentaciones cualitativas basadas en las analogías con el proceso de renormalización de $\gamma$ se propuso una $\gamma(k)$, su origen es, pues, diferente del expresado a lo largo de la memoria\cite{PhysRevA.39.6346}.\\

\subsection{Aplicación a la teoría de ondas capilares}
\label{sec:funcionesDistribucionAlturas}
\index{Ondas Capilares!Teoría}
La teoría de ondas capilares, \S\ref{Sec:Cap1OndasCapilares}, es una teoría gaussiana y podemos utilizar las expresiones anteriores para determinar el conjunto de funciones y promedios utilizados a lo largo de la memoria, estos se expresarán de modo general como,
\begin{equation}
<f(z,\xi(\vec{R}))>_{\xi}=\frac{1}{\mathcal{Z}_{I}}\int \mathcal{D}[\xi] e^{-\beta\mathcal{H}_{I}[\xi]}f(z,\xi(\vec{R}))
\label{eqn:promedioestadiscoGaussiano}
\end{equation}
Y por tanto el formalismo anterior es directamente aplicable a un hamiltoniano de la forma\footnote{Suponemos que $\xi(\vec{R})=\sum_{q}\xi_{q}e^{-\vec{q}\vec{R}}$}:
\begin{equation}
\mathcal{H}_{I}[\{\xi_{q}\}]=\frac{1}{2}L^{d}\sum_{q_{min}<q<q_{max}}|\xi_{q}|^{2}\left[h(q)\right]
\label{eqn:hamiltonianoGaussiano}
\end{equation}
que responde a la forma dada en ec. (\ref{eqn:GaussianoCONgammaq}). Esta expresión permite escribir, usando ec. (\ref{eqn:elibreGaussiana}) la corrección a energías como:
\begin{equation}
\beta\gamma=\beta\gamma_{0}-\frac{1}{2}L^{-d}\sum_{q_{min}<q<q_{max}}ln\frac{2\pi}{\beta h(q)L^{d}[\xi_{q}]^{2}}
\end{equation}
La desviación cuadrática media se puede escribir como:
\begin{equation}
<\xi_{q_{2}}\xi_{q_{1}}>=\delta_{q_{1},-q_{2}}<|\xi_{q_{1}}|^{2}>=\delta_{q_{1},-q_{2}}\frac{1}{\beta L^{d} h(q)}
\label{eqn:valorEsperadoXIq2}
\end{equation}
En consecuencia\footnote{En el caso del modelo clásico de ondas capilares esta función se puede determinar analiticamente para valores hasta d=4 y expresarse unificadamente mediante la función de Bessel de segunda especie $\mathcal{K}_{\nu}$,
\begin{equation}
S(|\vec{R}|)=\frac{\xi_{cw}^{2\nu}}{(2\pi)^{(d-1)/2}\beta\gamma}(|\vec{R}|^{2}+q_{max}^{-2})^{\nu/2}\mathcal{K}_{\nu}(|\vec{R}|^{2}+q_{max}^{-2})
\label{eqn:distrualturasSdeR}
\end{equation}
donde $2\nu=3-d$. Esto lleva a que para $d<3$ la anchura de la interfase es $\xi_{cw}^{3-d}/\beta\gamma$ y por tanto diverge. En el caso de $d=3$ tenemos una divergencia logarítmica mientras que en $d>3$ tenemos que la divergencia desaparece. },
\begin{equation}
S(\vec{R})\equiv<\xi(\vec{R})\xi(\vec{0})>=\frac{1}{(2\pi)^{d}}\int_{q_{min}<q<q_{max}} d^{d}\vec{q}\frac{e^{i\vec{q}\vec{R}}}{\beta h(q)}
\label{eqn:defdeSdeR}
\end{equation}
Y promedio de los perfiles de densidad en $\xi(\vec{R})$.
\begin{eqnarray}
\rho(z) &=& <\rho_{0}(z-\xi(\vec{R}))>_{\xi} \nonumber \\
 &=& \frac{1}{Z_{I}}\int \mathcal{D}[\xi]\left[\int dz_{1}\delta(\xi(\vec{R}-z_{1})\rho_{0}(z-z_{1})\right]e^{-\beta\mathcal{H}_{I}[\xi]} \nonumber \\
 &=& \frac{1}{Z_{I}}\int dz_{1}\rho_{0}(z-z_{1})\int \mathcal{D}[\xi]\left[\delta(\xi(\vec{R})-z_{1})\right]e^{-\beta \mathcal{H}_{I}[\xi]}  \nonumber \\
 &=&\frac{1}{Z_{I}}\int dz_{1}\rho_{0}(z-z_{1})\mathcal{P}(z_{1})\label{eqn:perfildensidadmedioCWT}
\end{eqnarray}
\index{Perfil de densidad!Medio}
Del mismo modo si escribo las correlaciones como:
\begin{eqnarray}
G^{(2)}(z_{1},z_{2},|\vec{R}_{12}|) = <(\rho_{0}(z_{1}-\xi(\vec{R_{1}}))-\rho(z_{1}))(\rho_{0}(z_{2}-\xi(\vec{R_{2}}))-\rho(z_{2}))>_{\xi}
\end{eqnarray}
Permite escribir:
\begin{eqnarray}
G^{(2)}(z_{1},z_{2},|\vec{R}_{12}|) &=& \rho(z_{1})\rho(z_{2})-\rho(z_{2})\int dz_{3}\rho_{0}(z_{3}-z_{1})\mathcal{P}(z_{3})\nonumber \\ 
&-&\rho(z_{1})\int dz_{3}\rho_{0}(z_{3}-z_{2})\mathcal{P}(z_{3})\nonumber \\&+&
\int dz_{3}\int dz_{4}\rho_{0}(z_{3}-z_{2})\rho_{0}(z_{3}-z_{1})\mathcal{P}(z_{1},z_{2},|\vec{R}_{12}|)
\end{eqnarray}
donde tenemos que:
\begin{eqnarray}
\mathcal{P}(z_{1})=\int \mathcal{D}[\xi]\left[\delta(\xi(\vec{R}_{1})-z_{1})e^{-\beta\mathcal{H}[\xi]}\right]
\end{eqnarray}
\begin{eqnarray}
\mathcal{P}(z_{1},z_{2},|\vec{R}_{12}|)=\int \mathcal{D}[\xi]\left[\delta(\xi(\vec{R}_{1})-z_{1})\delta(\xi(\vec{R}_{2})-z_{2})e^{-\beta\mathcal{H}[\xi]}\right]
\end{eqnarray}
Para determinar las expresiones anteriores escribimos la función $\delta(\xi(\vec{R}_{1})-z_{1})$ en el espacio Fourier así como la propia función $\xi(\vec{R})$. Las integraciones permite expresar,
\begin{eqnarray}
\mathcal{P}(z_{1})=\frac{1}{\sqrt{2\pi\Delta^{2}_{cw}}}e^{-\frac{z^{2}}{2\Delta^{2}_{cw}}}
\end{eqnarray}
mientras que:
\begin{eqnarray}
\Delta^{2}_{cw}=S(0)=\frac{1}{(2\pi)^{d}}\int_{q_{min}<q<q_{max}}d^{d}\vec{q}\frac{1}{\beta h(q)}
\label{eqn:DeltaCW}
\end{eqnarray}
En cuanto a la determinación de la función $\mathcal{P}(z_{1},z_{2},|\vec{R}_{12}|)$ es análogo, en este caso involucra a $S(0)$ y $S(\vec{R}_{12})$, y tenemos una distribución gaussina bivariante. La expresión formal resulta ser,
\begin{eqnarray}
\mathcal{P}(z_{1},z_{2},|\vec{R}_{12}|)=\frac{1}{\sqrt{2\pi(S(0)^{2}-S(R_{12})^{2})}}e^{-\frac{-S(0)(z_{1}^{2}+z_{2}^{2})+2S(R_{12})z_{1}z_{2}}{2(S(0)^{2}-S(R_{12})^{2})}}
\end{eqnarray}
Además en el se puede expresar $\mathcal{P}(z_{1},z_{2},|\vec{R}_{12}|)$ a partir de $\mathcal{P}(z_{1})$ y $S(\vec{R}_{12})$ mediante una serie, que a primer orden implica, y volvemos así a ec.(\ref{eqn:asintoticoG2cwt}),
\begin{eqnarray}
G^{(2)}(z_{1},z_{2},|\vec{R}_{12}|) \simeq S(R_{12})\int dz_{1}'\mathcal{P}(z_{1}')\rho'_{0}(z_{1}'-z_{1})\int dz_{2}'\mathcal{P}(z_{2}')\rho'_{0}(z_{2}'-z_{2})
\end{eqnarray}

Para resolver la expresión (\ref{eqn:denLxDESDEdenINT}) que determina el perfil de densidad $\rho(z,L_{x})$ cuando para el perfil intrínseco incluimos un término $\sum_{q}\xi_{q}e^{-\vec{q}\vec{R}}\rho_{q}(z)$. El cálculo es análogo a ec. (\ref{eqn:perfildensidadmedioCWT}) aunque en este caso el resultado es expresado como la transformada Fourier de la misma función exponencial involucrada en el caso anterior pero con un factor $iq$ que dará lugar al $\frac{d}{dz}\mathcal{P}(z)$.

\subsection{Expresión en el formalismo de proyectores de las funciones de distribución de alturas.}

Haciendo uso del formalismo del operador proyección podemos expresar la jerarquía de funciones de distribución de alturas del siguiente modo:
\begin{equation}
\mathcal{P}(z_{1})=\int d\mu(\omega)e^{-<\hat{\rho}|u>}\Lambda_{1}(\omega)
\end{equation}
\begin{equation}
\mathcal{P}(z_{1},z_{2},|\vec{R}_{12}|)=\int d\mu(\omega)e^{-<\hat{\rho}|u>}\Lambda_{12}(\omega)
\end{equation}
donde 
\begin{equation}
\Lambda_{1}(\omega)=\int \mathcal{D}[\xi]\Lambda(\xi,\omega)\delta(\xi(\vec{R}_{1})-z_{1})e^{-\beta\mathcal{H}[\xi]}
\end{equation}
\begin{equation}
\Lambda_{12}(\omega)=\int \mathcal{D}[\xi]\Lambda(\xi,\omega)\delta(\xi(\vec{R}_{1})-z_{1}))\delta(\xi(\vec{R}_{2})-z_{2})e^{-\beta\mathcal{H}[\xi]}
\end{equation}
Podemos obtener la jerarquía $\mathcal{P}^{(n)}$ de funciones de distribución de alturas desde la jerarquía de proyectores $\Lambda_{1...n}(\omega)$ que se obtiene únicamente promediando en el campo $\xi$ bajo una restricción que fija en el espacio un conjunto n de puntos para $\xi$ del operador proyección original.\\

Para terminar, en el contexto de las teorías de campo efectivo y relacionado con la interfase líquido-vapor se ha intentado indagar la relación entre las propiedades de la función de distribución $G^{(2)}(z_{1},z_{2},r_{12})$ y el perfil de densidad $\rho(z_{1})$, problema que ya se comentó en \S\ref{sec:correlacionesCWT} y constituye un ejemplo de aplicación de este formalismo de campos efectivos a la interfase líquido-vapor, véase \cite{PhysRevA.41.6732}.

\chapter{Cálculo de la función $c(r)$ en WDA}
\label{sec:apendiceCwda}
En el caso de WDA expresábamos la energía libre de un fluido de esferas duras mediante la expresión:
\begin{equation}
F_{HS}[\rho]=\int d\vec{r}\rho(\vec{r})\Delta\Psi(\bar{\rho}(\vec{r}))
\end{equation}
donde el funcional era construido a partir de tres funciones peso $\omega_{i}$. Aquí detallamos el cálculo de la función de correlación directa utilizando una expresión integral explicita (que además puede ser utilizada para obtener una expresión también explicita de las funciones peso en el cálculo de la función de distribución de pares por el método de la partícula test para el funcional FMT sin recurrir a métodos de Fourier para resolver las convoluciones). La definición en el formalismo del funcional de la densidad de la función de correlación directa me lleva directamente a la expresión:
\begin{eqnarray}
c^{(2)}(\vec{r}_{1},\vec{r}_{2};\rho_{B})&=&-\Delta\Phi'[\rho_{b}]\frac{\delta\bar{\rho}(\vec{r}_{1})}{\delta\rho(\vec{r}_{2})}\bigg\vert_{\rho_{B}}-\Delta\Phi'[\rho_{b}]\frac{\delta\bar{\rho}(\vec{r}_{1})}{\delta\rho(\vec{r}_{2})}\bigg\vert_{\rho_{B}}  \nonumber \\
&-&\Delta\Phi'[\rho_{b}]\int d\vec{r}\frac{\delta\bar{\rho}(\vec{r})}{\delta\rho(\vec{r}_{1})\delta\rho(\vec{r}_{2})}\bigg\vert_{\rho_{B}}\nonumber \\
&-&\Delta\Phi''[\rho_{b}]\int d\vec{r}\frac{\delta\bar{\rho}(\vec{r})}{\delta\rho(\vec{r}_{1})}\bigg\vert_{\rho_{B}}\frac{\delta\bar{\rho}(\vec{r})}{\delta\rho(\vec{r}_{2})}\bigg\vert_{\rho_{B}}
\end{eqnarray} 
donde las $\Delta\Phi'[\rho_{b}]$ son derivadas respecto de la densidad. Queda resolver las derivadas funcionales de la densidad promedio y evaluarlas en el caso uniforme. En nuestro caso tendremos,
\begin{equation}
\frac{\delta\bar{\rho}(\vec{r})}{\rho(\vec{r}_{1}}\bigg\vert_{\rho_{B}}=\omega(\vert\vec{r}-\vec{r}_{1})\vert,\rho_{B})
\end{equation} 
\begin{equation}
\frac{\delta\bar{\rho}(\vec{r})}{\delta\rho(\vec{r}_{1})\delta\rho(\vec{r}_{2})}\bigg\vert_{\rho_{B}}=\omega'(\vert\vec{r}-\vec{r}_{1}\vert,\rho_{B})\omega(\vert\vec{r}_{1}-\vec{r}_{2}\vert,\rho_{B})+\omega'(\vert\vec{r}-\vec{r}_{2}\vert,\rho_{B})\omega(\vert\vec{r}_{1}-\vec{r}_{2}\vert,\rho_{B})
\end{equation}
Al introducir estas expresiones en el caso uniforme podemos expresar,
\begin{eqnarray}
c^{(2)}(\vert\vec{r}_{12}\vert;\rho_{B})&=&-\Delta\Phi'[\rho_{B}]\omega(\vert\vec{r}_{12}\vert;\rho_{B})-\Delta\Phi'[\rho_{B}]\omega(\vert\vec{r}_{21}\vert;\rho_{B}) \nonumber \\
&-&\Delta\Phi'[\rho_{B}]\rho_{B}\int d\vec{x}\left[\omega'(\vert\vec{x}\vert;\rho_{B})\omega(\vert\vec{x}+\vec{r}_{12}\vert;\rho_{B})+\omega'(\vert\vec{x}+\vec{r}_{12}\vert;\rho_{B})\omega(\vert\vec{x}\vert;\rho_{B})\right] \nonumber \\
&-&\Delta\Phi''[\rho_{B}]\rho_{B}\int d\vec{x}[\omega(\vert\vec{x}\vert;\rho_{B})\omega(\vert\vec{x}+\vec{r}_{12}\vert;\rho_{B})]
\end{eqnarray}
Podemos expresar todo lo anterior como convoluciones de las funciones peso
\begin{eqnarray}
c^{(2)}(\vert\vec{r}_{12}\vert;\rho_{B}) &=& -\Delta\Phi'[\rho_{B}]\omega(\vert\vec{r}_{12}\vert;\rho_{B})-\Delta\Phi'[\rho_{B}]\omega(\vert\vec{r}_{21}\vert;\rho_{B}) \nonumber \\
&+&\Delta\Phi'[\rho_{B}]\rho_{B}\left[\omega'(\vert\vec{x}\vert;\rho_{B})\otimes\omega(\vert\vec{x}+\vec{r}_{12}\vert;\rho_{B})\right.\nonumber \\
&+&\left.\omega'(\vert\vec{x}+\vec{r}_{12}\vert;\rho_{B})\otimes\omega(\vert\vec{x}\vert;\rho_{B})\right] \nonumber \\
&+&\Delta\Phi''[\rho_{B}]\rho_{B}\omega(\vert\vec{x}\vert;\rho_{B})\otimes\omega(\vert\vec{x}+\vec{r}_{12}\vert;\rho_{B})]
\end{eqnarray}
como indicábamos para la resolución practica de las ecuaciones anteriores suele procederse bien mediante transformada de Fourier donde las convoluciones se transforman en productos de funciones, bien realizar un cambio de coordenadas que permite expresar,
\begin{equation}
\int d\vec{y}a(\vert\vec{y}\vert)b(\vert\vec{r}_{12}-\vec{y}\vert)=\frac{2\pi}{\vert\vec{r}_{12}\vert}\int dx x a(x)\int^{\vert r+x \vert}_{\vert r-x\vert } dr r b(r)
\end{equation}
Que completa los detalles específicos del cálculo usando que,
\begin{equation}
\omega=\omega_{0}+\omega_{1}\rho_{B}+\omega_{2}\rho_{B}^{2}
\end{equation}
de forma de convoluciones anteriores se expresan mediante convoluciones de las funciones peso elementales $\omega_{i}\otimes\omega_{j}$. Los detalles sobre las funciones $\omega_{i}$ pueden ser consultados en\cite{PhysRevA.31.2672}. Los resultados son similares a la función de correlación directa de Percus-Yevick, en un rango amplio de densidades sin embargo presenta algunas diferencias que en nuestros cálculos pueden ser bastante relevantes, una primera es la presencia de una cola oscilante para $\vec{r}_{12}>\sigma$ y la otra es la diferencia para $\vec{r}_{12}<\sigma$ para altas densidades. 

\newpage
\listoffigures
\newpage
\listoftables
\newpage
\bibliographystyle{utcaps}

\bibliography{articulos}

\end{document}